\documentclass[11pt, a4paper, oneside]{Thesis} 

\graphicspath{{Pictures/}} 

\usepackage[square,sort,compress,numbers]{natbib}
\usepackage{cite}
\usepackage{amsmath,latexsym} 
\usepackage{float}
\usepackage{multirow}
\usepackage{subfigure}
\usepackage{graphicx}
\usepackage{epigraph}
\usepackage{subcaption}   

\title{\ttitle} 
\DeclareUnicodeCharacter{2212}{-}
\DeclareUnicodeCharacter{2212}{+}
\begin{document}

\setstretch{1.3} 

\fancyhead{} 
\rhead{\thepage} 
\lhead{} 

%

\thesistitle{Qualitative Analysis of Cosmological Models Using Dynamical System Perspective}
\documenttype{Thesis}
\supervisor{Prof. Pradyumn Kumar Sahoo}
\supervisorposition{Professor}
\supervisorinstitute{BITS, Pilani, Hyderabad Campus}
\examiner{}
\degree{Ph.D. Research Scholar}
\coursecode{DOCTOR OF PHILOSOPHY}
\coursename{Thesis}
\authors{\textbf{SAYANTAN GHOSH}}
\IDNumber{2021PHXF0066H}
\addresses{}
\subject{}
\university{\texorpdfstring{\href{http://www.bits-pilani.ac.in/} 
                {Birla Institute of Technology and Science, Pilani}} 
                {Birla Institute of Technology and Science, Pilani}}
\UNIVERSITY{\texorpdfstring{\href{http://www.bits-pilani.ac.in/} 
                {BIRLA INSTITUTE OF TECHNOLOGY AND SCIENCE, PILANI}} 
                {BIRLA INSTITUTE OF TECHNOLOGY AND SCIENCE, PILANI}}



\department{\texorpdfstring{\href{http://www.bits-pilani.ac.in/pilani/Mathematics/Mathematics} 
                {Mathematics}} 
                {Mathematics}}
\DEPARTMENT{\texorpdfstring{\href{http://www.bits-pilani.ac.in/pilani/Mathematics/Mathematics} 
                {Mathematics}} 
                {Mathematics}}
\group{\texorpdfstring{\href{Research Group Web Site URL Here (include http://)}
                {Research Group Name}} 
                {Research Group Name}}
\GROUP{\texorpdfstring{\href{Research Group Web Site URL Here (include http://)}
                {RESEARCH GROUP NAME (IN BLOCK CAPITALS)}}
                {RESEARCH GROUP NAME (IN BLOCK CAPITALS)}}
\faculty{\texorpdfstring{\href{Faculty Web Site URL Here (include http://)}
                {Faculty Name}}
                {Faculty Name}}
\FACULTY{\texorpdfstring{\href{Faculty Web Site URL Here (include http://)}
                {FACULTY NAME (IN BLOCK CAPITALS)}}
                {FACULTY NAME (IN BLOCK CAPITALS)}}

\maketitle


\addtocontents{toc}{\vspace{2em}} 
\frontmatter 

\Declaration

\Certificate

\clearpage
\setstretch{1.3} 

\pagestyle{empty} 
\pagenumbering{gobble}
\Dedicatory{\bf \begin{LARGE}
Dedicated to
\end{LARGE} 
\\
\vspace{3cm}
To all my family and teachers\\
 for their continuous support}


\begin{acknowledgements}

It is a genuine pleasure to express my deep sense of thanks and gratitude to my mentor and guide, \textbf{Prof. Pradyumn Kumar Sahoo} (Prof. P. K. Sahoo), Professor, Department of Mathematics, BITS Pilani, Hyderabad Campus, Hyderabad, Telangana. His dedication and keen interest, above all his overwhelming attitude to helping his students, were solely responsible for completing my work. His timely advice, meticulous scrutiny, scholarly advice, and scientific approach have helped me immensely to accomplish this task.

I sincerely thank my Doctoral Advisory Committee (DAC) members, \textbf{Prof. Sashideep Gutti}, \textbf{Prof. Prasant Kumar Samantray}, for their valuable suggestions, theoretical assistance, and constant encouragement to improve my research work. Their advanced course in the Physics Department also helped me navigate through the research path.

It is my privilege to thank HoD, the DRC convener, faculty members, my colleagues, and the Department of Mathematics staff for supporting this amazing journey of my Ph.D. career.

I owe a deep sense of gratitude to all my co-authors for their valuable suggestions, discussions, encouragement, and help in working out my research works.

I gratefully acknowledge BITS Pilani, Hyderabad Campus, for providing me with the necessary facilities, and Council of Scientific and Industrial Research (CSIR), Government of India, Delhi, for providing CSIR JRF and SRF Fellowship (File no.09/1026(13105)/2022-EMR-I) to carry out my research work.

I would like to thank my lab-mates and colleagues for many helpful discussions and for creating a stimulating research environment. In particular, I would like to acknowledge Raja Solanki for guiding me through dynamical systems analysis; Prof. Genly Leon and Dr. Saikat Chakraborty for detailed explanations of dynamical systems techniques and MCMC analysis; Dr. Zinnat Hassan and Dr. Moreswar Tayde for their help on various wormhole projects; Debasmita Mohanty for collaboration on gravastars projects; Sneha Pradhan for assistance with compact object work; and Dr. Gaurav Gadbail for help with Gaussian process reconstruction. I have also greatly benefited from discussions with Dheeraj Singh Rana, Dr. Lakhan Jaybhaye, Dr. Simran Arora, Dr. Sanjay Mondal, and Dr. Sai Swagat Mishra during group meetings. 

I am thankful to many other faculty and researchers who offered helpful comments and discussions at various times: Prof. Rickmoy Samanta, Prof. Bivudutta Mishra, Prof. Sharmistha Banik, Prof. Rahul Nigam, Prof. Manabendra Kuri, Shivam Mishra, Priyobarta Singh,  Prof. Subhadeep Roy, Prof. Santanu Desai, Prof. Banibrata Mukhopadhyay, Dr. Apratim De, Dr. Pravakor Paul, Prof. Sujay Biswas, Dr. Nilanjandev Bhaumik, Dr. Parthiv Haldar, Dr. Sahel Dey,  Dr. Prerana Biswas, Partha Sarathi Rana, Dr. Anish Ghoshal, Dr. Swagat Saurav Sourav Mishra, Dr. Rajesh Ghosh and Prof. Burin Gumjudpai.

I am also grateful to the teachers who guided me before joining this institute and transformed my understanding, including Joydeep Ghosh (Rana Da), Arjun Sengupta (Arjun Da), Prof. Vijay Shenoy, Prof. Vamsi Pingali, Prof. Gadadhar Mishra, Prof. Govind Krishnaswamy, Prof. Alok Laddha, Prof. Dhiraj Kumar Hazra, Prof. L. Sriramkumar  and Prof. Suvrat Raju, just to name a few.

Special thanks to Timo Schulze for pointing out pitfalls in the calculation of field equations, and to Ameya Kolhatkar for help with starting the initial steps of this work. I also thank Raja Solanki and Zinnat Hassan for assistance with the modified Friedmann equations and wormhole solutions. Dr. Zinnat Hassan, Dr. Moreswar Tayde, Kartik Tathe, Dr. Hiren Gorai, Dr. Sushil Pathak, Pankaj Patel, Amit Pal, Jahir Sardar, Mintu Mondal, Soumya Kanta Bhoi, Avinash Gharde, Chandra Sekhar Mahaptra, Sidhant Kumar Panda, Kalpana Devi, and Suchita Patil have truly made the stay here worthwhile.

Throughout my tenure, I am grateful to ICTS for providing accommodation and travel support to attend the second IAGRG School (2023) and the Summer School in Gravitational Waves (2025). I am also grateful to the National Astronomical Research Institute / Mahidol University, Thailand for organizing cosmology activities and supporting local accommodation during our visit. I thank YITP for supporting my accommodation and local travel during the 4th Young Researchers' Workshop of the Extreme Universe Collaboration. Also, special thanks to the Tensor Society India for presenting me with the ``Young Relativist Award."

Finally, I am deeply grateful to my parents for their unwavering support and motivation, not just during this PhD but throughout my entire academic journey. Last but not least, I must thank my wife (Moli), who has not only stood steadfastly beside me during this tough PhD journey but, as a fellow cosmology researcher, has greatly benefited my work with her insights and technical support along the way.

\vspace{1em}
\noindent\textit{Any omission or inadvertent error in mentioning names is entirely my own, and to all those who have helped me in ways big and small, I remain deeply thankful.}

\vspace{1.2 cm}
Sayantan Ghosh,\\
ID: 2021PHXF0066H.

\end{acknowledgements}

\begin{abstract}

This thesis investigates theoretical and observational aspects of cosmic acceleration,
focusing on dark energy models and modified gravity frameworks. The goal is to analyse
their viability, dynamical behaviour, and perturbative stability in order to identify models
capable of describing the accelerated expansion of the Universe in a self-consistent way.

Chapter~\ref{Chapter1} reviews the essential background, the mathematical framework of
General Relativity, teleparallel and symmetric teleparallel geometries, basic cosmological
models, matter components, and the key observational probes. The chapter concludes with a summary
of modified gravity theories, including \(f(R)\), \(f(T)\), and \(f(Q)\).

Chapter~\ref{Chapter2} studies a dissipative Chaplygin gas cosmology in
\(f(Q)\) gravity. The model is constrained using the Cosmic Chronometer and
Pantheon+SH0ES dataset, and its performance is assessed through information criteria and
diagnostic tools such as \(Om\) and statefinder analysis.

Chapter~\ref{Chapter3} reconstructs the dynamics of Dirac-Born-Infeld (DBI) dark energy
using Hubble and DESI observations via Gaussian Processes. The reconstructed potential is
fitted to theoretical models using chi-square and MCMC techniques, providing constraints
on the DBI scalar field.

Chapter~\ref{Chapter4} analyses canonical scalar-field cosmology in coincident \(f(Q)\)
gravity using dynamical systems. Exponential and power-law potentials are examined,
their fixed points are classified, and the resulting cosmic evolution is interpreted.

Chapter~\ref{Chapter5} extends the dynamical analysis to the DBI scalar field in \(f(Q)\)
gravity. The system is studied for the same potentials, with emphasis on stability and the
distinct behavior introduced by the DBI kinetic structure.

Chapter~\ref{Chapter6} examines scalar-field evolution at both background and perturbation
levels. Perturbation equations for key gauge invariant variables are derived, and an
extended phase space combining background and perturbation dynamics is constructed.
Applications to matter perturbations in \(\Lambda\)CDM and quintessence settings are also
presented.

Finally, chapter~\ref{Chapter7} summarizes the main results and provides direction for
future research. \\

Further mathematical details and foundational derivations related to $f(Q)$ gravity are compiled in the following four sections: Appendix~\ref{AppendixA}, Appendix~\ref{AppendixB}, Appendix~\ref{AppendixC} and Appendix~\ref{AppendixD}.





 
\end{abstract}



\lhead{\emph{Contents}} 
\tableofcontents 
\addtocontents{toc}{\vspace{1em}}
\lhead{\emph{List of Figures}}
\listoffigures 
\addtocontents{toc}{\vspace{1em}}

\lhead{\emph{List of Tables}}
\listoftables 
\addtocontents{toc}{\vspace{1em}}





\lhead{\emph{List of symbols}}
\listofsymbols{ll}{
GR:\,\,\,\,\, \,\, General Relativity.\\
$\Lambda$CDM: $\Lambda$ Cold Dark Matter.\\
STGR:\,\,\,   Symmetric Teleparallel Gravity.\\
TGR: \,\, Teleparallel Gravity.\\
$g_{\mu\nu}:$ \,\,\,\,\,\,\, Lorentzian metric.\\
$g:$ \,\,\,\,\,\,\,\,\,\, Determinant of $g_{\mu\nu}$.\\
$\Gamma^{\lambda}_{\mu\nu}:$ \,\,\,\,\, General affine connection.\\
$\lbrace^{\lambda}_{\mu\nu}\rbrace $\,\,\,\,\,\,\,\, Levi-Civita connection.\\
$\nabla_{\lambda}$: \,\,\,\,\,\, Co-variant derivative w.r.t. Levi- Civita connection.\\
$(\mu\nu):$ \,\,\,\, Symmetrization over the indices $i$ and $j$.\\
$[\mu\nu]:$ \,\,\,\,\, Anti-symmetrization over the indices $i$ and $j$.\\
$R^{\lambda}_{\sigma \mu \nu}:$ \,\,\, Riemann tensor.\\
$R_{\mu\nu}:$\,\,\,\,\,\,\, Ricci tensor.\\
$R:$ \,\,\,\,\, \,\,\,  Ricci scalar.\\
$S_{M}:$ \,\,\,\,\, Matter action.\\
$T_{\mu\nu}:$ \,\,\,\,\,\, Stress-energy tensor.\\
$ S^{\mu \nu}_{\gamma}: $ \,\,\,\, Superpotential tensor.\\
$Q_{\gamma\mu\nu}: $\,\,\, Non-metricity tensor.\\
$Q: $\,\,\, Non-metricity scalar.\\
SCDM: Standard Cold Dark Matter.\\
ECs:\,\,\,\,\,  Energy Conditions.\\
EoS:\,\,\,\,\,    Equation of State parameter.\\
SNIa:\,\,\,   Type Ia Supernovae.\\
CMB: \,\,   Cosmic Microwave Background.\\
BAO: \,\,  Baryon Acoustic Oscillations.\\
DE:\,\,\,\,\, \,  Dark Energy.\\
DM:\,\,\,\,\, \,   Dark Matter.\\
HDE:\,\,\,\,\,    Holographic Dark Energy.\\
CG: \,\,\,\,\, \,\,  Chaplygin Gas.\\
GCG: \,\,\,\,\, \,\,  Generalized Chaplygin Gas.\\
GP: \,\,\,\,\, \,\,  Gaussian Process.\\
MCMC:\,   Markov Chain Monte Carlo.\\
DBI:\,\,\,\,\,   Dirac-Born-Infeld (field).\\
NH: \,\,  Non-Hyperbolic (fixed point).\\
CC:\,\,\,\,\, \,  Cosmic Chronometers (part of Hubble dataset).\\
AIC:\,\,\,\, Akaike Information Criterion.\\
BIC:\,\,\,\,\, Bayesian Information Criterion.\\
BEC: \,\,\,\,\,  Bose-Einstein Condensation (dark matter).}

\addtocontents{toc}{\vspace{2em}}

%
%


\clearpage 





\mainmatter 

\pagestyle{fancy} 


\chapter{Introduction} 
\label{Chapter1}

\lhead{Chapter 1. \emph{Introduction}} 

\clearpage
\pagebreak

\epigraph{``Perhaps there is a pattern set up in the heavens for one who desires to see it.''}{--- Plato, \textit{The Republic} (c.~375~BCE), Book~IX}


This thesis, entitled {\bf ``Qualitative Analysis of Cosmological Models Using Dynamical System Perspective"}, is devoted to the theoretical and phenomenological study of late--time cosmic acceleration within the framework of modified gravity and scalar field models. The primary focus is on understanding the dynamical behavior of the Universe through qualitative and quantitative analyses of cosmological models beyond the standard paradigm.

Before addressing the specific research problems, the present chapter provides an overview of the historical development of cosmological ideas, establishes the mathematical notation and conventions used throughout the thesis, and summarizes the essential elements of modern cosmology, fundamental theories of gravity, and key cosmological observations.

The subsequent chapters are organized as follows. One chapter is devoted to the dynamical-system analysis of Dirac-Born-Infeld (DBI) scalar-field cosmology, in which the phase-space structure and stability of cosmological solutions are investigated. Another chapter explores $f(Q)$ gravity, focusing on its cosmological dynamics and implications for late-time acceleration. The thesis further includes a reconstruction analysis of DBI scalar field models, aiming to connect theoretical predictions with observationally viable expansion histories. A separate chapter is dedicated to cosmological models involving the Chaplygin gas, examining their dynamical properties and their consistency with observational constraints. In addition, a comprehensive data analysis chapter confronts the theoretical models with recent cosmological observations. Finally, the thesis concludes with a covariant analysis of scalar field perturbations, providing insights into the stability and physical viability of the proposed cosmological scenarios.

\section{Historical overviews}
The questions of how we came to be and where we are headed remain among the most fundamental inquiries that have occupied human beings since the very beginning of cognitive development. Archaeological evidence, including prehistoric cave paintings depicting celestial bodies, suggests that curiosity about our place in the Universe predates even the origin of agriculture~\cite{harari:2014,marchant:2020}. Notable examples include El Castillo Cave in Spain ($\sim$40,000 years ago), where researchers have identified potential star symbols among Paleolithic hand prints and animal figures~\cite{rappengluck:2008}, and G\"obekli Tepe in Turkey ($\sim$11,600 years ago), where some researchers have interpreted carved pillars as depicting constellations associated with catastrophic astronomical events~\cite{piccardi:2007}. For most of human history, the age of the Universe was either considered eternal or determined through theological doctrine, with the most prominent calculation derived from biblical genealogies. Archbishop Ussher, in 1654, famously calculated the moment of Creation to be 4004 BC, a date that remained enshrined in religious orthodoxy until the rise of empirical science challenged the biblical timescale~\cite{usscher:1654}.

Outside of the Judeo-Christian tradition, people around the world have come up with a wide variety of stories to explain how the world began. Although these myths are unique to their own cultures, they often share the same big ideas. For instance, many ancient stories feature human-like characters to explain the Universe, making the gods' motives easier for us to understand. You often see a ``master builder" shaping the world's beauty or a story where a messy, chaotic start eventually turns into a clear, organized system. Take the Babylonian \textit{Enuma Elish} from around 1450 BCE. It describes the very beginning as a chaotic sea. In this story, the god Marduk has to defeat a chaos monster named Tiamat to finally bring order to the world~\cite{dalley:1998}. In Chinese tradition, the Pan Gu myth says that the Universe began as a giant egg. When Pan Gu woke up and cracked it open, the lighter parts rose to form the sky, while the heavier parts sank to become the Earth. Eventually, his own body transformed into stars and mountains~\cite{yang:2005}. African traditions are as deep. The Dogon people in Mali, for example, have creation myths that include surprisingly detailed knowledge of the Sirius star system~\cite{griaule:1986}. For the Yoruba of West Africa, the supreme god Olodumare is in charge, but he lets other spirits called Orishas do the actual work of building the Earth~\cite{idowu:1962}. Meanwhile, the San people in southern Africa have used rock art and spoken stories for thousands of years to explain how the first humans arrived and how the world was set right~\cite{lewis-williams:2002}. No matter where they come from, whether from Africa, Asia, or the Americas, these myths are structured so similarly that it suggests that humans everywhere have a common way of thinking about where we came from. Historically, human models of the divine have often been highly anthropomorphic; as the Greek philosopher Xenophanes pointed out, if animals could draw, they would invariably create gods in their own image. While Abrahamic traditions depart from this by emphasizing a formless, transcendent Creator, their accounts of the universe's origins are deeply tied to scriptural revelation. Consequently, though these narratives offer rich philosophical and spiritual meaning, they are methodologically distinct from modern scientific cosmology, which requires testable, purely physical frameworks rather than theological texts to explain the evolution of the cosmos.

Beyond these mythological narratives, other civilizations developed complex cosmological frameworks grounded in mathematical astronomy and natural observation. In ancient India, \textit{Rig Veda}, particularly \textit{Nasadiya Sukta}, pondered the origin of existence with profound ambiguity, suggesting a cyclic conception of time where Universes undergo repeated creation and dissolution over vast epochs known as \textit{Yugas}~\cite{rigveda:ancient}. This philosophical foundation was later complemented by rigorous mathematical astronomy. Aryabhata, in 499 CE, calculated the Earth's circumference as 4,967 yojanas and estimated the solar year as 365.25858 days, remarkably close to the modern value of 365.24219 days~\cite{aryabhata:499}. He proposed that the Earth rotates on its axis and used trigonometric methods to estimate the distance to the Sun as approximately 18 to 20 times the distance to the Moon~\cite{aryabhata:499,sarma:1997}. Although some scholars debate whether Aryabhata's model contained heliocentric elements, his systematic treatment of planetary positions represented a significant advance in quantitative cosmology~\cite{bag:1979}. Subsequent scholars built on this foundation. Brahmagupta, in the 7th century, developed mathematical techniques to predict planetary positions and the timing of solar and lunar eclipses~\cite{brahmagupta:628}. Bhaskara II, in the 12th century, further refined these planetary models and accurately defined numerous astronomical quantities, including the sidereal year~\cite{bhaskara:1150}. Similarly, during the Islamic Golden Age, scholars sought to reconcile scripture with natural observation. Polymaths such as Al-Biruni and Ibn Sina examined geological strata, noting that fossilized marine life found in mountains suggested that these formations were once near sea level, implying a dynamic Earth shaped over extensive periods rather than a static creation~\cite{albiruni:1000,ibnsina:1020}. These insights prefigured modern uniformitarianism, though they remained largely philosophical within their respective theological contexts.

Parallel to these theological chronologies, a structural revolution was unfolding within celestial mechanics that would eventually undermine the static Universe required by orthodoxy. For nearly 15 centuries, the Ptolemaic geocentric model remained an unchallenged dogma until Nicolaus Copernicus published \textit{De revolutionibus orbium coelestium} in 1543~\cite{copernicus:1543}. Although initially circulated with a cautious preface suggesting that the heliocentric model was merely a calculative convenience, Copernicus's axioms fundamentally displaced the Earth from the cosmic center~\cite{singh:2005}. This theoretical shift required empirical validation, which arrived with the advent of the telescope. In 1610, Galileo Galilei observed the phases of Venus and the satellites of Jupiter, providing irrefutable evidence that not all celestial bodies orbited the Earth~\cite{galileo:1610}. These observations directly contradicted the Ptolemaic prediction of the Venusian phases and demonstrated that the Earth was not the sole hub of orbital motion~\cite{singh:2005}. Concurrently, Johannes Kepler utilized Tycho Brahe's precise observational data to derive his laws of planetary motion, replacing circular orbits with ellipses and establishing a kinematic foundation for celestial dynamics~\cite{kepler:1609}. These developments transformed cosmology from a branch of natural philosophy into a quantitative physical science, culminating in Isaac Newton's synthesis of universal gravitation~\cite{newton:1687}. Thus, by the dawn of the 18th century, the architecture of the cosmos was subject to mathematical law, even if the temporal scale remained contested.

The first serious scientific challenge to the time scale came not from astronomy but from geology. James Hutton, in 1788, proposed the principle of uniformitarianism, suggesting that the Earth's features were shaped by slow, continuous processes rather than catastrophes, implying a timescale with ``no vestige of a beginning''~\cite{hutton:1788}. This view was later popularized by Charles Lyell and provided the necessary temporal framework for Charles Darwin's theory of evolution by natural selection, which required hundreds of millions of years to proceed~\cite{lyell:1830}. However, this geological timescale clashed violently with the laws of physics as understood in the nineteenth century. Lord Kelvin, applying the laws of thermodynamics to the cooling of the Earth and the gravitational contraction of the Sun, argued rigorously that the Earth and Sun could only be tens of millions of years old~\cite{kelvin:1862}. By 1897, Kelvin had set the upper limit on the lifetime of the Sun at 24 million years, creating a scientific impasse where biology and geology demanded billions of years, while classical physics permitted only millions~\cite{kelvin:1897}. The resolution of this conflict came with the discovery of radioactivity and the development of quantum mechanics. Ernest Rutherford and Arthur Holmes demonstrated that radioactive decay provided both a new source of energy to sustain the Sun and a precise clock to measure the age of rocks~\cite{rutherford:1905}. Radiometric dating pushed the age of the Earth back to billions of years, reconciling geological evidence with physical law. Similarly, Arthur Eddington applied the laws of physics to stellar structure, revealing that stars shine via nuclear fusion rather than gravitational contraction, allowing astrophysicists to model stellar evolution and determine that globular clusters were themselves over ten billion years old~\cite{eddingtion:1920}.

To measure the age of the Universe, astronomers first had to measure its size. The development of the cosmic distance ladder was pivotal in this expansion of scale. Henrietta Leavitt, working at the Harvard College Observatory, discovered a precise mathematical relationship between the brightness of a Cepheid variable star and its period in 1912~\cite{leavitt:1912}. This period-luminosity relationship provided a standard candle capable of probing extragalactic distances. This tool resolved the great debate between Harlow Shapley and Heber Curtis about the nature of spiral nebulae. Shapley, using Cepheids in globular clusters, determined that the Sun was not at the center of the Milky Way and that the Galaxy was much larger than previously thought, though he erroneously believed that the Milky Way constituted the entire Universe~\cite{shapley:1918}. Curtis argued that spiral nebulae were independent ``Island Universes,'' but lacked the distance indicators to prove it definitively~\cite{curtis:1920}. The final step in establishing a cosmic age came from the kinematics of these galaxies. Vesto Slipher had previously noted that most spiral nebulae exhibited significant redshifts, indicating high recession velocities~\cite{slipher:1917}. Building on theoretical frameworks proposed by Alexander Friedmann and Georges Lemaître, which suggested that Einstein's general theory of relativity allowed for an expanding spacetime, Edwin Hubble and Milton Humason correlated these redshifts with distance~\cite{lemaitre:1927}. Using the 100-inch Hooker Telescope at Mount Wilson, Hubble identified Cepheids in the Andromeda Nebula, conclusively proving that spiral nebulae were independent galaxies far beyond the Milky Way~\cite{hubble:1925}. Their 1929 publication established a linear relationship between recessional velocity and distance, now known as Hubble's Law~\cite{Hubble:1929}. This observation implied that the Universe is not static but expanding. If the galaxies are moving apart today, they must have been closer together in the past. Extrapolating backward implies a singular beginning, a moment of creation in which the density of the Universe was infinite.

This historical progression from the age of Earth to the age of the Universe represents the foundation upon which modern cosmology rests. The shift from Kelvin's thermodynamic limits to Hubble's expansion law demonstrates how cosmological parameters are inextricably linked to advances in fundamental physics and observational technology. Today, while the broad strokes of this history have settled, the determination of cosmological parameters remains a central challenge. Discrepancies in the measurement of the Hubble constant and the age of the Universe persist, driven by systematic uncertainties in distance indicators and stellar evolution models. This thesis contributes to this ongoing refinement, investigating theoretical and observational aspects of cosmic acceleration. Specifically, we focus on dark energy models and modified gravity frameworks, analyzing their viability, dynamical behavior, and perturbative stability. We explore dissipative Chaplygin gas cosmology and DBI dark energy within coincident $f(Q)$ gravity, utilizing dynamical system analysis to identify models capable of describing the accelerated expansion of the Universe in a self-consistent way. By building upon the legacy of Hubble, Humason, and the pioneers of radiometric dating, we continue the quest to measure the age of the Universe with ever-greater precision.
\section{Pre-relativity cosmology}

The quest to determine the age of the Universe did not begin with Einstein's field equations or the discovery of the Cosmic Microwave Background. Long before the advent of General Relativity, natural philosophers grappled with fundamental questions about the stability, structure, and temporal extent of the cosmos. The term \textit{cosmology} derives from the ancient Greek \textit{kosmos} ($\kappa$ó$\sigma\mu$os), 
meaning ``order'' or ``universe,'' combined with \textit{-logia} ($\lambda$o$\gamma$ía), 
denoting ``the study of''. The early investigations, though ultimately superseded by relativistic cosmology, established the conceptual framework within which modern questions remain posed. Understanding this historical progression is not merely an exercise in historiography; it reveals how deeply our current cosmological parameters are rooted in problems that troubled Newton, Kant, and their successors. The tensions between infinite and finite models, between static and dynamic Universes, and between local physics and global geometry all emerged in the pre-relativistic era and continue to resonate in contemporary debates about dark energy, inflation, and the Hubble tension.

Before Einstein's ``general theory of relativity", cosmology was constrained by the framework of Newtonian mechanics, in which space and time were treated as absolute, immutable backgrounds. The gravitational field obeys Poisson's equation,
\begin{equation}
\nabla^{2}\Phi = 4\pi G \rho \,\,\text{,}
\end{equation}
Yet applying this to the Universe as a whole reveals a fundamental instability. For a finite distribution of matter, mutual gravitational attraction inevitably leads to collapse. Newton himself recognized this paradox and proposed an infinite, homogeneous Universe where forces balance by symmetry \cite{newton:1687}. However, this resolution is mathematically precarious. The equilibrium corresponds to a saddle point in the potential landscape rather than a global minimum; any infinitesimal density perturbation breaks the symmetry and triggers runaway collapse or expansion. In dynamical systems language, the Laplacian vanishing ($\nabla^2 \Phi = 0$) does not guarantee stability; it merely indicates a critical point that is structurally unstable under perturbations \cite{kiessling/2012}. Newton himself entertained the idea of supplementing gravity with a  linear repulsive term to stabilize the Universe against collapse, a precursor to what would later become the cosmological constant \cite{Scali2025}.

This tension between mathematical idealization and physical reality was recognized even before the formal apparatus of instability analysis was developed. Immanuel Kant \cite{kant:1755}, in his \textit{Universal Natural History and Theory of the Heavens}, grappled with the cosmological implications of Newtonian gravity. Kant proposed that the Milky Way constitutes a rotating disk of stars and speculated that spiral nebulae might be independent ``Island Universes'' a remarkable anticipation of extragalactic astronomy. However, he acknowledged that a purely mechanical Newtonian system lacked an internal mechanism to explain the origin of cosmic structure without invoking initial conditions set by external agency. Kant's critique highlighted that Newtonian mechanics, while sufficient for local dynamics, remained incomplete as a global cosmology.

The mathematical incompleteness of Newtonian cosmology was further clarified by Hermann Weyl in his seminal work \textit{Space, Time, Matter} \cite{weyl:1918}. Weyl demonstrated that the gravitational potential in an infinite, homogeneous medium is ill-defined without specifying boundary conditions at infinity. The value of $\Phi$ depends on the order in which the limit is taken, rendering the force at any point indeterminate unless arbitrary constraints are imposed from ``outside'' the system. This circular dependency-requiring information from infinity to define physics locally-signaled that Newtonian gravity could not provide a self-contained description of the Universe. Weyl's analysis underscored the need for a theory in which spacetime geometry itself becomes dynamical and is locally determined by the matter content.

Parallel to these theoretical developments, astronomical observations were reshaping cosmological thinking. William Herschel's star counts suggested a flattened Milky Way structure, although he incorrectly placed the Sun near its center \cite{herschel:1785}. By the early twentieth century, the nature of spiral nebulae remained contested, with some astronomers viewing them as galactic substructures and others as independent systems. These observational ambiguities, combined with the theoretical instabilities inherent in Newtonian cosmology, created a situation in which neither finite nor infinite static Universes could be sustained within classical physics.

The pre-relativistic era thus concluded with a clear recognition: a static cosmos is incompatible with Newtonian gravity, whether finite or infinite. The Laplacian instability, the boundary-condition problem, and the lack of a mechanism to generate cosmic structure all pointed to the necessity of a new gravitational theory. This conceptual crisis set the stage for Einstein's general theory of relativity, which replaced absolute space and time with dynamical spacetime geometry and provided the first self-consistent framework for cosmology.

It is worth noting that the mathematical foundations of cosmology have long inspired profound reflections, notably from Alexander Grothendieck, one of the most transformative mathematicians of the 20th century. His playful yet deeply philosophical outlook is perfectly captured in his contribution to the 1968 mock obituary of Nicolas Bourbaki, where he wrote: “For God is the Alexandrov (one-point) compactification of the universe.”

\section{Relativistic cosmology}

The limitations of Newtonian gravity in describing the global structure of the Universe necessitated a more general theory of gravitation. The foundation of relativistic cosmology lies in Einstein's General Theory of Relativity (GR), which emerged from the synthesis of the principle of relativity and the equivalence principle. The journey from Special Relativity to a dynamic cosmological model was not linear; it involved a complex interplay between mathematical elegance, philosophical preconceptions about a static Universe, and eventually, observational evidence.

The principle of relativity states that the laws of physics are invariant under transformations between inertial frames. In Special Relativity, published by Einstein in 1905, this invariance is restricted to inertial observers and leads to the unification of space and time into a four-dimensional spacetime continuum \cite{einstein:1905}. However, Special Relativity does not include gravity, treating spacetime as a fixed Minkowski background. The equivalence principle provides the key conceptual step toward incorporating gravitation. In its simplest form, it states that locally, the effects of a gravitational field are indistinguishable from those of an accelerated reference frame. This implies that gravity is not a conventional force but rather a manifestation of spacetime geometry. Matter tells spacetime how to curve, and curved spacetime tells matter how to move. This conceptual leap was formalized in 1915, when Einstein presented the field equations of General Relativity (GR) \cite{einstein:1915}.

In Newtonian gravity, the gravitational potential satisfies Poisson's equation, yet in GR, this relation is generalized into Einstein's field equation as follows,

\begin{equation}
  G_{\mu\nu} = \frac{8\pi G}{c^4} T_{\mu\nu} \,\,\text{.}  
\end{equation}

Where $G_{\mu\nu}$ is the Einstein tensor representing the spacetime curvature, $T_{\mu\nu}$ is the energy-momentum tensor describing matter and energy content, and $g_{\mu\nu}$ is the metric tensor. We will use natural units where $8\pi G = c = 1$.  The inclusion of the cosmological constant $\Lambda$ was not part of the original 1915 formulation but was introduced by Einstein in 1917 when he attempted to apply his theory to the Universe as a whole \cite{Einstein:1917}. At the time, the prevailing astronomical view was that the Universe was static and eternal. Einstein found that his original equations demanded a dynamic Universe either expanding or contracting. To reconcile his theory with the static model, he introduced the $\Lambda$ term to provide a repulsive force that could balance gravitational attraction on cosmic scales. George Gamow later recalled that Einstein remarked to him that the introduction of the cosmological constant was the ``biggest blunder'' of his life \cite{gamow:1970}. However, this static solution was precarious; Eddington later demonstrated that the Einstein static Universe is unstable against small perturbations, much like a pencil balanced on its tip \cite{eddington:1930}.

Almost simultaneously, Willem de Sitter found another solution to the field equations \cite{desitter:1917}. The de Sitter Universe was empty of matter ($T_{\mu\nu}=0$) but possessed a cosmological constant. Remarkably, this model predicted that test particles would appear to recede from one another, producing a redshift effect purely due to the geometry of spacetime. Although the de Sitter model was not physically realistic because it lacked matter, it provided the first theoretical hint that GR naturally accommodates an expanding cosmos.

The first truly dynamic cosmological solutions were derived by Alexander Friedmann in 1922 \cite{friedmann:1922}. Friedmann showed that, without the cosmological constant, the field equations admit solutions in which the spatial scale of the Universe changes with time. He derived what are now known as the Friedmann equations, which govern the evolution of the scale factor $a(t)$. Initially, Einstein rejected Friedmann's work, believing it to be mathematically flawed, but later retracted his criticism and acknowledged the validity of the dynamic solutions.

The connection between these theoretical models and observation was established by Georges Lemaitre in 1927 \cite{lemaitre:1927}. Independent of Friedmann, Lemaitre derived the expanding Universe solutions and, crucially, linked them to the observed radial velocities of spiral nebulae measured by Vesto Slipher. Lemaitre proposed that the redshifts observed in distant nebulae were due to the expansion of space itself, deriving an initial estimate for the rate of expansion. While Lemaitre's work provided the theoretical framework for an expanding Universe, it was Edwin Hubble who provided the definitive observational confirmation.

In 1929, Hubble published his analysis of the distances and redshifts of extragalactic nebulae, demonstrating a linear relationship between recessional velocity and distance \cite{Hubble:1929}. This empirical relation, now known as Hubble's Law, vindicated the dynamic models of Friedmann and Lemaitre and rendered the Einstein static Universe obsolete.

The transition from a static to a dynamic understanding of the Universe necessitates a mathematical framework capable of describing evolving geometry. In Special Relativity, the metric is fixed (Minkowski), reflecting a static spacetime. However, if the Universe expands, the distance between comoving points changes with time. This requires a metric tensor $g_{\mu\nu}$ that is itself a function of cosmic time.

The metric encodes the invariant interval
\begin{equation}
    ds^2 = g_{\mu\nu} dx^\mu dx^\nu \,\,\text{.}
\end{equation}
defining how distances and time intervals are measured locally. Since the cosmological principle suggests that the Universe is homogeneous and isotropic on large scales, the metric must reflect these symmetries while allowing for temporal evolution. This motivation leads directly to the Friedmann-Lemaitre-Robertson-Walker (FLRW) metric, which provides the geometric backbone for modern cosmology by incorporating the scale factor $a(t)$ derived from the Friedmann equations into the spacetime interval.

\subsection{The FLRW metric and the cosmic geometry}

The standard model of cosmology is based on the cosmological principle. This principle states that the Universe looks the same at every point and in every direction on large scales. The mathematical formulation of this idea took shape in the 1920s and 1930s. Alexander Friedmann first found expanding solutions to Einstein's equations in 1922 \cite{friedmann:1922}. Georges Lemaitre derived similar results independently in 1927 and linked them to observational data \cite{lemaitre:1927}. Howard Robertson and Arthur Walker later proved that the metric form is the unique solution consistent with homogeneity and isotropy \cite{robertson:1935,walker:1937}. This is why we call it the Friedmann-Lemaitre-Robertson-Walker (FLRW) metric.

Homogeneity means that there exist Killing vectors that span the spatial hypersurfaces. Technically we require three spacelike Killing vectors satisfying specific commutation relations. These vectors generate translations that leave the metric invariant. Isotropy requires rotational symmetry around every point. Hawking and Ellis discuss the rigorous implications of these symmetries on the global structure \cite{hawking:1973}. Islam provides a detailed proof using the Frobenius theorem \cite{islam:2001}. The theorem ensures that the spacetime is foliated into maximally symmetric spatial slices. These slices can have positive, negative or zero curvature. Islam works through the geometry for all three cases explicitly \cite{islam:2001}.

 We note that the most general FLRW metric in spherical polar coordinates looks like,
\begin{equation}
ds^2 = -dt^2 + a^2(t)\left[\frac{dr^2}{1 - k r^2} + r^2 d\theta^2 + r^2 \sin^2\theta \, d\phi^2\right] \,\,\text{.}
\end{equation}
However, the latest observations from the Planck satellite suggest that the spatial curvature is very close to zero \cite{planck}. So we will assume a flat Universe for most of this work. The flat FLRW metric in cartesian coordinate is written as follows,
\begin{equation}
ds^2 = -dt^2 + a^2(t)\left[dx^2 + dy^2 + dz^2\right] \,\,\text{.}
\label{eq:flrw_metric}
\end{equation}

Here, $a(t)$ is the scale factor. It tells us how distances change with time. In natural units, this significantly simplifies the field equations. The parameter $k$ determines the spatial geometry in a general form. If $k = +1$ the space is closed and has positive curvature like a sphere. If $k = -1$ the space is open and has a negative curvature like a saddle. If $k = 0$ the space is flat and follows Euclidean geometry. This curvature is linked to the density of the Universe.

\subsection{Dynamics of cosmology}
We model matter as a perfect fluid with
\begin{equation}
T_{\mu\nu}=(\rho+p)u_\mu u_\nu + p g_{\mu\nu} \,\,\text{.}
\end{equation}
Using the FLRW metric in Einstein's equations yields the Friedmann equations
\begin{align}
H^2 &= \frac{\rho}{3} - \frac{k}{a^2} \,\,\text{,} \label{eq:F1}\\[4pt]
\frac{\ddot a}{a} &= -\frac{1}{6}(\rho+3p) \,\,\text{.} \label{eq:F2}
\end{align}
Here, \(H\equiv\dot a/a\). Define the critical density \(\rho_c=3H^2\) and
\(\Omega\equiv\rho/\rho_c\), so
\begin{equation}
\Omega-1=\frac{k}{a^2H^2},
\end{equation}
which links spatial curvature with the total energy content.

Energy conservation, \(\nabla_\mu T^{\mu\nu}=0\), gives the continuity equation
\begin{equation}
\dot\rho + 3H(\rho+p)=0 \,\,\text{,}
\end{equation}
and with \(p=w\rho\) the familiar scalings follow:
\[
\rho_m\propto a^{-3}\ (w=0),\qquad
\rho_r\propto a^{-4}\ (w=\tfrac{1}{3}),\qquad
\rho_\Lambda=\text{const}\ (w=-1) \,\,\text{.}
\]
Thus, the cosmic sequence radiation \(\to\) matter \(\to\) \(\Lambda\)-domination is a direct consequence of different dilution rates. Acceleration requires \(\rho+3p<0\), satisfied by a cosmological constant \(\Lambda\), which drives the observed late-time acceleration.

The deceleration parameter
\begin{equation}
q\equiv -\frac{\ddot a\,a}{\dot a^2} \,\,\text{.}
\end{equation}
is positive for deceleration and negative for acceleration; in a matter-dominated (\(w=0\)) Universe \(q=\Omega/2\).

This compact dynamical framework connects geometry, energy content, and expansion and forms the backbone of the standard cosmological model. Observational constraints (e.g., Planck full-mission results) place the contemporary energy budget at roughly \(\sim68\%\) dark energy, \(\sim27\%\) dark matter, and \(\sim5\%\) baryons, consistent with near flatness and with inflationary expectations \citep{planck}.

\begin{quote}
All of observational cosmology is the search for two numbers: the Hubble constant \(H_0\) and the deceleration parameter \(q_0\) \cite{sandage1970}.
\end{quote}

\section{Dark energy}
The discovery of a late-time acceleration of the cosmic expansion stands as one of the most profound challenges to our understanding of fundamental physics. Although the Friedmann equations successfully describe a Universe filled with matter and radiation, observational data suggest the existence of a dominant, yet mysterious, energy component that possesses negative pressure. This chapter explores the theoretical foundations, observational milestones, and the persistent conceptual difficulties surrounding this ``Dark Energy," beginning with its historical roots in Einstein's gravitational framework.
\subsection{Einstein's cosmological constant term and its reinterpretation}

The history of dark energy begins with Albert Einstein. In 1917, Einstein applied his general theory of relativity to the entire Universe \cite{Einstein:1917}. At that time, astronomers believed that the Universe was static. However, the original field equations predicted a dynamic Universe that would either expand or contract. To fix this, Einstein added a new term, the cosmological constant $\Lambda$, to allow for a static solution. The modified field equations are as follows.
\begin{equation}
G_{\mu\nu} + \Lambda g_{\mu\nu} =  T_{\mu\nu} \,\,\text{.}
\label{eq:einstein_lambda}
\end{equation}
Here, $\Lambda$ acts as a repulsive force that balances the attractive force of gravity. After Edwin Hubble discovered that the Universe is actually expanding in 1929 \cite{Hubble:1929}, Einstein removed the term, calling it his ``greatest blunder''. For decades, most cosmologists assumed $\Lambda$ was zero.

In the 1960s, the Russian physicist Yakov Zel'dovich reintroduced the idea \cite{ZEL}. He proposed that $\Lambda$ was not just a mathematical fix, but represented the energy of the physical vacuum. In Quantum Field Theory (QFT), an empty space is not truly empty. It contains quantum fields that fluctuate even in their lowest energy state. Zel'dovich suggested that these vacuum fluctuations have energy and gravity, just like matter.

We can describe this vacuum energy as a perfect fluid. Its stress--energy tensor is
\begin{equation}
T_{\mu\nu}(\Lambda) = - \rho_{\Lambda} g_{\mu\nu} \,\,\text{,}
\label{eq:vacuum_tensor}
\end{equation}
where $\rho_{\Lambda}$ is the energy density. The pressure $p_{\Lambda}$ is related to the density by
\begin{equation}
p_{\Lambda} = - \rho_{\Lambda} \,\,\text{.}
\label{eq:vacuum_eos}
\end{equation}
This negative pressure is the key. In the Friedmann equation for the Universe with cosmological constant can be given as follows,
\begin{equation}
\frac{\ddot{a}}{a} = - \frac{1}{6}(\rho + 3p) + \frac{\Lambda}{3} \,\,\text{,}
\label{eq:acceleration_lambda}
\end{equation}
where a positive $\Lambda$ creates a repulsive gravity effect. If this term dominates over matter, it causes the expansion of the Universe to accelerate.

\subsection{Observational evidences for acceleration}
Before 1998, there were some early hints that $\Lambda$ might not be zero. For example, some measurements of the ages of old stars seemed older than the age of the Universe without $\Lambda$ \cite{Efstathiou:1988,Efstathiou:1990}. Also, counts of distant galaxies suggested a geometry consistent with a non-zero cosmological constant \cite{Loh:1986}. However, these early signs were not conclusive.

The definitive discovery came in 1998 from observations of Type Ia supernovae. These exploding stars are very useful for cosmology because they act as ``standard candles.'' This means that they all have roughly the same intrinsic brightness. By comparing how bright they look to how bright they actually are, astronomers can measure their distance.

Two independent teams studied these supernovae at high redshifts (large distances). The Supernova Cosmology Project, led by Saul Perlmutter, and the High-$z$ Supernova Search Team, led by Adam Riess and Brian Schmidt, found surprising results \cite{late1, late2}. The distant supernovae were fainter than expected. This meant they were farther away than predicted by a Universe that was slowing down due to gravity. The only explanation was that the expansion of the Universe is speeding up. This acceleration requires a dominant energy component with negative pressure, now called ``Dark Energy".

Since this discovery, other methods have confirmed these results:

\begin{itemize}
\item \textbf{Cosmic Microwave Background (CMB):} Satellites like \textit{Planck} measure the leftover radiation from the Big Bang. The data show that the Universe is flat. Since matter only makes up about 30\% of the critical density, the remaining 70\% must be dark energy \cite{planck}. 
\item \textbf{Baryon Acoustic Oscillations (BAO):} This method uses the pattern of galaxy distribution as a standard ruler. Surveys like SDSS have measured this scale at different times in history, confirming the acceleration \cite{Eisenstein:2005}.
\item \textbf{Integrated Sachs-Wolfe Effect:} This is a specific imprint on the CMB caused by changing gravitational potentials in a dark energy dominated Universe \cite{Crittenden:1996}.
\end{itemize}

\subsection{The cosmological constant problem}
Although observations confirm that dark energy exists, its nature is a major mystery. If dark energy is the vacuum energy predicted by QFT, there is a huge problem.

In QFT, the vacuum energy density $\rho_{vac}$ is calculated by summing the zero-point energy of all field modes. This sum depends on the maximum energy scale (cutoff) $k_{max}$ where the theory works. The formula is as follows.
\begin{equation}
\rho_{vac} \sim \frac{\hbar c}{16 \pi^2} k_{max}^4  \,\,\text{.}
\label{eq:rho_vac}
\end{equation}
If we set the cutoff point on the Planck scale (where gravity becomes strong), the predicted energy density is enormous, and the discrepancy with the observation becomes of the order of $10^{120}$. Zel'dovich noted that even using the mass of a proton as a cutoff gives a value much larger than observed \cite{ZEL}. Using the Planck scale creates a mismatch of about 120 orders of magnitude compared to the observed value \cite{WENB}. This is known as the cosmological constant problem. It suggests that our understanding of how quantum vacuum energy gravitates is incomplete.

\subsection{The anthropic argument}
Because standard physics cannot explain why $\Lambda$ is so small, some physicists look to other explanations. Steven Weinberg proposed an argument based on the anthropic principle \cite{WENB}. He asked: What value of $\Lambda$ allows observers to exist?

For observers to exist, galaxies must form. Galaxies form when matter clumps together under gravity. If $\Lambda$ is too large and positive, the Universe expands too fast for matter to clump. If $\Lambda$ is too large and negative, the Universe collapses too quickly. Weinberg calculated that $\Lambda$ must be small enough so that vacuum energy does not dominate until after galaxies have formed.

This gives an upper bound
\begin{equation}
\rho_{\Lambda} \lesssim \rho_{m}(t_{gal}) \,\,\text{,}
\label{eq:anthropic_bound}
\end{equation}
where $\rho_{m}(t_{gal})$ is the matter density at the time of galaxy formation. Remarkably, this bound is close to the value we observe today. This suggests that the value of dark energy might not be determined by a fundamental symmetry but by the requirement that the Universe must allow for life.

\section{Modified gravity}

Modified gravity represents a prominent theoretical route to enhance our understanding of the evolution of the Universe. It incorporates early stages, such as inflation, and later phases marked by accelerated expansion, often attributed to dark energy \cite{CANT}. Additionally, these theories may address potential observational conflicts, such as the Hubble tension and structure growth discrepancies \cite{COSI}. Fundamentally, these theories involve the construction of extensions of GR that introduce additional degrees of freedom. These modifications can provide corrections at both the cosmological background and perturbation levels, offering a more comprehensive description of the behavior of the  Universe.

There exist various approaches to constructing such gravitational modifications, often categorized according to the geometric quantity used to describe the gravitational interaction. While standard GR relies on curvature, alternative formulations utilize torsion or non-metricity. One distinct class of gravitational theories emerges when employing the equivalent formulation of gravity based on non-metricity, commonly known as symmetric teleparallel gravity \cite{NEST}. In this formulation, a generic affine connection is used with vanishing curvature and torsion while relaxing the metric compatibility condition. This framework has recently been extended to $f(Q)$ gravity \cite{Jimenez/2018}. However, the more complete picture that we use in this paper is given in \cite{Harko/2018, Lazkoz/2019, Capozziello/2020}.

This extension of the symmetric teleparallel formalism has attracted considerable attention in cosmology as a possible avenue for physics beyond the standard $\Lambda$CDM model. Particular forms of the $f(Q)$ function have been shown to alleviate the $\sigma_8$ tension \cite{BARR}, while others provide improved fits to cosmological observations \cite{ANAG,ARORA,NUNES}. Interesting implications of $f(Q)$ gravity have also been explored in several contexts such as black hole physics \cite{RODR,LAVI-1}, neutrino physics \cite{NEOM}, quantum cosmology \cite{CAPE-1,PALIA}, bouncing cosmology \cite{GADB}, inflation \cite{CAPE-2}, phantom cosmology \cite{ANDER}, astrophysical objects \cite{SNEHA, Hassan/2022, Hassan/2023}, cosmological perturbations \cite{ET}, BBN constraints \cite{ANAG-2}, and several other applications \cite{DE-1,DE-2,PALIA-2,HOH,WOM-2}.

\subsection{$f(R)$ based modified gravity}

The late-time accelerated expansion of the Universe has motivated extensions of GR in which cosmic acceleration arises from modifications of gravity rather than from an explicit dark energy component. Among these approaches, $f(R)$ gravity represents one of the simplest and most widely studied generalizations.

Historically, the investigation of nonlinear functions of the Ricci scalar dates back to the work of Buchdahl in 1970 \cite{Buchdahl}. The idea gained further physical motivation with the Starobinsky model of inflation \cite{star}, where a quadratic correction $R^2$ naturally drives an inflationary phase without introducing additional scalar fields. From a more fundamental viewpoint, higher-order curvature terms are also expected from quantum corrections in effective field theory.

The action of $f(R)$ gravity is given by

\begin{equation}
S=\frac{1}{2}\int d^4x \sqrt{-g}\, f(R) + \int d^4x \sqrt{-g}\, \mathcal{L}_m \,\,\text{.}
\end{equation}

As usual, by the natural unit convention used in this thesis, we set $8\pi G = c = 1$.

Varying the action with respect to the metric yields modified field equations that are fourth-order in derivatives of the metric. For a spatially flat FLRW background, the cosmological evolution is governed by modified Friedmann equations, where additional curvature terms behave as an effective fluid driving cosmic expansion.

\subsection{$f(T)$ based teleparallel modified gravity}

An alternative geometric description of gravity relies on torsion rather than curvature. This approach originates from the work of Elie Cartan in the 1920s, who introduced torsion as a generalization of Riemannian geometry \cite{Cartan}. Later, Sciama proposed that intrinsic spin could act as a source of torsion in analogy to how mass generates curvature \cite{Sciama}. It is also historically notable that Einstein himself explored teleparallel formulations while attempting to unify gravity and electromagnetism \cite{Einstein-Tele}.

In this framework, the spacetime metric is constructed from tetrad fields $e^i_{\ \mu}$ as

\begin{equation}
    g_{\mu\nu}=\eta_{ij} e^i_{\ \mu} e^j_{\ \nu} \,\,\text{.}
\end{equation}

The gravitational dynamics are encoded in the torsion scalar $T$, constructed from the torsion tensor and the contorsion tensor. The action of $f(T)$ gravity is given by \cite{Bengochea/2009,Baojiu/2011}

\begin{equation}
S=\int d^4x\, e \left[\frac{1}{2}f(T)+\mathcal{L}_{matter}\right] \,\,\text{,}
\end{equation}

where $e=\det(e^i_{\ \mu})=\sqrt{-g}$ \,\,\text{.}

Unlike $f(R)$ gravity, the resulting field equations remain second order in derivatives of the tetrad fields. However, generic $f(T)$ theories can break local Lorentz invariance depending on the choice of tetrad. The field equations are obtained by varying the action with respect to the tetrad \cite{Bengochea/2009,Baojiu/2011,Bohmer/2011}.

\begin{figure}[H]
{\includegraphics[width=8cm,height=3.5cm]{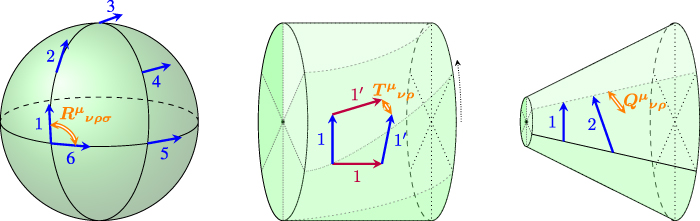}}
{\includegraphics[width=7.5cm,height=5cm]{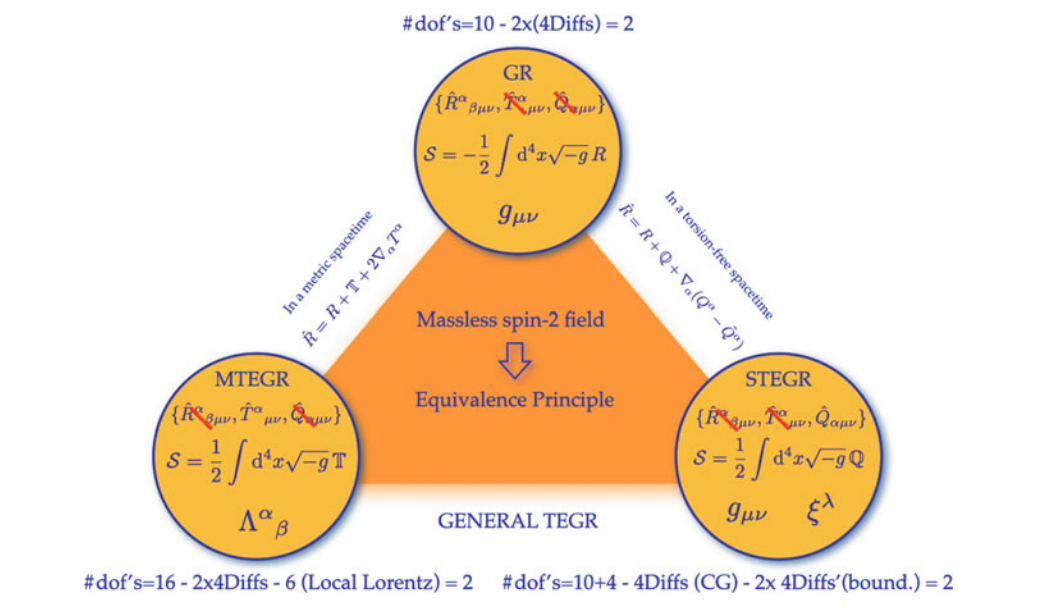}}
\caption{On the left side, we compare three different types of parallel transportation: Affine, Torsion-based, and non-metricity-based, respectively \cite{Jarv2020}. The right-hand side provides a pictorial representation of all three formulations of GR and their equivalence \cite{Jimenez/2019}.}\label{fig_1.1} 
\end{figure}

\subsection{$f(Q)$ based symmetric teleparallel gravity}
We begin by noting that a connection plays a vital role in the transport of tensors across a manifold. In the realm of GR, which is built on Riemannian geometry, gravitational interactions are governed by a symmetric connection known as the Levi-Civita connection. Nevertheless, a more generic connection comprises two components: an antisymmetric part and another component exhibiting the non-metricity condition. This extended affine connection can be expressed as follows \cite{Jimenez/2019},
\begin{equation}\label{2a}
\Upsilon^\alpha_{\ \mu\nu}=\Gamma^\alpha_{\ \mu\nu}+K^\alpha_{\ \mu\nu}+L^\alpha_{\ \mu\nu} \,\,\text{.}
\end{equation}
where the first term denotes the metric-compatible Levi-Civita connection,
\begin{equation}\label{2b}
\Gamma^\alpha_{\ \mu\nu}\equiv\frac{1}{2}g^{\alpha\lambda}(g_{\mu\lambda,\nu}+g_{\lambda\nu,\mu}-g_{\mu\nu,\lambda}) \,\,\text{.}
\end{equation}
The second is the contortion tensor that represents the antisymmetric part of the affine connection, and that can be estimated in terms of the torsion tensor $ T^\alpha_{\ \mu\nu}\equiv \Upsilon^\alpha_{\ \mu\nu}-\Upsilon^\alpha_{\ \nu\mu}$ as follows.
\begin{equation}\label{2c}
K^\alpha_{\ \mu\nu}\equiv\frac{1}{2}(T^{\alpha}_{\ \mu\nu}+T_{\mu \ \nu}^{\ \alpha}+T_{\nu \ \mu}^{\ \alpha}) \,\,\text{.}
\end{equation}
and the last one is the distortion tensor,
\begin{equation}\label{2d}
L^\alpha_{\ \mu\nu}\equiv\frac{1}{2}(Q^{\alpha}_{\ \mu\nu}-Q_{\mu \ \nu}^{\ \alpha}-Q_{\nu \ \mu}^{\ \alpha}) \,\,\text{.}
\end{equation}	
which is expressed in terms of the non-metricity tensor, 
\begin{equation}\label{2e}
Q_{\alpha\mu\nu}\equiv\nabla_\alpha g_{\mu\nu} = \partial_\alpha g_{\mu\nu}-\Upsilon^\beta_{\,\,\,\alpha \mu}g_{\beta \nu}-\Upsilon^\beta_{\,\,\,\alpha \nu}g_{\mu \beta} \,\,\text{.}
\end{equation}

The geometric significance of the non-metricity tensor $Q_{\alpha\mu\nu}$ becomes evident when considering the behavior of a vector $v^\mu$ under parallel transport along a curve $\gamma$ with tangent vector $u^\alpha$. Unlike in Riemannian geometry, where the length of a vector is preserved, the presence of non-metricity implies a change in the magnitude $\|v\|^2 = g_{\mu\nu}v^\mu v^\nu$. Applying the parallel transport equation $u^\alpha \nabla_\alpha v^\mu = 0$ and the Leibniz rule, the variation of the squared magnitude along the curve is given by \cite{Heisenberg/2024}
\begin{equation}\label{2e-a}
\frac{d}{dt} \|v\|^2 = u^\alpha \nabla_\alpha (g_{\mu\nu} v^\mu v^\nu) = (u^\alpha \nabla_\alpha g_{\mu\nu}) v^\mu v^\nu = Q_{\alpha\mu\nu} u^\alpha v^\mu v^\nu \,\,\text{.}
\end{equation}
Eq.~\eqref{2e-a} demonstrates that $Q_{\alpha\mu\nu}$ acts as a direct measure of how the metric—and consequently the lengths and angles of vectors—deforms during transport. A connection is said to be metric-compatible if and only if $Q_{\alpha\mu\nu} = 0$, a condition strictly enforced in GR but relaxed in the symmetric teleparallel framework to allow the non-metricity to carry the gravitational information.

Beyond individual vectors, non-metricity also governs the evolution of the $n$-dimensional volume elements. For a region $\Omega \subset \mathcal{M}$ with volume $\text{Vol}(\Omega) = \int_\Omega \sqrt{-g} \, d^nx$, the rate of change of the volume under parallel transport along the flow of $u^\alpha$ is determined by the trace of the non-metricity tensor given as follows \cite{Heisenberg/2024}
\begin{equation}\label{2e-b}
\frac{d}{dt} \text{Vol}(\Omega) = \int_\Omega u^\alpha \left( \nabla_\alpha \sqrt{-g} \right) d^nx = \frac{1}{2} \int_\Omega \sqrt{-g} u^\alpha Q_\alpha \, d^nx \,\,\text{,}
\end{equation}
where $Q_\alpha = g^{\mu\nu}Q_{\alpha\mu\nu}$ is the non-metricity trace. This relationship highlights that non-metricity does not merely affect local vector alignment but fundamentally alters the global geometric properties of the manifold, such as the expansion or contraction of space-time volumes independently of the curvature.

In addition, we define the superpotential tensor as,
\begin{equation}\label{2f}
4P^\lambda\:_{\mu\nu} = -Q^\lambda\:_{\mu\nu} + 2Q_{(\mu}\:^\lambda\:_{\nu)} + (Q^\lambda - \tilde{Q}^\lambda) g_{\mu\nu} - \delta^\lambda_{(\mu}Q_{\nu)} \,\,\text{.}
\end{equation}
where $Q_\alpha = Q_\alpha\:^\mu\:_\mu $ and $ \tilde{Q}_\alpha = Q^\mu\:_{\alpha\mu} $ are non-metricity vectors.
Now by contracting the superpotential tensor with the non-metricity tensor, we obtain the non-metricity scalar in a more confined form,
\begin{equation}\label{2g}
Q = -Q_{\lambda\mu\nu}P^{\lambda\mu\nu}. 
\end{equation}
We know that the curvature tensor can be estimated as
\begin{equation}\label{2h}
R^\alpha_{\: \beta\mu\nu} = 2\partial_{[\mu} \Upsilon^\alpha_{\: \nu]\beta} + 2\Upsilon^\alpha_{\: [\mu \mid \lambda \mid}\Upsilon^\lambda_{\nu]\beta} \,\,\text{.}
\end{equation} 
Now, using the affine connection Eq.~\eqref{2a}, one can have
\begin{equation}\label{2i}
R^\alpha_{\: \beta\mu\nu} = \mathring{R}^\alpha_{\: \beta\mu\nu} + \mathring{\nabla}_\mu X^\alpha_{\: \nu \beta} - \mathring{\nabla}_\nu X^\alpha_{\: \mu \beta} + X^\alpha_{\: \mu\rho} X^\rho_{\: \nu\beta} - X^\alpha_{\: \nu \rho} X^\rho_{\: \mu\beta} \,\,\text{.}
\end{equation}
Here, $\mathring{R}^\alpha_{\: \beta\mu\nu}$ and $\mathring{\nabla}$ are described in terms of the Levi-Civita connection Eq.~\eqref{2b}, and $X^\alpha_{\ \mu\nu}=K^\alpha_{\ \mu\nu}+L^\alpha_{\ \mu\nu}$.
 Applying suitable contractions on the curvature term and torsion-free constraint $ T^\alpha_{\ \mu\nu}=0$ in the Eq.~\eqref{2i}, we have
\begin{equation}\label{2j}
R=\mathring{R}-Q + \mathring{\nabla}_\alpha \left(Q^\alpha-\tilde{Q}^\alpha \right) \,\,\text{,} 
\end{equation}
where $\mathring{R}$ is the usual Ricci scalar evaluated in terms of the Levi-Civita connection.
 Furthermore, employing a teleparallel constraint $R=0$, we acquire curvature-free teleparallel geometries, and hence the relation Eq.~\eqref{2j} becomes
\begin{equation}\label{2k}
\mathring{R}=Q - \mathring{\nabla}_\alpha \left(Q^\alpha-\tilde{Q}^\alpha \right).   
\end{equation}
The relation obtained in Eq.~\eqref{2k} indicates that the Ricci scalar curvature differs from the non-metricity scalar by a boundary term. The Eq.~\eqref{2k} reveals that the gravity theory incorporates only a non-metricity scalar $Q$ in the action, differs from Einstein's GR by a boundary term. This indicates that STEGR presents an equivalent formulation to GR, and hence the theory is known as symmetric teleparallel equivalent to GR \cite{KUHN}. 

Now, we present the action for $f(Q)$ gravity which is a generalization to STEGR theory, in the presence of a scalar field, 
\begin{equation}\label{2l}
\mathcal{S}=\int\frac{1}{2}\,f(Q)\sqrt{-g}\,d^4x+\int \mathcal{L}_{\phi}\,\sqrt{-g}\,d^4x\, \,\,\text{,}
\end{equation}
where $g=\text{det}(g_{\mu\nu})$, $f(Q)$ is an arbitrary function of the non-metricity scalar $Q$, and $\mathcal{L}_{\phi}$ is the Lagrangian density of a scalar field $\phi$ given by \cite{BAHA},

\begin{equation}\label{2m}
\mathcal{L}_{\phi} = -\frac{1}{2} g^{\mu \nu} \partial_\mu \phi  \partial_\nu \phi -V(\phi) \,\,\text{.}
\end{equation}
Here, $V(\phi)$ represents a potential for the field $\phi$. We obtain the following governing field equation by varying the action in Eq.~\eqref{2l} with respect to the metric,

\begin{equation}\label{2n}
\frac{2}{\sqrt{-g}}\nabla_\lambda (\sqrt{-g}f_Q P^\lambda\:_{\mu\nu}) + \frac{1}{2}g_{\mu\nu}f+f_Q(P_{\mu\lambda\beta}Q_\nu\:^{\lambda\beta} - 2Q_{\lambda\beta\mu}P^{\lambda\beta}\:_\nu) = -T_{\mu\nu}^{\phi} \,\,\text{,}
\end{equation}

where $f_Q=\frac{df}{dQ}$ and $T_{\mu\nu}^{\phi}$ represents the stress-energy tensor of the scalar field given as
\begin{equation}\label{2o}
 T_{\mu\nu}^{\phi}= \partial_\mu \phi  \partial_\nu \phi -\frac{1}{2} g_{\mu \nu} g_{\alpha \beta} \partial^\alpha \phi  \partial^\beta \phi -  g_{\mu \nu} V(\phi) \,\,\text{.}
\end{equation}
Moreover, we obtain the following equation of motion for the scalar field, i.e., the Klein-Gordon equation, from the Euler-Lagrangian equation for the Lagrangian density given by Eq.~\eqref{2m}
\begin{equation}\label{2p}
\square \phi - V,_\phi =0.
\end{equation}
Here, $\square$ denotes the d'Alembertian and $V,_\phi = \frac{\partial V}{\partial \phi}$. Furthermore, on varying the action in Eq.~\eqref{2l} with respect to the connection (similar to Palatini prescription), we have 
\begin{equation}\label{2q}
\nabla_\mu \nabla_\nu (\sqrt{-g}f_Q P^{\mu\nu}\:_\lambda) =  0 \,\,\text{.}
\end{equation}
To ensure the main text remains focused on the primary physical results and cosmological implications, we have compiled the more rigorous mathematical derivations and theoretical consistency checks in a series of appendices. 

We begin in Appendix~\ref{AppendixA}, where we provide a comprehensive account of the variational procedure used to derive the field equations. This includes the explicit expansion of the non-metricity scalar within the coincident gauge, ensuring that the transition from the action to the equations of motion is transparent and reproducible. Following this, Appendix~\ref{AppendixB} specializes these general results to the FLRW metric. This allows us to extract the specific cosmological field equations that govern the background evolution of the Universe, which serve as the basis for our subsequent analysis.

Beyond the classical evolution of the background, it is vital to verify the theoretical viability of the symmetric teleparallel framework. To this end, Appendix~\ref{AppendixC} is dedicated to a formal degree-of-freedom analysis. Our objective is to demonstrate that the Symmetric Teleparallel Equivalent of General Relativity (STEGR) is well-behaved, propagating exactly the two physical modes associated with a massless spin-2 graviton. We establish this through two complementary paths: 
\begin{enumerate}
    \item A Hamiltonian constraint analysis following the Dirac-Bergmann algorithm, building upon foundational work \cite{Jimenez/2018, Jimenez/2019} and incorporating recent developments in the covariant Hamiltonian formulation and stability of $f(Q)$ gravity \cite{Pati:2023, Hu:2022, Guzman:2024}.
    \item A linear perturbation approach utilizing scalar-vector-tensor (SVT) decomposition \cite{Mukhanov, Bardeen:1980kt}, which confirms the absence of unwanted scalar or vector modes that often plague more general metric-affine theories.
\end{enumerate}

Finally, Appendix~\ref{AppendixD} addresses the construction of the STEGR Lagrangian from first principles. We derive the necessary coefficients by requiring consistency with the Fierz-Pauli action for a massless spin-2 field in the weak-field limit \cite{Fierz:1939ix, Hinterbichler:2011tt}. Furthermore, we impose nonlinear consistency via the Bianchi identity to ensure that the resulting field equations remain divergence-free. Together, these appendices provide the robust geometric and theoretical bedrock upon which the $f(Q)$ framework employed in this thesis is built.

\section{Cosmology in $f(Q)$ gravity under scalar field}

In this chapter, we assume that our Universe is homogeneous and isotropic, which is evident from the large galaxy survey \cite{2df}. We would also like to note that the observation \cite{planck} suggests that the Universe is also flat to a very good approximation, so the line element in which we are interested is given by the FLRW metric. It can be shown \cite{hawking:1973} that for a homogeneous and isotropic Riemannian manifold, the FLRW metric is the unique metric. Thus, we consider the standard FLRW metric given by,
\begin{equation}\label{3a}
ds^2= -dt^2 + a^2(t)[dx^2+dy^2+dz^2].    
\end{equation}
Here, $a(t)$ is a measure of the expansion of the Universe (scale factor). Beginning with the teleparallel constraint that corresponds to a flat geometry characterizing a pure inertial connection, one can execute a gauge transformation parameterized by $\Lambda^\alpha_\mu$  \cite{Hehl1995, JIM-2},
\begin{equation}\label{3b}
 \Upsilon^\alpha_{\: \mu \nu}  = (\Lambda^{-1})^\alpha_{\:\: \beta} \partial_{[ \mu}\Lambda^\beta_{\: \: \nu ]} \,\,\text{.}
\end{equation}
Consequently, the generic affine connection can be expressed as follows, using the general element of $ GL(4,\mathbb{R}) $ characterized by the transformation $ \Lambda^\alpha_{\: \: \mu}=\partial_\mu \zeta^\alpha$, where $ \zeta^\alpha $ is an arbitrary vector field,
\begin{equation}\label{3c}
\Upsilon^\alpha_{\: \mu \nu} = \frac{\partial x^\alpha}{\partial \zeta^\rho} \partial_\mu \partial_\nu \zeta^\rho \,\,\text{.}
\end{equation}
This reveals the possibility of eliminating the connection through a coordinate transformation. The coordinate transformation responsible for eliminating the connection Eq.~\eqref{3c} is termed gauge coincident. We utilize the coincident gauge in the present thesis. Hence, the non-metricity scalar corresponds to the metric Eq.~\eqref{3a} becoming $Q=6H^2$.\\
The stress-energy tensor for the perfect fluid distribution reads as follows.
\begin{equation}\label{3d}
T_{\mu\nu}=(\rho+p)u_\mu u_\nu + pg_{\mu\nu} \,\,\text{,}
\end{equation}
where $u^\mu=(1,0,0,0)$ are components of the four velocities. Upon comparing Eq.~\eqref{3d} and \eqref{2o}, we have
\begin{equation}\label{3e}
\rho=-\frac{1}{2}g_{\alpha \beta}\partial^\alpha \phi \partial^\beta \phi +V(\phi) \,\,\text{,}
\end{equation}
\begin{equation}\label{3f}
p=-\frac{1}{2}g_{\alpha \beta}\partial^\alpha \phi \partial^\beta \phi -V(\phi) \,\,\text{.}
\end{equation}
As the scalar field considered here does not depend on the spatial coordinates, we have the following expressions for the energy density and pressure component of the scalar field,
\begin{equation}\label{3g}
\rho_{\phi}=\frac{1}{2}\dot{\phi}^2+V(\phi) \,\,\text{,}
\end{equation}
\begin{equation}\label{3h}
p_{\phi}=\frac{1}{2}\dot{\phi}^2-V(\phi) \,\,\text{,}
\end{equation}
and the corresponding equation of the state parameter can be written as,
\begin{equation}\label{3i}
\omega_{\phi}=\frac{p_{\phi}}{\rho_{\phi}}=\frac{\frac{1}{2}\dot{\phi}^2-V(\phi)}{\frac{1}{2}\dot{\phi}^2+V(\phi)} \,\,\text{.}
\end{equation}
Moreover, corresponding to the metric Eq.~\eqref{3a}, the Klein-Gordon equation given in Eq.~\eqref{2p} becomes
\begin{equation}\label{3j}
\ddot{\phi}+3H\dot{\phi}+V_{,\phi}=0 \,\,\text{.}
\end{equation}
We obtain the following Friedmann-like equations governing the gravitational interactions under the $f(Q)$ gravity background in the presence of scalar field,
\begin{equation}\label{3k}
3H^2=\frac{1}{2f_Q} \left( -\rho_{\phi}+\frac{f}{2}  \right) \,\,\text{,a }
\end{equation}
\begin{equation}\label{3l}
    \dot{H}+3H^2+ \frac{\dot{f_Q}}{f_Q}H = \frac{1}{2f_Q} \left( p_{\phi}+\frac{f}{2} \right) \,\,\text{.}
\end{equation}
For the $f(Q)$ functional $f(Q)=-Q+\Psi(Q)$, we can rewrite the Friedmann equations becomes Eq.~\eqref{3k}-\eqref{3l} as (where we can recover ordinary GR by putting $\Psi=0$)
\begin{equation}\label{3m}
    3H^2= \rho_\phi + \rho_{de} \,\,\text{,}
\end{equation}
\begin{equation}\label{3n}
    \dot{H}=-\frac{1}{2} [\rho_\phi + p_\phi+\rho_{de}+p_{de}] \,\,\text{,}
\end{equation}
where $\rho_{de}$ and $p_{de}$ represent the energy density and pressure of the dark energy component evolving due to the geometry of spacetime,
\begin{equation}\label{3o}
    \rho_{de}=-\frac{\Psi}{2}+ Q\Psi_Q \,\,\text{,}
\end{equation}
\begin{equation}\label{3p}
    p_{de}=-\rho_{de}-2\dot{H} \left( \Psi_Q+2Q\Psi_{QQ} \right) \,\,\text{.}
\end{equation}

\section{Scalar fields}
Scalar fields play a significant role in describing the physical properties of the Universe, particularly in the context of the inflationary scenario \cite{ALAN, Linde}, and can explain the cosmic late-time acceleration. Although the $\Lambda$CDM model is highly consistent with observational data, successfully describing structure formation, it has yet to effectively quantify quantum vacuum fluctuations \cite{ZEL,WENB}. This is the key inspiration for proposing the dark energy candidate as an alternative to $\Lambda$. Various examples include the quintessence field \cite{Ratra:1988,quint2}, a quintom scalar field \cite{quintom1,Guo/2005}, a phantom scalar field \cite{phant}, and multi-scalar field models \cite{mult}. If one assumes that the cosmological constant $\Lambda$ is coming from a single (or multiple) scalar field, then one can solve the cosmological constant problem, as the scalar field can decay the vacuum energy to stop the exponential increase. However, there exists a tension between the Hubble constant values measured from early observations (such as Planck \cite{planck}) and estimated via local observations (such as SH0ES \cite{Riess}). One potential solution to this tension involves assuming extensions beyond the $\Lambda$CDM. This article explores scalar field cosmology in the modified symmetric teleparallel gravity background, utilizing the dynamical system approach. The modified $f(Q)$ function is now responsible for the observed accelerating scenario, whereas the quantum field theory prediction of the tiny value of $\Lambda$ ($\sim 10^{-120}$) cannot be described by such gravitational modification and thus the addition of a scalar field takes into account the quintessence scenario.\\
Note that it is prominent to choose both modified gravity and scalar field, such as in previous work on $f(R,\phi)$ \cite{fr}, $f(T,\phi) \cite{ft}$ and $f(\mathcal{G},\phi)$ \cite{fg}. There are many reasons to believe that the effect of both scalar field and modified gravity can describe the late time acceleration efficiently. One of the reasons is the cosmological constant problem that we discussed in the previous section; the second reason is from various calculations of QFT in curved spacetime, which indicate that the inflation field could indeed survive up to today to act as the Quintessence field \cite{Ratra:1988}. For a detailed discussion of how the early Universe scalar field could survive even in the late Universe, we recommend a review by Peebles and Ratra \cite{ratrareview}. Another motivation to take the exponential and polynomial potential is that the background perturbation (Bardeen potential) behaves very well in these two types of scalar field, which we have already explored in our work \cite{rajame}. Some works have been done to suggest that torsion-based gravity could explain the Hubble tension \cite{tension}. In this thesis, our motivation is to present the dynamical system of $f(Q)$ in the presence of a scalar field in full generality. 
\subsection{Inflation}

Inflation is a paradigm of the early Universe that postulates a period of accelerated expansion occurring shortly after the Big Bang. It was originally proposed to address several conceptual problems of standard hot Big Bang cosmology, including the horizon problem, the flatness problem, and the absence of unwanted relics such as magnetic monopoles. During inflation, the Universe expands quasi--exponentially, allowing regions that are currently causally disconnected to have been in causal contact in the past and dynamically driving the spatial geometry toward near flatness.

In most models, inflation is driven by one or more scalar fields, collectively referred to as inflaton fields. The dynamics of the inflaton is governed by its potential energy, which dominates its kinetic energy during inflation, leading to a negative pressure and accelerated expansion. For a homogeneous scalar field $\phi$ minimally coupled to gravity, the energy density and pressure are given by
\begin{equation}
\rho_\phi = \frac{1}{2}\dot{\phi}^2 + V(\phi) \,\,\text{,} \qquad
p_\phi = \frac{1}{2}\dot{\phi}^2 - V(\phi) \,\,\text{,}
\end{equation}
where $V(\phi)$ is the scalar field potential. When $V(\phi)$ dominates, the equation of state approaches $p_\phi \simeq -\rho_\phi$, providing the necessary condition for accelerated expansion.

Inflation also provides a natural mechanism for the origin of the cosmic structure. Quantum fluctuations of the scalar field and the spacetime metric are stretched to cosmological scales during inflation, seeding the primordial density perturbations that later grow into galaxies and large-scale structure. The success of inflation is strongly supported by observations of the cosmic microwave background, which reveal a nearly scale-invariant, Gaussian spectrum of primordial fluctuations. Consequently, inflation has become a central component of modern cosmology, linking high-energy field theory with observational data.
\subsection{Quintessence}

Quintessence is a dynamical scalar field model proposed to explain the late-time accelerated expansion of the Universe as an alternative to a strictly constant cosmological constant. Unlike $\Lambda$, which corresponds to a fixed vacuum energy density, quintessence allows the dark energy component to evolve with cosmic time, potentially alleviating some conceptual issues associated with the cosmological constant.

In quintessence models, cosmic acceleration is driven by a canonical scalar field $\phi$ minimally coupled to gravity, with dynamics governed by a potential $V(\phi)$. The energy density and pressure of the field are given by,
\begin{equation}
\rho_\phi = \frac{1}{2}\dot{\phi}^2 + V(\phi) \,\,\text{,} \qquad
p_\phi = \frac{1}{2}\dot{\phi}^2 - V(\phi) \,\,\text{.}
\end{equation}
The equation of state parameter is defined as,
\begin{equation}
\omega_\phi \equiv \frac{p_\phi}{\rho_\phi}
= \frac{\frac{1}{2}\dot{\phi}^2 - V(\phi)}
{\frac{1}{2}\dot{\phi}^2 + V(\phi)} \,\,\text{.}
\end{equation}
When the potential energy dominates the kinetic term, $\omega_\phi$ approaches $-1$, mimicking the behavior of a cosmological constant and leading to accelerated expansion.

A key advantage of quintessence is its dynamical nature. Depending on the choice of the potential, the scalar field can exhibit tracking or scaling behavior, where its energy density evolves alongside matter or radiation for much of cosmic history before becoming dominant at late times. This feature offers a possible explanation for the observed coincidence between the dark energy and matter densities today.

In the limit where the scalar field is frozen, $\dot{\phi}\rightarrow 0$ and $V(\phi)=\text{constant}$, quintessence is exactly reduced to a cosmological constant with $\omega_\phi = -1$. Thus, quintessence provides a unifying framework in which $\Lambda$CDM emerges as a special case, while more general dynamics allows departures from $\omega=-1$ that can be tested observationally.

\section{Dirac-Born-Infeld scalar field}
It is well known that Einstein's general theory of relativity is not renormalizable in the context of quantum field theory. There have been several attempts to find a renormalizable theory of quantum gravity, and string theory offers one such unification. It is well known that even in bosonic string theory, quantization of the Polyakov action (conformal transformation of the Nambu-Goto action) gives a Tachyon-like field which soon decays via spontaneous symmetry breaking \cite{green,polchinski}. It was first observed by Mazumdar et al. \cite{mazumdar} that the decay of a non-BPS $D4$ brane to a stable $D3$ brane can give rise to a Tachyon field, which can act as an inflation field in the cosmological context. In 2002, a series of three papers by Sen \cite{sen1,sen2,sen3} showed how, in string theory, as well as in string field theory, Tachyons occur naturally, and in \cite{sen3} it has been shown that the effective field of such Tachyons can be viewed as DBI scalar field theory.\\
Soon after these proposals, Padmanabhan \cite{paddy} and Gibbons \cite{gibbons1} showed how these DBI-type fields could be used in the FLRW background to give inflation field-like behavior. Alternative ways of obtaining the DBI field from other forms of string theory have been reviewed by Gibbons \cite{gibbons2}. The study of the DBI field in the late time acceleration context has been done by Bhagla et al. \cite{bhagla}, while Gorini et al. \cite{gorini}
offered an alternative way of visualizing the DBI field as a modified Chaplygin gas. We also note that the DBI field has been proposed as an alternative to dark matter by Padmanabhan \cite{paddy2}, which shows that the DBI field could indeed affect the late-time cosmology.\\
 Copeland \cite{copeland1} and Aguirregabiria \cite{aguirregabiria} first studied the study of the DBI field in the dynamical system setting. In this thesis, we also closely follow the treatment given in \cite{copeland1}. Soon after that, Fang and Lu \cite{fang2010a} considered a much more general type of potential beyond the inverse square potential, the work later extended by Quiros et al. \cite{Quiros} to include much more general potentials, and they have given an exact treatment of the $sinh(\phi)$ potential. Guo \cite{guoexp} has chosen an exponential potential for dynamical system analysis which we have used here to make an autonomous dynamical system. It is also worth noting that as Silverstein and Tong \cite{tong} have shown, if one considers a D3-brane moving towards the horizon of AdS space, one can get a generalized DBI field in a strong coupling limit (as opposed to a weak coupling limit where the previous work has been done). In the strong coupling limit, it can be shown that the DBI field receives additional contributions from the movement of the D3 brain and the Lagrangian becomes $\mathcal{L}_{GDBI}=\frac{1}{f(\phi)}(\sqrt{1+f(\phi)\partial \phi^2}-1)-V(\phi)$.\\
The conventional concept of relativity, especially in GR, which interprets gravity as the curvature of spacetime, may not offer the definitive solution to elucidate dark energy. This encourages the exploration of alternative theoretical frameworks in cosmology that can effectively address cosmic acceleration while remaining consistent with observational data. GR and its curvature-based extensions have been formulated and thoroughly examined in previous research \cite{CANT, R15}. Recently, alternative theories of gravitation based solely on non-metricity and flat spacetime geometry have been established and extensively explored \cite{NEST, Jimenez/2018}. The $f(Q)$ gravity, with its various astrophysical and cosmological implications, has been widely investigated \cite{R18, JIM-2, LAVI-1, ANAG, KUHN, R24, R25, R26, R27, WOM-2, R29}. In this thesis, we have shown that even with modified $f(Q)$ gravity, we obtain a similar late-time accelerating behavior with $ q =-1$, as expected from de Sitter-like expansion. We also note that in our investigation, the present value of the deceleration parameter is approaching $-0.8$, which is quite consistent with the observed value of $-0.55$.

\section{Chaplygin gas and generalized Chaplygin gas}
The Chaplygin gas was originally proposed as a unified candidate for dark matter and dark energy, characterized by an exotic equation of state of the form
\begin{equation}
p = -\frac{A}{\rho},
\end{equation}
where $A$ is a positive constant. This model naturally interpolates between a dust--dominated phase ($\omega\simeq 0$) at early times and a cosmological constant--like phase ($\omega\simeq -1$) at late times. At high redshift, the energy density behaves as $\rho \propto a^{-3}$, mimicking pressure-less matter, while at late times it asymptotically approaches a constant value, driving accelerated expansion.

The model was later generalized to the Generalized Chaplygin Gas (GCG), defined by
\begin{equation}
p = -\frac{A}{\rho^{\alpha}} \,\,\text{,}
\end{equation}
where $\alpha$ is a dimensionless parameter. This generalization provides additional flexibility in fitting observational data and retains the desirable feature of smoothly connecting matter and dark energy epochs within a single fluid description.

An important aspect of the Chaplygin gas is that it can be realized from fundamental field theories. For a minimally coupled scalar field $\phi$ with a Lagrangian
\begin{equation}
\mathcal{L} = -\frac{1}{2}g^{\mu\nu}\partial_\mu \phi \partial_\nu \phi - V(\phi) \,\,\text{.}
\end{equation}
The corresponding energy density and pressure in a spatially flat FLRW background are
\begin{equation}
\rho_\phi = \frac{1}{2}\dot{\phi}^2 + V(\phi) \,\,\text{,}
\qquad
p_\phi = \frac{1}{2}\dot{\phi}^2 - V(\phi) \,\,\text{.}
\end{equation}
By appropriately reconstructing the potential $V(\phi)$, one can reproduce the Chaplygin gas equation of state. In this way, the Chaplygin gas can be interpreted as an effective scalar field model.

Furthermore, the Chaplygin equation of state arises naturally in DBI scalar field or Tachyonic field theories with action given by,
\begin{equation}
S = -\int d^4x \sqrt{-g}\, V(T)\sqrt{1 - g^{\mu\nu}\partial_\mu T \partial_\nu T} \,\,\text{.}
\end{equation}
For a homogeneous Tachyon field, the energy density and pressure become
\begin{equation}
\rho_T = \frac{V(T)}{\sqrt{1-\dot{T}^2}} \,\,\text{,}
\qquad
p_T = -V(T)\sqrt{1-\dot{T}^2} \,\,\text{,}
\end{equation}
which can directly lead to a Chaplygin-type relation between $p$ and $\rho$. 

Beyond phenomenology, the Chaplygin gas also appears in higher-dimensional and brane-world scenarios, such as in the Randall--Sundrum framework, where effective four-dimensional cosmological dynamics can generate similar exotic equations of state. These theoretical connections suggest that the Chaplygin gas is not merely an ad hoc phenomenological model but may have deeper origins in high-energy physics.

Observationally, Chaplygin gas models have been tested against Type~Ia supernovae, cosmic microwave background data, gravitational lensing, and cluster observations, showing consistency with late-time acceleration within certain parameter ranges. Owing to its ability to unify dark matter and dark energy behavior and its natural appearance in scalar and DBI field theories, the Chaplygin gas remains an interesting alternative framework for describing cosmic acceleration.
\subsection{Bose-Einstein condensate dark matter}
The notion of unseen mass in the Universe was first raised by Fritz Zwicky 
in 1933, who coined the german term \textit{dunkle Materie} dark matter
to account for the anomalous velocity dispersion of galaxies in the Coma 
cluster \cite{Zwicky1933}. Yet for decades, the idea remained on the 
fringes of mainstream astronomy, treated more as a curiosity than a 
physical reality. It was Vera Rubin who changed that. Through painstaking 
observations of galactic rotation curves with Kent Ford in the 1970s, 
Rubin demonstrated that stars in the outer reaches of spiral galaxies orbit 
far too quickly to be accounted for by visible matter alone, a result so 
unambiguous, it forced the astronomical community to take dark matter 
seriously \cite{Rubin1980}. Her philosophy on scientific discovery perhaps 
best explains why she looked where others did not bother: \textit{``Don't 
shoot for the stars, we already know what's there. Shoot for the space in 
between because that's where the real mystery lies''} 
\cite{Yeager2022}. It is in that same spirit, probing what cannot be 
seen, only inferred, that the dark energy sector of the Universe 
continues to motivate the present work.

Bose-Einstein Condensation (BEC) is a purely quantum phenomenon that arises from Bose--Einstein statistics, originally predicted by Einstein following the work of Bose. Since bosonic wave functions are symmetric under particle exchange, a large number of particles can occupy the same quantum ground state at sufficiently low temperatures, leading to macroscopic quantum coherence. This phenomenon was experimentally confirmed in 1995 and has since played a central role in condensed matter physics.

In a cosmological context, BEC has been proposed as a viable candidate for dark matter if the dark sector is composed of ultra-light bosonic particles. The critical temperature for condensation depends on the particle mass and number density and can be comparable to temperatures achieved during cosmic structure formation. This idea was systematically explored by Böhmer and Harko \cite{Boehmer/2007}, who showed that galactic dark matter halos could be modeled as self-gravitating Bose--Einstein condensates. Such models naturally predict smooth, cored density profiles, potentially alleviating the cusp problem encountered in standard cold dark matter simulations.

The dynamics of weakly interacting bosonic condensates is governed by the Gross--Pitaevskii equation, which includes kinetic energy, short-range interactions, and gravitational self-attraction. In the hydrodynamic representation, the condensate behaves as a quantum fluid with effective pressure arising from particle interactions. In the Thomas--Fermi regime, where interaction energy dominates over quantum pressure, the equation of state of BEC dark matter takes a quadratic form,
\[
p \propto \rho^2 \,\,\text{,}
\]
where the proportionality constant depends on the particle mass and the length of scattering.

More generally, one may consider an extended phenomenological equation of state,
\[
p = \alpha \rho + \beta \rho^2 \,\,\text{,}
\]
where the linear term represents ordinary barotropic behavior, while the quadratic term captures two-body interaction effects characteristic of a condensate. Special cases of this equation recover cold dark matter, normal bosonic dark matter, or pure BEC dark matter scenarios.

Overall, Bose-Einstein condensate dark matter provides a compelling framework in which microscopic quantum effects manifest at galactic scales, offering an alternative to conventional cold dark matter while remaining consistent with a wide range of astrophysical observations.

\section{Datasets for statistical analysis}

In this thesis, theoretical predictions are confronted with observational data to constrain the free parameters of the cosmological models. A joint analysis is performed using Cosmic Chronometers, Type~Ia Supernovae from the Pantheon+SH0ES compilation, and large-scale structure observations. Parameter estimation is carried out within a Bayesian framework using likelihood-based methods and Markov Chain Monte Carlo (MCMC) sampling.

\subsection{Cosmic Chronometers (CC)}

Cosmic chronometers are massive, passively evolving galaxies that allow a model-independent determination of the Hubble parameter through differential age measurements,
\begin{equation}
H(z) = -\frac{1}{1+z}\frac{dz}{dt} \,\,\text{.}
\end{equation}
In this work, $31$ independent $H(z)$ measurements covering the redshift range $0.07 \leq z \leq 2.41$ are used. The corresponding chi-square is defined as
\begin{equation}
\chi^2_{\mathrm{CC}} = \sum_{k=1}^{31}
\frac{\left[H_{\mathrm{th}}(z_k)-H_{\mathrm{obs}}(z_k)\right]^2}{\sigma_{H,k}^2} \,\,\text{.}
\end{equation}

\subsection{Type~Ia Supernovae: Pantheon+SH0ES}

Type~Ia supernovae act as standardizable candles and constrain the expansion history of the Universe through luminosity distance measurements. We used the Pantheon+SH0ES compilation, consisting of $1701$ supernovae in the redshift range $0.001 \leq z \leq 2.3$. The theoretical distance modulus is given by
\begin{equation}
\mu^{\mathrm{th}}(z) = 5\log_{10}\!\left(\frac{D_L(z)}{\mathrm{Mpc}}\right)+25 \,\,\text{,}
\end{equation}
with the luminosity distance
\begin{equation}
D_L(z) = c(1+z)\int_0^z \frac{dx}{H(x)} \,\,\text{.}
\end{equation}
Cepheid-calibrated distances are incorporated to break the degeneracy between the absolute magnitude and the Hubble constant. The likelihood is constructed using the full covariance matrix, including systematic uncertainties.

\subsection{Baryon Acoustic Oscillations (BAO)}

Baryon Acoustic Oscillations (BAO) provide a standard ruler for probing the geometry of the Universe. BAO measurements constrain combinations of the angular diameter distance, the Hubble parameter, and the sound horizon scale. These data offer robust and largely model-independent information on the expansion history and play a key role in breaking parameter degeneracies when combined with other probes.

\subsection{DESI}
We also used five data points from the latest DESI survey. The Dark Energy Spectroscopic Instrument (DESI) employs a multi-tracer strategy that combines galaxy surveys (including the Bright Galaxy Survey, Luminous Red Galaxies, and Emission Line Galaxies) covering redshifts up to $z\simeq1.6$, quasars extending to $z\simeq2.1$, and the Lyman--$\alpha$ forest, which probes the large-scale structure up to $z\simeq2.33$. 

By analyzing more than $5.7$ million objects over approximately $7500$ square degrees, DESI achieves sub-percent precision in BAO measurements, with a highest detection significance of $9.1\sigma$ at an effective redshift of $z_{\rm eff}=0.93$. In addition, DESI provides constraints on both the transverse comoving distance and the Hubble parameter at high redshift. The availability of these complementary measurements makes DESI a powerful probe of the cosmic distance ladder, helping break parameter degeneracies in cosmological analyses.

\section{Bayesian analysis in cosmology}

\subsection{Statistical modeling of the Universe}

Once a cosmological model is constructed at the theoretical level, it must be confronted with observational data. The connection between theory and observation is established through statistical inference. In modern cosmology, Bayesian statistics provides the standard framework for parameter estimation and model comparison.

The general procedure for statistical modeling in cosmology may be summarized as follows:

\begin{itemize}
    \item \textbf{Specification of parameters:} 
    After solving the cosmological field equations, the model typically contains free parameters such as $H_0$, $\Omega_m$, $\Omega_\Lambda$, coupling constants, or parameters of modified gravity. These must be constrained using observational data.

    \item \textbf{Construction of observables:}
    The theoretical model predicts observable quantities such as Hubble parameter $H(z)$, luminosity distance $d_L(z)$, angular diameter distance $d_A(z)$, growth factor $f\sigma_8$, etc., which can be directly compared with data from supernovae, BAO, CMB, cosmic chronometers, and other surveys.

    \item \textbf{Likelihood function:}
    For a given dataset, we construct a likelihood function $\mathcal{L}(\theta)$, which quantifies the probability of observing the data given the model parameters $\theta$.
\end{itemize}

\subsection{Basics of Bayesian inference}

Bayesian statistics is based on Bayes' theorem:
\[
P(\theta|D) = \frac{P(D|\theta)\,P(\theta)}{P(D)} \,\,\text{.}
\]

Here:
\begin{itemize}
    \item $P(\theta|D)$ is the \textit{posterior probability distribution} of parameters $\theta$ given the data $D$.
    \item $P(D|\theta)$ is the \textit{likelihood function}.
    \item $P(\theta)$ is the \textit{prior distribution}, representing our prior knowledge about the parameters.
    \item $P(D)$ is the Bayesian evidence (a normalization constant).
\end{itemize}

In cosmology, the posterior distribution is the primary object of interest, as it provides the most probable parameter values along with credible intervals.

\subsection{Chi-Square minimization}

For Gaussian-distributed observational errors, the likelihood function is commonly written in terms of the chi-square statistic:

\[
\chi^2(\theta) = \sum_{i,j} \left[ O_i^{\text{obs}} - O_i^{\text{th}}(\theta) \right]
C^{-1}_{ij}
\left[ O_j^{\text{obs}} - O_j^{\text{th}}(\theta) \right] \,\,\text{,}
\]

where $O^{\text{obs}}$ denotes observational data, $O^{\text{th}}$ represents theoretical predictions, and $C^{-1}_{ij}$ is the inverse covariance matrix.

For Gaussian errors, the likelihood is related to the chi-square by,
\[
\mathcal{L}(\theta) \propto \exp\left(-\frac{\chi^2(\theta)}{2}\right) \,\,\text{.}
\]

The traditional frequentest approach determines the best-fit parameters by minimizing $\chi^2$. The minimum value $\chi^2_{\min}$ corresponds to the maximum likelihood estimate. The confidence intervals are obtained from the constant contours $\Delta \chi^2$.

\subsection{Markov Chain Monte Carlo (MCMC) methods}

In realistic cosmological models, the parameter space is multi-dimensional and highly correlated. Analytical maximization of the posterior becomes impractical. Therefore, numerical sampling techniques such as Markov Chain Monte Carlo (MCMC) are employed.

MCMC methods generate a sequence (chain) of parameter values whose distribution converges to the posterior distribution. The most commonly used algorithm is the Metropolis--Hastings algorithm, which proceeds as follows:

\begin{enumerate}
    \item Choose an initial parameter set $\theta_0$.
    \item Propose a new point $\theta'$ from a distribution of proposals.
    \item Compute the acceptance probability:
    \[
    A = \min\left(1, \frac{P(\theta'|D)}{P(\theta|D)}\right) \,\,\text{.}
    \]
    \item Accept or reject the proposed point based on $A$.
\end{enumerate}

After a sufficient number of iterations (burn-in phase removed), the chain samples the posterior distribution. From this chain, one computes:

\begin{itemize}
    \item Mean parameter values,
    \item Marginalized distributions,
    \item Credible intervals (e.g., 68\%, 95\%),
    \item Parameter correlations.
\end{itemize}

Widely used cosmological MCMC codes include \texttt{CosmoMC}, \texttt{MontePython}, and \texttt{emcee}.

\subsection{Model comparison and Bayesian evidence}

Beyond parameter estimation, Bayesian analysis allows model comparison through Bayesian evidence.
\[
\mathcal{Z} = \int \mathcal{L}(\theta) P(\theta) d\theta.
\]

The ratio of evidence between two models (Bayes factor) quantifies which model is statistically preferred by the data. This approach naturally penalizes overly complex models, implementing Occam’s razor.

\subsection{Cosmological interpretation}

Once the parameters are constrained using MCMC analysis, one studies the cosmological implications of the best-fit model. Quantities such as the deceleration parameter $q(z)$, the effective equation of state $\omega(z)$, statefinder diagnostics, or growth parameters are evaluated using the values of posterior-constrained parameters. These predictions are then compared with the standard $\Lambda$CDM model or other competing cosmological scenarios.

In summary, Bayesian inference combined with MCMC sampling provides a robust and systematic framework for confronting theoretical cosmological models with observational data, allowing reliable parameter estimation, uncertainty quantification, and model selection.
\subsection{AIC and BIC}

To assess the goodness of fit and compare different cosmological models while accounting for model complexity, we employ the Akaike Information Criterion (AIC) and the Bayesian Information Criterion (BIC). These information criteria provide a quantitative balance between the quality of the fit and the number of free parameters, thus penalizing overly complex models.

The AIC measures the relative predictive accuracy of a model and is defined as
\begin{equation}
\text{AIC} = \chi^2_{\min} + 2d \,\,\text{,}
\end{equation}
where $\chi^2_{\min}$ is the minimum chi-square value obtained from the fit and $d$ denotes the number of free parameters. Model comparison is performed using the difference
\begin{equation}
\Delta \text{AIC} = \text{AIC}_{\text{model}} - \text{AIC}_{\min} \,\,\text{,}
\end{equation}
with $\Delta \text{AIC} < 2$ indicating strong support, $4 < \Delta \text{AIC} \leq 7$ moderate support, and $\Delta \text{AIC} > 10$ little or no support for a given model.

The BIC provides an approximation to Bayesian evidence and introduces a stronger penalty for additional parameters, especially for large datasets. It is defined as
\begin{equation}
\text{BIC} = \chi^2_{\min} + d \ln N  \,\,\text{,}
\end{equation}
where $N$ is the total number of observational data points. The interpretation of $\Delta \text{BIC}$ follows similar guidelines: $\Delta \text{BIC} < 2$ corresponds to strong evidence, $2 \leq \Delta \text{BIC} < 6$ to moderate evidence, and $\Delta \text{BIC} > 6$ to weak or no evidence in favor of the model.

In this work, both AIC and BIC are computed from the best-fit $\chi^2$ values obtained through MCMC analysis. The resulting criteria allow for a direct comparison of the proposed models with the standard $\Lambda$CDM scenario and provide a robust statistical measure of their relative viability.
\section{Dark energy null tests}
In order to check the deviation from the $\Lambda$CDM model, quite a few tests have been proposed, roughly known as the null test for standard cosmological models. Basically, the idea is to construct a scalar quantity (real numbers) that gives a particular value for $\Lambda$CDM and only deviates when the model is not $\Lambda$CDM. Ideally, these quantities should also distinguish between two different types of nonstandard cosmology as well, such as quintessence and phantom, etc.\\
In this thesis, we have taken mainly two diagnoses, that is, Om and the statefinder diagnosis, to test the deviation from $\Lambda$CDM. Even though, through the MCMC analysis, we have shown that the best fit model parameters with the latest observational data, such as Hubble and Pantheon+SH0ES datasets, it is still not obvious what the late-time behavior of our model is, that is, whether it is approaching the de Sitter model from the Phantom or Quintessence side. In addition, these diagnoses provide a very good consistency check, namely, whether we are obtaining late-time de Sitter solutions or not, and also how severe the deviation from $\Lambda$CDM is.
\subsection{Om diagnostics}
Om diagnostics is the simplest form of diagnosis available for classification of the various dark energy-based cosmological models and their deviation from $\Lambda$CDM.\\
It was first proposed by Sahni et al. \cite{Sahni/2008}, noting that the matter density falls as a third power of the scalar factor while $\Lambda$ remains constant. The formula for $Om(z)$ is given as
\begin{equation}
    Om(z)=\frac{\left(\frac{H(z)}{H_0}\right)^2-1}{(1+z)^3-1}\,\,\text{.}
\end{equation}
\subsection{State finder diagnostics}

 Although diagnosis $Om$ is an excellent null test, there are several problems; for example, it does not distinguish between various types of quintessence and phantom models. There is also the fact that the formula is too simple, so it does not take into account the nuances of the other type of late cosmology and how much they differ from the $\Lambda$CDM.\\
  In order to circumvent these things, Sahni et al. \cite{Sahni/2003} have proposed a null test based on two parameters different from $r,s$, where they are given by the formula:
  \begin{equation}
      r=\frac{\dddot{a}}{aH^3}\,\,\text{,}
  \end{equation}
  and 
  \begin{equation}
      s=\frac{r-1}{3\left(q-\frac{1}{2}\right)}\,\,\text{.}
  \end{equation}
One first note is that for the standard $\Lambda$CDM model, $q=-1$, so $r=0$, so one of the advantages of $r$ is that even though many different models give similar $H$ and $q$, they differ by the third derivative, that is, $r$. $s$ is defined to distinguish between various forms of dark energy models. 
\section{Gaussian process}
Gaussian Processes (GPs) are widely used in machine learning as a non-parametric Bayesian method to reconstruct continuous functions and their derivatives directly from noisy data \citep{Rasmussen2005}. The key strength of GP lies in its model-independent nature: instead of assuming a specific functional form to fit the data, GP infers the function by placing a prior over functions, guided by a covariance kernel. This makes it particularly suitable for cosmology, where one often aims to reconstruct quantities such as the Hubble parameter $H(z)$, its derivatives, or the DE equation of state without assuming any specific cosmological model \citep{Seikel2012, Seikel2013, Mehrabi2021}.

In this framework, the unknown function $f(x) \equiv H(x)$ is modeled as a GP
$$
f(x)\sim\mathcal{GP}\bigl(m(x),\,k(x,x')\bigr) \,\,\text{,}
$$
where $m(x) = \mathbb{E}[f(x)]$ is the prior mean function, often chosen as a constant or low-order polynomial, and $k(x,x')=\text{cov}(f(x),f(x'))$ is the kernel function that determines the covariance between observations at points $x$ and $x'$. There are many covariance functions that we can choose \citep{Rasmussen2005,Seikel2013}. Since the Hubble parameter is expected to evolve smoothly with redshift, it is natural to choose a covariance function that depends only on the separation between data points. For this reason, and following its successful use in similar cosmological reconstruction studies \citep{Seikel2012, Busti:2014, Sun:2021}, we adopt the squared exponential covariance function, which is given by
$$
k(x,x') = \sigma_f^2\exp\!\left[-\frac{(x - x')^2}{2\ell^2}\right] \,\,\text{.}
$$
This covariance function depends on two hyperparameters, $\ell$ and $\sigma_f^2$, which encode the smoothness of the reconstructed function through a characteristic length scale and the output variance, respectively. This approach enables a robust, data-driven reconstruction of cosmological quantities, making GP a powerful tool for probing the expansion history of the Universe and the nature of DE.

Importantly, derivatives of $H(z)$ are obtained analytically by differentiating the kernel
$$
\frac{\mathrm{d}^m}{\mathrm{d}x^m}\frac{\mathrm{d}^n}{\mathrm{d}x'^n}k(x,x')\,\,\text{.}
$$
Since the kernel maintains its Gaussian form even after taking derivatives, we can easily reconstruct $H'(z)$ and $H''(z)$, together with their uncertainties \citep{Seikel2013}. This feature is crucial for diagnostics of cosmic acceleration and model selection.
\section{Swampland conjecture}
It should also be noted that the interest in fixing the parameters of the potentials from the data is not only important for phenomenological purposes but also has a deeper theoretical necessity. It is well known that string theory is one of the most promising theories of quantum gravity, yet there is no consistent de Sitter vacuum solution in string theory in the context of cosmology. Such an unexpected result has been handled by C. Vafa's Swampland conjectures \citep{vafa}. The main idea of Swampland is that there are certain gravity theories that are consistent with the quantum theory (that the low-energy gravity theory has a full UV-complete theory); these belong to the ``landscape". However, there are some other theories, such as theories based on the de Sitter vacuum belonging in the ``swampland", in a sense, these theories cannot have a reasonable UV-complete quantum theory. Based on this paradigm, Vafa has given some conjectures, such as the Swampland Distance Conjecture (SDC), the de Sitter Conjecture (dSC) and the refined de Sitter Conjecture (RdSC), which in principle gives a very tight bound on \( \left|\frac{\nabla_{\phi}V}{V}\right| \). Elizalde and Khurshudyan \citep{elizaldeswampland1} have shown that one can indeed verify the Swampland conjecture or place bounds on the constants of the potentials, using GPs. Later, they have also extended this verification to modified \( f(R) \) gravity formulations \citep{elizaldeswampland2}.
\section{Dynamical system approach in cosmology}

The dynamical system approach provides a powerful and systematic framework for studying the qualitative evolution of cosmological models without requiring exact analytical solutions. In this method, the cosmological field equations are rewritten as an autonomous system of first-order differential equations by introducing suitable dimensionless variables. The evolution of the Universe is then described as trajectories in a finite-dimensional phase space. Fixed (critical) points of the system correspond to cosmological epochs with well-defined physical behavior, such as matter-dominated, radiation-dominated, or dark-energy-dominated phases. The stability of these fixed points is determined by linearizing the system around them and analyzing the eigenvalues of the Jacobian matrix. Negative real parts of the eigenvalues indicate stable attractors, while positive real parts correspond to unstable or saddle points. This analysis allows one to determine the asymptotic behavior of the Universe, particularly the late-time attractors responsible for accelerated expansion, without solving the equations exactly. Consequently, dynamical system techniques are especially useful in cosmology for understanding generic features of cosmic evolution, robustness of solutions, and model viability at early and late-time limits.
\section{Curvature perturbation}

\subsection{Theory}

Cosmological perturbation theory provides a systematic framework for studying small inhomogeneities around an exactly homogeneous and isotropic FLRW background spacetime. These perturbations are believed to originate from quantum fluctuations in the early Universe and later grow to form the large-scale structures observed today. The spacetime metric is decomposed as
\[
g_{\mu\nu} = g^{(0)}_{\mu\nu} + \delta g_{\mu\nu} \,\,\text{,}
\]
where $g^{(0)}_{\mu\nu}$ denotes the FLRW background metric and $\delta g_{\mu\nu}$ represents small perturbations.

In linear order, perturbations can be uniquely decomposed into three mutually decoupled sectors based on their transformation properties under spatial rotations: scalar, vector, and tensor perturbations. Scalar perturbations are responsible for density fluctuations and curvature inhomogeneities, vector perturbations describe vorticity and typically decay with cosmic expansion, and tensor perturbations correspond to gravitational waves. Since scalar perturbations dominate structure formation and CMB anisotropies, they are of primary importance in cosmology.

\subsection{Choice of gauge}

A key subtlety in cosmological perturbation theory is the presence of gauge freedom, which arises due to the invariance of GR under infinitesimal coordinate transformations. Different choices of spacetime coordinates can yield distinct mathematical representations of the same physical perturbation, leading to spurious gauge artifacts if not handled carefully. A gauge choice corresponds to fixing this coordinate freedom by imposing specific conditions on the perturbation variables.

To avoid ambiguities associated with gauge dependence, one can either work in a specific convenient gauge or construct gauge-invariant combinations of perturbation variables. Gauge-invariant quantities are particularly useful since they represent true physical degrees of freedom and allow for direct comparison with observations.

\subsection{Three types of gauge}

For scalar perturbations, the most general perturbed FLRW metric in conformal time $\eta$ can be written as,
\[
ds^2 = a^2(\eta)\left[
(1+2\Phi)d\eta^2 - 2\partial_i B\, d\eta\, dx^i
- \left((1-2\Psi)\delta_{ij} + 2\partial_i\partial_j E\right) dx^i dx^j
\right] \,\,\text{.}
\]
where $\Phi$, $\Psi$, $B$, and $E$ are scalar perturbation variables.

Commonly used gauges include the following.
\begin{itemize}
    \item \textit{Newtonian (longitudinal) gauge}: Defined by setting $B=E=0$, leaving only two scalar potentials $\Phi$ and $\Psi$. This gauge is particularly intuitive, as it closely resembles Newtonian gravity.
    \item \textit{Synchronous gauge}: Characterized by $\Phi=B=0$, often used in numerical Boltzmann codes but requiring careful treatment of residual gauge freedom.
    \item \textit{Comoving gauge}: Defined so that the velocity perturbation vanishes, making it useful for tracking matter perturbations.
\end{itemize}

Gauge-invariant combinations of the scalar perturbations were first constructed by Bardeen and are known as the Bardeen potentials:
\[
\Phi_B = \Phi - \mathcal{H}(B - E') - (B - E')'\,\,\text{,}
\qquad
\Psi_B = \Psi + \mathcal{H}(B - E') \,\,\text{,}
\]
where $\mathcal{H}=a'/a$ is the conformal Hubble parameter and a prime denotes differentiation with respect to conformal time.

In the absence of anisotropic stress, the Einstein field equations imply $\Phi_B = \Psi_B$. The evolution equation for the Bardeen potential in a Universe filled with a perfect fluid can be derived from the perturbed Einstein equations and takes the form
\[
\Phi_B'' + 3\mathcal{H}(1+c_s^2)\Phi_B' 
+ \left[2\mathcal{H}' + (1+3c_s^2)\mathcal{H}^2 - c_s^2 \nabla^2 \right]\Phi_B = 0 \,\,\text{,}
\]
where $c_s^2 = \delta p / \delta \rho$ is the sound speed of the perturbations. This equation governs the evolution of curvature perturbations and plays a central role in connecting early-Universe physics with late-time cosmological observations.
\section{Matter perturbations}

To complement the background and dynamical analysis, we investigate the evolution of cosmological perturbations in a Universe containing two non-interacting matter components, namely pressureless dust matter and a dynamical dark energy component modeled either as a perfect fluid or an effective scalar field. While many studies in the literature focus on a single dominant component during a specific cosmological epoch, a multi-component treatment is more general and allows one to capture the interplay between matter and dark energy perturbations. Such an analysis is particularly relevant during scalar-field-dominated eras, for instance, during early-time inflation or the late-time accelerated expansion.

We consider scalar perturbations around an FLRW background in the Newtonian (longitudinal) gauge, assuming vanishing anisotropic stress so that the two Bardeen potentials coincide. In Fourier space, the evolution equations for matter and dark energy perturbations in a general non-interacting scenario are given by
\begin{subequations}
\begin{align}
&\dot{\delta}_m + \frac{\theta_m}{a} = 0 \,\,\text{,} \label{eq:pert_m1} \\
&\dot{\delta}_\phi + (1+w_\phi)\frac{\theta_\phi}{a}
+ 3H\left(c_{\mathrm{eff}}^{2}-w_\phi\right)\delta_\phi = 0 \,\,\text{,} \label{eq:pert_phi1} \\
&\dot{\theta}_m + H\theta_m - \frac{k^{2}\Phi}{a} = 0 \,\,\text{,} \label{eq:pert_m2} \\
&\dot{\theta}_\phi + H\theta_\phi
- \frac{k^{2}c_{\mathrm{eff}}^{2}\delta_\phi}{(1+w_\phi)a}
- \frac{k^{2}\Phi}{a} = 0 \,\,\text{,} \label{eq:pert_phi2}
\end{align}
\end{subequations}
where $k$ is the comoving wavenumber, $\Phi$ is the scalar gravitational potential, and an overdot denotes differentiation with respect to cosmic time. Density contrasts are defined as $\delta_i \equiv \delta\rho_i/\rho_i$, while $\theta_i$ represents the velocity divergence of each component of the fluid, with $i\in\{m,\phi\}$.

On sub-horizon scales, the system can be closed using the Poisson equation, which relates the gravitational potential to the matter and dark energy perturbations,
\begin{equation}
-\frac{k^{2}}{a^{2}}\Phi =
\frac{3}{2}H^{2}\left[
\Omega_m \delta_m
+ \left(1+3c_{\mathrm{eff}}^{2}\right)\Omega_\phi \delta_\phi
\right] \,\,\text{,}
\label{eq:poisson}
\end{equation}
where the density parameters are defined as
\begin{equation}
\Omega_i = \frac{\rho_i}{3H^2} \,\,\text{.}
\end{equation}

The clustering properties of dark energy are governed by its effective sound speed $c_{\mathrm{eff}}^{2}$. For $c_{\mathrm{eff}}^{2}=0$, dark energy clusters in a manner similar to pressureless matter, while for larger values of $c_{\mathrm{eff}}^{2}$ its perturbations are suppressed on sub-horizon scales due to pressure support. Even in the clustering case, the amplitude of dark energy perturbations is typically smaller than that of matter because of its negative pressure. In this work, we restrict ourselves to the case $c_{\mathrm{eff}}^{2}=0$ for simplicity.

Furthermore, treating dark energy as a perfect fluid implies that its effective sound speed coincides with the adiabatic sound speed, given by
\begin{equation}
c_{\mathrm{a}}^{2}
= w_\phi - \frac{a\,dw_\phi/da}{3(1+w_\phi)} \,\,\text{.}
\label{eq:ca}
\end{equation}
The coupled system of perturbation equations described above plays a crucial role in understanding the growth of cosmic structures and provides a direct link between theoretical models of dark energy and observational probes such as large-scale structure and weak gravitational lensing.

\section{Conclusion}

In this chapter, we discuss the evolution of the understanding of the Universe from our ancestors to modern times. But the proper geometrical representation of the Universe started with Einstein's general theory of relativity. Later in this chapter, we present the fundamental mathematical frameworks, their applications to cosmology, and cosmological observations that help validate cosmological models. The fundamental theory of gravity, like GR, fails to address some fundamental issues of the Universe, and it seems incomplete. Therefore, its modifications and generalizations are more successful at addressing this issue, and we conclude this chapter by reviewing the modified theories of gravity. In the upcoming chapters, some problems are investigated by applying the modified theories of gravity mentioned above.




\chapter{Observational Constraints on Dissipative Chaplygin Gas Cosmology in Coincident $f(Q)$ Gravity} 

 \label{Chapter2}
\lhead{Chapter 2. \emph{Observational Constraints on Dissipative Chaplygin Gas Cosmology in Coincident $f(Q)$ Gravity}} 


\vspace{10 cm}
* The work, in this chapter, is covered by the following publication:

\textit{Observational Constraints on Dissipative Chaplygin Gas Cosmology in the Framework of Coincident $f(Q)$ Gravity}, Chinese Physics C, DOI:10.1088/1674-1137/ae50e5 (2026)

\clearpage

\epigraph{``But if there is no solace in the fruits of our research, there is at least some consolation in the research itself\dots\ The effort to understand the Universe is one of the very few things that lifts human life a little above the level of farce, and gives it some of the grace of tragedy.''}{--- Steven Weinberg, \textit{The First Three Minutes: A Modern View of the Origin of the Universe} (1977), Ch.~1}

In this chapter, we shed light on the unified approach to both dark energy and dark matter via the generalized Chaplygin gas model in STGR. We have taken the equation of state given by the generalized Chaplygin gas, which naturally arises in string theory, tachyonic field theory, and Rundall Sundaram type brane world solutions. We show that such a generalized Chaplygin gas can not just give a lucrative candidate to dark energy but also a viable candidate for dark matter via Bose-Einstein Condensation (BEC). We have also taken the interaction between dark matter and dark energy to give a more realistic point of view. We performed MCMC analysis with combined Hubble and Pantheon datasets. We have also done the Om-diagnostics and $r-s$ plot to comment on the late behavior of our model. We have also found that the values are in Phantom regions, and we have given physical reasons why this is expected. Finally, we give some future directions where our work can be carried out. 

\section{Introduction}\label{sec_2.1}
Since the discovery of late-time acceleration in the late 90s, by Riess et al. \cite{late1} and Perlmutter et al. \cite{late2}, it has been clear that Einstein's cosmological constant has to be non-zero in order to explain the late-time de Sitter type expansion. It has also been seen from observation of galactic halo and structure formation, and BAO data that there is dark matter which is weakly interacting and also cold and non-baryonic in nature. So, from the late 90s to the early 2000s was dominated by $\Lambda$CDM cosmology, that is, the dark energy (responsible for the late time acceleration) is given by the cosmological constant ($\Lambda$) and the dark matter (matter responsible for galactic rotation curve anomaly) is cold and non-baryonic in nature. \\
The standard cosmology model, $\Lambda$CDM, is one of the most successful models that can explain almost all present and past observations with only six free parameters. Although the $\Lambda$CDM paradigm is successful, there are still unanswered questions that make people question standard cosmology. For starters, there is no way to explain $\Lambda$ in the standard cosmological model, although identifying it with dark energy gives remarkable observational tests. For example, if one tries to explain $\Lambda$ (the cosmological constant) as just vacuum energy, one runs into a serious problem: the observed value of $\Lambda$ versus the theoretical prediction using one-loop corrections in QFT shows a discrepancy of order $10^{120}$. In order to explain dark energy, either one has to use some sort of scalar field (a quintessence field), which can explain the origin of $\Lambda$ and hence why the cosmological constant is so small. There is an alternative way to explain dark energy or late-time acceleration via modifying Einstein's relativity. It is well known that standard GR is a classical theory, so it breaks down in extreme-curvature regions such as near singularities. Thus, to avoid this, one must quantize the theory, just like electromagnetism, to find a quantum theory of gravity. However, it is also well known that even after many proposals, such as string theory and loop quantum gravity (just to name a few), a full theory of quantum gravity (which can be tested via experiment or observation) is still out of reach. So, noting that even though we cannot find a full quantum gravity theory, one can still make some reasonable guesses about how the Einstein-Hilbert action would change while maintaining diffeomorphism symmetry. As for CDM, we do not have any clear idea what exactly constitutes dark matter. (In this chapter, we have taken the Chaplygin gas, which can explain both dark energy and dark matter in unified matter) In addition to these, there is also a growing concern regarding contemporary cosmology about the Hubble tension ($H_0$) \cite{Simpson/2025} and $\sigma_8$ \cite{Kazantzidis/2018}. Roughly speaking, these tensions show that our understanding of the full time scale of the Universe is limited; that is, the contemporary observation does not match the early Universe and late Universe, which challenges the current $\Lambda$CDM model in a profound way.\\
It is well understood that one reason for such an anomaly is that the quantum effects become quite important in explaining the cosmological constant or dark matter. However, since we do not have a full quantum gravity theory, we can roughly argue by noting how the infrared behavior of quantum gravity theory might be. It can be seen that one loop correction of the graviton-graviton interaction can give rise to the $R^2$ term in the Einstein-Hilbert action, which has been used by Starobinsky \cite{star} to predict inflation in the early Universe. So, modified gravity is a very lucrative alternative to explain not just dark energy, but it can also be used for explaining various other astrophysical phenomena (such as the super-Chandrasekhar mass limit \cite{Das/2013}, black hole accretion \cite{Pun/2008}, etc.). When it comes to modifying gravity, the usual approach is to modify the Einstein-Hilbert action with a suitable diffeomorphic scale and then find the modified equation by varying the metric. From there, one can put a suitable metric (in this case, the FLRW metric) into the field equation to get the modified Einstein equations.\\
   In this chapter, we have taken $f(Q)$  gravity, a version of modified gravity that is based on non-metricity ($\nabla_{\lambda}g_{\mu\nu}\neq0$) and reduces to GR in the linear limit. Although the actual origin of using non-metricity to unify electromagnetism and gravity goes back to the idea of Weyl, the modern version of $f(Q)$ cosmology was first written by Jimenez et al. \cite{Jimenez/2018} and later expanded by them \cite{Jimenez/JCAP2018, Jimenez/2019}. Since then, in the last half a decade, a large number of articles have been published on modified $f(Q)$ gravity. Some of the researches focus on dynamic system analysis using scalar field in $f(Q)$ gravity site, and also some articles have also claimed to have solved the Hubble and $\sigma_8$ tension. It is also worth noting that the modified $f(Q)$ gravity theory has shown remarkable consistency in various astrophysical objects, such as wormholes \cite{Hassan/2022, Hassan/2023} and gravastars \cite{Mohanty/2024, Mohanty/2025}, as well. In this chapter, we will consider a particular form of $f(Q)$, namely $f(Q)=\gamma\left(\frac{Q}{Q_0}\right)^n$, with $\gamma=-Q_0$ and $ n=1$, which reduces to standard GR. This form of $f(Q)$ is widely used in the literature; in fact, it is the most common form, and we can integrate the equation precisely to find $H(z)$ as a function of $z$ that fits the data.\\
In this chapter, we have taken the Chaplygin gas as our matter source. Chaplygin gas has been one of the most prominent candidates for dark energy for a very long time. There are several reasons for that, as we discuss in Section-\ref{sec_2.2}, such as the microscopically Chaplygin gas that arises very naturally in string theory. There is also a very natural way to construct the Chaplygin gas equation of state using a scalar field and a DBI field, and also by Brane compactification. On the phenomenological side, the Chaplygin gas is not just consistent with the current observation, but it can also give a very natural transition from matter to a dark-energy-dominated Universe. So, here we have taken the most general equation of state of the Chaplygin gas, which is given by $p=\alpha \rho+ \beta\rho^m$. We note that typically one takes $m<0$ to model it as dark energy.\\
There has been a lot of study regarding the interaction (non-gravitational) between dark energy and dark matter over the past few decades. One of the standard quantum field theoretic approaches to this is to note that the quintessence scalar field, whose effective mass becomes comparable to that of dark matter particles, could lead to an exchange of energy between the two components \cite{Amendola/2000, FarrarPeebles/2004}. There are several ways to model this interaction in the continuity equation, but the most simplistic and phenomenologically consistent way to model this interaction is via $  \mathcal{I} \propto H\rho$, as this encodes the FLRW geometry, which is dilution via a factor of $H$, and also is consistent with the standard continuity equation. This model of interaction between dark sectors has been widely used in the standard literature, as seen in references such as \cite{ZimdahlPavon/2007,Bolotin/2015,Wang/2016}, among others. The interaction term relieves the tension between dark matter and dark energy, as the analysis with only BEC dark matter should yield only $m=2$, which has been studied in detail in the following article by Bhat et al. \cite{Bhat/2024}. However, there is also an explicitly dark energy component which is very much consistent with the generalized Chaplygin gas given as $m<0$. The presence of the interaction term therefore provides a unified and physically motivated mechanism through which both branches emerge as viable solutions, with their relative relevance determined by observational constraints rather than imposed assumptions.\\
 As mentioned earlier, even though the Chaplygin gas can explain dark energy very well, it falls short in explaining dark matter. So, in this chapter, we have taken dark matter as an effective Bose-Einstein condensate scale field. There are two primary reasons for this, as explained in Sec.~\ref{sec_2.3}. First, of course, Bose-Einstein condensation is a very lucrative option for dark matter as it does not have any viscosity; so it explains why the galactic halos are of uniform density. It is also very much visible from the microscopic point of view as the Gross-Pitaevskii equation for a moderately interacting dilute Bose gas can naturally give an equation of state of the form $ p=\alpha\rho+\beta\rho^2$, which is basically a Chaplygin gas with $m=2$. So, apart from microscopic and phenomenological motivation, our work gives a unified framework that we can use to explain both dark energy and dark matter. Apart from that, we have also included an interacting term $Q=3H^2\rho$ which can make a transition between dark matter and dark energy.\\
   We have performed the MCMC analysis using Hubble and Pantheon+ SH0ES data and found that the $\Lambda$CDM transition is coming for both $m\approx2,-5$. This gives concrete evidence that even in the $f(Q)$ gravity model, one can consider both dark energy and dark matter in a unified manner using the Chaplygin gas.\\
   After plotting the 3-$\sigma$ contour, we have tested the model using two of the most popular tests, such as AIC and BIC. The result shows that model-B that is the We have also done the so-called null test for $\Lambda$CDM cosmology by statefinder and $Om$ diagnostics. Both show that our model eventually converges to the $\Lambda$CDM model through a phantom region. This is very natural in the Chaplygin gas models, which reproduce phantom behavior. 

\section{Chaplygin gas in cosmology}\label{sec_2.2}
Chaplygin gas was first proposed as a dark energy candidate by Kamenshchik et al. \cite{Kamenshchik/2001} with the relation $p\propto \rho^{-1}$, which can smoothly make a transition from matter-dominated ($\omega=0$) to dark energy-dominated epochs ($\omega=-1$). Later, it was generalized by Bento et al. \cite{Bento/2002} with the relation $p\propto \rho^{-\alpha}$ where $\alpha$ is not necessarily one. Both the Chaplygin and the generalized Chaplygin gas models give remarkable constant solutions that can explain the late-time acceleration and are also consistent with the observations. Even though the Chaplygin gas has been mainly utilized as a candidate for dark energy, it appears very naturally in quantum gravity theories, such as in the context of Brane stabilization for AdS-Schwarzschild black holes to study the critical horizon \cite{Kamenshchik/2000} or in the study of 2+1 black holes in string theory \cite{Kama/1998}. Apart from string theory, there are other conventional ways to obtain the Chaplygin equation of state, for example, following the DBI field equation \cite{Novello/2005} or scalar fields \cite{Benaoum/2022} or even via the Brane world scenario \cite{Kamenshchik/2000}.\\
 Even though the Chaplygin gas is a very natural candidate for dark energy because it not only gives matter-dominated and dark-energy-dominated solutions, but also smoothly makes a transition between these two regions, although there are other detailed studies that show that the Chaplygin gas is not just a good candidate for phenomenology but is also extremely consistent with the observational datasets. For example, one can look at the review article \cite{Carturan/2003} for a full analysis with an observational test. Note that the consistency with dark energy using type IA supernova has been done in \cite{Bertolami/2004,Makler/2003}, and the consistency using the CMB anisotropy data has also been in favor of the Chaplygin gas theory \cite{Bento/2003}. There are also gravitational lensing data \cite{Dev/2003}, age measurement of high-redshift object data \cite{Alcaniz/2003}, and last but not least, the X-ray gas mass fraction of clusters data \cite{Cunha/2004}, which shows a good agreement with the Chaplygin gas model as dark energy.

Here, we give a very short derivation of the scalar field, which can be responsible for the Chaplygin gas under the influence of a scalar field (FLRW metric).\\
It is well known that the Lagrangian density of a scalar field can be written like this,
\begin{equation} \label{eq_2.1} 
    L(\phi)=-\frac{1}{2} g^{\mu\nu}\partial_\mu \phi\partial_\nu \phi - V(\phi) \,\,\text{.}
\end{equation}
The stress-energy tensor is given by,
\begin{equation}
    T^\phi_{\mu\nu}= \partial_\mu \phi\partial_\nu\phi-\frac{1}{2} g_{\mu\nu}g_{\alpha\beta}\partial^\alpha\phi\partial^\beta\phi-g_{\mu\nu}V(\phi) \,\,\text{.}
\end{equation}
Also, the variation of the Lagrangian with respect to $\phi$ would give the Klein-Gordon equation, given as,
\begin{equation}
    \Box\phi-V_{,\phi}(\phi)=0 \,\,\text{.}
\end{equation}
where $V_{,\phi}=\frac{\partial V}{\partial \phi}$.

If one tries to calculate the stress-energy tensor ($T_{\mu\nu}$) with such a Lagrangian and equates with $\rho_{\phi}$ and $p_{\phi}$ as a fluid analogue, one can get an expression for density and pressure on a FLRW background with metric $ds^2=dt^2-a^2(t)dl^2$, where $dl^2$ is the usual spherically symmetric line element. Then we can get the expressions of $\rho_{\phi}$ and $p_{\phi}$ as follows,

\begin{equation} \label{eq_2.4} 
\rho_{\phi}=\frac{1}{2} {\dot{\phi}}^2 +V(\phi)=\sqrt{A+\frac{B}{a^6}}  \,\,\text{.}
\end{equation}
and,
\begin{equation} \label{eq_2.5} 
    p_{\phi}=\frac{1}{2} {\dot{\phi}}^2-V(\phi)=-\frac{A}{\sqrt{A+\frac{B}{a^6}}} \,\,\text{.}
\end{equation}
From Eq.~(\ref{eq_2.4}) and (\ref{eq_2.5}) we get,
\begin{equation} \label{eq_2.6} 
    {\dot{\phi}}^2=\frac{B}{a^6 \sqrt{A+\frac{B}{a^6}}} \,\,\text{.}
\end{equation}
and, 
\begin{equation} \label{eq_2.7} 
    V(\phi)= \frac{2a^6(A+\frac{B}{a^6})-B}{2a^6 \sqrt{A+\frac{B}{a^6}}} \,\,\text{.}
\end{equation}
We also note that taking the derivative with respect to $a$ we get,
\begin{equation}\label{eq_2.8} 
    \phi'=\frac{\sqrt{B}}{a(Aa^6+B)^{\frac{1}{2}}} \,\,\text{.}
\end{equation}
where $``\prime"$ denotes the derivative with respect to $a$.\\
One can integrate the above expression and get,
\begin{equation}\label{eq_2.9} 
    a^6 =\frac{4B \exp(6 \phi)}{A(1-\exp(6\phi))^2} \,\,\text{.}
\end{equation}
Putting in the expression Eq.~(\ref{eq_2.7}), we get the expression for $V(\phi)$ as follows,
\begin{equation} \label{eq_2.10} 
    V(\phi)=\frac{1}{2} \sqrt{A}\left(\cosh{3\phi}+\frac{1}{\cosh{3\phi}}\right) \,\,\text{.}
\end{equation}
We have shown above that a potential scalar field is given by $ V(\phi)=\frac{1}{2} \sqrt{A}\left(\cosh{3\phi}+\frac{1}{\cosh{3\phi}}\right)$ (here $B$ is just an integration constant which is obtained by solving the Friedmann equation).\\
Now we note how one can recover the Chaplygin gas equation of state from the Tachyons.\\
We first note that action in Tachyon-based action or (DBI field action) can be written like,

\begin{equation}\label{eq_2.11} 
    S= - \int d^4 x \sqrt{-g} V(T) \sqrt{1-g^{\mu \nu} T_{,\mu} T_{, \nu}} \,\,\text{.}
\end{equation}
We will write the Lagrangian density as follows,
\begin{equation}\label{eq_2.12} 
    L=-V(T) \sqrt{1-\dot{T}^2} \,\,\text{.}
\end{equation}
Here, we have defined $\dot{T}^2=g^{\mu \nu} T_{,\mu} T_{, \nu}$ the kinetic term of the Lagrangian.\\
Like in the previous case, if we calculate the stress energy tensor ($T_{\mu\nu}$) and equate it with the fluid stress energy tensor, we can get an expression for $\rho_{Tac}$ and $p_{Tac}$ as follows,
\begin{equation} \label{eq_2.13} 
    \rho_{Tac} = \frac{V(T)}{\sqrt{1-\dot{T}^2}} \,\,\text{,}
\end{equation}
and,
\begin{equation}\label{eq_2.14} 
p_{Tac}=-V(T)\sqrt{1-\dot{T}^2} \,\,\text{.}
\end{equation}







From the previous expressions of $\rho_{Tac}, p_{Tac}$ using the Friedmann equations, one can derive the potential as \cite{Benaoum/2022},
\begin{equation} \label{eq_2.15} 
V(T) = \frac{\Lambda}{\sin^{2}\!\left(\tfrac{3\sqrt{\Lambda(1+k)}T}{2}\right)} 
\left( \sqrt{1 - (1+k)\cos^{2}\!\left(\tfrac{3\sqrt{\Lambda(1+k)}T}{2}\right)} \right) \,\,\text{.}
\end{equation}
Also, the kinetic energy term is given by,
\begin{equation}\label{eq_2.16} 
    T(t)= \frac{2}{3\sqrt{\lambda(1+k)}} tan^{-1} \left(\sinh \frac{3\sqrt{\Lambda}(1+k)t}{2}\right) \,\,\text{.}
\end{equation}
which is obtained by integrating \(\dot T = \sqrt{1+p/\rho}\) with the Chaplygin solution for the scale factor \(a(t)\propto \sinh^{2/3(1+k)}\!\big(\tfrac{3\Lambda(1+k)t}{2}\big)\).

Similarly, the generalized Chaplygin gas equation of state can be obtained from Tachyon reconstruction by the same procedure (solve continuity $\to$ obtain \(\rho(a)\) $\to$ \(a(t)\) $\to$ \(\rho(t)\) $\to$ reconstruct \(V(T)\)).

Now, we move to the generalized Chaplygin gas equation of state and show that it can also be derived from a scalar field in an FLRW background.\\
We note that the general equation of state for the generalized Chaplygin gas is given as follows,
\begin{equation} \label{eq_2.17} 
    p= A \rho- \frac{B}{\rho^\chi} \,\,\text{.}
\end{equation}
Substituting into the continuity equation \(\dot\rho+3H(\rho+p)=0\) gives
\[
\dot\rho + 3H\big((1+A)\rho - B\rho^{-\chi}\big)=0 \,\,\text{,}
\]
or, in scale-factor variables,
\[
a\frac{d\rho}{da} + 3\big((1+A)\rho - B\rho^{-\chi}\big)=0 \,\,\text{.}
\]
This can be solved by the substitution \(u=\rho^{\chi+1}\), yielding the solution
\[
\rho(a)=\left(\frac{B}{1+A} + C a^{-3(1+A)(1+\chi)}\right)^{\tfrac{1}{1+\chi}} \,\,\text{,}
\]
where \(C\) is an integration constant.
If we take the stress-energy tensor and equate it with fluid, we can obtain the expression for $U(\varphi)$ as follows \cite{Benaoum/2022},
\begin{equation} \label{eq_2.18} 
U(\varphi) = \frac{1}{2} \left(\frac{B}{A+1} \right)^{\tfrac{1}{\chi+1}}
\Bigg[
    \frac{1+A}{\cosh^{\tfrac{2\chi}{\chi+1}}\!\big(\varpi(\chi+1)\Delta \varphi\big)} + (1-A)\cosh^{\tfrac{2}{\chi+1}}\!\big(\varpi(\chi+1)\Delta \varphi\big)
\Bigg] \,\,\text{.}
\end{equation}
Where $\Delta\varphi=\varphi-\varphi_0$(where $\varphi_0$ is an integration constant), $\varpi=\sqrt{\frac{(D-1)(A+1)}{2(D-2)}}$, (in our case $D=4$).
Similarly, one can show that the generalized Chaplygin gas equation of state \cite{Benaoum/2022} can also be derived from the Tachyon field.\\
We also note that from the Randall-Sundrum model \cite{Randall/1999}, where a dimension is wrapped, the metric can be written as,
\begin{equation} \label{eq_2.19} 
    ds^2=e^{\frac{-2|y|}{l}} (dt^2-{dx_1}^2-{dx_2}^2-{dx_3}^2)-dy^2 \,\,\text{.}
\end{equation}  
where $y$ is the coordinate for the extra dimension and $l$ is often called the ``wrapping factor". We also note that at $y=0$, there is a singularity (the derivative is discontinuous), which implies a brane-like object embedded in the fifth dimension. 

The 5D action is
\[
S_5=\int d^5x\,\sqrt{-g_5}\left(\frac{1}{2\kappa_5^2}R_5 - \Lambda_5\right) + S_{\rm brane} \,\,\text{,}
\]
and imposing \(Z_2\) symmetry and using the Israel junction conditions \cite{Israel:1966} relates jumps in the extrinsic curvature to the brane stress tensor. Projecting the 5D equations onto the 4D brane yields modified Friedmann equations \cite{Shiromizu:2000} of the form,
\[
H^2 = \frac{\rho}{3}\Big(1+\frac{\rho}{2\lambda}\Big) + \frac{\Lambda_4}{3} - \frac{C}{a^4} \,\,\text{,}
\]
where \(\lambda\) is the tension of the brane, \(\Lambda_4\) the effective 4D cosmological constant, and \(C\) an integration constant (dark radiation).

One can show that for such a metric, one can get the equation of state as \cite{Kamenshchik/2000},
\begin{equation}\label{eq_2.20} 
    p=\frac{-(n-1)\rho}{n}-\frac{4n}{\rho l^2} \,\,\text{.}
\end{equation}
Here, $n$ denotes the dimension of branes. We have assumed that locally the manifold (orbifold) has $R\times S^n$ topology (Chaplygin gas-dominated anisotropic brane world cosmological models).\\ 
So, as we can see, the Randall-Sundrum model can naturally give rise to the generalized Chaplygin gas equations.\\
We would like to note that for the remainder of our chapter, we have taken the Chaplygin gas equation of state as $p=\alpha\rho+\beta\rho^m$. Here, we have used $A,B,\chi$ in order to be consistent with the literature; however, since these are used, for example, $\chi^2$ and models A and B, we have taken the equation of state to be $p=\alpha\rho+\beta\rho^m$. Also, the purpose of this section is to show that one can not just get the Chaplygin gas equation of state from a scalar field, but one can completely reverse equations 13 or 16 to reconstruct the appropriate scalar field potential or DBI field potential from the known parameters of the Chaplygin gas equation.

\section{Dark matter from BEC}\label{sec_2.3}
Bose-Einstein condensation is an effect of Bose-Einstein statistics that was predicted by Einstein. The idea is that because Boson wave functions are symmetric under interchange, there is a possibility of a large degeneracy at the lowest temperature; such a degeneracy could give rise to condensation at very low temperatures. It was first experimentally verified by Bradley et al. \cite{Bradley/1995}. Even though BEC is a very important concept in condensed matter physics to give a partial explanation for superfluidity and superconductivity, it has been shown that BEC could be used as an effective theory for dark matter due to its noninteractive nature.\\
 Although there are many papers that have conjectured BEC as a dark-matter candidate, it was Boehmer and Harko \cite{Boehmer/2007} who gave a solid reasoning why BEC should be taken seriously as a dark-matter candidate. The argument is that since the critical temperature for BEC to occur is very close to the interstellar medium, the scalar fields responsible for dark matter could also condense. This could naturally explain the uniformity of the dark-matter halo that is observed in the galactic center. Later, Harko \cite{Harko/2011} expanded this idea, noting how the early Universe temperature dependence could make the scalar field condense to BEC. The above paper also discussed how the scalar field that satisfies the Gross-Pitaevskii equation for a weakly interacting Bose gas could give rise to superfluidity and, in principle, could explain the uniform density of the galactic halo. In the later paper by Harko and Lobo \cite{Harko/2015}, they have expanded the analysis to include gravitational lensing analysis and have shown that it is remarkably consistent with the current observation.\\
  We would also like to note that it can be shown that a similar type of treatment can be carried out for other models of dark energies as well, such as dark matter using the Lagrangian having global symmetry \cite{Mambrini/2016} or extra dimension (brane world) responsible for dark matter \cite{Hooper/2007} or primordial black hole as a candidate for dark matter \cite{Mishra/2020}. It has also been shown that the axion dark matter \cite{Odintsov/2019} also gives a similar equation of state after condensation. Das et al. \cite{Das/2018,Das/2023} have shown that the quantum potential can be used to obtain a similar equation of state. \\
Finally, last but not least, the BEC equation of state was used by Mahichi et al. \cite{Mahichi/2021,Mahichi/2022,Mahichi/2023} in various modified gravity scenarios, such as $f(T,B)$ gravity or Gauss-Bonnet gravity.\\
Here, we will give a brief outline about how one can tackle weakly interacting gravitationally bound boson particles, as discussed in the work of Harko \cite{Harko/2011}. One first note is that Bose-Einstein condensation occurs at very low temperatures and when quantum scales are important. More precisely, the weakly interacting particles would go under a Bose-Einstein condensation provided that this condition is satisfied, that is, $T_{\text{cr}} \approx 2\pi \times \hbar^2 \rho^{2/3}/ m^{5/3} k_B$, such conditions are physically possible during the structure formation time in cosmology. Now, in order to include the weak interactions between bosons, we take the potential form as $V(r'-r)=\lambda\delta(r'-r)$, where $\lambda$ is a length scale associated with the interaction and is dimensionally consistent. Now, it can be shown that the functional energy for such weakly interacting bosons can be written by the Gross-Pitaevskii formula given as,
\begin{equation}
E[\psi] = \int \left[ \frac{\hbar^2}{2m_\chi} |\nabla \psi(\vec{r})|^2 + \frac{U_0}{2} |\psi(\vec{r})|^4 \right] d\vec{r}
- \frac{1}{2} Gm_\chi^2 \int \int \frac{|\psi(\vec{r})|^2 |\psi(\vec{r}')|^2}{|\vec{r} - \vec{r}'|} d\vec{r} d\vec{r}' \,\,\text{.}
\end{equation}
Here, $\psi$ is closely analogous to the entire wave function for the many-body system. So, following this, one can define the mass density as,
\begin{equation}
\rho_\chi(\vec{r}) = m_\chi |\psi(\vec{r})|^2 = m_\chi \rho(\vec{r}) \,\,\text{.}
\end{equation}
In addition, the total number of dark matter particles can be given as $N=\int |\psi(\vec{r})|^2 d^3x$. From the above energy functional, one can use the variational principle to reach the Schrodinger-like equation for $\psi$ for weakly interacting Bose particles under gravitational attraction as follows,
\begin{equation}
- \frac{\hbar^2}{2m_\chi} \nabla^2 \psi(\vec{r}) + m_\chi V(\vec{r}) \psi(\vec{r}) + U_0 |\psi(\vec{r})|^2 \psi(\vec{r}) = \mu \psi(\vec{r}) \,\,\text{.}
\end{equation}
Finally, in order to obtain the desired equation of state one has to choose the trial wave function, the plane polar coordinate as $\psi(\vec{r}, t) = \sqrt{\rho(\vec{r}, t)} \exp \left[ \frac{i}{\hbar} S(\vec{r}, t) \right]$, just like the Grizberg-Landau ansatz to obtain the uncertainty relation in phase or chemical potential.


Substituting the Madelung ansatz $\psi=\sqrt{\rho/m_\chi}e^{iS/\hbar}$ into the Gross-Pitaevskii equation and separating the real and imaginary components yields the continuity equation and the quantum Euler equation. The latter governs the momentum density and is expressed as
\begin{equation*}
\frac{\partial\mathbf v}{\partial t}+(\mathbf v\cdot\nabla)\mathbf v = -\nabla\Phi - \frac{1}{\rho}\nabla P_{\rm int} + \nabla Q \,\,\text{,}
\end{equation*}
where $\mathbf v=\nabla S/m_\chi$ represents the velocity of the fluid and $Q = \frac{\hbar^2}{2m_\chi^2}\frac{\nabla^2\sqrt{\rho}}{\sqrt{\rho}}$ is the quantum potential. The identification of the interaction-pressure gradient $P_{\rm int}$ follows from the mean-field nonlinear term via the relation $\frac{1}{\rho}\nabla P_{\rm int} = \frac{1}{m_\chi}\nabla(U_0 |\psi|^2)$. By substituting the density definition $\rho = m_\chi |\psi|^2$, the gradient term becomes $\frac{U_0}{m_\chi^2}\nabla\rho$. The quadratic density dependence $\rho^2$ of the pressure arises directly and unambiguously from the integration of this term with respect to $\rho$. Specifically, we find $\nabla P_{\rm int} = \frac{U_0}{m_\chi^2} \rho \nabla \rho$, which integrates into $P_{\rm int} = \frac{U_0}{2m_\chi^2}\rho^2 \equiv \beta\rho^2$. Employing the standard contact coupling $U_0 = 4\pi\hbar^2 a/m_\chi$, the interaction coefficient is defined as $\beta = 2\pi\hbar^2 a/m_\chi^3$, where the factor of $1/2$ is a necessary consequence of the fluid-dynamical mapping of two-body interactions.

To incorporate residual finite-temperature velocity dispersion alongside these quantum self-interactions, we adopt an Extended Bose-Einstein Condensate (EBEC) equation of state of the form $p = \alpha\rho + \beta\rho^2$. In this construction, the linear term $\alpha\rho$ (with $\alpha = k_B T/m_\chi$) accounts for thermal effects, while the quadratic term $\beta\rho^2$ reflects repulsive two-body interactions. This structure ensures dynamical stability against gravitational collapse, as the adiabatic sound speed $c_s^2 = \partial p/\partial \rho = \alpha + 2\beta\rho$ remains non-zero even at high central densities. Following the standard Jeans criterion, the scale of structure suppression is determined by the Jeans length
\begin{equation*}
\lambda_J = \frac{\pi c_s}{\sqrt{G\rho}} = \pi \sqrt{\frac{\alpha + 2\beta\rho}{G\rho}} \,\,\text{.}
\end{equation*}
This generalized model is highly versatile, recovering pressureless CDM in the limit $\alpha = \beta = 0$, and ordinary thermal dark matter with linear pressure when $\beta = 0$. In the zero-temperature limit where $\alpha = 0$, the model isolates the pure BEC contribution, where the purely two-body interaction pressure $p = \beta\rho^2$ typically yields the core density profiles observed in dark matter halo cores.


In this analysis, we consider dark matter as a bosonic substance whose number density follows Bose-Einstein statistics. This statistical behavior suggests that such particles emerged from the thermal decoupling of the Early-Universe plasma. The energy density of standard bosonic dark matter is defined by the product of its number density and particle mass. Its pressure, according to Bose-Einstein statistics, can be described by a sphere defined by the momentum radius of the particles. Under these assumptions, the pressure of normal dark matter is found to vary linearly with the energy density, leading to the equation of state (EoS),
\[
p = \alpha \rho \,\,\text{,}
\]
where $\alpha$ denotes the proportionality constant associated with single-body interactions in the dark-matter medium.\\
We now extend this description to a BEC form of dark matter, comprising non-relativistic bosons undergoing two-body interactions at extremely low temperatures. As the temperature approaches absolute zero, quantum effects dominate and individual-particle wave functions overlap, leading to condensation into a single quantum state. The dynamics of such a condensate is governed by the Gross-Pitaevskii equation. Within a gravitational context, the pressure of BEC dark matter is shown to follow a quadratic dependence on energy density,
\[
p = \beta \rho^2 \,\,\text{,}
\]
where $\beta$ encapsulates the interaction strength through the mass of the particles and the length of the scattering. To incorporate both conventional and BEC-like behavior, an extended form of dark matter EoS, referred to as Extended Bose-Einstein Condensate (EBEC), is proposed by the following equation,
\[
p = \alpha \rho + \beta \rho^2 \,\,\text{,}
\]
here, the linear term $\alpha \rho$ accounts for single-particle effects, while the nonlinear term $\beta \rho^2$ reflects two-body interactions. Special cases of this model recover various dark matter scenarios: $\alpha = \beta = 0$ corresponds to cold dark matter, $\beta = 0$ recovers normal dark matter, and $\alpha = 0$ isolates the BEC contribution from dark matter halos.

\section{An uniform framework for dark energy and dark matter}\label{sec_2.4}

In this work, we have used the generalized Chaplygin gas model to describe both dark energy and dark matter with a single equation of state, $p=\alpha \rho+\beta \rho^m$. As we have seen from previous sections, for $m<0$ we get the ``pure" Chaplygin gas, which is responsible for the dark energy, and for $m=2$ we will get $p=\alpha \rho + \beta \rho^2$, which is just the equation of state for BEC dark matter. At this point, we bifurcate the models; that is, we have one model in which we obtain $m\approx-5$, which is very consistent with the Chaplygin gas as dark energy. However, we also obtain $ma\approx2$ for another MCMC run with different priors. This is extremely consistent with the idea that this can be a BEC dark-matter equation of state. This also gives us the idea that we need to add an interacting term between dark matter and dark energy, such as $\mathcal{Q}$, which can take care of the fact that when $\mathcal{Q}>0$ there is a positive energy transfer from dark matter to dark energy and vise versa for $\mathcal{Q}<0$. In order to fix a form of $\mathcal{Q}$, we note that from the dimension analysis, there is a natural form in terms of $[\rho][T^{-1}]$, so the most obvious choice is $\mathcal{Q} \propto \rho H$. So, following convention, we take the proportional constant to be $3b^2$ ($b^2$ to make the expression positive). It should also be noted that there are other alternatives to take the interacting term, also discussed in \cite{Setare/2009,Chimento/2003,Guo/2005}.\\
We also would like to note that we have taken the model for $f(Q)$ gravity as $f(Q)=\gamma\left(\frac{Q}{Q_0}\right)^n$. As is well known in our convention, $f(Q)$ gravity reduces the Einstein GR when $f(Q)=-Q$ so our model reduces to GR when $\gamma=-Q_0$ and $n=1$. In the MCMC analysis, we can see that there is a significant deviation of $n$ from 1, which hints that the observation-wise model is only consistent when gravity is ``modified".


\section{Formulation of $f(Q)$ gravity}\label{sec_2.5}
Even though in the original formulation of GR, Einstein used the Levi-Civita connection to construct the field equation, it was soon clear that one can (uniquely) split the most general connection on a tangent bundle over a manifold into three different parts: Levi-Civita, anti-symmetric, and non-metricity. The proof and motivation for this can be found in the review article by Heisenberg \cite{Heisenberg/2024}. So in principle, the most general affine connection can be written in the form \cite{Jimenez/2019},


\begin{equation}\label{eq_2.24} 
\Upsilon^\alpha_{\ \mu\nu}=\Gamma^\alpha_{\ \mu\nu}+K^\alpha_{\ \mu\nu}+L^\alpha_{\ \mu\nu} \,\,\text{.}
\end{equation}
Here, the first term $\Gamma^\alpha_{\mu\nu}$ denotes the usual Levi-Civita connection,
\begin{equation}\label{eq_2.25} 
\Gamma^\alpha_{\ \mu\nu}\equiv\frac{1}{2}g^{\alpha\lambda}(g_{\mu\lambda,\nu}+g_{\lambda\nu,\mu}-g_{\mu\nu,\lambda}) \,\,\text{.}
\end{equation}
The second term $K^\alpha_{\ \mu\nu}$ is known as the contortion tensor. The formula can be written in the form of a torsion tensor ($T^\alpha_{\ \mu\nu}\equiv \Upsilon^\alpha_{\ \mu\nu}-\Upsilon^\alpha_{\ \nu\mu}$) as follows,
\begin{equation}\label{eq_2.26} 
K^\alpha_{\ \mu\nu}\equiv\frac{1}{2}(T^{\alpha}_{\ \mu\nu}+T_{\mu \ \nu}^{\ \alpha}+T_{\nu \ \mu}^{\ \alpha}) \,\,\text{.}
\end{equation}
Finally, the last term is known as the distortion tensor, which is the most relevant for our current chapter. The formula given in the form of the non-metricity tensor is given as follows,
\begin{equation}\label{eq_2.27} 
L^\alpha_{\ \mu\nu}\equiv\frac{1}{2}(Q^{\alpha}_{\ \mu\nu}-Q_{\mu \ \nu}^{\ \alpha}-Q_{\nu \ \mu}^{\ \alpha}) \,\,\text{.}
\end{equation}	
The expression of the non-metricity tensor is given as,
\begin{equation}\label{eq_2.28} 
Q_{\alpha\mu\nu}\equiv\nabla_\alpha g_{\mu\nu} = \partial_\alpha g_{\mu\nu}-\Upsilon^\beta_{\,\,\,\alpha \mu}g_{\beta \nu}-\Upsilon^\beta_{\,\,\,\alpha \nu}g_{\mu \beta} \,\,\text{.}
\end{equation} 
We can also define the superpotential tensor as follows,
\begin{equation}\label{eq_2.29} 
4P^\lambda\:_{\mu\nu} = -Q^\lambda\:_{\mu\nu} + 2Q_{(\mu}\:^\lambda\:_{\nu)} + (Q^\lambda - \tilde{Q}^\lambda) g_{\mu\nu} - \delta^\lambda_{(\mu}Q_{\nu)} \,\,\text{.}
\end{equation}
where $Q_\alpha = Q_\alpha\:^\mu\:_\mu $ and $ \tilde{Q}_\alpha = Q^\mu\:_{\alpha\mu} $ are non-metricity vectors. If one contracts the non-metricity tensor with the superpotential tensor, one can get the non-metricity scalar ($Q$) as follows,
\begin{equation}\label{eq_2.30} 
Q = -Q_{\lambda\mu\nu}P^{\lambda\mu\nu} \,\,\text{.}
\end{equation}
It can be shown that the Riemann curvature tensor is given as follows,
\begin{equation}\label{eq_2.31} 
R^\alpha_{\: \beta\mu\nu} = 2\partial_{[\mu} \Upsilon^\alpha_{\: \nu]\beta} + 2\Upsilon^\alpha_{\: [\mu \mid \lambda \mid}\Upsilon^\lambda_{\nu]\beta} \,\,\text{.}
\end{equation} 
Now, using the affine connection Eq.~\eqref{2a}, one can have,
\begin{equation}\label{eq_2.32} 
R^\alpha_{\: \beta\mu\nu} = \mathring{R}^\alpha_{\: \beta\mu\nu} + \mathring{\nabla}_\mu X^\alpha_{\: \nu \beta} - \mathring{\nabla}_\nu X^\alpha_{\: \mu \beta} + X^\alpha_{\: \mu\rho} X^\rho_{\: \nu\beta} - X^\alpha_{\: \nu \rho} X^\rho_{\: \mu\beta} \,\,\text{.}
\end{equation}
Here, $\mathring{R}^\alpha_{\: \beta\mu\nu}$ and $\mathring{\nabla}$ are described in terms of the Levi-Civita connection Eq.~\eqref{2b}. We also note that $X^\alpha_{\ \mu\nu}=K^\alpha_{\ \mu\nu}+L^\alpha_{\ \mu\nu}$. If we use the contraction on the Riemann curvature tensor using the torsion-free constraint $ T^\alpha_{\ \mu\nu}=0$ in the equation Eq.~\eqref{2i}, we get the following,
\begin{equation}\label{2j}
R=\mathring{R}-Q + \mathring{\nabla}_\alpha \left(Q^\alpha-\tilde{Q}^\alpha \right) \,\,\text{.}
\end{equation}
Here, $\mathring{R}$ is the usual Ricci scalar evaluated with respect to the Levi-Civita connection. We also use the teleparallel constraint (the choice for such a gauge is given in detail in \cite{Heisenberg/2024}), i.e., $R=0$. Using the teleparallel constraint, the relation Eq.~\eqref{2j} becomes,
\begin{equation}\label{2k}
\mathring{R}=Q - \mathring{\nabla}_\alpha \left(Q^\alpha-\tilde{Q}^\alpha \right) \,\,\text{.}  
\end{equation}
From the above equation Eq.~\eqref{2k}, it can be seen that the Ricci scalar (using the Levi-Civita connection) differs from the non-metricity scalar ($Q$) by a total derivative. Using a generalized Stokes theorem, one can transform this total derivative into a boundary term. So we can see that the Lagrangian density changes by a boundary term, and as far as the action is concerned, $Q$ is equivalent to $\mathring{R}$. As we can see, $Q$ gives a comparable description of GR with curvature. We also note that since we have taken the torsion to be zero, the theory is known as STEGR \cite{KUHN}.\\
Now, we offer a general form of the STEGR theory in the presence of matter using a general form of $f(Q)$ in the action as follows,
\begin{equation}\label{2l2}
\mathcal{S}=\int\frac{1}{2}\,f(Q)\sqrt{-g}\,d^4x+\int \mathcal{L}_{m}\,\sqrt{-g}\,d^4x\, \,\,\text{.}
\end{equation}
where $g=\text{det}(g_{\mu\nu})$, $f(Q)$ is a function of the non-metricity scalar $Q$.
\section{Cosmology with Chaplygin gas in $f(Q)$ gravity}\label{sec_2.6}
In this chapter, we first take $f(Q)$ as,
\begin{equation}
    f(Q)=\gamma\left(\frac{Q}{Q_0}\right)^n \,\,\text{.}
\end{equation}
So we can rewrite the Friedmann equations Eq.~\eqref{3k}-\eqref{3l} as follows,
\begin{equation}\label{3m}
    3H^2= \rho + \rho_{de} \,\,\text{,}
\end{equation}
\begin{equation}\label{3n}
    \dot{H}=-\frac{1}{2} [\rho + p+\rho_{de}+p_{de}] \,\,\text{.}
\end{equation}
and the $\rho_{de}$ and $p_{de}$ become,
\begin{equation}
    \rho_{de}=\frac{1}{2}(Q-f)+Qf_Q \,\,\text{,}
\end{equation}
\begin{equation}
    p_{de}= -\rho_{de}-2\dot{H}(1+f_Q+2Qf_{QQ}) \,\,\text{.}
\end{equation}
We note that the continuity equation holds as a full set of $\rho$ and $p$.\\
That is,
\begin{equation}
(\dot{\rho}+\dot{\rho}_{de})+3H(\rho+p+\rho_{de}+p_{de})=0 \,\,\text{.}
\end{equation}
However, in the quintessence scenario, we can separate this equation by noting that there may be an interacting term.\\
The modified conservation equation with interacting terms becomes,
\begin{equation}
(\dot{\rho}_{de})+3H(\rho_{de}+p_{de})=\mathcal{Q} \,\,\text{,}
\end{equation}
\begin{equation}
(\dot{\rho})+3H(\rho+p)=-\mathcal{Q} \,\,\text{.}
\end{equation}
We note that $\mathcal{Q}=3b^2H\rho$ \cite{Setare/2009,Chimento/2003,Guo/2005} is an interacting term motivating form quintessence. Here $\rho_{de}$ and $p_{de}$ are the quantities arising due to $f(Q)$ geometry, and $\rho_{int}$  and $p_{int}$ account for the energy exchange arising due to interaction of dark energy and dark matter. Observe that, in the present framework, the effective dark energy arises from contributions of both the geometric sector through the $f(Q)$ gravity function and the generalized Chaplygin gas matter sector. Therefore, the late-time acceleration in our model is not solely geometric in origin, but rather results from a combined contribution of modified gravity and the exotic matter sector. The interaction term further couples these two effective components, leading to richer cosmological dynamics. Also note that the continuity equation holds as a full set of $\rho$ and $p$ as follows,\\
\begin{equation}
(\dot{\rho}+\dot{\rho}_{de}^{eff})+3H(\rho+p+\rho_{de}^{eff}+p_{de}^{eff})=0
\end{equation}
We take the form of $f(Q)$ as a monomial, that is,
\begin{equation}
    f(Q)=\gamma\left(\frac{Q}{Q_0}\right)^n,
\end{equation}
and from the first Friedman equation, we get the following,
\begin{equation}
    \rho=\frac{1-2n}{2}\left(\frac{H}{H_0}\right)^{2n} \,\,\text{.}
\end{equation}
Taking $\rho_0$ as present $\rho$, we get,
\begin{equation}
    \rho=\rho_0\left(\frac{H}{H_0}\right)^{2n} \,\,\text{.}
\end{equation}
Now, taking the generalized Chaplygin gas, we get the density as,
\begin{equation}
    \rho=\rho_0\left(\frac{c\eta a_0^{3\eta(m-1)}-\beta}{c\eta a^{3\eta(m-1)}-\beta}\right)^{\frac{1}{(m-1)}} \,\,\text{.}
\end{equation}
As a consequence, we get the scale factor as follows.
\begin{equation}
    H=H_0\left(\frac{c\eta a_0^{3\eta(m-1)}-\beta}{c\eta a^{3\eta(m-1)}-\beta}\right)^{\frac{1}{2n(m-1)}} \,\,\text{.}
\end{equation}

\section{Statistical analysis and datasets}\label{sec_2.7}

In this section, we compare the predictions of the theoretical model with observational data with the aim of constraining the free parameters. We used a joint sample consisting of 31 cosmic chronometer measurements and the Pantheon+SH0ES compilation of 1701 Type~Ia supernovae. Bayesian statistical methods are employed, using a likelihood-based approach and MCMC sampling, to estimate the posterior distributions of the parameters.

\subsection{Cosmic Chronometers (CC)}\label{2.7.1}

Cosmic chronometers are massive, passively evolving galaxies whose star formation has ceased \cite{Moresco/2016}. The age differences of these galaxies at nearby redshifts provide a direct estimate of the Hubble parameter $H(z)$ through,
\begin{equation}
    H(z) = -\frac{1}{1+z}\,\frac{dz}{dt} \,\,\text{.}
\end{equation}
We used 31 independent $H(z)$ measurements in the range $0.07 \leq z \leq 2.41$ \cite{Solanki2021,Solanki2022}. For each measurement $k$, with observed value $H_{\text{obs},k}$, uncertainty $\sigma_{H,k}$, and model prediction $H_{\text{th}}(z_k)$, the chi-square is,
\begin{equation}
    \chi^2_{\text{CC}} = \sum_{k=1}^{31}\frac{\left[H_{\text{th}}(z_k)-H_{\text{obs},k}\right]^2}{\sigma_{H,k}^2} \,\,\text{.}
\end{equation}

\subsection{Pantheon+SH0ES Supernovae (SN)}\label{2.7.2}

The Pantheon+SH0ES compilation contains $1701$ Type~Ia supernovae that span redshifts $0.001 \leq z \leq 2.3$. SNe Ia act as standardizable candles via the distance modulus,
\begin{equation}
    \mu^{\text{th}}(z) = 5\log_{10}\!\left(\frac{D_L(z)}{\text{Mpc}}\right)+25 \,\,\text{.}
\end{equation}
where the luminosity distance is given by,
\begin{equation}
    D_L(z) = c(1+z)\int_0^z \frac{dx}{H(x,\theta)} \,\,\text{.}
\end{equation}
During the past two decades, several Type Ia supernova compilations have been developed, beginning with the Union sample~\cite{Kowalski/2008}, followed by the Union2~\cite{Amanullah/2010}, Union2.1~\cite{Suzuki/2012}, the JLA compilation~\cite{Betoule/2014}, Pantheon~\cite{Scolnic/2018}, and most recently the Pantheon+SH0ES dataset~\cite{Scolnic/2022}.

For each supernova with observed peak magnitude $m_{B,i}$ and absolute magnitude $M$, the residual vector is,
\begin{equation}
    D_i = m_{B,i} - M - \mu^{\text{th}}(z_i) \,\,\text{.}
\end{equation}
To break the $M$--$H_0$ degeneracy, Pantheon+SH0ES replaces $\mu^{\text{th}}$ with Cepheid-calibrated distances $\mu^{\text{Ceph}}_i$ for host galaxies with independent calibrations,
\begin{equation}
    \bar{D}_i =
    \begin{cases}
        m_{B,i}-M-\mu^{\text{Ceph}}_i, & \text{Cepheid hosts} \,\,\text{,}\\
        m_{B,i}-M-\mu^{\text{th}}(z_i), & \text{otherwise} \,\,\text{.}
    \end{cases}
\end{equation}
With covariance matrix $C_{\text{SN}}$ (including systematics), the SN chi-square is given by,
\begin{equation}
    \chi^2_{\text{SN}} = \bar{\mathbf{D}}^\mathsf{T} C_{\text{SN}}^{-1} \bar{\mathbf{D}} \,\,\text{.}
\end{equation}

\subsection{Joint likelihood}\label{2.7.3}

The combined chi-square minimized in the MCMC analysis is,
\begin{equation}
    \chi^2_{\text{total}} = \chi^2_{\text{CC}} + \chi^2_{\text{SN}} \,\,\text{.}
\end{equation}
We adopt Gaussian priors on the parameters,
$H_0 \in [50,100]$, $n \in [-5,0]$, $\beta \in [0,5]$, $\eta \in [-5,0]$, and $c \in [0,1]$.

\section{Results and constraints}\label{sec_2.8}
\subsection{Model-I}\label{2.8.1}
For Chaplygin gas, that is, when $m\approx-5$, the best-fit values are:
\\
From the combined CC+Pantheon+SH0ES dataset, the best-fit parameters with $68\%$ confidence limits are:
\begin{table}[h]
    \centering
    \caption{Best-fit cosmological parameters with $68\%$ confidence intervals.}
    \begin{tabular}{lc}
        \toprule
        Parameter & Best fit $\pm 1\sigma$ \\
        \midrule
        $H_0$ & $73^{+0.19}_{-0.18}$ \\
        $n$   & $0.52^{+0.03}_{-0.03}$ \\
        $\beta$ & $0.86^{+0.1}_{-0.1}$ \\
        $\eta$ & $0.45^{+0.02}_{-0.021}$ \\
        $c$ & $-0.62^{+0.046}_{-0.046}$ \\
        $m$ & $-5^{+0.098}_{-0.098}$ \\
        \midrule
        $\chi^2_{\min}$ & $1644.828$ \\
        \bottomrule
    \end{tabular}
    \label{tab:results}
\end{table}
The minimum chi-square for the fit is
\begin{equation}
    \chi^2_{\min} = 1644.828 \,\,\text{.}
\end{equation}

\subsection{Model-II}\label{2.8.2}

For BEC ($m\approx2$), the best-fit values are given as:
\begin{table}[H]
    \centering
    \caption{Best-fit cosmological parameters with $68\%$ confidence intervals.}
    \begin{tabular}{lc}
        \toprule
        Parameter & Best fit $\pm 1\sigma$ \\
        \midrule
        $H_0$ & $73^{+0.19}_{-0.2}$ \\
        $n$   & $-2^{+0.005}_{-0.0049}$ \\
        $\beta$ & $0.51^{+0.0047}_{-0.0045}$ \\
        $\eta$ & $-1.7^{+0.042}_{-0.042}$ \\
        $c$ & $-0.019^{+0.0024}_{-0.0022}$ \\
        $m$ & $1.8^{+0.051}_{-0.052}$ \\
        \midrule
        $\chi^2_{\min}$ & $1634.029$ \\
        \bottomrule
    \end{tabular}
    \label{tab:results}
\end{table}
The minimum chi-square for the fit is,
\begin{equation}
    \chi^2_{\min} = 1634.029 \,\,\text{.}
\end{equation}

\begin{figure}[H]
\centering
\includegraphics[scale=0.50]{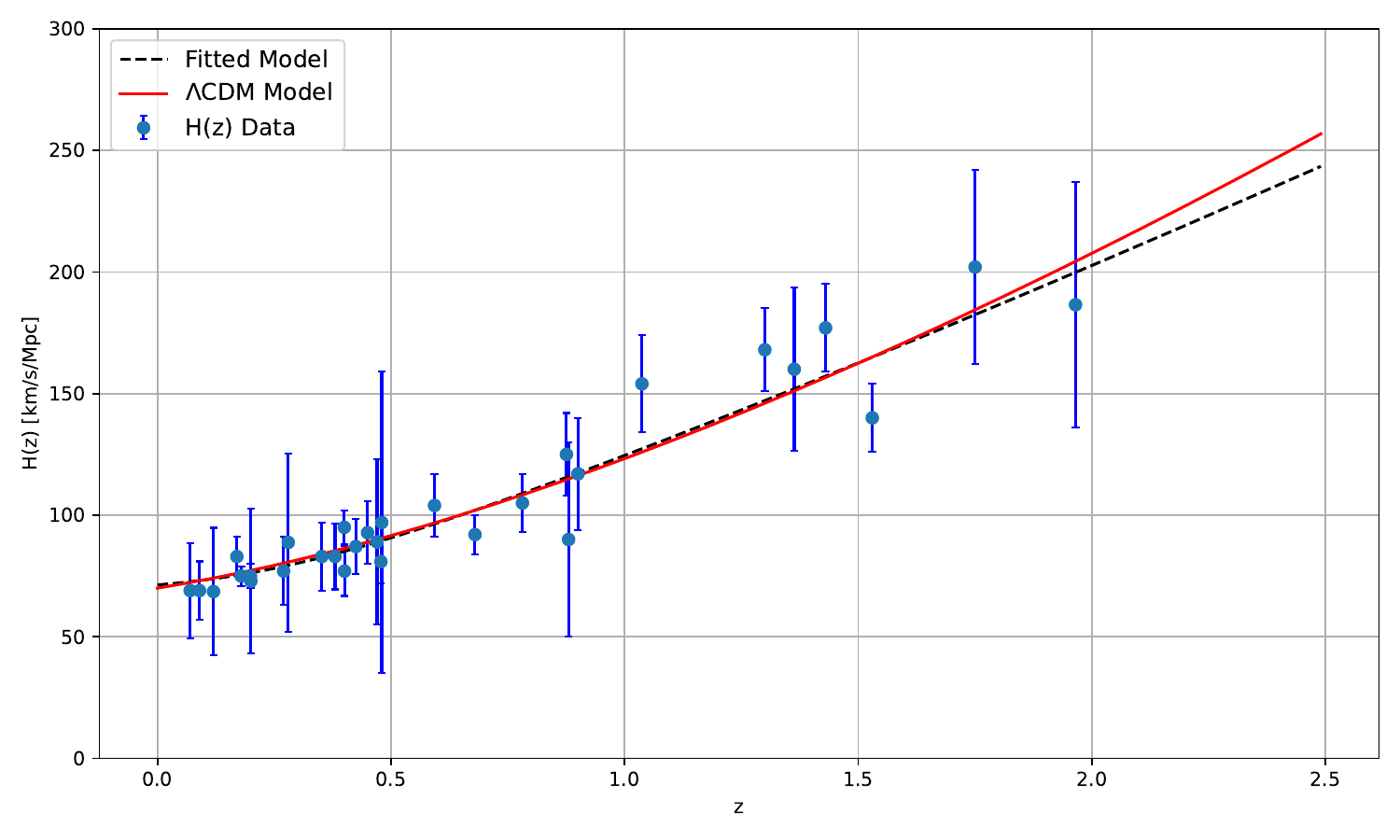}
\caption{The least square fitting for the Chaplygin gas model (as dark energy $m\approx-5$). Compared with the standard $\Lambda$CDM model (black dashed line), the figure shows the fit of the Hubble function $H(z)$ versus redshift $z$ for our proposed model (red line). An error bar plot showing the 31 CC dataset points used in the analysis is also included.}\label{fig_2.1} 
\end{figure}

\begin{figure}[H]   
\centering
\includegraphics[scale=0.50]{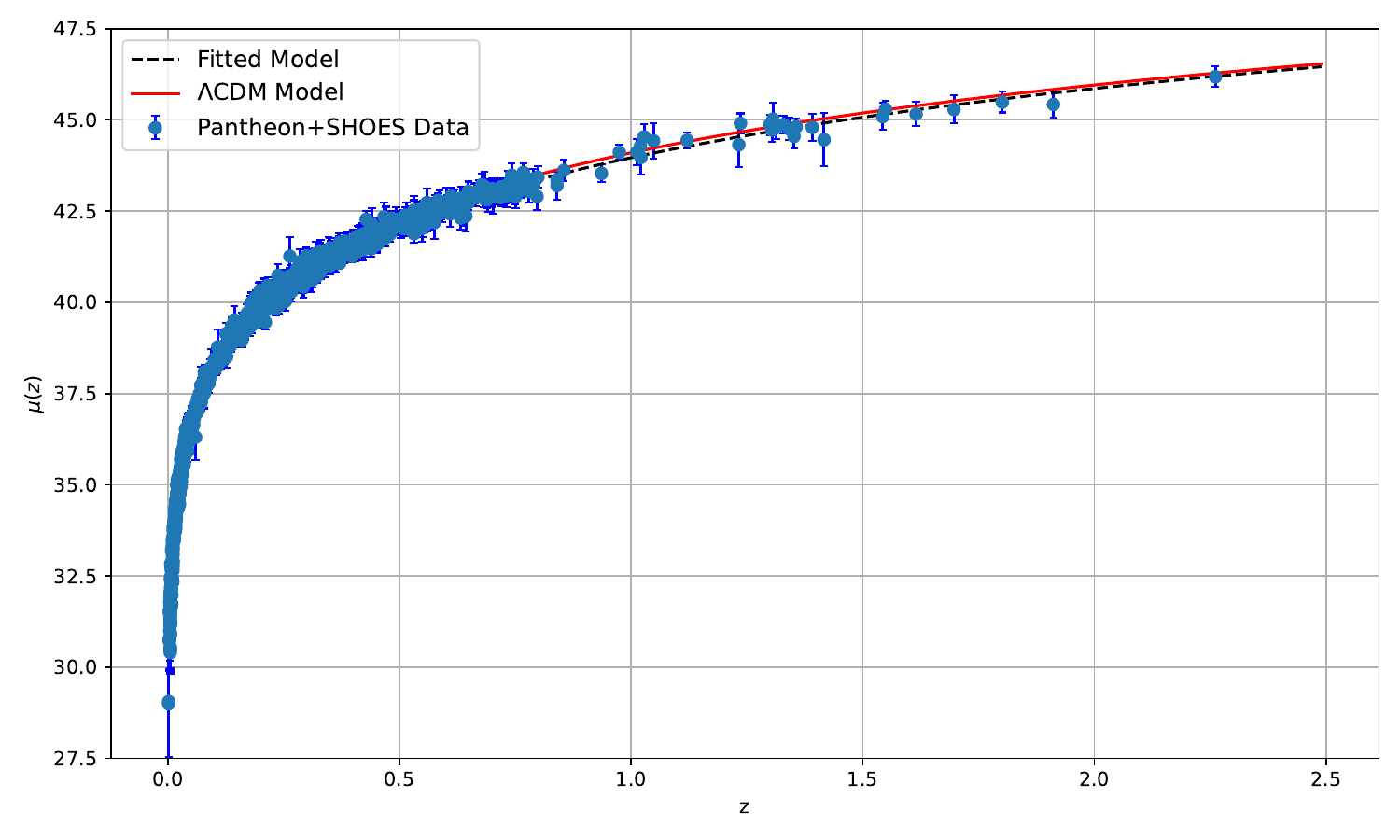}
\caption{The least square fitting for the Chaplygin gas model (as dark energy $m\approx-5$). Compared with the standard $\Lambda$CDM model (black dashed line), the figure shows the fit of the function $\mu(z)$ versus redshift $z$ for our proposed model (red line). An error bar plot representing the 1701 points of the Pantheon+SH0ES dataset used in the analysis is also included.}\label{fig_2.2} 
\end{figure}

\begin{figure}[H]
\includegraphics[scale=0.51]{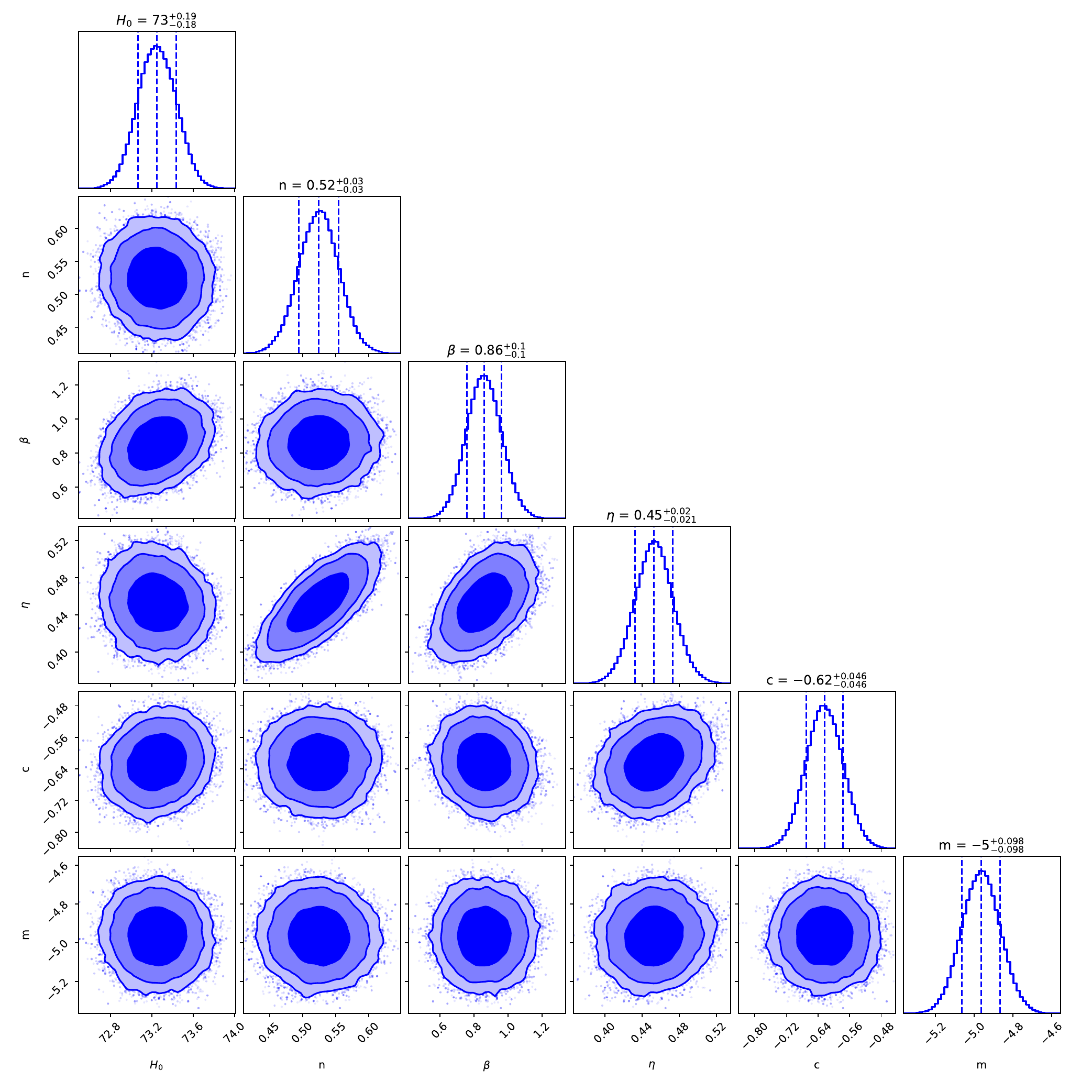}
\caption{The MCMC analysis for the Chaplygin gas model (as dark energy $m\approx-5$). Based on a combined examination of the CC, BAO, and Pantheon+SH0ES datasets, the 2D-contour plot of the model parameters $m$, $c$, $\eta$,$\beta$, $n$ and $H_0$ displays the most likely values and the confidence areas up to 3$-\sigma$.}\label{fig_2.3} 
\end{figure}

\begin{figure}[H]
{\includegraphics[width=7.5cm,height=5cm]{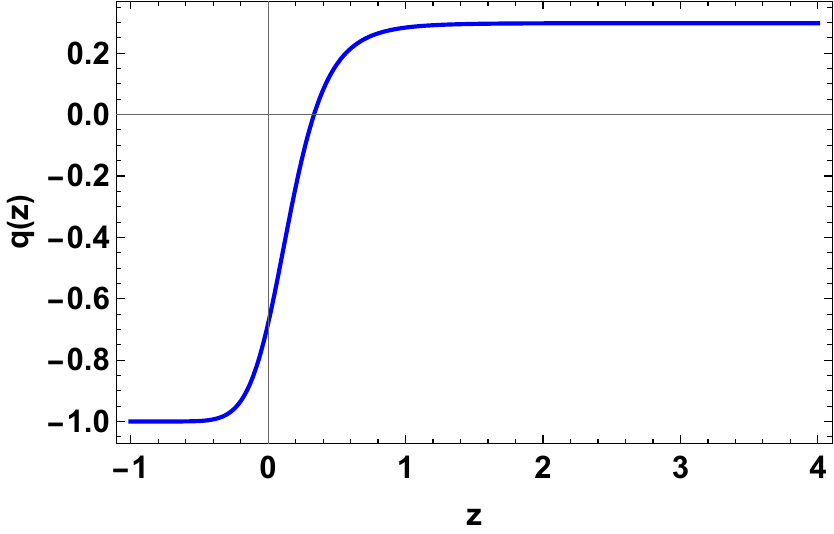}}
{\includegraphics[width=7.5cm,height=5cm]{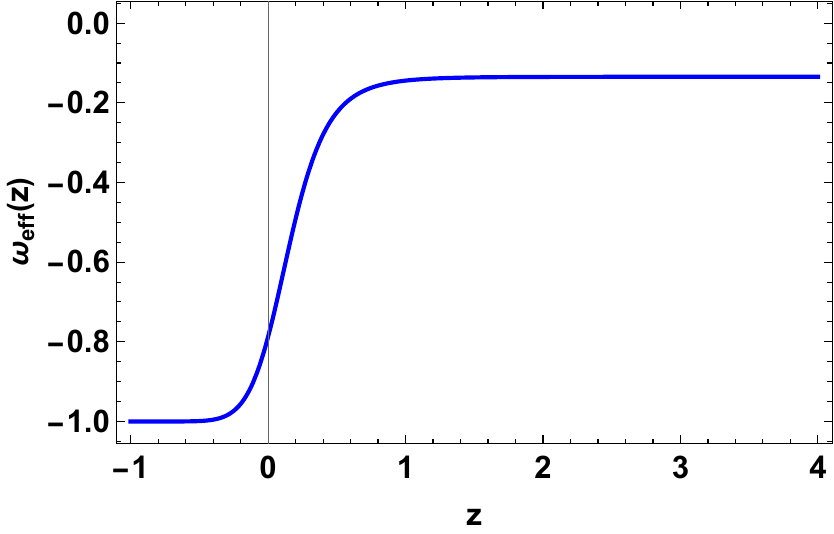}}
\caption{The deceleration parameter $q(z)$ and effective equations of state $\omega(z)$ for the Chaplygin gas model (as dark energy $m\approx-5$). As both blots show at $z\xrightarrow{}-\infty$, both of them converge to $-1$ as expected and also it is very consistent with the current values of the observed $q$ and $\omega$.}\label{fig_2.4} 
\end{figure}

\begin{figure}[H]
{\includegraphics[width=7.5cm,height=5cm]{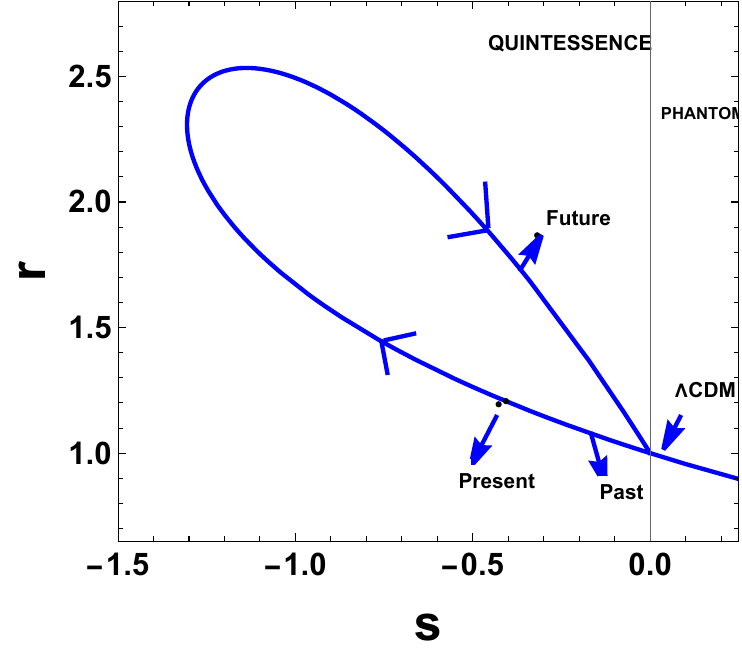}}
{\includegraphics[width=7.5cm,height=5cm]{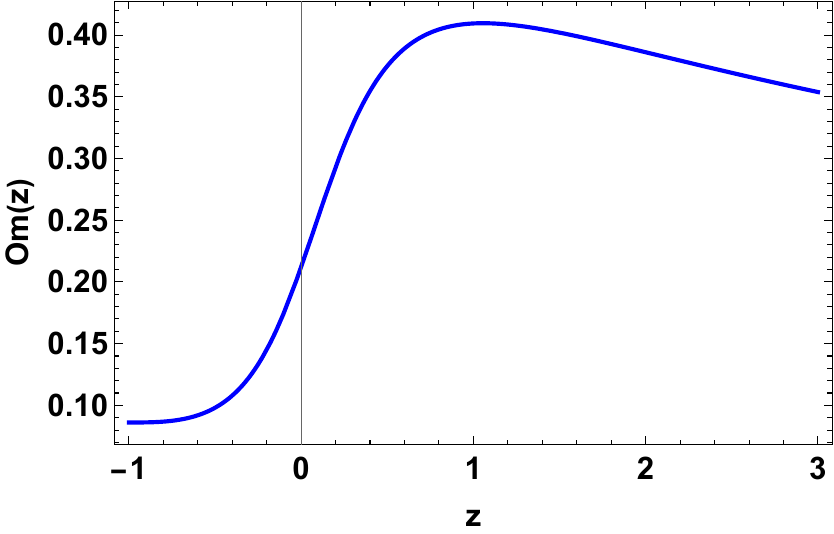}}
\caption{To check the models, we have used two very popular null tests for the $\Lambda$CDM model. First, we have done the $r-s$ plot, which shows that at late time it is converging to $\Lambda$CDM model, and we have also used $Om$ diagnostics for testing variation from $\Lambda$CDM for the Chaplygin gas model (as dark energy $m\approx-5$). }\label{fig_2.5} 
\end{figure}


\begin{figure}[H]
\centering
\includegraphics[scale=0.5]{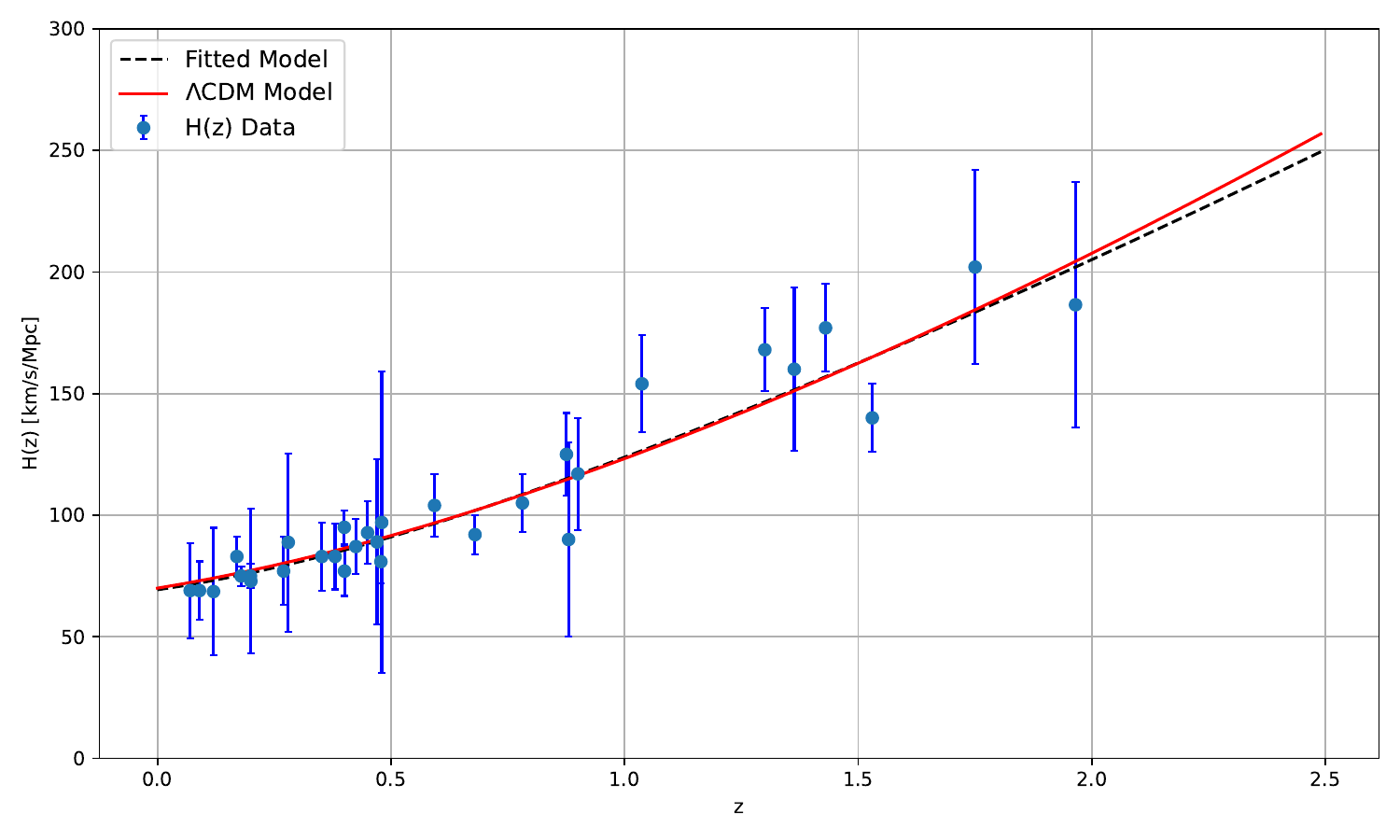}
\caption{The least square fitting for the Chaplygin gas model (as BEC dark matter $m\approx2$). Compared with the standard $\Lambda$CDM model (black dashed line), the figure shows the fit of the Hubble function $H(z)$ versus redshift $z$ for our proposed model (red line). An error bar plot showing the 31 CC dataset points used in the analysis is also included.}\label{fig_2.6} 
\end{figure}

\begin{figure}[H]
\centering
\includegraphics[scale=0.5]{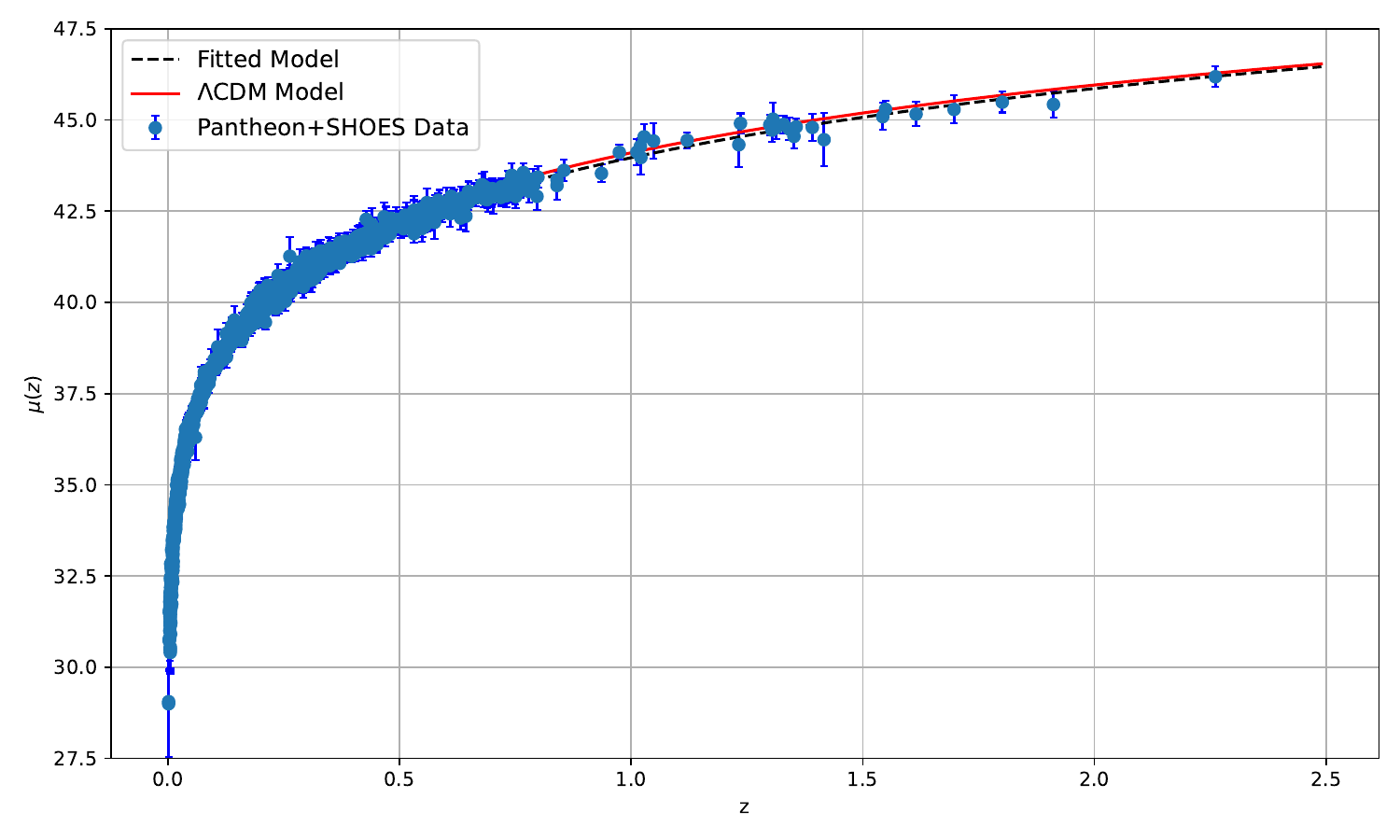}
\caption{The least square fitting for the Chaplygin gas model (as BEC dark matter $m\approx2$). Compared with the standard $\Lambda$CDM model (black dashed line), the figure shows the fit of the function $\mu(z)$ versus redshift $z$ for our proposed model (red line). An error bar plot representing the 1701 points of the Pantheon+SH0ES dataset used in the analysis is also included.}\label{fig_2.7} 
\end{figure}

\begin{figure}[H]
\includegraphics[scale=0.51]{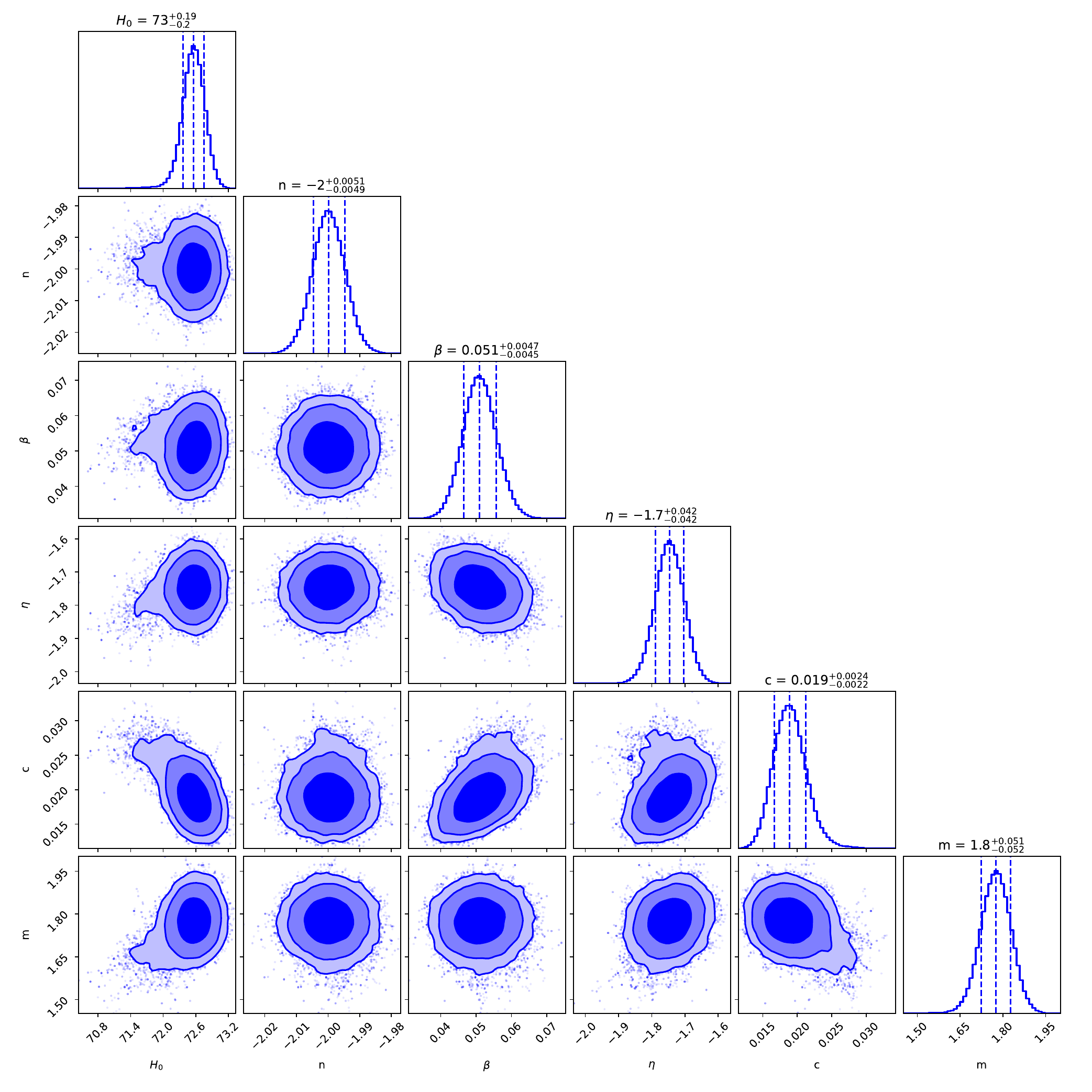}
\caption{The MCMC analysis for the Chaplygin gas model (as BEC dark matter $m\approx2$). Based on a combined examination of the CC, BAO, and Pantheon+SH0ES datasets, the 2D-contour plot of the model parameters $m$, $c$, $\eta$,$\beta$, $n$, and $H_0$ displays the most likely values and the confidence areas up to 3$-\sigma$.}\label{fig_2.8} 
\end{figure}

\begin{figure}[H]
{\includegraphics[width=7.5cm,height=5cm]{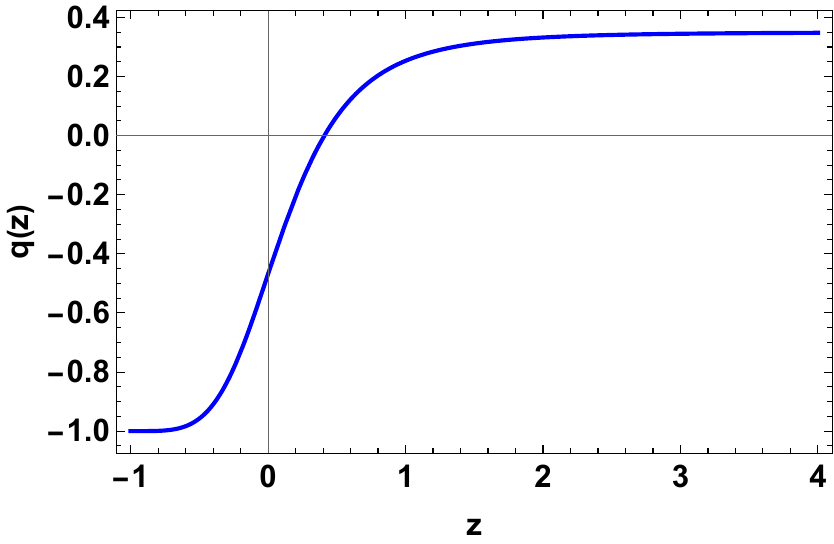}}
{\includegraphics[width=7.5cm,height=5cm] {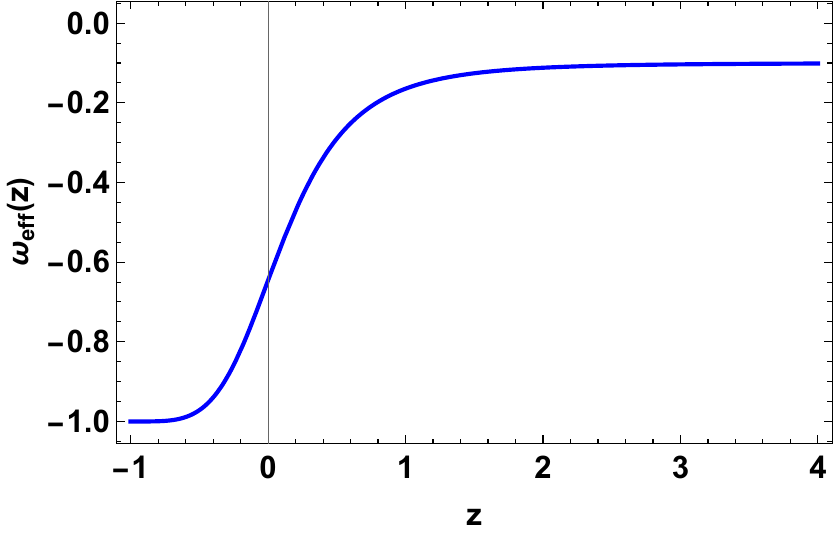}}
\caption{The deceleration parameter $q(z)$ and effective equations of state $\omega(z)$ for the Chaplygin gas model (as dark matter $m\approx2$). As both blots show at $z\xrightarrow{}-\infty$, both converge to $-1$ as expected, and it is also very consistent with the current values of the observed $q$ and $\omega$.}\label{fig_2.9} 
\end{figure}

\begin{figure}[H]
{\includegraphics[width=7.5cm,height=5cm]{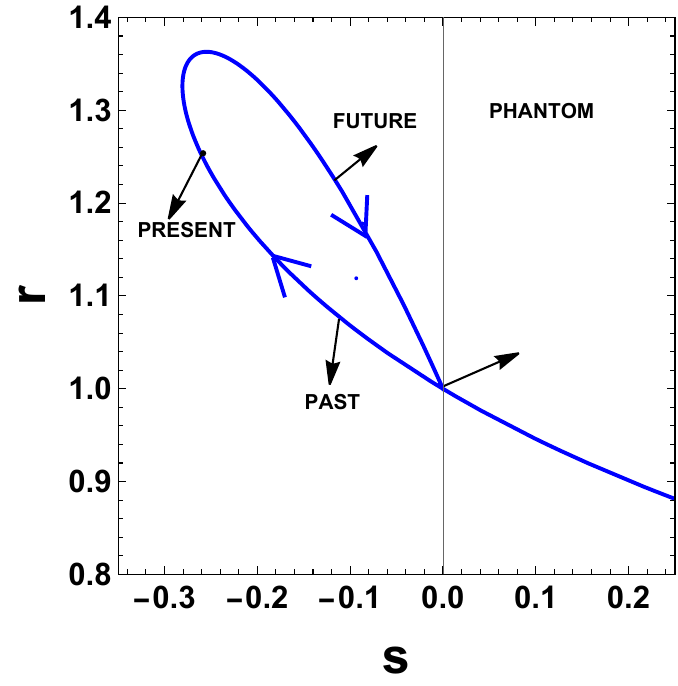}}
{\includegraphics [width=7.5cm,height=5cm]{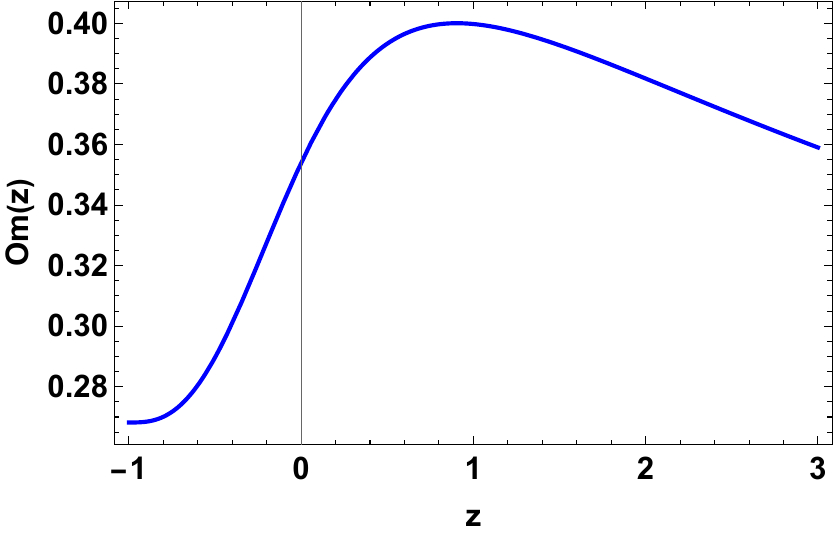}}
\caption{To check the models, we have used two very popular null tests for the $\Lambda$CDM model. First, we have done the $r-s$ plot, which shows that at late time it is converging to $\Lambda$CDM model, and we have also used $Om$ diagnostics for testing variation from $\Lambda$CDM for the Chaplygin gas model (as BEC dark matter $m\approx2$).}\label{fig_2.10} 
\end{figure}
\subsection{Model comparison via AIC and BIC}\label{2.8.3}
To evaluate the effectiveness of our MCMC analysis, we perform a statistical comparison using the AIC \cite{Akaike/1974} and the BIC \cite{Schwarz/1978}. The purpose of these tests is to make sure that one does not "overfit" the data by introducing a large number of free parameters. Although AIC and BIC are very common practices in general Bayesian analysis, they were first systematically introduced in the context of cosmology by Liddle \cite{Liddle/2007}. In this thesis, we have used only Liddle's approach for the model comparison.

The AIC is defined as follows,
\begin{equation}
\text{AIC} = \chi^2_{\text{min}} + 2d \,\,\text{.}
\end{equation}
where \( d \) is the number of free parameters in the model. To compare our results with the standard \(\Lambda\)CDM model, we compute the difference, that is,
\begin{equation}
\Delta \text{AIC} = |\text{AIC}_{\text{MOG}} - \text{AIC}_{\Lambda\text{CDM}}| \,\,\text{.}
\end{equation}
A value of \( \Delta \text{AIC} < 2 \) indicates strong support for the MOG model, while \( 4 < \Delta \text{AIC} \leq 7 \) suggests moderate evidence. If \( \Delta \text{AIC} > 10 \), there is no significant support for the MOG model.

The BIC is defined as follows,
\begin{equation}
\text{BIC} = \chi^2_{\text{min}} + d \ln(N) \,\,\text{.}
\end{equation}
where \( N \) is the number of data points used in the MCMC analysis. The interpretations are similar to above and roughly given as follows,
\begin{itemize}
    \item \( \Delta \text{BIC} < 2 \): strong support for MOG,
    \item \( 2 \leq \Delta \text{BIC} < 6 \): moderate support,
    \item \( \Delta \text{BIC} > 6 \): weak or no support.
\end{itemize}

The computed AIC and BIC values for the viscous modified gravity (MOG) models considered are summarized in Table -III. Based on these results, there is strong evidence in favor of the proposed models across all three datasets. In particular, model I demonstrates the closest agreement with the \(\Lambda\)CDM model.


\begingroup
  \renewcommand{\arraystretch}{0.85}%
  \setlength{\tabcolsep}{4pt}%
  \small
  \begin{table}[ht]
    \centering
    \caption{Minimum $\chi^2$ values and the corresponding AIC and BIC values for both Chaplygin gas (model-I) and BEC dark matter (model-II)}
    \begin{tabular}{@{} c c cc cc cc cc @{}} 
      \toprule
      & & \multicolumn{2}{c}{\(\chi^2_{\min}\)} 
          & \multicolumn{2}{c}{AIC} 
          & \multicolumn{2}{c}{BIC} 
          & \(\Delta\)AIC 
          & \(\Delta\)BIC \\
      \cmidrule(lr){3-4}\cmidrule(lr){5-6}\cmidrule(lr){7-8}
      Model 
        & Dataset 
        & MOG 
        & \(\Lambda\)CDM 
        & MOG 
        & \(\Lambda\)CDM 
        & MOG 
        & \(\Lambda\)CDM 
        & 
        & \\
      \midrule
      \multirow{3}{*}{I}
        & CC             & 28.001  & 32.132 & 40.001  & 38.132 & 48.604  & 42.431 & 1.869 & 6.173 \\
        & Pantheon+      & 1616.827 & 1609.917 & 1628.827 & 1615.917 & 1661.460 & 1632.231 & 12.91 & 29.22 \\
        & CC+Pantheon+   & 1644.828 & 1642.044  & 1656.828 & 1648.044  & 1689.570 & 1664.204  & 8.784 & 25.366 \\
      \midrule
      \multirow{3}{*}{II}
        & CC             & 38.912  & 32.132 & 50.912  & 38.132 & 59.515   & 42.431 & 12.78 & 17.08 \\
        & Pantheon+      & 1595.117  & 1609.917 & 1607.117  & 1615.917 & 1639.750 & 1632.231 & 8.8 & 7.519 \\
        & CC+Pantheon+   & 1634.029 & 1642.044  & 1646.029 & 1648.044  & 1678.771 & 1664.204  & 2.015 & 14.567 \\
      \bottomrule
    \end{tabular}
  \end{table}\label{table_3}
\endgroup

\section{Om diagnostics and state finder analysis}\label{sec_2.9}
To check deviations from the $\Lambda$CDM model, several tests have been proposed, roughly categorized as null tests for standard cosmological models. Basically, the idea is to construct a scalar quantity (a real number) that gives a particular value for $\Lambda$CDM and only deviates from it when the model is not $\Lambda$CDM. Ideally, these quantities should also distinguish between two different types of nonstandard cosmology as well, such as quintessence and phantom, etc.\\
In this chapter, we have taken mainly two diagnoses, that is Om and the statefinder diagnosis, to test the deviation from $\Lambda$CDM. Even though, through the MCMC analysis, we have shown that the best fit model parameters with the latest observational data, such as Hubble and Pantheon+SH0ES datasets, it is still not obvious what the late-time behavior of our model is, that is, whether it is approaching the de Sitter model from the Phantom or Quintessence side. In addition, these diagnoses give a very good consistency check, that is, whether we are getting late-time de Sitter solutions or not, and also how severe the deviation from $\Lambda$CDM is.
\subsection{Om}\label{2.9.1}
Om diagnostics is the simplest form of null-diagnosis available for classification of the various dark energy-based cosmological models and their deviation from the $\Lambda$CDM.\\
It was first proposed by Sahni et al.\cite{Sahni/2008} noting that the matter density falls as a third power of the scalar factor while the $\Lambda$ remains constant. The formula for $Om(z)$ is given as,
\begin{equation}
    Om(z)=\frac{\left(\frac{H(z)}{H_0}\right)^2-1}{(1+z)^3-1} \,\,\text{.}
\end{equation}
It should be noted that the expression depends solely on the values of $z$ and $H(z)$, so it can be easily calculated using any data point.\\
In our work, we have plotted this in the figure and it shows that for approximately $z>1$ the slope is negative, indicating quintessence scenarios, but for $-1<z<1$ the slope is positive, giving the phantom scenario. So, we can say for both of our cases that is both for the Chaplygin gas ($m\approx -5$) (Fig.~\ref{fig_2.5}) and the BEC (Figure-$m\approx 2$) (Fig.~\ref{fig_2.8}), we first obtain the quintessence scenario up to $z=1,$ then from $z=1$ to present ($z=0$) and future ($z<0$) it follows phantom behavior to reach the de Sitter solution or $\Lambda$CDM.\\
 So, in general, both models show that globally there are in the Phantom region.
\subsection{Statefinder}\label{2.9.2}
  Although diagnosis $Om$ is an excellent null test, there are several problems; for example, it does not distinguish between various types of quintessence and phantom models. There is also the fact that the formula is too simple, so it does not take into account the nuances of the other type of late cosmology and how much they differ from the $\Lambda$CDM.\\
  In order to circumvent these things, Sahni et al. \cite{Sahni/2003} have proposed a null test based on two different parameters that $r,s$, where they are given by the formula,
  \begin{equation}
      r=\frac{\dddot{a}}{aH^3} \,\,\text{.}
  \end{equation}
  and,
  \begin{equation}
      s=\frac{r-1}{3\left(q-\frac{1}{2}\right)} \,\,\text{.}
  \end{equation}
One first note is that for the standard $\Lambda$CDM model, $q=-1$, so $r=0$, so one of the advantages of $r$ is that even though many different models give similar $H$ and $q$, they differ by the third derivative, that is, $r$. $s$ is defined to distinguish between various forms of dark energy models.  Together, $r$ and $s$ can distinguish between Chaplygin gas, Phantom, Quintessence, etc. So, it can be easily shown that for $\Lambda$CDM ($\omega=-1$) $s=0,r=1$; however, $s<0$ and $r>1$ (for both of our models), they are phantom in nature. However, it also shows that our current Universe ($z\approx 0$) lies in the phantom region and will eventually converge to $\Lambda$CDM.

\section{Conclusion}\label{sec_2.10}
In this chapter, we have presented a unified model of both dark energy and dark matter based on the (Generalized) Chaplygin gas. We have shown that the MCMC analysis with both the Hubble and Pantheon+SH0ES datasets favors a modified-gravity dark-energy ($m\approx-5$) or dark-matter ($m\approx2$)- dominated Universe (dark energy arising from the modified gravity). To systematically account for the transition, we have used a standard interaction term, $\mathcal{Q}$, which can serve as a dark matter-to-dark energy transformation. Overall, we have shown that data analysis ($\Delta$AIC and $\Delta$BIC tests) favors the BEC dark-matter model over dark energy purely by the Chaplygin gas model.\\
Here, is a quick overview of the final results which have been presented throughout the chapter. First of all, in Fig.~\ref{fig_2.1} and Fig.~\ref{fig_2.6}, we have given the least square fitting for the Hubble dataset and compared our model Chaplygin Gas ($m\approx-5$) and BEC ($m\approx2$) respectively, with the standard $\Lambda$CDM model to show that there is not much significant deviation between the two, which shows that at least phenomenologically, our model is consistent with the current observation. In Fig.~\ref{fig_2.2} and Fig.~\ref{fig_2.7}, the same thing has been done for the Pantheon+ dataset for Chaplygin gas ($m\approx-5$) and BEC ($m\approx2$), respectively. Like the previous Hubble dataset, it also shows remarkable consistency with the $\Lambda$CDM cosmology. For Fig.~\ref{fig_2.3}, we have done the MCMC analysis and plotted the $3\sigma$ contour plot using the corner plot in the EMCEE package in Python. All the free parameters can be read off the plot; however, the most important point is that we obtain $m\approx-5$ for the Chaplygin gas model, which shows it is remarkably consistent with the dark energy model. Similarly, in Fig.~\ref{fig_2.8} we have done the same for BEC dark matter and shown that $m\approx2$ is also a valid solution that produces a late-time de Sitter solution. In Fig.~\ref{fig_2.4} and Fig.~\ref{fig_2.9}, we have shown the behaviors for phenomenological quantities such as $q$ and $\omega_{eff}$ for both $m\approx-5$ and $m=2$, respectively. As expected, both are showing late time de Sitter behavior (i.e., at $z\rightarrow -\infty$, $q,\omega_{eff}\rightarrow -1$). Finally, in Fig.~\ref{fig_2.5} and Fig.~\ref{fig_2.10} we have done the two most famous Null-tests for the $\Lambda$CDM model, that is, the statefinder and Om diagnostics, both of which show the Phantom behavior in the current time and late-time de Sitter transition, showing the consistency of our model with the current paradigm.\\
It should also be noted from Table~\ref{table_3} that, after qualitatively examining $\Delta$AIC and $\Delta$BIC, our model generally favors the BEC model over the Chaplygin Gas model, indicating that there is indeed a nontrivial interaction term between the two. \\
Here, we provide a brief outline of the entire chapter.\\
First, we start with an introduction (Sec.~\ref{sec_2.1}) where we provide a detailed motivation for thinking of the Chaplygin gas equation of state as a unification of both dark matter and dark energy. We also give a rough idea of why Chaplygin gas models are appealing: they naturally arise in string theory and naturally connect a matter-dominated Universe to a dark-energy-dominated Universe. We also discussed how Bose-Einstein Condensation could be responsible for the dark-matter candidate. We also discussed the motivation for including the interaction term between dark matter and dark energy. After giving a brief outline of the symmetric teleparallel-based gravity, we close with the data analysis and the null-tests, such as Om and statefinder diagnosis.\\
In Sec.~\ref{sec_2.2}, we give a more detailed discussion on the Chaplygin gas, and we also show how various microscopic theories, such as the scalar field, DBI field, and Brane world scenarios, can give rise to the Chaplygin gas equation of state. This section also serves as a primer showing that, given the best fit to the Chaplygin gas model, one can indeed reconstruct the scalar field potential, the DBI field potential, or even the wrapping factor for the Branes. So in a sense, we have shown the Universality of the Chaplygin gas model from both a microscopical and a phenomenological point of view. \\
 In Sec.~\ref{sec_2.3} we discuss how dark matter can be modeled by BEC. We have given a brief physical motivation: the Universe, after the state of formation, could indeed create the ideal environment for BEC to occur. The BEC could explain why the density of dark-matter halos is so uniform, and it is also an effective theory, in the sense that no matter what the underlying bosons are, as long as they are massive, BEC is possible. We have also discussed the microscopic origin of BEC by the Gross-Pitaevskii equations for a weakly interacting Bose gas. Then we argue that the variation of such an energy function could indeed lead to a form similar to the Chaplygin gas equation. Overall, this gives a microscopic and observationally sound motivation for why dark matter can be modeled as BEC.\\
 In Sec.~\ref{sec_2.4}, we advocate for a uniform framework for dark matter and dark energy. We argue that, given the values of $m$, one can easily check whether the model is dominated by dark energy, as in the Chaplygin gas equation, or by BEC dark matter. We also argued that it is important to include an interacting term that is consistent with the observational standard model of cosmology and maintains the ratios of dark matter and dark energy. \\
 In Sec.~\ref{sec_2.5}, we give a brief description of the $f(Q)$ gravity. We discuss which gauge to choose and why the boundary term does not affect the formulation. We also give a brief detour on how to calculate the non-metric scalar from $Q_{\alpha\mu\nu}$. In Sec.~\ref{sec_2.6}, we continue the discussion for $f(Q)$ gravity under the FLEW metric. We have used a flat FLRW metric with a signature $(- + + +)$ and given a way to calculate the non-metric scalar, which turns out to be $6H^2$. We also argue how $f(Q)=-Q$ could give the Einstein GR, providing a consistency check between modified gravity and standard Einstein GR.\\
 In Sec.~\ref{sec_2.7} we give the complete formula for $H(z)$ as a function of $z$ (redshift). We use the continuity equation and the Friedmann  equation to arrive at the exact formula. This would enable us to do the data analysis and constrain the free parameters.\\
  In Sec.~\ref{sec_2.8}, we perform a complete data analysis using the formula $H(z)$. We have found that there are two models that are model I ($m\approx-5$) and model II ($m\approx2$), both of which give constant observational values. We also see that both models have similar $q$ and $\omega$, which closely resemble our current observational data. Even though both models give observably constant results and satisfy the null test, one can see that model II, which is the model that is dark matter dominates, it gives the lowest $\Delta$ AIC and $\Delta$ BIC, which shows that the Chaplygin gas is more biased toward dark matter modeled by BEC, and also there is non-trivial (as $b\neq 0$) interaction between bark matter and dark energy, that is, dark matter is transformed into dark energy. \\
  From the MCMC analysis presented in Sec.~\ref{sec_2.8}, we observe that the exponent parameter takes two distinct values depending on the chosen priors, which phenomenologically connects our unified equation of state to two distinct physical pictures discussed earlier in the chapter. For model-I, we obtain , consistent with the generalized Chaplygin gas behavior as originally discussed by Kamenshchik et al. \cite{Kamenshchik/2001} and Bento et al. \cite{Bento/2002}, and as shown in Sec.~\ref{sec_2.2}, such negative exponents can be reconstructed from a scalar field with potential (Eq.~\ref{eq_2.10}), where our fitted value sets the normalization scale; the gravity parameter indicates departure from GR, which phenomenologically could arise from brane-world corrections of the type discussed in Eq.~(\ref{eq_2.20}), where the parameter corresponds to the brane tension scale, while the positive interaction parameter suggests energy transfer from dark matter to dark energy, consistent with the interacting dark energy frameworks of Setare \cite{Setare/2009} and Guo \cite{Guo/2005}, or alternatively, following the Tachyon/DBI reconstruction of Benaoum \cite{Benaoum/2022} (Eq.~\ref{eq_2.15}), our constraint corresponds to a parameter , yielding a well-behaved Tachyonic potential supporting late-time acceleration. For model-II, we find , lying close to the value expected for Bose-Einstein Condensate dark matter as formulated by Boehmer and Harko \cite{Boehmer/2007} and Harko \cite{Harko/2011,Harko/2015}, where the parameter encodes the two-body interaction strength of the condensate, relating through the Gross-Pitaevskii formalism (Sec.~\ref{sec_2.3}) to the boson mass and scattering length; the parameter differs significantly from model-I, suggesting stronger gravitational modification if the BEC interpretation holds, while the near-zero value indicates minimal brane-world contributions, making the BEC description primarily a four-dimensional effective theory consistent with Mahichi et al. \cite{Mahichi/2021,Mahichi/2022,Mahichi/2023}, and the negative interaction parameter reverses the energy flow direction compared to model-I, now indicating the transfer from dark energy to dark matter, supporting continued structure formation at late times and aligning with the quantum potential interpretations of Das et al. \cite{Das/2018,Das/2023}. The slightly lower for model-II (1634.029 versus 1644.828), together with the IC tests from Table~\ref{tab:results}, provides marginal statistical preference for the BEC scenario, and this bimodality in phenomenologically suggests that the Chaplygin gas framework may admit multiple physical interpretations depending on the epoch or scale considered, although distinguishing between these scenarios conclusively will require next-generation observations.\\
In Sec.~\ref{sec_2.9}, we have done the two popular null tests for the $\Lambda$CDM model, that is, Om and the statefinder design. We have shown that both models under both diagnoses approach the standard $\Lambda$CDM model while currently passing through the Phantom epoch. This is hardly surprising, as the Chaplygin gas equation of state typically exhibits Phantom behavior, as we have shown in Sec.~\ref{sec_2.2}. Indeed, one needs a non-standard kinetic energy term to arrive at the Chaplygin gas equation of state.\\
 Finally, we end the chapter with the conclusion (Sec.~\ref{sec_2.10}). Overall, this chapter shows that one can obtain a unified picture of dark matter and dark energy via the Chaplygin gas model in modified $f(Q)$ gravity, and that data analysis yields very tight observational constraints on the free parameters. We can further investigate this study by taking more general modified gravity models, such as $f(Q,T),f(Q,B),f(Q,T)$ or general $f(Q,T,L_m)$ models, to see whether the conclusions of our models are empirical or not. \\



\chapter{Reconstruction of Dark Energy Using Hubble and DESI Data for Dirac-Born-Infeld Scalar Field via Gaussian Process} 

 \label{Chapter3}
\lhead{Chapter 3. \emph{Reconstruction of Dark Energy Using Hubble and DESI data for Dirac-Born-Infeld scalar field via Gaussian Process}} 

\vspace{8 cm}
* The work, in this chapter, is covered by the following publication: \\
 
\textit{Reconstruction of Dark Energy Using Hubble and DESI data for Dirac-Born-Infeld scalar field via Gaussian Process}, currently under review, (arXiv:2607.09731).

\clearpage

\epigraph{``Two things fill the mind with ever new and increasing admiration and awe\dots\ the starry heavens above and the moral law within.''}{--- Immanuel Kant, \textit{Critique of Practical Reason} (1788), Conclusion}

In this chapter, we have reconstructed the dark energy as a Dirac-Born-Infeld scalar field from the Hubble dataset (32 CC + 26 BAO) and the DESI dataset using the Gaussian process. As the Gaussian process is a non-parametric and model-independent way to reconstruct a function and its derivative using the data, our reconstruction of $\omega_\phi$, $\Omega_\phi$, and $\mathcal{V}(\phi)$ does not assume any particular model of cosmology.
Using the reconstructed profiles of the scalar potential $\mathcal{V}(\phi)$ as a function of the field $\phi$, along with their associated uncertainties, we performed a chi-square curve fitting procedure to assess the viability of four different scalar field potentials such as exponential, power-law, free field (quadratic), and Higgs-like. This allowed us to identify which potential best fits the reconstructed data. Furthermore, we used MCMC analysis to place quantitative constraints on the model parameters associated with each potential.
Furthermore, we do a $\chi^2$ analysis for all four potentials and comment on the goodness of the fit for each of them. Finally, we conclude with the potential generalization of our model and the phenomenological implications of our study.

\section{Introduction}\label{sec_3.1}  
Since the discovery of CMB radiation in 1965 \cite{cmb}, it has been more or less confirmed that our Universe originated from a very hot, dense state, popularly known as the hot Big Bang model or just the Big Bang model. It was soon confirmed that CMB radiation is around $3K$, which is remarkably consistent with the decade-old calculation carried out by Gamow and his collaborators \cite{gamow}. Also, it has been shown by them that the Big Bang Nucleosynthesis (BBN) exactly predicts the abundance of heavier materials and also predicts the Hydrogen-to-Helium ratio in the Universe with remarkable accuracy. However, there have been some problems with the standard Big Bang model of cosmology, such as the Horizon problems, the flatness problem, and the monopole problem. It was Starobinsky \cite{star} and Guth \cite{ALAN} who independently proposed the ``inflation theory" as a way to avoid these problems. Later, in 1983, Linde \cite{Linde} solved some of the shortcomings of the inflationary model (mainly the initial value problem) and introduced ``chaotic inflation," which makes a broader class of scalar field potentials applicable for the inflationary paradigm. One can think of inflation as a scalar field that acts for a very short time, gives rise to a very rapid expansion (de Sitter-like solution) for a very short time, and decays. Such a rapid expansion in the very early Universe could naturally solve all the above-mentioned problems. Even though there are plenty of scalar field potentials and other types of non-canonical scalar fields, such as K-essence field, DBI field, generalized DBI field, it has not been possible to pinpoint one single type of scale field or mechanism that is responsible for the inflation. The best one can do is give a bound on the potential forms via various phenomenological constraints such as the scalar-to-tensor ratio or anisotropies in the CMB.\\
  The second paradigm change in cosmology came with the discovery of late-time acceleration, as noted independently by Riess \cite{late1} and Perlmutter \cite{late2}. By analyzing high-redshift Type Ia supernovae, they have shown that the Universe is not just expanding, but is accelerating. An obvious explanation for such a situation is that the cosmological constant ($\Lambda$) is responsible for the de Sitter-like expansion. Although the $\Lambda$CDM model is extremely phenomenologically sound and can explain a wide range of phenomenological and observable results (including the latest Planck satellite observations \cite{planck}) based on very few free parameters. However, it has some serious downsides, especially since no explanation is given of how the term $\Lambda$ originates. First and foremost, it was argued by Zel'dovich \cite{ZEL} that maybe the cosmological constant $\Lambda$ is just the zero point energy of the vacuum in the quantum fields, however, it has been soon realized a the energy contribution of the first loop correction using plank energy cut off would give discrepancy of $\Lambda$ with the observation of the order of $10^{120}$. There is no natural explanation for such a huge discrepancy in contemporary quantum field theory. One natural way to bypass this issue is given by Weinberg \cite{WENB} that perhaps the cosmological constant has not been uniform throughout the age of the Universe. There is a secondary ``quintessence" scalar field which can naturally explain the late time acceleration and such a tiny value of observed $\Lambda$.\\ 
  In this chapter, our focus is on the DBI scale field, a special type of $k$-essence field, which naturally occurs in open string theory when one takes the low-energy effective Lagrangian of D-Branes \cite{green,polchinski}. In particular, it has been noted that \cite{mazumdar,sen1,sen2,sen3} Tachyon condensation of the non-BPS Branes leads to such a DBI potential, which can be used in the context of cosmology as the energy scale is very similar to that of pre-inflationary scenarios. The detailed study of the DBI field as inflation has been done by Padmanabhan \cite{paddy} and Gibbons \cite{gibbons1}. Moreover, the DBI field as a candidate for quintessence field has also been explored by various authors, as it can not just reproduce the late-time acceleration but also give a value that is very consistent with the observation, and last but not least, it has a very sound physical ground. Alternative ways of obtaining the DBI field from other forms of string theory have been reviewed by Gibbons \cite{gibbons2}. The study of the DBI field in late time acceleration context has been done by Bhagla et al. \cite{bhagla}, while Gorini et al. \cite{gorini}
offered an alternative way of visualizing the DBI field as a modified Chaplygin gas. It has also been noted by Padmanabhan and Roy Choudhury \cite{paddy2} that the DBI field can not just explain DE but could also be used to explain DM, making it an ideal candidate for the study of late-time cosmology. It should also be noted that apart from string theory, one can get the DBI field using the generalised Chaplygin gas equation of state as shown by Gorini \cite{gorini}.\\
The analytical study of the DBI or Tachyon fields is usually done using dynamical system analysis. As the Klein-Gordon equation for DBI fields is highly nonlinear in nature, so it is very difficult to study using the exact solutions, so one incorporates the fixed point analysis using the dynamical system method. The first study using dynamical system analysis was conducted by Copeland \cite{copeland1} and Aguirregabiria \cite{aguirregabiria}. However, they only used the inverse square type potential in order to find the tracker solution. Soon after that, it was extended to more general potentials beyond the inverse square potentials by Fang and Lu \cite{fang2010a}. Quiros \cite{Quiros} first tackled even more general functional forms, including the $sinh(\phi)$ type potential. Guo \cite{guoexp} has studied the exponential potentials in detail and given a very solid motivation as to why exponential potentials, apart from the power law potentials, are a very good choice for such cases. We have followed these leads to choose the power law and exponential potentials, and we have included the more general potentials such as free field and Higgs potentials.\\
It is also noteworthy that, as demonstrated by Silverstein and Tong~\cite{tong}, when a D3-brane is considered moving toward the horizon of AdS space, one obtains a generalized DBI action in the strong coupling regime, contrasting with earlier analyses conducted in the weak coupling limit. In this strong coupling scenario, the DBI field receives additional contributions due to the motion of the D3-brane, leading to a modified Lagrangian of the form
$\mathcal{L}_{\text{GDBI}} = -\frac{1}{f(\phi)} \left( \sqrt{1 + f(\phi) \, \partial_\mu \phi \partial^\mu \phi} - 1 \right) - V(\phi)$. This shows that the DBI fields are a good candidate to study cosmology from string theory, as both high- and low coupling ones can get the DBI field. This gives us a wide range of flexibility in choosing the model that is consistent with the data as well as phenomenologically sound.
We should also mention that the DBI field or Tachyon fields have already been explored in the context of modified gravity such as minimally coupled \cite{mincouple} and non-minimally coupled \cite{nonmincouple} $f(R)$ gravity as well as in the context of $f(Q)$ gravity \cite{medbi}.\\
In this work, we used the Gaussian process algorithm to construct a model-independent scalar field potential profile for the DBI field. To do this, we used both the combined Hubble dataset \cite{Jimenez2002,Moresco2016,Gaztanaga2009,Alam2017} and the latest realized DESI dataset \cite{desi,desi1} to constrain the DBI field potential using the GP. GP is a non-parametric algorithm that can enable the reconstruction of a function using the data instead of assuming a specific functional form. The detailed motivation for the algorithm comes from the machine learning algorithm, which is discussed in detail in \cite{Rasmussen2005}. For the cosmological context, it is a very powerful tool, as it is well known that one can reconstruct the form of DE in the form of $H(z)$ and its derivatives via Sahni and Starobinsky's work \cite{sahnireconstruction}. Even though constraining the form of DE or quintessence scalar field potential from data is very common via MCMC or other Bayesian-type analysis, the problem lies with the fact that almost all the Bayesian algorithms choose a form of the potential and find the constraint on the free parameters from the data by performing MCMC analysis. The advantage of GP is that, by design, it is nonparametric, so it does not rely on any prior knowledge of cosmographic quantities such as $\omega$ or $V(\phi)$, etc. The study of cosmography, that is, the time evolution of the cosmic expansion in a model-independent way using GP, was first done by Shafieloo, Kim, and Linder \cite{lindergaussian}. The reconstruction of the model-independent DE profile using GP was done by Seikel et al. \cite{Seikel2012} and was later optimized and expanded to include various priors \cite{Seikel2013}. Mehrabi et al. \cite{Mehrabi2021} have used the GP and more datasets from Supernova to study the cosmography, showing the robustness of the GP. The reconstruction of model-independent quintessence scalar field potential using GP was performed by Jesus et al. \cite{jesusgaussian}. Niu et al. \cite{niugaussian} have expanded on this paradigm and, apart from reconstructing the scalar field potential, have taken exponential and power-law potentials and have performed MCMC analysis in order to bound the free parameters of the exponential and power-law potential. It is also worth noting that in recent years, considerable efforts have been made to validate the swampland conjecture of cosmology using GP. Elizalde et al. have shown \cite{elizaldeswampland1,elizaldeswampland2} that one can verify the consistency of the swampland conjecture \cite{vafa} with the scalar field potential reconstructed from the GP. Significant work is also done on reconstruction of DE and scalar field potentials using GP in the context of modified symmetric teleparallel gravity by Gadbail et al. \cite{gaurav1,gaurav2}.\\
We have constructed the kinetic term ($\dot{\phi}^2$) and the potential ($V(\phi)$) from the cosmographic parameters ($H(z)$ and its derivatives). Then we have numerically solved the ODE and used GP to reconstruct the potential. We have also taken four different types of DBI potentials such as exponential, power-law, free field, and Higgs field, and have done an MCMC analysis to constrain the free parameters of the potential. Finally, we used the $\chi^2$ test to comment on the goodness of the analysis. In general, this work provides a complete and systematic model-independent analysis of the DBI scalar field and constraints on potentials using the GP.\\
It should be noted that the latest released DESI dataset has already been used to reconstruct $f(T)$ gravity using Genetic Algorithms by Ouardi et al. \cite{genetic}, and to reconstruct the Om-diagnostic using the GP of Mukherjee et al. \cite{gaussianom}. However, to the best of our knowledge, this is the first thesis that uses DESI data to tackle the DBI field and find constraints on the coefficients of the DBI potential.\\
It should also be noted that the interest in fixing the parameters of the potentials from the data is not only important for phenomenological purposes but also has a deeper theoretical necessity. It is well known that string theory is one of the most promising theories of quantum gravity, yet there is no consistent de Sitter, vacuum solution in string theory in the context of cosmology. Such an unexpected result has been handled by Vafa's Swampland conjectures \cite{vafa}. The main idea of Swampland is that there are certain gravity theories that are consistent with the quantum theory (that the low-energy gravity theory has a full UV-complete theory), these belong to the ``landscape". However, there are some other theories such as theories based on de Sitter vacuum belonging in the ``swampland", in a sense these theories cannot have a reasonable UV- complete quantum theory. Based on this paradigm Vafa has given some conjectures such as Swampland Distance Conjecture (SDC), de Sitter Conjecture (dSC) and the refined de Sitter Conjecture (RdSC), which in principle gives a very tight bound on \( \left|\frac{\nabla_{\phi}V}{V}\right| \). Elizalde and Khurshudyan \cite{elizaldeswampland1} have shown that one can indeed verify the Swampland conjecture or place bounds on the constants of the potentials, using GPs. Later, they have also extended this verification to modified \( f(R) \) gravity formulations \cite{elizaldeswampland2}.\\
The structure of this chapter is as follows: In Sec.~\ref{sec_3.2}, we briefly discuss the physical motivation behind the DBI field. Sec.~\ref{sec_3.3} presents the basic mathematical formalism of DBI field theory. In Sec.~\ref{sec_3.4}, we perform a model-independent reconstruction of the Hubble function and its derivatives using the CC+BAO+DESI datasets. Using this reconstruction, we further reconstruct the DE equation of state and the DE density parameter. Sec.~\ref{sec_3.5} focuses on reconstructing the scalar field potential using the GP, where we also employ chi-square analysis for model selection and MCMC to constrain the model parameters. Finally, we summarize and conclude our results in Sec.~\ref{sec_3.6}.

\section{Physical motivation for Dirac-Born-Infeld (DBI) field}\label{sec_3.2} 

The DBI action arises in open string theory as an effective low-energy description of D-brane dynamics \cite{green,polchinski}. In particular, Tachyon condensation in non-BPS D-branes \cite{mazumdar,sen1,sen2,sen3} generates a DBI-type scalar field with potential applications to cosmology. Padmanabhan \cite{paddy} and Gibbons \cite{gibbons1} demonstrated that DBI fields in FLRW backgrounds exhibit inflationary behavior, while subsequent studies explored their role in late-time acceleration \cite{bhagla,gorini,paddy2}. These works suggest DBI fields could provide a string-theoretic origin for cosmic acceleration without ad hoc scalar-field introduction.

Dynamical systems analysis has proven invaluable for studying cosmological scalar field dynamics \cite{copeland1,aguirregabiria}. By recasting the evolution equations as autonomous systems, one can identify critical points corresponding to cosmological solutions (e.g., matter domination, de Sitter expansion) and analyze their stability. Previous studies considered DBI fields with inverse square \cite{copeland1}, exponential \cite{guoexp}, and hyperbolic potentials \cite{Quiros}, revealing rich phase space structures. In particular, Silverstein and Tong \cite{tong} showed that D3 brain motion in warped throats generates generalized DBI actions with non-canonical kinetic terms, offering new phenomenological possibilities.\\
In the context of cosmological models that address the acceleration of the Universe, scalar fields play an integral role. Specifically, the D-brane inflation (DBI) model has garnered considerable attention due to its ability to provide a self-consistent framework for scalar field dynamics in a high-energy regime. The DBI scalar field, when coupled with various types of potentials, can describe a wide range of cosmological phenomena, including inflationary dynamics and DE. \\
In this work, we have taken four types of DBI field potential, that is, exponential, power-law, free field (quadratic), and Higgs-like field, all of which naturally arise from sound microscopic descriptions, such as the effective field theory of string theory or other UV-complete theories. Even from a phenomenological point of view, these potentials have been widely studied as candidates for dark energy and are extremely consistent with current observations.

\section{Basic mathematical formalism of the DBI field theory}\label{sec_3.3} 

One of the main troubles of using string theory in cosmology directly is the so-called no-go theorem \cite{hao,chingangbam}, for wrapped products by compactifying the extra dimensions. \\
In string theory, it was predicted by Sen \cite{sen1,sen2,sen3} that there are 
Tachyon fields in both open and closed string theory. For more on open and closed string theory, one can refer to \cite{polchinski}.  Even though for closed string theory, the Tachyon fields are projected out in open string, they remain. Although one can use a spontaneous symmetry-breaking argument to get rid of Tachyon modes, one can still fully explain the reason for their existence. In bosonic string theory, if one uses Nambu- Goto action then it is almost impossible to quantize, in order to get meaningful quantization rules one has to invoke the conformal invariant Polyakov action, using the conformal field theory techniques one can quantize such an action which leads to the undesirable Tachyon modes, even though they violate casualty it can be shown that they are unstable. So, Tachyon modes are typically given by the DBI Lagrangian density, which has the following form,
\begin{equation}\label{4a}
\mathcal{L}_{DBI}=-V(\phi)\sqrt{1+\partial \phi^2} \,\,\text{,}
\end{equation}
where $V(\phi)$ is a potential function for the scalar field and  $\partial\phi^2=\partial^{\mu}\phi\partial_{\mu}\phi$ denotes the kinetic term for Tachyon fields.\\
From the Lagrangian, one can find the field equation for the Tachyon field from the Euler-Lagrangian equation as,
\begin{equation}\label{4b}
    \frac{\Ddot{\phi}}{1-\dot{\phi}^2}+3H\dot{\phi}+\frac{V_{,\phi}}{V}=0 \,\,\text{.}
\end{equation}
which is called the modified Klein-Gordon equation for the DBI field.\\
For the full Lagrangian density in natural units becomes,
\begin{equation}
    \mathcal{L}_{tot}= \frac{1}{2}R+\mathcal{L}_{DBI} \,\,\text{.}
\end{equation}
So, the Einstein-Hilbert action becomes,
\begin{equation}
    S= \int \left[\frac{1}{2} R + \mathcal{L}_{DBI}\right] \sqrt{-g} d^4x \,\,\text{.}
\end{equation}
Now, one can vary this action with respect to the metric $g_{\mu\nu}$ and recover Einstein's field equations;
\begin{equation}
    G_{\mu\nu}= R_{\mu\nu}-\frac{1}{2}R g_{\mu\nu}= T_{\mu\nu} \,\,\text{.}
\end{equation}
If one puts the flat FLRW metric, which is given as,
\begin{equation}
    ds^2= -dt^2+a(t)^2[dx^2+dy^2+dz^2] \,\,\text{.}
\end{equation}
One recovers the following Friedman equations given as,
\begin{equation}\label{4c}
    3H^2= \rho_{\phi} + \rho_{m} \,\,\text{.}
\end{equation}
and,
\begin{equation}\label{4d}
    \dot{H}=-\frac{1}{2} [\rho_{\phi} + p_{\phi}+\rho_{m}+p_{m}] \,\,\text{.}
\end{equation}
The energy density $\rho_{\phi}$ and pressure $p_{\phi}$ for the DBI scalar field are given by,
\begin{equation}\label{4e}
    \rho_{\phi}=\frac{V}{\sqrt{1-\dot{\phi}^2}} \,\,\text{,}
\end{equation}
and,
\begin{equation}\label{4f}
    p_{\phi}=-V\sqrt{1-\dot{\phi}^2} \,\,\text{.}
\end{equation}
respectively. In addition, the corresponding equation for the state parameter $\omega_{\phi}$ and the density parameter $\Omega_{\phi}$ for the DBI scalar field is defined as
\begin{equation}\label{w}
    \omega_{\phi}=\frac{p_{\phi}}{\rho_{\phi}}=\dot{\phi}^2-1 \,\,\text{.}
\end{equation}
and,
\begin{equation}\label{density}
    \Omega_{\phi}=\frac{V}{3H^2\,\sqrt{1-\dot{\phi}^2}} \,\,\text{.}
\end{equation}
respectively.\\
Further, by rewriting the Friedmann equations in the presence of a DBI scalar field and pressure-less matter ($p_m=0$), we obtain,
\begin{equation}
\label{RFE1}
    3H^2-\rho_m=\rho_{\phi}=\frac{V}{\sqrt{1-\dot{\phi}^2}} \,\,\text{.}
\end{equation}
and,
\begin{equation}
\label{RFE2}
    2\dot{H}+3H^2=-p_{\phi}-p_m=-p_{\phi}=V \sqrt{1-\dot{\phi}^2} \,\,\text{.}
\end{equation}
Using Eq.~\eqref{RFE1} and Eq.~\eqref{RFE2}, the scalar field potential can be expressed as
\begin{equation}\label{V}
    V^2=(2\dot{H}+3H^2)(3H^2-\rho_m) \,\,\text{,}
\end{equation}
and the kinetic term satisfies the following relation,
\begin{equation}\label{KE}
    (1-\dot{\phi}^2)=\frac{(2\dot{H}+3H^2)}{(3H^2-\rho_m)} \,\,\text{.}
\end{equation}
For the case of pressureless matter $p_m=0$, the matter density is given by $\rho_m(z)=3H_0\Omega_{m,0}(1+z)^3$, where the subscript zero refers to values measured at the present epoch. Here, $z$ represents the redshift, which is defined as $z=\frac{1}{a}-1$, with $a(t)$ being the scale factor. \\
In this study, we need to express the time dependence of the equation in terms of redshift $z$ to investigate the observational study. This can be achieved using the relation,
\begin{equation}\label{relation}
    \frac{d}{dt}=-(1+z)H(z)\frac{d}{dz} \,\,\text{.}
\end{equation}

\section{Model-independent reconstructions}\label{sec_3.4} 
\subsection{Observational data and priors}
For the Gaussian process reconstruction, we employ two datasets. The first consists of 58 Observational Hubble Data (OHD) points, comprising 26 points from radial BAO measurements and 32 points from CC. The OHD is taken in the redshift interval $0.07 \le z \le 2.42$. The second dataset includes 5 recently released data points from the DESI .\\
  The CC data consist of 32 measurements derived from the differential age evolution of passively evolving galaxies, which yield $H(z)$ in a model-independent fashion \cite{Jimenez2002, Moresco2016}. The rest is radial BAO data, consisting of 26 measurements based on the position of the baryon acoustic oscillation peak as a standard ruler, calibrated by the sound horizon scale \cite{Gaztanaga2009, Alam2017}. It should be noted that the CC data are based solely on the distance ladder, so no assumptions have been made about cosmological models, but it cannot place constraints on the $H(z)$ function. However, for the radial BAO data, one must assume the position of the sound horizon to obtain the pick position. So, this makes the 26 radial BAO data model dependent. So, our combined OHD datasets break the model's degeneracy by noting that CC data are based solely on the distance ladder. The full 58 data points with proper citation can be found in the paper by Gadbail et al. \cite{gaurav1}.\\ 
  We also used the five data points from the latest DESI survey \cite{desi,desi1}. DESI employs a multi-tracer approach, which includes: 1. the Galaxy Survey (Including the Bright Galaxy Survey (BGS), Luminous Red Galaxies (LRGs), and Emission Line Galaxies (ELGs), covering redshifts up to 1.6); 2. Quasars (Distant, luminous objects with redshifts up to 2.1) and finally 3. Lyman-$\alpha$ Forest: The absorption features in the spectra of quasars caused by intervening hydrogen clouds, allowing measurements up to redshift 2.33.\\
  By analyzing more than 5.7 million objects in a 7500-square-degree area, DESI has achieved a combined precision of approximately 0.52 percent in BAO measurements. The highest significance of BAO detection is 9.1$\sigma$ at an effective redshift of 0.93, with a constraint of 0.86 percent placed on the BAO scale. One can also note that, at redshift 2.33, DESI measures the transverse comoving distance $(D_L)$ and the Hubble parameter $(H_0(z))$. These make it an ideal candidate for the distance ladder, as two independent observations of the cosmological parameters break the degeneracy in the measurements.\\ 
  To initiate GP reconstruction, suitable priors are required. We adopt the following values based on the Planck 2018 results \cite{planck}:
\begin{equation}
H_0 = 67.4 \pm 0.5 \, \text{km s}^{-1} \text{Mpc}^{-1}, \quad \Omega_{m,0} = 0.315 \pm 0.007 \,\,\text{.}
\end{equation}
These priors provide a well-constrained baseline for the iterative reconstruction process.  

\subsection{Gaussian process}
Gaussian Processes (GPs) are widely used in machine learning as a non-parametric Bayesian method to reconstruct continuous functions and their derivatives directly from noisy data \cite{Rasmussen2005}. The key strength of GP lies in its model-independent nature: instead of assuming a specific functional form to fit the data, GP infers the function by placing a prior over functions, guided by a covariance kernel. This makes it particularly suitable for cosmology, where one often aims to reconstruct quantities such as the Hubble parameter $H(z)$, its derivatives, or the DE equation of state without assuming any specific cosmological model \cite{Seikel2012, Seikel2013, Mehrabi2021}.

Reconstructing the DBI scalar field potential directly from observational 
data cosmic chronometers (CC), baryon acoustic oscillations (BAO), and 
DESI requires a method that imposes no prior functional form on 
$V(\phi)$, since theoretical constraints on the potential shape remain 
largely unconstrained for exotic dark energy models. Gaussian process (GP) 
regression provides precisely such a non-parametric framework, allowing the 
data themselves to drive the reconstruction. In this framework, the unknown 
function $f(x) \equiv H(x)$ is modeledThe  as a Gaussian process,

\begin{equation}
    f(x) \sim \mathcal{GP}\bigl(m(x),\, k(x,x')\bigr),
\end{equation}

\noindent where $m(x) = \mathbb{E}[f(x)]$ is the prior mean function, 
typically chosen as a constant or low-order polynomial, and $k(x,x')$ is 
the covariance kernel. A standard choice is the squared-exponential form,

\begin{equation}
    k(x,x') = \sigma_f^2\exp\!\left[-\frac{(x-x')^2}{2\ell^2}\right],
\end{equation}

\noindent where $\ell$ controls the correlation length scale and $\sigma_f^2$ 
the output variance, together encoding the smoothness of the reconstructed 
function. When $\ell$ is too small, the kernel approaches a Dirac delta, 
forcing exact interpolation of every training point and capturing 
observational noise rather than the true underlying trend. This overfitting 
is naturally suppressed by maximising the marginal log-likelihood,

\begin{equation}
    \log\mathcal{L} = -\frac{1}{2}\mathbf{y}^\top(K+C)^{-1}\mathbf{y}
    -\frac{1}{2}\log|K+C| + \mathrm{const},
\end{equation}

\noindent where the log-determinant term acts as an automatic complexity 
penalty proportional to the effective volume of the function space, 
implementing Occam's razor without manual regularisation. Once the 
hyperparameters $\{\ell, \sigma_f\}$ are optimised via likelihood 
maximisation, the posterior predictive mean 
$f_* = \mathbf{k}_*^\top(K+C)^{-1}\mathbf{y}$ and variance 
$\sigma_*^2 = k(x_*,x_*) - \mathbf{k}_*^\top(K+C)^{-1}\mathbf{k}_*$ 
deliver uncertainty-quantified reconstructions at arbitrary redshifts, 
making GP a robust and principled tool for probing the expansion history 
of the Universe and the nature of dark energy.

\begin{figure}[H]
\centering
\includegraphics[width=0.47\textwidth]{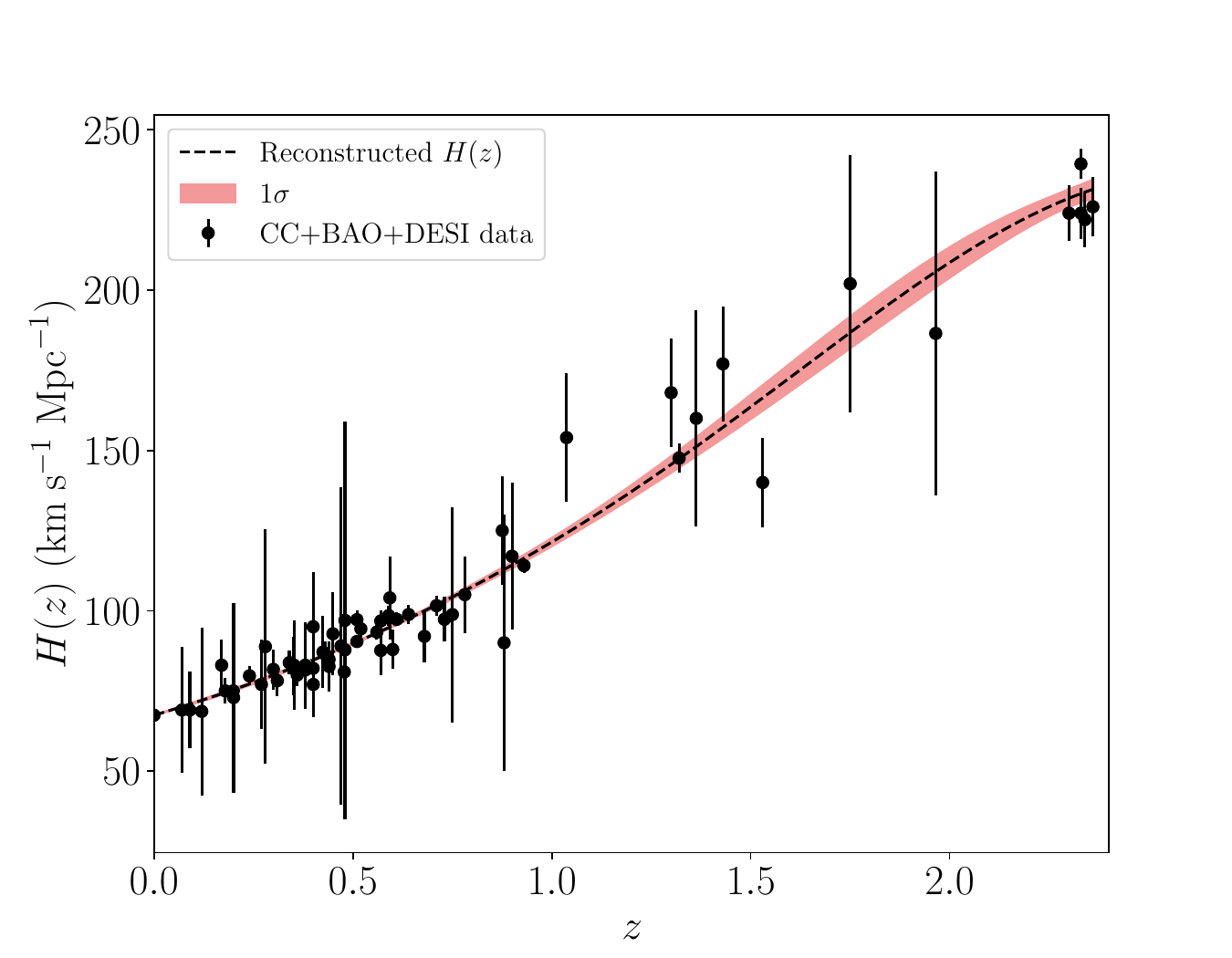}
 \hspace{0.15in} 
\includegraphics[width=0.47\textwidth]{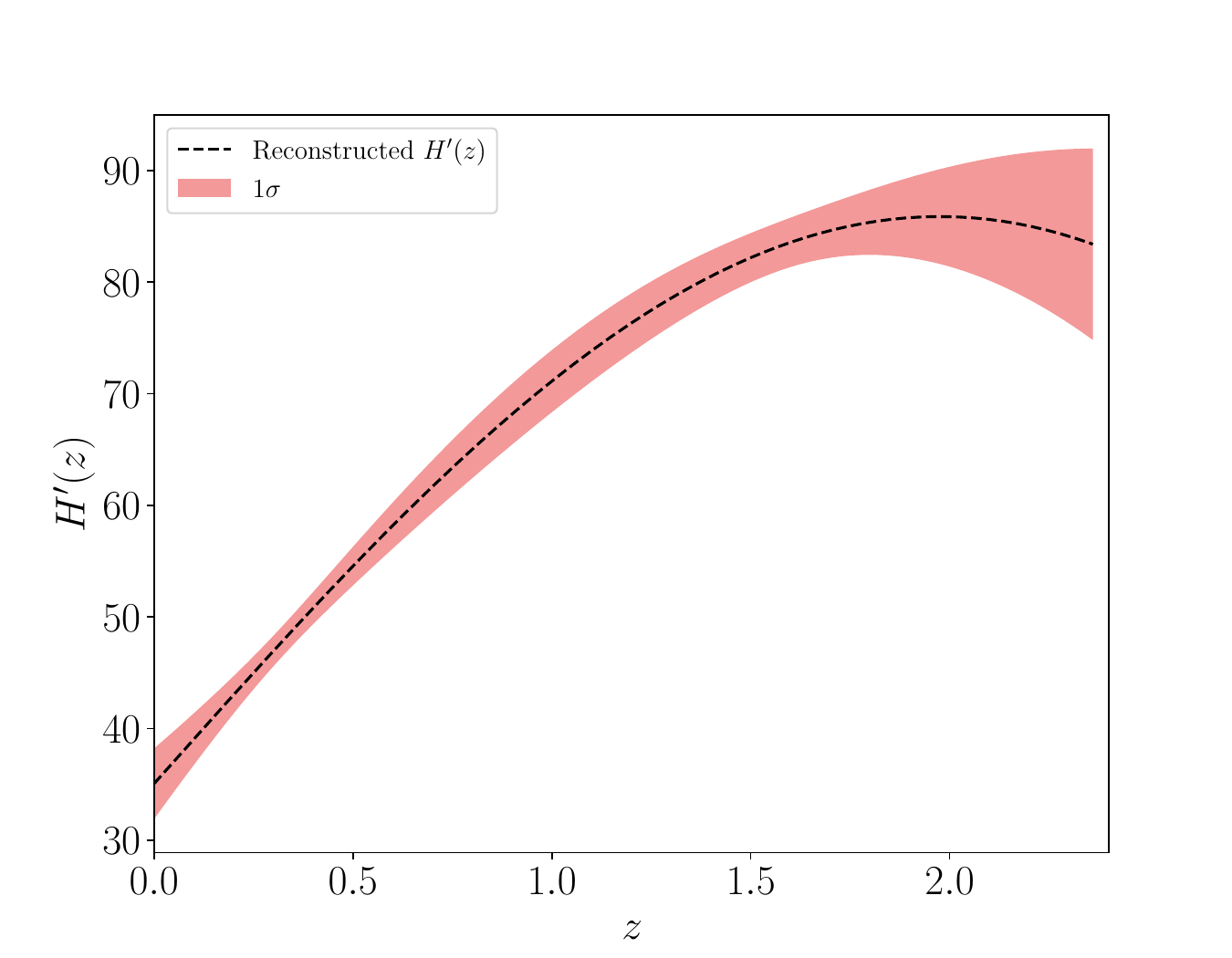}
\\ 
\caption{In the left panel, we have plotted the reconstructed mean $H(z)$ (the black dashed graph) using 32 CC + 26 radial BAO + 5 DESI data points, the black solid lines show the errors of data points, and the light red region shows the $1\sigma$ uncertaintfieldy associated with the GP. In the right panel, we have reconstructed the mean $H'(z)$ (black dashed line) with the light red region showing the $1\sigma$ uncertainty associated with the GP.}
\label{fig_3.1} 
\end{figure}

Importantly, derivatives of $H(z)$ are obtained analytically by differentiating the kernel,
\begin{equation*}
    \frac{\mathrm{d}^m}{\mathrm{d}x^m}\frac{\mathrm{d}^n}{\mathrm{d}x'^n}k(x,x') \,\,\text{,}
\end{equation*}
retains Gaussian form, allowing for direct reconstruction of $H'(z)$ and $H''(z)$ with uncertainty estimates \cite{Seikel2013}. This feature is crucial for diagnosing cosmic acceleration and for model selection.\\ 
Finally, we implement this methodology using the publicly available \texttt{GaPP} code \cite{Seikel2012}, which efficiently handles matrix inversions ($\mathcal{O}(N^3)$) and hyperparameter optimization to reconstruct $H(z)$ and its redshift derivative from the combined CC+BAO + DESI data. The results, along with the corresponding $1\sigma$ confidence bands, are presented in Fig.~\ref{fig_3.1}.\\  

\subsection{Reconstructed $w_{\phi}$ and $\Omega_{\phi}$}
\begin{figure}[H]
\centering
\includegraphics[width=0.47\textwidth]{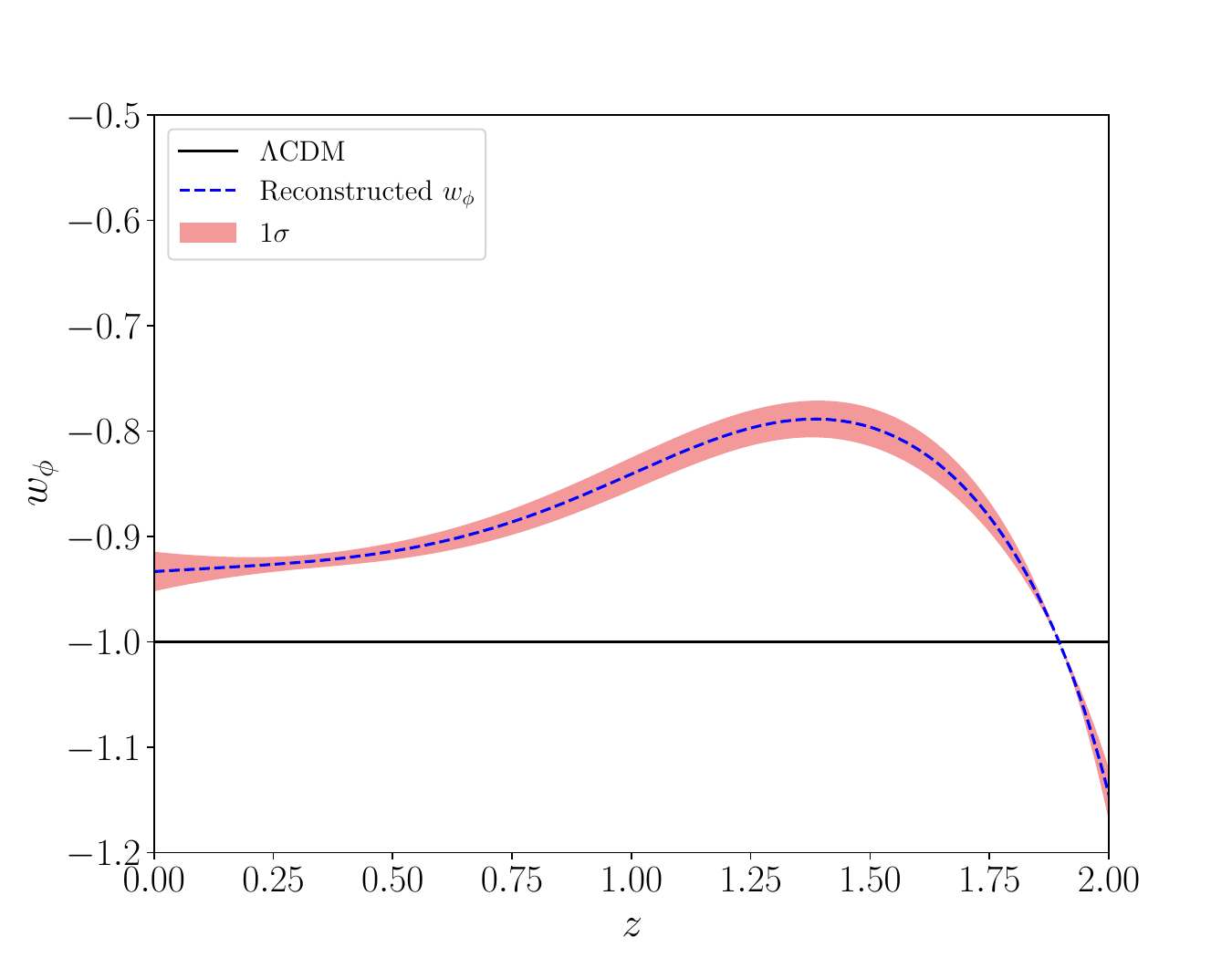}
 \hspace{0.15in} 
\includegraphics[width=0.47\textwidth]{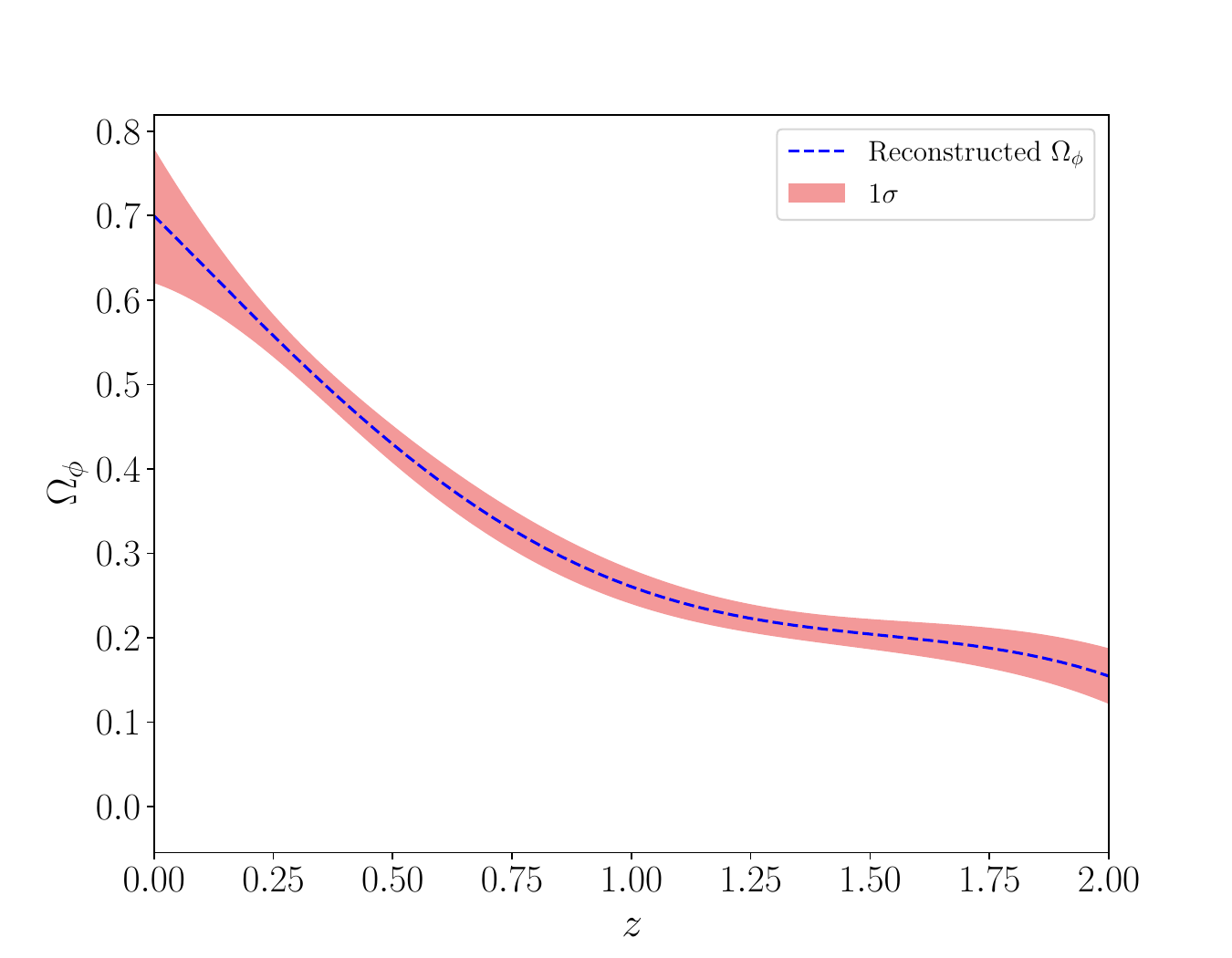}
\\ 
\caption{In the left panel, we have plotted the reconstructed mean equation of state $\omega_\phi$ (the black dashed graph) using 32 CC + 26 radial BAO + 5 DESI data points, the light red region shows the $1\sigma$ uncertainty associated with the GP. We have also plotted the $\Lambda$CDM EoS, which is -1 for comparison purposes. In the right panel, we have reconstructed the mean total DE density $\Omega_\phi$ (black dashed line) with the light red region showing the $1\sigma$ uncertainty associated with the GP. }
\label{fig_3.3} 
\end{figure}

In this subsection, we reconstructed the DE EoS parameter $w_{\phi}$ and the DE density parameter $\Omega_{\phi}$ using the above reconstructed Hubble function and its derivative.\\
Using the value of the matter density parameter and Eq.~\eqref{KE} in Eq.~\eqref{w} and Eq.~\eqref{density}, we get the simplified $w_{\phi}$ and $\Omega_{\phi}$ as
\begin{equation}
    w_{\phi}=\frac{-2(1+z)H(z)\frac{dH}{dz}+3H^2}{3H^2-3H_0^2\,\Omega_{m,0}(1+z)^3} \,\,\text{,}
\end{equation}
and,
\begin{equation}
    \Omega_{\phi}=1-\frac{H_0^2\,\Omega_{m,0}(1+z)^3}{H^2} \,\,\text{.}
\end{equation}
To plot the above quantities, we utilized the reconstructed functions $H(z)$ and $H'(z)$, together with the assumed prior values for $H_0$ and $\Omega_{m,0}$. The resulting reconstructions of $w_{\phi}(z)$ and $\Omega_{\phi}(z)$ with 1$\sigma$ error are presented in Fig.~\ref{fig_3.3}. From the analysis, we obtain the current value of the DE equation of state as $w_{\phi0} = -0.9333 \pm 0.0062$, and the present value of the DE density parameter as $\Omega_{\phi0} = 0.6998 \pm 0.1593$. These results are in good agreement with the standard $\Lambda$CDM model, where $w_{\Lambda0} = -1$ and $\Omega_{\Lambda0} \approx 0.7$, indicating the consistency of our reconstruction approach with the current cosmological paradigm. Even though, as we have seen from Eq.~\eqref{w}, in the slow-roll approximation, that is, $\dot{\phi} \ll 1$, $\omega_{\phi}$ always becomes $\omega_{\phi} \approx -1$. What our analysis shows is how fast it reaches $-1$ without invoking any prior model.

\begin{figure}[H]
\centering
\includegraphics[width=0.39\textwidth]{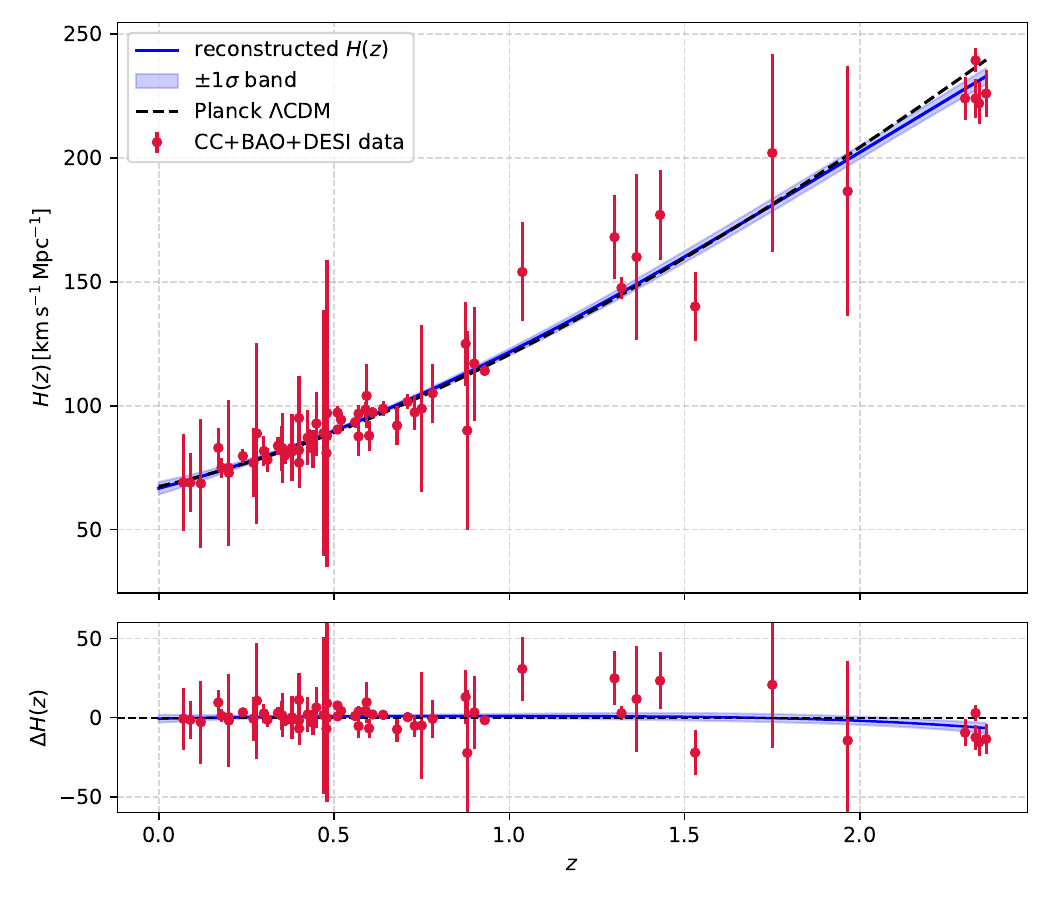}
 \hspace{0.15in} 
\includegraphics[width=0.56\textwidth]{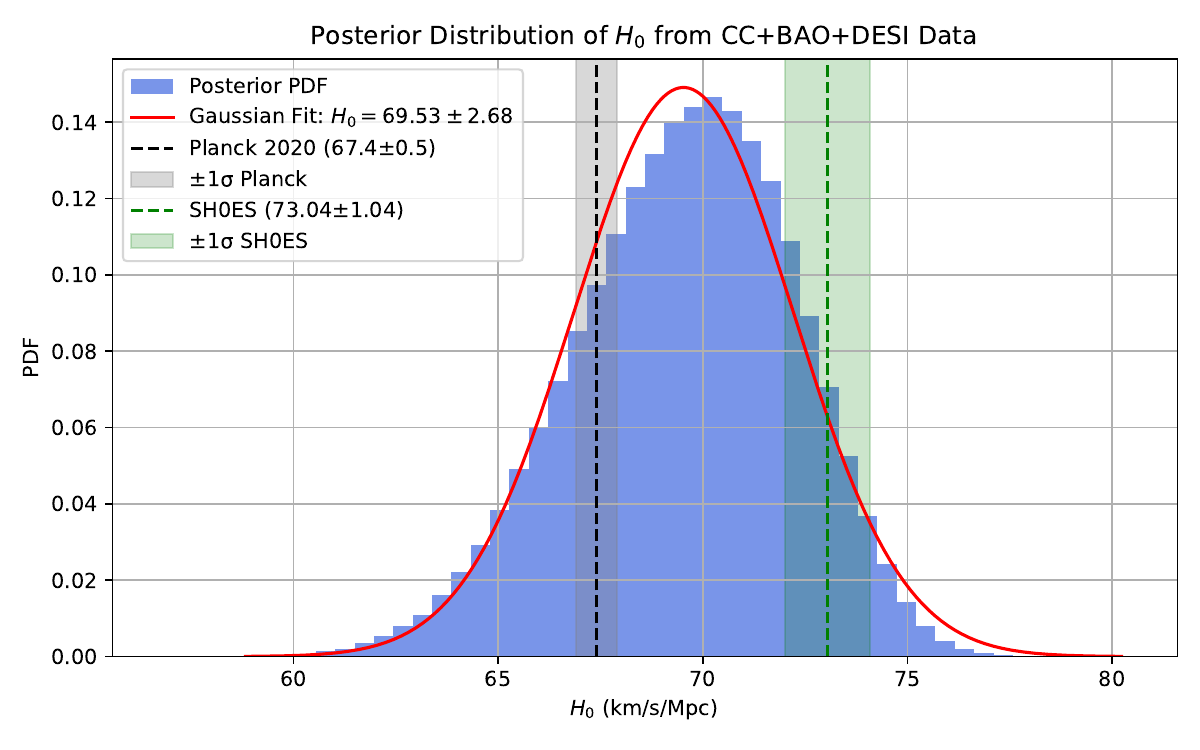}
\\ 
\caption{In the left panel, we have plotted the reconstructed mean $H(z)$ (the blue line) using 32 CC + 26 radial BAO + 5 DESI data points, the red solid points show the errors of data points, and the light blue region shown the $1\sigma$ uncertainty associated with the GP. In the right panel, the posterior probability distribution of the Hubble constant $H_0$ is reconstructed from the combined CC + BAO + DESI dataset. The histogram represents the Monte Carlo realizations of $H_0$, while the solid curve denotes the best-fit Gaussian with $H_0 = 69.53 \pm 2.68$ km s$^{-1}$ Mpc$^{-1}$. The vertical shaded regions correspond to the $1\sigma$ intervals from the Planck 2020 ($67.4 \pm 0.5$) and SH0ES ($73.04 \pm 1.04$) measurements, respectively, shown for comparison.}
\label{fig_3.2} 
\end{figure}

\section{Reconstruction of DE scalar field potential}\label{sec_3.5} 
In this section, we reconstruct the DE scalar field potential using the GP framework described above. To perform this reconstruction, we first outline the methodological setup; subsequently, we apply the chi-square fitting method for model selection; and finally, we constrain the model parameters using the MCMC technique as detailed in the following subsections.
\subsection{Method setup}
Now, using the relation Eq.~\eqref{relation} and the value of matter density $\rho_m(z)$, we rewrite the scalar field potential Eq.~\eqref{V} in terms of redshift $z$ as,
\begin{equation}\label{V(z)}
    \left(V(z)\right)^2=-6\,H(1+z)\frac{dH}{dz}\left(H^2-H_0^2\,\Omega_{m,0}(1+z)^3\right)+9H^2\left(H^2-H_0^2\,\Omega_{m,0}(1+z)^3\right) \,\,\text{,}
\end{equation}
and the kinetic term of the scalar field Eq.~\eqref{KE} becomes,
\begin{equation}
    (\dot{\phi}(z))^2=\frac{-2H(1+z)\frac{dH}{dz}+3H_0^2\Omega_m(1+z)^3}{3H_0^2\Omega_m(1+z)^3-3H^2} \,\,\text{.}
\end{equation}

Using relation Eq.~\eqref{relation}, the kinetic term can be rewritten as,
\begin{align}
 \frac{d\phi}{dz}&=\frac{1}{(1+z)\,H}\left(\frac{-2H(1+z)\frac{dH}{dz}+3H_0^2\Omega_{m,0}(1+z)^3}{3H_0^2\Omega_{m,0}(1+z)^3-3H^2}\right)^{\frac{1}{2}} \,\,\text{.}
 \end{align}
Since an analytic solution to the above differential equation is difficult to obtain, we proceed by seeking a numerical solution. To facilitate numerical integration, we derive a recursive relation between consecutive redshift points $z_i$ and $z_{i+1}$. This allows us to express $\phi(z_{i+1})$ in terms of $\phi(z_i)$, along with the values of $H(z_i)$ and $H'(z_i)$, as following,
 \begin{equation}\label{Recc}
   \phi(z_{i+1})=\phi(z_i)+\frac{(z_{i+1}-z_i)}{(1+z_i)\,H(z_i)}\left(\frac{-2H(1+z_i)\frac{dH}{dz}\big|_{z_i}+3H_0^2\,\Omega_{m,0}(1+z_i)^3}{3H_0^2\,\Omega_{m,0}(1+z_i)^3-3H(z_i)^2}\right)^{\frac{1}{2}} \,\,\text{,}
\end{equation}
where
\begin{equation*}
    \phi'(z) = \frac{\phi(z + \Delta z) - \phi(z)}{\Delta z} \,\,\text{,}
\end{equation*}
for small $\Delta z$.\\
According to Eq.~\eqref{Recc}, in order to determine the values of $\phi(z)$, it is necessary to know the values of $H(z)$, $H'(z)$, $H_0$, and $\Omega_{m,0}$. To obtain $H(z)$ and $H'(z)$ in the redshift range $z \in [0,\,2.5]$, we employed the GP reconstruction method using the combined CC+BAO + DESI dataset. Parameters $H_0 = 67.4 \pm 0.5 \, \text{km s}^{-1} \text{Mpc}^{-1}$ and $\Omega_{m,0} = 0.315 \pm 0.007$ are adopted as priors, consistent with recent observational constraints.
\subsection{Model selection and chi-square fitting}
Using Eq.~\eqref{V(z)} and Eq.~\eqref{Recc}, we reconstruct the dimensionless scalar field potential $\mathcal{V}(\phi)=V/3H_0^2$ of DE and obtain the dataset in the form of $\mathcal{V}(\phi)$ versus $\phi$, along with its associated uncertainties $\sigma_{\mathcal{V}}$. To identify the most suitable model corresponding to our reconstructed potential, we performed a chi-square fitting analysis using standard theoretical scalar field models. The chi-square statistic is given by,
\begin{equation}
\chi^2 = \sum_i \frac{\left(\mathcal{V}_{SM}(\theta_s,\phi_i) - \mathcal{V}_{obs}(\phi_i)\right)^2}{\sigma_{\mathcal{V}}^2}\,\,\text{,}
\end{equation}  
where $\mathcal{V}_{SM}$ is the standard theoretical scalar field potential model, $\theta_s$ is the model parameter space, and $\mathcal{V}_{obs}$ are the reconstructed scalar field potential data.\\
 Moreover, to assess which standard theoretical model provides a better fit to reconstruction $\mathcal{V}(\phi)$, we calculated the reduced values of chi-square ($\chi_{\text{red}}^2$) between the reconstructed $\mathcal{V}(\phi)$ and various standard theoretical models. The $\chi_{\text{red}}$ is computed as,
 \begin{equation}
     \chi^2_{\text{red}} =\frac{1}{N-P} \sum_i \frac{\left(\mathcal{V}_{SM}(\theta_s,\phi_i) - \mathcal{V}_{obs}(\phi_i)\right)^2}{\sigma_{\mathcal{V}}^2} \,\,\text{,}
 \end{equation}
where $N$ represents the number of data points and $P$ represents the number of parameters in the model.\\
In this work, we consider four different standard scalar field potentials, namely, the exponential, power-law, free field, and Higgs potentials, to compare against our reconstructed dataset. Each potential is fitted to the reconstructed $\mathcal{V}(\phi)$ data, and the quality of fit is assessed through the reduced chi-square values. In the following, we summarize the properties and interpretations of each model:
\begin{enumerate}
    \item \textbf{Exponential potential ($\mathcal{V}_0+\mathcal{V}_1e^{-\lambda\phi}$):} The exponential potential is one of the most studied scalar field potentials, as this potential is known for its tracker solution-like behavior, and in recent years it also provides justification for the swampland conjecture \cite{expnentiallavinia}. In our work, the coefficients of the exponential potential give a very good estimation with the observational dataset. In addition, we find that the exponential potential provides a good fit to the reconstructed data, with a reduced chi-square value of $0.7694$. This indicates that the model is statistically consistent with the data and captures its behavior without significant deviations. However, we note that an additional one $\mathcal{V}_0$ is needed for the data to fit. It shows that not just quintessence potential but a nonzero "zero-point" energy is needed for observational consistency and $\mathcal{V}_0$ just mimics the $\Lambda$ term in $\Lambda$CDM cosmology while the $\mathcal{V}_1e^{-\lambda \phi}$ denotes the quintessence potential which is responsible for such a low value of cosmological constant. It is also evident from the MCMC analysis that there is a negative correlation between $\mathcal{V}_0$ vs $\mathcal{V}_1$ and $\lambda$ which is reasonable for a certain cosmological constant if $\mathcal{V}_0$ is higher than $\mathcal{V}_1$ it would be lower and $\lambda$ would be higher so that overall potential does not change observational $\Lambda$. However, there is an additional positive correlation between $\mathcal{V}_1$ and $\lambda$ which gives tight constraints on the values of $\lambda$ and $\mathcal{V}_1$.
\item  \textbf{Power-law potential ($\mathcal{V}_0+\mathcal{V}_1 \phi^{\lambda}$):} Power-law potential is another widely studied potential in the context of cosmology, as the study of power law potential is very easy in quantum field theory and as Peebles and Ratra have shown \cite{Peebles:1988,Ratra:1988}, power-law potential in the de Sitter background could be a very good candidate for quintessence. Although this potential, which does not rely on a tracker solution-like behavior, yields an excellent fit to the reconstructed data, with $\chi^2_{\text{red}}=0.9888$. As in the exponential case, the constant term $\mathcal{V}_0$ plays a crucial role in matching the overall energy scale of the potential, again reflecting the need for a nonzero ``zero-point" energy. For the power-law potential, as we can see, there is a negative correlation between $\mathcal{V}_1$, $\mathcal{V}_0$, and $\lambda$, which is again resolvable as the $\Lambda$ for the current Universe is fixed from the observation. However, there is a positive correlation between $\mathcal{V}_0$ and $\lambda$, which gives us the constraints on the $\lambda$ for a fixed $\mathcal{V}_0$ and $\mathcal{V}_1$.
\item \textbf{Free field potential ($\mathcal{V}_0+ \frac{m^2}{2}\phi^2$):} This potential was proposed by Linde \cite{Linde} as a candidate for chaotic inflation. Later, the study for the free field has been extended by Urna-Lopez et al. \cite{freefieldlopez} to study the details of the phenomenology for the late time cosmology. This simple quadratic form also provides a good fit to the reconstructed data, with a reduced chi-square value of $\chi^2_{\text{red}} = 1.2150$. Here too, the additional $\mathcal{V}_0$ term is needed to reconcile the model with observations, highlighting the role of vacuum energy in driving late-time cosmic acceleration. In this context, we are not concerned with the initial conditions during inflation--as emphasized by Linde \cite{Linde} and others--but rather focus on the late-time dynamics of the Universe. There we can see that there is a strong negative correlation between $\mathcal{V}_0$ and $m$, which is very consistent with the fact that $\Lambda$ at present is fixed. 
\item \textbf{Higgs scalar field potential (\(\mathcal{V}_0 + \frac{m^2}{2} \phi^2 + \frac{\lambda}{4} \phi^4\)):} This potential, which includes both a quadratic and quartic term, naturally arises from the spontaneous symmetry breaking of the $Z_2$ group \cite{higgsamin}. This potential has a feature very similar to Higgs' mechanism \cite{higgs} in the weak interaction, where $U(1)$ gauge symmetry breaking gives the masses of the $ W$ and $Z$ bosons. This potential also has a spontaneous symmetry-breaking mechanism, which can lead to preferred vacuum solutions. With $\chi^2_{\text{red}}=0.8120$, it also fits the reconstructed data reasonably well. This potential is particularly significant in describing the transition of the Universe from one unstable false vacuum to the true vacuum via spontaneous symmetry breaking. As with the previous potentials, an additional \(\mathcal{V}_0\) term is necessary to be consistent with observational data, again indicating the need for a nonzero "zero-point" energy. The \(\mathcal{V}_0\) term mimics the \(\Lambda\) term in \(\Lambda\)CDM cosmology, while the \(\frac{m^2}{2} \phi^2 + \frac{\lambda}{4} \phi^4\) terms govern the dynamics of the scalar field, describing the transition between vacuum and spontaneous symmetry breaking. For the Higgs-like field, we see a negative correlation between $m$ and $\mathcal{V}_0$ and between $ m$ and $\lambda$, which one would again expect given that the $\Lambda$ of the current Universe is fixed. However, here too we can see a positive correlation between $\mathcal{V}_0$ and $\lambda$, which imposes a constraint on $\lambda$ for the Higgs field. 
\end{enumerate}
\begin{table}[]
\begin{center}
  \caption{Reduced chi-square ($\chi^2_{\text{red}}$) analysis of reconstructed $\mathcal{V}(\phi)$ against standard scalar field models.}
    \label{Table 2}
    \begin{tabular}{l c c }
\hline\hline 
Model      &  $\chi^2_{red}$ & Interpretation     \\[1ex] \hline\hline
Exponential: $\mathcal{V}_0+\mathcal{V}_1e^{-\lambda\phi}$  & $0.7694$ & Good fit  \\[1ex]
   Power-law: $\mathcal{V}_0+\mathcal{V}_1 \phi^{\lambda}$   & $0.9888$  & Excellent fit  \\[1ex] 
Free field: $\mathcal{V}_0+ \frac{m^2}{2}\phi^2$ & $1.2150$  & Good fit  \\[1ex] 
Higgs scalar field: $\mathcal{V}_0+\frac{m^2}{2}\phi^2+\frac{\lambda}{4}\phi^4$ & $0.8120$ & Reasonably good fit \\[1ex]
\hline
\end{tabular}
\end{center}
\end{table}

\begin{figure}[htbp]
\centering

\begin{minipage}{0.45\textwidth}
  \centering
  \includegraphics[width=\linewidth]{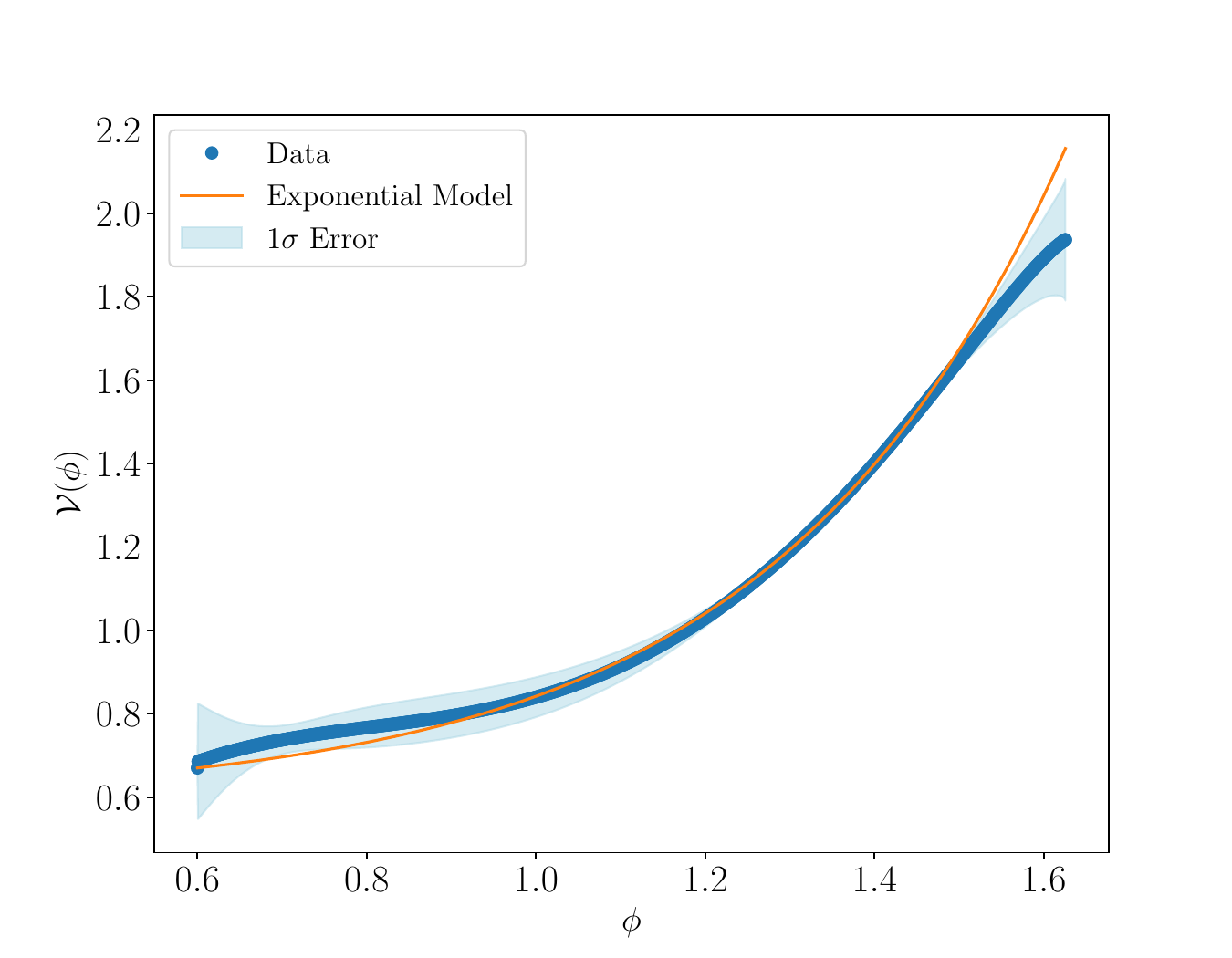}
  \vskip1ex
  \small\textbf{Exponential Potential}
\end{minipage}\hspace{0.15in}%
\begin{minipage}{0.45\textwidth}
  \centering
  \includegraphics[width=\linewidth]{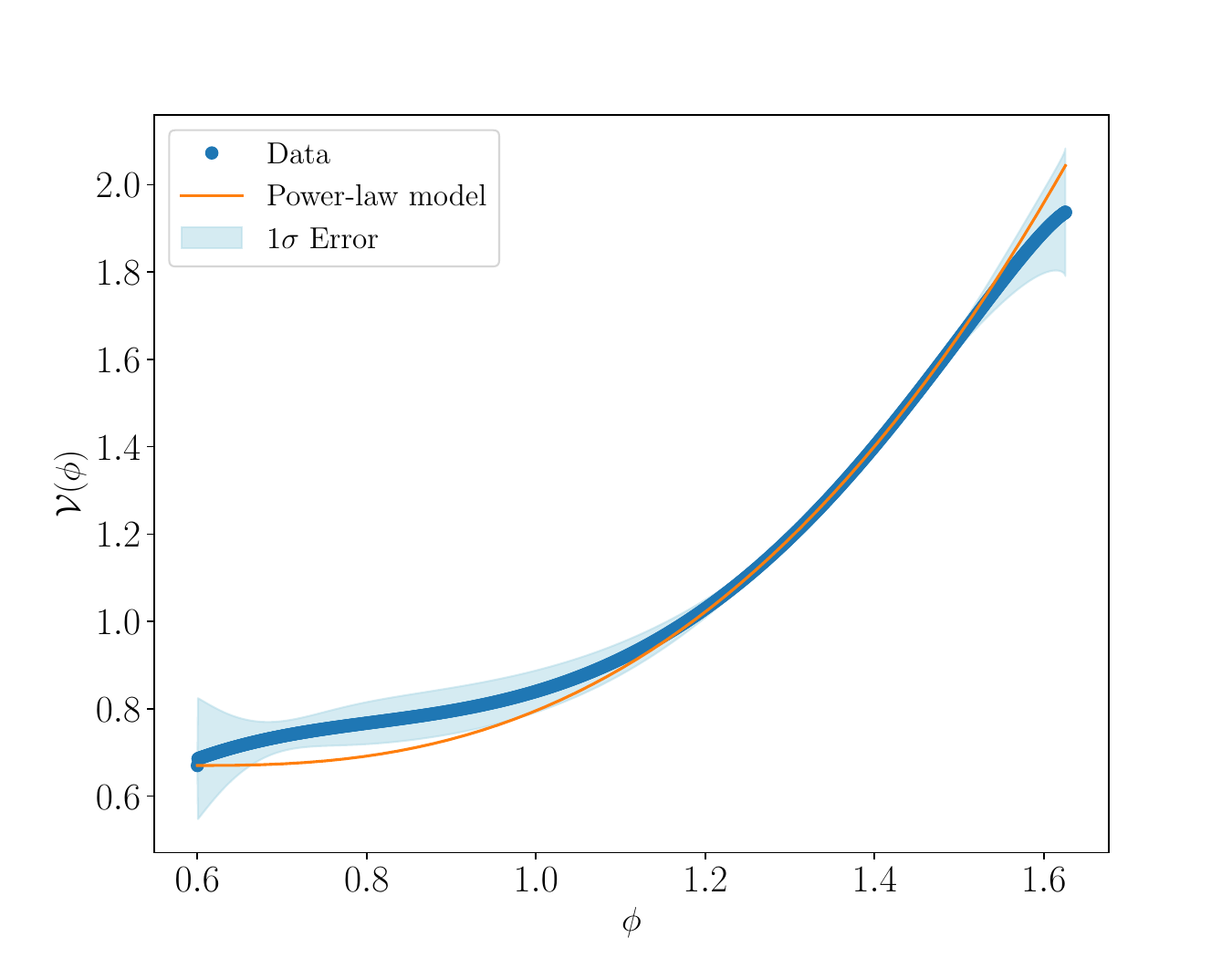}
  \vskip1ex
  \small\textbf{Power-law Potential}
\end{minipage}

\vspace{0.2cm}

\begin{minipage}{0.45\textwidth}
  \centering
  \includegraphics[width=\linewidth]{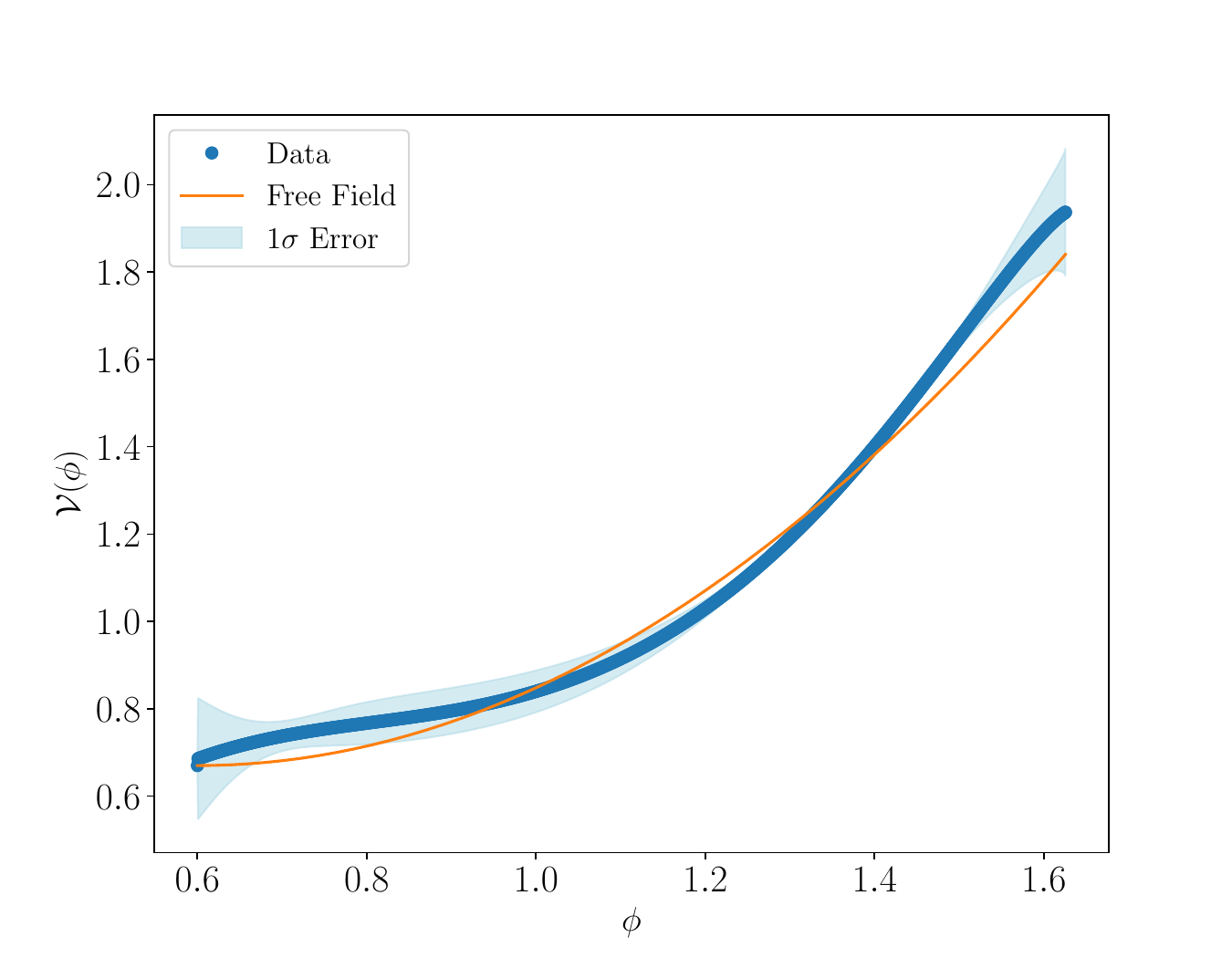}
  \vskip1ex
  \small\textbf{Free field Potential}
\end{minipage}\hspace{0.15in}%
\begin{minipage}{0.45\textwidth}
  \centering
  \includegraphics[width=\linewidth]{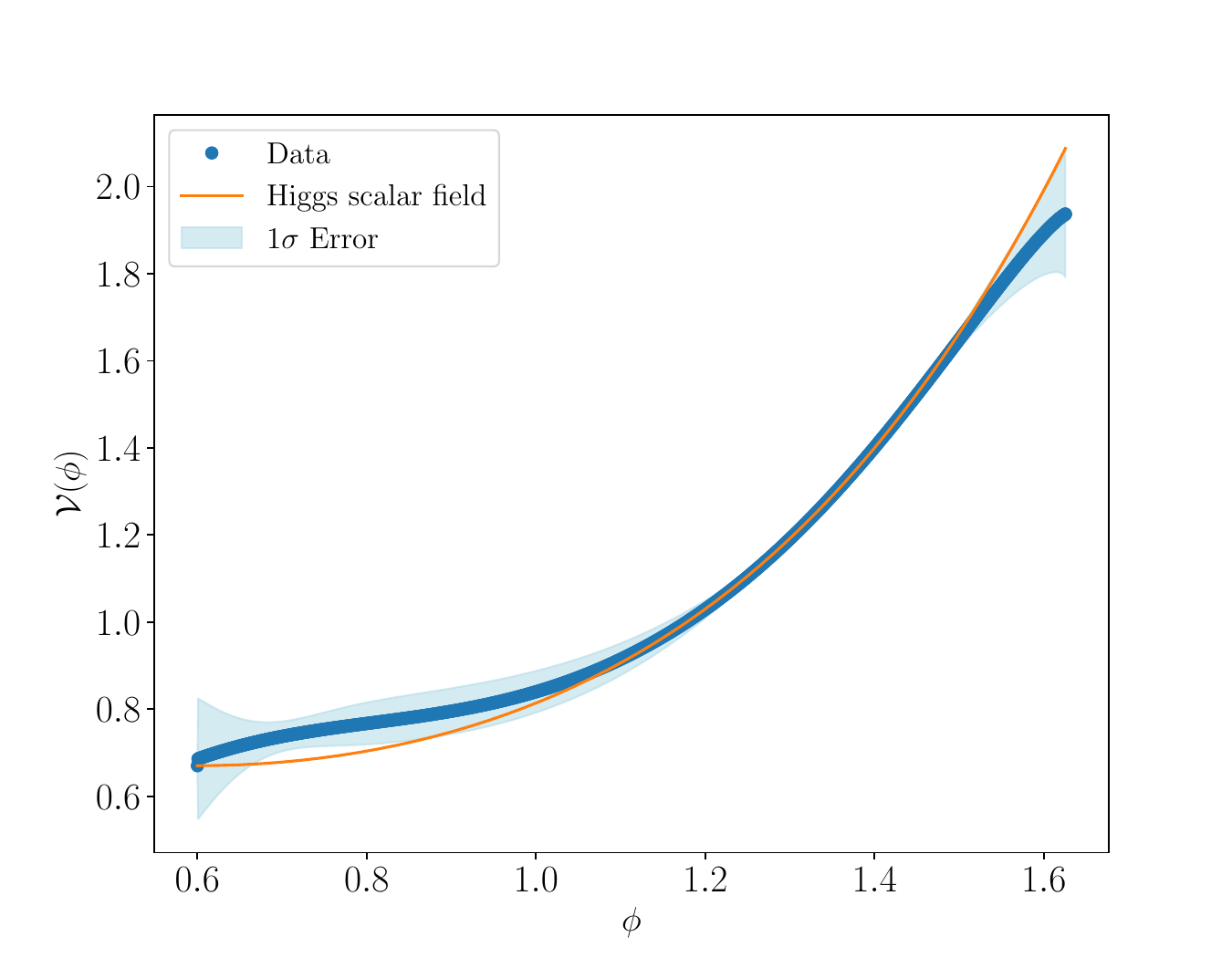}
  \vskip1ex
  \small\textbf{Higgs-like Potential}
\end{minipage}

\caption{\justifying This figure shows the reconstructed scalar field potential with $1\sigma$ uncertainty using the CC+BAO+DESI datasets via a GP. Standard theoretical scalar field potential models are fitted to the reconstructed data using chi-square curve fitting. The orange line represents the theoretical model, while the thick blue dots with error bars (light blue) denote the reconstructed data with $1\sigma$ uncertainty.}
\label{fig_3.4} 
\end{figure}

\subsection{MCMC analysis}
We have employed the MCMC technique as a robust Bayesian inference tool to constrain the free parameters of the theoretical model under consideration. Specifically, MCMC is used to determine the best-fit values of the scalar field potential parameters, with the potential reconstruction guided by GP techniques applied to observational data. The MCMC algorithm allows us to efficiently sample the posterior distribution of the model parameters by constructing a Markov chain whose equilibrium distribution corresponds to the posterior probability \( P(\theta | D) \propto \mathcal{L}(D | \theta)\pi(\theta) \), where \( \mathcal{L}(D | \theta) \) is the likelihood of the data \( D \) given the parameters \( \theta \), and \( \pi(\theta) \) denotes the prior distribution. 
The MCMC sampling is performed using the \texttt{emcee} ensemble sampler, enabling efficient exploration of the parameter space of the reconstructed scalar field potentials. This methodology ensures a statistically consistent estimation of the most probable model parameters compatible with current cosmological observations.\\
For this analysis, we use the reconstructed $\mathcal{V}(\phi)$ as a function of $\phi$, along with its associated uncertainties.
To set constants for the potential coefficient, we have relied on the reconstructed data and their error values from the GP rather than selecting particular dataset, which is standard practice. There are two key advantages to this approach. First, it does not rely directly on the observational data in the MCMC analysis. Instead, data uncertainties are effectively smoothed out by GP reconstruction, leading to a more stable and controlled input. Second, and more importantly, this method is completely non-parametric and model-independent. Unlike BAO data, which require an assumed value of the sound horizon for calibration, our approach avoids such assumptions. The reconstructed data are obtained purely from the GP, which yields a much smoother and well-behaved dataset compared to the raw measurements. This is evident from the $\chi^2$ analysis, which shows that the reconstructed data provide an excellent fit for all four types of potentials considered.



\begin{figure}[htbp]
\centering

\begin{minipage}{0.45\textwidth}
  \centering
  \includegraphics[width=\linewidth]{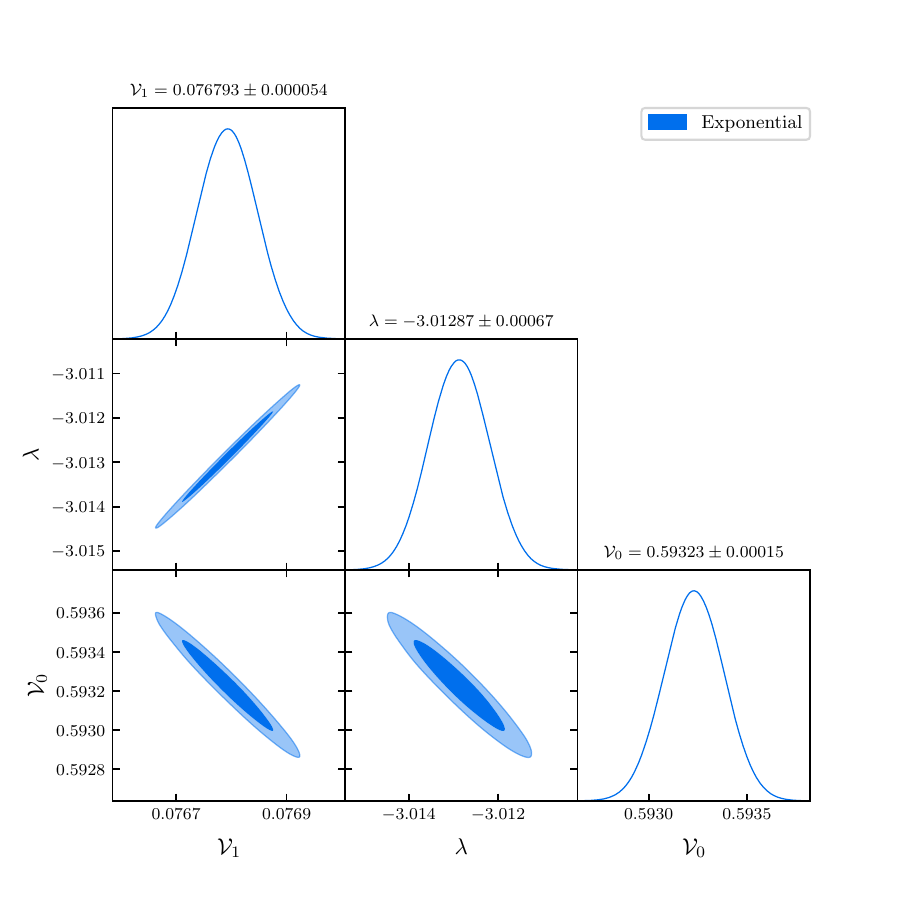}
\end{minipage}\hspace{0.15in}%
\begin{minipage}{0.45\textwidth}
  \centering
  \includegraphics[width=\linewidth]{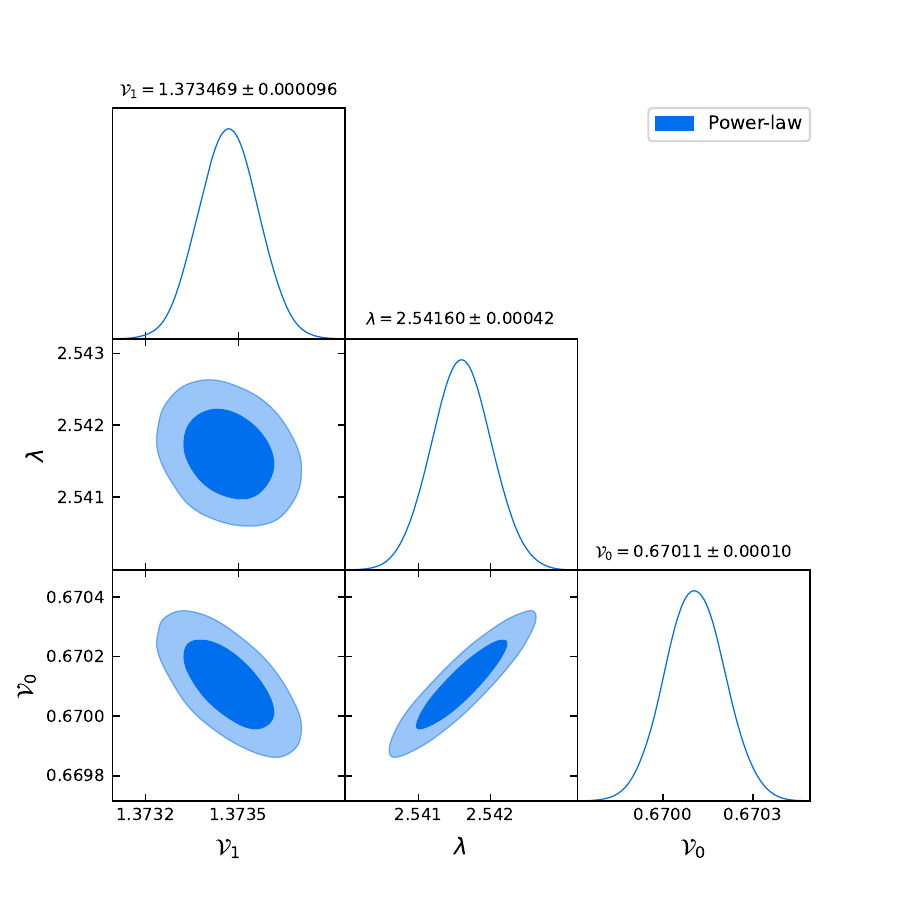}
\end{minipage}

\vspace{0.15in}

\begin{minipage}{0.45\textwidth}
  \centering
  \includegraphics[width=\linewidth]{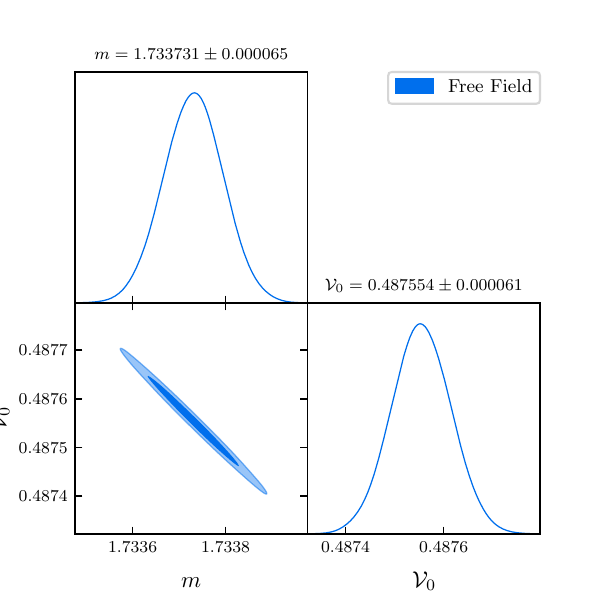}
\end{minipage}\hspace{0.15in}%
\begin{minipage}{0.45\textwidth}
  \centering
  \includegraphics[width=\linewidth]{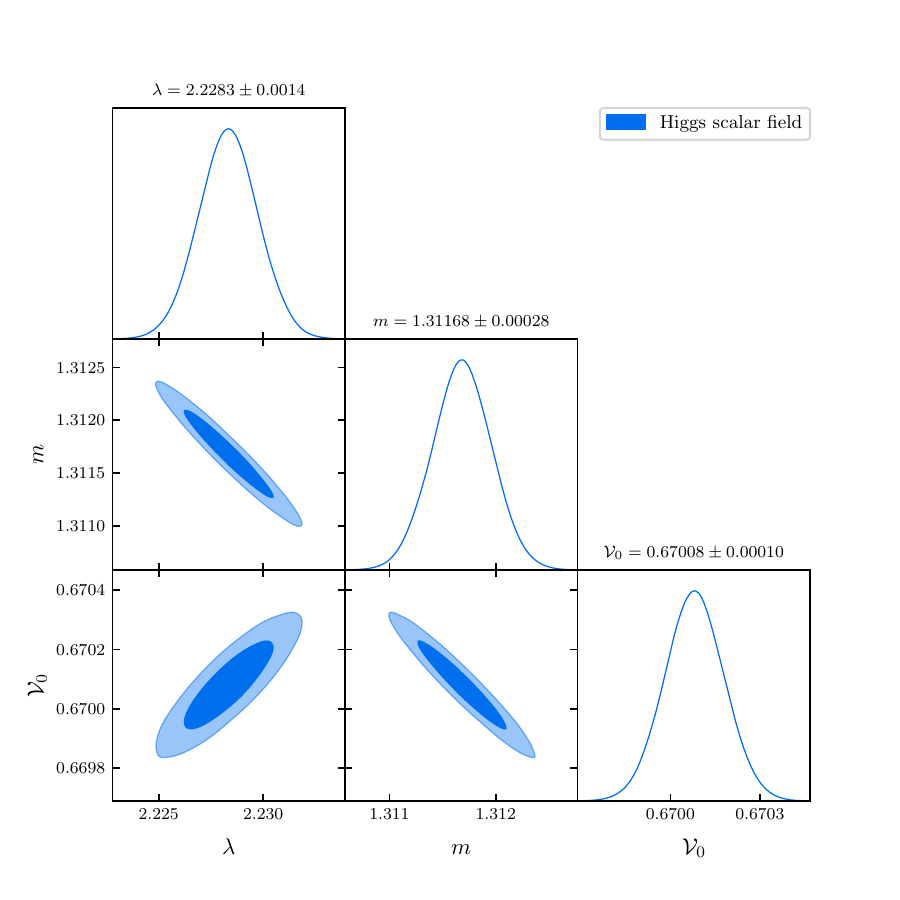}
\end{minipage}

\caption{\justifying Corner plots showing the correlations and posterior distributions of the free parameters for the exponential, power-law, free field, and Higgs-like scalar field potentials. The contours represent the $1\sigma$ and $2\sigma$ confidence regions, derived from the reconstructed scalar field potential data.}
\label{fig_3.5} 
\end{figure}

    

\section{Conclusion}\label{sec_3.6} 

In this study, we reconstruct the DE scalar field potential using the Gaussian process method within the DBI framework. To examine the behavior of DE, we also reconstruct the DE EoS parameter $\omega(z)$ and the DE density parameter $\Omega(z)$. The key strength of the GP method lies in its model independence; unlike traditional Bayesian analyses, our reconstruction does not rely on any assumptions about the underlying cosmological model. We would also like to note that the $\omega_{\phi}$ we get from the GP reconstruction is very close to -0.95, which is not just consistent with the $\Lambda$CDM model but phenomenologically justifies our choice of DBI field, as $\omega_{\phi}\approx-1$ when the field is very slow rolling.\\
To reconstruct the scalar field potential, we require both observational data and a prior, along with a choice of covariance function for the GP. For the dataset, we utilize a compilation of Hubble parameter measurements consisting of 32 Cosmic Chronometers, 26 Baryon Acoustic Oscillation points, and 5 data points from the DESI survey. As priors, we adopt the Planck 2018 results: $H_0 = 67.4 \pm 0.5 , \text{km s}^{-1} \text{Mpc}^{-1}$ and $\Omega_{m,0} = 0.315 \pm 0.007$, to study their influence on the reconstructed scalar potential. For the GP kernel, we employ the squared exponential covariance function.\\
After running the GP reconstruction, we obtain the reconstructed scalar field potential $\mathcal{V}(\phi)$ in the form of dataset $\mathcal{V}(\phi)$ versus $\phi$, along with its uncertainty $\sigma_V$. To identify the best-fit model for the reconstructed potential, we consider four widely used and physically motivated forms of the potential: the exponential potential, the power-law potential, the free field (quadratic) potential, and the Higgs-like potential. Using the reconstructed $\mathcal{V}(\phi)$ vs. $\phi$ data, we perform a MCMC analysis for each of these potentials.\\
This approach ensures that the fitted parameters for each potential are free from any model-dependent biases that would otherwise arise from the assumption of a specific cosmological model. Such model-independent reconstructions are particularly relevant in the current cosmological context, where tensions such as the Hubble tension highlight the challenges in identifying the most accurate and predictive models for our Universe.\\
The four scalar field potentials considered in this study are chosen based on their relevance in current cosmological research and their ability to describe the evolution of DE across cosmic time. These potentials include the \textbf{exponential potential} given by \( \mathcal{V}(\phi) = \mathcal{V}_0 + \mathcal{V}_1 e^{-\lambda \phi} \), which provides an excellent fit with \( \chi^2_{\text{red}} = 0.7694 \), suggesting that even dynamical DE models require a baseline cosmological constant-like component for observational consistency. Moreover, this potential aligns well with the Swampland conjecture, indicating that it may be consistent with quantum gravity constraints. The \textbf{power-law potential}, \( \mathcal{V}(\phi) = \mathcal{V}_0 + \mathcal{V}_1 \phi^{\lambda} \), yields the best statistical agreement with the reconstruction (\( \chi^2_{\text{red}} = 0.9888 \)). Although it does not rely on a tracker solution, it still captures the key features of cosmic acceleration and DE dynamics, requiring the addition of the \( \mathcal{V}_0 \) term for consistency with the data. The \textbf{free field (quadratic) Potential}, \( \mathcal{V}(\phi) = \mathcal{V}_0 + \frac{1}{2} m^2 \phi^2 \) provides a good fit with \( \chi^2_{\text{red}} = 1.2150 \), although it requires an additional \( \mathcal{V}_0 \) term to match the observational data, highlighting the need for a vacuum energy contribution in late-time cosmology. Lastly, the \textbf{Higgs-Like potential}, \( \mathcal{V}(\phi) = \mathcal{V}_0 + \frac{1}{2} m^2 \phi^2 + \frac{\lambda}{4} \phi^4 \) fits the data well with \( \chi^2_{\text{red}} = 0.8120 \). This potential is significant in describing transitions between vacua, and, as in the other models, requires the inclusion of \( \mathcal{V}_0 \) to account for the vacuum energy and ensure consistency with observations.
Based on the reduced values of the chi-square ($\chi^2_{\text{red}}$), the potential of the power-law exhibits the closest agreement with the reconstructed data, indicating an excellent fit. The exponential and Higgs-like potentials also show good consistency with the reconstruction. Although the free field potential yields a slightly higher $\chi^2_{\text{red}}$ value, it still lies within the acceptable range for a good fit. Finally, the analysis shows that all four potential models are compatible with the reconstructed data, with the power-law potential providing the best match. The inclusion of these four well-motivated potentials enables a comprehensive exploration of how various scalar field dynamics correspond to model-independent reconstructions of the DE potential.\\
It is worth emphasizing that finding the parameter for the DBI filed potential is not just of phenomenological interest, but also carries deep theoretical significance. As mentioned in the introduction, how ``swampland conjectures" \cite{vafa} try to constrain all the cosmological models, especially the scale field models, by giving a very robust bound on the potential and derivative of the potential. As mentioned, a GP is non-parametric; the potential reconstruction by the GP has been of great interest in the string theory community, as it can be used as a proof to verify the swampland bound for the scalar field potentials as given by the ``string landscape" (particularly dSC or RdSC). So, one can also extend our work regarding the constraints on the DBI field potential and could check the validity of the ``swampland criteria" as done in \cite{elizaldeswampland1,elizaldeswampland2}.\\
  As far as future work is concerned, one can extend this chapter in various directions, such as using a more general K-essence field beyond the DBI field to constrain the potentials using GP. One can also verify the swampland-like conjectures in DBI fields and other K-essence fields to comment on the consistency of the observation and string theory prediction. Last but not least, one can also extend this study to general teleparallel gravity theories, such as $f(T)$ and $f(Q)$ gravity, in order to check the consistency of such modified gravity frameworks with current observational data.   

\chapter{Dynamical System Analysis of Scalar Field Cosmology in Coincident $f(Q)$ Gravity} 

 \label{Chapter4}
\lhead{Chapter 4. \emph{Dynamical System Analysis of Scalar Field Cosmology in Coincident $f(Q)$ \\ Gravity}} 
\vspace{10 cm}
* The work presented in this chapter is covered by the following publication: \\
 
\textit{Dynamical system analysis of scalar field cosmology in coincident $f(Q)$ gravity}, 
Physica Scripta, \textbf{99}, 055021 (2024).

\clearpage

\epigraph{``What is it that breathes fire into the equations and makes a Universe for them to describe?''}{--- Stephen Hawking, \textit{A Brief History of Time} (1988), Ch.~12}

In this chapter, we investigate scalar field cosmology in the coincident $f(Q)$ gravity formalism. We calculate the motion equations of $f(Q)$ gravity under the flat FLRW background in the presence of a scalar field. We consider a non-linear $f(Q)$ model, particularly $f(Q)=-Q+\alpha Q^n$, which is nothing but a polynomial correction to the STEGR case. In addition, we assumed two well-known specific forms of the potential function, specifically the exponential form from $V(\phi)= V_0 e^{-\beta \phi}$ and the power-law form $V(\phi)= V_0\phi^{-k}$. We employ some phase-space variables and transform the cosmological field equations into an autonomous system. We calculate the critical points of the corresponding autonomous systems and examine their stability behaviors. We discuss the physical significance corresponding to the exponential case for parameter values $n=2$ and $n=-1$ with $\beta=1$, and $n=-1$ with $\beta=\sqrt{3}$. Moreover, we discuss the same for the power-law case with parameter values $n=-2$ and $k=0.16$. We also analyze the behavior of the corresponding cosmological parameters, such as the scalar field and dark energy density, deceleration, and the effective equation of state parameter. As in the exponential case, we find that the results for the parameter constraints in case III are better than those in the other two cases, and this reflects the evolution of the Universe from a decelerated stiff era to an accelerated de Sitter era via a matter-dominated epoch. Further, in the power-law case, we find that all trajectories exhibit identical behavior, representing the evolution of the Universe from a decelerated stiff era to an accelerated de Sitter era. Lastly, we conclude that the exponential case shows better evolution than the power-law case.

\section{Introduction}\label{sec_4.1} 

Modified gravity is a prominent route to enhance our understanding of the evolution of the Universe, incorporating its early stages (inflation) and later phases (dark energy) marked by accelerated expansion \cite{CANT}. In addition, it may address potential observational conflicts \cite{COSI}. These theories involve constructing extensions and modifications of GR that introduce additional degrees of freedom. These modifications can potentially provide corrections at both the cosmological background and perturbation levels, offering a more comprehensive description of the behavior of the Universe. There are various methods for constructing such gravitational modifications. One different category of gravitational modifications emerges when utilizing the equivalent formulation of gravity that incorporates non-metricity, which is widely known as symmetric teleparallel gravity \cite{NEST}. This approach employs a generic affine connection with vanishing torsion and vanishing curvature with respect to the Levi-Civita connection, while relaxing the ``metric compatibility" condition with respect to the generic connection. This formulation has recently been extended to $f(Q)$ gravity \cite{Jimenez/2018}. This extension of the symmetric teleparallel formalism has gained attention from the cosmology community as a potential avenue for exploring new physics beyond conventional $\Lambda$CDM cosmology. Particular forms of the $f(Q)$ function have been demonstrated to alleviate the $\sigma_8$ tension \cite{BARR}, whereas others enable a more accurate representation of cosmological dataset \cite{ANAG,ARORA,NUNES}. Recently, some interesting cosmological implications of the $f(Q)$ gravity in different context have appeared, for instance, black hole physics \cite{RODR,LAVI-1}, neutrino physics \cite{NEOM}, quantum cosmology \cite{CAPE-1,PALIA}, bouncing cosmology \cite{GADB}, inflation \cite{CAPE-2}, phantom cosmology \cite{ANDER}, astrophysical objects \cite{SNEHA,Hassan/2022}, cosmological perturbations \cite{ET}, BBN constraints \cite{ANAG-2}, and many others \cite{DE-1,DE-2,PALIA-2,HOH,WOM-2}.

Scalar fields play a significant role in describing the physical properties of the Universe, particularly in the context of the inflationary scenario \cite{ALAN}, and they can offer explanations for the cosmic late-time acceleration. Although the $\Lambda$CDM model is highly consistent with observational data, successfully describing structure formation, it has yet to effectively quantify quantum vacuum fluctuations \cite{ZEL,WENB}. This is the key inspiration for proposing the dark energy candidate as an alternative to $\Lambda$. Various examples include the quintessence field \cite{Ratra:1988,quint2}, a quintom scalar field \cite{quintom1,Guo/2005}, a phantom scalar field \cite{phant}, and multi-scalar field models \cite{mult}. If one assumes that the cosmological constant $\Lambda$ is coming from a single (or multiple) scalar field, then one can solve the cosmological constant problem, as the scalar field can decay the vacuum energy to stop the exponential increase. However, there exists a tension between the Hubble constant values measured from early observations (such as Planck \cite{planck}) and estimated via local observations (such as SH0ES \cite{Riess}). One potential solution to this tension involves assuming extensions beyond the $\Lambda$CDM. This chapter explores scalar field cosmology in the modified symmetric teleparallel gravity background, utilizing the dynamical system approach. The modified $f(Q)$ function is now responsible for the observed accelerating scenario, whereas the quantum field theory prediction of the tiny value of $\Lambda$ ($\sim 10^{-120}$) cannot be described by such gravitational modification and thus the addition of a scalar field takes into account the quintessence scenario.\\
Note that it is prominent to choose both modified gravity and scalar field, such as in previous work on $f(R,\phi)$ \cite{fr}, $f(T,\phi)$ \cite{ft}, and $f(\mathcal{G},\phi)$ \cite{fg}. There are many reasons to believe that the effect of both scalar field and modified gravity can describe the late time acceleration efficiently. One of the reasons is the cosmological constant problem that we have discussed in the previous section; the second reason is from various calculations of QFT in curved spacetime, which indicate that the inflation field could indeed survive up to today to act as the quintessence field \cite{Ratra:1988}. For a detailed discussion on how an early Universe scalar field could survive even in the late Universe, we recommend the review by Peebles and Ratra \cite{ratrareview}. Another motivation to take the exponential and polynomial potential is that the background perturbation (Bardeen potential) behaves very well in these two types of scalar field, which have been explored in the chapter~\ref{Chapter6}. Some works have been done to suggest that torsion-based gravity could explain the Hubble tension \cite{tension}. In this chapter our motivation is to present a dynamical system of $f(Q)$ in the presence of a scalar field in full generality. \\
It is widely recognized that when investigating cosmological models, introducing auxiliary variables allows one to transform cosmological equations into an autonomous dynamical system \cite{COPE}. This results in a system $X' =f(X)$, where $X$ represents the column vector consisting of auxiliary variables and $f(X)$ denotes the vector field. The stability analysis of the given autonomous entity involves several steps. Initially, critical points (or equilibrium points) $X_c$ are identified by setting $X'=0$.  Then, we consider linear perturbations around the critical point $X_c$ as $X=X_c+P$, where $P$ represents the column vector of perturbed auxiliary variables. As a result, one can have (up to first order) $P'=AP$, where $A$ is the matrix consisting of coefficients of the perturbed equations. Finally, the eigenvalues of the coefficient matrix $A$ determine the stability behavior of each hyperbolic critical point. A critical point $X_c$ is considered stable (unstable), or a saddle if the real parts of the corresponding eigenvalues are negative (positive), or have real parts with different signs. Several interesting outcomes in the context of modified gravity utilizing the dynamical system method can be found in \cite{DE-3,WOM,Mishra-1,Mishra-2,HAMID,APLL}. The present chapter is organized as follows. In Sec.~\ref{sec_4.2}, we present the mathematical background of extended symmetric teleparallel gravity with the Lagrangian density of a scalar field. In Sec.~\ref{sec_4.3}, we present governing equations of motion under the FLRW background in the presence of a scalar field. Further in Sec.~\ref{sec_4.4}, we perform a detailed dynamical analysis of two well-known scalar field potentials along with a non-linear $f(Q)$ cosmological background model. Finally, in Sec.~\ref{sec_4.5}, we highlight our findings.

\section{$f(Q)$ gravity formulation with scalar field}\label{sec_4.2} 

Following \eqref{2l2} in Chapter~\ref{Chapter2} here we present the action for $f(Q)$ gravity, which is a generalization of the STEGR theory, in the presence of a scalar field,
\begin{equation}\label{2l4}
\mathcal{S}=\int\frac{1}{2}\,f(Q)\sqrt{-g}\,d^4x+\int \mathcal{L}_{\phi}\,\sqrt{-g}\,d^4x\, .
\end{equation}
where $g=\text{det}(g_{\mu\nu})$ and wec have also taken the natural units that is $8\pi G=c=1$, $f(Q)$ is an arbitrary function of the non-metricity scalar $Q$, and $\mathcal{L}_{\phi}$ is the Lagrangian density of a scalar field $\phi$ given by \cite{BAHA},
\begin{equation}\label{2m4}
\mathcal{L}_{\phi} = -\frac{1}{2} g^{\mu \nu} \partial_\mu \phi  \partial_\nu \phi -V(\phi) \,\,\text{.}
\end{equation}
Here, $V(\phi)$ represents a potential for the field $\phi$. We obtain the following governing field equation by varying the action Eq.~\eqref{2l4} with respect to the metric,
\begin{equation}\label{2n}
\frac{2}{\sqrt{-g}}\nabla_\lambda (\sqrt{-g}f_Q P^\lambda\:_{\mu\nu}) + \frac{1}{2}g_{\mu\nu}f+f_Q(P_{\mu\lambda\beta}Q_\nu\:^{\lambda\beta} - 2Q_{\lambda\beta\mu}P^{\lambda\beta}\:_\nu) = -T_{\mu\nu}^{\phi} \,\,\text{.}
\end{equation}
where $f_Q=\frac{df}{dQ}$ and $T_{\mu\nu}^{\phi}$ represents the stress-energy tensor of the scalar field given as
\begin{equation}\label{2o4}
 T_{\mu\nu}^{\phi}= \partial_\mu \phi  \partial_\nu \phi -\frac{1}{2} g_{\mu \nu} g_{\alpha \beta} \partial^\alpha \phi  \partial^\beta \phi -  g_{\mu \nu} V(\phi).
\end{equation}
Moreover, we obtain the following equation of motion for the scalar field, i.e., the Klein-Gordon equation from the Euler-Lagrangian equation for the Lagrangian density given by Eq.~\eqref{2m4},
\begin{equation}\label{2p4}
\square \phi - V,_\phi =0 \,\,\text{.}
\end{equation}
Here, $\square$ denotes the d'Alembertian and $V,_\phi = \frac{\partial V}{\partial \phi}$. Furthermore, on varying the action Eq.~\eqref{2l4} with respect to the connection (similar to Palatini prescription), we have 
\begin{equation}\label{2q}
\nabla_\mu \nabla_\nu (\sqrt{-g}f_Q P^{\mu\nu}\:_\lambda) =  0 \,\,\text{.}
\end{equation}

\section{Equations of motion}\label{sec_4.3} 

We begin with the following flat FLRW line element in order to probe the cosmological implications under the assumption of spatial isotropy and homogeneity of the Universe,
\begin{equation}\label{3a4}
ds^2= -dt^2 + a^2(t)[dx^2+dy^2+dz^2] \,\,\text{.}   
\end{equation}
Here, $a(t)$ is a measure of the expansion of the Universe. Beginning with the teleparallel constraint that corresponds to a flat geometry characterizing a pure inertial connection, one can execute a gauge transformation parameterized by $\Lambda^\alpha_\mu$  \cite{JIM-2},
\begin{equation}\label{3b}
 \Upsilon^\alpha_{\: \mu \nu}  = (\Lambda^{-1})^\alpha_{\:\: \beta} \partial_{[ \mu}\Lambda^\beta_{\: \: \nu ]} \,\,\text{.}
\end{equation}
Consequently, the generic affine connection can be expressed as follows, using the general element of $ GL(4,\mathbb{R}) $ characterized by the transformation $ \Lambda^\alpha_{\: \: \mu}=\partial_\mu \zeta^\alpha$, where $ \zeta^\alpha $ is an arbitrary vector field,
\begin{equation}\label{3c4}
\Upsilon^\alpha_{\: \mu \nu} = \frac{\partial x^\alpha}{\partial \zeta^\rho} \partial_\mu \partial_\nu \zeta^\rho \,\,\text{.}
\end{equation}
This reveals the possibility of eliminating the connection through a coordinate transformation. The coordinate transformation is responsible for eliminating the connection Eq.~\eqref{3c4} is termed gauge coincident. We utilize the coincident gauge in the present chapter. Hence, the non-metricity scalar corresponds to the metric Eq.~\eqref{3a4} becoming $Q=6H^2$.\\
The stress-energy tensor for the perfect fluid distribution reads as,
\begin{equation}\label{3d4}
T_{\mu\nu}=(\rho+p)u_\mu u_\nu + pg_{\mu\nu} \,\,\text{.}
\end{equation}
where $u^\mu=(1,0,0,0)$ are components of the four velocities. Upon comparing Eq.~\eqref{3d4} and Eq.~\eqref{2o4}, we have
\begin{equation}\label{3e}
\rho_{\phi}=-\frac{1}{2}g_{\alpha \beta}\partial^\alpha \phi \partial^\beta \phi +V(\phi) \,\,\text{,}
\end{equation}
\begin{equation}\label{3f}
p_{\phi}=-\frac{1}{2}g_{\alpha \beta}\partial^\alpha \phi \partial^\beta \phi -V(\phi) \,\,\text{.}
\end{equation}
As the scalar field considered here does not depend on the spatial coordinates, we have the following expressions for the energy density and pressure component of the scalar field,
\begin{equation}\label{3g}
\rho_{\phi}=\frac{1}{2}\dot{\phi}^2+V(\phi) \,\,\text{,}
\end{equation}
and,
\begin{equation}\label{3h}
p_{\phi}=\frac{1}{2}\dot{\phi}^2-V(\phi) \,\,\text{.}
\end{equation}
and the corresponding equation of state parameter can be written as,
\begin{equation}\label{3i}
\omega_{\phi}=\frac{p_{\phi}}{\rho_{\phi}}=\frac{\frac{1}{2}\dot{\phi}^2-V(\phi)}{\frac{1}{2}\dot{\phi}^2+V(\phi)} \,\,\text{.}
\end{equation}
Moreover, corresponding to the metric Eq.~\eqref{3a4}, the Klein-Gordon equation which is given by Eq.~\eqref{2p4} becomes
\begin{equation}\label{3j}
\ddot{\phi}+3H\dot{\phi}+V_{,\phi}=0 \,\,\text{.}
\end{equation}
We obtain the following Friedmann like equations governing the gravitational interactions under the $f(Q)$ gravity background in the presence of scalar field,
\begin{equation}\label{3k4}
3H^2=\frac{1}{2f_Q} \left( -\rho_{\phi}+\frac{f}{2}  \right) \,\,\text{,}
\end{equation}
\begin{equation}\label{3l4}
    \dot{H}+3H^2+ \frac{\dot{f_Q}}{f_Q}H = \frac{1}{2f_Q} \left( p_{\phi}+\frac{f}{2} \right) \,\,\text{.}
\end{equation}
For the $f(Q)$ functional $f(Q)=-Q+\Psi(Q)$, we can rewrite the Friedmann equations which are given by Eq.~\eqref{3k4}-\eqref{3l4} as (where we can recover ordinary GR by putting $\Psi=0$),
\begin{equation}\label{3m4}
    3H^2= \rho_\phi + \rho_{Q} \,\,\text{,}
\end{equation}
\begin{equation}\label{3n4}
    \dot{H}=-\frac{1}{2} [\rho_\phi + p_\phi+\rho_{Q}+p_{Q}] \,\,\text{.}
\end{equation}
where $\rho_{Q}$ and $p_{Q}$ represent the energy density and pressure of the dark energy component evolving due to the geometry of spacetime,
\begin{equation}\label{3o}
    \rho_{Q}=-\frac{\Psi}{2}+ Q\Psi_Q \,\,\text{,}
\end{equation}
\begin{equation}\label{3p}
    p_{Q}=-\rho_{de}-2\dot{H} \left( \Psi_Q+2Q\Psi_{QQ} \right) \,\,\text{.}
\end{equation}

\section{The cosmological model and dynamical system analysis}\label{sec_4.4} 

In this section, we attempt to examine some specific class of scalar field potentials and $f(Q)$ functional forms, incorporating phase-space techniques. In order to do so, we transform motion equations of considered cosmological settings, discussed in the previous section, into an autonomous system with the help of phase-space variables. We define the following phase-space variables,
\begin{equation}\label{4a}
x^2=\frac{\dot{\phi}^2}{6H^2}, \:\: y^2=\frac{V}{3H^2}, \:\: \text{and} \:\: s^2=\Omega_{de}=\frac{\rho_{de}}{3H^2} \,\,\text{.}
\end{equation}
Now, using the fact that $\Omega_\phi=\frac{\rho_\phi}{3H^2}=\frac{\dot{\phi}^2}{6H^2}+\frac{V}{3H^2}=x^2+y^2$ and the Eq.~\eqref{3m4}, we have the following constraint,
\begin{equation}\label{4b}
x^2+y^2+s^2=1 \,\,\text{.}
\end{equation}
Hence, it follows that $\Omega_{de}=s^2=1-x^2-y^2=1-\Omega_\phi$ and the constraint $0\leq x^2+y^2 \leq 1$.

In order to obtain the closed form of the required autonomous system with general potential, we define another phase-space variable as follows,
\begin{equation}\label{4c4}
\lambda=-\frac{V_{,\phi}}{V} \,\,\text{.} 
\end{equation}
We obtain the following dynamical system  with respect to the e-folding time $N=ln(a)$,
\begin{equation}\label{4d4}
x'=-3x-x \frac{\dot{H}}{H^2} +\sqrt{\frac{3}{2}} \lambda y^2 \,\,\text{,} 
\end{equation}
\begin{equation}\label{4e}
y'=\frac{1}{2y} \left[ \frac{\dot{V}}{3H^2} - y^2 \frac{2\dot{H}}{H^2} \right] \,\,\text{,}
\end{equation}
\begin{equation}\label{4f4}
\lambda'=-\sqrt{6}\lambda^2 x (\Gamma-1) \,\,\text{,}
\end{equation}
where $\Gamma=\frac{VV_{,\phi \phi}}{V_{,\phi}^2}$. Here, using Eq.~\eqref{3n4}, we have
\begin{equation}\label{4g4}
\frac{\dot{H}}{H^2}=\frac{3x^2}{\left( \Psi_Q+2Q\Psi_{QQ}-1 \right)} \,\,\text{.}
\end{equation}
Note that the dynamical system Eq.~\eqref{4d4}-\eqref{4f4} is still not in its closed form, as it depends on the choice of $f(Q)$ functional form. We also note that Eq.~\eqref{4d4}-\eqref{4f4} and Eq.~\eqref{4g4} are in the most general form (as long as $\Psi$ and $V$ are analytic or at least twice continuously differentiable.) and to create a closed autonomous and exactly analytically solved answer we choose some forms of $\Psi(Q)$ and $V(\phi)$. Further, it also appears that we gain nothing from this extra phase-space variable $\lambda$, due to the presence of a quantity $\Gamma$ in the given dynamical system, which directly relies on the scalar field $\phi$. Nevertheless, considering that both $\lambda$ and $\Gamma$ are functions dependent on the scalar field $\phi$, there exists an explicit relation between them, i.e. if the function $\lambda(\phi)$ is invertible, allowing us to derive $\phi(\lambda)$, and thus we have $\Gamma$ as a function of $\lambda$.

We consider a power-law functional form $\Psi(Q)=\alpha Q^n$, i.e. $f(Q)=-Q+\alpha Q^n$, where $\alpha$ and $n$ are free parameters. The choice of the considered functional form is appropriate to close the obtained dynamical system.
We would like to note that the dimensional consistency requires $[\alpha] = 2(1-n)$ in mass units. However, in the autonomous formulation, $\alpha$ is absorbed into the dimensionless phase-space variables $(x, y)$ and drops out of the evolution equations entirely. As a result, the qualitative dynamics and the expressions for the deceleration parameter $q$ and the total equation of state $\omega_{\mathrm{total}}$ depend only on $n$ and the state variables $(x, y)$. The explicit mass dimension of $\alpha$ is therefore irrelevant for the stability analysis, though it must be restored when mapping phase-space trajectories back to physical cosmological scales.\\
Now, for the assumed $f(Q)$ function, the expression Eq.~\eqref{4g4} becomes,
\begin{equation}\label{4h}
    \frac{\dot{H}}{H^2}=\frac{3x^2}{n \left( 1-x^2-y^2 \right) -1} \,\,\text{.}
\end{equation}
Moreover, we have the following expressions for two cosmological parameters that play a crucial role in characterizing the evolutionary phase of expansion of the Universe, namely, deceleration and equation of state parameter.
\begin{equation}\label{4i}
q=-1- \frac{3x^2}{n \left( 1-x^2-y^2 \right) -1} \,\,\text{,}
\end{equation}
\begin{equation}\label{4j}
\omega_{total}=-1- \frac{2x^2}{n \left( 1-x^2-y^2 \right) -1} \,\,\text{.}
\end{equation}
Now, in order to probe the cosmological implications of the considered scenario, we choose two specific forms of potential function, namely exponential and power-law potentials, which are widely discussed in the literature in different cosmological contexts.

\subsection{Exponential potential}
\justifying
We assume the following form of an exponential potential,
\begin{equation}\label{s14}
V(\phi)= V_0 e^{-\beta \phi} \,\,\text{.}
\end{equation}
The assumed exponential potential represents the simplest illustration of quintessence scenarios \cite{LAS, TBR}, as it is a slowly varying scalar field that could drive the cosmological constant $\Lambda$ to zero. Hence, it can be a suitable candidate to address the cosmological constant problem. This potential arises naturally from the string theory compactifications discussed in \cite{DBL}. In addition, this type of potential naturally arises from Kaluza-Klein type compactifications, and their effect on anisotropic product cosmology has been investigated in the reference \cite{JYJ}. We note that the dynamical system analysis of scale-invariant solutions of the exponential scalar field coupled with a barotropic fluid system has been studied, and it has been shown that such a system can have a stable scale-invariant fixed point \cite{COPE}. Now, using Eq.~\eqref{s14}, the dynamical variable in Eq.~\eqref{4c4} becomes, 
\begin{equation}\label{s2}
\lambda=\beta \:\: \text{with} \:\: \Gamma=1 \,\,\text{.} 
\end{equation}
Hence, the system Eq.~\eqref{4d4}-\eqref{4f4} reduces to the following autonomous system,
\begin{equation}\label{s3}
x'=-3x \left[ 1+ \frac{x^2}{n \left( 1-x^2-y^2 \right)-1}  \right] + \sqrt{\frac{3}{2}} \beta y^2 \,\,\text{,}
\end{equation}
\begin{equation}\label{s4}
y'= -xy \left[ \sqrt{\frac{3}{2}} \beta + \frac{3x}{n \left( 1-x^2-y^2 \right)-1}  \right] \,\,\text{.}
\end{equation}
The critical points and their physical behavior, corresponding to autonomous system Eq.~\eqref{s3}-\eqref{s4}, are presented in the Table~\eqref{4Table-1}.

\begin{table}[htbp]
\centering
\caption{Critical points and their behavior for $f(Q)=-Q+\alpha Q^n$ with exponential potential $V(\phi)=V_0 e^{-\beta\phi}$.}
\resizebox{1.10\textwidth}{!}{%
\begin{tabular}{|c|c|c|c|c|}
\hline
Critical Points $(x_c,y_c)$ & Eigenvalues $\lambda_1,\lambda_2$ & Nature of critical point & $q$ & $\omega$ \\
\hline
$O(0,0)$ & $-3,\;0$ & Stable & $-1$ & $-1$ \\
$A(-1,0)$ & $6-6n,\;3+\sqrt{\tfrac{3}{2}}\beta$ & Stable for $(n>1 \ \&\ \beta<-\sqrt{6})$ & $2$ & $1$ \\
$B(1,0)$  & $6-6n,\;3-\sqrt{\tfrac{3}{2}}\beta$ & Stable for $(n>1 \ \&\ \beta>\sqrt{6})$ & $2$ & $1$ \\
$C\!\left(\tfrac{\beta}{\sqrt{6}},\sqrt{1-\tfrac{\beta^2}{6}}\right)$ & $\tfrac{1}{2}(\beta^2-6),\;(1-n)\beta^2$ & Stable for $(-\sqrt{6}<\beta<0\ \&\ n>1)$ or $(0<\beta<\sqrt{6}\ \&\ n>1)$ & $-1+\tfrac{\beta^2}{2}$ & $-1+\tfrac{\beta^2}{3}$ \\
$D\!\left(\tfrac{\beta}{\sqrt{6}},-\sqrt{1-\tfrac{\beta^2}{6}}\right)$ & $\tfrac{1}{2}(\beta^2-6),\;(1-n)\beta^2$ & Stable for $(-\sqrt{6}<\beta<0\ \&\ n>1)$ or $(0<\beta<\sqrt{6}\ \&\ n>1)$ & $-1+\tfrac{\beta^2}{2}$ & $-1+\tfrac{\beta^2}{3}$ \\
\hline
\end{tabular}%
}
\label{4Table-1}
\end{table}

\begin{figure}[h]
{\includegraphics[scale=0.37]{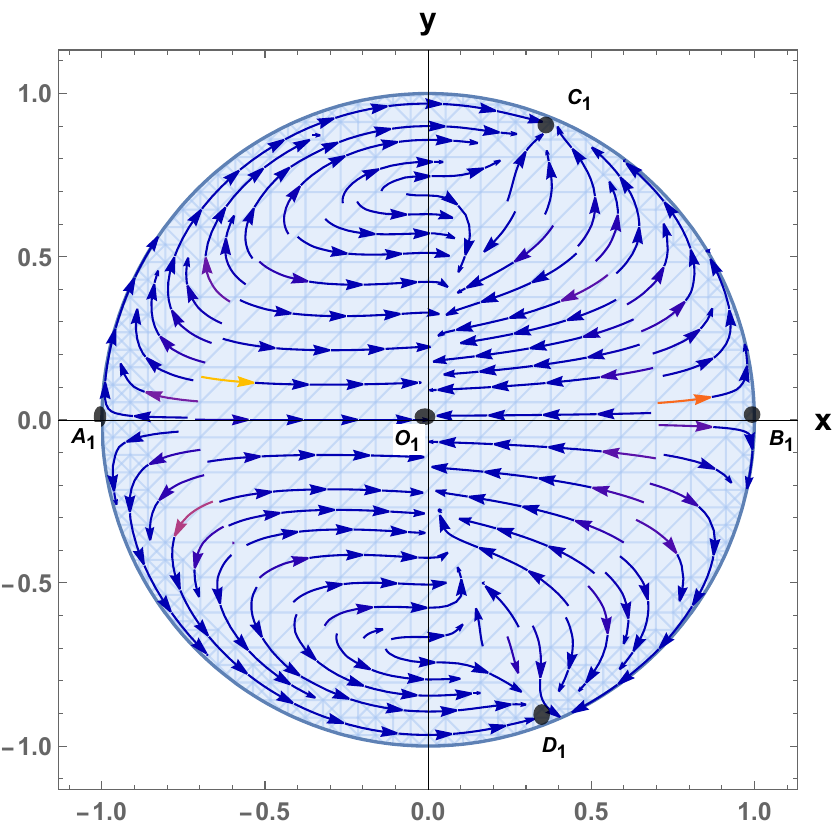}}
{\includegraphics[scale=0.37]{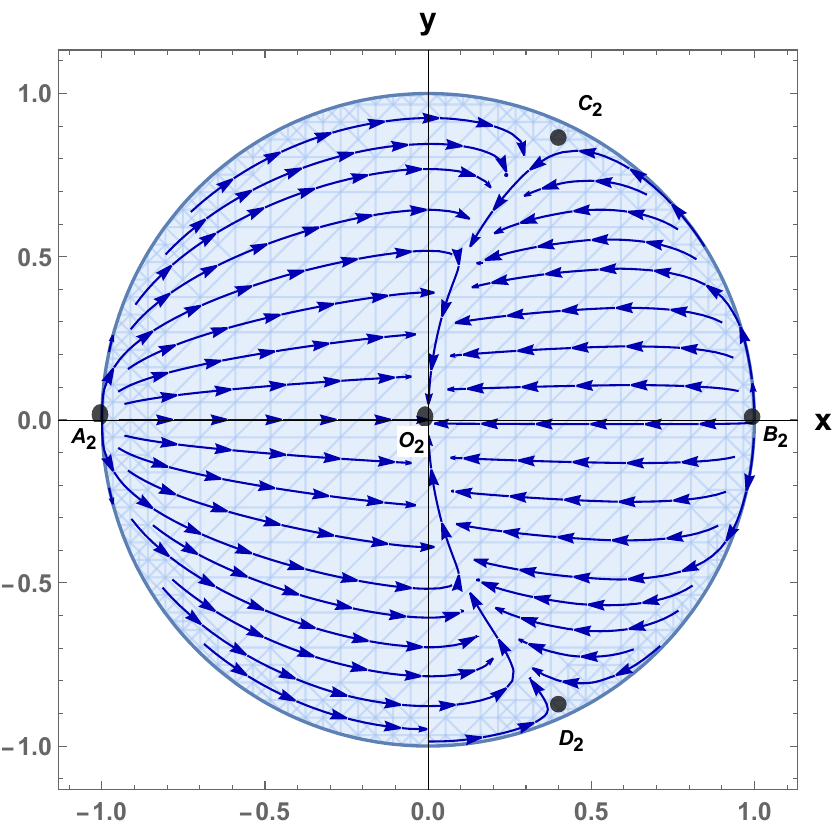}}
{\includegraphics[scale=0.37]{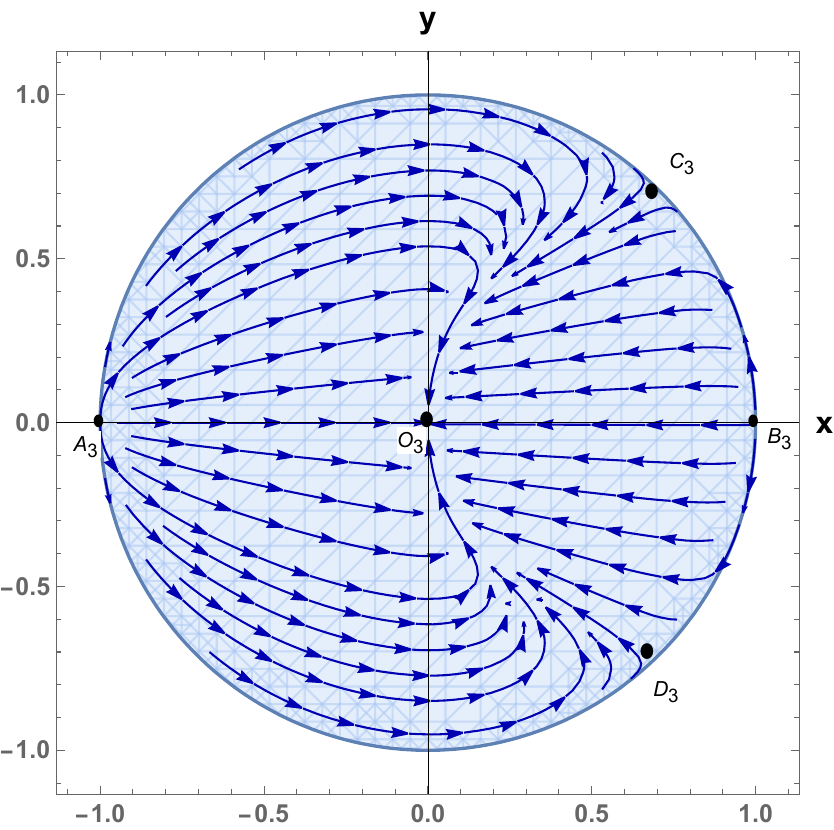}}
\caption{Phase-space plots for the case $n=2$ (left panel above) and $n=-1$ (right panel above), with $\beta=1$ , and for the case $n=-1$ with $\beta=\sqrt{3}$ (below) corresponding to the exponential potential.}\label{fig_4.1}
\end{figure}

\begin{figure}[H]
{\includegraphics[scale=0.28]{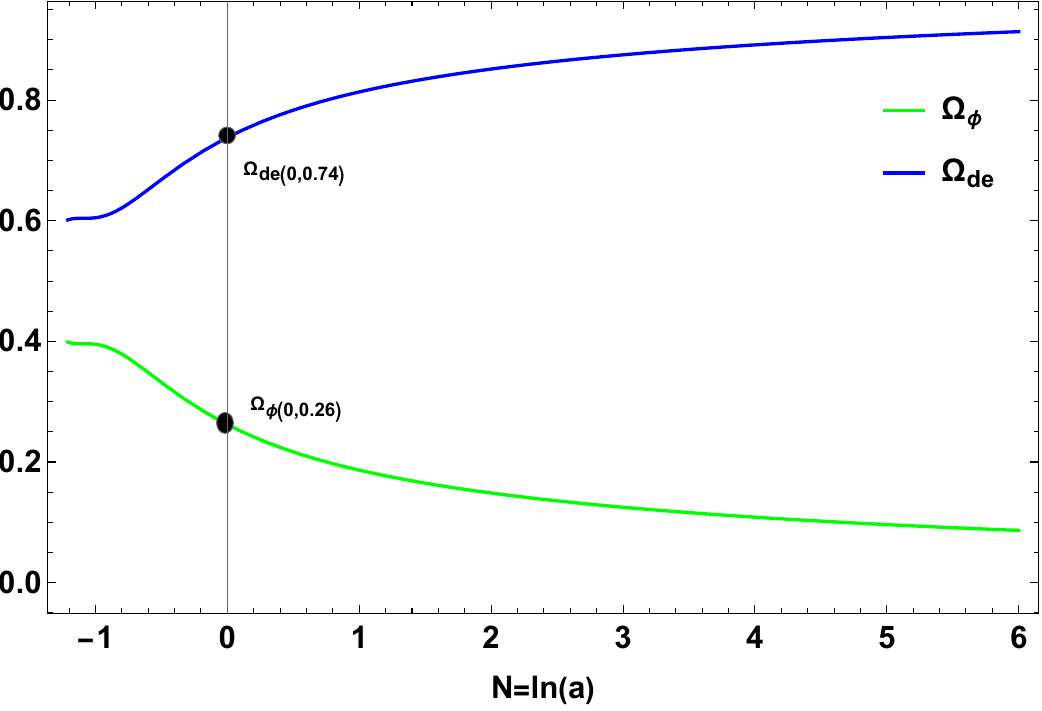}}
{\includegraphics[scale=0.28]{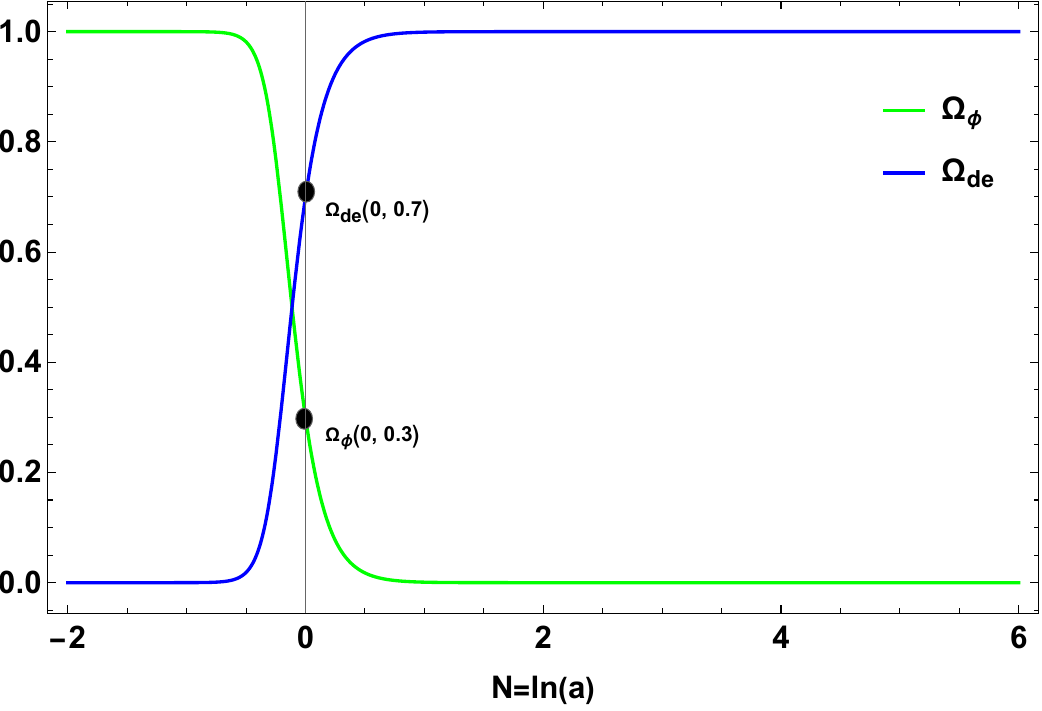}}
{\includegraphics[scale=0.28]{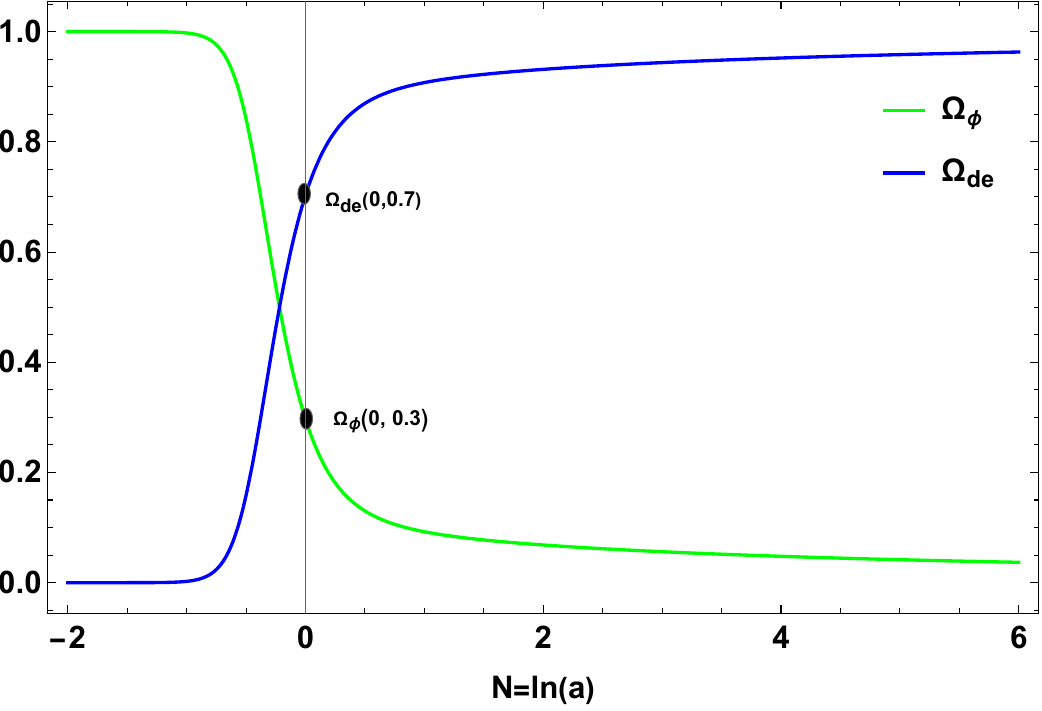}}
\caption{Evolutionary profile of the scalar field density and dark energy density for the case $n=2$ (left panel) and $n=-1$ (middle panel), with $\beta=1$, and for the case $n=-1$ with $\beta=\sqrt{3}$ (right panel) corresponding to the exponential potential.}\label{fig_4.2}
\end{figure}

\begin{figure}[H]
{\includegraphics[scale=0.28]{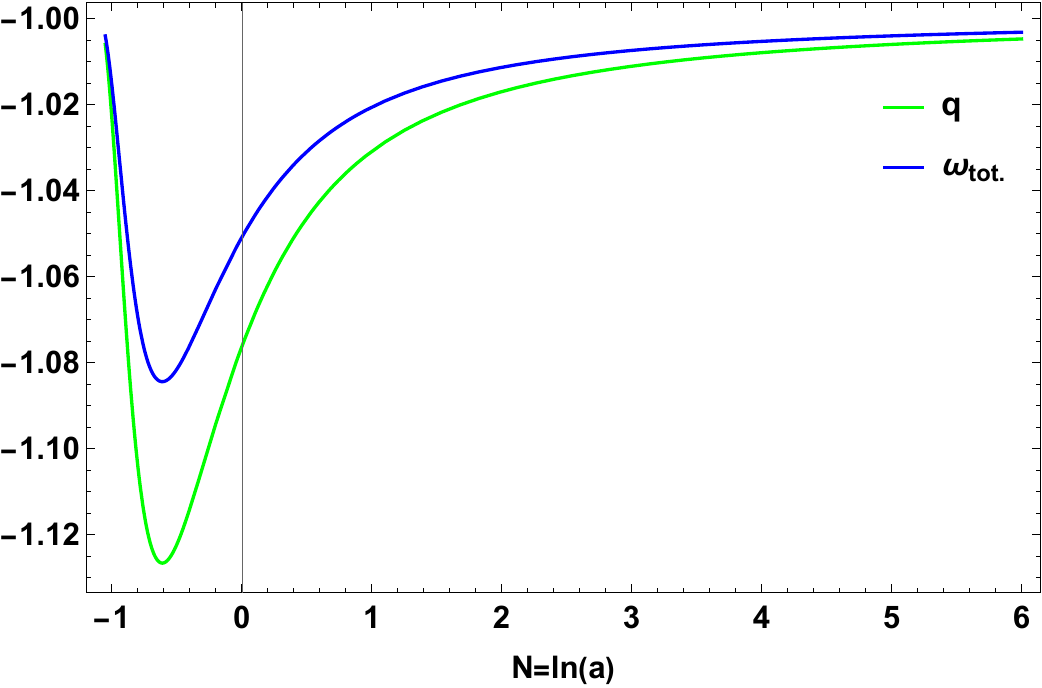}}
{\includegraphics[scale=0.28]{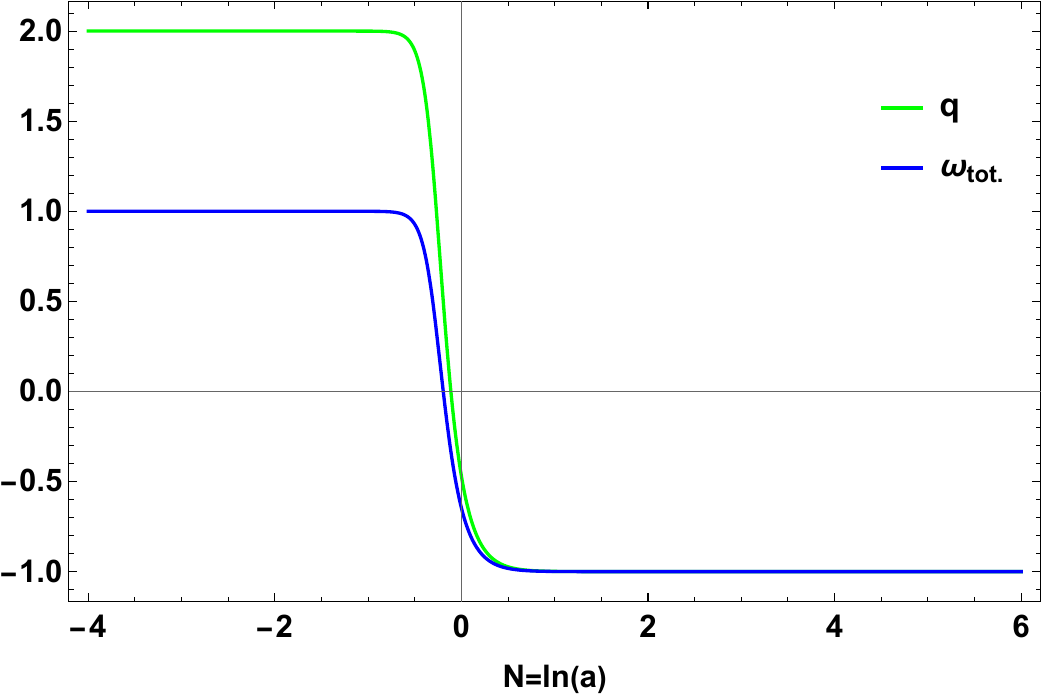}}
{\includegraphics[scale=0.28]{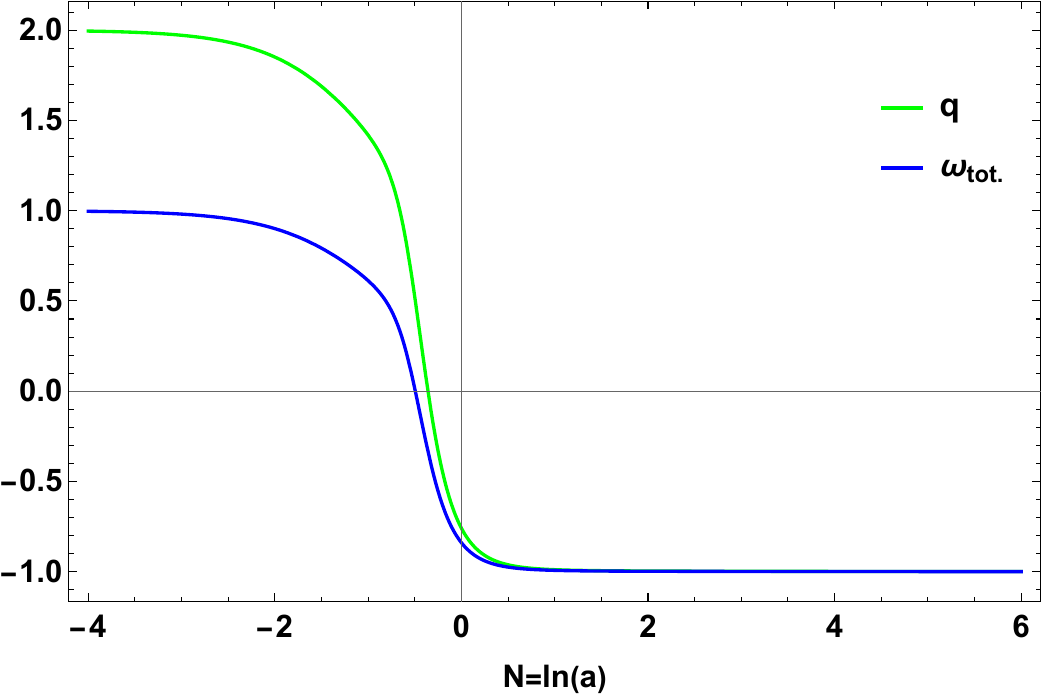}}
\caption{Evolutionary profile of the deceleration and the equation of state parameter for the case $n=2$ (left panel) and $n=-1$ (middle panel), with $\beta=1$, and for the case $n=-1$ with $\beta=\sqrt{3}$ (right panel) corresponding to the exponential potential.}\label{fig_4.3}
\end{figure}

We present the asymptotic behavior of some prominent cosmological parameters such as the scale factor, the Hubble parameter, and the scalar field corresponding to the obtained set of critical points,
\begin{itemize}
\item Similarly to the cases $x_c=0$ and $y_c=0$, we obtain the cosmic scale factor as $a(t)=e^{H_0(t-t_0)}$, the Hubble parameter as $H(t)=H(t_0)=H_0$, and the scalar field $\phi(t)=\phi(t_0)=\phi_0$, where $t_0$ being the current time, and $a(t_0)=1$ and $H_0$ are the present values of the scale factor and the Hubble parameter.
\item Similarly to the cases $x_c=\pm 1$ and $y_c=0$, we obtain the cosmic scalar factor as $a(t)=\left[ 3H_0(t-t_0)+1\right]^\frac{1}{3}$, the Hubble parameter as $H(t)=\frac{H_0}{3H_0(t-t_0)+1}$, and the scalar field as $\phi(t)=\phi_0 \pm \frac{\sqrt{6}}{3} ln\left[ 3H_0(t-t_0)+1\right] $.
\item Similarly to the cases $x_c=\frac{\beta}{\sqrt{6}}$ and $y_c= \pm \sqrt{1-\frac{\beta^2}{6}}$, we obtain the cosmic scalar factor as $a(t)=\left[ \frac{\beta^2}{2}H_0(t-t_0)+1\right]^\frac{2}{\beta^2}$, the Hubble parameter as $H(t)=\frac{H_0}{\frac{\beta^2}{2}H_0(t-t_0)+1}$, and the scalar field as $\phi(t)=\phi_0 + \left[ \frac{\beta^2}{2}H_0(t-t_0)+1\right]^\frac{2}{\beta} $.
\end{itemize}
The 2-D phase-space diagrams corresponding to the autonomous system Eq.~\eqref{s3}-\eqref{s4} for some specific parameter values are presented in Fig.~\ref{fig_4.1}. In addition, the corresponding evolutionary profile of the scalar field and dark energy density is presented in Figure \ref{fig_4.2}. In further detail in Fig.~\ref{fig_4.3}, we present the corresponding evolutionary profile of the deceleration and the effective equation of state parameter.

Now we discuss the physical significance and stability of the critical points obtained for the specific parameter values as follows:
\begin{itemize}
\item \textbf{Case I ($n=2$ and $\beta=1$) :} In this case, the set of critical points obtained is $O_1(0,0)$, $A_1(-1,0)$, $B_1(1,0)$, $C_1\left( \frac{1}{\sqrt{6}}, \sqrt{\frac{5}{6}} \right)$ and $D_1\left( \frac{1}{\sqrt{6}}, -\sqrt{\frac{5}{6}} \right)$ with the corresponding eigenvalues $(\lambda_1,\lambda_2)=(-3,0)$, $\left(-6,3+\sqrt{\frac{3}{2}}\right)$, $\left(-6,3-\sqrt{\frac{3}{2}}\right)$, $\left( -\frac{5}{2},-1 \right)$, and $\left( -\frac{5}{2},-1 \right)$, respectively. Moreover, the corresponding pairs of values $(q,\omega)$ are $(-1,-1)$, $(2,1)$, $(2,1)$, $\left( -\frac{1}{2},-\frac{2}{3} \right)$, and $\left( -\frac{1}{2},-\frac{2}{3} \right)$. Therefore, the critical points $O_1$, $C_1$, and $D_1$ are stable, while the points $A_1$ and $B_1$ are saddle. It should also be noted that one can neglect the flow of trajectories below the X-axis in the phase-space diagrams, since the results are the same as those of the above X-axis due to the symmetry in the positive and negative axes of Y. Moreover, from the flow of phase-space trajectories in Fig.~\ref{fig_4.1} (left panel above), it is clear that we obtained a stable de Sitter accelerated Universe (represented by $O_1$) without any transition epoch. The same is reflected in the evolutionary profiles of the corresponding cosmological parameters presented in Fig.~\ref{fig_4.2} and \ref{fig_4.3} (left panel).

\item \textbf{Case II ($n=-1$ and $\beta=1$) :} In this case, the set of critical points obtained is $O_2(0,0)$, $A_2(-1,0)$, $B_2(1,0)$, $C_2\left( \frac{1}{\sqrt{6}}, \sqrt{\frac{5}{6}} \right)$ and $D_2\left( \frac{1}{\sqrt{6}}, -\sqrt{\frac{5}{6}} \right)$ with the corresponding eigenvalues $(\lambda_1,\lambda_2)=(-3,0)$, $\left(12,3+\sqrt{\frac{3}{2}}\right)$, $\left(12,3-\sqrt{\frac{3}{2}}\right)$, $\left( -\frac{5}{2},2\right)$, and $\left( -\frac{5}{2},2 \right)$, respectively. Moreover, the corresponding pairs of values $(q,\omega)$ are $(-1,-1)$, $(2,1)$, $(2,1)$, $\left( -\frac{1}{2},-\frac{2}{3} \right)$, and $\left( -\frac{1}{2},-\frac{2}{3} \right)$. Therefore, the critical point $O_2$ is stable, $A_2$ and $B_2$ are unstable, and the points $C_2$ and $D_2$ are saddles. Furthermore, from the flow of phase-space trajectories in Fig.~\ref{fig_4.1} (right panel above), we obtained the evolutionary phase $A_2 \rightarrow O_2$ and $B_2 \rightarrow O_2$, both representing the evolution of the Universe from a stiff fluid-dominated decelerated epoch to an accelerated de Sitter epoch. The same is reflected in the evolutionary profiles of the corresponding cosmological parameters presented in Fig.~\ref{fig_4.2} and \ref{fig_4.3} (middle panel). The results obtained for the parameter constraints in case-II are better than those in case-I.

\item \textbf{Case III ($n=-1$ and $\beta=\sqrt{3}$) :} In this case, the set of critical points obtained is $O_3(0,0)$, $A_3(-1,0)$, $B_3(1,0)$, $C_3\left( \frac{1}{\sqrt{2}}, \sqrt{\frac{1}{2}} \right)$ and $D_3\left( \frac{1}{\sqrt{2}}, -\sqrt{\frac{1}{2}} \right)$ with the corresponding eigenvalues $(\lambda_1,\lambda_2)=(-3,0)$, $\left(12,3+\frac{3}{\sqrt{2}}\right)$, $\left(12,3-\frac{3}{\sqrt{2}}\right)$, $\left( -\frac{3}{2},6 \right)$, and $\left( -\frac{3}{2},6 \right)$, respectively. Moreover, the corresponding pairs of values $(q,\omega)$ are $(-1,-1)$, $(2,1)$, $(2,1)$, $\left( \frac{1}{2},0 \right)$, and $\left( \frac{1}{2},0 \right)$. Therefore, the critical point $O_3$ is stable, $A_3$ and $B_3$ are unstable, and the points $C_3$ and $D_3$ are saddles. Furthermore, from the flow of the phase-space trajectories in Fig.~\ref{fig_4.1} (below), we obtained the evolutionary phase $A_3 \rightarrow O_3$, $B_3 \rightarrow O_3$, and $A_3 \rightarrow C_3 \rightarrow O_3$. The evolutionary trajectory $A_3 \rightarrow C_3 \rightarrow O_3$ is quite interesting, representing the evolution of the Universe from a decelerated stiff era to an accelerated de Sitter era via matter dominated epoch. The same is reflected in the evolutionary profile of the corresponding cosmological parameters presented in Fig.~\ref{fig_4.2} and \ref{fig_4.3} (right panel). The results for the parameter constraints in case-III are better than those in the other two cases.
\end{itemize}
We note that the critical point $O(0,0)$ possesses eigenvalues $(-3, 0)$, making it a non-hyperbolic fixed point since one eigenvalue vanishes. For such points, the Hartman-Grobman theorem does not apply directly, and the linear stability analysis is inconclusive \cite{Perko:2001}. To rigorously determine stability, we employ the center manifold theory. Linearization on $O(0,0)$ produces a stable subspace $E^s$ along the $x$-axis (corresponding to $\lambda_1 = -3$) and a center subspace $E^c$ along the $y$-axis (corresponding to $\lambda_2 = 0$). According to the Center Manifold Theorem \cite{Wiggins:2003}, there exists a local center manifold $W^c$ tangent to $E^c$ at the origin, described by $x = h(y)$ with $h(0) = h'(0) = 0$. Approximating the system Eq.~\eqref{s3}-\eqref{s4} near the origin as $x' \approx -3x + \sqrt{\frac{3}{2}}\beta y^2$ and $y' \approx -\sqrt{\frac{3}{2}}\beta xy$, and assuming an expansion $h(y) = ay^2 + O(y^3)$, the invariance condition $h'(y)y' = x'$ produces the coefficient $a = \beta/\sqrt{6}$. The flow restricted to this manifold is governed by $\dot{y} \approx -(\beta^2/2)y^3$, which indicates asymptotic stability along the center direction since $\dot{y}$ opposes $y$. Combined with the transverse attraction of $\lambda_1 = -3$, this confirms that $O(0,0)$ is a stable attractor, consistent with our numerical phase-space analysis (Fig.~\ref{fig_4.1}) within the physical constraint $x^2 + y^2 \leq 1$ \cite{Carr:2012}.

\subsection{Power-law potential}
\begin{figure}[H]
\centering
\includegraphics[scale=0.48]{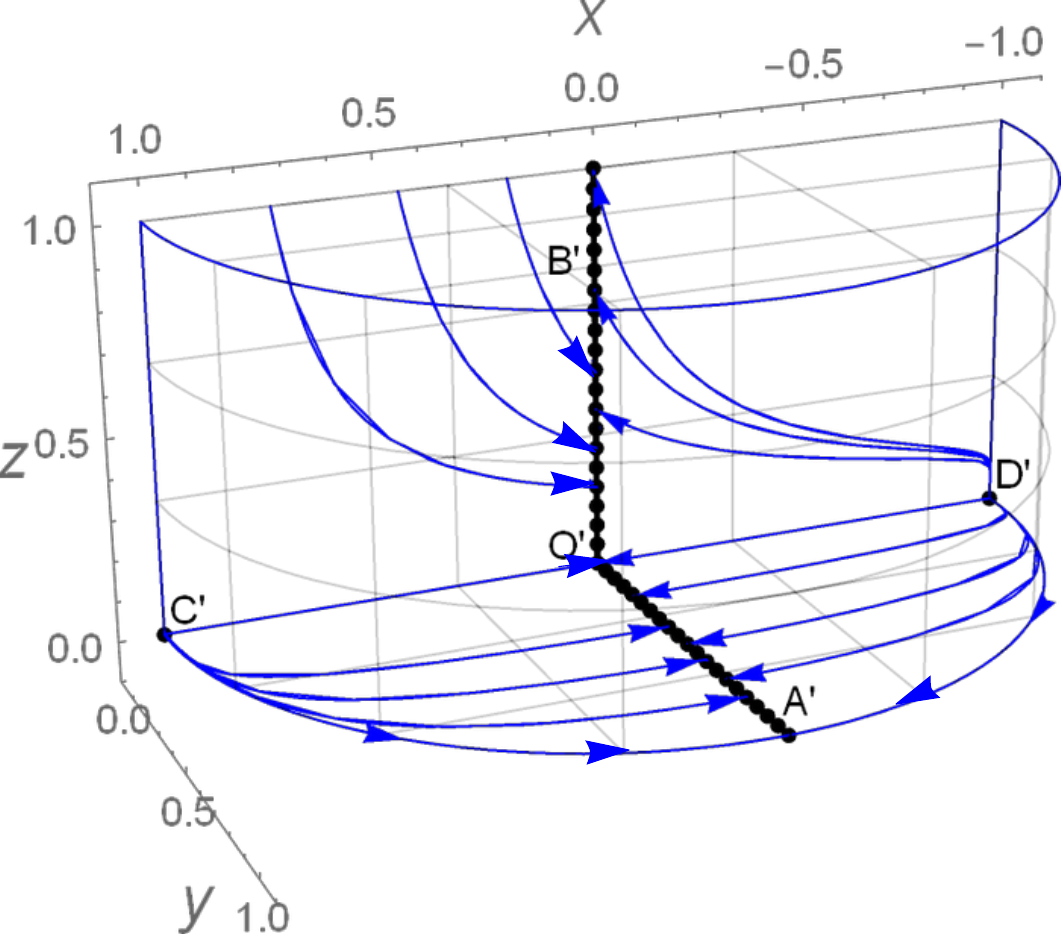}
\caption{The 3-D phase-space trajectories plotted for a set of solutions to the autonomous system for the parameter value  $n=-2$ and $k=0.16$ corresponding to the power-law potential.}\label{fig_4.4}
\end{figure}

\begin{figure}[H]
{\includegraphics[scale=0.41]{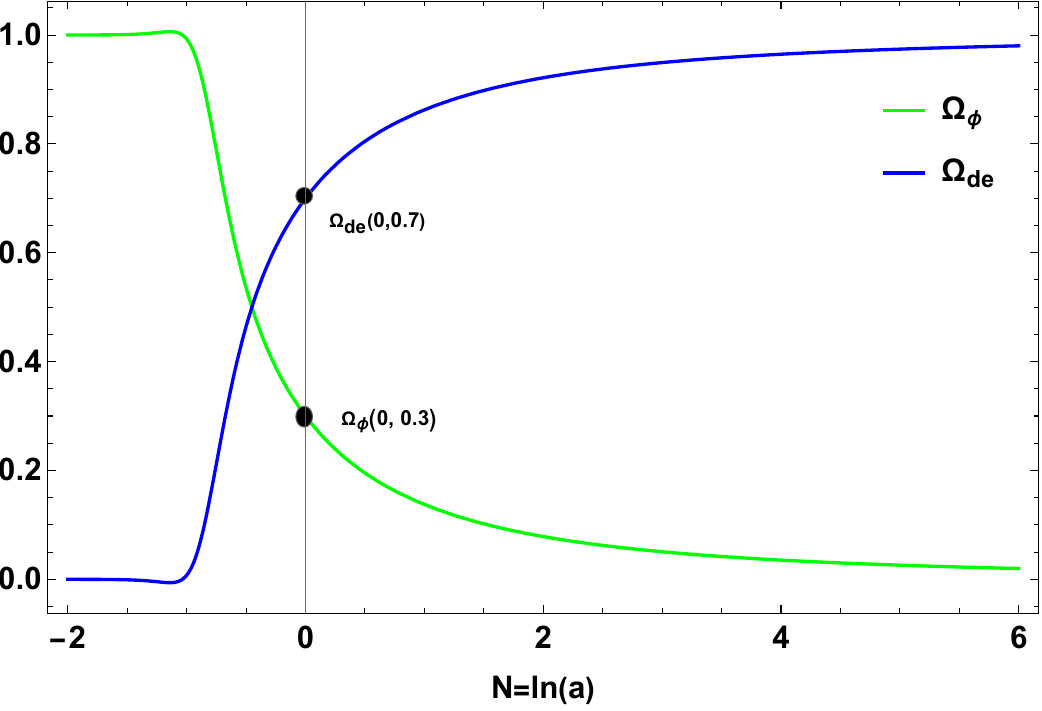}}
{\includegraphics[scale=0.41]{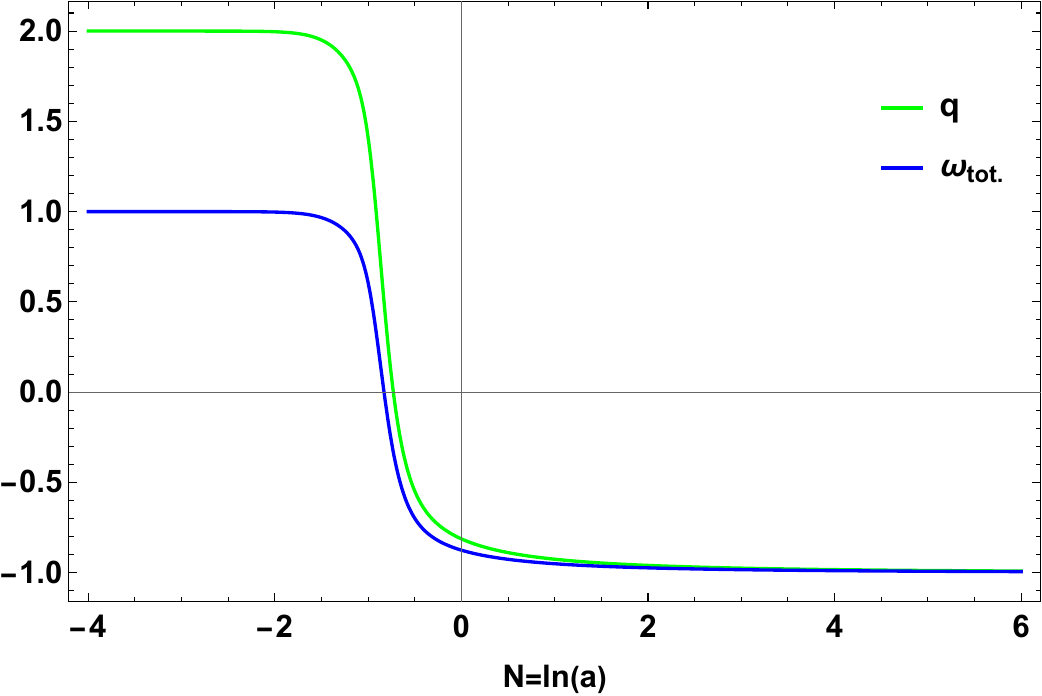}}
\caption{Evolutionary profile of scalar field density, dark energy density, deceleration, and the equation of state parameter for the parameter value $n=-2$ and $k=0.16$ corresponding to the power-law potential.}\label{fig_4.5}
\end{figure}

\justifying
We assume the following form of power-law potential,
\begin{equation}\label{r1}
V(\phi)= V_0\phi^{-k} \,\,\text{.}
\end{equation}
The power-law-type scalar field potentials have already been investigated in the context of both inflation \cite{Ratra:1988} and to address the cosmological constant problem \cite{Peebles:1988}. Such a potential naturally arises in string theory, such as in low-energy states of D-branes, which give an effective Tachyon field \cite{sen1}, and in a Born-Infeld-inspired Lagrangian \cite{born}. In addition, such potential has appeared in Tachyon-dominated backgrounds to probe cosmological perturbations \cite{LRA}. Now, using Eq.~\eqref{r1}, the dynamical variable Eq.~\eqref{4c4} becomes 
\begin{equation}\label{r2}
\lambda=\frac{k}{\phi} \:\: \text{with} \:\: \Gamma= \frac{k+1}{k} \,\,\text{,}
\end{equation}
hence, the system Eq.~\eqref{4d4}-\eqref{4f4} reduces to the following autonomous system,
\begin{equation}\label{r3}
x'=-3x \left[ 1+ \frac{x^2}{n \left( 1-x^2-y^2 \right)-1}  \right] + \sqrt{\frac{3}{2}} \lambda y^2 \,\,\text{,}
\end{equation}
\begin{equation}\label{r4}
y'=  \sqrt{\frac{3}{2}} \lambda xy \left[ 1- \frac{ \sqrt{6}x}{\lambda \{ n \left( 1-x^2-y^2 \right)-1 \} }  \right] \,\,\text{,}
\end{equation}
\begin{equation}\label{r5}
\lambda'= - \frac{\sqrt{6}}{k} \lambda^2 x \,\,\text{.}
\end{equation}
Note that the autonomous system Eq.~\eqref{r3}-\eqref{r5} is invariant under transformation $y \rightarrow -y$, and hence the flow of trajectories obtained in the negative region of $y$ would be a copy of the positive region. Moreover, the system Eq.~\eqref{r3}-\eqref{r5} is also invariant under simultaneous transformations $x \rightarrow -x$ and $\lambda \rightarrow -\lambda$. Therefore, the physical region of the given dynamical system is represented by the positively half cylinder with infinite length from $\lambda=0$ to $\lambda= +\infty$. Hence, we define the following phase-space variable to compactify the variable $\lambda$,
\begin{equation}\label{r6}
z=\frac{\lambda}{\lambda+1}  \:\: \text{or} \:\: \lambda=\frac{z}{1-z} \,\,\text{.}
\end{equation}
The new phase-space variable $z$ is bounded as $0 \leq z \leq 1$ and follows $z=0$ when $\lambda=0$ and $z=1$ when $\lambda \rightarrow + \infty$. Now, the autonomous system Eq.~\eqref{r3}-\eqref{r5} reduces to
\begin{equation}\label{r7}
x'=-3x \left[ 1+ \frac{x^2}{n \left( 1-x^2-y^2 \right)-1}  \right] + \sqrt{\frac{3}{2}} \frac{z}{(1-z)}y^2 \,\,\text{,}
\end{equation}
\begin{equation}\label{r8}
y'=  \sqrt{\frac{3}{2}} \frac{z}{(1-z)} xy \left[ 1- \frac{ \sqrt{6}x(1-z)}{z \{ n \left( 1-x^2-y^2 \right)-1 \} }  \right] \,\,\text{,}
\end{equation}
\begin{equation}\label{r9}
z'= - \frac{\sqrt{6}}{k} z^2 x \,\,\text{.}
\end{equation}
The critical points and their physical behavior, corresponding to autonomous system Eq.~\eqref{r7}-\eqref{r9}, are presented in the Table~\eqref{Table-2}.
\begin{table}[H]
\begin{center}\caption{Table shows the critical points and their behavior corresponding to the model $f(Q)=-Q+ \alpha Q^n$ with potential $V(\phi)= V_0 \phi^{-k}$.}
\begin{tabular}{|c|c|c|c|c|c|}
\hline
 Critical Points $(x_c,y_c,z_c)$ & Eigenvalues ($\lambda_1$, $\lambda_2$,  $\lambda_3$) & Nature of critical point  & $q$ & $\omega$ \\
\hline 
$O'(0,0,0)$ & $ (-3,0,0) $ & Stable  & $-1$ & $-1$ \\
$A'(0,y,0)$ & $ (-3,0,0)  $ & Stable  & $-1$ & $-1$ \\
$B'(0,0,z)$ & $ (0,0,-3) $ & Stable  & $-1$ & $-1$ \\
$C'(1,0,0)$ & $ (3,0,6-6n) $ & Unstable for $ n \leq 1 $ and N.H. for $ n > 1 $   & $2$ & $1$ \\
$D'(-1,0,0)$ & $ (3,0,6-6n) $ & Unstable for $ n \leq 1 $  and N.H. for $ n > 1 $   & $2$ & $1$ \\
\hline
\end{tabular}\label{Table-2}
\end{center}
\end{table}   
We present the asymptotic behavior of some prominent cosmological parameters such as the scale factor, the Hubble parameter, and the scalar field corresponding to the obtained set of critical points.
\begin{itemize}
\item Similarly to case $x_c=0$, we obtain the cosmic scalar factor as $a(t)=e^{H_0(t-t_0)}$, the Hubble parameter as $H(t)=H(t_0)=H_0$, and the scalar field $\phi(t)=\phi(t_0)=\phi_0$, where $t_0$ being the present time, and $a(t_0)=1$ and $H_0$ are the present values of the scale factor and the Hubble parameter.
\item Similarly to the cases $x_c=\pm 1$ and $y_c=0$, we obtain the cosmic scalar factor as $a(t)=\left[ 3H_0(t-t_0)+1\right]^\frac{1}{3}$, the Hubble parameter as $H(t)=\frac{H_0}{3H_0(t-t_0)+1}$, and the scalar field as $\phi(t)=\phi_0 \pm \frac{\sqrt{6}}{3} ln\left[ 3H_0(t-t_0)+1\right] $.
\end{itemize}
The 3-D phase-space diagram plotted for a set of solutions to the autonomous system Eq.~\eqref{r7}-\eqref{r9} that utilize the numerical approach to the value of specific parameters is presented in Fig.~\ref{fig_4.4}. Furthermore, the corresponding evolutionary profile of the scalar field and dark energy density with deceleration and the effective equation of state parameter are presented in Fig.~\ref{fig_4.5}. 

Now we discuss the physical significance and stability of the critical points obtained for the specific parameter value $n=-2$ and $k=0.16$. In this case, the set of critical points obtained is $O'(0,0,0)$, $A'(0,y,0)$ where $0 \leq y \leq 1$, $B'(0,0,z)$ where $0 \leq z \leq 1$, $C'(1,0,0)$, and $D'(-1,0,0)$ with the corresponding eigenvalues $(\lambda_1,\lambda_2,\lambda_3)= (-3,0,0) $, $(-3,0,0) $, $(0,0,-3)$, $(3,0,18)$ and $(3,0,18)$, respectively. Moreover, the corresponding pairs of values $(q,\omega)$ are $(-1,-1)$, $(-1,-1)$, $(-1,-1)$, $(2,1)$, and $(2,1)$. Therefore, the critical points $O'$, $A'$, and $B'$ are stable, while the points $C'$ and $D'$ are unstable. Furthermore, from the flow of the trajectories in the 3-D phase-diagram presented in Fig.~\ref{fig_4.4}, we obtained the evolutionary phase $C' \rightarrow O'$, $D' \rightarrow O'$, $C' \rightarrow A'$, $D' \rightarrow A'$ and $D' \rightarrow B'$. All the presented flow of trajectories presented exhibit identical behavior, representing the evolution of the Universe from a decelerated stiff era to an accelerated de Sitter era. The same is reflected in the evolutionary profile of the corresponding cosmological parameters presented in Fig.~\ref{fig_4.5}. The left panel of Figure \ref{fig_4.5} indicates that the scalar field density vanishes with the expansion of the universe, whereas the dark energy density dominates completely at a late time. The right panel of Fig.~\ref{fig_4.5} indicates the transition from a decelerated epoch to an accelerated epoch of the universe in the recent past, and that ends up with a de Sitter universe at late times. \\
 We observe from Table~\ref{Table-2} that multiple critical points ($O', A', B', C', D'$) possess zero eigenvalues, rendering them non-hyperbolic. Specifically, $O'(0,0,0)$ has eigenvalues $(-3, 0, 0)$, while $A'$ and $B'$ represent lines of fixed points with eigenvalues $(-3, 0, 0)$ and $(0, 0, -3)$ respectively. For fixed point lines ($A', B'$), one zero eigenvalue corresponds to the tangent direction along the continuous set of fixed points (neutral stability along the line), while the negative eigenvalue ensures attraction in the transverse directions. For the isolated point $O'$, the Hartman-Grobman theorem is inconclusive due to the two-dimensional center subspace spanned by the $y$ and $z$ axes. According to Center Manifold Theory \cite{Wiggins:2003}, the dynamics near $O'$ reduce to a 2D center manifold $x = h(y,z)$ tangent to the $y$-$z$ plane. Linearization of Eq.~\eqref{r7}-\eqref{r9} near the origin yields $x' \approx -3x$, indicating rapid contraction along the $x$-direction onto the center manifold. The stability is then determined by the higher-order terms on the manifold. Although analytical reduction is complex due to the compactification variable $z$, our numerical phase-space analysis (Fig.~\ref{fig_4.4}) shows that the trajectories converge to $O'$ along the stable manifold and remain bounded in the center subspace, confirming its stability. Similar reductions in center manifolds apply to $C'$ and $D'$, where the single zero eigenvalue is outweighed by positive eigenvalues, confirming their unstable nature \cite{Perko:2001, Carr:2012}.

\section{Conclusions}\label{sec_4.5} 
In this chapter, we studied scalar field cosmology in the coincident $f(Q)$ gravity formalism. Modified gravity with non-metricity has been extensively investigated recently in different contexts such as late-time observational constraints, black holes, wormholes, and dynamical analysis, whereas the scalar field cosmology has been prominent in describing inflation as well as late-time acceleration. The modified $f(Q)$ function is now responsible for the observed accelerating expansion, whereas the quantum field theory prediction of a tiny value of $\Lambda$ cannot be described by such a modified gravity scenario, and hence the addition of a scalar field takes into account the quintessence scenario. In Sec.~\ref{sec_4.2}, we presented the mathematical formulation of $f(Q)$ gravity with the Lagrangian density of the scalar field. Further, in Sec.~\ref{sec_4.3}, we presented Friedmann-like equations of $f(Q)$ gravity ruling the gravitational interactions under the FLRW background in the presence of a scalar field. In Sec.~\ref{sec_4.4}, we begin with a non-linear $f(Q)$ functional, specifically $f(Q)=-Q+\Psi(Q)=-Q+\alpha Q^n$, where $\alpha$ and $n$ are free model parameters. The assumed functional form is a polynomial correction to the STEGR case and is of great significance in early- and late-time cosmology. The considered $f(Q)$ function with $n > 1$ can potentially apply to the inflationary scenario, whereas the case $n < 1$ corrects the late-time cosmology. Furthermore, we have defined phase-space variables (presented in equations Eq.~\eqref{4a} and Eq.~\eqref{4c4}), and then we expressed the deceleration and the effective equation of state parameter in terms of phase-space variables. Now to probe the cosmological implications of the considered scenario, we assumed two specific forms of the potential function, specifically the exponential one $V(\phi)= V_0 e^{-\beta \phi}$ and the power-law $V(\phi)= V_0\phi^{-k}$, which are widely discussed in the literature. The corresponding critical points and their behaviors for the autonomous systems obtained are presented in Tables~\eqref{4Table-1} and \eqref{Table-2}. Moreover, for both cases, the asymptotic behavior of the scale factor, the Hubble parameter, and the scalar field corresponding to the obtained set of critical points have been presented. In addition, we presented the 2-D phase-space diagrams corresponding to the exponential case for some parameter values, specifically, $n=2$ and $n=-1$ with $\beta=1$ and $n=-1$ with $\beta=\sqrt{3}$, in Fig.~\ref{fig_4.1}. The behavior of corresponding cosmological parameters, such as scalar field and dark energy density, deceleration, and the effective equation of state parameter, is presented in  Fig.~\ref{fig_4.2} and \ref{fig_4.3}. We then discussed the physical significance and stability of the critical points obtained for all three cases. We found that the results for the parameter constraints in case-III are better than those in the other two cases. We obtained an evolutionary phase $A_3 \rightarrow C_3 \rightarrow O_3$ representing the evolution of the Universe from a decelerated stiff era to an accelerated de Sitter era via a matter-dominated epoch. In addition, we have presented the 3-D phase-space diagram plotted for a set of solutions to the autonomous system corresponding to the power-law case for the parameter value $n=-2$ and $k=0.16$ (see Fig.~\ref{fig_4.4}) and the behavior of the corresponding cosmological parameters presented in Fig.~\ref{fig_4.5}. In this case, all trajectories exhibit identical behavior, representing the universe's evolution from a decelerated stiff era to an accelerated de Sitter era. Thus, we can conclude that the exponential case shows better evolution than the power-law case, and hence the present study successfully describes the different cosmological epochs of the Universe.

\chapter{Dynamical System Analysis of Dirac-Born-Infeld Scalar Field Cosmology in Coincident $f(Q)$ Gravity} 

 \label{Chapter5}
\lhead{Chapter 5. \emph{Dynamical System Analysis of Dirac-Born-Infeld Scalar Field Cosmology in Coincident $f(Q)$ Gravity}} 

\vspace{8 cm}
* The work, in this chapter, is covered by the following publications: \\
 
\textit{Dynamical system analysis of Dirac-Born-Infeld scalar field cosmology in coincident $f(Q)$ gravity}, Chinese Physics C, \textbf{48}, 095102 (2024).

\clearpage

\epigraph{``The researches of many commentators have already thrown much darkness on this subject, and it is probable that, if they continue, we shall soon know nothing at all about it.''}{--- Mark Twain, \textit{The Innocents Abroad} (1869), Ch.~26}

In this chapter, we offer the dynamical system analysis of the DBI scalar field in a modified $f(Q)$ gravity context. We have taken a polynomial form of modified gravity and used two different kinds of scalar potential, i.e., polynomial and exponential, and found a closed autonomous dynamical system of equations. We have analyzed the fixed points of such a system and commented on the conditions under which deceleration to late-time acceleration happens in this model. We have noted the similarity of the two models and have also shown that our result is indeed consistent with the previous work done on Einstein's gravity. We have also investigated the phenomenological implications of our models by plotting the EoS ($\omega$), the energy density ($\Omega$), and the deceleration parameter ($q$) w.r.t. to e-fold time and comparing with the present value. Finally, we conclude this chapter by observing how the analysis of the dynamical system differs in modified $f(Q)$ gravity, and we also provide some of the future scope of our work. 

\section{Introduction}\label{sec_5.1}
After the discovery of CMB in 1965 \cite{cmb}, it became clear that our Universe started in a very hot, dense state called ``Hot Big Bang" \cite{gamow} and has evolved in its current form. This is known as the standard Big Bang theory of cosmology. After the discovery of late time acceleration \cite{late1,late2} and observation from the galaxy rotation curve, it has been clear that there are other objects in our Universe, besides baryonic matter \cite{planck}. In the standard $\Lambda$CDM paradigm, which is probably the most successful theory of the current state of the Universe, one takes dark energy (which is responsible for late-time acceleration) to be the cosmological constant $\Lambda$ and dark matter to be cold (non-radiative). Although the $\Lambda$CDM model is so successful with phenomenological predictions and observational evidence, it has some serious problems. One of the main problems is the nature of dark energy. If one assumes that the cosmological constant ($\Lambda$) is solely responsible for dark energy, then the calculations from QFT can be shown to have a discrepancy of order ($10^{120}$) \cite{WENB}. One natural way to explain this is by introducing the scalar field (quintessence field \cite{Carroll:2000}), which can explain why the current value of the cosmological constant is so low. The scalar field also appears quite naturally in the early inflation scenarios, which can naturally explain the horizon problem and the flatness problem, etc.  \\
Although the scalar field can explain both early inflation and late-time acceleration, the exact form or origin of the scalar field is not known. There are many such candidates for the origin of the inflation or quintessence fields. In this chapter, we take DBI as the origin of the scalar field, which naturally comes from string theory. We have also performed a dynamical system analysis in the flat FLRW background and provided phenomenological predictions (evolutionary graphs of $q$, $\Omega$, and $\omega$) based on the fixed-point analysis.\\
It is well known that Einstein's general theory of relativity is not renormalizable in the context of quantum field theory. There have been several attempts to find a renormalizable theory of quantum gravity, and string theory offers one such unification. It is well known that even in bosonic string theory, quantization of the Polyakov action (conformal transformation of the Nambu-Goto action) gives a Tachyon-like field which soon decays via spontaneous symmetry breaking \cite{green,polchinski}. It was first observed by Mazumdar et al. \cite{mazumdar} that the decay of a non-BPS $D4$ brane to a stable $D3$ brane can give rise to a Tachyon field, which can act as an inflation field in the cosmological context. In 2002, a series of three papers by Sen \cite{sen1,sen2,sen3} showed how, in string theory, as well as in string field theory, Tachyons occur naturally, and in \cite{sen3} it has been shown that the effective field of such Tachyons can be viewed as DBI scalar field theory.\\
Soon after these proposals, Padmanabhan \cite{paddy} and Gibbons \cite{gibbons1} showed how these DBI-type fields could be used in the FLRW background to give inflation field-like behaviour. Alternative ways of obtaining the DBI field from other forms of string theory have been reviewed by Gibbons \cite{gibbons2}. The study of the DBI field in late time acceleration context has been done by Bhagla et al. \cite{bhagla}, while Gorini et al.\cite{gorini}
offered an alternative way to visualize the DBI field as a modified Chaplygin gas. We also note that the DBI field has been proposed as an alternative to dark matter by Padmanabhan \cite{paddy2}. This shows that the DBI field could indeed affect the late-time cosmology.\\
 Copeland \cite{copeland1} and Aguirregabiria \cite{aguirregabiria} first studied the study of the DBI field in the dynamical system setting. In this paper, we are also closely following the treatment given in \cite{copeland1}. Soon after that, Fang and Lu \cite{fang2010a} considered a much more general type of potential beyond the inverse square potential, the work later extended by Quiros et al. \cite{Quiros} to include much more general potentials, and they have given an exact treatment of the $sinh(\phi)$ potential. Guo \cite{guoexp} has chosen an exponential potential for dynamical system analysis, which we have used here to make an autonomous dynamical system. It is also worth noting that as Silverstein and Tong \cite{tong} have shown, if one considers a D3-brane moving towards the horizon of AdS space, one can get a generalized DBI field in a strong coupling limit (as opposed to a weak coupling limit where the previous work has been done). In the strong coupling limit, it can be shown that the DBI field gets extra contributions from the movement of the D3-brane and the lagrangian becomes $\mathcal{L}_{GDBI}=\frac{1}{f(\phi)}(\sqrt{1+f(\phi)\partial \phi^2}-1)-V(\phi)$.\\
The conventional concept of relativity, especially GR, which interprets gravity as the curvature of spacetime, may not offer the definitive solution to elucidate dark energy. This encourages the exploration of alternative theoretical frameworks in cosmology that can effectively address cosmic acceleration while remaining consistent with observational data. GR and its curvature-based extensions have been formulated and thoroughly examined in previous research \cite{CANT, R15}. Recently, alternative theories of gravitation based on a flat spacetime geometry, relying solely on non-metricity, have been established and extensively explored \cite{NEST, Jimenez/2018}. The $f(Q)$ gravity, with its various astrophysical and cosmological implications, has been widely investigated \cite{R18, LAVI-1, JIM-2, Jimenez/JCAP2018, ANAG, KUHN, R24, R25, R26, R27, WOM-2, R29}. In this chapter, we have shown that even with modified $f(Q)$ gravity, we are getting a similar kind of late-time accelerating behavior where $q$ is $-1$ as expected from de Sitter-like expansion. We also note that in our investigation, the present value of the deceleration parameter is reaching nearly $-0.8$, which is quite consistent with the observed value $-0.55$.\\
We initiate our exploration by introducing a set of dimensionless variables that encapsulate the complete evolution of the system's phase space. These variables facilitate transforming the system's dynamics into an autonomous structure, thereby enhancing our understanding of its behavior. Several noteworthy findings within the context of modified gravity utilizing dynamical system techniques have appeared in references \cite{DE-3,WOM,HAMID,APLL,OO1,DiValentino:2021izs}. We note that in this chapter we have taken both the DBI scalar field (for the quintessence field )and $f(Q)$ gravity to discuss the late-time acceleration phase. We are taking the DBI field as a quintessence field to explain late-time acceleration, which is reasonable, as the inflationary field could indeed be responsible for late-time acceleration as discussed in \cite{Ratra:1988,Peebles:1988}. Also, in those energy scales, it is reasonable to expect that $f(Q)$ gravity would emerge as an effective field theory of higher order corrections to graviton-graviton interactions \cite{effective}. The general criterion for the DBI field to give a later acceleration similar to that of de Sitter is given by the theorem of Hao \cite{hao} and Chingangbam \cite{chingangbam}. In this chapter, we have used the results from the chapter~\ref{Chapter2}, especially from Sec.~\ref{sec_2.5}, where the detailed formulation of $f(Q)$ gravity is given and from Sec.~\ref{sec_2.6} where we have applied flat FLRW metric in the $f(Q)$ gravity to get modified Friedmann equations. In Sec.~\ref{sec_5.4}, we invoke the phase-space variables and perform the complete dynamical system analysis for the exponential and power-law potentials under the $f(Q)$ gravity formalism in the presence of a DBI scalar field. Finally, in Sec.~\ref{sec_5.5}, we present our findings of the investigation.

\section{The dynamical system analysis in the presence of DBI field}\label{sec_5.4}
One of the main troubles of using string theory in cosmology directly is the so-called no-go theorem \cite{hao,chingangbam}, for wrapped products by compactifying the extra dimensions. We note that from the equations below the dynamical system equations we can see that it does not get closed for the generalized DBI field but is closed for ordinary DBI. We also have compactified the phase space (the $\lambda$ axis) using Eq.~\ref{5g}. Using compactification, we have drawn the 3D phase space given in Figure \ref{fig_5.1}. \\
In string theory, it was predicted by Sen \cite{sen1,sen2,sen3} that there are 
Tachyon fields in both open and closed string theory. For more on open and closed string theory, one can refer to \cite{polchinski}.  Even though for closed string theory, the Tachyon fields are projected out in open string, they remain. Although one can use a spontaneous symmetry-breaking argument to get rid of tachyon modes, one can still fully explain the reason for their existence. In bosonic string theory, if one uses the Nambu-Goto action, then it is almost impossible to quantize. In order to obtain meaningful quantization rules, one has to invoke the conformal invariant Polyakov action. Using the conformal field theory techniques, one can quantize such an action, which leads to the undesirable Tachyon modes. Although they violate causality, they can be shown to be unstable. So, Tachyon modes are typically given by the Dirac-Born-Infeld (DBI) Lagrangian, which has the following form,
\begin{equation}\label{4a4}
\mathcal{L}_{DBI}= - V(\phi)\sqrt{1+\partial \phi^2} \,\,\text{.} 
\end{equation}
Here, $\partial\phi^2=\partial^{\mu}\phi\partial_{\mu}\phi$ and $V(\phi)$ is a potential function of the scalar field and $\partial^{\mu}\phi\partial_{\mu}\phi$ denotes the kinetic term for Tachyon fields.\\
From the Lagrangian, one can find the field equation for the DBI field from the Euler-Lagrangian equation as,
\begin{equation}\label{4b}
    \frac{\Ddot{\phi}}{1-\dot{\phi}^2}+3H\dot{\phi}+\frac{V_{,\phi}}{V}=0 \,\,\text{.} 
\end{equation}
This is the modified Klein-Gordon equation for the DBI field.\\
The Friedmann equations in  Eq.~\eqref{3m4}-\eqref{3n4} becomes,
\begin{equation}\label{4c}
    3H^2= \rho_{DBI} + \rho_{Q} \,\,\text{,} 
\end{equation}
and,
\begin{equation}\label{4d}
    \dot{H}=-\frac{1}{2} [\rho_{DBI} + p_{DBI}+\rho_{Q}+p_{Q}] \,\,\text{.} 
\end{equation}
Here, $\rho_{Q}$ and $p_{Q}$ is given by Eq.~\eqref{3o}-\eqref{3p}.\\
We also note that for such cases the energy density ($\rho_{DBI}$) and pressure ($p_{DBI}$) are given by ,
\begin{equation}\label{4e}
    \rho_{DBI}=\frac{V}{\sqrt{1-\dot{\phi}^2}} \,\,\text{,} 
\end{equation}
and,
\begin{equation}\label{4f}
    p_{DBI}=-V\sqrt{1-\dot{\phi}^2} \,\,\text{.} 
\end{equation}
So, the equation of state ($\omega_{DBI}$) is given by,
\begin{equation}\label{4g}
    \omega_{DBI}=\frac{p_{DBI}}{\rho_{DBI}}=\dot{\phi}^2-1 \,\,\text{.} 
\end{equation}
To construct the autonomous dynamical system, we use the following. We can define the variables as $x=\dot{\phi}$ and $y=\frac{\sqrt{V}}{\sqrt{3}H}$, so we can have $x^2=\dot{\phi}^2$ and $y^2=\frac{V}{3H^2}$ and $s^2=\Omega_{de}=\frac{\rho_{de}}{3H^2}$. Now, Eq.~\eqref{4c} becomes, 
\begin{equation}\label{4h}
 s^2 = 1-\frac{y^2}{\sqrt{1-x^2}} \,\,\text{.}  
\end{equation}
We note that to form the dynamical system, it is more convenient to take the ``e-folding" timing defined as $N=\ln a$, so we get $\frac{d}{dt}=H\frac{d}{dN} $.\\
As $\dot{x}=\ddot{\phi}$ we can write $x^{\prime}=\frac{\Ddot{\phi}}{H}$ (where $\prime$ denotes the derivative with respect to ``e-folding" time and $\dot{}$ denotes the derivative with respect to ordinary time). Using these expressions in the Klein-Gordon equation for the DBI field given in Eq.~\eqref{4b}, we get,
\begin{equation}\label{4i}
 \Ddot{\phi}= (1-x^2)[\lambda V^{\frac{1}{2}}-3Hx] \,\,\text{.}    
\end{equation}
Here, we have defined the variable $\lambda$ as $\lambda = -\frac{V_{,\phi}}{V^{\frac{3}{2}}}$.
Now using Eq.~\eqref{4i} and the fact that $y=\frac{\sqrt{V}}{\sqrt{3}H}$ in the expression $x^{\prime}=\frac{\Ddot{\phi}}{H}$, we obtain the following,
\begin{equation}\label{4j}
 x^{\prime}=(x^2-1)[3x-\sqrt{3}\lambda y] \,\,\text{.}  
\end{equation}
Furthermore, in differentiating the variable $y$ from the e-folding time $N$, we obtain,  
\begin{equation}\label{4k}
 y^{\prime}= -\frac{1}{2}y\left[\sqrt{3}\lambda xy+2\frac{\dot{H}}{H^2}\right]  \,\,\text{.}    
\end{equation}
Now, utilizing Eq.~\eqref{3o}-\eqref{3p} and Eq.~\eqref{4e}-\eqref{4f} in Eq.~\eqref{4d}, we have,
\begin{equation}\label{4l}
   \frac{\dot{H}}{H^2}= \frac{3x^2y^2}{2\sqrt{1-x^2}[\Psi_Q+2Q\Psi_{QQ}-1]} \,\,\text{.}   
\end{equation}    
Hence, Eq.~\eqref{4k} becomes,
\begin{equation}\label{4m}
  y^{\prime}= -\frac{1}{2}y\left[\sqrt{3}\lambda xy+ \frac{3x^2y^2}{\sqrt{1-x^2}[\Psi_Q+2Q\Psi_{QQ}-1]}\right] \,\,\text{.} 
\end{equation}
Now, in order to obtain the closed form of the variable $\lambda$, we define another quantity $\Gamma$ as $\Gamma=\frac{VV_{,\phi\phi}}{V_{,\phi}}$. Differentiating the variable $\lambda$ with respect to e-folding time $N$ with the quantity $\Gamma$, we obtain the following,
\begin{equation}\label{4n}
  \lambda^{\prime}= \sqrt{3} xy\lambda^2\left[\frac{3}{2}-\Gamma\right] \,\,\text{.}   
\end{equation}
In our analysis, we consider the cosmological model $f(Q)=-Q+\Psi(Q)=-Q+\alpha Q^n$ to be of great significance. A power-law correction to the STEGR will give rise to branches of solution applicable either to the early Universe or to late-time cosmic acceleration. The model characterized by value $n < 1$ can describe late-time cosmology, potentially influencing the emergence of dark energy, whereas the model characterized by value $n > 1$ can describe the early Universe phenomenon \cite{JIM-2}.  Moreover, one can observe that the case $\alpha=0$, i.e., $\Psi=0 \Rightarrow f(Q)=-Q$, recovers the GR.\\
Hence, using $\Psi(Q)=\alpha Q^n$, we have $[\Psi_Q+2Q\Psi_{QQ}-1]=(2n-1)\alpha n Q^{n-1}-1$.\\
Moreover, $s^2=(-\frac{\Psi}{2}+Q\Psi_Q)\frac{1}{3H^2}= (2n-1)\alpha Q^{n-1}$. Thus, we have $[\Psi_Q+2Q\Psi_{QQ}-1]=ns^2-1 $. Therefore, the expressions Eq.~\eqref{4l} and Eq.~\eqref{4m} become,
\begin{equation}\label{4o}
  \frac{\dot{H}}{H^2}= \frac{3x^2y^2}{[(n-1)\sqrt{1-x^2}-ny^2]} \,\,\text{,}    
\end{equation}
and,
\begin{equation}\label{4p}
  y^{\prime}= -\frac{1}{2}y\left[\sqrt{3}\lambda xy+ \frac{3x^2y^2}{[(n-1)\sqrt{1-x^2}-ny^2]}\right]  \,\,\text{.}    
\end{equation}
   
\subsection{Exponential potential}
\begin{figure}[H]
\centering
\includegraphics[scale=0.52]{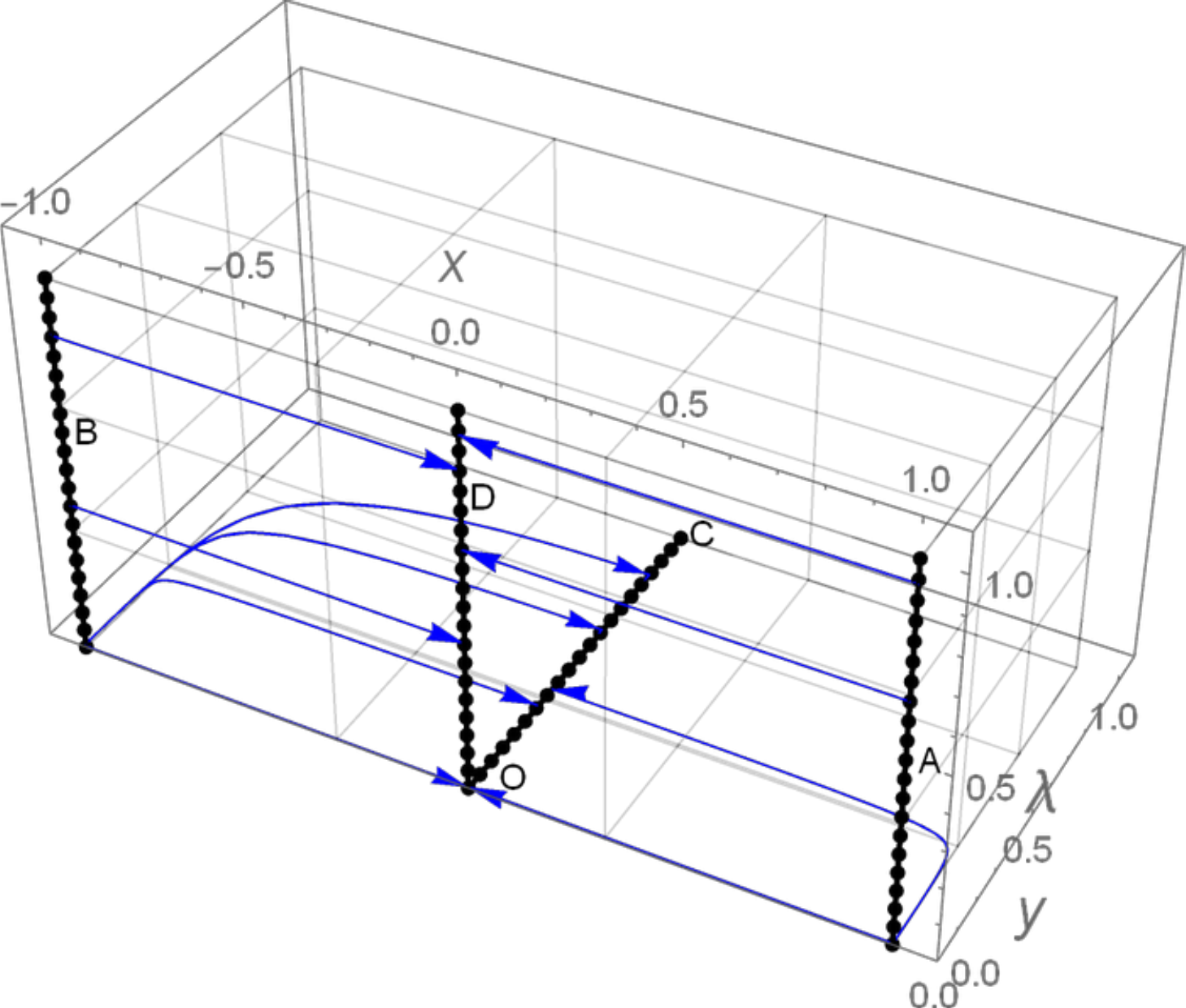}
\caption{The 3-D phase-space trajectories plotted for a set of solutions to the autonomous system Eq.~\eqref{5b}-\eqref{5d} corresponding to the exponential potential.}\label{fig_5.1}
\end{figure}

\begin{figure}[H]
{\includegraphics[scale=0.52]{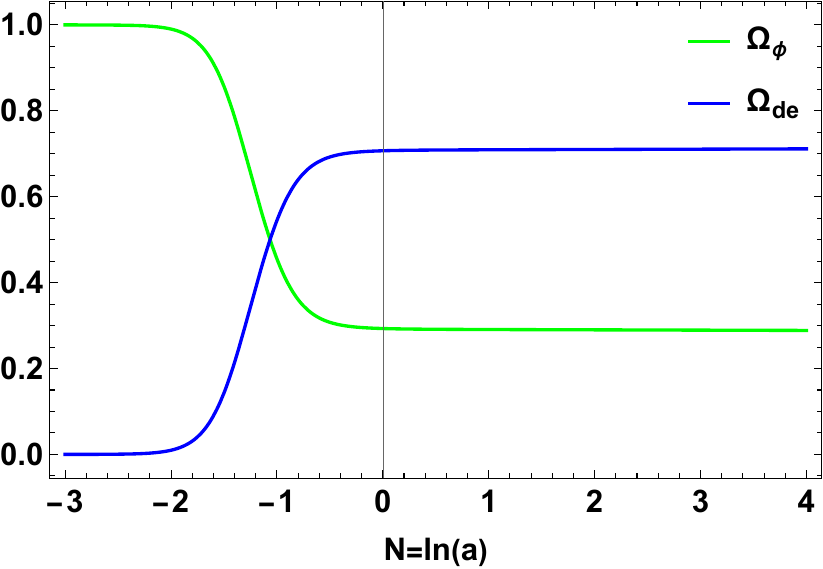}}
{\includegraphics[scale=0.56]{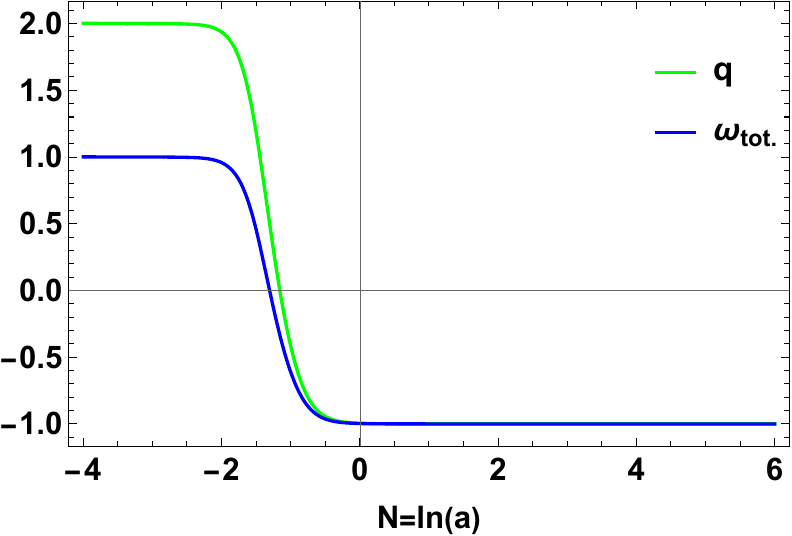}}
\caption{Evolutionary profile of scalar field density, dark energy density, deceleration, and the equation of state parameter for the exponential potential in DBI scalar field.}\label{fig_5.2} 
\end{figure}

\justifying
We note that an exponential potential in the DBI field can arise from the fact that if one takes the dark matter with the phantom field (which can naturally arise from string theory) and applies this fact in the present epoch, the dark matter energy density and phantom energy density are comparable (coincidence problem). One can show that it would indeed give rise to an exponential potential \cite{guoexp}. \\
Therefore, we take the following form of the exponential potential, 
\begin{equation}\label{5a}
V(\phi)= V_0 e^{-\beta \phi} \,\,\text{.} 
\end{equation}
Then for this choice of potential, we obtain $\lambda=-\frac{V_{,\phi}}{V^{\frac{3}{2}}}=\frac{\beta}{\sqrt{V_0e^{-\beta\phi}}}$, and $\Gamma=\frac{VV_{,\phi\phi}}{V_{,\phi}^2}=1$.\\
Therefore, the complete autonomous form of dynamical Eq.~\eqref{4j}, Eq.~\eqref{4n}, and Eq.~\eqref{4p} given as
\begin{equation}\label{5b}
  x^{\prime}=(x^2-1)(3x-\sqrt{3}\lambda y) \,\,\text{,} 
\end{equation}
\begin{equation}\label{5c}
  y^{\prime}= -\frac{1}{2}y\left[\sqrt{3}\lambda xy+ \frac{3x^2y^2}{[(n-1)\sqrt{1-x^2}-ny^2]}\right] \,\,\text{,}    
\end{equation}
\begin{equation}\label{5d}
\lambda^{\prime}= \frac{\sqrt{3}}{2} xy\lambda^2  \,\,\text{.} 
\end{equation}
Using Eq.~\eqref{4o} and the definition of the deceleration parameter $q=-1-\frac{\dot{H}}{H^2}$, we obtain the following expression corresponding to the model parameter $n=-1$,
\begin{equation}\label{5e}
q=-1-\frac{3x^2y^2}{-2\sqrt{1-x^2}+y^2} \,\,\text{,} 
\end{equation}
and the effective equation of state parameter is given by,
\begin{equation}\label{5f}
\omega= \omega_{total}= \frac{p_{eff}}{\rho_{eff}}  = \frac{p_{DBI} + p_{Q}}{\rho_{DBI} + \rho_{Q}}= -1-\frac{2\dot{H}}{3H^2} = -1-\frac{2x^2y^2}{-2\sqrt{1-x^2}+y^2} \,\,\text{.} 
\end{equation}
We present the critical points and their behavior (see Table~\eqref{Table-1}) for the autonomous system Eq.~\eqref{5b}-\eqref{5d} corresponding to the model parameter $n=-1$.
\begin{table}[H]
\begin{center}\caption{Table shows the critical points and their behavior corresponding to the model parameter $n=-1$ and potential $V(\phi)= V_0 e^{-\beta\phi}$.}
\begin{tabular}{|c|c|c|c|c|c|}
\hline
 Critical Points $(x_c,y_c,z_c)$ & Eigenvalues ($\lambda_1$, $\lambda_2$,  $\lambda_3$) & Nature of critical point  & $q$ & $\omega$ \\
\hline 
$O(0,0,0)$ & $ (-3,0,0) $ & Nonhyperbolic (stable)  & $-1$ & $-1$ \\
$A(1,0,\lambda)$ & $ (6,0,0)  $ & Nonhyperbolic  & $-1$ & $-1$ \\
$B(-1,0,\lambda)$ & $ (6,0,0) $ & Nonhyperbolic  & $-1$ & $-1$ \\
$C(0,y,0)$ & $ (-3,0,0) $ & Nonhyperbolic (stable)  & $-1$ & $-1$ \\
$D(0,0,\lambda)$ & $ (-3,0,0) $ & Nonhyperbolic (stable)  & $-1$ & $-1$ \\
\hline
\end{tabular}\label{Table-1}
\end{center}
\end{table} 
We note that since $\lambda$ can take any value and is an even function (as the equation remains the same when $\lambda\rightarrow -\lambda$), we can take the physical region of the given dynamical system as the positive half cylinder with infinite length from $\lambda=0$ to $\lambda= +\infty$. Hence, we compactify the variable by defining the following phase-space variable $z$ as follows \cite{BAHA},
\begin{equation}\label{5g}
z=\frac{\lambda}{\lambda+1}  \:\: \text{or} \:\: \lambda=\frac{z}{1-z} \,\,\text{.} 
\end{equation}
Now, the evolutionary trajectories of the autonomous system Eq.~\eqref{5b}-\eqref{5d} utilizing the above compactified variable are presented in Fig.~\ref{fig_5.1}.
The evolutionary profile of the scalar field density,
dark energy density, deceleration, and the equation of state parameter for the exponential potential have been presented in Fig.~\ref{fig_5.2}.
Note that we have taken the entire plot on the $ln(a)$ axis (see Fig.~\ref{fig_5.2}) and the present value of the scale factor is taken to be $a=1$, that is, $N=ln(1)=0$ is the present time. Furthermore, the value $a < 1$ i.e., $N=ln(a) < 0$ represents a distant past, while the value $a > 1$ i.e., $N=ln(a) > 0$ represents a distant future.

\subsubsection{Limit point analysis for $(\pm1,0,\lambda)$}
Note that, for the critical point $A(1,0,\lambda)$ (obtained in Table~\eqref{Table-1}), we find that $q$ is taking an undefined form (see Eq.~\eqref{5e}), so we circumvent the problem by taking the appropriate limit of that fixed point.\\
 We first take $x=1-\epsilon_1$ and $y=\epsilon_2$ in the expression given in Eq.~\eqref{4o} (for the $n=-1$ case) where $\epsilon_1,\epsilon_2>0$ and when we take $\epsilon_1,\epsilon_2\rightarrow 0$ we recover the original fixed points.\\
 Also, $\frac{\dot{H}}{H^2}$ becomes,
 \begin{equation}\label{5h}
\frac{\dot{H}}{H^2} =\frac{3(1-\epsilon_1)^2\epsilon_2^2}{\epsilon_2^2-2\sqrt{2\epsilon_1-\epsilon_1^2}}
     = \frac{3(1-\epsilon_1)^2}{1-2\sqrt{2\frac{\epsilon_1}{\epsilon_2^4}-\frac{\epsilon_1^2}{\epsilon_2^4}}}.   
 \end{equation}
 If we happen to take the limit in such a way that $\frac{\epsilon_1}{\epsilon_2^4}\rightarrow 1$, that is, $\frac{1-x}{y^4}\rightarrow1$, then, as we note that $\frac{\epsilon_1^2}{\epsilon_2^4}\rightarrow0$, we get the following expression.
\begin{equation}\label{5i}
\frac{\dot{H}}{H^2}=\frac{3}{1-2\sqrt{2}}\approx -1.64<-1 \,\,\text{.} 
\end{equation}
Hence we get
\begin{equation}\label{5j}
q=-1-\frac{\dot{H}}{H^2}\approx 0.64 \,\,\text{.} 
\end{equation}
We note that for a matter dominated Universe ($a=t^{\frac{2}{3}}$), we know $q=0.5$, as we can see in that our limit along that particular trajectory when $\frac{1-x}{y^4}\rightarrow1$ gives 0.64, which is quite consistent with the observation from matter dominated to late time acceleration phase.\\
In the same limit, we also note that $\omega=-1-\frac{2x^2y^2}{-2\sqrt{1-x^2}+y^2}\approx0.09$.\\
The deceleration parameter plays a vital role in describing the expansion phase of the Universe. The value $q < 0$ depicts the accelerating behavior, whereas the transition from acceleration phase to deceleration phase admits the value $q \geq 0$. Thus, we get
\begin{equation}\label{5k}
q \geq 0 \iff  \frac{\dot{H}}{H^2} \leq -1 \iff  \frac{3}{2\sqrt{2\frac{\epsilon_1}{\epsilon_2^4}-\frac{\epsilon_1^2}{\epsilon_2^4}}-1} \geq 1 \iff 2 \geq \sqrt{2\frac{\epsilon_1}{\epsilon_2^4}-\frac{\epsilon_1^2}{\epsilon_2^4}} \iff  2 \geq\frac{\epsilon_1}{\epsilon_2^4} \iff 2 \geq \frac{1-x}{y^4} \,\,\text{.} 
\end{equation}
We note that in the above equations, we have used the facts $x\rightarrow1$ and $y\rightarrow0$. In addition, the equality holds when $a\propto t$.\\
Similarly, the criteria Eq.~\eqref{5k} gives $\omega\geq -\frac{1}{3}$. In addition, for arbitrarily $n$, the criterion Eq.~\eqref{5k} for the transition from acceleration phase to deceleration phase becomes $\frac{1-x}{y^4}\leq \frac{8}{(1-n)^2}$.
We also note that $\frac{\dot{H}}{H^2}=\frac{3}{1-2\sqrt{2}}$ gives $a\propto t^{\frac{2\sqrt{2}-1}{3}}\approx t^{0.609}$, we note that for matter dominated Universe $a\propto t^{\frac{2}{3}}\approx t^{0.66}$. Similarly, for the case of $B(-1,0,\lambda)$ we can get identical expressions with similar expressions for $q$ and $\omega$, as in the previous case, taking $\epsilon_1=x+1$ (noting that $x\geq-1$ and close to $(-1,0,0)$).\\
\begin{figure}[H]
{\includegraphics[scale=0.36]{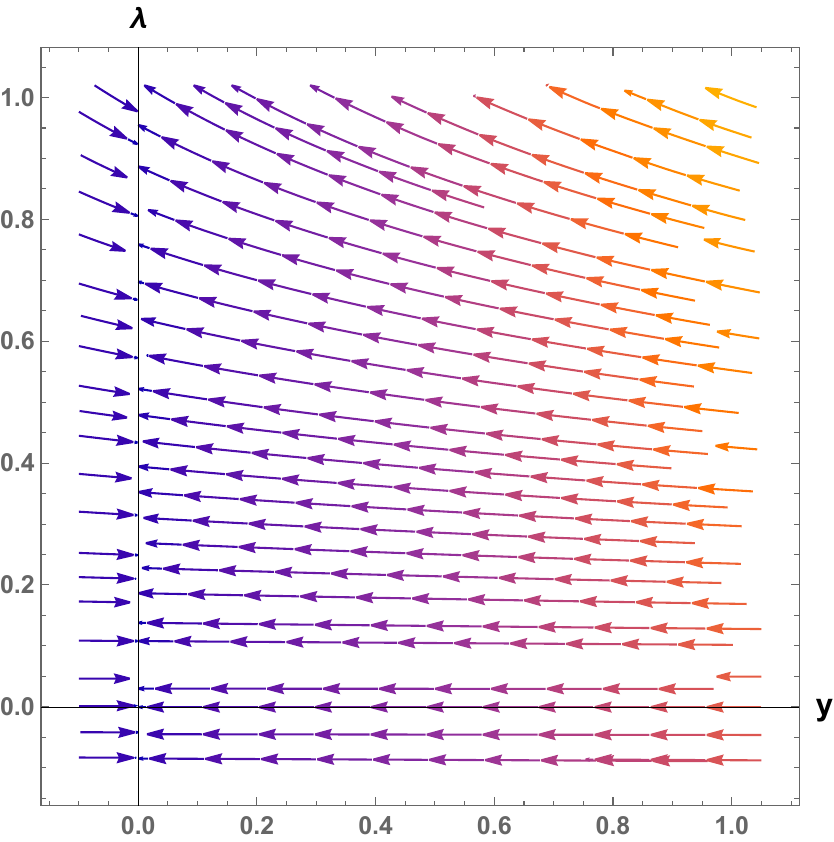}}
{\includegraphics[scale=0.36]{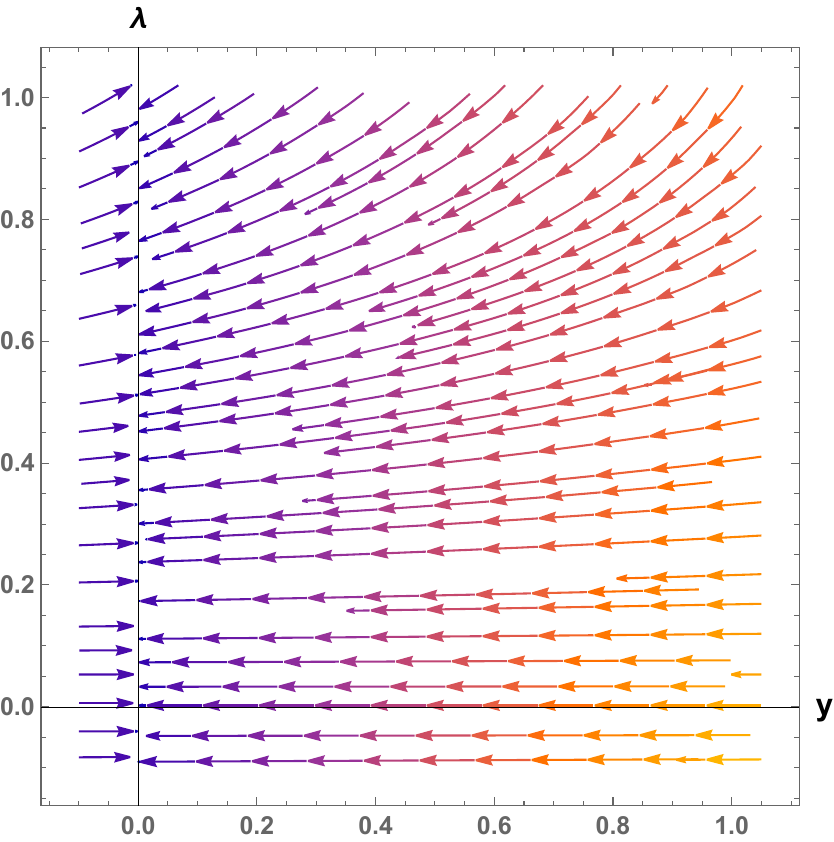}}
{\includegraphics[scale=0.50]{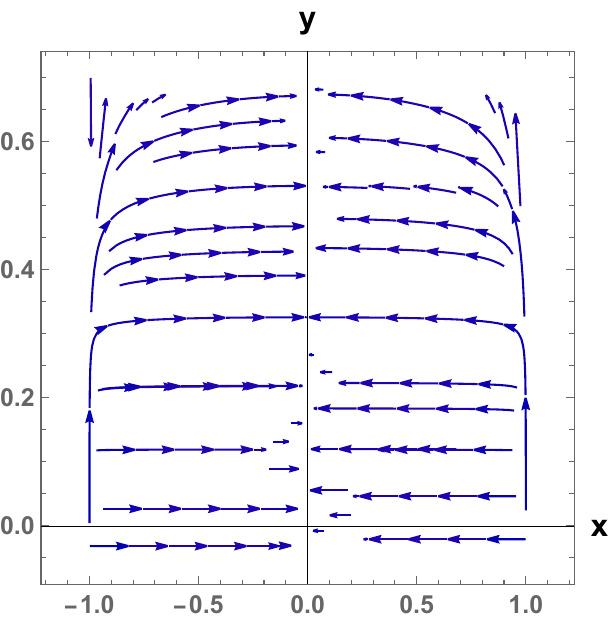}}
\caption{2D phase portrait for the $x=1,-1$ and $xy$ planes and shows that even though the critical points are non-hyperbolic, one can still take some particular limits to see the matter-dominated to de Sitter transition.}\label{fig_5.3}
\end{figure}
We see in Fig.~\ref{fig_5.3} that we get a 2D phase portrait for $x=1$, as we can see that the corresponding eigenvalue for the critical point $A(1,0,\lambda)$, as we can see from the nature it is unstable, also $q=-1$ and $\omega=-1$ so it is giving a de Sitter type solution. Indeed, there is a general theorem of \cite{hao} and \cite{chingangbam} that this type of potential with a well-defined minimum would always lead to de Sitter-type solutions. We can partially verify this by noting that the assertion holds even in modified $f(Q)$ gravity. \\
We observe from Table~\ref{Table-1} that all critical points ($O, A, B, C, D$) possess zero eigenvalues, rendering them non-hyperbolic. Specifically, the points $O, C, D$ have eigenvalues $(-3, 0, 0)$ with one negative eigenvalue and two zero eigenvalues, while the points $A, B$ have eigenvalues $(6, 0, 0)$ with one positive and two zero eigenvalues. For such points, the Hartman-Grobman theorem does not apply and linear stability analysis is inconclusive \cite{Perko:2001}. The zero eigenvalues arise from the continuous parameter $\lambda$ (compactified as $z$), which creates lines of fixed points rather than isolated points. For points $O, C, D$, the negative eigenvalue ensures attraction along the transverse $x$-direction, while the center manifold dynamics along the $y$-$z$ plane determine the ultimate stability. For points $A, B$, the positive eigenvalue ensures repulsion along the transverse direction, making them unstable regardless of the center manifold behavior \cite{Wiggins:2003}. Since the analysis of the central manifold reduction is complex for this compactification 3D DBI system, we rely on numerical phase-space analysis (Fig.~\ref{fig_5.1}-\ref{fig_5.3}), which confirms that the trajectories converge to $O, C, D$ and diverge from $A, B$, supporting our stability classification \cite{Carr:2012}.

\subsection{Power-law potential}
\justifying
In this subsection, we take the power-law potential, noting that the power-law potential most naturally gives global attractor solutions \cite{copeland1}. In addition, these solutions have been shown to be stable under perturbation \cite{fang2010a}.
We assume the following form of power-law potential,
\begin{equation}\label{6a}
V(\phi)= V_0\phi^{-k} \,\,\text{.} 
\end{equation}
Then for this choice of potential, we obtain $\lambda=-\frac{V_{,\phi}}{V^{\frac{3}{2}}}=\frac{V_0k\phi^{-k-1}}{V_0\phi^{-k}(V_0\phi^{-k})^{\frac{1}{2}}}=\frac{k}{\sqrt{V_0}}\phi^{-1+\frac{k}{2}}$.\\
In particular, for $k=2$ we get $\lambda=\frac{2}{\sqrt{V_0}}$.\\
In addition, the quantity $\Gamma$ for the power-law potential considered becomes $\Gamma=\frac{VV_{,\phi\phi}}{V_{,\phi}^2}=\frac{k+1}{k}$,\\
Therefore, the complete autonomous form of dynamical Eq.~\eqref{4j}, Eq.~\eqref{4n}, and Eq.~\eqref{4p} for the power-law potential becomes
\begin{equation}\label{6b}
x^{\prime}= (x^2-1)(3x-\sqrt{3}\lambda y)  \,\,\text{,} 
\end{equation}
\begin{equation}\label{6c}
y^{\prime}= -\frac{1}{2}y^2\left[\sqrt{3}\lambda xy+ \frac{3x^2y}{[(n-1)\sqrt{1-x^2}-ny^2]}\right]  \,\,\text{,}  
\end{equation}
\begin{equation}\label{6d}
\lambda^{\prime}= \frac{\sqrt{3}(k-2)}{2k} xy\lambda^2  \,\,\text{.}  
\end{equation}
We present the critical points and their behavior (see Table~\eqref{4Table-2}) for the autonomous system in Eq.~\eqref{6b}-\eqref{6d} with model parameter $n=-1$.
\begin{table}[H]
\begin{center}\caption{Table shows the critical points and their behaviour corresponding to the model parameter $n=-1$ and potential $V(\phi)= V_0\phi^{-k}$ with $k\neq2$.}
\begin{tabular}{|c|c|c|c|c|c|}
\hline
 Critical Points $(x_c,y_c,z_c)$ & Eigenvalues ($\lambda_1$, $\lambda_2$,  $\lambda_3$) & Nature of critical point  & $q$ & $\omega$ \\
\hline 
$O'(0,0,\lambda)$ & $ (-3,\sqrt{3}\lambda,0) $ & $\lambda\leq0$ (NH(stable)), $\lambda>0$ (saddle)  & $-1$ & $-1$ \\
$A'(x,y,0)$ & $ (-3,0,0)  $ & Non-hyperbolic (stable)  & $-1$ & $-1$ \\
$B'(0,y,0)$ & $ (-3,0,0) $ & Non-hyperbolic (stable)  & $-1$ & $-1$ \\
\hline
\end{tabular}\label{4Table-2}
\end{center}
\end{table}   
Note that the dynamical equations presented for the power-law case in Eq.~\eqref{6b}-\eqref{6d} are identical to those presented for the exponential case in Eq.~\eqref{5b}-\eqref{5d}, just differ by a constant in the $\lambda'$ equation, and hence the further analysis, i.e., evolutionary trajectories, are identical.\\
We observe from Table~\ref{4Table-2} that all critical points ($O', A', B'$) possess zero eigenvalues, rendering them non-hyperbolic. Specifically, $O'(0,0,\lambda)$ represents a line of fixed points along the $\lambda$-axis with eigenvalues $(-3, \sqrt{3}\lambda, 0)$, where the zero eigenvalue corresponds to the tangential direction along the line. For $\lambda \leq 0$, the non-zero eigenvalues are non-positive, indicating stability transverse to the line. The points $A'$ and $B'$ possess eigenvalues $(-3, 0, 0)$, with a negative eigenvalue that ensures attraction along the $x$-direction, while the two zero eigenvalues correspond to the center subspace in the $y$-$\lambda$ plane. As the analysis of the reduction of the center manifold is complex for this DBI system, we rely on the numerical analysis of the phase-space (similar to the exponential case), which confirms that the trajectories converge to these points for the relevant parameter ranges, supporting their classification as stable attractors \cite{Perko:2001, Wiggins:2003}.

In particular, for $k=2$, the power-law case in Eq.~\eqref{6b}-\eqref{6d} differs from the exponential one, as it reduces to the following 2-dimensional dynamical system, since $ \lambda'=0$ for $ k=2$,
\begin{equation}\label{6e}
x^{\prime}= (x^2-1)(3x-2\sqrt{3} y)  ,
\end{equation}
\begin{equation}\label{6f}
y^{\prime}= -\frac{1}{2}y^2\left[2\sqrt{3} xy+ \frac{3x^2y}{[(n-1)\sqrt{1-x^2}-ny^2]}\right]  \,\,\text{.}  
\end{equation}
The unique role of the $k=2$ power-law potential, $V(\phi) = V_0 \phi^{-k}$, arises from a dimensional balancing that enforces scale invariance within the system. As previously shown, the steepness parameter $\lambda$ is governed by the ratio $V_{,\phi}/V^{3/2}$; for the specific case of an inverse-square potential ($k=2$), the field $\phi$ cancels out entirely, rendering $\lambda$ a constant. This ensures the potential's slope remains identical regardless of the field's magnitude, making the dynamics "blind" to the absolute scale of $\phi$. Complementing this, the DBI radical $\sqrt{1-\dot{\phi}^2}$ introduces a non-canonical kinetic structure that imposes a relativistic speed limit on the field's evolution, functioning much like the Lorentz factor in special relativity. The synergy between this square-root term and the inverse-square potential allows the warped geometry and the potential's gradient to perfectly compensate for one another, preserving the form of the equations of motion under simultaneous rescaling of time and field. Consequently, $\lambda$ decouples from the dynamical flow, reducing the original three-dimensional system $(x,y,\lambda)$ to a closed two-dimensional subsystem characterized by exact scaling solutions. This reduction confirms that the underlying conformal symmetry effectively renders the global phase-space dynamics independent of the specific field amplitude.

\begin{figure}[H]
\centering
\includegraphics[scale=0.52]{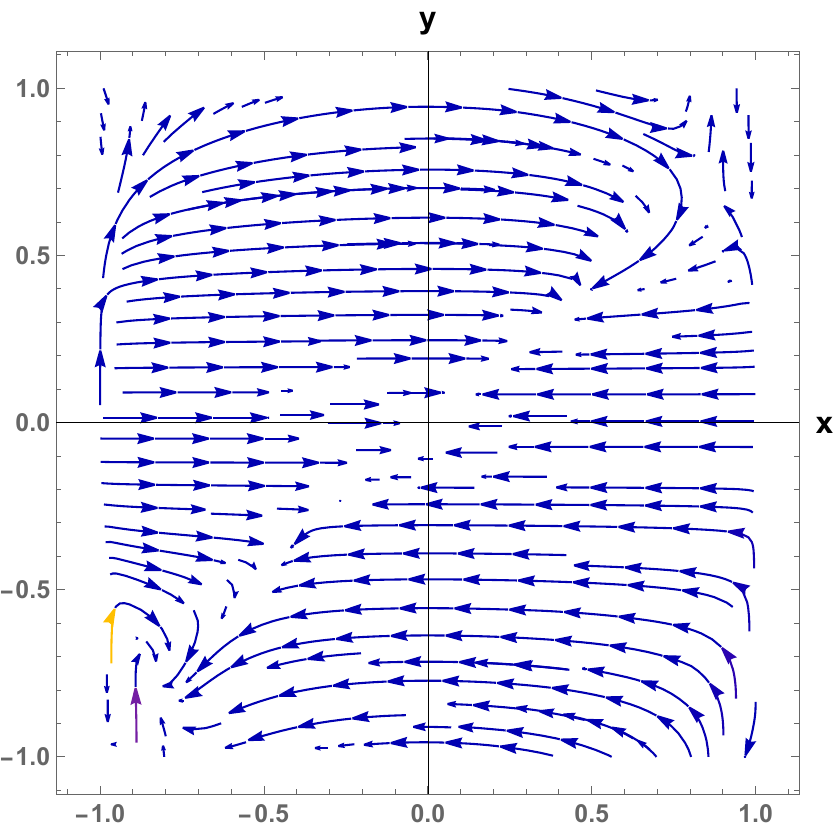}
\caption{2D phase portrait for polynomial potential when $k=2$}\label{fig_5.4}
\end{figure}

Here, without loss of generality, we assume $V_0=1$ and hence $\lambda=\frac{2}{\sqrt{V_0}}=2$. We present the critical points and their behavior (see Table~\eqref{Table-3}) for the autonomous system given by Eq.~\eqref{6e}-\eqref{6f} with model parameter $n=-1$.
\begin{table}[H]
\begin{center}\caption{Table shows the critical points and their behavior corresponding to the model parameter $n=-1$ with potential $V(\phi)= V_0\phi^{-k}$ having $k=2$ and $V_0=1$.}
\begin{tabular}{|c|c|c|c|c|}
\hline
 Critical Points $(x_c,y_c)$  & Nature of critical point  & $q$ & $\omega$ \\
\hline 
$O''(0,0)$ &  Stable (NH)  & $-1$ & $-1$ \\
$A''(0.806,0.698)$ & Saddle  & $0.362$ & $-0.092$ \\
\hline
\end{tabular}\label{Table-3}
\end{center}
\end{table}   
The evolutionary trajectories of the autonomous system Eq.~\eqref{6e}-\eqref{6f} are presented in Fig.~\ref{fig_5.4}.
From the phase-space trajectories obtained in Fig.~\ref{fig_5.4}, it is evident that the solution trajectory indicates the evolution from the saddle point $A''$ representing matter-like behavior to the stable point $O''$ representing the accelerated expansion phase of type de Sitter. Moreover, it matches the previous analysis by Copeland et al.\cite{copeland1}.\\


We note that the critical point $O''(0,0)$ is non-hyperbolic. The Jacobian matrix evaluated at $O''(0,0)$ for the system Eq.~\eqref{6e}-\eqref{6f} is given by
\begin{equation*}
J_{O''} = \begin{pmatrix} 
\frac{\partial x'}{\partial x} & \frac{\partial x'}{\partial y} \\
\frac{\partial y'}{\partial x} & \frac{\partial y'}{\partial y}
\end{pmatrix}_{(0,0)} = \begin{pmatrix} 
-3 & 2\sqrt{3} \\
0 & 0
\end{pmatrix} \,\,\text{,} 
\end{equation*}
which yields eigenvalues $\lambda_1 = -3$ and $\lambda_2 = 0$. The zero eigenvalue arises from Eq.~\eqref{6f} where $y' \propto y^2$, so $\frac{\partial y'}{\partial y}|_{(0,0)} = 0$. For such non-hyperbolic points, linear stability analysis is inconclusive \cite{Perko:2001}. However, the negative eigenvalue ($\lambda_1 = -3$) ensures attraction along the corresponding eigen direction, and since $y' < 0$ for $y > 0$ near the origin (trajectories approach $y=0$ from above), the dynamics of the center manifold also leads to convergence. This is confirmed by our numerical phase-space analysis (Fig.~\ref{fig_5.4}), where all trajectories in the physical region converge to $O''$, supporting its classification as a stable attractor \cite{Wiggins:2003}.
Here, without loss of generality, we assume $V_0=1$ and hence $\lambda=\frac{2}{\sqrt{V_0}}=2$. 
Note that $\lambda$ appears in the dynamical equations as $\sqrt{3}\lambda = 2\sqrt{3}$.

The reduction to a two-dimensional phase space at $k=2$ is rooted in the relativistic kinetic structure of the DBI action, which enforces a strict speed limit $\dot{\phi}^2<1$ through the Lorentz-like factor $\gamma=(1-\dot{\phi}^2)^{-1/2}$. For a scaling solution where $w_\phi$ and $\dot{\phi}$ remain constant, the acceleration term $\ddot{\phi}$ vanishes, reducing the field equation to a steady-state balance between Hubble friction and the potential gradient, $3H\dot{\phi}\propto V_{,\phi}$. Consistency with the Friedmann equation ($H^2\propto\rho\propto V$, since $\gamma$ is fixed) requires $H\propto\sqrt{V}$, which combined with the friction balance yields the differential condition $V^{3/2}\propto V_{,\phi}$. The unique solution is the inverse-square potential $V(\phi)\propto\phi^{-2}$. In the autonomous formulation, this scaling is monitored by the steepness parameter $\lambda\equiv-V_{,\phi}/V^{3/2}$, which for $V=V_0\phi^{-k}$ evaluates to $\lambda=k V_0^{-1/2}\phi^{k/2-1}$. At the critical exponent $k=2$, the power of $\phi$ vanishes and $\lambda$ becomes a fixed constant, signaling an emergent conformal symmetry. Under the joint rescaling $\phi\to c\phi$ and $t\to ct$, the potential's mass dimension exactly matches the scaling weight of the DBI kinetic sector, rendering the equations of motion form-invariant. The dynamics thereby become blind to the absolute field value, with Hubble friction and the potential force diluting at identical rates. Consequently, $\lambda'\equiv0$, the steepness parameter decouples from the flow, and the original three-dimensional system collapses into a closed two-dimensional phase space governed solely by the algebraic friction--slope balance.

\section{Conclusions} \label{sec_5.5}
In this chapter, we have studied the DBI scalar field (the corresponding Lagrangian density is presented in Eq.~\eqref{4a4}) and its effect on cosmology via dynamical system analysis. The corresponding Klein-Gordon equation with Friedmann-like equations under the $f(Q)$ gravity formalism has been presented in equations given in Eq.~\eqref{4b}, Eq.~\eqref{4c}, and Eq.~\eqref{4d}. For our analysis, we considered the cosmological model $f(Q)=-Q+\Psi(Q)=-Q+\alpha Q^n$, which is nothing but a power-law correction to the STEGR case, which will give rise to branches of solution applicable either to the early Universe or to late-time cosmic acceleration. The model characterized by $n < 1$ can describe late-time cosmology, potentially influencing the emergence of dark energy, whereas the model characterized by $n > 1$ can describe early Universe phenomena \cite{JIM-2}. We obtained a set of dynamical equations (i.e., Eq.~\eqref{4j}, Eq.~\eqref{4n}, and Eq.~\eqref{4p}) corresponding to the choice of our $f(Q)$ function. Further, in order to obtain the closed form (i.e. autonomous form) of the system, we investigated two specific forms of the potential function, specifically the exponential one $V(\phi)= V_0 e^{-\beta \phi}$ and the power-law $V(\phi)= V_0\phi^{-k}$, which have been extensively studied in the literature in the GR context. We obtained the corresponding autonomous systems Eq.~\eqref{5b}-\eqref{5d} and Eq.~\eqref{6b}-\eqref{6d}, and their stability analysis has been presented in Table~\eqref{Table-1} and Table~\eqref{4Table-2}. Moreover, the 3-D phase-space trajectories of the autonomous system corresponding to an exponential potential have been presented in Fig.~\ref{fig_5.1}, and the behavior of cosmological parameters such as deceleration, energy density, and effective equation of state parameter have been presented in Fig.~\ref{fig_5.2}. Note that the autonomous systems presented for the power-law case in Eq.~\eqref{6b}-\eqref{6d} are identical to those presented for the exponential case in Eq.~\eqref{5b}-\eqref{5d}, they just differ by a constant in the equation $\lambda'$, and hence the further analysis, i.e., evolutionary trajectories, are identical. We also observe that our fixed points have a late time de Sitter type solution, and that is somewhat what we expected from the general theorem given by Hao et al. \cite{hao} and Chingangbam et al.\cite{chingangbam} about the sufficient condition for a de Sitter type solution. In addition, we analyze the fixed point $(\pm1,0,\lambda)$ to show the transition from matter-dominated to de Sitter-type solutions. Also, we found that the necessary and sufficient condition for the fixed points $(\pm1,0,\lambda)$ to show a matter-dominated phase to de Sitter phase is given by $\frac{1-x}{y^4}\leq \frac{8}{(1-n)^2}$ (where in the limit $x\rightarrow 1$ and $y\rightarrow 0$) corresponding to the generic model parameter $n$. For the other fixed point $(-1,0,\lambda)$, we get similar results. In addition, we plotted the 2D phase portrait in Fig.~\ref{fig_5.3} for the $x=1,-1$ and $xy$ planes, which shows that even though the critical points are non-hyperbolic, one can still take some particular limits to see the matter-dominated to de Sitter transition. We also note that in the current time ($N=0$), we obtained $q\approx -0.8$, which is somewhat consistent with the present observed value (-0.55). The discrepancy in the value of $q$ arises because there is no dark matter or ordinary matter in our calculations. We further note that the power-law potential case for $k=2$ is different from the exponential one, since $k=2$ gives $\lambda^{\prime}=0$, making it a 2D autonomous system, which we have presented in Eq.~\eqref{6e}-\eqref{6f}. The corresponding stability analysis and phase portrait are presented in Table~\eqref{Table-3} and Fig.~\ref{fig_5.4}. Moreover, the case $k=2$ matches the previous studies done in detail by \cite{bhagla,copeland1}. Thus, we conclude that the investigation presented in this chapter successfully describes the late-time epochs of the Universe, particularly de Sitter expansion and the observed transition epoch, along with effects of the DBI field under modified $f(Q)$ cosmology.

\chapter{Scalar Field Evolution at Background and Perturbation Levels For a Broad Class of Potentials} 

 \label{Chapter6}
\lhead{Chapter 6. \emph{Scalar Field Evolution at Background and Perturbation Levels For a Broad Class of Potentials}} 

\vspace{10 cm}
* The work, in this chapter, is covered by the following publications: \\
 
\textit{Scalar Field Evolution at Background and Perturbation Levels for a Broad Class of Potentials}, Fortschritte der Physik, \textbf{71}, 2300006 (2023).

\clearpage
\pagebreak

\epigraph{``Nature uses only the longest threads to weave her patterns, so that each small piece of her fabric reveals the organization of the entire tapestry.''}{--- Richard P. Feynman, \textit{The Character of Physical Law} (1965), Ch.~1}

In this chapter, we investigate a non-interacting scalar field cosmology with an arbitrary potential using the $f$-deviser method that relies on the differentiability properties of the potential. Using this alternative mathematical approach, we present a unified dynamical system analysis at a scalar field's background and perturbation levels with arbitrary potentials. For illustration, we consider a monomial and double exponential potential. These two classes of potentials comprise the asymptotic behaviour of several classes of scalar field potentials, and, therefore, they provide the skeleton for the typical behaviour of arbitrary potentials. Moreover, we analyse the linear cosmological perturbations in the matter less case by considering three scalar perturbations: the evolution of the Bardeen potentials, the comoving curvature perturbation, the so-called Sasaki-Mukhanov variable, or the scalar field perturbation in uniform curvature gauge. Finally, an exhaustive dynamical system analysis for each scalar perturbation is presented, including the evolution of Bardeen potentials in the presence of matter.


\section{Introduction}

Scalar fields are prominent in the physical description of the Universe in the inflationary scenario \cite{ALAN} and can be used to explain the late-time acceleration of the Universe. Although $\Lambda$CDM has excellent concordance with observations, describes structure formation, and successfully provides a late-time acceleration \cite{Carroll:2000}, $\Lambda$ has yet to succeed in quantifying quantum vacuum fluctuations \cite{ZEL, WENB}. That is the primary motivation for introducing dark energy as an alternative to $\Lambda$CDM. Some examples are quintessence field \cite{Ratra:1988, quint2, Parsons:1995kt, Barrow:2016qkh}, a phantom scalar field (which, however, suffers ghost instabilities \cite{phant}),  a quintom scalar field model  \cite{quintom1, Guo/2005, Feng:2004ff, Zhang:2005eg, Zhang:2005kj, Lazkoz:2006pa, Lazkoz:2007mx, Setare:2008pz, Setare:2008pc, Leon:2012vt, Leon:2018lnd, Mishra:2018dzq, Marciu:2020vve, Dimakis:2020tzc},   a chiral cosmology \cite{Dimakis:2020tzc, Chervon:2013btx, Paliathanasis:2020wjl}, or multi-scalar field models. The latter describes various epochs in the cosmological history \cite{mult, Elizalde:2008yf, Paliathanasis:2018vru}. However, the Hubble constant value measured with local observations (see SH0ES \cite{Riess}) is in tension with that estimated from early observations (see Planck \cite{planck}). A possible alternative to solve this tension is considering extensions beyond $\Lambda$CDM \cite{DiValentino:2021izs}. There could be other reasons for the $H_0$ tension, e.g., incomprehension between the SnIa absolute magnitude and the Cepheid-based distance ladder, rather than an exotic late-time physics  \cite{Efstathiou:2021ocp}. Even more,  $H_0$ tension seems to permeate dark energy models (including quintessence), whereby $H_0$ is sent to lower values by any dark energy model with $w_{DE}(z) > -1$, whereas local (model-independent) $H_0$ determinations are biased to more significant values than Planck-$\Lambda$CDM \cite{Banerjee:2020xcn, Lee:2022cyh}. 
Although the exploration of scalar field models has the attention of several researchers, such that 
scalar field evolution at the background level has been studied in several works, say \cite{Capozziello:2005tf, Nojiri:2005pu, Briscese:2006xu, Nojiri:2006ww, Capozziello:2005mj, Astashenok:2012kb, Astashenok:2012tv, Ito:2011ae, Frampton:2011rh, Bamba:2014daa, Odintsov:2018zai, Odintsov:2018uaw}. 
To analyze the early and late-time dynamics of cosmological problems, the perturbation and averaging methods \cite{verhulst2006method, fenichel, Fusco1989SlowmotionMD, Berglund, holmes2012introduction, kevorkian2013perturbation} were used in \cite{Rendall:2006cq, Alho:2015cza, Alho:2019pku, Fajman:2020yjb, Fajman:2021cli, Leon:2021lct, Leon:2021rcx} for single field scalar field cosmologies and for scalar field cosmologies with two scalar fields which interact only gravitationally with matter in \cite{Chakraborty:2021vcr}. In reference \cite{Leon:2020pvt}, scalar field cosmology with a generalized harmonic potential was investigated in FLRW with flat and negative spatial curvature and for Bianchi I metrics. In addition, an interaction between the scalar field and matter was considered in the conservation equations. In these references, asymptotic methods and the theory of averaging in nonlinear dynamical systems are essential tools to obtain relevant information about the solution space of a scalar field with generalized harmonic potential in a vacuum and adding matter, \cite{Leon:2021lct, Leon:2021rcx, Chakraborty:2021vcr, Leon:2020pvt, Leon:2019iwj, Leon:2020pfy, Leon:2020ovw,   Leon:2021hxc}. The amplitude-angle transformation  was used in \cite{Fajman:2020yjb, Leon:2020pvt,  Leon:2020pfy, Leon:2020ovw, Llibre:2012zz}. In \cite{Leon:2020pfy, Leon:2020ovw}, scalar field cosmologies with generalized harmonic potentials and exponential couplings to matter in the sense of \cite{Leon:2008de, Giambo:2009byn, Tzanni:2014eja} were examined.   In  \cite{Fajman:2021cli},  a theorem about the large-time behavior of solutions of spatial homogeneous cosmology with oscillatory behavior was presented. Moreover, slow-fast methods were used, for example, in analyzing theories based on a Generalized Uncertainty Principle (GUP), say in \cite{Paliathanasis:2015cza, Paliathanasis:2021egx}. In \cite{Paliathanasis:2021egx}, a preliminary study of linear perturbations in the matter-dominated phase was presented in the context of GUP was presented. More precisely, the dynamical equations for linear cosmological perturbations were derived, forming a singular differential equation system. In contrast to the usual quintessence, one can explicitly write the perturbed equations' solution in fast and slow manifolds. The extra components enhance the growth of the scalar perturbations due to the higher-order derivative terms of the GUP in the fast manifold. However, scalar perturbations either decay, grow, or describe an oscillatory solution in the slow manifold. Consequently, the perturbation equations are also affected by the minimum length  \cite{Paliathanasis:2021egx}. 

Similarly, dynamical system methods are useful for investigating scalar field cosmologies for a wide class of potentials. To use this procedure and handle the differentiation involved, it is necessary to determine a specific potential form
$V(\phi)$ of the scalar field $\phi$. This procedure has the disadvantage that for each different potential, one must repeat all the calculations
from the beginning. Therefore, developing an
extended method that could handle the potential differentiation in
a unified way would be beneficial, without the need for any ``apriori" specification. That is the method of $f$-devisers improved in \cite{Escobar:2013js}  and
applied in \cite{delCampo:2013vka} for scalar-field FLRW cosmologies in the
presence of a Generalized Chaplygin  Gas. With this method, the classes of models discussed in \cite{Ratra:1988, Wetterich:1987fm, JYJ, Sahni:1999qe, Sahni:1999gb, Urena-Lopez:2000ewq, Matos:2000ng, Cardenas:2002np, Matos:2009hf, Copeland:2009be, Leyva:2009zz, delCampo:2013vka, Lidsey:2001nj, Pavluchenko:2003ge, TBR, Gonzalez:2007hw, Gonzalez:2006cj, Peebles:1988, LRA, aguirregabiria, copeland1, Saridakis:2009pj, Saridakis:2009ej, Leon:2009dt, Chang:2013cba, Skugoreva:2013ooa, Pavlov:2013nra} can be studied.

However, there is an interest in simultaneously investigating cosmological linear perturbations and background equations. One can obtain a unified dynamical system analysis at a scalar field cosmology's background and perturbation levels in cosmological studies. That can be done using the methods of \cite{Bardeen:1980kt, Mukhanov, Brandenberger:1992dw, Brandenberger:1993zc, Brandenberger:1992qj, dunsby:1997, amendola_tsujikawa_2010, Amendola:1999dr, Basilakos:2019dof, Alho:2019jho, Alho:2020cdg} (see references for the notation as well as for the theory to improve the background analysis of a cosmological model). Generally, one can investigate the dynamical system for the model consisting of a system of autonomous nonlinear first-order ordinary differential equations. The state space $S$ has a product structure $S = B \times P$,  
where $B$ is the background state space, which describes the dynamics of a FLRW background, and $P$ are the perturbation state space. This space contains Fourier decomposed
gauge-invariant variables that describe linear cosmological perturbations. In this way, the background dynamics determines the dynamics of the perturbations. Several recent studies examine the stability of cosmological perturbations on top or in an extended phase space that incorporates both perturbed scalar quantities and normalised (background) phase space variables \cite{dunsby:1997, amendola_tsujikawa_2010, Amendola:1999dr, Basilakos:2019dof, Alho:2019jho, Alho:2020cdg, Uggla:2011jn, Uggla:2011hs, Uggla:2012gg, Uggla:2013paa, Uggla:2013kya, Uggla:2014hva, Uggla:2018fiy, Uggla:2018cct, Uggla:2019rho, Landim:2019lvl, Uggla:2019zdm, Khyllep:2021wjd, Khyllep:2022spx}. 

In \cite{Basilakos:2019dof}, the authors performed this dynamical system analysis of the background and perturbation equations for a $\Lambda$CDM cosmology and quintessence scenario with exponential potential in a unified way. For the $\Lambda$CDM cosmology, the perturbations do not change the stability of the late-time attractor of the background equations, and the system still results in the dark energy-dominated de Sitter solution, with a transition by a dark matter era with growth index $\gamma\approx 6/11$. Here, $\gamma$ is defined through the relation $d \ln \delta_m/d \ln a \approx \Omega_m^{\gamma}$, where $\delta_m$ is the contrast of matter, and $\Omega_m$, the fractional energy density of matter.   In the case of quintessence, incorporating linear perturbations results in a change in the stability and properties of the background evolution. The only conditionally stable points present either an exponentially increasing matter clustering not
favoured by observations or suffering Laplacian instabilities and, thus, are not of physical interest. This result
severely disadvantages quintessence cosmology compared to the $\Lambda$CDM paradigm. In this vein, the work \cite{Alho:2019jho} introduced a dynamical system method to describe linear scalar
and tensor perturbations of the $\Lambda$CDM model. That provided pedagogical examples showing the global illustrative powers of dynamical systems in cosmological perturbations. It discussed the validity of the perturbations as approximations to the Einstein field equations. Furthermore, the linear growth rate $d \ln \delta_m/d \ln a \approx \Omega_m^{\gamma}$ was corrected to $d \ln \delta_m/d \ln a \approx \Omega_m^{\frac{6}{11}} -\frac{1}{70}(1-\Omega_m)^{\frac{5}{2}}$ and showed that it is much more accurate than the previous ones in the literature. That was the starting point of a series of technical papers. For example, in \cite{Alho:2020cdg}, a new regular dynamical system was derived on a compact three-dimensional state space that describes linear scalar perturbations of spatially flat RW geometries for relativistic models with a minimally coupled scalar field with exponential potential. That enables them to construct the global solution space, where known solutions are shown to reside on some invariant sets. They use their dynamical systems approach to obtain new results about comoving and uniform density curvature perturbations.
Finally, they show how to extend this approach to more general scalar field potentials. That leads to state spaces where the state space of the models with an exponential potential appears as invariant boundary sets, thereby illustrating their role as building blocks in a hierarchy of increasingly complex cosmological models. 
More generalisations appeared in \cite{Landim:2019lvl} and \cite{Sharma:2021ivo, Sharma:2021ayk}, which examined the imprints of interacting dark energy in linear scalar field perturbations. These results extend the analysis of \cite{Basilakos:2019dof}. Moreover, in reference, \cite{Tot:2022dpr} investigated the linear cosmological perturbations for a two-field quintom model that interacts through the kinetic terms, following the results of  \cite{Chervon:2013btx} for N-field chiral action.

In \cite{Khyllep:2021wjd}, the authors applied the formalism of \cite{Basilakos:2019dof} to investigate interacting dark energy scenarios at the background and perturbation levels in a unified way. An extra perturbation variable related to the matter over-density was introduced. The combined analysis found critical points describing the non-accelerating matter-dominated epoch with the proper growth of matter structure. These saddles provide the natural exit from this phase. Furthermore, late-time stable attractors correspond to dark energy-dominated accelerated solutions with constant matter perturbations. It is claimed that interacting cosmology describes the matter and dark energy epochs correctly, both at the background and perturbation levels, which reveals the capabilities of the interaction. 

In \cite{Khyllep:2022spx}, the authors studied cosmological models based on $f(Q)$  gravity, which is based on the non-metricity scalar $Q$ \cite{Jimenez/2018}. The systems were analyzed for background and perturbation levels using a dynamical system analysis. Two $f(Q)$  models of the literature are examined: the power law and the exponential ones. Both cases obtained a matter-dominated saddle with the correct growth rate of matter perturbations. This epoch is followed by the transition to a stable dark-energy-dominated accelerated Universe in which matter perturbations remain constant. Furthermore, analyzing the behavior of $f\sigma_8$ was deduced that the models fit the
observational data successfully, obtaining a behavior similar to that of the $\Lambda$CDM scenario. However, the exponential model does not possess  $\Lambda$CDM  as a limit. That is, through the independent approach of dynamical systems, it was verified that $f(Q)$ gravity can be considered an emerging alternative to the $\Lambda$CDM concordance model.

This chapter investigates a non-interacting scalar field cosmology with an arbitrary potential using the $f$-deviser method. We present a unified dynamical system analysis at a scalar field's background and perturbation levels with arbitrary potentials using this alternative mathematical approach. Using this procedure, we perform a dynamical system analysis of Background quantities using Hubble-normalised variables. For simplicity, we assume the matter less case for analysing linear cosmological perturbations.
Nevertheless, our analysis with perturbation will be perfectly viable during scalar field-dominated epochs of the Universe, e.g. inflation and late-time acceleration. Following the line of \cite{Alho:2020cdg}, we investigated the dynamics of linear scalar cosmological perturbations for a generic scalar field model using dynamical systems methods. We considered three types of gauge-invariant scalar perturbation quantities. For the case of a single scalar field, we investigate the Bardeen potentials \cite{Bardeen:1980kt, Mukhanov, Brandenberger:1992dw, Brandenberger:1993zc, Brandenberger:1992qj}, the comoving curvature perturbation  \cite{fr}, and the so-called Sasaki-Mukhanov variable or the scalar field perturbation in uniform curvature gauge \cite{Kodama:1984ziu, Mukhanov:1988jd}. An exhaustive dynamical system analysis for each scalar perturbation will be presented.

The chapter is organised as follows: In Sec.~\ref{sec2}, we present the field equations for a scalar field minimally coupled to gravity, with an arbitrary potential $V(\phi)$ in the presence of matter. 
We discuss there the $f$-devisers method. In Sec .~\ref {sect:3}, we perform a dynamical system analysis of background quantities using Hubble-normalized variables and the method of $f$-devisers. For illustration, we consider the monomial potential in Subsec.~\ref{sect:3-1} as a first example. This potential $V(\phi)=\left|\frac{\mu}{n}\right|\phi^{n}$ has been investigated in \cite{Ratra:1988, Peebles:1988, LRA, aguirregabiria,  Saridakis:2009pj, Saridakis:2009ej, Leon:2009dt, Chang:2013cba, Skugoreva:2013ooa, Pavlov:2013nra}. As a second example, we study the usual exponential potential in Sec.~\ref{sect:3-2}. For $f=0$ and $\lambda$ constant, we recover the quintessence scenario with an exponential potential $V=V_0 e^{-\lambda\phi}$ as studied in \cite{COPE}. As a final example, in Sec .~\ref{sect:3-3}, we investigate the double exponential, say 
$V(\phi)=V_1 e^{\alpha \phi}+ V_2 e^{\beta \phi}$  \cite{TBR, Gonzalez:2007hw, Gonzalez:2006cj}. This example contains the particular case of hyperbolic cosine $V(\phi)=\frac{1}{2} \left(e^{\alpha \phi}+ e^{-\alpha \phi}\right)$ setting $V_1=
 V_2=1/2$ and $\beta=-\alpha$. When one of the exponents is zero, this corresponds to the exponential potential plus a cosmological constant \cite{JYJ, Pavluchenko:2003ge, Cardenas:2002np}. The potentials that are sums of two exponents are interesting in the context of $f(R)$ gravity because the conformal transformation of a metric gives
$f(R)$ in analytic form \cite{Vernov:2019ubo}.  These two classes of potentials, monomial (power-law) and exponential (double or single exponential plus a cosmological constant), comprise the asymptotic behavior of several classes of scalar field potentials. Therefore, they provide the skeleton for the typical behavior of arbitrary potentials. For simplicity, we assume the matterless case for analyzing linear cosmological perturbations in Sec.~\ref {sec3}. An exhaustive dynamical system analysis for three types of gauge-invariant scalar perturbation quantities, the Bardeen potentials, the comoving curvature perturbation, and the Sasaki-Mukhanov variable, is presented in Sec.~\ref{sect:4}. In Sec.~\ref {sect:new}, we investigate cosmological perturbations in the presence of two matter components, e.g., a perfect fluid plus a cosmological constant or a perfect fluid plus a scalar field with exponential potential. A widespread practice in the literature concentrates on a particular cosmological epoch when only one matter component is dominant. In that sense, even though not generic, our subsequent analysis is still relevant when the Universe is a scalar field-dominated, e.g., during the early inflationary epoch or the late-time acceleration. Conclusions are presented in Sec.~\ref{sect:6}.

\section{Equations}\label{sec2}
The action we are working with is
\begin{equation}
    \mathcal{S} = \int d^{4}x\sqrt{-g}\left[\frac{R}{2} -\frac{1}{2}\phi_{\mu}\phi^{\mu} -V(\phi)\right] + \mathcal{S}_m \,\,\text{,} 
\end{equation}
where we denote $\phi$ as the scalar field, $R$ is the Ricci scalar, $\mathcal{S}_m$ denotes the matter action, and as usual we are using the natural units. Now, for a scalar field $\phi$ with self-interacting potential $V(\phi)$, we see that their energy density and pressure are given by
\begin{align}
   \rho_{\phi}=\frac{\dot \phi^2}{2}+V(\phi) \,\,\text{,} \label{(2)}
\end{align} 
\begin{align}
    p_{\phi}=\frac{\dot \phi^2}{2}-V(\phi) \,\,\text{,}\label{(3)}
\end{align}
respectively. Also, for a pressure-less matter, we can write the Friedman equation as follows:
\begin{align}
    3H^2= \rho_m+\rho_{\phi} \,\,\text{,}  \label{(4)}
\end{align}
\begin{align}
   \dot H=-\frac{1}{2}(\rho_m+\dot \phi^2)\,\,\text{,} \label{(5)}
\end{align}
where $H$ is the Hubble parameter defined as $H=\frac{\dot{a}}{a}$, $a$ is the scale factor, and $\rho_m$ is the matter-energy density, whose corresponding conservation equation is given by
\begin{align}
    \dot \rho_m+3H\rho_m=0 \,\,\text{.}  \label{(6)}
\end{align}

On the other hand, given the scalar field Lagrangian, we can get the Klein-Gordon equation as follows
\begin{align}
    \ddot \phi= -3H\dot\phi - \frac{d V}{d\phi} \,\,\text{.}   \label{(7)}
\end{align}

To extend the standard dynamical analysis method to generic classes of potentials, one uses the  method of $f$-devisers in which two new dynamical variables are introduced, namely $\lambda$ and $f$, as 
\begin{eqnarray}
\label{sdef}
\lambda&\equiv&-\frac{V^{\prime}(\phi )}{  V(\phi)},\\
f&\equiv& \frac{V^{\prime \prime}(\phi )}{V(\phi )}-\frac{V^{\prime}(\phi )^2}{V(\phi )^2} \,\,\text{,}
\label{fdef}
\end{eqnarray}
such that
\begin{eqnarray}   
V^{\prime}(\phi ) &=& -  \lambda  V(\phi), \label{eq11}\\
    V^{\prime \prime}(\phi )&=&   \left(f+\lambda ^2\right) V(\phi) \,\,\text{.}  \label{eq10}
\end{eqnarray}
The only requirement is that  $f$ can be expressed as an explicit function of $\lambda$, that is, $f=f(\lambda)$. Following the above procedure, one can transform a cosmological system into a closed dynamical system for a set of auxiliary normalized variables and the new one $\lambda$. Then, using this procedure, one can investigate a wide range of potentials. In particular, the usual ansatzes of the cosmological literature can be covered by simple forms for $f$, as seen in Table~\ref{fsform}. Note that the variable $\lambda$ is not required for the single exponential potential since it is a constant, i.e., $f$ is automatically zero. 

\begin{table}[!ht]
\centering
\caption{The function $f(\lambda)$ for the most common quintessence potentials
\protect.}
\begin{tabular*}{\columnwidth}{@{\extracolsep{\fill}}ll@{}}
\hline
\multicolumn{1}{@{}l}{Potential  $V(\phi)$}  & $f(\lambda)$\\
\hline
$\left|\frac{\mu}{n}\right|\phi^{n}$ \cite{LRA, aguirregabiria, Saridakis:2009pj, Saridakis:2009ej, Leon:2009dt, Chang:2013cba, Skugoreva:2013ooa, Pavlov:2013nra} & $-\frac{\lambda ^2}{n}$ \\
$V_{0}e^{-\alpha \phi}+V_1$  \cite{JYJ, Pavluchenko:2003ge, Cardenas:2002np} & $-\lambda(\lambda-\alpha)$ \\ 
$V_1 e^{\alpha \phi}+ V_2 e^{\beta \phi}$  \cite{TBR, Gonzalez:2007hw, Gonzalez:2006cj} & $-(\lambda+\alpha)(\lambda+\beta)$\\ 
$V_{0}/\sinh^{\alpha}(\beta \phi)$  \cite{Wetterich:1987fm, Copeland:2009be, Leyva:2009zz, Pavluchenko:2003ge, Sahni:1999gb, Urena-Lopez:2000ewq}
& $\frac{\lambda^2}{\alpha}-\alpha\beta^2$\\
$V_{0}\left[\cosh\left( \xi \phi\right)-1\right]$  \cite{Wetterich:1987fm, Matos:2009hf, Copeland:2009be, Leyva:2009zz, Pavluchenko:2003ge, delCampo:2013vka, Sahni:1999qe, Sahni:1999gb, Lidsey:2001nj, Matos:2000ng}
& $-\frac{1}{2}(\lambda^2-\xi^2)$ \\
\hline
\end{tabular*}
\label{fsform}
\end{table}
On the other hand, when the function $f(\lambda)$ is given, we can straightforwardly reconstruct
the corresponding potential form starting with
\begin{eqnarray}
&& \frac{d\lambda}{d\phi}=-  f, \quad  \frac{dV}{d\phi}=-\lambda  V,
\label{dV-dphi}
\end{eqnarray}
which leads to
\begin{eqnarray}
\phi(\lambda)&=&-\int \frac{1}{f} \, d\lambda,\label{quadphi}\\
V(\lambda)&=&V_0 e^{\int \frac{\lambda}{f} \, d\lambda}\label{quadV}.
\end{eqnarray} 
Note that the relations
(\ref{quadphi}) and (\ref{quadV}) are always valid, giving the
potential in an implicit form. However, for the usual cosmological cases of
Table~\ref{fsform}  we can additionally eliminate  $\lambda$  between
Eq.~\eqref{quadphi} and Eq.~\eqref{quadV}, and write the potential explicitly as
$V=V(\phi)$.
Finally, note that the $f$-devisers method  
also allows   reconstructing a scalar field potential from a model with
stable equilibrium points. In particular, choosing a function $f$ with
the requested properties (existence of minimum, intervals of monotony,
differentiability) to have stable late-time attractors,
one uses Eq.~\eqref{quadphi} and Eq.~\eqref{quadV} to explicitly obtain  
$V(\phi)$. That is similar to the superpotential construction
method  \cite{Arefeva:2009tkq}, which allows the construction of stable
kink-type solutions in scalar-field cosmological models, starting from the
dynamics, and specifically for the Lyapunov stability. A field model with a stable kink solution was considered
earlier in \cite{Arefeva:2004odl}. 

Nevertheless, this method is not universal. That means that it cannot be applied to any arbitrary potential. The procedure can be fully implemented only when  $f$ is an explicit function of
$\lambda$. For example, in some specific forms in the inflationary context, such as 
$V(\phi)\propto \phi ^p \ln ^q(\phi )$ \cite{Barrow:1995xb} and  $V(\phi)\propto \phi^n e^{-q\phi^m}$
\cite{Parsons:1995ew}, the expression $f$ cannot be expressed as a 
single-valued function of $\lambda$. In general, for a wide
range of potentials, the introduction of the
variables $f$ and $\lambda$ add an extra direction in the phase space, whose
neighboring points correspond to ``neighboring'' potentials.

\section{Dynamical system in terms of background quantities}
\label{sect:3}

It is well-known that for the investigation of cosmological models, one can introduce auxiliary
variables that transform the cosmological equations into an autonomous
dynamical system
\cite{Burd:1988ss, tavakol_1997, Coley:2003mj, Leon2012CosmologicalDS, Gong:2006sp, Setare:2008sf, Chen:2008ft, Gupta:2009kk, Farajollahi:2011ym, Urena-Lopez:2011gxx, Escobar:2011cz, Escobar:2012cq, Xu:2012jf, Leon:2013qh}. Hence, we obtain a system of the form  $\textbf{X}'=\textbf{f(X)}$, where $\textbf{X}$ is the
column vector of the auxiliary variables and $\textbf{f(X)}$ is a vector field for autonomous equations. The prime denotes the differentiation with respect to a logarithmic time scale. The stability analysis comprises several steps. First, the critical points $\bf{X_c}$ are extracted under the requirement  of $\bf{X}'=0$. Then, one considers linear perturbations around $\bf{X_c}$ as $\bf{X}=\bf{X_c}+\bf{U}$, with $\textbf{U}$ the column vector of the  auxiliary variable's perturbations. Therefore, up
to first order we obtain $\textbf{U}'={\bf{\Xi}}\cdot \textbf{U}$, where  the matrix ${\bf {\Xi}}$ contains coefficients of the perturbed equations. Finally, the type and stability of each hyperbolic critical point are determined by the eigenvalues of ${\bf {\Xi}}$. The point is stable (unstable) if the real parts of the eigenvalues are negative (positive) or a saddle
if the eigenvalues have real parts with different signs.

To proceed, we can take the equation in Eq.~\eqref{(4)} and divide them by $3H^2$, and also putting the value of $\rho_{\phi}$ in Eq.~\eqref{(2)}, we get 
\begin{align}
    1=\frac{\rho_m}{3H^2}+\frac{ \dot \phi^2}{6H^2}+\frac{  V}{3H^2} \,\,\text{.}  \label{(8)}
\end{align}
Now we denote the following
\begin{align}
    x^2=\frac{ \dot \phi^2}{6H^2},\quad y^2=\frac{  V}{3H^2},\quad \Omega_m=\frac{ \rho_m}{3H^2} \,\,\text{,} \label{(9)}
\end{align}
So, Eq.~\eqref{(8)} becomes
\begin{align}
    1=\Omega_m+x^2+y^2 \quad \textrm{or}\quad 1-x^2-y^2=\Omega_m \,\,\text{.}  \label{(10)}
\end{align}
As we see from the  Eq.~\eqref{(10)}, $x^2+y^2\leq 1$ and $x^2+y^2\geq 0$, i.e., the system is bounded for a non-negative fluid density $ \rho_m\geq 0$. Then, the evolution of this system is completely described by trajectories within the unit disc, where the lower half-disc, $y<0$, corresponds to contracting Universes. As the system is symmetric under the reflection $(x,y)\mapsto (x, -y)$ and time reversal $t\mapsto -t $, we only consider the
upper half-disc, $y\geq 0 $ in the following discussion.

Now we write a dynamical equation for each of the variables. 
Using the dynamical variable $N=\ln(a)$ with $dN=Hdt$, we write our dynamical system for $(x, y,\lambda)$ as a system of first-order equations. 
\begin{align}
x^{\prime}& =-\frac{3}{2}x\left(y^2-x^2+1\right)+\sqrt{\frac{3}{2}}\lambda y^2 \,\,\text{,} \label{(18)}
\\
y^{\prime}& =-\sqrt{\frac{3}{2}}\lambda x y-\frac{3}{2}y\left(y^2-x^2-1\right) \,\,\text{,} \label{(19)}
\\
\lambda^{\prime}& =-\sqrt{6} x f \,\,\text{,} \label{(20)}
\end{align}
where to close the system, we assume that  
$f$ can be written as an implicit function of $\lambda$. That is, $f(\lambda)$ can be explicitly obtained by inverting Eq.~\eqref{sdef} and 
Eq.~\eqref{fdef}. This procedure only gives a closed dynamical system when we can explicitly obtain $f=f(\lambda)$. In Table~\ref{fsform}, we present the cases in which this approach can be completely implemented.

An important cosmological parameter is the deceleration parameter, which can be written in terms of the dynamical variables as follows.
\begin{equation}\label{dec_param}
    q \equiv -1-\frac{\dot{H}}{H^2} = \frac{1}{2}\left(1 + 3x^2 - 3y^2\right) \,\,\text{.}
\end{equation}
From the above equation, we can see that, at the equilibrium points, the deceleration parameter is constant. Then, we can obtain an expression for the scale factor $a(t)$ that is asymptotically valid according to whether the constant   $q=-1$ or $q\neq -1$. Indeed, for the $q$ constant, integrating the expression
\begin{equation}
    \frac{a\ddot{a}}{\dot{a}^2}=-q \,\,\text{,}
\end{equation}
with the initial values $a(t_U)=1$, $\dot{a}(t_U)=H_0$, where $t_U$ is the age of the Universe and $H(t_U)=H_0$ is the current value of the Hubble parameter, we can obtain $a(t)$. Then, by definition, we obtain $H(t)= \dot{a}(t)/a(t)$. Summarizing,
\begin{align}
  a(t)& = \left\{ \begin{array}{cc}
      \left(1+H_0 \left(q+1\right) \left(t-t_U\right)\right)^{\frac{1}{q+1}}, &  q\neq -1 \\
      e^{H_0 \left(t-t_U\right)}, & q=-1
  \end{array} \right.,\\
  H(t)& = \left\{ \begin{array}{cc}\frac{H_0}{H_0 \left(q+1\right) \left(t-t_U\right)+1}, & q\neq -1 \\
      H_0, & q=-1
  \end{array} \right..
\end{align}
Finally, because $x$ is a constant at the equilibrium points, we have 
\begin{equation}
\phi(t)= \phi_0 + \sqrt{6} x_c \int_{t_U}^t H(s) ds =  \phi_0 + \left\{\begin{array}{cc}                \ln \left(\left(H_0 \left(q+1\right)
   \left(t-t_U\right)+1)\right)^{\frac{\sqrt{6}x_c}{(1+q)}}\right), & x_c \neq 0\\         
    0, &  x_c= 0      \end{array}\right.
\end{equation}

The asymptotic behaviors of the scale factor, the Hubble scalar, and the scalar field as a function of $t$ depend on the value of $x_c$ at the equilibrium points. 
\begin{itemize}
\item Case $x_c= \lambda^*/\sqrt{6}$ for any $\lambda^*$ that satisfies $f(\lambda^*)=0, -\sqrt{6}<\lambda^*<\sqrt{6}$:      
\begin{align}
    & a(t)=\left\{\begin{array}{cc}                \left(\frac{H_0}{2}{\lambda^*}^2 \left(t-t_U\right)+1\right)^{\frac{2}{{\lambda_*}^2}}, & \lambda^* \neq 0\\                e^{H_0\left(t-t_U\right)}, &  \lambda^*= 0      \end{array}\right., \label{A1}
    \\
    & H(t)=\left\{\begin{array}{cc}  \frac{H_0}{\frac{H_0}{2}{\lambda^*}^2 \left(t-t_U\right)+1},  & \lambda^* \neq 0\\               H_0, &  \lambda^*= 0      \end{array}\right., \label{A1b}\\
    & \phi(t)=   \phi_0 + \left\{\begin{array}{cc}                \ln \left(\left(\frac{1}{2} H_0 {\lambda^*}^2 \left(t-t_U\right)+1\right)^{\frac{2}{|\lambda^*|}}\right), & \lambda^* \neq 0\\        0, &  \lambda^*= 0      \end{array}\right.. \label{A1c}
\end{align}
\item Case $x_c= \pm 1$: 
\begin{align}
    & a(t)=\left(3 H_0  \left(t-t_U\right)+1\right)^{\frac{1}{3}}, \label{A2}\\
    & H(t)=\frac{H_0}{3 H_0 \left(t-t_U\right)+1}, \label{A2b}\\
    & \phi(t)=   \phi_0 \pm \frac{\sqrt{6}}{3} \ln \left(3 H_0 \left(t-t_U\right)+1\right). \label{A2c}
\end{align} 
\item Case $-1< x_c<1$:  \begin{align}
    & a(t)=\left\{\begin{array}{cc}     \left(3 H_0 {x_c}^2 \left(t-t_U\right)+1\right)^{\frac{1}{3{x_c}^2}},  & x_c \neq 0\\                e^{H_0\left(t-t_U\right)}, &  x_c= 0      \end{array}\right., \label{CASE-A}\\ 
    & H(t)=\left\{\begin{array}{cc}  \frac{H_0}{3 H_0 {x_c}^2 \left(t-t_U\right)+1},  & x_c \neq 0\\               H_0, &  x_c= 0      \end{array}\right.,\label{CASE-Ab}\\
    & \phi(t)= \phi_0 + \sqrt{6} x_c \int_{t_U}^t H(s) ds =  \phi_0 + \left\{\begin{array}{cc}                \ln \left(\left(3 H_0 {x_c}^2 \left(t-t_U\right)+1\right)^{\frac{\sqrt{6}}{3|x_c|}}\right), & x_c \neq 0\\         
    0, &  x_c= 0      \end{array}\right.. \label{CASE-Ac}
\end{align}
\item Case $x_c= \pm \sqrt{3}/3$: 
\begin{align}
    & a(t)=H_0 \left(t-t_U\right)+1, \label{A3}\\
    & H(t)=\frac{H_0}{  H_0   \left(t-t_U\right)+1},    \label{A3b} \\
    & \phi(t)=  \phi_0 \pm  \sqrt{2} \ln \left( H_0   \left(t-t_U\right)+1\right). \label{A3c}
\end{align}
\end{itemize}

\subsection{Physical interpretation and stability of the equilibrium points}

The equilibrium points of the system Eq.~\eqref{(18)}, Eq.~\eqref{(19)}, and Eq.~\eqref{(20)},  in the finite region for an arbitrary function $f(\lambda)$ are presented in Table~\ref{Background_a}. For arbitrary potentials, we have the following. 

\begin{enumerate}
    \item The set of equilibrium points $O$ corresponds to matter-dominated solutions which, as expected, are saddles, i.e., intermediate cosmological epochs. The deceleration parameter is $q= \frac{1}{2}$. Then we have the asymptotic solutions 
   $a(t)=\left(\frac{3}{2} H_0 \left(t-t_U\right)+1\right)^{2/3}$, 
 $ H(t)  =\frac{H_0}{\frac{3}{2} H_0 \left(t-t_U\right)+1}$,   $\rho_m(t)= \rho_{m0} \left(\frac{3}{2} H_0    \left(t-t_U\right)+1\right)^{-2}$, and  $\phi(t)= 0$.

\item $K_\pm(\lambda^*)$  exist for $f({\lambda^*})=0$, and they represent kinetic-dominated solutions. They are associated with the early stages of the Universe and correspond to stiff solutions. 
    
\item $K_-(\lambda^*)$  is a source for  $\lambda^*>-\sqrt{6}, \; f'({\lambda^*})>0$. It is a saddle for $\lambda^*<-\sqrt{6}$ or $ f'({\lambda^*})<0$. Non-hyperbolic for  $\lambda^*=-\sqrt{6}$ or $ f'({\lambda^*})=0$. 
 
\item $K_+(\lambda^*)$  is a source for  $\lambda^*<\sqrt{6}, \; f'({\lambda^*})<0$. It is a saddle for $\lambda^*>\sqrt{6}$ or $ f'({\lambda^*})>0$. Non-hyperbolic for  $\lambda^*= \sqrt{6}$ or $ f'({\lambda^*})=0$.
    
   For these solutions, the value of the deceleration parameter is $q=2$. Then we have the same asymptotic behaviour   $a(t)=(3 H_0  \left(t-t_U\right)+1)^{\frac{1}{3}}$,  $H(t)=\frac{H_0}{3 H_0 \left(t-t_U\right)+1}$, $\phi(t)=   \phi_0 \pm \frac{\sqrt{6}}{3} \ln \left(3 H_0 \left(t-t_U\right)+1\right)$,  and $\rho_m(t)=0$. 
 
\item $MS_{-}(\lambda^*)$ exists for  $f({\lambda^*})=0, \lambda^*<-\sqrt{3}$. It represents a matter-scalar field scaling solution where neither the scalar field nor the matter field dominates.   \newline It  is a sink for ${\lambda^*}\leq -2 \sqrt{\frac{6}{7}}, f'({\lambda^*})<0$ (stable spiral) or $ -2 \sqrt{\frac{6}{7}}<{\lambda^*}<-\sqrt{3}, f'({\lambda^*})<0$ (stable node). It is non-hyperbolic for ${\lambda^*}=-2 \sqrt{\frac{6}{7}}$ or $f'({\lambda^*})=0$. It is a saddle otherwise. 
   
\item  $MS_{+}(\lambda^*)$ exists for  $f({\lambda^*})=0, \lambda^*>\sqrt{3}$. It represents a matter-scalar field scaling solution where neither the scalar field nor the matter field dominates. \newline It  is a sink for ${\lambda^*}\geq 2 \sqrt{\frac{6}{7}}, f'({\lambda^*})>0$ (stable spiral) or $\sqrt{3}<{\lambda^*}<2 \sqrt{\frac{6}{7}}, f'({\lambda^*})>0$ (stable node). It is non-hyperbolic for ${\lambda^*}=2 \sqrt{\frac{6}{7}}$ or $f'({\lambda^*})=0$. It is a saddle otherwise.
    
  For these solutions, the deceleration parameter is $q= \frac{1}{2}$. Then, we have asymptotic solutions
 $\rho_m(t)= 0$, $a(t)=\left(\frac{3}{2} H_0 \left(t-t_U\right)+1\right)^{2/3}$, 
 $ H(t)  =\frac{H_0}{\frac{3}{2} H_0 \left(t-t_U\right)+1}$. Since $x_c=\frac{\sqrt{\frac{3}{2}}}{{\lambda^*}}$, we have $\phi(t)=   \phi_0 + \ln \left(\left(\frac{3}{2} H_0
   \left(t-t_U\right)+1\right)^{2/\lambda^* }\right)$. For $\lambda=\lambda^*$ and $f(\lambda^*)=0$, the potential asymptotically behaves as $V(\phi)= 3 y_c^2 H(t)^2= \frac{3 H_0^2}{2 {\lambda^{*2}} \left(\frac{3}{2} H_0
   \left(t-t_U\right)+1\right)^2}\sim  e^{-\lambda^*(\phi-\phi_0)}$. 

\item $Sf(\lambda^*)$ exists $-\sqrt{6}<\lambda^*<\sqrt{6}$. It represents an scalar-field dominated solution. It is a sink for $-\sqrt{3}<{\lambda^*}<0,  f'({\lambda^*})<0$ or $0<{\lambda^*}<\sqrt{3}, f'({\lambda^*})>0$. It is non-hyperbolic for  ${\lambda^*}\in \left\{-\sqrt{3}, 0, \sqrt{3}\right\}$ or   $f'({\lambda^*})=0$. It is a saddle otherwise. 
    For this solution, the deceleration parameter is 
   $q= \frac{1}{2} \left({\lambda^{*2}}-2\right)$. Then $a(t)=\left(\frac{1}{2} H_0\lambda^{*2}
   \left(t-t_U\right)+1\right)^{\frac{2}{\lambda^{*2}}}$, $H(t)= \frac{2 H_0}{H_0 \lambda^{*2} \left(t-t_U\right)+2}$,  $\phi(t)=  \phi_0 + \ln \left(\left(\frac{1}{2} H_0 {\lambda^{*2}}
   \left(t-t_U\right)+1\right)^{2/{\lambda^*} }\right)$, and $\rho_m(t)=0$.  For $\lambda=\lambda^*$ and $f(\lambda^*)=0$, the potential asymptotically behaves as $V(\phi)= 3 y_c^2 H(t)^2= {2 H_0^2 \left(6- \lambda^{*2}\right)}/{\left(H_0 \lambda^{*2} \left(t-t_U\right)+2\right)^2} \sim  e^{-\lambda^*(\phi-\phi_0)}$. 

\item $dS$ is a potential dominated solution that represents the de Sitter solutions. It is stable for $f(0)> 0$ or a saddle for $f(0)<0$.  For this solution, the deceleration parameter is 
   $q=-1$. Then $a(t)= e^{H_0 \left(t-t_U\right)}$, $H(t)= H_0$,  $\phi(t)=\phi_0$, $V(\phi)= 3 H_0^2$, and $\rho_m(t)=0$. 
   
\end{enumerate}

\begin{table}[]
    \centering
    \caption{Equilibrium points of the system Eq.~\eqref{(18)}, Eq.~\eqref{(19)}, and Eq.~\eqref{(20)}, and their eigenvalues $k_1$, $k_2$ and $k_3$, in the finite region for an arbitrary function $f(\lambda)$. $\lambda^*$ represent zeros of the function $f(\lambda)$. Note that $O$ is actually a line of fixed points whereas all the others are isolated fixed points.}
            \resizebox{\textwidth}{!}{%
    \begin{tabular}{|c|c|c|c|c|c|c|c|}
    \hline 
Label &    $x$ & $y$  & $\lambda$ & Existence & $k_1$ & $k_2$ & $k_3$ \\\hline
$O$ & $ 0$ & $0$ & $\lambda_c$ & $\lambda_c\in \mathbb{R}$  & $-\frac{3}{2}$ & $\frac{3}{2}$ & $0$ \\\hline
$K_-(\lambda^*)$ & $-1$ & $0$ & ${\lambda^*}$ & $f({\lambda^*})=0$ & $3$ & $\sqrt{\frac{3}{2}} {\lambda^*}+3$ & $\sqrt{6} f'({\lambda^*})$ \\\hline
$K_+(\lambda^*)$ & $ 1$ &$ 0$ & ${\lambda^*}$ & $f({\lambda^*})=0$ &$ 3$ & $3-\sqrt{\frac{3}{2}} {\lambda^*}$ &$ -\sqrt{6} f'({\lambda^*})$ \\\hline
$MS_{-}(\lambda^*)$ & $ \frac{\sqrt{\frac{3}{2}}}{{\lambda^*}}$ &$ -\frac{\sqrt{\frac{3}{2}}}{{\lambda^*}}$ &$ {\lambda^*}$ & $f({\lambda^*})=0, \lambda^*<-\sqrt{3}$ &$ -\frac{3 \left(\sqrt{24-7 {\lambda^*}^2}+{\lambda^*}\right)}{4 {\lambda^*}}$ &$ \frac{3}{4} \left(\frac{\sqrt{24-7 {\lambda^*}^2}}{{\lambda^*}}-1\right) $&$ -\frac{3 f'({\lambda^*})}{{\lambda^*}}$ \\\hline
$MS_{+}(\lambda^*)$  & $ \frac{\sqrt{\frac{3}{2}}}{{\lambda^*}}$ & $\frac{\sqrt{\frac{3}{2}}}{{\lambda^*}}$ & ${\lambda^*}$ & $f({\lambda^*})=0, \lambda^*>\sqrt{3}$ &$ -\frac{3 \left(\sqrt{24-7 {\lambda^*}^2}+{\lambda^*}\right)}{4 {\lambda^*}} $& $\frac{3}{4} \left(\frac{\sqrt{24-7 {\lambda^*}^2}}{{\lambda^*}}-1\right)$ & $-\frac{3 f'({\lambda^*})}{{\lambda^*}}$ \\\hline
$Sf(\lambda^*)$ &  $\frac{{\lambda^*}}{\sqrt{6}}$ &$ \sqrt{1-\frac{{\lambda^*}^2}{6}}$ & ${\lambda^*}$& $f({\lambda^*})=0,-\sqrt{6}<\lambda^*<\sqrt{6}$ & $\frac{1}{2} \left({\lambda^*}^2-6\right)$ & ${\lambda^*}^2-3 $&
  $ -{\lambda^*} f'({\lambda^*}) $\\\hline
$dS$ &  $0$ &$ 1$ & $0$ & always & $-3$ &$ \frac{1}{2} \left(-3-\sqrt{9-12 f(0)}\right)$ & $\frac{1}{2} \left(-3+ \sqrt{9-12 f(0)}\right)$\\\hline
    \end{tabular}}
    \label{Background_a}
\end{table}

To illustrate the potentiality of the $f$-devisers method, we consider some examples. 

\subsection{First example: monomial potential}
\label{sect:3-1}

Consider the potential $V(\phi)=\left|\frac{\mu}{n}\right|\phi^{n}$ \cite{Ratra:1988, Peebles:1988, LRA, aguirregabiria, copeland1, Saridakis:2009pj, Saridakis:2009ej, Leon:2009dt, Chang:2013cba, Skugoreva:2013ooa, Pavlov:2013nra}, which produces the function $f(\lambda)=-\frac{\lambda ^2}{n}$. For this potential, the evolution equations are given by Eq.~\eqref{(18)}, Eq.~\eqref{(19)}, together with 
\begin{align}
\lambda^{\prime}& = \frac{\sqrt{6}}{n} x \lambda ^2 \,\,\text{.} \label{(20b)}
\end{align}
For the function $f(\lambda)$ we have $f^{\prime}(\lambda)= -\frac{2\lambda}{n}$ and $f(\lambda)=0 \Longleftrightarrow\lambda=0$. Then $\lambda^*=0$ and $f'(\lambda^*)=0$. The equilibrium points of this example are the following, summarised in Table~\ref{Background_a-powerlaw}.

\begin{table}[]
    \centering
    \caption{Equilibrium points of the system Eq.~\eqref{(18)}, Eq.~\eqref{(19)}, and Eq.~\eqref{(20)},  in the finite region for $f(\lambda)=-\frac{\lambda ^2}{n}$.}
    \begin{tabular}{|c|c|c|c|c|c|c|c|c|}
    \hline 
Label &    $x$ & $y$  & $\lambda$ & Existence & $k_1$ & $k_2$ & $k_3$ & Stability \\\hline
$O$ & $ 0$ & $0$ & $\lambda_c$ & $\lambda_c\in \mathbb{R}$  & $-\frac{3}{2}$ & $\frac{3}{2}$ & $0$ & saddle \\\hline
$K_-(0)$ & $-1$ & $0$ & ${0}$ & always & $3$ & $3$ & $0$ & unstable \\\hline
$K_+(0)$ & $ 1$ &$ 0$ & ${0}$ & always  &$ 3$ & $3$ & $0$ & unstable \\\hline
$dS$ &  $0$ &$ 1$ & $0$ & always & $-3$ &$ -3$ & $0$ & saddle ($n>0$); sink ($n<0$)  \\\hline
    \end{tabular}
    \label{Background_a-powerlaw}
\end{table}

Analyzing the case of the de Sitter solution $dS$ in detail, note that the eigenvalues are $-3, -3, 0$, i.e., non-hyperbolic. Using the Center Manifold theorem, we find that the graph locally gives the center manifold of the origin
\begin{align}
& \Big\{(x, y, \lambda)\in [-1,1] \times [0,1] \times \mathbb{R}: x=\frac{\lambda}{\sqrt{6}}+h_1(\lambda), y=1+ h_2(\lambda),  \nonumber \\
& h_1(0)=0, h_2(0)=0,  h_1'(0)=0, h_2'(0)=0, |\lambda|<\delta\Big\},
\end{align}
for a small enough $\delta$, where the functions $h_1$ and $h_2$ satisfy the differential equations

\begin{align}
   & -24 \lambda ^2 \left(\left(\sqrt{6} h_1(\lambda )+\lambda \right) h_1'(\lambda )+h_1(\lambda
   )\right) +6 n \Big(\sqrt{6} \lambda  \left(3 h_1(\lambda )^2+h_2(\lambda ) (h_2(\lambda
   )+2)\right) \nonumber \\
   & +6 h_1(\lambda ) \left(h_1(\lambda )^2-h_2(\lambda ) (h_2(\lambda
   )+2)-2\right)+3 \lambda ^2 h_1(\lambda )\Big)+\sqrt{6} \lambda ^3 (n-4)=0,
\\
   & -\frac{\lambda ^2 \left(\sqrt{6}
   h_1(\lambda )+\lambda \right) h_2'(\lambda )}{n}-\frac{1}{4} (h_2(\lambda )+1) \left(-6
   h_1(\lambda )^2+6 h_2(\lambda ) (h_2(\lambda )+2)+\lambda ^2\right)=0 
\end{align}
Then, using the Taylor series, we have the solutions
\begin{align}
 x(\lambda) & =  \frac{\lambda }{\sqrt{6}}-\frac{\lambda ^3}{3 \left(\sqrt{6} n\right)}-\frac{(n-8) \lambda ^5}{18
   \left(\sqrt{6} n^2\right)}-\frac{((n-18) (n-6)) \lambda ^7}{108 \left(\sqrt{6} n^3\right)}\nonumber \\
   &+\frac{(1984-n ((n-48)
   n+592)) \lambda ^9}{648 \sqrt{6} n^4} +\frac{(-n (n ((n-80) n+1880)-16320)-45280) \lambda ^{11}}{3888 \sqrt{6}
   n^5}\nonumber \\
   & +\frac{(1223424-n (n (n ((n-120) n+4560)-72896)+504624)) \lambda ^{13}}{23328 \sqrt{6} n^6}+O\left(\lambda
   ^{14}\right), \label{expansion-x-t}
\\
y(\lambda) & =1-\frac{\lambda ^2}{12}-\frac{(n-16) \lambda ^4}{288 n}-\frac{((n-48) n+288) \lambda ^6}{3456
   n^2} \nonumber \\
   &+\frac{(n (-5 (n-96) n-8064)+31744) \lambda ^8}{165888 n^3} +\frac{(n (376832-7 n ((n-160) n+5184))-1159168)
   \lambda ^{10}}{1990656 n^4} \nonumber \\
   &+\left(\frac{n (5147648-21 n ((n-240) n+12672))-40069120}{47775744 n^4}+\frac{59}{27
   n^5}\right) \lambda ^{12}+O\left(\lambda ^{14}\right) \,\,\text{.}
\end{align}

The 1D dynamical system dictates that the dynamics at the center manifold is given by,
\begin{align}
    \frac{d \lambda}{d N} & =- U'(\lambda) \,\,\text{,}
\end{align}
which corresponds to a gradient-like equation with potential 
\begin{align}
 U(\lambda)& =-\frac{\lambda ^4}{4 n}+\frac{\lambda ^6}{18 n^2}+\frac{\lambda ^8 (n-8)}{144 n^3}+\frac{\lambda ^{10} (n-18)
   (n-6)}{1080 n^4}   -\frac{\lambda ^{12} (1984-n ((n-48) n+592))}{7776 n^5} \nonumber \\
   & +\frac{\lambda ^{14} (n (n ((n-80)
   n+1880)-16320)+45280)}{54432 n^6}\nonumber \\
   & -\frac{\lambda ^{16} (1223424-n (n (n ((n-120) n+4560)-72896)+504624))}{373248
   n^7}+O\left(\lambda ^{17}\right) \,\,\text{.} \label{Gradient-Potential}
\end{align}
Therefore, since $U^{(4)}(0)=-6/n\neq 0$, the origin is a degenerate maximum of the potential for $n>0$, and the center manifold of the origin and the origin are unstable (saddle). At the same time, it is stable if $n<0$. In this example, the points $MS_{-}(\lambda^*)$ and 
$MS_{+}(\lambda^*)$  do not exist and $Sf(\lambda^*)$ reduces to $dS$. 
However, there are equilibrium points at the invariant sets 
$\lambda= \pm\infty$, where the dynamics are given, under a time re-scaling which does not affect the orbits of the phase space, by  
\begin{align}
\frac{d x}{d \tau}& = \pm \sqrt{\frac{3}{2}} y^2 \,\,\text{,}  \quad
\frac{d y}{d \tau} =\mp \sqrt{\frac{3}{2}}  x y \,\,\text{.} \label{PowerlawInf-b}
\end{align}
The orbits as $\lambda= \pm\infty$ are semicircles $x^2+y^2= x_0^2+y_0^2$. 

One can define, 
\begin{equation}
u=\frac{2 \tan ^{-1}(\lambda )}{\pi } \,\,\text{,} -1<u<1 \,\,\text{,} \label{eqU}
\end{equation}
obtaining a compactification of the phase space and the vector field, which defines a global phase space that  comprises the dynamics at finite $\lambda$, and, i.e., 
\begin{align}
\frac{d x}{d N}& = \left\{\begin{array}{cc}
      \sqrt{\frac{3}{2}} y^2 \,\,\text{,} & u=1\\
    \sqrt{\frac{3}{2}} y^2 \tan \left(\frac{\pi  u}{2}\right)+\frac{3}{2} x \left(x^2-y^2-1\right)\,\,\text{,} &  -1<u<1 \\
     -\sqrt{\frac{3}{2}} y^2 \,\,\text{,} & u=-1
     \end{array} \right.,  \label{Powerlaw-2a}
\\
\frac{d y}{d N}& =  \left\{\begin{array}{cc}
     - \sqrt{\frac{3}{2}}  x y \,\,\text{,} & u=1\\
    -\frac{1}{2} y \left(\sqrt{6} x \tan \left(\frac{\pi  u}{2}\right)-3 x^2+3 y^2-3\right) \,\,\text{,}&  -1<u<1 \\
      \sqrt{\frac{3}{2}}  x y \,\,\text{,} & u=-1
\end{array} \right., \label{Powerlaw-2b}
\\
\frac{d u}{d N}& = -\frac{\sqrt{6} x (\cos (\pi  u)-1)}{\pi  n} \,\,\text{,} \label{Powerlaw-2c}
\end{align}

\begin{figure}[]
    \centering
    \includegraphics[scale=0.65]{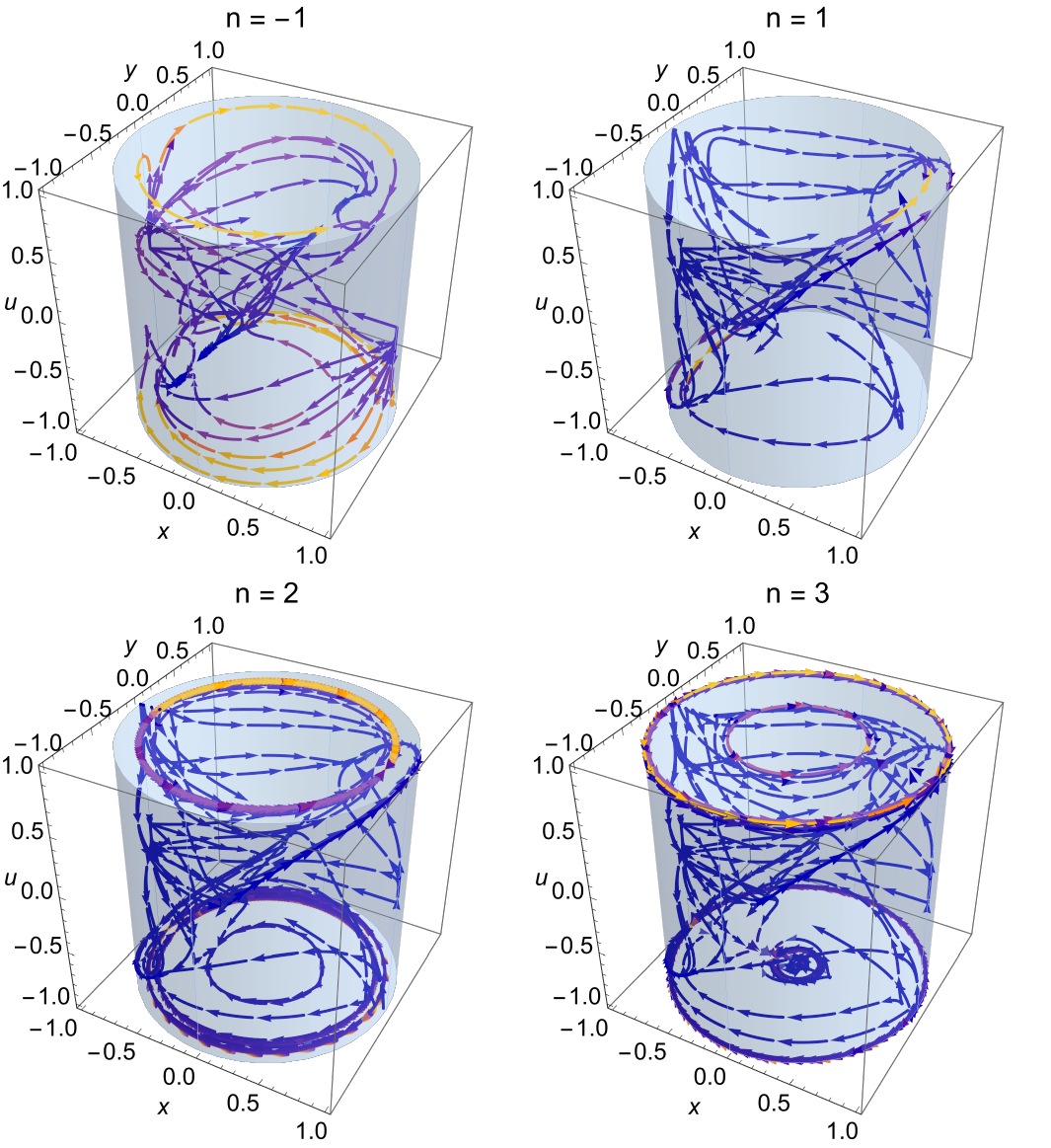}
    \caption{Compact 3D phase space of the system Eq.~\eqref{Powerlaw-2a}, Eq.~\eqref{Powerlaw-2b}, and Eq.~\eqref{Powerlaw-2c}, for $n=-1$, $1$, $2$, and $3$.}
    \label{fig:PL2}
\end{figure}

In Fig.~\ref{fig:PL2} is represented the flow of the system Eq.~\eqref{Powerlaw-2a}, Eq.~\eqref{Powerlaw-2b}, and Eq.~\eqref{Powerlaw-2c}, for  $n=-1$, $n=1$, $2$, and $3$. The dynamics as   $\lambda \rightarrow \pm \infty$ is represented in the top and bottom disks ($u=\pm 1, x^2+y^2 \leq 1$), where the orbits are concentric semicircles.   The planes $u=\pm1$ correspond to the limiting cases when the scalar field potential has an infinite negative/positive slope (see the definition of $\lambda$ in Eq.~\eqref{sdef}). One half of this plane acts as an attracting invariant sub-manifold, while the other half acts as a repelling invariant sub-manifold, with $x=0$ separating the two regions (see Eq.~\eqref{Powerlaw-2c}).

\subsection{Second example: exponential potential}
\label{sect:3-2}
For $f=0$ and $\lambda$ constant, we recover the quintessence scenario with an exponential potential $V=V_0 e^{-\lambda\phi}$ as studied in \cite{COPE},  where for generality,  we have considered a perfect fluid with linear equation of state $p_m=  w_m \rho_m$, and barotropic index $\gamma_m\equiv w_m+1$. 

The equations of the dynamical system are  
\begin{align}
   \frac{d x}{d N} & =   -3x + \lambda \sqrt{\frac{3}{2}} y^2
 + \frac{3}{2} x \left[ 2x^2 + \gamma_m \left( 1 - x^2 - y^2 \right)
\right] \,\,\text{,} \label{eomx}\\
\frac{d y}{d N} & =  - \lambda \sqrt{\frac{3}{2}} xy
 + \frac{3}{2} y \left[ 2x^2 + \gamma_m \left( 1 - x^2 - y^2 \right)
\right] \,\,\text{,} \label{eomy}
\end{align}
defined in the phase space 
\begin{equation}
\left\{(x,y)\in \mathbb{R}^2: x^2+y^2\leq 1, y\geq 0\right\} \,\,\text{.}
\end{equation}

\begin{table}[]
\begin{center}
\caption{The critical points, their stability conditions (the
corresponding eigenvalues are given in \protect\cite{COPE}), and
the values of $\Omega_\phi$ and $w_\phi$, for the quintessence scenario, with $%
\protect\gamma_m\equiv w_m+1$ and $w_d=\gamma_\phi -1$. }
\label{Quintback}
\resizebox{\textwidth}{!}{
\begin{tabular}{|c|c|c|c|c|c|c|}
\hline
C.P. & $x$ & $y$ & Existence & Stability & $\Omega_d$ & $w_d$ \\ \hline
A & $0$ & $0$ & Always & Saddle for $0 < \gamma_m < 2$ & $0$ & Undefined \\ \hline
B & $1$ & $0$ & Always & Unstable node for $\lambda < \sqrt{6}$ & $1$ & $1$ \\ 
&  &  &  & Saddle for $\lambda > \sqrt{6}$ &  &  \\ \hline
C & $-1$ & $0$ & Always & Unstable node for $\lambda > -\sqrt{6}$ & $1$ & $1 $\\ 
&  &  &  & Saddle for $\lambda < -\sqrt{6}$ &  &  \\ \hline
D & $\lambda/\sqrt{6}$ & $[1-\lambda^2/6]^{1/2}$ & $\lambda^2 < 6$ & Stable
node for $\lambda^2 < 3\gamma_m$ & 1 & $\frac{\lambda^2}{3}-1$ \\ 
&  &  &  & Saddle for $3\gamma_m < \lambda^2 < 6$ &  &  \\ \hline
E & $(3/2)^{1/2} \, \gamma_m/\lambda$ & $[3(2-\gamma_m)\gamma_m/2%
\lambda^2]^{1/2}$ & $\lambda^2 > 3\gamma_m$ & Stable node for $3\gamma_m <
\lambda^2 < 24 \gamma_m^2/(9\gamma_m -2)$ & $3\gamma_m/\lambda^2$ & $w_m$ \\ 
&  &  &  & Stable spiral for $\lambda^2 > 24 \gamma_m^2/(9\gamma_m -2)$ &  & 
\\ \hline
\end{tabular}}%
\end{center}
\end{table}
In Table~\ref{Quintback} the critical points, their stability conditions (the
corresponding eigenvalues are given in \protect\cite{COPE}), and
the values of $\Omega_\phi$ and $w_\phi$, for the quintessence scenario, with a perfect fluid with linear equation of state $p_m=  (\gamma_m-1) \rho_m$ and $\gamma_\phi \equiv {\rho_\phi+p_\phi \over \rho_\phi}  = {\dot\phi^2 \over V + \dot\phi^2/2}
 = {2x^2 \over x^2 + y^2}$.

\begin{figure}[]
    \centering
    \includegraphics[scale=0.8]{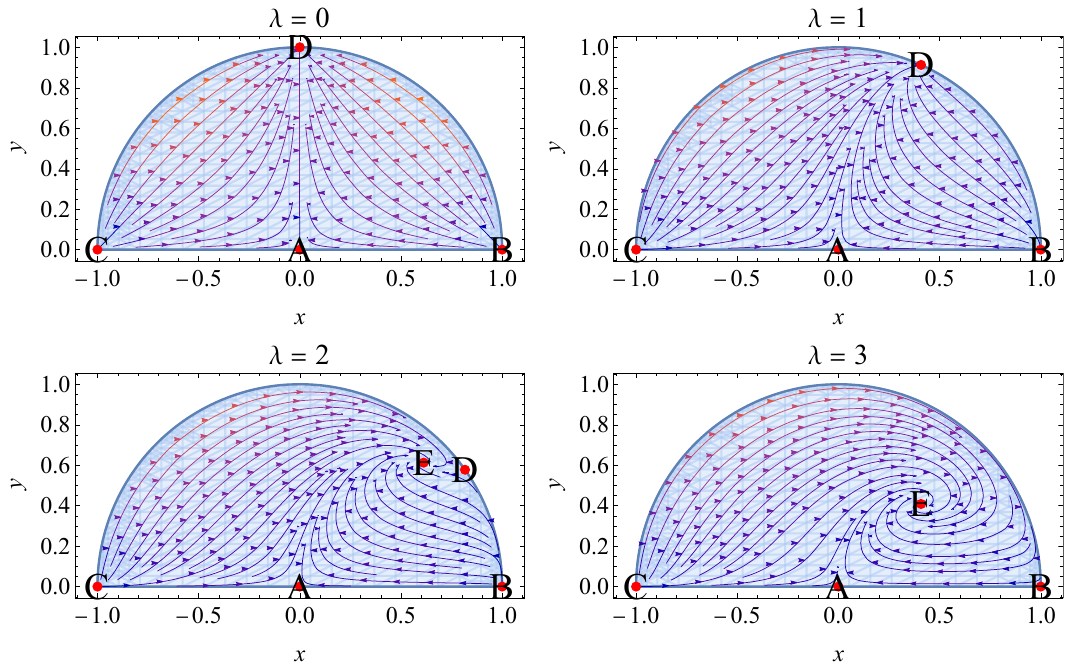}
    \caption{Compact 2D phase space of the system Eq.~\eqref{eomx} and Eq.~\eqref{eomy}, for different values of $\lambda$ and $w_m=0$. }
    \label{fig:exp}
\end{figure}
In Fig.~\ref{fig:exp} presents a compact 2D phase space of the system Eq.~\eqref{eomx} and Eq.~\eqref{eomy}, for different values of $\lambda$ and $w_m=0$.

The equilibrium points are at finite values of $x$ and $y$ in the phase-plane
correspond to solutions where the scalar field has a barotropic
equation of state and the scale factor of the Universe evolve as
$a\propto t^p$ where $p={2/3\gamma_\phi}$.

Two of the fixed points ($B$ and $C$) correspond to solutions where the Friedman constraint
Eq.~\eqref{(10)} is dominated by the kinetic energy of the
scalar field with a stiff equation of state, $\gamma_\phi=2$. As
expected, these solutions are unstable and are only expected to be
relevant at early times.

Moreover, we find that the barotropic
fluid dominated solution (A) where $\Omega_\phi=0$ is a saddle for all values of $\gamma_m>0$ (saddle). For any $\gamma_m>0$, and however steep
the potential (i.e.~whatever the value of $\lambda$), the energy density of
the scalar field never vanishes with respect to the other matter
in the Universe. Generally, the system accepts only two possible late-time attractor solutions. One
of these is the well-known scalar field-dominated solution
($\Omega_\phi=1$) that exists for sufficiently flat potentials,
$\lambda^2<6$. The scalar field has an effective barotropic index
$\gamma_\phi=\lambda^2/3$, giving rise to a power-law inflationary
expansion~\cite{Lucchin:1984yf,Kitada:1992uh} ($\ddot{a}>0$) for
$\lambda^2<2$. Previous phase-plane
analyses to \cite{COPE}, as \cite{Halliwell:1986ja, Burd:1988ss,Coley:1997nk} have shown that a wide class of
homogeneous vacuum models approach the spatially-flat FRW model for
$\lambda^2<2$. This scalar field-dominated solution
is a late-time attractor in the presence of a barotropic fluid when  $\lambda^2<3\gamma_m$.

However, for $\lambda^2>3\gamma_m$, we find a different late-time attractor
where neither the scalar field nor the barotropic fluid entirely
dominates the evolution. Instead, there is a scaling solution in which the
energy density of the scalar field remains proportional to that of the
barotropic fluid with $\Omega_\phi=3\gamma_m/\lambda^2$. This solution was
first found by Wetterich~\cite{Wetterich:1987fm} and was shown to be the global
attractor solution for $\lambda^2>3\gamma_m$ in ~\cite{Wands:1993zm}.

\begin{enumerate}
\item $\lambda^2<3\gamma_m$. 
Both kinetic-dominated solutions are unstable nodes.
The fluid-dominated solution is a saddle point.
The scalar field-dominated solution is the late-time attractor and is 
inflationary in the parameter region~$\lambda^2<\min\{2, 3\gamma_m\}$ and non-inflationary in the region~$2<\lambda^2<3\gamma_m$.
\item $3\gamma_m<\lambda^2<6$. 
Both kinetic-dominated solutions are unstable nodes.
The fluid-dominated solution is a saddle point.
The scalar field-dominated solution is a saddle point.
The scaling solution is a stable node/spiral.
\item $6<\lambda^2$. 
The kinetic-dominated solution with $\lambda x<0$ is an unstable
node.
A saddle point is a kinetic-dominated solution with $\lambda x>0$.
The fluid-dominated solution is a saddle point.
The scaling solution is a stable spiral.
\end{enumerate}

The bifurcation value $\gamma=0$ was studied in \cite{COPE}, where the largest eigenvalue for linear perturbations
vanishes. Thus,  higher-order perturbations about the
critical point are involved in determining its stability. The result is that $x=y=0$ is a stable attractor, but that trajectory only approaches this as the
logarithm of the scale factor, $N$. The late-time evolution is 
given by $y^2 = {\sqrt{6}\over \lambda} x \approx {1 \over \lambda^2 N} $.

\subsection{Third example: double exponential potential}
\label{sect:3-3}

Consider the potential
$V(\phi)=V_1 e^{\alpha \phi}+ V_2 e^{\beta \phi}$ \cite{TBR, Gonzalez:2007hw, Gonzalez:2006cj} that provides the function  $f(\lambda)=-(\lambda+\alpha)(\lambda+\beta)$. This example contains the particular case of hyperbolic cosine $V(\phi)=\frac{1}{2} \left(e^{\alpha \phi}+ e^{-\alpha \phi}\right)$ setting $V_1=
 V_2=1/2$ and $\beta=-\alpha$. 
 
 For this potential, we have $f^{\prime}(\lambda)=-\alpha -\beta -2 \lambda$ and $f(\lambda)=0 \Longleftrightarrow\lambda\in\left\{-\alpha, -\beta\right\}$, with $ f'(-\alpha)=\alpha-\beta$ and $ f'(-\beta)=-(\alpha-\beta)$. Moreover, we have $f(0)=-\alpha \beta$ and $f'(0)=-\alpha- \beta$. Without losing generality, we can assume $\alpha<\beta$. The equilibrium points of this example are the following, summarised in  Table~\ref{Background_a-double-exp}.

 \begin{table}[]
    \centering
     \caption{Equilibrium points of the system Eq.~\eqref{(18)}, Eq.~\eqref{(19)}, and Eq.~\eqref{(20)},  in the finite region for $f(\lambda)=-(\lambda+\alpha)(\lambda+\beta), \; \alpha \neq \beta$. Without loss of generality, we can assume $\alpha<\beta$. }
        \resizebox{\textwidth}{!}{%
    \begin{tabular}{|c|c|c|c|c|c|c|c|c|c|c|}
    \hline 
Label &    $x$ & $y$  & $\lambda$ & Existence & $k_1$ & $k_2$ & $k_3$ & Stability \\\hline
$O$ & $ 0$ & $0$ & $\lambda_c$ & $\lambda_c\in \mathbb{R}$  & $-\frac{3}{2}$ & $\frac{3}{2}$ & $0$ & saddle \\\hline
$K_-(-\alpha)$ & $-1$ & $0$ & $-\alpha$ & always & $3$ & $3 -\sqrt{\frac{3}{2}}\alpha$ & $\sqrt{6} (\alpha -\beta )$ & saddle \\\hline
$K_-(-\beta)$ & $-1$ & $0$ & $-\beta$ & always & $3$ & $3 -\sqrt{\frac{3}{2}}\beta$ & $-\sqrt{6} (\alpha -\beta )$ & source for \\
 &&&&&&&& $\alpha<\beta<\sqrt{6}$\\
 &&&&&&&& saddle for \\
 &&&&&&&& $\alpha<\beta, \beta> \sqrt{6}$ \\\hline
$K_+(-\alpha)$ & $ 1$ &$ 0$ & $-\alpha$ & always &$ 3$ & $3 + \sqrt{\frac{3}{2}}\alpha$ &$ -\sqrt{6} (\alpha -\beta )$  & source for \\
 &&&&&&&& $-\sqrt{6}<\alpha<\beta$\\
 &&&&&&&& saddle for  \\
 &&&&&&&& $\alpha<\beta, \alpha < -\sqrt{6}$ \\\hline
$K_+(-\beta)$ & $ 1$ &$ 0$ & $-\beta$ & always &$ 3$ & $3 + \sqrt{\frac{3}{2}} \beta$ &$ \sqrt{6} (\alpha -\beta )$ & saddle \\\hline
$MS_{-}(-\alpha)$ & $ -\frac{\sqrt{\frac{3}{2}}}{{\alpha}}$ &$ \frac{\sqrt{\frac{3}{2}}}{{\alpha}}$ &$-\alpha$ & $ \alpha>\sqrt{3}$ &$ -\frac{3 \left(\alpha-\sqrt{24-7 {\alpha}^2}\right)}{4 {\alpha}}$ &$ -\frac{3 \left(\alpha+\sqrt{24-7 {\alpha}^2}\right)}{4 {\alpha}}$&$ 3 \left(1-\frac{\beta }{\alpha }\right)$ & sink for \\
 &&&&&&&& $ \beta >\alpha  >\sqrt{3}$\\\hline
$MS_{-}(-\beta)$ & $ -\frac{\sqrt{\frac{3}{2}}}{{\beta}}$ &$ \frac{\sqrt{\frac{3}{2}}}{\beta}$ &$-\beta$ & $\beta>\sqrt{3}$ &$ -\frac{3 \left(\beta-\sqrt{24-7 {\beta}^2}\right)}{4 {\beta}}$ &$ -\frac{3 \left(\beta+\sqrt{24-7 {\beta}^2}\right)}{4 {\beta}}$&$ 3 \left(1-\frac{\alpha }{\beta }\right)$ & saddle \\\hline
$MS_{+}(-\alpha)$  & $- \frac{\sqrt{\frac{3}{2}}}{{\alpha}}$ & $-\frac{\sqrt{\frac{3}{2}}}{{\alpha}}$ & $-\alpha$ & $\alpha<-\sqrt{3}$ &$ -\frac{3 \left(\alpha-\sqrt{24-7 {\alpha}^2}\right)}{4 {\alpha}}$ &$ -\frac{3 \left(\alpha+\sqrt{24-7 {\alpha}^2}\right)}{4 {\alpha}}$ & $3 \left(1-\frac{\beta }{\alpha }\right)$ & saddle \\\hline
$MS_{+}(-\beta)$  & $-\frac{\sqrt{\frac{3}{2}}}{{\beta}}$ & $-\frac{\sqrt{\frac{3}{2}}}{{\beta}}$ & $-\beta$ & $\beta <-\sqrt{3}$ &$ -\frac{3 \left(\beta-\sqrt{24-7 {\beta}^2}\right)}{4 {\beta}}$ &$ -\frac{3 \left(\beta+\sqrt{24-7 {\beta}^2}\right)}{4 {\beta}}$& $3 \left(1-\frac{\alpha }{\beta }\right)$ & sink for \\
 &&&&&&&& $\alpha<\beta < -\sqrt{3}$ \\\hline
$Sf(-\alpha)$ &  $-\frac{{\alpha}}{\sqrt{6}}$ &$ \sqrt{1-\frac{{\alpha}^2}{6}}$ & $-\alpha$& $-\sqrt{6}<\alpha<\sqrt{6}$ & $\frac{1}{2} \left({\alpha}^2-6\right)$ & ${\alpha}^2-3 $&
  $\alpha  (\alpha -\beta )$ & sink for \\
 &&&&&&&&  $0<\alpha <\sqrt{3}, \beta >\alpha$\\
 &&&&&&&&  saddle otherwise\\\hline
  $Sf(-\beta)$ &  $-\frac{{\beta}}{\sqrt{6}}$ &$ \sqrt{1-\frac{{\beta}^2}{6}}$ & $-\beta$& $-\sqrt{6}<\beta<\sqrt{6}$ & $\frac{1}{2} \left({\beta}^2-6\right)$ & ${\beta}^2-3 $&
  $-\beta (\alpha -\beta ) $ & sink for \\
 &&&&&&&& $-\sqrt{3}<\beta <0, \alpha <\beta$\\
 &&&&&&&&  saddle otherwise\\\hline
$dS$ &  $0$ &$ 1$ & $0$ & always & $-3$ &$ \frac{1}{2} \left(-3-\sqrt{9+12 \alpha \beta}\right)$ & $\frac{1}{2} \left(-3+ \sqrt{9+12 \alpha \beta}\right)$ & stable for $\alpha \beta< 0$\\\hline
    \end{tabular}}
    \label{Background_a-double-exp}
\end{table}

As in Sec.~\ref{sect:3-1}, using the same compact variable as defined in Eq.~\eqref{eqU}, we obtain a compactification of the phase space and the vector field, which defines a global phase space that  comprises the dynamics at finite $\lambda$, and the dynamics at infinity under a time re-scaling, which does not affect the orbits of the phase space, i.e., 
 \begin{align}
\frac{d x}{d N}& = \left\{\begin{array}{cc}
    \sqrt{\frac{3}{2}} y^2 \,\,\text{,} & u=1\\
    \sqrt{\frac{3}{2}} y^2 \tan \left(\frac{\pi  u}{2}\right)+\frac{3}{2} x \left(x^2-y^2-1\right) \,\,\text{,} &  -1<u<1 \\
     -\sqrt{\frac{3}{2}} y^2, & u=-1
\end{array} \right. \,\,\text{,}  \label{double-exp-2a}
\\
\frac{d y}{d N}& =  \left\{\begin{array}{cc}
      - \sqrt{\frac{3}{2}}  x y \,\,\text{,} & u=1\\
    -\frac{1}{2} y \left(\sqrt{6} x \tan \left(\frac{\pi  u}{2}\right)-3 x^2+3 y^2-3\right) \,\,\text{,}&  -1<u<1 \\
      \sqrt{\frac{3}{2}}  x y \,\,\text{,} & u=-1
\end{array} \right., \label{double-exp-2b}
\\
\frac{d u}{d N}& =  \frac{2 \sqrt{6} x \cos ^2\left(\frac{\pi  u}{2}\right) \left(\alpha +\tan \left(\frac{\pi  u}{2}\right)\right) \left(\beta +\tan \left(\frac{\pi  u}{2}\right)\right)}{\pi} \,\,\text{.} \label{double-exp-2c}
\end{align}
\begin{figure}[]
    \centering
    \includegraphics[scale=0.65]{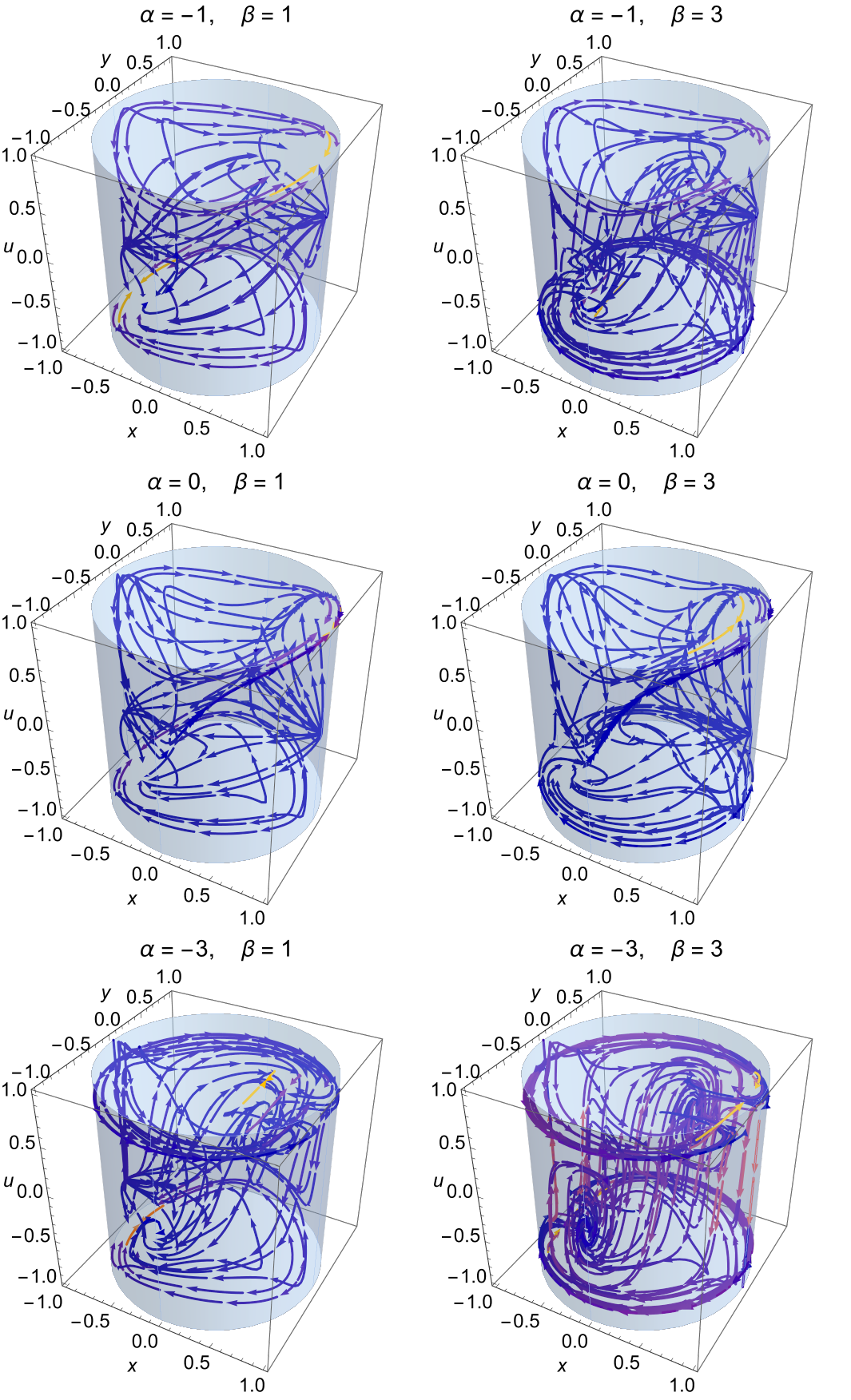}
    \caption{Compact 3D phase space of the system Eq.~\eqref{double-exp-2a}, Eq.~\eqref{double-exp-2b}, and Eq.~\eqref{double-exp-2c}, for different values of $\alpha$ and $\beta$.   }
    \label{fig:double-exp}
\end{figure}

In Fig.~\ref{fig:double-exp} is represented the flow of the system Eq.~\eqref{double-exp-2a}, Eq.~\eqref{double-exp-2b}, and Eq.~\eqref{double-exp-2c}, for different values of $\alpha$ and $\beta$.  The planes $u=\pm1$ correspond to the limiting cases when the scalar field potential has an infinite negative/positive slope (see the definition of $\lambda$ in Eq.~\eqref{sdef}).

When $\beta=-\alpha$, the potential reduces to a hyperbolic cosine, and when one parameter is zero, it reduces to an exponential potential plus a cosmological constant. Therefore, we cover three of the most common quintessence potentials shown in Table~\ref{fsform}. 
\section{Evolution of cosmological perturbations}\label{sec3}

In this section, following the line of \cite{Alho:2020cdg}, we investigate the dynamics of linear scalar cosmological perturbations for a generic scalar field model using the methods of dynamical systems. We used the perturbation of a scalar field $\phi_0$ in the background. The most generic scalar perturbed FLRW metric can be written as \cite{Bardeen:1980kt},
\begin{equation}
    ds^2 = - \left(1+\alpha\right)dt^2 - 2a(t)\left(\beta,_{i}-S_{i}\right)dtdx^i + a^{2}(t)\left[\left(1+2\psi\right)\delta_{ij} + 2\partial_{i}\partial_{j}\gamma + 2\partial_{(i}F_{j)} + h_{ij}\right]dx^{i}dx^{j} \,\,\text{,}
\end{equation}
where the inhomogeneous perturbation quantities $\alpha,\,\beta,\,\psi,\,\gamma,\,F_i,\,h_{ij}$ are functions of both $t$ and $\bar{x}$. The quantity $\psi(t,\bar{x})$ is directly related to the 3-curvature of the spatial hyper-surface,
\begin{equation}
    ^{(3)}R = - \frac{4}{a^2}\nabla^{2}\psi \,\,\text{.}
\end{equation}
Also, $h_{ij}$ satisfies $\partial^i h_{ij}=0$ (transverse) and $h^i_i=0$ (traceless), leaving exactly 2 polarization states that are automatically gauge-invariant at linear order.\\
Note that the total 4 scalars, 4 vectors (two divergence-free vectors), and 2 tensors degrees of freedom ensure no physical modes are lost or over counted ($\delta g_{ij}$ had 4 dof.).\\
For a scalar field, one also needs to take into account the perturbation of the scalar field $\delta\phi(t,\bar{x})$ and, for a perfect fluid, the perturbed energy-momentum tensor is,
\begin{equation}
    T^0_0 = -\left(\rho(t) + \delta\rho(t,\bar{x})\right) \,\,\text{,} \quad
    T^0_i = -\left(\rho(t) + P(t)\right)\partial_{i}v(t,\bar{x})\,\,\text{,} \quad
    T^i_j = \left(P(t) + \delta P(t,\bar{x})\right)\delta^i_j\,\,\text{,}
\end{equation}
being $v(t,\bar{x})$ the velocity potential. In what follows, we will restrict ourselves to the case where there is no matter but only a scalar field is present. The reason is simplicity.\\
The scalar perturbations introduced in the metric decomposition are not directly observable, as their values shift under infinitesimal coordinate transformations $x^\mu \to x^\mu + \xi^\mu$. To resolve this gauge ambiguity, one typically imposes conditions that align the coordinate system with a specific physical picture. In the Newtonian (or longitudinal) gauge, setting $\beta=\gamma=0$ removes scalar shear and time-space mixing, leaving $\alpha$ to play the role of a Newtonian gravitational potential; this makes the gauge particularly suited for sub-horizon dynamics, structure formation, and gravitational lensing. Alternatively, the comoving gauge fixes the fluid peculiar velocity to zero ($v=0$) alongside $\gamma=0$, effectively placing the observer in the rest frame of the cosmic fluid. Here, spatial curvature carries the full perturbation signal, and the resulting comoving curvature perturbation $\mathcal{R}$ remains conserved on super-horizon scales, providing a direct link between inflationary initial conditions and late-time CMB anisotropies. For early-universe calculations, the spatially flat gauge ($\psi=\gamma=0$) forces the spatial metric to remain unperturbed, which isolates the scalar field dynamics and greatly simplifies the quantization of primordial fluctuations. Because each choice is optimized for a different physical regime, cosmological predictions are ultimately expressed in terms of gauge-invariant combinations such as the Bardeen potential $\Phi$, the comoving curvature $\mathcal{R}$, or the Sasaki--Mukhanov variable $Q$. These quantities reduce to simple metric or field perturbations in their respective gauges but remain invariant under coordinate transformations, ensuring that theoretical results correspond to measurable physical observables rather than coordinate artifacts.\\
Suppose that one wants to investigate cosmological perturbations in the presence of two matter components, e.g., a perfect fluid and a scalar field. In that case, one needs to consider entropy perturbations as well. A widespread practice in the literature concentrates on a particular cosmological epoch when only one matter component is dominant. In that sense, even though not generic, our subsequent analysis is still relevant when the Universe is a scalar field-dominated, e.g., during the early inflationary epoch or the late-time acceleration. 
Of course, there is the gauge issue; the perturbation quantities defined above are not gauge-invariant \cite{Mukhanov, Brandenberger:1992dw, Brandenberger:1993zc, Brandenberger:1992qj, fr}. In this sense, various gauge-invariant perturbation quantities have been introduced in the literature. In this chapter, we will consider the following three gauge-invariant perturbation quantities.
\begin{itemize}

\item \textbf{Bardeen potential $\Phi$}:  
James Bardeen introduced Bardeen potentials \cite{Bardeen:1980kt}, who gave the first-ever gauge-invariant formulation for cosmological perturbations. These quantities are gauge-invariant perturbation quantities constructed solely out of metric perturbations. There are two such quantities,
\begin{equation}
    \Phi \equiv \alpha - \frac{d}{dt}\left[a\left(\beta+a\gamma\right)\right] \,\,\text{,} \qquad \Psi \equiv - \psi + aH\left(\beta+a\dot{\gamma}\right) \,\,\text{.}
\end{equation}
It can be shown that for the case of a single scalar field, both Bardeen potentials are equal and follow the equation \cite{Bardeen:1980kt, Mukhanov, Brandenberger:1992dw, Brandenberger:1993zc, Brandenberger:1992qj},
\begin{align}\label{1h}
    \Phi^{\prime\prime} + 2\left(\mathcal{H}-\frac{\phi_0^{\prime\prime}}{\phi_0^{\prime}}\right)\Phi^{\prime} + 2\left(\mathcal{H}^{\prime}-\mathcal{H}\frac{\phi_0^{\prime\prime}}{\phi_0^{\prime}}\right)\phi - {\mathcal{H}^{-2}}\nabla^2  \Phi = 0 \,\,\text{,}
\end{align}
where $\mathcal{H} = a H$, and $'$ denotes the derivative with respect to the conformal time $\eta$ given by,
 \begin{equation}\label{1b}
      d\eta =\frac{dt}{a(t)} \,\,\text{.}
 \end{equation}
The time derivatives with respect to $t$ and $\eta$ are related as, 
 \begin{equation}\label{1d}
     \frac{d}{d\eta} =a \frac{d}{dt} \,\,\text{,} \qquad \frac{d^2}{d\eta^2}=a^2H\frac{d}{dt}+a^2\frac{d^2}{dt^2} \,\,\text{.}
 \end{equation}
 The variable $\psi$ gives the 3-curvature perturbation of the otherwise spatially flat constant time slice: $^{(3)}R=-\frac{4}{a^2}\nabla^{2}\psi$.
 
\item \textbf{Comoving curvature perturbation $\mathcal{R}$}:
For single scalar field models, the comoving curvature perturbation is defined as,
\begin{equation}
    \mathcal{R} \equiv \psi - \frac{H}{\dot{\phi}}\delta\phi \,\,\text{.}
\end{equation}
The name comes from the fact that this variable coincides with the 3-curvature perturbation of the spatial slice in the \emph{comoving} gauge, which, for single scalar field models, is given by $\delta\phi=0$. At the linear level, comoving curvature perturbation evolves according to the following equation \cite{fr},
\begin{align}\label{2h}
    \ddot{\mathcal{R}} + \frac{\left(a^{3}\frac{\dot{\phi}^2}{H^2}\right)^.}{\left(a^{3}\frac{\dot{\phi}^2}{H^2}\right)}\dot{\mathcal{R}} - \frac{1}{a^2}\nabla^2\mathcal{R}=0 \,\,\text{.}
\end{align}

\item \textbf{Sasaki-Mukhanov variable $\varphi_c$}:
Another gauge-invariant perturbation variable that we will consider is the so-called Sasaki-Mukhanov variable \cite{Kodama:1984ziu, Mukhanov:1988jd}, or the scalar field perturbation in uniform curvature gauge, defined as,
\begin{equation}
    \varphi_c \equiv \delta\phi - \frac{\dot{\phi}}{H}\psi \,\,\text{.}
\end{equation}
At the linear level, this variable follows the perturbation equation,
\begin{align}\label{3h}
& \frac{d^2\varphi_c}{dN^2}+\frac{d\varphi_c}{dN}\left(\frac{V}{H^2}\right) + \left( \frac{V_{,\phi \phi} + 2\frac{\dot{\phi}}{H} V_{,\phi}+\left(\frac{\dot{\phi}}{H}\right)^2 V}{H^2}\right)\varphi_c - {\mathcal{H}^{-2}}\nabla^2 \varphi_c = 0 \,\,\text{.}
 \end{align}
\end{itemize}
The perturbation equations given in Eq.~\eqref{1h}, Eq.~\eqref{2h}, and Eq.~\eqref{3h} are strictly valid only in the absence of matter.

To obtain a dynamical system that describes the evolution of perturbations, we first introduce Cartesian spatial coordinates and take the Fourier transform of the perturbation variables. This results in, 
\begin{equation}
   \mathcal{H}^{-2} \nabla^2 \longrightarrow - k^2 \mathcal{H}^{-2} \,\,\text{,}
\end{equation}
We now consider the evolution of the three perturbation quantities in separate sections.

\subsection{Evolution of perturbed quantities}
\label{Section-6.3} 

We have  confirmed that, generically, for $q\neq -1$, the scale factor $a$ has a power law
dependence on conformal/cosmic time and thereby a constant deceleration parameter. 

In analyzing the solutions, we need the following properties of conformal time that follow
from the assumption that $q$ is a non-zero constant (and different from $-1$). 

For fixed $x_c\neq 0$, $a(t)= (3 H_0 {x_c}^2 \left(t-t_U\right)+1)^{\frac{1}{3{x_c}^2}}$ and $q=-1+3 x_*^2\neq 0$ and $q\neq -1$. Moreover, from Eq.~\eqref{1d} it follows that,
     \begin{align}
    \eta & = \int \frac{d\eta}{dt} dt = \int a^{-1} d t= \int \left(3 H_0 {x_c}^2 \left(t-t_U\right)+1\right)^{-\frac{1}{3{x_c}^2}} dt \nonumber \\
    & = \frac{\left(3 {H_0} {x_c}^2
   \left(t-t_U\right)+1\right)^{1-\frac{1}{3 {x_c}^2}}}{{H_0}
   \left(3 {x_c}^2-1\right)}  = \frac{\left({H_0} (q+1)
   \left(t-t_U\right)+1\right)^{\frac{q}{q+1}}}{{H_0} q} \,\,\text{.}
     \end{align}
     On the other hand, 
 \begin{align}
   a= \left({H_0} (q+1) \left(t-t_U\right)+1\right)^{\frac{1}{q+1}} \,\,\text{,} \quad
   H= \frac{{H_0}}{{H_0} (q+1) \left(t-t_U\right)+1} \,\,\text{.}
 \end{align}
Then 
\begin{align}
  \mathcal{H}= a H=  {H_0} \left({H_0} (q+1) \left(t-t_U\right)+1\right)^{-\frac{q}{q+1}}= H_0 a^{-q} \,\,\text{.}
\end{align}
Finally,
\begin{equation}
  \mathcal{H} \eta=  \frac{1}{q} \,\,\text{,} \quad  \eta =  \frac{a^{q}}{{H_0} q} = \eta_0 e^{q N} \,\,\text{,} \quad  \mathcal{H}=\mathcal{H}_0 e^{-q N} \,\,\text{.}
\end{equation}
where $\eta_0=\frac{1}{{H_0} q}$, $\mathcal{H}_0= a_0 H_0=H_0$, recall that we have taken $a_0=1$ such that $N=\ln a$ .

In this case, it is convenient to introduce a new variable
\begin{equation}
    \nu  = a^p \times \text{Perturbed quantity} \,\,\text{,}
\end{equation}
(where $p$ is chosen to remove the first order derivative of  $\nu$) 
and to use conformal time $\eta$ instead of e-fold time $N$.  In making the transition from N to $\eta$, we use the relations given by,
\begin{align}
 \frac{d}{dN} & = \mathcal{H}^{-1}  \frac{d}{d\eta}  \,\,\text{,} \quad 
   \frac{d^2}{dN^2} =   \mathcal{H}^{-2} \frac{d^2}{d\eta^2} +  q  \mathcal{H}^{-1}  \frac{d}{d\eta} \,\,\text{.}
\end{align}

\subsubsection{Bardeen potential}\label{Section-6.3.1}

Using the dynamical variables, we can write the perturbation equation given in Eq.~\eqref{1h} as,
\begin{align}\label{ptbn_bardeen3}
& \frac{d^2\Phi_k}{dN^2} + \left[7-3x^{2}+\sqrt{6}\lambda\left(\frac{1-x^2}{x}\right)\right]\frac{d\Phi_k}{dN} + \left[6\left(1-x^2\right)+\sqrt{\frac{3}{2}}\lambda\left(\frac{1-x^2}{x}\right) + \frac{k^2}{a^2H^2}\right]\Phi_k=0 \,\,\text{.}
\end{align} 

When $q$ is a constant, $q\notin\{0,-1\}$, $dq/dN=0$, $dx/dN=0$, hence, 
$\left(6x - \sqrt{6} \lambda\right)\left(1-x^2\right)=0$. Therefore, either $x=\lambda/\sqrt{6}$ or $x=\pm 1$, such $x\neq 0$. Then 
$6x \left(1-x^2\right)= \sqrt{6} \lambda\left(1-x^2\right)$ implies $\lambda\left(\frac{1-x^2}{x}\right)=\sqrt{6} \left(1-x^2\right)$. Then, Eq.~\eqref{ptbn_bardeen3} becomes
\begin{align}
& \frac{d^2\Phi_k}{dN^2} + \left(13-9 x^2\right)\frac{d\Phi_k}{dN} + \left[9\left(1-x^2\right) + \frac{k^2}{a^2H^2}\right]\Phi_k=0  \,\,\text{.}
\end{align}
But $q=-1+3 x^2$ implies this.
\begin{align}\label{ptbn_bardeen33}
& \frac{d^2\Phi_k}{dN^2} + \left[10-3 q\right]\frac{d\Phi_k}{dN} + \left[6 -3 q + \frac{k^2}{a^2H^2}\right]\Phi_k=0  \,\,\text{.}
\end{align}
Then, passing to the variable $\eta$, and using the relation $  \mathcal{H} \eta=  1/{q}$, Eq.~\eqref{ptbn_bardeen33} becomes 
\begin{align}
&    \frac{d^2\Phi_k}{d\eta^2} +  \left[10-2 q\right](q \eta)^{-1}  \frac{d\Phi_k}{d\eta} + \left[(6 -3 q)(q \eta)^{-2} + k^2\right]\Phi_k=0  \,\,\text{.}
\end{align}

Defining \begin{equation}
    v_k = a^p \Phi_{k}  \,\,\text{,}
\end{equation}
we have
\begin{equation}
\frac{d^2\Phi_k}{d\eta^2}=\frac{p v_k a^{-p}   (p+q+1)}{\eta ^2 q^2}-\frac{2 p  a^{-p}}{\eta  q} \frac{d  v_k}{d\eta} +a^{-p} \frac{d^2 v_k}{d\eta^2},
\end{equation}
and
\begin{equation}
  \frac{d  \Phi_k}{d\eta}=a^{-p}  \frac{d  v_k}{d\eta}-\frac{p v_k   a^{-p}}{\eta  q}  \,\,\text{.}
\end{equation}
Then
\begin{align}
& \frac{d^2 v_k}{d\eta^2} -\frac{2 (p+q-5)}{\eta  q} \frac{d  v_k}{d\eta} + v_k \left(k^2+\frac{p (p+3 q-9)-3 q+6}{\eta ^2 q^2}\right)=0  \,\,\text{.}
\end{align}
Because $q$ is constant at fixed points, we can eliminate the first-order derivative of $v_k$ by defining the constant $p$ such that 
\begin{equation}
  (p+q-5)=0  \,\,\text{.}
\end{equation}
Then we obtain the Bessel equation for the function $v_k$, 
\begin{align}
 \frac{d^2 v_k}{d\eta^2} +    v_k \left(k^2-\frac{(q-2) (2 q-7)}{\eta ^2 q^2}\right)=0  \,\,\text{,}
\end{align}
This equation can be written as, 
\begin{equation}
  \frac{d^2 v_k}{d\eta^2} +    v_k \left(k^2-\left({\nu ^2-\frac{1}{4}}\right){\eta ^{-2}}\right)=0  \,\,\text{.}
\end{equation}
by defining
\begin{align}
    \nu^2= \frac{14-11 q}{q^2}+\frac{9}{4}  \,\,\text{.}
\end{align}
The resulting equation admits the solution given by,
\begin{align}
  v_k(\eta )= C_+ \sqrt{\eta } J_{\nu }(k \eta )+C_- \sqrt{\eta } Y_{\nu }(k \eta)  \,\,\text{.} 
\end{align}
 $C_+$ and $C_-$ are complex constants depending on $k$. 
 
In the special case $x_c = \pm \sqrt{3}/3$, $q=0$, the previous asymptotic analysis fails. 
To analyze this case, we use the relation $ \mathcal{H}= a H=  {H_0}$ (constant), and find that $\lambda$ is constant at the equilibrium point. Then, Eq.~\eqref{ptbn_bardeen3} becomes 
\begin{align}\label{ptbn_bardeen3-special}
& \frac{d^2 \Phi_k}{d\eta^2} + 2\left[3 \pm \sqrt{2}\lambda\right] \mathcal{H}  \frac{d\Phi_k}{d\eta} + \left[ \left(4\pm \sqrt{2}\lambda\right)\mathcal{H}^{2}+  k^2\right]\Phi_k=0  \,\,\text{.}
\end{align}
Defining 
\begin{equation}\label{2ptbn_bardeen3-special} 
 \Phi_k(\eta)=  v_k(\eta )   e^{-H_0\eta   \left(3 \pm \sqrt{2} \lambda\right)}  \,\,\text{,}
\end{equation}
Eq.~\eqref{ptbn_bardeen3-special} becomes, 
\begin{align}\label{3ptbn_bardeen3-special} 
 \frac{d^2 v_k}{d\eta^2}  + v_k  \left(k^2-H_0^2 \left(2 \lambda ^2+5 \sqrt{2} \lambda  \epsilon +5\right)\right)=0  \,\,\text{,}
\end{align}
with $\epsilon=\pm 1$. The solution
is as follows,
\begin{equation} \label{4ptbn_bardeen3-special}
 v_k(\eta)=   C_+ e^{\eta  \sqrt{H_0^2 \left(2 \lambda ^2+5 \sqrt{2} \lambda  \epsilon +5\right)-k^2}}+C_- e^{-\eta 
   \sqrt{H_0^2 \left(2 \lambda ^2+5 \sqrt{2} \lambda  \epsilon +5\right)-k^2}}  \,\,\text{,}
\end{equation}
where  $C_+$ and $C_-$ are complex constants depending on $k$.

\subsubsection{Comoving curvature perturbation}
\label{Section-6.3.2}

Using the dynamical variables, we can write the perturbation Eq.~\eqref{2h} as
 \begin{align}\label{ptbn_comov}
 & \frac{d^2\mathcal{R}_k}{dN^2} + \sqrt{6}\lambda\left(\frac{1-x^2}{x}\right)\frac{d\mathcal{R}_k}{dN} + \left(\frac{k^2}{a^2H^2}\right)\mathcal{R}_k = 0  \,\,\text{.}
\end{align}

Using the relation $\lambda\left(\frac{1-x^2}{x}\right)=\sqrt{6} \left(1-x^2\right)$, valid for constant $q$, Eq.~\eqref{ptbn_comov} becomes 

\begin{align}
 &  \frac{d^2 \mathcal{R}_k}{d\eta^2} +  \left(4-q\right) \frac{1}{q \eta}  \frac{d  \mathcal{R}_k}{d\eta} +   k^2 \mathcal{R}_k = 0  \,\,\text{.}
\end{align}
Defining \begin{equation}
    v_k = a^p \mathcal{R}_k  \,\,\text{,}
\end{equation}
we have
\begin{align}
\frac{d^2 v_k}{d\eta^2} -\frac{(2 p+q-4)}{\eta  q} \frac{dv_k}{d\eta} +  v_k \left(k^2+\frac{p (p+2 q-3)}{\eta ^2 q^2}\right)=0  \,\,\text{.}
\end{align}
Defining 
\begin{equation}
    p=  2-\frac{q}{2}  \,\,\text{,}
\end{equation}
we have 
\begin{equation}
 \frac{d^2 v_k}{d\eta^2} +  v_k \left(k^2-\frac{(q-4) (3 q-2)}{4 \eta ^2 q^2}\right)=0  \,\,\text{.}
\end{equation}
Defining 
\begin{equation}
    \nu^2=-\frac{7}{2 | q| }+\frac{2}{q^2}+1  \,\,\text{,}
\end{equation}
the equation can be written as 
\begin{equation}
  \frac{d^2 v_k}{d\eta^2} +    v_k \left(k^2-\left({\nu ^2-\frac{1}{4}}\right){\eta ^{-2}}\right)=0  \,\,\text{,}
\end{equation}
that admits the solution; 
\begin{align}
  v_k(\eta )= C_+ \sqrt{\eta } J_{\nu }(k \eta )+C_- \sqrt{\eta } Y_{\nu }(k \eta)  \,\,\text{.}
\end{align}
 $C_+$ and $C_-$ are complex constants depending on $k$.

In the special case $x_c = \pm \sqrt{3}/3$, $q=0$, the previous asymptotic analysis fails. 
To analyse this case, we use the relation $ \mathcal{H}= a H=  {H_0}$ (constant), and find that $\lambda$ is constant at the equilibrium point. In this case, Eq.~\eqref{ptbn_comov} becomes 
\begin{align}\label{ptbn_comov3-special}
& \frac{d^2 \mathcal{R}_k}{d\eta^2}\pm  2\sqrt{2}\lambda \mathcal{H}  \frac{d\mathcal{R}_k}{d\eta} +   k^2\mathcal{R}_k=0  \,\,\text{.}
\end{align}
Defining 
\begin{equation}\label{2ptbn_comov3-special} 
 \mathcal{R}_k(\eta)=  v_k(\eta )   e^{\mp H_0  \eta \sqrt{2} \lambda}  \,\,\text{,}
\end{equation}
then Eq.~\eqref{ptbn_comov3-special} becomes 
\begin{align}\label{3ptbn_comov33-special} 
 \frac{d^2 v_k}{d\eta^2}  + v_k  \left(k^2-2H_0^2 \lambda ^2\right)=0  \,\,\text{.}
\end{align}
The solution is as follows.
\begin{equation} \label{4ptbn_comov3-special}
 v_k(\eta)=   C_+ e^{\eta  \sqrt{2 H_0^2 \lambda ^2-k^2}}+C_- e^{-\eta  \sqrt{2 H_0^2 \lambda ^2-k^2}}  \,\,\text{,}
\end{equation}
where  $C_+$ and $C_-$ are complex constants depending on $k$. 
\subsubsection{Sasaki-Mukhanov variable}
\label{Section-6.3.3}
Using the dynamical variables, we can write the perturbation equation given in Eq.~\eqref{3h} as,
\begin{align}
    \frac{d^2\varphi_{ck}}{dN^2} + 3 \left( 1-x^2\right)\frac{d\varphi_{ck}}{dN} + \left[ 18 \left(1-x^2\right)\left( \frac{f}{6}+\left(x-\frac{\lambda }{\sqrt{6}}\right)^2\right) +\frac{k^2}{a^2H^2}\right]\varphi_{ck} = 0  \,\,\text{.} \label{eq:Uggla} 
\end{align}

As before, when $q$ is constant,  $dx/dN=0$, hence 
$\left(6x - \sqrt{6} \lambda\right)\left(1-x^2\right)=0$. Therefore, $x=\lambda/\sqrt{6}$ or $x=\pm 1$ (note that $x\neq 0$). Using $x\neq 0$ in Eq.~\eqref{(20)}, it follows at a fixed point that $\lambda$ is constant and $f(\lambda)=0$. At equilibrium, $\left(1-x^2\right) \left(\frac{f}{6}+\left(x-\frac{\lambda }{\sqrt{6}}\right)^2\right)=0$. Then, passing to the time variable $\eta$, Eq.~\eqref{eq:Uggla}  becomes, 
\begin{align}
 \frac{d^2 \varphi_{ck}}{d\eta^2} +  2 (\eta q)^{-1} \frac{d\varphi_{ck} }{d\eta} +  k^2\varphi_{ck} = 0  \,\,\text{.}
\end{align}

Defining \begin{equation}
    v_k = a \varphi_{ck},
\end{equation}
we have
\begin{align}
 \frac{d^2 v_k}{d\eta^2}+    v_k  \left(k^2+\eta ^{-2} |q|^{-1}\right)=0  \,\,\text{,}
 \end{align}
with the solution 
\begin{align}
  v_k(\eta )= C_+ \sqrt{\eta } J_{\nu}(k \eta )+C_- \sqrt{\eta } Y_{\nu}(k \eta )  \,\,\text{,}
\end{align}
where 
 \begin{equation}
   \nu=\frac{1}{2} \sqrt{1-4 |q|^{-1}}  \,\,\text{.}
 \end{equation}
 $C_+$ and $C_-$ are complex constants depending on $k$.

In the special case $x_c = \pm \sqrt{3}/3, \lambda=\pm \sqrt{2}$, with $f(\pm \sqrt{2})=0$ and $q=0$, the previous asymptotic analysis fails. 
To analyse this case, we use the relation $ \mathcal{H}= a H=  {H_0}$ (constant). In this case, Eq.~\eqref{eq:Uggla} becomes 
\begin{align}\label{eq:Uggla-special}
&  \frac{d^2\varphi_{ck}}{d\eta^2} + 2H_0\frac{d\varphi_{ck}}{d\eta} + k^2\varphi_{ck} = 0  \,\,\text{.}
\end{align}
Defining 
\begin{equation}\label{2eq:Uggla-special} 
\varphi_{ck}(\eta)=  v_k(\eta )   e^{-H_0 \eta}  \,\,\text{,}
\end{equation}
so Eq.~\eqref{eq:Uggla-special} becomes 
\begin{align}\label{3eq:Ugglaspecial} 
 \frac{d^2 v_k}{d\eta^2}  + v_k  \left(k^2-H_0^2\right)=0  \,\,\text{,}
\end{align}
with $\epsilon=\pm 1$. The solution
is as follows.
\begin{equation} \label{4eq:Uggla-special}
 v_k(\eta)=   C_+ e^{\eta  \sqrt{H_0^2-k^2}}+C_- e^{-\eta  \sqrt{H_0^2 -k^2}}  \,\,\text{,}
\end{equation}
where  $C_+$ and $C_-$ are complex constants depending on $k$. 

\section{Dynamical system analysis at background and perturbation levels for the matterless case}
\label{sect:4}

Defining
\begin{align}
    Z=k^2(aH)^{-2} \,\,\text{,} \label{EQ:(63)}
\end{align}
then $Z^{\prime}$ satisfies the following equation via the deceleration parameter $q$
\begin{align}
    Z^{\prime}=2qZ \,\,\text{,}
\end{align}
The latter is related to the Hubble parameter through the expression $\dot{H}=-(1+q)H^2$ or, as $\frac{d}{dN}=H^{-1}\frac{d}{dt}$,
\begin{align}
    H^{\prime}=-(1+q)H \,\,\text{.}
\end{align}
Finally, the above expression can be written as
\begin{align}
    1+q = \frac{1}{2}\frac{\dot{\phi}^2}{H^2} = 3x^2 \,\,\text{.}
\end{align}

For $q\neq -1$ and $x\neq 0$, we have at the equilibrium points 
\begin{align}
    1+q_*  = 3{{x}^*}^2 \,\,\text{.}
\end{align}
Then 
\begin{align}
    & \frac{d \ln H}{d \ln a}=-3{{x}^*}^2 \implies H=H_0 a^{-3{{x}^*}^2}  \nonumber \\
    & \implies a(t)=\left(3 H_0 {{x}^*}^2  \left(t-t_U\right)+1\right)^{\frac{1}{3{{x}^*}^2}}, \;  H(t)=\frac{H_0}{3 H_0 {{x}^*}^2 \left(t-t_U\right)+1} \,\,\text{,}
\end{align}
where $t_U$ is the age of the Universe, and we have assumed $H(t_U)=H_0$, and  $a(t_U)=1$. 

When ${{x}^*}=0$, $q_*=-1$ we have a de Sitter expansion with 
\begin{align}
a(t)= e^{H_0 \left(t-t_U\right)} \,\,\text{,} \;  H(t)= H_0 \,\,\text{.} 
\end{align}
Deepening into the interpretation of the variable $Z= \frac{k^2}{\mathcal{H}^2}$, with $\mathcal{H}= a H$. Perturbations with $k^2 \mathcal{H}^{-2}\ll 1$ are called long wavelength or super-horizon. Those with $k^2 \mathcal{H}^{-2}\gg 1$ are considered short wavelengths or sub-horizon. Long wavelength perturbations are usually studied by choosing the idealized limiting value $k=0$, corresponding to $Z = 0$.
On the other hand, short wavelength perturbations correspond to $Z\rightarrow \infty$. We also note that in
choosing $Z= \frac{k^2}{\mathcal{H}^2}$ as a dynamical variable, we see that the wave number
$k$ is absorbed in the definition when formulating the dynamical system. However, if we choose the reference time $t=t_U$ (i.e.,
when $N:= \ln a = 0$) to be the time to set the initial data in the state space, then different choices of $Z_0= \frac{k^2}{H_0^2}$ for a given $H_0$ (we assume $a(t_U)=1$) yield solutions with different wave
numbers $k$ \cite{Alho:2020cdg}.

The evolution of the background quantities leads to the (not bounded) dynamical system 
\begin{align}
& x^{\prime} = -\left(3x - \sqrt{\frac{3}{2}} \lambda\right)\left(1-x^2\right), \quad  \lambda^{\prime} =-\sqrt{6} x f, \quad  Z^{\prime} = 2(3x^2 -1)Z \,\,\text{,}
\end{align}
To do the compactification,   we denote
\begin{align}
    \bar{Z}=\frac{Z}{1+Z}=\frac{k^2}{k^2+(aH)^2},\quad 
    Z=\frac{\bar{Z}}{1-\bar{Z}} \,\,\text{.}
\end{align}
We note that
\begin{align}
    \bar{Z}^{\prime} = 2(3x^2 -1)\bar{Z}\left(1-\bar{Z}\right) \,\,\text{.}
\end{align}

However, as we can see from that, when $\bar{Z}\to 1$, we have a singularity, so we change the ``$e$ folding time'' from $N$ to $\bar{N}$ through
\begin{align}
    \frac{d\bar{N}}{dN}=\frac{1}{1-\bar{Z}}=1+Z \,\,\text{.}
\end{align}

Recall that we refer to the invariant set $\bar{Z}= 0$ as the long
wavelength boundary (or the super-horizon boundary), and $\bar{Z} = 1$ as the short wavelength boundary (or sub-horizon boundary). 
After compactification of $Z$ to $\bar{Z}$,  we get the final set of equations  as
\begin{subequations}
\label{(eq:99)}
\begin{align}
& \frac{dx}{d\bar{N}} =-\left(3x - \sqrt{\frac{3}{2}} \lambda\right)\left(1-x^2\right)\left(1-\bar{Z}\right) \,\,\text{,} \\
& \frac{d\lambda}{d\bar{N}} = - \sqrt{6}xf\left(1-\bar{Z}\right) \,\,\text{,} \\
& \frac{d\bar{Z}}{d\bar{N}} = 2(3x^2 - 1)\bar{Z}\left(1-\bar{Z}\right)^2 \,\,\text{.} 
\end{align}
\end{subequations}

The function $\lambda$ can be negative, zero, positive, or unbounded. However, in some exceptional cases,  say, when $f(\lambda)$ is an even function, $f(\lambda)= f(-\lambda)$, there is no loss of generality in assuming that $\lambda$ is non-negative. In this case, the field equations are invariant under the transformation $(x, \phi) \rightarrow -(x, \phi)$ and $\lambda \rightarrow -\lambda$. That is, the case of the exponential potential where $f\equiv 0$. 

For an arbitrary potential, given $\lambda^*$, with $f(\lambda^*)=0$, the equilibrium point  $Sf(\lambda^*)$, with ${{x}^*}= \lambda^*/\sqrt{6}$ exists for $-\sqrt{6}<\lambda^*<\sqrt{6}$ and is a sink for $-\sqrt{3}<{\lambda^*}<0,  f'({\lambda^*})<0$ or $0<{\lambda^*}<\sqrt{3}, f'({\lambda^*})>0$. In addition, 
$dS$ is  stable for $f(0)> 0$. That is, at the future attractor, we have  ${{x}^*}= \lambda^*/\sqrt{6}$ or zero at the 
background state space. At these equilibrium points, the deceleration parameter $q= -1+ {{\lambda}^*}^2/2$ or $q=-1$ is thus constant. The range
$-\sqrt{6} \leq \lambda^* \leq \sqrt{6}$ corresponds to the range $-1 \leq q \leq 2$ for $q$. This range of $q$ also describes a space-time with a perfect fluid with a linear equation of state $p = w \rho$, with $w:= (2 q-1)/3$ in the range $-1\leq w \leq 1$. Thus, ${\lambda}^*=0$ corresponds to a cosmological
constant while the bifurcation value $\lambda^*= \pm \sqrt{6}$ corresponds to a stiff fluid with the speed of sound equal to that of light. On the other hand, ${\lambda^*}^2>6$ yields an equation of state with superluminal speed. Therefore, at the physically interesting late-time attractors, $\lambda$ is bounded. However, we can handle the cases $\lambda \rightarrow \pm \infty$ using the new variable Eq.~\eqref{eqU}.

\subsection{Stability analysis of the fixed points on the background space $B$}
\label{sect:A.1}

The dynamics at the background space  $B = \left\{\left(x, \lambda, \bar{Z}\right)\in [-1,1]\times \mathbb{R} \times [0,1]\right\}$ is  given by 
\begin{align}
\label{background(eq:99)}
& \frac{dx}{d {N}} =-\left(3x - \sqrt{\frac{3}{2}} \lambda\right)\left(1-x^2\right) \,\,\text{,} \quad  \frac{d\lambda}{d {N}} = - \sqrt{6}xf \,\,\text{,} \quad  \frac{d\bar{Z}}{d {N}} = 2(3x^2 - 1)\bar{Z}\left(1-\bar{Z}\right) \,\,\text{,} 
\end{align}
where it is convenient to use the e-folding variable as the time variable.

\begin{table}[]
    \centering
    \caption{Equilibrium points of system Eq.~\eqref{background(eq:99)} in the finite region for an arbitrary function $f(\lambda)$. N. H. stands for Non-hyperbolic.}
               \resizebox{\textwidth}{!}{
    \begin{tabular}{|c|c|c|c|c|c|c|c|c|}
    \hline 
Label &    $x$ &$\lambda$ & $\bar{Z}$ & Existence & $k_1$ & $k_2$ & $k_3$ & Stability \\\hline
$P_1({\lambda^*})$ & $\frac{{{\lambda^*}}}{\sqrt{6}}$ & ${{\lambda^*}}$ & $0$ & $-\sqrt{6} \leq \lambda^*\leq \sqrt{6}$ &  $\frac{1}{2} \left({{\lambda^*}}^2-6\right)$ & ${{\lambda^*}}^2-2$ &  $-{{\lambda^*}} f'({{\lambda^*}})$&  Saddle  for \\
 &&&&&&&& $f'(\lambda^*)<0, -\sqrt{2}<\lambda^*<0$, \\
 &&&&&&&& or $f'(\lambda^*)>0, 0<\lambda^*<\sqrt{2}$, \\
 &&&&&&&& or $2 <\lambda^{*^2}<6$, \\
 &&&&&&&& or ${\lambda^{*}} f'({\lambda^{*}})<0$. \\
 &&&&&&&& N. H.  otherwise. \\
\hline
$P_2({\lambda^*})$ & $-1$ & ${{\lambda^*}}$ & $0$ & always & $4$ & $\sqrt{6} {{\lambda^*}}+6$ & $\sqrt{6} f'({{\lambda^*}})$ & Source for \\
&&&&&&&& $ {{\lambda^*}}>- \sqrt{6}, f'({{\lambda^*}})>0$. \\
&&&&&&&& Saddle for  $ {{\lambda^*}}<- \sqrt{6}$ \\
&&&&&&&& or  $f'({{\lambda^*}})<0$.\\
&&&&&&&& N. H.  otherwise. \\\hline  
$P_3({\lambda^*})$ & $1$ & $ {{\lambda^*}}$ & $ 0$ & always & $4$ & $6-\sqrt{6} {{\lambda^*}}$ & $-\sqrt{6} f'({{\lambda^*}})$ & Source for \\
&&&&&&&& $ {{\lambda^*}}< \sqrt{6},  f'({{\lambda^*}})<0$, \\
&&&&&&&& saddle for  $ {{\lambda^*}}> \sqrt{6}$, \\
&&&&&&&& or $f'({{\lambda^*}})>0$. \\
&&&&&&&& N. H.  otherwise.\\\hline 
$P_4({\lambda^*})$ & $\frac{{{\lambda^*}}}{\sqrt{6}}$ & ${{\lambda^*}}$ & $1$ & $-\sqrt{6} \leq\lambda^*\leq \sqrt{6}$ &  $\frac{1}{2} \left({{\lambda^*}}^2-6\right)$ & $ 2-{{\lambda^*}}^2$ & $ -{{\lambda^*}} f'({{\lambda^*}})$& sink for \\
&&&&&&&& $2 <\lambda^{*^2}<6, {\lambda^{*}} f'({\lambda^{*}})>0$. \\
&&&&&&&& saddle for $0\leq \lambda^{*^2}<2$, \\
&&&&&&&& or ${\lambda^{*}} f'({\lambda^{*}})<0$. \\
&&&&&&&& N. H.  otherwise. \\\hline
$P_5({\lambda^*})$ & $-1$ & ${{\lambda^*}}$ & $ 1$ & always & $-4$ & $\sqrt{6} {\lambda^{*}}+6$ & $\sqrt{6} f'({\lambda^{*}})$ &  sink for \\
&&&&&&&& ${\lambda^{*}}<-\sqrt{6}, f'({\lambda^{*}})<0$. \\
&&&&&&&& saddle for ${\lambda^{*}}<-\sqrt{6} $, \\
&&&&&&&& or $f'({\lambda^{*}})>0$.  \\
&&&&&&&& N. H.  otherwise. \\\hline 
$P_6({\lambda^*})$ & $1$ & $ {{\lambda^*}}$ & $ 1$ & always & $-4$ & $6-\sqrt{6} {\lambda^{*}}$ & $-\sqrt{6} f'({\lambda^{*}})$ & sink for \\
&&&&&&&& ${\lambda^{*}}>\sqrt{6}, f'({\lambda^{*}})>0$.\\
&&&&&&&& saddle for ${\lambda^{*}}<\sqrt{6} $.\\
&&&&&&&& or $f'({\lambda^{*}})<0$ \\
&&&&&&&& N. H.  otherwise \\\hline 
$P_7$ & $-\frac{1}{\sqrt{3}}$ & $ -\sqrt{2}$ & $ \bar{Z}_c$ & $\begin{array}{c}
   f(-\sqrt{2})=0, \\
   0\leq \bar{Z}_c \leq 1
\end{array}$ & $-2 $ & $0$ & $\sqrt{2} f'\left(-\sqrt{2}\right)$ & saddle if $f'\left(-\sqrt{2}\right)>0$ \\
&&&&&&&& sink for $f'\left(-\sqrt{2}\right)<0$\\\hline
$P_8$ & $\frac{1}{\sqrt{3}}$ & $ \sqrt{2}$ & $ \bar{Z}_c$ & $\begin{array}{c}f(\sqrt{2})=0,\\ 0\leq \bar{Z}_c \leq 1
\end{array}$ & $-2 $ &  $0$ & $-\sqrt{2} f'\left(\sqrt{2}\right)$.  &  saddle if $f'\left(\sqrt{2}\right)<0$. \\
&&&&&&&& sink for $f'\left(\sqrt{2}\right)>0$.\\\hline
$P_9$ & $0$ & $0$ & $0$ & always & $-2$ & $-\frac{1}{2} \left(3+\sqrt{9-12 f(0)}\right)$ & $ -\frac{1}{2} \left(3-\sqrt{9-12 f(0)}\right)$ & sink for $f(0)> 0$.\\
&&&&&&&& saddle for $f(0)<0$.\\\hline 
$P_{10}$ & $0$ & $0$ & $1$ & always & $2$  & $-\frac{1}{2} \left(3+\sqrt{9-12 f(0)}\right)$ & $ -\frac{1}{2} \left(3-\sqrt{9-12 f(0)}\right)$ & saddle. \\\hline
\end{tabular}}
    \label{Background_b}
\end{table}

The equilibrium points in the background space are the following. 
\begin{enumerate}
 \item  $P_1({\lambda^*}): \left(x, \lambda, \bar{Z}\right)=\left(\frac{{{\lambda^*}}}{\sqrt{6}}, {{\lambda^*}}, 0\right)$ that  exists for $-\sqrt{6} \leq \lambda^*\leq \sqrt{6}$. The eigenvalues are $\frac{1}{2} \left({{\lambda^*}}^2-6\right), {{\lambda^*}}^2-2, -{{\lambda^*}} f'({{\lambda^*}})$. It is a saddle  for $f'(\lambda^*)<0, -\sqrt{2}<\lambda^*<0$, or $f'(\lambda^*)>0, 0<\lambda^*<\sqrt{2}$, or $2 <\lambda^{*^2}<6$ or ${\lambda^{*}} f'({\lambda^{*}})<0$. It is non-hyperbolic otherwise.

 \item $P_2({\lambda^*}): \left(x, \lambda, \bar{Z}\right)=\left(-1, {{\lambda^*}}, 0\right)$ that always exists. The eigenvalues are \newline $4, \sqrt{6} {{\lambda^*}}+6, \sqrt{6} f'({{\lambda^*}})$. It is a source for $ {{\lambda^*}}>- \sqrt{6}, f'({{\lambda^*}})>0$. It is a saddle for  $ {{\lambda^*}}<- \sqrt{6}$ or  $f'({{\lambda^*}})<0$.  It is non-hyperbolic otherwise.

 \item $P_3({\lambda^*}): \left(x, \lambda, \bar{Z}\right)=\left(1, {{\lambda^*}}, 0\right)$ that always exists. The eigenvalues are \newline $4, 6-\sqrt{6} {{\lambda^*}}, -\sqrt{6} f'({{\lambda^*}})$. It is a source for $ {{\lambda^*}}< \sqrt{6},  f'({{\lambda^*}})<0$. It is a saddle for  $ {{\lambda^*}}> \sqrt{6}$ or  
   $f'({{\lambda^*}})>0$.  It is non-hyperbolic otherwise.

\item $P_4({\lambda^*}): \left(x, \lambda, \bar{Z}\right)=\left(\frac{{{\lambda^*}}}{\sqrt{6}}, {{\lambda^*}}, 1\right)$ that exists for $-\sqrt{6} \leq\lambda^*\leq \sqrt{6}$. The eigenvalues are $\frac{1}{2} \left({{\lambda^*}}^2-6\right), 2-{{\lambda^*}}^2, -{{\lambda^*}} f'({{\lambda^*}})$. It is a sink for $2 <\lambda^{*^2}<6$ and ${\lambda^{*}} f'({\lambda^{*}})>0$. It is a saddle for $0\leq \lambda^{*^2}<2$ or ${\lambda^{*}} f'({\lambda^{*}})<0$. It is non-hyperbolic otherwise. 

 \item  $P_5({\lambda^*}): \left(x, \lambda, \bar{Z}\right)=\left(-1, {{\lambda^*}}, 1\right)$ that always exists. The eigenvalues are \newline $-4, \sqrt{6} {\lambda^{*}}+6, \sqrt{6} f'({\lambda^{*}})$.  It is a sink for ${\lambda^{*}}<-\sqrt{6} $ and $f'({\lambda^{*}})<0$. It is a saddle for ${\lambda^{*}}<-\sqrt{6} $ or $f'({\lambda^{*}})>0$.  It is non-hyperbolic otherwise.

 \item  $P_6({\lambda^*}): \left(x, \lambda, \bar{Z}\right)=\left(1, {{\lambda^*}}, 1\right)$ that always exists. The eigenvalues are \newline $-4, 6-\sqrt{6} {\lambda^{*}}, -\sqrt{6} f'({\lambda^{*}})$. It is a sink for ${\lambda^{*}}>\sqrt{6} $ and $f'({\lambda^{*}})>0$. It is a saddle for ${\lambda^{*}}<\sqrt{6} $ or $f'({\lambda^{*}})<0$.  It is non-hyperbolic otherwise. 

\item  The line
$P_7: \left(x, \lambda, \bar{Z}\right)=\left(-\frac{1}{\sqrt{3}}, -\sqrt{2}, \bar{Z}_c\right)$ exists for $f(-\sqrt{2})=0$ and $0\leq \bar{Z}_c \leq 1$. 
The eigenvalues are $-2, 0, \sqrt{2} f'\left(-\sqrt{2}\right)$. The eigenvector associated with the zero eigenvalues is tangent to the line. Then, it is normally hyperbolic. This implies that it is a  saddle if $f'\left(-\sqrt{2}\right)>0$ or a sink for $f'\left(-\sqrt{2}\right)<0$.

\item  The line 
$P_8: \left(x, \lambda, \bar{Z}\right)=\left(\frac{1}{\sqrt{3}}, \sqrt{2}, \bar{Z}_c\right)$ exists for $f(\sqrt{2})=0$ and $0\leq \bar{Z}_c \leq 1$. The eigenvalues are $-2, 0, -\sqrt{2} f'\left(\sqrt{2}\right)$. The eigenvector associated with the zero eigenvalues is tangent to the line. Then, it is normally hyperbolic. This implies that it is a  saddle if $f'\left(\sqrt{2}\right)<0$ or a sink for $f'\left(\sqrt{2}\right)>0$.

\item $P_9: \left(x, \lambda, \bar{Z}\right)=\left(0,0,0\right)$.  The eigenvalues are \newline $-2, -\frac{1}{2} \left(3+\sqrt{9-12 f(0)}\right), -\frac{1}{2} \left(3-\sqrt{9-12 f(0)}\right)$. It is a sink for $f(0)>0$ or a saddle for $f(0)<0$.

\item $P_{10}: \left(x, \lambda, \bar{Z}\right)=\left(0,0,1\right)$.  The eigenvalues are \newline $2, -\frac{1}{2} \left(3+\sqrt{9-12 f(0)}\right), -\frac{1}{2} \left(3-\sqrt{9-12 f(0)}\right)$. It is a saddle.

\end{enumerate}

Now we present some numerical solutions. As we commented on, $\lambda$ is generically bounded at late-time attractors. However,  we handle the cases $\lambda \rightarrow \pm \infty$ using the new variable Eq.~\eqref{eqU}.

\subsubsection{First example: monomial potential}

Substituting the function $f(\lambda)=-\frac{\lambda ^2}{n}$ in Eq.~\eqref{background(eq:99)}, we obtain
\begin{subequations}
\label{background(eq:99)-monomial}
\begin{align}
\frac{dx}{d {N}} &=-\left(3x - \sqrt{\frac{3}{2}} \lambda\right)\left(1-x^2\right) \,\,\text{,} \\ \frac{d\lambda}{d N} &= \frac{\sqrt{6}}{n} x \lambda ^2 \,\,\text{,} \\  \frac{d\bar{Z}}{d {N}} &= 2(3x^2 - 1)\bar{Z}\left(1-\bar{Z}\right) \,\,\text{,}
\end{align}
\end{subequations}
defined on the background space  $B = \left\{\left(x, \lambda, \bar{Z}\right)\in [-1,1]\times \mathbb{R} \times [0,1]\right\}$.

\begin{table}[]
    \centering
      \caption{Equilibrium points of system Eq.~\eqref{background(eq:99)-monomial}   in the finite region for $f(\lambda)=-\frac{\lambda ^2}{n}$.}  
    \begin{tabular}{|c|c|c|c|c|c|c|c|c|}\hline
 Label & $x$ & $\lambda$ & $\bar{Z}$ & Existence & $k_1$ & $k_2$ & $k_3$ & Stability \\    \hline 
 $P_1(0)$ & $0$ & $0$ & $0$ & always & $-3$ & $-2$ & $0$ & saddle ($n>0$); sink ($n<0$) \\\hline
 $P_2(0)$ & $-1$ & $ 0$ & $ 0$ & always & $4$ & $ 6$ & $0$ & unstable  \\\hline
 $P_3(0)$ & $1$ & $ 0$ & $ 0$  & always  & $4$ & $6$ & $ 0$ & unstable \\\hline
 $P_4(0)$ & $0$ & $0$ & $1$ & always  & $-6$ & $ 2$ & $0$ &  saddle \\\hline
 $P_5(0)$ & $-1$ & $ 0$ & $ 1$& always & $-4$ & $ 6$ & $ 0$ &  saddle \\\hline 
 $P_6(0)$ & $1$ & $0$ & $ 1$ & always & $-4$ & $ 6$ & $ 0$ & saddle \\\hline
\end{tabular}
    \label{Background_a-powerlawb}
    \end{table}

Table~\ref{Background_a-powerlawb} presents the equilibrium points of system Eq.~\eqref{background(eq:99)-monomial} in the finite region.

Interestingly, the de Sitter point $P_1(0): \left(x, \lambda, \bar{Z}\right)=\left(0, 0, 0\right)$ always exists. The eigenvalues are $-3, -2, 0$. It is non-hyperbolic. Using the Centre Manifold theorem, we find that the graph locally gives the centre manifold of the origin
\begin{align}
& \Big\{\left(x, \lambda, \bar{Z}\right)\in [-1,1]\times \mathbb{R} \times [0,1]: x=\frac{\lambda}{\sqrt{6}}+h_1(\lambda), \bar{Z}=h_2(\lambda), \nonumber \\
& h_1(0)=0, h_2(0)=0,  h_1'(0)=0, h_2'(0)=0, |\lambda|<\delta\Big\}
\end{align}
for a small enough $\delta$. The functions $h_1$ and $h_2$ satisfy the differential equations
\begin{align}
   & 6 \lambda ^2 \left(\left(\sqrt{6} h_1(\lambda )+\lambda \right) h_1'(\lambda )+h_1(\lambda )\right)  -3
   n h_1(\lambda ) \left(2 \sqrt{6} \lambda  h_1(\lambda )+6 h_1(\lambda )^2+\lambda
   ^2-6\right)+\sqrt{6} \lambda ^3=0,
\\
   & -\frac{\lambda ^2 \left(\sqrt{6} h_1(\lambda )+\lambda \right) h_2'(\lambda )}{n}-\left(2 \sqrt{6} \lambda 
   h_1(\lambda )+6 h_1(\lambda )^2+\lambda ^2-2\right) h_2(\lambda )^2 \nonumber \\
   & +\left(2 \sqrt{6} \lambda 
   h_1(\lambda )+6 h_1(\lambda )^2+\lambda ^2-2\right) h_2(\lambda )=0.
\end{align}
Using the Taylor series, we have the solution for $ x(\lambda)$ given by Eq.~\eqref{expansion-x-t} and 
\begin{align}
Z(\lambda) & =   O\left(\lambda ^{14}\right). 
\end{align}

The 1D dynamical system dictates the dynamics at the centre manifold 
\begin{align}
    \frac{d \lambda}{d N} & =- U'(\lambda) \,\,\text{.}
\end{align}
That is a gradient-like equation with potential 
$U(\lambda)$ defined  through Eq.~\eqref{Gradient-Potential}. 
Since $U^{(4)}(0)=-6/n\neq 0$, the origin is a degenerate maximum of the potential for $n>0$. Therefore, the centre manifold of the origin and the origin are unstable (saddle), and if $n<0$, it is stable. 

\begin{figure}
    \centering
    \includegraphics[scale=0.6]{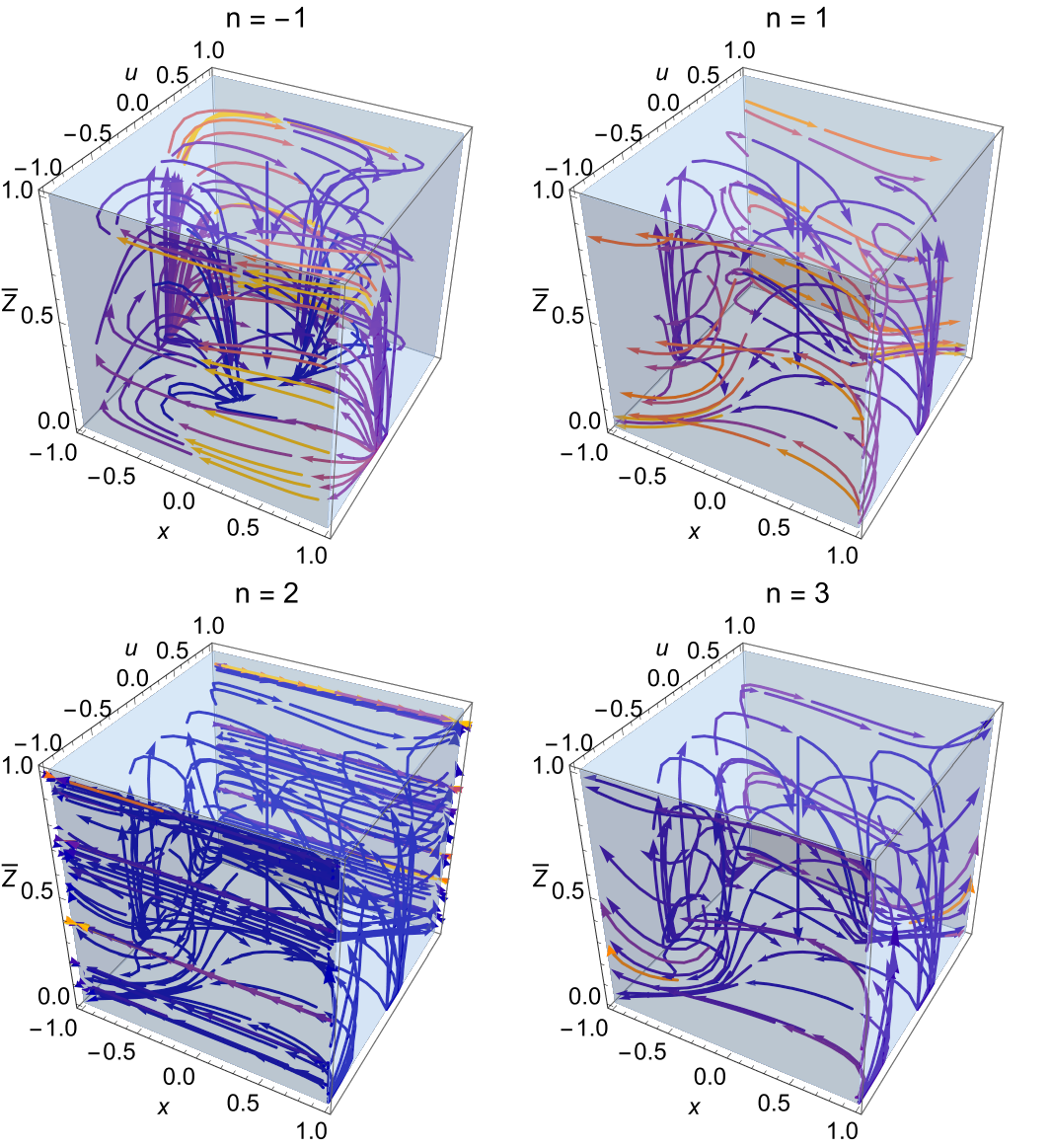}
    \caption{Flow of  the system Eq.~\eqref{background(eq:99)-monomial} in the representations   $\left(x, u\right)$ for $n=-1,1,2,3$.}
    \label{fig:Powerlaw-background}
\end{figure}
In Fig.~\ref{fig:Powerlaw-background} is represented the flow of the system Eq.~\eqref{background(eq:99)-monomial} in the phase space $\left(x, u, \bar{Z}\right)$ for $n=-1,1,2,3$. 

\begin{figure}
    \centering
    \includegraphics[scale=0.6]{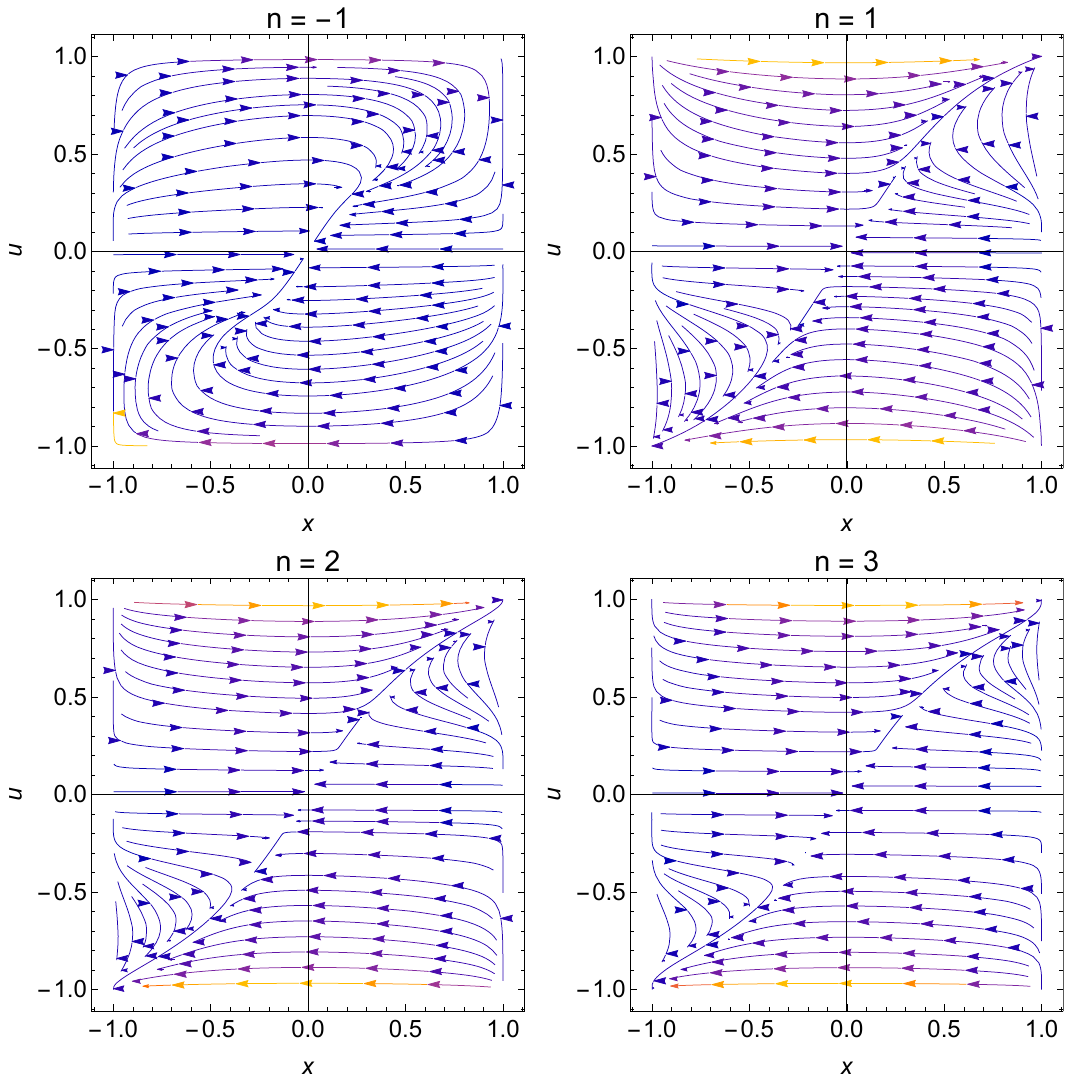}
    \caption{Flow of the system Eq.~\eqref{background(eq:99)-monomial} in the plane $\left(x, u\right)$ for $n=-1,1,2,3$.
}
    \label{fig:powerlaw-background2D}
\end{figure}
In Fig.~\ref{fig:powerlaw-background2D} is represented the flow of the system Eq.~\eqref{background(eq:99)-monomial} restricted to the plane $\left(x, u\right)$ for $n=-1,1,2,3$.

As shown in these Figs.~\ref{fig:Powerlaw-background} and
\ref{fig:powerlaw-background2D}, for $n>0$ the late time attractor corresponds to $\lambda\rightarrow \pm\infty$ (along the centre manifold of the origin). This corresponds to $(x, \lambda,\bar{Z})=\left(\pm 1, \pm \infty, 1\right)$. For $n<0$, the attractor is the origin. For $n>0$, the origin is a saddle point. 
These numerical results illustrate the analytical results in Table~\ref{Background_a-powerlawb}, therefore, the stability analysis is numerically confirmed.   
\subsubsection{Second example: double exponential}

Substituting the function $f(\lambda)=-\left(\lambda+\alpha\right)\left(\lambda+\beta\right)$ into Eq.~\eqref{background(eq:99)} we obtain
\begin{subequations}
\label{background(eq:99)-double-exp}
\begin{align}
\frac{dx}{d {N}} & =-\left(3x - \sqrt{\frac{3}{2}} \lambda\right)\left(1-x^2\right) \,\,\text{,} \\  \frac{d\lambda}{d {N}} & =  \sqrt{6}x \left(\lambda+\alpha\right)\left(\lambda+\beta\right) \,\,\text{,} \\  \frac{d\bar{Z}}{d {N}} & = 2\left(3x^2 - 1\right)\bar{Z}\left(1-\bar{Z}\right) \,\,\text{,} 
\end{align}
\end{subequations}
defined on the background space  $B = \left\{\left(x, \lambda, \bar{Z}\right)\in [-1,1]\times \mathbb{R} \times [0,1]\right\}$.

Recall that $f^{\prime}(\lambda)=-\alpha -\beta -2 \lambda$, and $f(\lambda)=0 \Longleftrightarrow\lambda\in\left\{-\alpha, -\beta\right\}$ and $ f'(-\alpha)=\alpha-\beta$,  $ f'(-\beta)=-\left(\alpha-\beta\right)$. Moreover, we have $f(0)=-\alpha \beta$, and $f'(0)=-\alpha- \beta$. Without losing generality, we can assume $\alpha<\beta$. 

\begin{table}[]
    \centering
     \caption{Equilibrium points of system Eq.~\eqref{background(eq:99)-double-exp}  in the finite region for $f(\lambda)=-(\lambda+\alpha)(\lambda+\beta), \; \alpha \neq \beta$. Without losing generality, we can assume $\alpha<\beta$. }
               \resizebox{\textwidth}{!}{
    \begin{tabular}{|c|c|c|c|c|c|c|c|c|}\hline
 Label & $x$ & $\lambda$ & $\bar{Z}$ & Existence & $k_1$ & $k_2$ & $k_3$ & Stability \\\hline
$P_1(-\alpha)$ & $-\frac{\alpha }{\sqrt{6}}$ & $-\alpha $  &$ 0$ & $-\sqrt{6}<\alpha < \sqrt{6}$ & $\frac{1}{2} \left(\alpha ^2-6\right) $& $\alpha ^2-2$ &$ \alpha  (\alpha  -\beta )$ &   N. H.  for  \\
 &&&&&&&& $\alpha \in\left\{-\sqrt{2}, 0, \sqrt{2}\right\}$\\
 &&&&&&&&  sink for \\
 &&&&&&&&  $ 0<\alpha <\sqrt{2}, 
      \beta >\alpha$\\
 &&&&&&&&  saddle otherwise\\\hline
$P_1(-\beta)$ &  $-\frac{\beta }{\sqrt{6}}$ & $-\beta $ & $0$ & $-\sqrt{6}<\beta < \sqrt{6}$  & $\frac{1}{2} \left(\beta ^2-6\right)$& $\beta ^2-2$ & $\beta  (\beta -\alpha )$ &    N. H.  for \\
 &&&&&&&& $\beta \in\left\{-\sqrt{2}, 0, \sqrt{2}\right\}$\\
  &&&&&&&&  sink for \\
 &&&&&&&& $-\sqrt{2}<\beta <0, \alpha <\beta$\\
 &&&&&&&&  saddle otherwise\\\hline
$P_2(-\alpha)$ &  $-1 $& $-\alpha$  & $0 $ & always & $4$ &$ 6-\sqrt{6} \alpha$  & $\sqrt{6} (\alpha -\beta ) $ & N. H.  for $\alpha=\sqrt{6}$\\
  &&&&&&&&  saddle otherwise\\\hline
$P_2(-\beta)$ &$ -1$ & $-\beta$  & $0 $ & always & $4$  & $6-\sqrt{6} \beta $ & $\sqrt{6} (\beta -\alpha )$  &  
N. H.  for $\beta=\sqrt{6}$\\
 &&&&&&&& source for $\alpha <\beta <\sqrt{6}$\\
  &&&&&&&&  saddle otherwise\\\hline
$P_3(-\alpha)$ & $1$ & $-\alpha $ &$ 0$ & always & $4 $&$ \sqrt{6} \alpha +6$ & $\sqrt{6} (\beta -\alpha ) $ & N. H.  for $\alpha=-\sqrt{6}$\\
 &&&&&&&& source for $-\sqrt{6}<\alpha <\beta$\\
  &&&&&&&&  saddle otherwise\\\hline
$P_3(-\beta)$ & $ 1$ & $-\beta$  &$ 0$ & always &$ 4$ & $\sqrt{6} \beta +6 $ &$ \sqrt{6} (\alpha -\beta ) $ & N. H.  for $\beta=-\sqrt{6}$\\
  &&&&&&&&  saddle otherwise\\\hline
$P_4(-\alpha)$ &$ -\frac{\alpha }{\sqrt{6}}$ & $-\alpha$  & $1 $  & $-\sqrt{6}<\alpha < \sqrt{6}$  & $\frac{1}{2} \left(\alpha ^2-6\right)$ &$ 2-\alpha ^2$ & $\alpha  (\alpha -\beta )$  &  N. H.  for \\
 &&&&&&&& $\alpha \in\left\{-\sqrt{2}, 0, \sqrt{2}\right\}$\\
 &&&&&&&& sink for \\
 &&&&&&&& $\sqrt{2}<\alpha <\sqrt{6}, \beta >\alpha$\\
 &&&&&&&&  saddle otherwise\\\hline
$P_4(-\beta)$ &  $-\frac{\beta }{\sqrt{6}}$ &$ -\beta $ & $1$ & $-\sqrt{6}<\beta < \sqrt{6}$   & $\frac{1}{2} \left(\beta ^2-6\right)$ & $2-\beta ^2$ & $\beta  (\beta -\alpha )$  &  N. H.  for \\
 &&&&&&&& $\beta \in\left\{-\sqrt{2}, 0, \sqrt{2}\right\}$\\
 &&&&&&&& sink for \\
 &&&&&&&& $-\sqrt{6}<\beta <-\sqrt{2}, \alpha <\beta$ \\
 &&&&&&&&  saddle otherwise\\\hline
$P_5(-\alpha)$ & $ -1$ &$ -\alpha$  & $1$ & always & $-4 $& $6-\sqrt{6} \alpha$  & $\sqrt{6} (\alpha -\beta ) $ &  N. H.  for $\alpha= \sqrt{6}$\\
 &&&&&&&& sink for $\sqrt{6}<\alpha <\beta$\\
  &&&&&&&&  saddle otherwise\\\hline
$P_5(-\beta)$ &  $-1 $&$ -\beta$  &$ 1$ & always & $ -4 $  & $6-\sqrt{6} \beta  $ & $\sqrt{6} (\beta -\alpha ) $  & N. H.  for  $\beta=\sqrt{6}$\\
  &&&&&&&&  saddle otherwise\\\hline
$P_6(-\alpha)$ &$ 1 $& $-\alpha$  &$ 1 $& always & $-4 $&$ \sqrt{6} \alpha +6 $&$ \sqrt{6} (\beta -\alpha ) $ & N. H.  for $\alpha=-\sqrt{6}$\\
  &&&&&&&&  saddle otherwise\\\hline
$P_6(-\beta)$ & $1 $&$ -\beta $ &$ 1$ & always & $-4$  &$ \sqrt{6} \beta +6$ & $\sqrt{6} (\alpha -\beta )$ &  N. H.  for $\beta=-\sqrt{6}$\\
  &&&&&&&&  sink for $\alpha <\beta <-\sqrt{6}$ \\
  &&&&&&&& saddle otherwise\\\hline
 $P_7$ & $-\frac{1}{\sqrt{3}}$ & $ -\sqrt{2}$ & $ \bar{Z}_c$ & $\begin{array}{c}
  \beta>\alpha=\sqrt{2}, \\
   0\leq \bar{Z}_c \leq 1
\end{array}$ & $-2 $ & $0$ & $\sqrt{2} \left(\sqrt{2}-\beta \right)$ & sink \\ 
&  & &   & $\begin{array}{c} \alpha<\beta=\sqrt{2}, \\
   0\leq \bar{Z}_c \leq 1
\end{array}$ & $-2 $ & $0$ & $\sqrt{2} \left(\sqrt{2}-\alpha \right)$ & saddle\\\hline
$P_8$ & $\frac{1}{\sqrt{3}}$ & $ \sqrt{2}$ & $ \bar{Z}_c$ & $\begin{array}{c}
  \beta>\alpha=-\sqrt{2}, \\
   0\leq \bar{Z}_c \leq 1
\end{array}$ & $-2 $ &  $0$ & $\sqrt{2} \left(\sqrt{2}+\beta\right)$  &  saddle \\
&  & &   & $\begin{array}{c} \alpha<\beta=-\sqrt{2}, \\
   0\leq \bar{Z}_c \leq 1
\end{array}$ &&&  $\sqrt{2} \left(\sqrt{2}+\alpha\right)$ & sink\\\hline
$P_{9}$ & $0$ & $0$ & $0$ & always & $-2$ & $\frac{1}{2} \left(-\sqrt{12 \alpha  \beta +9}-3\right)$ & $\frac{1}{2} \left(\sqrt{12 \alpha  \beta
   +9}-3\right)$ & stable for $\alpha \beta< 0$ \\\hline
$P_{10}$ & $0$ & $0$ & $1$ & always  &$ 2$ & $\frac{1}{2} \left(-\sqrt{12 \alpha  \beta +9}-3\right)$ &$ \frac{1}{2} \left(\sqrt{12 \alpha  \beta
   +9}-3\right)$ & saddle \\\hline
\end{tabular}}
    \label{Background_Zb}
\end{table}
Table~\ref{Background_Zb} presents the equilibrium points of system Eq.~\eqref{background(eq:99)-double-exp}  in the finite region.

\begin{figure}
    \centering
    \includegraphics[scale=0.6]{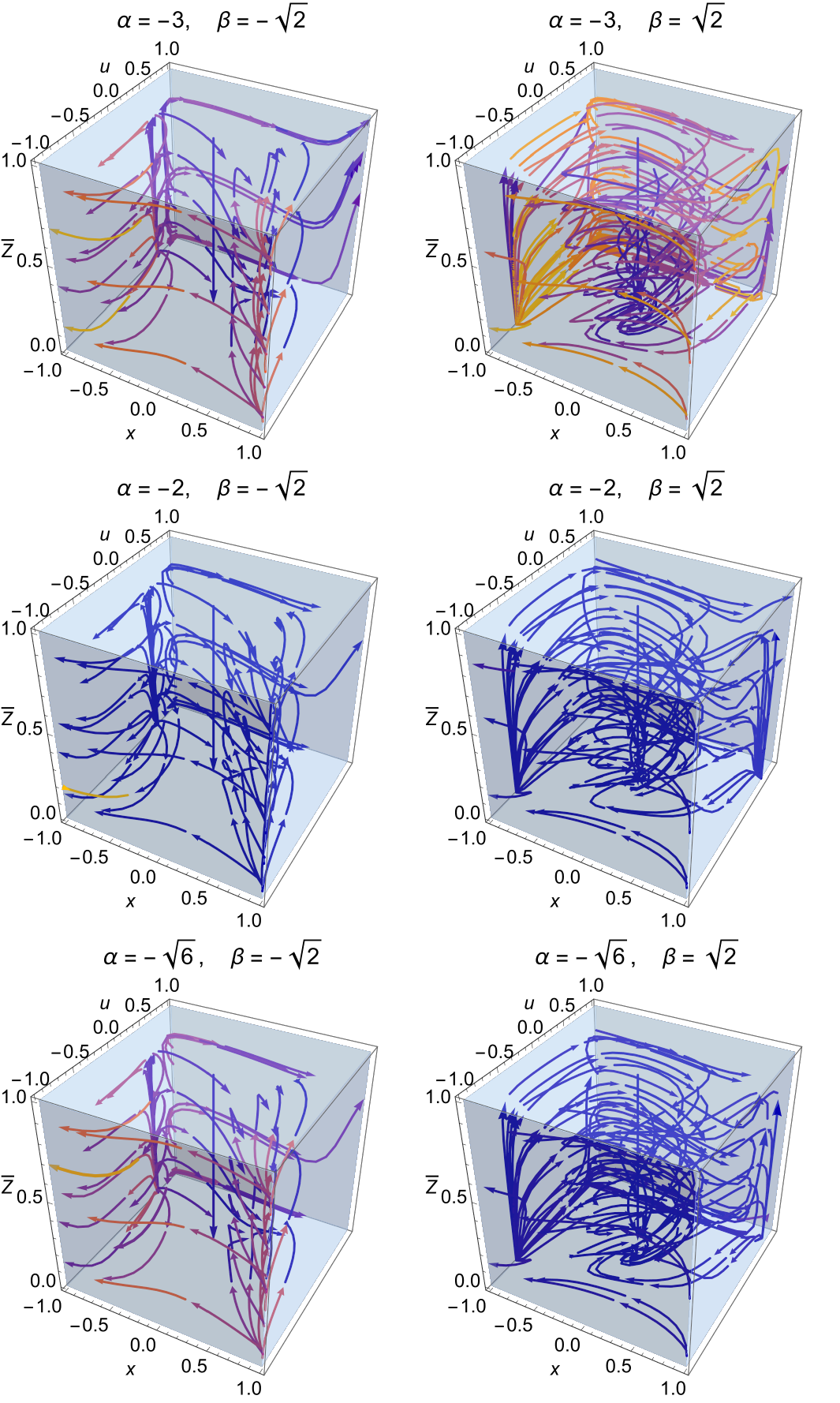}
    \caption{Flow of the system Eq.~\eqref{background(eq:99)-double-exp} for different values of $\alpha$ and $\beta$.}
    \label{fig:double-exp-background}
\end{figure}

\begin{figure}
    \centering
    \includegraphics[scale=0.6]{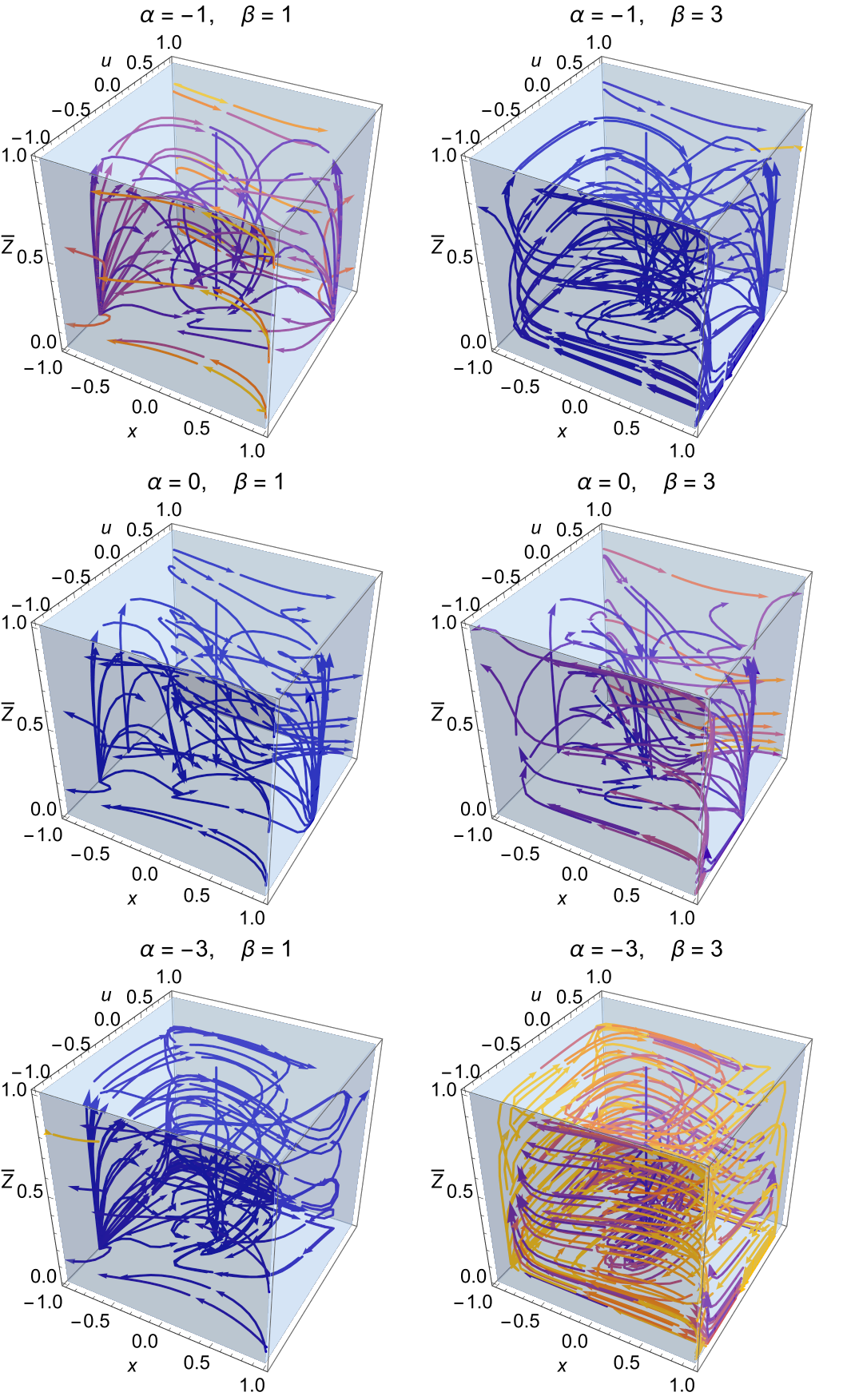}
    \caption{Flow of the system Eq.~\eqref{background(eq:99)-double-exp} for different values of $\alpha$ and $\beta$.}
    \label{fig:double-exp-background2}
\end{figure}
Figs.~\ref{fig:double-exp-background} and  \ref{fig:double-exp-background2}  represent the flow of the system Eq.~\eqref{background(eq:99)-double-exp} for different values of $\alpha$ and $\beta$ in the phase space $(x, u, \bar{Z})$. 

\begin{figure}
    \centering
    \includegraphics[scale=0.6]{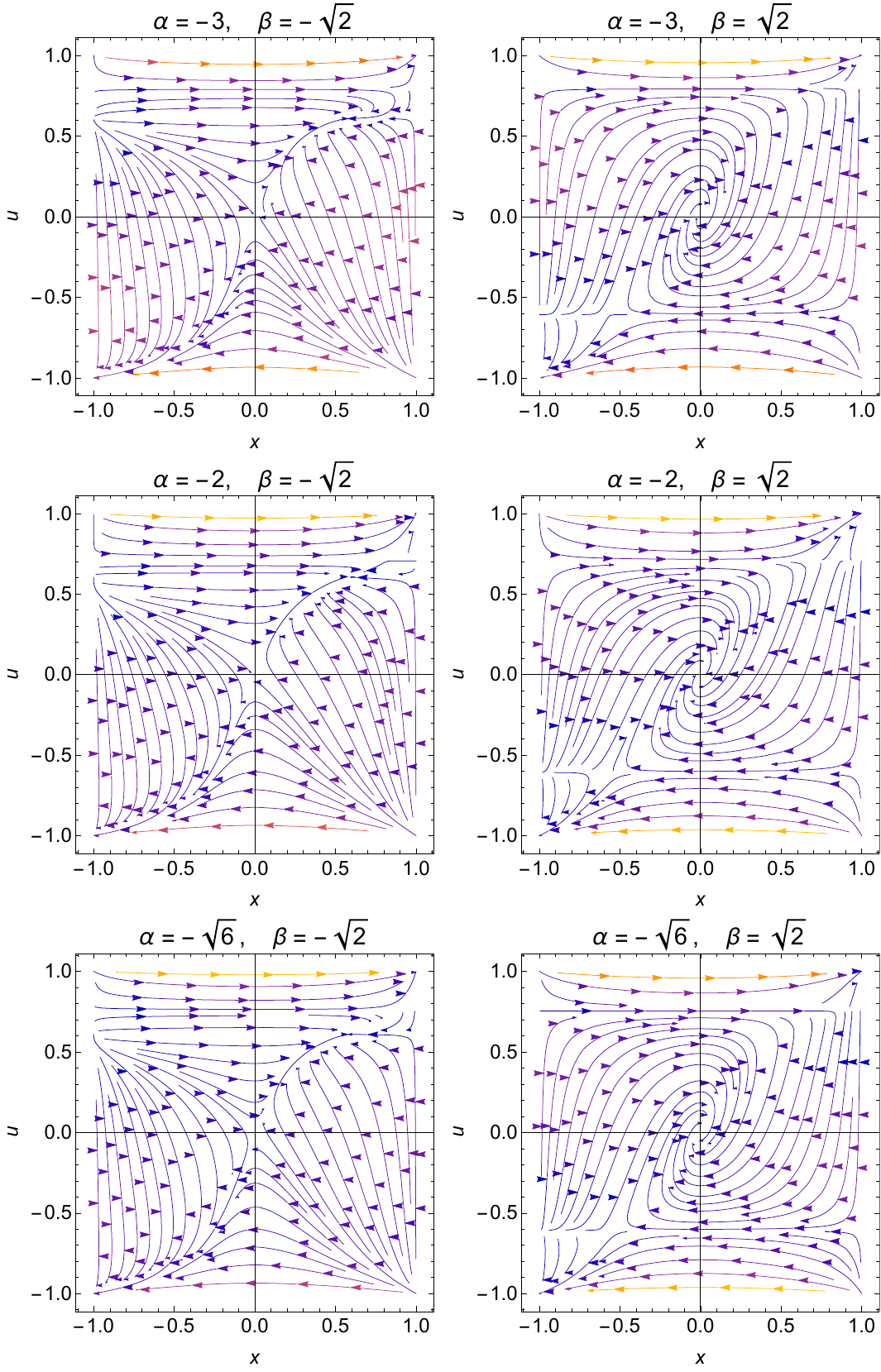}
    \caption{Flow of the system Eq.~\eqref{background(eq:99)-double-exp} in the plane $\left(x, u\right)$ for different values of $\alpha$ and $\beta$.}
    \label{fig:double-exp-background2D}
\end{figure}

\begin{figure}
    \centering
    \includegraphics[scale=0.6]{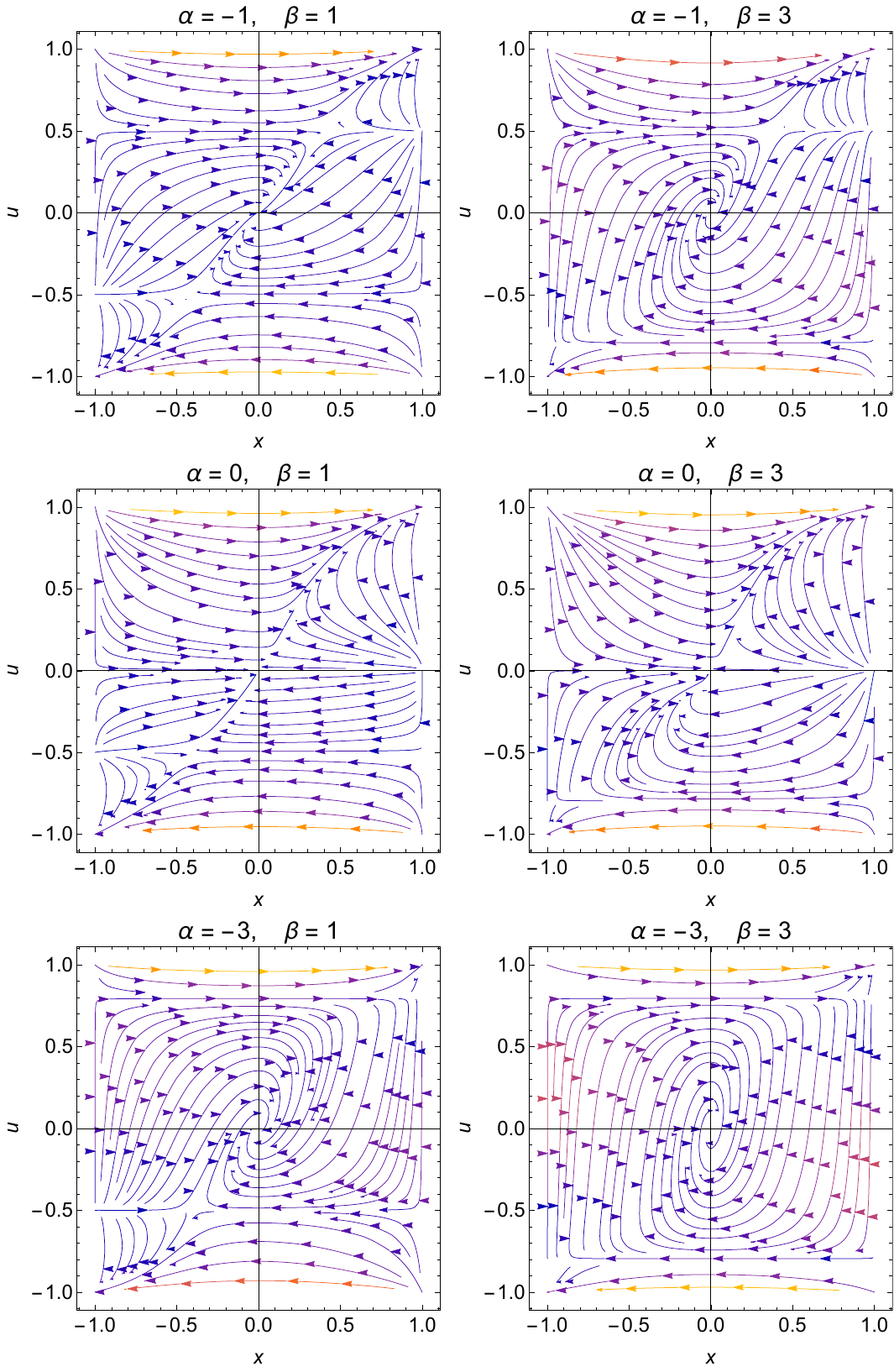}
    \caption{Flow of the system Eq.~\eqref{background(eq:99)-double-exp} in the plane $\left(x, u\right)$ for different values of $\alpha$ and $\beta$.}
    \label{fig:double-exp-background22D}
\end{figure} 
Fig.~\ref{fig:double-exp-background2D} and \ref{fig:double-exp-background22D} represent the flow of the system Eq.~\eqref{background(eq:99)-double-exp} in the plane $\left(x, u\right)$ for different values of $\alpha$ and $\beta$.
These numerical results illustrate the analytical results in Table~\ref{Background_Zb}, therefore, the stability analysis is numerically confirmed.

\subsection{Extended phase space at background and perturbation levels: Bardeen Potential}

In Eq.~\eqref{ptbn_bardeen3}, we first note that ${\Phi}_k$ is generally complex (as it came from a Fourier transformation). 
So, we write ${\Phi}_k=F_1+iF_2$, where $F_1$ and $F_2$ are the real and imaginary parts of ${\Phi}_k$, respectively. Moreover, the resulting equation   has the structure 
\begin{equation}
  F{^{\prime \prime}}+P F^{\prime}+Q F =0 \,\,\text{,} \label{pert-eq}
\end{equation}
where 
\begin{align}
    & P= \left[7-3x^{2}+\sqrt{6}\lambda\left(\frac{1-x^2}{x}\right)\right], \quad  Q= \left[6\left(1-x^2\right)+\sqrt{\frac{3}{2}}\lambda\left(\frac{1-x^2}{x}\right) + \frac{k^2}{a^2H^2}\right],
\end{align}
that is the same for $F_1$ and $F_2$. Generically, we denote $F_i=r_i\cos\theta_i$ and $F^{\prime}_i=r_i\sin\theta_i$ where $i=1,2$. So,
\begin{align}
   F^{\prime}=F  \tan\theta \,\,\text{,} \label{def-theta}
\end{align}
where $\frac{F^{\prime}}{F}=\mathcal{Y}$ has a period of $\pi$. Hence, the mapping $\mathcal{Y}=\tan\theta$ is two-to-one and, therefore, when $\theta$ makes one revolution ($0 \rightarrow  2\pi$)
$\mathcal{Y}$ has to be traversed twice $-\infty \rightarrow +\infty$ \cite{Alho:2020cdg}. Following this line, Eq.~\eqref{pert-eq} can be expressed as 
\begin{equation}
{\mathcal{Y}}{^{\prime}}=-{\mathcal{Y}} ^2  -P {\mathcal{Y}} -Q \label{pert-eq_Y} \,\,\text{,}
\end{equation}
or 
\begin{equation}
 \theta^{\prime} =-\sin^2\theta-P \sin\theta \cos\theta -Q\cos^2\theta \,\,\text{.}
\end{equation}
We also note that it is possible to get $F_i$ from $\mathcal{Y}_i$ through the expression
\begin{align}
    F_i(N)=F_i(0)\exp\left(\int_0^N \mathcal{Y}_i(\Tilde{N})d\Tilde{N}\right) \,\,\text{.}
\end{align}
The sign of $\tan \theta$ denotes whether $F_i(N)$ (for $i=1$ it is the real part) will grow or decay as $\theta$ ranges from $(-\pi,\pi]$.

Therefore, Eq.~\eqref{EQ:(63)}  becomes 
\begin{align}
    Z^{\prime} = 2\left(3x^2 - 1\right)Z \,\,\text{,}
\end{align}
and Eq.~\eqref{ptbn_bardeen3} lead to the \textbf{Bardeen potential $\Phi$} 
\begin{align}\label{theta_bardeen}
   \theta^{\prime} &  = - \sin^{2}\theta - \left[7-3x^{2}+\sqrt{6}\lambda\left(\frac{1-x^2}{x}\right)\right] \sin\theta\cos\theta \nonumber \\
   & - \left[6\left(1-x^2\right)+\sqrt{\frac{3}{2}}\lambda\left(\frac{1-x^2}{x}\right) + Z\right]\cos^{2}\theta \,\,\text{.}
\end{align}

The final equations for the Bardeen potential are the background Eq.~\eqref{(eq:99)} with the perturbation equation
\begin{align}
&\frac{d\theta}{d\bar{N}} = - \Bigg[\sin^2\theta + \left(7-3x^{2}+\sqrt{6}\lambda\left(\frac{1-x^2}{x}\right)\right) \sin\theta\cos\theta
    \nonumber\\
   & \qquad\qquad +\left(6\left(1-x^2\right)+\sqrt{\frac{3}{2}}\lambda\left(\frac{1-x^2}{x}\right)\right)\cos^2\theta\Bigg]\left(1-\bar{Z}\right)  - \bar{Z}\cos^2\theta \,\,\text{,} \label{(eq:99b)}
\end{align}
defined in the phase-space $B\times P$
modulo $n\pi, n\in\mathbb{Z}$, where  the background space is 
\begin{equation}
  B = \left\{\left(x, \lambda, \bar{Z}\right)\in [-1,1]\times \mathbb{R} \times [0,1]\right\}, \label{B-space}
\end{equation}
and the perturbation space is 
\begin{equation}
  P =  \left\{ \theta\in  [-\pi, \pi]\right\}. \label{P-space}
\end{equation}
The scalar field perturbations from Bardeen potentials, the comoving curvature perturbation, and the Sasaki-Mukhanov variable are growing up, decaying or frozen, according to the $\theta$-values. They are classified as super- or sub-horizon perturbations according to whether $\bar{Z}\rightarrow 0$ or  $\bar{Z}\rightarrow 1$. Therefore, we omitted the analysis of non-hyperbolic equilibrium points; the analysis of those points can be done numerically. 

\subsubsection{Sub-horizon boundary}
Recall that the limit $k^2 \mathcal{H}^{-2}\gg 1$ corresponds to the short wavelength or sub-horizon boundary. It is related to the limit $\bar{Z}=1$.  
In this limit Eq.~\eqref{(eq:99)} and Eq.~\eqref{(eq:99b)} become, 
\begin{align}
& \frac{dx}{d\bar{N}} =0 \,\,\text{,}  \;  \frac{d\lambda}{d\bar{N}} = 0 \,\,\text{,} \;    \frac{d \theta}{d \bar{N}} = -\cos^2 \theta \,\,\text{.}\label{(eq:99c)}
\end{align}
Then, we have two asymptotic behaviours as $k^2 \mathcal{H}^{-2}\gg 1$, say there are two sets of equilibrium points with constant $x, \lambda$ and $\theta= \pi/2 + n \pi, n=-1,0$. When $\cos^2 \theta>0$, $\theta$ is monotonically decreasing at constant $x, \lambda$. Then, the invariant set is spanned by a family of heteroclinic cycles with constant $x, \lambda$. They are denoted by $A_{19}$  and $A_{20}$ in Table~\ref{tab:III}. Due to their physical importance, we have distinguished some special points, say $A_{21}(\lambda^*)$, from $A_{30}$. 

At $\theta= \pi/2 + n \pi, n=-1,0$ we have
\begin{equation}
    \frac{d \theta}{d \bar{N}}|_{\cos\theta=0}= -\left(1-\bar{Z}\right), \quad   \frac{d \theta}{d N}|_{\cos\theta=0}= -1 \,\,\text{.}
\end{equation}
which implies that the orbits near $\bar{Z}=1$ shadow the heteroclinic cycles and do not end at the equilibrium points in these cycles. 

Now, passing to the  e-folding time, the stability of the set of equilibrium points on $\bar{Z}=1$ can be examined by analyzing the  fast-slow system 
\begin{subequations}
\label{(eq:99sub-horizon)}
\begin{align}
 \frac{dx}{d{N}}& =-\left(3x - \sqrt{\frac{3}{2}} \lambda\right)\left(1-x^2\right) \,\,\text{,} \\ 
 \frac{d\lambda}{d{N}} & = - \sqrt{6}xf \,\,\text{,} \\
 \frac{d\varepsilon}{d{N}} & = - 2\left(3x^2 - 1\right)\varepsilon \left(1-\varepsilon\right) \,\,\text{,} \\
\varepsilon\frac{d\theta}{d{N}}& = -\cos^2\theta \,\,\text{.}
\end{align}
\end{subequations}
where $0<\left(1-\bar{Z}\right):=\varepsilon \leq 1$. We see that, whenever $q:=(3x^2 - 1)>0$ the perturbation variable $\varepsilon$ monotonically tends to zero, so the surface $\bar{Z}=1$ is approached as $q^*:=\left(3{x^*}^2 - 1\right)>0$ for a fixed value of $x$. On the other hand, for  $q^*<0$ for a fixed value of $x={x^*}$ the perturbation $\varepsilon$ is enhanced and  $\varepsilon\rightarrow 1$, whence, $\bar{Z}\rightarrow 0$. 

The angular variable produces an eigenvalue $-2\left(\cos\theta \sin\theta\right)/\varepsilon$ along the $\theta$-axis, which is zero at $\theta= \pi/2 + n \pi, n=-1,0$ as $\varepsilon\rightarrow 0$. Therefore, at the points with $\bar{Z}=1$, we have, in addition to the eigenvalues presented in Sec .~\ref{sect:A.1}, a zero eigenvalue corresponding to $\theta$. The stability conditions in the background space are the building blocks for the analysis of the extended phase space $B\times P$, modulo $n\pi, n\in\mathbb{Z}$, where  the background space is Eq.~\eqref{B-space}
and the perturbation space is 
Eq.~\eqref{P-space}.

\subsubsection{Stability analysis of the fixed points on the space $B\times P$}

\begin{table}[!ht]
    \centering
    \caption{Equilibrium points of the system Eq.~\eqref{(eq:99)} and Eq.~\eqref{(eq:99b)}. }
      \resizebox{\textwidth}{!}{%
    \begin{tabular}{|c|c|c|c|c|c|c|c|c|c|}
    \hline
    Label & $x$ & $\lambda$ & $\bar{Z}$ &$\theta$ & $k_1$ & $k_2$ & $k_3$ &$k_4$ & $a(t), H(t), \phi(t)$ \\\hline
    $A_{1}(\lambda^{*})$ &$ \frac{{{\lambda^*}}}{\sqrt{6}}$ &$ {{\lambda^*}}$ &$ 0$ & $-\cos ^{-1}(- \Delta_{1} )$ & $ \frac{1}{2} \left(\lambda^{*^2}-6\right)$ &$ \lambda^{*^2}-2 $& $ \Gamma_{1} +\left(8-\frac{3 \lambda^{*^2}}{2}\right) \sin \left(2 \sin ^{-1}( \Delta_{1} )\right) $& $ -{{\lambda^*}} f'({{\lambda^*}})$ & Eq.~\eqref{A1},  Eq.~\eqref{A1b}, Eq.~\eqref{A1c} \\\hline
    $A_{2}(\lambda^{*})$ & $ \frac{{{\lambda^*}}}{\sqrt{6}}$ & ${{\lambda^*}}$ & $0$ & $\cos ^{-1}( \Delta_{1} )$ & $\frac{1}{2} \left(\lambda^{*^2}-6\right)$ & $\lambda^{*^2}-2$ & $ \Gamma_{1} +\left(8-\frac{3 \lambda^{*^2}}{2}\right) \sin \left(2 \cos ^{-1}( \Delta_{1} )\right)$ & $-{{\lambda^*}} f'({{\lambda^*}})$ & Eq.~\eqref{A1},  Eq.~\eqref{A1b}, Eq.~\eqref{A1c}\\\hline
    $A_{3}(\lambda^{*})$ & $ \frac{{{\lambda^*}}}{\sqrt{6}}$ & ${{\lambda^*}}$ & $0$ &$ -\cos ^{-1}( \Delta_{1} )$ & $\frac{1}{2}\left(\lambda^{*^2}-6\right)$ &$ \lambda^{*^2}-2 $& $ \Gamma_{1} +\left(\frac{3 \lambda^{*^2}}{2}-8\right) \sin \left(2 \cos ^{-1}( \Delta_{1} )\right)$ & $-{{\lambda^*}} f'({{\lambda^*}}) $& Eq.~\eqref{A1},  Eq.~\eqref{A1b}, Eq.~\eqref{A1c}\\\hline
    $A_{4}(\lambda^{*})$ &  $\frac{{{\lambda^*}}}{\sqrt{6}}$ & $ {{\lambda^*}}$ & $0$ & $\cos ^{-1}(- \Delta_{1} )$ & $\frac{1}{2}
   \left(\lambda^{*^2}-6\right)$ &$ \lambda^{*^2}-2$ & $ \Gamma_{1} +\left(\frac{3 \lambda^{*^2}}{2}-8\right) \sin \left(2 \sin ^{-1}( \Delta_{1} )\right) $& $ -{{\lambda^*}} f'({{\lambda^*}})$ & Eq.~\eqref{A1},  Eq.~\eqref{A1b}, Eq.~\eqref{A1c} \\\hline
    $A_{5}(\lambda^{*})$ &  $\frac{{{\lambda^*}}}{\sqrt{6}}$ & ${{\lambda^*}}$ &$ 0$ & $-\cos ^{-1}(- \Delta_{2} )$ & $\frac{1}{2}
   \left(\lambda^{*^2}-6\right)$ & $\lambda^{*^2}-2$ &$  \Gamma_{2} +\left(8-\frac{3 \lambda^{*^2}}{2}\right) \sin \left(2 \sin ^{-1}( \Delta_{2} )\right)$ & $-{{\lambda^*}} f'({{\lambda^*}})$ & Eq.~\eqref{A1},  Eq.~\eqref{A1b}, Eq.~\eqref{A1c}\\\hline
    $A_{6}(\lambda^{*})$ & $\frac{{{\lambda^*}}}{\sqrt{6}} $& ${{\lambda^*}}$ & $0$ &$ \cos ^{-1}( \Delta_{2} )$ & $\frac{1}{2}
   \left(\lambda^{*^2}-6\right)$ &$ \lambda^{*^2}-2$ & $ \Gamma_{2} +\left(8-\frac{3 \lambda^{*^2}}{2}\right) \sin \left(2 \cos ^{-1}( \Delta_{2} )\right)$ & $ -{{\lambda^*}} f'({{\lambda^*}}) $ & Eq.~\eqref{A1},  Eq.~\eqref{A1b}, Eq.~\eqref{A1c} \\\hline
    $A_{7}(\lambda^{*})$ &$ \frac{{{\lambda^*}}}{\sqrt{6}}$ &$ {{\lambda^*}} $&$ 0$ & $-\cos ^{-1}( \Delta_{2} )$ & $\frac{1}{2}
   \left(\lambda^{*^2}-6\right)$ & $\lambda^{*^2}-2$ & $ \Gamma_{2} +\left(\frac{3 \lambda^{*^2}}{2}-8\right) \sin \left(2 \cos ^{-1}( \Delta_{2} )\right)$ &$ -{{\lambda^*}} f'({{\lambda^*}})$ & Eq.~\eqref{A1},  Eq.~\eqref{A1b}, Eq.~\eqref{A1c}\\\hline
    $A_{8}(\lambda^{*})$ & $\frac{{{\lambda^*}}}{\sqrt{6}} $& ${{\lambda^*}}$ & $0 $&$ \cos ^{-1}(- \Delta_{2} )$ & $\frac{1}{2}
   \left(\lambda^{*^2}-6\right)$ & $\lambda^{*^2}-2$ &$  \Gamma_{2} +\left(\frac{3 \lambda^{*^2}}{2}-8\right) \sin \left(2 \sin ^{-1}( \Delta_{2} )\right)$ &$ -{{\lambda^*}} f'({{\lambda^*}})$ & Eq.~\eqref{A1}, Eq.~\eqref{A1b}, Eq.~\eqref{A1c}\\\hline
    $A_{9}(\lambda^{*})$ & $-1$ &$ {{\lambda^*}}$ & $0$ & $0$ & $-4$ & $4$ & $\sqrt{6} {{\lambda^*}}+6$ & $\sqrt{6} f'({{\lambda^*}})$ & Eq.~\eqref{A2}, Eq.~\eqref{A2b}, Eq.~\eqref{A2c}\\\hline
    $A_{10}(\lambda^{*})$ & $-1$ & ${{\lambda^*}}$ & $0 $&$ -\pi$  & $-4$ & $4$ & $\sqrt{6} {{\lambda^*}}+6$ & $\sqrt{6} f'({{\lambda^*}})$ & Eq.~\eqref{A2}, Eq.~\eqref{A2b}, Eq.~\eqref{A2c}\\\hline
    $A_{11}(\lambda^{*})$ & $-1$ & ${{\lambda^*}}$ & $0$ & $\pi$  &$ -4$ &$ 4$ & $\sqrt{6} {{\lambda^*}}+6$ & $\sqrt{6} f'({{\lambda^*}})$ & Eq.~\eqref{A2}, Eq.~\eqref{A2b}, Eq.~\eqref{A2c}\\\hline
    $A_{12}(\lambda^{*})$ & $-1$ & ${{\lambda^*}}$ & $0$ & $\sec ^{-1}\left(-\sqrt{17}\right)$ & $4$ & $4$ & $\sqrt{6} {{\lambda^*}}+6$ & $\sqrt{6}
   f'({{\lambda^*}})$ & Eq.~\eqref{A2}, Eq.~\eqref{A2b}, Eq.~\eqref{A2c}\\\hline
    $A_{13}(\lambda^{*})$ & $ -1$ & ${{\lambda^*}}$ & 0 &$ -\sec ^{-1}\left(\sqrt{17}\right)$ & $4$ &$ 4$ & $\sqrt{6} {{\lambda^*}}+6$ & $\sqrt{6}
   f'({{\lambda^*}})$ & Eq.~\eqref{A2}, Eq.~\eqref{A2b}, Eq.~\eqref{A2c}\\\hline
    $A_{14}(\lambda^{*})$ &  $1$ &$ {{\lambda^*}}$ & $0$ & $0$ &$ -4$ & $4$ & $6-\sqrt{6} {{\lambda^*}}$ & $-\sqrt{6} f'({{\lambda^*}})$& Eq.~\eqref{A2}, Eq.~\eqref{A2b}, Eq.~\eqref{A2c}\\\hline
    $A_{15}(\lambda^{*})$ &  $1$ & ${{\lambda^*}}$ & $0$ &$ -\pi$  & $-4$ & $4$ & $6-\sqrt{6} {{\lambda^*}}$ & $-\sqrt{6} f'({{\lambda^*}})$ & Eq.~\eqref{A2}, Eq.~\eqref{A2b}, Eq.~\eqref{A2c}\\\hline
    $A_{16}(\lambda^{*})$ &  $1$ & ${{\lambda^*}}$ &$ 0$ & $\pi$  &$ -4$ & $4 $&$ 6-\sqrt{6} {{\lambda^*}}$ & $-\sqrt{6} f'({{\lambda^*}}) $ & Eq.~\eqref{A2}, Eq.~\eqref{A2b}, Eq.~\eqref{A2c}\\\hline
    $A_{17}(\lambda^{*})$ &  $1$ &$ {{\lambda^*}}$ & $0$ & $\sec ^{-1}\left(-\sqrt{17}\right)$ &$ 4$ & $4$ & $6-\sqrt{6} {{\lambda^*}} $&$ -\sqrt{6}
   f'({{\lambda^*}})$ & Eq.~\eqref{A2}, Eq.~\eqref{A2b}, Eq.~\eqref{A2c}\\\hline
    $A_{18}(\lambda^{*})$ &  $1$ & ${{\lambda^*}}$ &$ 0 $& $-\sec ^{-1}\left(\sqrt{17}\right)$ & $4 $& $4 $& $6-\sqrt{6} {{\lambda^*}}$ &$ -\sqrt{6}
   f'({{\lambda^*}})$ & Eq.~\eqref{A2}, Eq.~\eqref{A2b}, Eq.~\eqref{A2c}\\\hline
    $A_{19}$&  $x_c$ & $\lambda_c$  & $1$ & $-\frac{\pi }{2}$ & $0$ & $0$ & $0$ & $0$ & Eq.~\eqref{CASE-A}, Eq.~\eqref{CASE-Ab}, Eq.~\eqref{CASE-Ac} \\\hline
    $A_{20}$ & $x_c$ & $\lambda_c$ & $1$ & $\frac{\pi }{2}$ & $0$ & $0$ & $0$ & $0$ & Eq.~\eqref{CASE-A}, Eq.~\eqref{CASE-Ab}, Eq.~\eqref{CASE-Ac}\\\hline   
    $A_{21}(\lambda^{*})$ & $\frac{{{\lambda^*}}}{\sqrt{6}}$ & ${{\lambda^*}}$ & $1$ & $-\frac{\pi }{2}$ & $0$ &$0$ &$0$ &$0$ & Eq.~\eqref{A1},  Eq.~\eqref{A1b}, Eq.~\eqref{A1c}\\\hline
    $A_{22}(\lambda^{*})$ & $\frac{{{\lambda^*}}}{\sqrt{6}}$ & ${{\lambda^*}}$ &$ 1$ & $\frac{\pi }{2}$ & $0$ &$0$ &$0$ & $0$ & Eq.~\eqref{A1},  Eq.~\eqref{A1b}, Eq.~\eqref{A1c}\\\hline
    $A_{23}$ &$ -1 $& $\lambda_c $ & $1$ &$ -\frac{\pi }{2}$ & $0$ &$ 0 $& $ 0$ & $0$  & Eq.~\eqref{A2}, Eq.~\eqref{A2b}, Eq.~\eqref{A2c} \\\hline
    $A_{24}$ & $1$ & $\lambda_c$  & $1$ & $-\frac{\pi }{2}$ & $0$ & $0$ & $0 $& $0$ & Eq.~\eqref{A2}, Eq.~\eqref{A2b}, Eq.~\eqref{A2c} \\\hline
    $A_{25}$ &  $ -1$ & $\lambda_c$  & $1$ & $\frac{\pi }{2}$ & $0$ & $0$ & $0$ & $0$ & Eq.~\eqref{A2}, Eq.~\eqref{A2b}, Eq.~\eqref{A2c} \\\hline
    $A_{26}$  & $1$ & $\lambda_c$  & $1$ & $\frac{\pi }{2}$ & $0$ & $0$ &$ 0 $& $0$ & Eq.~\eqref{A2}, Eq.~\eqref{A2b}, Eq.~\eqref{A2c} \\\hline
    $A_{27}$   & $-\frac{1}{\sqrt{3}}$ & $\lambda_c$  & $1 $& $-\frac{\pi }{2}$ & $0$ &$0$ &$0$ &$0$ & Eq.~\eqref{A3}, Eq.~\eqref{A3b}, Eq.~\eqref{A3c}\\\hline
    $A_{28}$ & $\frac{1}{\sqrt{3}}$ & $\lambda_c$  &  $1$ & $-\frac{\pi }{2}$ & $0$ &$0$ &$0$ &$0$ & Eq.~\eqref{A3}, Eq.~\eqref{A3b}, Eq.~\eqref{A3c}\\\hline
    $A_{29}$ & $-\frac{1}{\sqrt{3}}$ & $\lambda_c $ & $1$ & $\frac{\pi }{2}$ & $0$ &$0$ &$0$ &$0$ & Eq.~\eqref{A3}, Eq.~\eqref{A3b}, Eq.~\eqref{A3c}\\\hline
    $A_{30}$ & $\frac{1}{\sqrt{3}}$ & $\lambda_c $ & $1$ & $\frac{\pi }{2}$ & $0$ &$0$ &$0$ &$0$ & Eq.~\eqref{A3}, Eq.~\eqref{A3b}, Eq.~\eqref{A3c}\\\hline
    \end{tabular}}
    \label{tab:III}
\end{table}
In Table~\ref{tab:III} the equilibrium points of the system Eq.~\eqref{(eq:99)} and Eq.~\eqref{(eq:99b)}  are presented, where we denote by $\lambda^*$ any zeroes of $f(\lambda)$ and we define the quantities

\begin{align}
   \Delta_{1,2} & = \frac{2 \sqrt{2}}{\sqrt{9 {{\lambda^*}}^4\pm 3 \left(\sqrt{9 {{\lambda^*}}^4-132 {{\lambda^*}}^2+532} \mp 48\right) {{\lambda^*}}^2\mp 26 \sqrt{9 {{\lambda^*}}^4-132
   {{\lambda^*}}^2+532}+612}} \,\,\text{,}
\end{align}
and 
\begin{align}
 \Gamma_{1,2}&= \frac{6 \left(3
   {{\lambda^*}}^2-26\right) \left(3 {{\lambda^*}}^2-23\right)}{-9 {{\lambda^*}}^4+108 {\lambda^{*}}^2\pm 3 {{\lambda^*}}^2\sqrt{9 {{\lambda^*}}^4-132 {{\lambda^*}}^2+532} \mp 26 \sqrt{9 {\lambda^{*}}^4-132 {{\lambda^*}}^2+532}-320} \,\,\text{.}
\end{align}

These equilibrium points and the stability conditions are summarized as follows.

 $A_{1,2}(\lambda^{*}):\left(\frac{{{\lambda^*}}}{\sqrt{6}}, {{\lambda^*}}, 0, \mp\cos ^{-1}(\mp\Delta_{1} )\right)$ exist for $-\sqrt{6}\leq \lambda^{*}\leq \sqrt{6}$, they are saddles  for $f'(\lambda^*)<0, -\sqrt{2}<\lambda^*<0$, or $f'(\lambda^*)>0, 0<\lambda^*<\sqrt{2}$, or $2 <\lambda^{*^2}<6$ or ${\lambda^{*}} f'({\lambda^{*}})<0$. They are non-hyperbolic otherwise. We have a cosmological solution for these equilibrium points with an asymptotic scale factor Eq.~\eqref{A1}. Since $\frac{{\Phi}'}{\Phi}=\frac{\sqrt{1- \Delta_{1} ^2}}{ \Delta_{1} }$, and $ \Delta_{1}>0$, the amplitude of super-horizon Bardeen potential perturbation grows exponentially. Using the procedures of section \ref{Section-6.3.1}, that is, under the transformation 
\begin{equation}
    \label{transform}
    \Phi_k = a^{-\left(6-\frac{\lambda^{*^2}}{2}\right)}v_k \,\,\text{,}
\end{equation} we obtain the equation  
\begin{equation}
\label{Bessel}
    \frac{d^2 v_k}{d\eta^2} +    v_k\left(k^2-\frac{2 ({\lambda^{*}} -3) ({\lambda^{*}} +3) \left({\lambda^{*}} ^2-6\right)}{\eta ^2 \left({\lambda^{*}} ^2-2\right)^2}\right)=0 \,\,\text{,}
\end{equation}
with solution
\begin{equation}
    \label{sol}
 v_k(\eta)=  C_+ \sqrt{\eta } J_{\nu}(k \eta )+C_- \sqrt{\eta
   } Y_{\nu}(k \eta ) \,\,\text{,}
\end{equation}
where 
\begin{equation}\label{nu}
 \nu=   \frac{\sqrt{9 {\lambda^{*}} ^4-124 {\lambda^{*}} ^2+436}}{2 \left({\lambda^{*}} ^2-2\right)} \,\,\text{,}
\end{equation} and  $C_+$ and $C_-$ are complex constants depending on $k$.

 $A_{3,4}(\lambda^{*}):\left(\frac{{{\lambda^*}}}{\sqrt{6}}, {{\lambda^*}}, 0, \mp\cos ^{-1}(\pm \Delta_{1} )\right)$, with $-\sqrt{6}\leq \lambda^{*}\leq \sqrt{6}$, are sinks for $f'(\lambda^*)<0, -\sqrt{2}<\lambda^*<0$, or $f'(\lambda^*)>0, 0<\lambda^*<\sqrt{2}$.  For the range $2 <\lambda^{*^2}<6$ or ${\lambda^{*}} f'({\lambda^{*}})<0$  they are saddles. They are non-hyperbolic otherwise.      We have a cosmological solution for these equilibrium points with an asymptotic scale factor Eq.~\eqref{A1}. Since $\frac{{\Phi}'}{\Phi}=-\frac{\sqrt{1- \Delta_{1} ^2}}{ \Delta_{1} }$, and $ \Delta_{1}>0$, the amplitude of the super-horizon Bardeen potential perturbation decays exponentially. Introducing the transformation Eq.~\eqref{transform}, we acquire the Bessel equation Eq.~\eqref{Bessel} with solution Eq.~\eqref{sol} where the parameter $\nu$ is defined by Eq.~\eqref{nu}, and $C_+$ and $C_-$ are complex constants depending on $k$.

 $A_{5,6}(\lambda^{*}):\left(\frac{{{\lambda^*}}}{\sqrt{6}}, {{\lambda^*}}, 0, \mp\cos ^{-1}(\mp\Delta_{2} )\right)$,  with $-\sqrt{6}\leq \lambda^{*}\leq \sqrt{6}$, are saddles for $f'(\lambda^*)<0, -\sqrt{2}<\lambda^*<0$, or $f'(\lambda^*)>0, 0<\lambda^*<\sqrt{2}$, or   $2 <\lambda^{*^2}<6$ or ${\lambda^{*}} f'({\lambda^{*}})<0$. They are non-hyperbolic otherwise. We have a cosmological solution for these equilibrium points with an asymptotic scale factor Eq.~\eqref{A1}. Since $\frac{{\Phi}'}{\Phi}=\frac{\sqrt{1- \Delta_{2} ^2}}{ \Delta_{2} }$, and $ \Delta_{2}>0$, the amplitude of super-horizon Bardeen potential perturbation grows exponentially.   Introducing the transformation Eq.~\eqref{transform}, we acquire the Bessel equation given by Eq.~\eqref{Bessel} with solution Eq.~\eqref{sol} where the parameter $\nu$ is defined by Eq.~\eqref{nu}, and $C_+$ and $C_-$ are complex constants depending on $k$. 
   
 $A_{7,8}(\lambda^{*}):\left(\frac{{{\lambda^*}}}{\sqrt{6}}, {{\lambda^*}}, 0, \mp\cos ^{-1}( \pm\Delta_{2} )\right)$,  with $-\sqrt{6}\leq \lambda^{*}\leq \sqrt{6}$, are saddles for $f'(\lambda^*)<0, -\sqrt{2}<\lambda^*<0$, or $f'(\lambda^*)>0, 0<\lambda^*<\sqrt{2}$, or $2 <\lambda^{*^2}<6$ or ${\lambda^{*}} f'({\lambda^{*}})<0$.  They are non-hyperbolic otherwise. We have a cosmological solution for these equilibrium points with an asymptotic scale factor Eq.~\eqref{A1}. Since $\frac{{\Phi}'}{\Phi}=-\frac{\sqrt{1- \Delta_{2} ^2}}{ \Delta_{2} }$, and $ \Delta_{2}>0$, the amplitude of the super-horizon Bardeen potential perturbation decays exponentially. Introducing the transformation Eq.~\eqref{transform}, we acquire the Bessel equation in Eq.~\eqref{Bessel} with solution Eq.~\eqref{sol} where the parameter $\nu$ is defined by Eq.~\eqref{nu}, and $C_+$ and $C_-$ are complex constants depending on $k$.

  $A_{9}(\lambda^*):\left(-1, {{\lambda^*}},0,0\right)$,  $A_{10}(\lambda^{*}):\left(-1, {{\lambda^*}}, 0, -\pi\right)$ and 
  $A_{11}(\lambda^{*}):\left(-1, {\lambda^*}, 0,\pi\right)$ are saddles. We have a cosmological solution for these equilibrium points with an asymptotic scale factor 
Eq.~\eqref{A2}. Since $\frac{{\Phi}'}{\Phi}=0$, the amplitude of the super-horizon Bardeen potential perturbation is frozen. 
Under the transformation 
\begin{equation}
    \label{transform1b}
    \Phi_k = a^{-3}v_k \,\,\text{,}
\end{equation} we obtain the equation  
\begin{equation}
\label{Bessel1b}
    \frac{d^2 v_k}{d\eta^2} +  k^2  v_k =0 \,\,\text{,}
\end{equation}
with solution are given by,
\begin{equation}
    \label{sol1b}
 v_k(\eta)=  C_+ \cos(k \eta )+C_- \sin(k \eta ) \,\,\text{.}
\end{equation}
Where  $C_+$ and $C_-$ are complex constants depending on $k$. 

  $A_{12,13}(\lambda^{*}):\left(-1, {{\lambda^*}}, 0, \pm\sec ^{-1}\left(\mp\sqrt{17}\right)\right)$  are sources for $ {{\lambda^*}}>- \sqrt{6}, 
   f'({{\lambda^*}})>0$. They are saddles for  $ {{\lambda^*}}<- \sqrt{6}$ or  
   $f'({{\lambda^*}})<0$. They are non-hyperbolic otherwise. We have a cosmological solution for this equilibrium point with an asymptotic scale factor
Eq.~\eqref{A2}. Since $\frac{{\Phi}'}{\Phi}=-4$,  the amplitude of the super-horizon Bardeen potential perturbation exponentially decays. Introducing the transformation Eq.~\eqref{transform1b}, we acquire the Eq.~\eqref{Bessel1b} with solution Eq.~\eqref{sol1b} where  $C_+$ and $C_-$ are complex constants depending on $k$.  
    
  $A_{14}(\lambda^{*}):\left(1, {{\lambda^*}}, 0,0\right)$,  $A_{15}(\lambda^{*}):\left(1, {{\lambda^*}}, 0,  -\pi\right)$
 and $A_{16}(\lambda^{*}):\left(1, {{\lambda^*}},  0, \pi\right)$ are saddles. We have a cosmological solution for these equilibrium points with an asymptotic scale factor  
Eq.~\eqref{A2}. Since $\frac{{\Phi}'}{\Phi}=0$, the amplitude of the super-horizon Bardeen potential perturbation is frozen. Introducing the transformation Eq.~\eqref{transform1b}, we acquire the Eq.~\eqref{Bessel1b} with solution Eq.~\eqref{sol1b} where  $C_+$ and $C_-$ are complex constants depending on $k$.     
 
  $A_{17,18}(\lambda^{*}):\left(1,  {{\lambda^*}}, 0, \pm\sec ^{-1}\left(\mp\sqrt{17}\right)\right)$  are sources for $ {{\lambda^*}}< \sqrt{6}, 
   f'({{\lambda^*}})<0$. They are saddles for  $ {{\lambda^*}}> \sqrt{6}$ or  
   $f'({{\lambda^*}})>0$.  They are non-hyperbolic otherwise. We have a cosmological solution for these equilibrium points with an asymptotic scale factor Eq.~\eqref{A2}. Since $\frac{{\Phi}'}{\Phi}=-4$,  the amplitude of the super-horizon Bardeen potential perturbation exponentially decays. Introducing the transformation Eq.~\eqref{transform1b}, we acquire the equation given by Eq.~\eqref{Bessel1b} with solution given by Eq.~\eqref{sol1b} where  $C_+$ and $C_-$ are complex constants depending on $k$.

As we commented,  the invariant set $\bar{Z}=1$ is spanned by a family of heteroclinic cycles with constant $x, \lambda$. They are denoted by       $A_{19}$  and $A_{20}$ in Table~\ref{tab:III}. Due to their physical importance, we have distinguished some special points, say $A_{21}(\lambda^*)$, from $A_{30}$. The eigenvalues of the linearization of system Eq.~\eqref{(eq:99)} are  $0, 0, 0, 0$ at these equilibrium points. Therefore, they are non-hyperbolic.

  $A_{19}:\left(x_c, \lambda_c, 1, -\frac{\pi }{2}\right)$, with $-1\leq x_c\leq 1$. We have a cosmological solution for this equilibrium point with an asymptotic scale factor Eq.~\eqref{CASE-A}. For  ${x_c}=0$ we have a de Sitter expansion with $a(t)= e^{H_0 \left(t-t_U\right)}$.  Since $\frac{{\Phi}'}{\Phi}\rightarrow-\infty$, the amplitude of sub-horizon Bardeen potential perturbation quickly decays.  

Assume $x_c\notin\{0, \pm \sqrt{3}/3\}$, then, under the transformation 
\begin{equation}
    \label{transform19}
    \Phi_k = a^{-\left(6-3 x_c^2\right)}v_k \,\,\text{,}
\end{equation} we obtain the equation,  
\begin{equation}
\label{Bessel19}
    \frac{d^2 v_k}{d\eta^2} +    v_k\left(k^2-\frac{9 (x_c-1) (x_c+1) \left(2 x_c^2-3\right)}{\eta ^2 \left(1-3 x_c^2\right)^2}\right)=0 \,\,\text{,}
\end{equation}
with solution
\begin{equation}
    \label{sol19}
 v_k(\eta)=  C_+ \sqrt{\eta } J_{\nu}(k \eta )+C_- \sqrt{\eta
   } Y_{\nu}(k \eta ) \,\,\text{,}
\end{equation}
where 
\begin{equation}\label{nu19}
 \nu= \frac{\sqrt{81 x_c^4-186 x_c^2+109}}{2-6 x_c^2} \,\,\text{,}
\end{equation} and  $C_+$ and $C_-$ are complex constants depending on $k$.

 $A_{20}:\left(x_c, \lambda_c,1, \frac{\pi }{2}\right)$.   For this equilibrium point, we have a cosmological solution with an asymptotic scale factor 
 Eq.~\eqref{CASE-A}. For  ${x_c}=0$ we have a de Sitter expansion with $a(t)= e^{H_0 \left(t-t_U\right)}$.  Since $\frac{{\Phi}'}{\Phi}\rightarrow \infty$, the amplitude of sub-horizon Bardeen potential perturbation quickly diverges. Introducing the transformation Eq.~\eqref{transform19}, we acquire the Bessel equation given in  Eq.~\eqref{Bessel19} with solution Eq.~\eqref{sol19} where the parameter $\nu$ is defined by Eq.~\eqref{nu19}. 

  $A_{21}(\lambda^{*}):\left(\frac{{{\lambda^*}}}{\sqrt{6}}, {{\lambda^*}}, 1, -\frac{\pi }{2}\right)$,  with $-\sqrt{6}\leq \lambda^{*}\leq \sqrt{6}$. We have a cosmological solution for this equilibrium point with an asymptotic scale factor Eq.~\eqref{A1}. Since $\frac{{\Phi}'}{\Phi}\rightarrow-\infty$, the amplitude of sub-horizon Bardeen potential perturbation quickly decays. Introducing the transformation Eq.~\eqref{transform}, we acquire the Bessel equation in Eq.~\eqref{Bessel} with solution Eq.~\eqref{sol} where the parameter $\nu$ is defined by Eq.~\eqref{nu}, and $C_+$ and $C_-$ are complex constants depending on $k$.  

  $A_{22}(\lambda^{*}):\left(\frac{{{\lambda^*}}}{\sqrt{6}}, {{\lambda^*}}, 1,\frac{\pi }{2}\right)$,  with $-\sqrt{6}\leq \lambda^{*}\leq \sqrt{6}$.   We have a cosmological solution for this equilibrium point with an asymptotic scale factor Eq.~\eqref{A1}. Since $\frac{{\Phi}'}{\Phi}\rightarrow \infty$, the amplitude of sub-horizon Bardeen potential perturbation quickly diverges.   Introducing the transformation Eq.~\eqref{transform}, we acquire the Bessel equation in Eq.~\eqref{Bessel} with solution Eq.~\eqref{sol} where the parameter $\nu$ is defined by Eq.~\eqref{nu}, and $C_+$ and $C_-$ are complex constants depending on $k$.  

 $A_{23}:\left( -1,\lambda_c,1, -\frac{\pi }{2}\right)$ and $A_{24}:\left(1,\lambda_c, 1,-\frac{\pi }{2}\right)$ always exist. We have a cosmological solution for these lines of equilibrium points with an asymptotic scale factor 
Eq.~\eqref{A2}. Since $\frac{{\Phi}'}{\Phi}\rightarrow-\infty$, the amplitude of sub-horizon Bardeen potential perturbation quickly decays.   Introducing the transformation Eq.~\eqref{transform1b}, we acquire the equation in Eq.~\eqref{Bessel1b} with solution Eq.~\eqref{sol1b} where  $C_+$ and $C_-$ are complex constants depending on $k$.  

 $A_{25}:\left(-1,\lambda_c,1,\frac{\pi }{2}\right)$ and $A_{26}:\left(1,\lambda_c,1,\frac{\pi }{2}\right)$ always exist.  We have a cosmological solution for these lines of equilibrium points with an asymptotic scale factor Eq.~\eqref{A2}. Since $\frac{{\Phi}'}{\Phi}\rightarrow \infty$, the amplitude of sub-horizon Bardeen potential perturbation quickly diverges.   Introducing the transformation Eq.~\eqref{transform1b}, we acquire the equation in Eq.~\eqref{Bessel1b} with solution Eq.~\eqref{sol1b} where  $C_+$ and $C_-$ are complex constants depending on $k$.  
 
 $A_{27}:\left(-\frac{1}{\sqrt{3}}, \lambda_c,1, -\frac{\pi }{2}\right)$ and $A_{28}:\left(\frac{1}{\sqrt{3}}, \lambda_c, 1, -\frac{\pi }{2}\right)$ always exist. We have a cosmological solution for these lines of equilibrium points with an asymptotic scale factor Eq.~\eqref{A3}. Since $\frac{{\Phi}'}{\Phi}\rightarrow-\infty$, the amplitude of sub-horizon Bardeen potential perturbation quickly decays. Introducing the new variable Eq.~\eqref{2ptbn_bardeen3-special}, equation in Eq.~\eqref{ptbn_bardeen3-special} becomes Eq.~\eqref{3ptbn_bardeen3-special} with solution Eq.~\eqref{4ptbn_bardeen3-special} where $\epsilon =-1, \lambda=\lambda_c$, and   $C_+$ and $C_-$ are complex constants depending on $k$.  

 $A_{29}:\left(-\frac{1}{\sqrt{3}},\lambda_c ,1,\frac{\pi }{2}\right)$ and $A_{30}:\left(\frac{1}{\sqrt{3}}, \lambda_c, 1,\frac{\pi }{2}\right)$ always exist. We have a cosmological solution for these lines of equilibrium points with an asymptotic scale factor \eqref{A3}. Since $\frac{{\Phi}'}{\Phi}=\tan\left(\frac{\pi}{2}\right)\rightarrow \infty$, the amplitude of sub-horizon Bardeen potential perturbation quickly diverges. 
Introducing the new variable Eq.~\eqref{2ptbn_bardeen3-special}, equation in Eq.~\eqref{ptbn_bardeen3-special} becomes Eq.~\eqref{3ptbn_bardeen3-special} with solution Eq.~\eqref{4ptbn_bardeen3-special} where $\epsilon =1, \lambda=\lambda_c$, and   $C_+$ and $C_-$ are complex constants depending on $k$.

\subsection{Extended phase space at background and perturbation levels: comoving curvature perturbation} 
  
First, note that in the equation in Eq.~\eqref{ptbn_comov}, $\mathcal{R}_k$ is generally complex (as it came from the Fourier transformation). 
So, we write $\mathcal{R}_k=F_1+iF_2$, where $F_1$ and $F_2$ are the real and imaginary parts of $\mathcal{R}_k$, respectively. Substituting  $\mathcal{R}_k=F_1+iF_2$ in  Eq.~\eqref{ptbn_comov}, we would get the same equation for both real and imaginary parts. 
Denoting by $F$ in the equation given by Eq.~\eqref{ptbn_comov},  we get the following,
\begin{align}
& F^{\prime\prime} + \sqrt{6}\lambda\left(\frac{1-x^2}{x}\right)F^{\prime}
+ \left(\frac{k^2}{a^2H^2}\right)F = 0 \,\,\text{.} \label{eq:95}
\end{align}
This equation has the structure of equation  Eq.~\eqref{pert-eq},
where, 
\begin{align}
    & P=\sqrt{6}\lambda\left(\frac{1-x^2}{x}\right), \quad Q=\left(\frac{k^2}{a^2H^2}\right) \,\,\text{.}
\end{align}
As before, note that Eq.~\eqref{eq:95} is a two-degree equation of $f$. We can expect to use phase-space type analysis for this equation.
Using Eq.~\eqref{def-theta} we obtain, 
\begin{align}
   \theta^{\prime} =-\sin^2\theta - P \sin\theta \cos\theta -Q\cos^2\theta \,\,\text{.}
\end{align}
Replacing equation given in Eq.~\eqref{EQ:(63)} in equation $Q$ and considering equation in the Eq.~\eqref{ptbn_comov}, we obtain for the \textbf{Comoving curvature perturbation $\mathcal{R}$} the expression
\begin{equation}\label{theta_comov}
    \theta^{\prime} = - \sin^2\theta - \sqrt{6}\lambda\left(\frac{1-x^2}{x}\right)\sin\theta\cos\theta - Z\cos^2 \theta \,\,\text{.}
\end{equation}

The final equations for the comoving curvature perturbation are given by the background equations given by Eq.~\eqref{(eq:99)} and the perturbation equation, 
\begin{align}
& \frac{d\theta}{d\bar{N}} = - \left[\sin^2\theta + \sqrt{6}\lambda\left(\frac{1-x^2}{x}\right)\sin\theta\cos\theta\right]\left(1-\bar{Z}\right) - \bar{Z}\cos^2 \theta \,\,\text{,} \label{eq100}
\end{align}
defined in the phase-space $B\times P$, modulo $n\pi, n\in\mathbb{Z}$, where  the background space is Eq.~\eqref{B-space}
and the perturbation space is 
Eq.~\eqref{P-space}.

\subsubsection{Sub-horizon boundary}

In  the limit $\bar{Z}=1$, Eq.~\eqref{(eq:99)} and Eq.~\eqref{eq100} becomes Eq.~\eqref{(eq:99c)}. As before,  we have two asymptotic behaviors as $k^2 \mathcal{H}^{-2}\gg 1$, say there are two set of equilibrium points with constant $x, \lambda$ and $\theta= \pi/2 + n \pi, n=-1,0$. When $\cos^2 \theta>0$, $\theta$ is monotonically decreasing at constant $x, \lambda$. Then, the invariant set is spanned by a family of heteroclinic cycles with constant $x, \lambda$. They are denoted by       $B_{14}$  and $B_{15}$ in Table~\ref{tab:IV}. Due to their physical importance, we have distinguished some special points from these sets of equilibrium points ($B_{16}(\lambda^*)$  to $B_{25}$).

\subsubsection{Stability analysis of the fixed points on the space $B\times P$}

\begin{table}[!ht]
    \centering
    \caption{Equilibrium points of system Eq.~\eqref{(eq:99)} and Eq.~\eqref{eq100}.}
      \resizebox{\textwidth}{!}{%
    \begin{tabular}{|c|c|c|c|c|c|c|c|c|c|}
    \hline
 Label & $x$ & $\lambda$ & $\bar{Z}$ &$\theta$ & $k_1$ & $k_2$ & $k_3$ &$k_4$ & $a(t), H(t), \phi(t)$\\\hline
$B_1({\lambda^{*}})$ & $ \frac{{\lambda^{*}}}{\sqrt{6}}$ & ${\lambda^{*}}$ &$ 0$& $-\cos ^{-1}\left(-\frac{1}{\sqrt{(\lambda^{*^2}-6)^2 +1}}\right)$ & $\frac{1}{2}
   \left(\lambda^{*^2}-6\right)$ & $6-\lambda^{*^2}$ & $\lambda^{*^2}-2$ & $-{\lambda^{*}} f'({\lambda^{*}}) $& Eq.~\eqref{A1},  Eq.~\eqref{A1b}, Eq.~\eqref{A1c}\\\hline
$B_2({\lambda^{*}})$ & $\frac{{\lambda^{*}}}{\sqrt{6}}$ & ${\lambda^{*}}$ & $0$ & $\cos ^{-1}\left(\frac{1}{\sqrt{(\lambda^{*^2}-6)^2 +1}}\right)$ & $\frac{1}{2} 
 \left(\lambda^{*^2}-6\right)$ & $6-\lambda^{*^2}$ & $\lambda^{*^2}-2$  & $-{\lambda^{*}} f'({\lambda^{*}})$ & Eq.~\eqref{A1},  Eq.~\eqref{A1b}, Eq.~\eqref{A1c}\\\hline
$B_3({\lambda^{*}})$ &   $ \frac{{\lambda^{*}}}{\sqrt{6}}$ &$ {\lambda^{*}}$ & $0$ & $-\cos ^{-1}\left(\frac{1}{\sqrt{(\lambda^{*^2}-6)^2 +1}}\right)$ &$ \frac{1}{2} \left(\lambda^{*^2}-6\right)$ & $(\lambda^{*^2}-6)\left[\frac{4}{(\lambda^{*^2}-6)^2 +1} -1\right]$ &$ \lambda^{*^2}-2$ &
  $ -{\lambda^{*}} f'({\lambda^{*}}) $& Eq.~\eqref{A1},  Eq.~\eqref{A1b}, Eq.~\eqref{A1c}\\\hline
$B_4({\lambda^{*}})$ & $ \frac{{\lambda^{*}}}{\sqrt{6}}$ & ${\lambda^{*}}$ & $0$ & $\cos ^{-1}\left(-\frac{1}{\sqrt{(\lambda^{*^2}-6)^2 +1}}\right)$ & $\frac{1}{2}
   \left(\lambda^{*^2}-6\right)$ &$ (\lambda^{*^2}-6)\left[\frac{4}{(\lambda^{*^2}-6)^2 +1} -1\right]$  & $\lambda^{*^2}-2$& $-{\lambda^{*}} f'({\lambda^{*}})$ & Eq.~\eqref{A1},  Eq.~\eqref{A1b}, Eq.~\eqref{A1c}\\\hline
$B_5({\lambda^{*}})$ &  $\frac{{\lambda^{*}}}{\sqrt{6}}$ & ${\lambda^{*}}$ & $0$ &$ 0$ &$ \frac{1}{2} \left(\lambda^{*^2}-6\right)$ & $\lambda^{*^2}-6$ & $\lambda^{*^2}-2$ &
   $-{\lambda^{*}} f'({\lambda^{*}})$ & Eq.~\eqref{A1},  Eq.~\eqref{A1b}, Eq.~\eqref{A1c}\\\hline
$B_6({\lambda^{*}})$ &  $\frac{{\lambda^{*}}}{\sqrt{6}}$ & ${\lambda^{*}}$ & $0$ & $-\pi$  & $\frac{1}{2} \left(\lambda^{*^2}-6\right) $& $\lambda^{*^2}-6$ &$ \lambda^{*^2}-2$ &
   $-{\lambda^{*}} f'({\lambda^{*}})$ & Eq.~\eqref{A1},  Eq.~\eqref{A1b}, Eq.~\eqref{A1c}\\\hline
$B_7({\lambda^{*}})$ & $\frac{{\lambda^{*}}}{\sqrt{6}}$ & ${\lambda^{*}}$ &$ 0 $& $\pi$  & $\frac{1}{2} \left(\lambda^{*^2}-6\right)$ & $\lambda^{*^2}-6$ & $\lambda^{*^2}-2$ &
   $-{\lambda^{*}} f'({\lambda^{*}})$ & Eq.~\eqref{A1},  Eq.~\eqref{A1b}, Eq.~\eqref{A1c}\\\hline
$B_8({\lambda^{*}})$ &  $-1$ & ${\lambda^{*}}$ &$ 0 $& $0$ & $4$ &$ 0$ & $\sqrt{6} {\lambda^{*}}+6$ & $ \sqrt{6} f'({\lambda^{*}}) $  & Eq.~\eqref{A2}, Eq.~\eqref{A2b}, Eq.~\eqref{A2c} \\\hline
$B_9({\lambda^{*}})$ &  $-1$ & ${\lambda^{*}}$ & $0$ & $-\pi$  & $4$ & $0$ & $\sqrt{6} {\lambda^{*}}+6 $& $\sqrt{6} f'({\lambda^{*}})$  & Eq.~\eqref{A2}, Eq.~\eqref{A2b}, Eq.~\eqref{A2c} \\\hline
$B_{10}({\lambda^{*}})$ &  $-1$ & ${\lambda^{*}}$ &$ 0$ &$ \pi$  & $4$ &$ 0$ & $\sqrt{6} {\lambda^{*}}+6$ & $\sqrt{6} f'({\lambda^{*}})$  & Eq.~\eqref{A2}, Eq.~\eqref{A2b}, Eq.~\eqref{A2c} \\\hline
$B_{11}({\lambda^{*}})$ & $ 1$ & ${\lambda^{*}}$ & $0$ &$ 0$ &$4 $& $0$ &$ 6-\sqrt{6} {\lambda^{*}} $& $-\sqrt{6} f'({\lambda^{*}}) $ & Eq.~\eqref{A2}, Eq.~\eqref{A2b}, Eq.~\eqref{A2c} \\\hline
$B_{12}({\lambda^{*}})$ & $1$ & ${\lambda^{*}}$ & $0$ & $-\pi$  & $4$ &$ 0$ & $6-\sqrt{6} {\lambda^{*}}$ & $-\sqrt{6} f'({\lambda^{*}})$  & Eq.~\eqref{A2}, Eq.~\eqref{A2b}, Eq.~\eqref{A2c} \\\hline
$B_{13}({\lambda^{*}})$ & $ 1 $& ${\lambda^{*}}$ & $0$ & $\pi$  & $4$ & $0$ & $6-\sqrt{6} {\lambda^{*}}$ & $-\sqrt{6} f'({\lambda^{*}})$ & Eq.~\eqref{A2}, Eq.~\eqref{A2b}, Eq.~\eqref{A2c}  \\\hline
$B_{14}$ &$x_c$ & $\lambda_c$  & $1$ & $-\frac{\pi }{2}$ & $0$ & $0$ & $0$ & $0$ & Eq.~\eqref{CASE-A}, Eq.~\eqref{CASE-Ab}, Eq.~\eqref{CASE-Ac}\\\hline
$B_{15}$ &$ x_c $& $\lambda_c$  & $1$ & $\frac{\pi }{2}$ & $0$ & $0$ & $0$ & $0$ & Eq.~\eqref{CASE-A}, Eq.~\eqref{CASE-Ab}, Eq.~\eqref{CASE-Ac}\\\hline
$B_{16}({\lambda^{*}})$ & $ \frac{{\lambda^{*}}}{\sqrt{6}}$ & ${\lambda^{*}}$ &$ 1$ & $-\frac{\pi }{2}$ & $0$ &$0$& $0$ &$0$ & Eq.~\eqref{A1},  Eq.~\eqref{A1b}, Eq.~\eqref{A1c}\\\hline
$B_{17}({\lambda^{*}})$ & $ \frac{{\lambda^{*}}}{\sqrt{6}}$ & $ {\lambda^{*}}$ & $1$ & $\frac{\pi}{2}$ & $0$ & $0$ & $0$ & $0$ & Eq.~\eqref{A1},  Eq.~\eqref{A1b}, Eq.~\eqref{A1c}\\\hline
$B_{18}$ & $-1$ & $\lambda_c$  & $1$ &$ -\frac{\pi }{2}$ & $0$ &$ 0$ & $0$ &$ 0$ & Eq.~\eqref{A2}, Eq.~\eqref{A2b}, Eq.~\eqref{A2c} \\\hline
$B_{19}$ & $1 $& $\lambda_c$  & $1$ & $-\frac{\pi }{2}$ & $0$ & $0$ & $0$ &$ 0$ & Eq.~\eqref{A2}, Eq.~\eqref{A2b}, Eq.~\eqref{A2c} \\\hline
$B_{20}$ & $-1$ & $\lambda_c$  & $1$ & $\frac{\pi }{2}$ & $0$ &$ 0$ &$ 0$ &$ 0$ & Eq.~\eqref{A2}, Eq.~\eqref{A2b}, Eq.~\eqref{A2c} \\\hline
$B_{21}$ & $1$ & $\lambda_c$  & $1$ & $\frac{\pi }{2}$ & $0$ & $0$ &$ 0$ &$ 0$ & Eq.~\eqref{A2}, Eq.~\eqref{A2b}, Eq.~\eqref{A2c} \\\hline
$B_{22}$ & $ -\frac{1}{\sqrt{3}}$ & $\lambda_c $ &$1$ &$ -\frac{\pi }{2}$ & $0 $& $0$ & $0$ &$ 0$ & Eq.~\eqref{A3}, Eq.~\eqref{A3b}, Eq.~\eqref{A3c}\\\hline
$B_{23}$ & $ \frac{1}{\sqrt{3}}$ & $\lambda_c$  & $1$ & $-\frac{\pi }{2}$ &$0$ & $0$ &$ 0 $& $0$ & Eq.~\eqref{A3}, Eq.~\eqref{A3b}, Eq.~\eqref{A3c}\\\hline
$B_{24}$ & $-\frac{1}{\sqrt{3}}$ & $\lambda_c$  & $1$ & $\frac{\pi }{2}$ & $0 $& $0$ & $0 $& $0$ & Eq.~\eqref{A3}, Eq.~\eqref{A3b}, Eq.~\eqref{A3c}\\\hline
$B_{25}$ & $ \frac{1}{\sqrt{3}}$ & $\lambda_c $ & $1 $& $\frac{\pi }{2}$ & $0$ &$0$ & $0$ & $0$ & Eq.~\eqref{A3}, Eq.~\eqref{A3b}, Eq.~\eqref{A3c}\\\hline
    \end{tabular}}
    \label{tab:IV}
 \end{table}

In Table~\ref{tab:IV}, the equilibrium points of system Eq.~\eqref{(eq:99)} and Eq.~\eqref{eq100} are presented.

These equilibrium points and the stability conditions are summarized as follows.

 $B_{1,2}({\lambda^{*}}):\left(\frac{{\lambda^{*}}}{\sqrt{6}}, {\lambda^{*}}, 0, \mp\cos ^{-1}\left(\mp\frac{1}{\sqrt{(\lambda^*-6)^2 +1}}\right)\right)$ exist for $-\sqrt{6}\leq \lambda^{*}\leq \sqrt{6}$. They are saddles. For a perturbation $k$-mode, this corresponds to the super-horizon limit of a cosmology with an asymptotic scale factor Eq.~\eqref{A1}. 
   Since $\frac{\mathcal{R}'}{\mathcal{R}}=|\lambda^*-6|$, the amplitude of super-horizon comoving curvature perturbation is exponentially increasing.
   
    Using the procedures of section \ref{Section-6.3.2}, that is, under the transformation 
\begin{equation}
    \label{transform2}
   \mathcal{R}_k = a^{-\left(\frac{5}{2}-\frac{1}{4}{\lambda^{*^2}}\right)}v_k \,\,\text{,}
\end{equation} we obtain the equation, 
\begin{equation}
\label{Bessel2}
    \frac{d^2 v_k}{d\eta^2} +    v_k\left(k^2-\frac{\left({\lambda^{*}}^2-10\right) \left(3 {\lambda^{*}}^2-10\right)}{4 \eta ^2 \left({\lambda^{*}}^2-2\right)^2}\right)=0 \,\,\text{,}
\end{equation}
with solution,
\begin{equation}
    \label{sol2}
 v_k(\eta)=  C_+ \sqrt{\eta } J_{\nu}(k \eta )+C_- \sqrt{\eta
   } Y_{\nu}(k \eta ) \,\,\text{,}
\end{equation}
where, 
\begin{equation}\label{nu2}
 \nu=  \frac{\sqrt{{\lambda^{*}} ^4-11 {\lambda^{*}} ^2+26}}{{\lambda^{*}} ^2-2} \,\,\text{,}
\end{equation}
and $C_+$ and $C_-$ are complex constants depending on $k$. 

 $B_{3,4}({\lambda^{*}}):\left(\frac{{\lambda^{*}}}{\sqrt{6}},   {\lambda^{*}}, 0, \mp\cos^{-1}\left(\pm \frac{1}{\sqrt{(\lambda^*-6)^2 +1}}\right)\right)$ exist for $-\sqrt{6}\leq \lambda^{*}\leq \sqrt{6}$. They are saddles for $f'(\lambda^*)<0, -\sqrt{2}<\lambda^*<0$, or $f'(\lambda^*)>0, 0<\lambda^*<\sqrt{2}$ or $2 <\lambda^{*^2}<6$ or ${\lambda^{*}} f'({\lambda^{*}})<0$. They are non-hyperbolic otherwise. For a perturbation $k$-mode, this corresponds to the super-horizon limit of a cosmology with an asymptotic scale factor Eq.~\eqref{A1}.    
   Since $\frac{\mathcal{R}'}{\mathcal{R}}=-|\lambda^*-6|$, the amplitude of super-horizon comoving curvature perturbation is exponentially decreasing. Introducing the transformation Eq.~\eqref{transform2}, we acquire the Bessel equation in Eq.~\eqref{Bessel2} with solution in Eq.~\eqref{sol2} where the parameter $\nu$ is defined by Eq.~\eqref{nu2}, and $C_+$ and $C_-$ are complex constants depending on $k$. 

 $B_5({\lambda^{*}}):\left(\frac{{\lambda^{*}}}{\sqrt{6}}, {\lambda^{*}}, 0,  0\right)$, $B_6({\lambda^{*}}):\left(\frac{{\lambda^{*}}}{\sqrt{6}},{\lambda^{*}},0, -\pi\right)$ and $B_7({\lambda^{*}}):\left(\frac{{\lambda^{*}}}{\sqrt{6}}, {\lambda^{*}}, 0, \pi\right)$, exist for $-\sqrt{6}\leq \lambda^{*}\leq \sqrt{6}$. They are sinks for $f'(\lambda^*)<0, -\sqrt{2}<\lambda^*<0$, or $f'(\lambda^*)>0, 0<\lambda^*<\sqrt{2}$. They are saddles for $2 <\lambda^{*^2}<6$ or ${\lambda^{*}} f'({\lambda^{*}})<0$.   They are non-hyperbolic otherwise.  
   For a perturbation $k$-mode, this corresponds to the super-horizon limit of cosmology with an asymptotic scale factor Eq.~\eqref{A1}. Since $\frac{\mathcal{R}'}{\mathcal{R}}=0$, the amplitude of super-horizon comoving curvature perturbation is frozen. Introducing the transformation Eq.~\eqref{transform2}, we acquire the Bessel equation in Eq.~\eqref{Bessel2} with solution in Eq.~\eqref{sol2} where the parameter $\nu$ is defined by Eq.~\eqref{nu2}, and $C_+$ and $C_-$ are complex constants depending on $k$.
  
 $B_8({\lambda^{*}}): \left(-1, {\lambda^{*}},   0 ,  0\right)$, $B_9({\lambda^{*}}):\left(-1, {\lambda^{*}}, 0, -\pi\right)$ 
and $B_{10}({\lambda^{*}}):\left(-1, {\lambda^{*}},   0,   \pi\right)$  are non-hyperbolic with a three-dimensional unstable manifold for  $\lambda^{*}>-\sqrt{6}$ and $f'({\lambda^{*}})>0$. For a perturbation $k$-mode, this corresponds to the super-horizon limit with an asymptotic scale factor Eq.~\eqref{A2}. Since $\frac{\mathcal{R}'}{\mathcal{R}}=0$, the amplitude of super-horizon comoving curvature perturbation is frozen. Under the transformation, 
\begin{equation}
    \label{transform18}
     \mathcal{R}_k = a^{-1}v_k \,\,\text{,}
\end{equation} we obtain the equation,  
\begin{equation}
\label{Bessel18}
    \frac{d^2 v_k}{d\eta^2} +    v_k\left(k^2+\frac{1}{2 \eta ^2}\right)=0 \,\,\text{,}
\end{equation}
with solution,
\begin{equation}
    \label{sol18}
 v_k(\eta)=  \sqrt{\eta } \left(C_+ J_{\frac{i}{2}}(k \eta )+C_- Y_{\frac{i}{2}}(k \eta )\right) \,\,\text{,}
\end{equation}
where  $C_+$ and $C_-$ are complex constants depending on $k$.
 
 $B_{11}({\lambda^{*}}):\left(1, {\lambda^{*}}, 0,   0\right)$,  $B_{12}({\lambda^{*}}):\left(1, {\lambda^{*}}, 0, -\pi\right)$ 
 and $B_{13}({\lambda^{*}}):\left(1 ,  {\lambda^{*}}, 0, \pi\right)$ are non-hyperbolic with a three-dimensional unstable manifold for  $\lambda^{*}<\sqrt{6}$ and $f'({\lambda^{*}})<0$. For a perturbation $k$-mode, this corresponds to the super-horizon limit of a cosmology of the form Eq.~\eqref{A2}. Since $\frac{\mathcal{R}'}{\mathcal{R}}=0$, the amplitude of super-horizon comoving curvature perturbation is frozen. Introducing the transformation Eq.~\eqref{transform18}, we acquire the Bessel equation in Eq.~\eqref{Bessel18} with solution in Eq.~\eqref{sol18}, and $C_+$ and $C_-$ are complex constants depending on $k$.

As we commented,  the invariant set $\bar{Z}=1$ is spanned by a family of heteroclinic cycles with constant $x, \lambda$. They are denoted by       $B_{14}$  and $B_{15}$ in Table~\ref{tab:IV}. Due to their physical importance, we have distinguished some special points from these sets of equilibrium points ($B_{16}(\lambda^*)$  to $B_{25}$). The eigenvalues of the linearisation of system Eq.~\eqref{eq100} are  $0, 0, 0, 0$ at those equilibrium points. Therefore, they are non-hyperbolic.
 
 $B_{14}:(x_c,\lambda_c,1,-\frac{\pi}{2})$, with $-1\leq x_c\leq 1$. For a perturbation $k$-mode, this corresponds to the sub-horizon limit with an asymptotic scale factor Eq.~\eqref{CASE-A}. For  ${x_c}=0$ we have a de Sitter expansion with $a(t)= e^{H_0 \left(t-t_U\right)}$.  Since $\frac{\mathcal{R}'}{\mathcal{R}}\rightarrow-\infty$, the amplitude of sub-horizon comoving curvature perturbation decays quickly. 
Assume $x_c\notin\{0, \pm \sqrt{3}/3\}$, then, under the transformation, 
\begin{equation}
    \label{transform14}
     \mathcal{R}_k = a^{-\frac{1}{2} \left(5-3 x_c^2\right)}v_k \,\,\text{,}
\end{equation} we obtain the equation,  
\begin{equation}
\label{Bessel14}
    \frac{d^2 v_k}{d\eta^2} +    v_k\left(k^2-\frac{\left(3 x_c^2-5\right) \left(9 x_c^2-5\right)}{4 \eta ^2 \left(1-3 x_c^2\right)^2}\right)=0 \,\,\text{,}
\end{equation}
with solution,
\begin{equation}
    \label{sol14}
 v_k(\eta)=  C_+ \sqrt{\eta } J_{\nu}(k \eta )+C_- \sqrt{\eta
   } Y_{\nu}(k \eta ) \,\,\text{,}
\end{equation}
where, 
\begin{equation}\label{nu14}
 \nu=\frac{\sqrt{9 x_c^4-\frac{33 x_c^2}{2}+\frac{13}{2}}}{1-3 x_c^2} \,\,\text{,}
\end{equation} and  $C_+$ and $C_-$ are complex constants depending on $k$.

   $B_{15}:(x_c,\lambda_c,1,\frac{\pi}{2})$, with $-1\leq x_c\leq 1$. For a perturbation $k$-mode, this corresponds to the sub-horizon limit with an asymptotic scale factor Eq.~\eqref{CASE-A}. For  ${x_c}=0$ we have a de Sitter expansion with $a(t)= e^{H_0 \left(t-t_U\right)}$. Since $\frac{\mathcal{R}'}{\mathcal{R}}\rightarrow\infty$, the amplitude of sub-horizon comoving curvature perturbation diverges quickly. Introducing the transformation Eq.~\eqref{transform14}, we acquire the Bessel equation in Eq.~\eqref{Bessel14} with solution in Eq.~\eqref{sol14} where the parameter $\nu$ is defined in Eq.~\eqref{nu14}. 
 
 $B_{16}({\lambda^{*}}):\left(\frac{{\lambda^{*}}}{\sqrt{6}}, {\lambda^{*}},   1, -\frac{\pi }{2}\right)$, with $-\sqrt{6}\leq \lambda^{*}\leq \sqrt{6}$. For a perturbation $k$-mode, this corresponds to the sub-horizon limit with an asymptotic scale factor Eq.~\eqref{A1}. Since $\frac{\mathcal{R}'}{\mathcal{R}}\rightarrow-\infty$, the amplitude of sub-horizon comoving curvature perturbation decays quickly. Introducing the transformation Eq.~\eqref{transform2}, we acquire the Bessel equation in Eq.~\eqref{Bessel2} with solution in Eq.~\eqref{sol2} where the parameter $\nu$ is defined in Eq.~\eqref{nu2}, and $C_+$ and $C_-$ are complex constants depending on $k$.
 
$B_{17}({\lambda^{*}}):\left(\frac{{\lambda^{*}}}{\sqrt{6}},  {\lambda^{*}}, 1, \frac{\pi}{2}\right)$, with $-\sqrt{6}\leq \lambda^{*}\leq \sqrt{6}$. For a perturbation $k$-mode, this corresponds to the sub-horizon limit with an asymptotic scale factor Eq.~\eqref{A1}. Since $\frac{\mathcal{R}'}{\mathcal{R}}\rightarrow\infty$, the amplitude of sub-horizon comoving curvature perturbation diverges quickly. Introducing the transformation Eq.~\eqref{transform2}, we acquire the Bessel equation in Eq.~\eqref{Bessel2} with solution in Eq.~\eqref{sol2} where the parameter $\nu$ is defined in Eq.~\eqref{nu2}, and $C_+$ and $C_-$ are complex constants depending on $k$.
 
  $B_{18}:(-1,\lambda_c,1,-\frac{\pi}{2})$ and  $B_{19}:(1,\lambda_c,1,-\frac{\pi}{2})$. For a perturbation $k$-mode, this corresponds to the sub-horizon limit with an asymptotic scale factor Eq.~\eqref{A2}. Since $\frac{\mathcal{R}'}{\mathcal{R}}\rightarrow-\infty$, the amplitude of sub-horizon comoving curvature perturbation decays quickly. Introducing the transformation given in Eq.~\eqref{transform18}, we acquire the Bessel equation in Eq.~\eqref{Bessel18} with solution in Eq.~\eqref{sol18}, and $C_+$ and $C_-$ are complex constants depending on $k$.
 
 $B_{20}:(-1,\lambda_c,1,\frac{\pi}{2})$ and  $B_{21}:(1,\lambda_c,1,\frac{\pi}{2})$. For a perturbation $k$-mode, this corresponds to the sub-horizon limit with an asymptotic scale factor Eq.~\eqref{A2}. Since $\frac{\mathcal{R}'}{\mathcal{R}}\rightarrow\infty$, the amplitude of sub-horizon comoving curvature perturbation diverges quickly. Introducing the transformation in Eq.~\eqref{transform18}, we acquire the Bessel equation in Eq.~\eqref{Bessel18} with solution in Eq.~\eqref{sol18}, and $C_+$ and $C_-$ are complex constants depending on $k$.
 
 $B_{22}:(-\frac{1}{\sqrt{3}},\lambda_c,1,-\frac{\pi}{2})$ and  $B_{23}:(\frac{1}{\sqrt{3}},\lambda_c,1,-\frac{\pi}{2})$. For a perturbation $k$-mode, this corresponds to the sub-horizon limit with an asymptotic scale factor Eq.~\eqref{A3}. Since $\frac{\mathcal{R}'}{\mathcal{R}}\rightarrow-\infty$, the amplitude of sub-horizon comoving curvature perturbation decays quickly. 
Introducing the new variable in Eq.~\eqref{2ptbn_comov3-special}, equation in Eq.~\eqref{ptbn_comov3-special} becomes Eq.~\eqref{3ptbn_comov33-special} with solution Eq.~\eqref{4ptbn_comov3-special} where  $C_+$ and $C_-$ are complex constants depending on $k$.  
 
 $B_{24}:(-\frac{1}{\sqrt{3}},\lambda_c,1,\frac{\pi}{2})$ and  $B_{25}:(\frac{1}{\sqrt{3}},\lambda_c,1,\frac{\pi}{2})$. For a perturbation $k$-mode, this corresponds to the sub-horizon limit with an asymptotic scale factor Eq.~\eqref{A3}. Since $\frac{\mathcal{R}'}{\mathcal{R}}\rightarrow\infty$, the amplitude of sub-horizon comoving curvature perturbation diverges quickly. Introducing the new variable Eq.~\eqref{2ptbn_comov3-special}, equation Eq.~\eqref{ptbn_comov3-special} becomes Eq.~\eqref{3ptbn_comov33-special} with solution Eq.~\eqref{4ptbn_comov3-special} where  $C_+$ and $C_-$ are complex constants depending on $k$.

\subsection{Extended phase space at background and perturbation levels: Sasaki-Mukhanov variable}

First, note that $\varphi_{ck}$ in Eq.~\eqref{eq:Uggla} is generally complex (since it comes from the Fourier transformation). 
So, we write $\varphi_{ck}=F_1+iF_2$, where $F_1$ and $F_2$ are the real and imaginary parts of ${\Phi}_k$, respectively. Following the same procedures as before, the resulting equation  has the structure of equation Eq.~\eqref{pert-eq}, where 
\begin{equation}
 P=3 \left( 1-x^2\right) \,\,\text{,}  \quad  Q=18\left(1-x^2\right) \left[  \frac{f}{6}+\left(x-\frac{\lambda }{\sqrt{6}}\right)^2\right] +\frac{k^2}{a^2H^2} \,\,\text{,}
\end{equation}
that is the same for $F_1$ and $F_2$.

As before, note that Eq.~\eqref{eq:Uggla} is a two-degree equation of $f$, so we can expect to use phase space type analysis for this equation. Using Eq.~\eqref{def-theta} 
we obtain 
\begin{align}
   \theta^{\prime} =-\sin^2\theta - P \sin\theta \cos\theta -Q\cos^2\theta \,\,\text{,}
\end{align}
where the replacement of Eq.~\eqref{EQ:(63)} in $Q$ leads to
\begin{align}
     \theta^{\prime} & =-\sin^2\theta - 3\left(1-x^2\right)\sin\theta\cos\theta \nonumber \\
     & - 18\left(1-x^2\right) \left[  \frac{f}{6}+\left(x-\frac{\lambda }{\sqrt{6}}\right)^2\right]\cos^2\theta - Z
\cos^2\theta \,\,\text{,}
\end{align}
with $f\equiv \lambda^2(\Gamma-1)$. For the exponential potential $f\equiv 0$, and with the re-definitions  $\left(\frac{\lambda }{\sqrt{6}}, x\right)\mapsto \left(\lambda, \Sigma_\varphi\right)$ we recover Eq. (30) of \cite{Alho:2020cdg}.

For the scalar field perturbation in the uniform curvature gauge, the evolution of background quantities and perturbations leads to a dynamical system given by 
the background equations Eq.~\eqref{(eq:99)} and  the perturbation equation 

\begin{align}
& \frac{d\theta}{d\bar{N}} =- \Bigg[ \sin^2\theta + 3\left(1-x^2\right)\sin\theta\cos\theta \nonumber \\
& + 18\left(1-x^2\right) \left(  \frac{f}{6}+\left(x-\frac{\lambda }{\sqrt{6}}\right)^2 \right)\cos^2\theta\Bigg]\left(1-\bar{Z}\right)   -\bar{Z} \cos^2\theta \,\,\text{,}\label{(eq:103)}
\end{align}

defined in the phase-space $B\times P$, modulo $n\pi, n\in\mathbb{Z}$, where  the background space is Eq.~\eqref{B-space}
and the perturbation space is 
Eq.~\eqref{P-space}.

\subsubsection{Sub-horizon boundary}

Recall that the limit $k^2 \mathcal{H}^{-2}\gg 1$ corresponds to the short wavelength or sub-horizon boundary. It is related to the limit $\bar{Z}=1$.  
In this  limit Eq.~\eqref{(eq:99)} and Eq.~\eqref{(eq:103)} become Eq.~\eqref{(eq:99c)}. As before,  we have two asymptotic behaviours as $k^2 \mathcal{H}^{-2}\gg 1$, say there are two set of equilibrium points with constant $x, \lambda$ and $\theta= \pi/2 + n \pi, n=-1,0$. When $\cos^2 \theta>0$, $\theta$ is monotonically decreasing at constant $x, \lambda$. Then, the invariant set is spanned by a family of heteroclinic cycles with constant $x, \lambda$. They are denoted by $C_{14}$  and $C_{15}$ in Table~\ref{tab:V}. Due to their physical importance, we have distinguished some special points from these sets of equilibrium points ($C_{16}(\lambda^*)$  to $C_{21}$).  

\subsubsection{Stability analysis of the fixed points on the space $B\times P$}

\begin{table}[!ht]
    \centering
    \caption{Equilibrium points of system Eq.~\eqref{(eq:99)} and Eq.~\eqref{(eq:103)}.}
      \resizebox{\textwidth}{!}{%
    \begin{tabular}{|c|c|c|c|c|c|c|c|c|c|}
    \hline
 Label & $x$ & $\lambda$ & $\bar{Z}$ &$\theta$ & $k_1$ & $k_2$ & $k_3$ &$k_4$ & $a(t), H(t), \phi(t)$\\\hline
$C_{1}({\lambda^{*}})$&$ \frac{\lambda ^*}{\sqrt{6}}$ & $\lambda ^* $&$ 0$ &$ -\cos ^{-1}\left(-\frac{2}{\sqrt{\lambda ^{*^4}-12 \lambda ^{*^2}+40}}\right)$ &$ \frac{1}{2}
   \left(\lambda ^{*^2}-6\right)$ & $\lambda ^{*^2}-2 $& $3-\frac{\lambda ^{*^2}}{2}$ & $-\lambda ^* f'\left(\lambda ^*\right) $& Eq.~\eqref{A1},  Eq.~\eqref{A1b}, Eq.~\eqref{A1c}\\\hline
$C_{2}({\lambda^{*}})$ &$ \frac{\lambda ^*}{\sqrt{6}}$ &$ \lambda ^*$ & $0$ &$ \cos ^{-1}\left(\frac{2}{\sqrt{\lambda ^{*^4}-12 \lambda ^{*^2}+40}}\right)$ & $\frac{1}{2} \left(\lambda ^{*^2}-6\right)$ & $\lambda ^{*^2}-2$ & $3-\frac{\lambda ^{*^2}}{2} $& $-\lambda ^* f'\left(\lambda ^*\right)$ & Eq.~\eqref{A1},  Eq.~\eqref{A1b}, Eq.~\eqref{A1c}\\\hline
$C_{3}({\lambda^{*}})$ & $\frac{\lambda ^*}{\sqrt{6}}$ & $\lambda ^*$ & $0 $&$ -\cos ^{-1}\left(\frac{2}{\sqrt{\lambda ^{*^4}-12 \lambda ^{*^2}+40}}\right)$ &$ \frac{1}{2} \left(\lambda ^{*^2}-6\right)$ & $\lambda ^{*^2}-2$ & $-\frac{1}{2} \lambda ^{*^2}+\frac{8 \left(\lambda ^{*^2}-6\right)}{\lambda ^{*^4}-12 \lambda ^{*^2}+40}+3 $& $-\lambda ^* f'\left(\lambda ^*\right)$& Eq.~\eqref{A1},  Eq.~\eqref{A1b}, Eq.~\eqref{A1c} \\\hline
$C_{4}({\lambda^{*}})$& $\frac{\lambda ^*}{\sqrt{6}}$ &$\lambda ^* $& $0$ & $\cos ^{-1}\left(-\frac{2}{\sqrt{\lambda ^{*^4}-12 \lambda ^{*^2}+40}}\right)$ &$ \frac{1}{2} \left(\lambda ^{*^2}-6\right)$ & $\lambda ^{*^2}-2$ & $-\frac{1}{2} \lambda ^{*^2}+\frac{8 \left(\lambda ^{*^2}-6\right)}{\lambda ^{*^4}-12 \lambda ^{*^2}+40}+3$ &$ -\lambda ^* f'\left(\lambda ^*\right)$ & Eq.~\eqref{A1},  Eq.~\eqref{A1b}, Eq.~\eqref{A1c}\\\hline
$C_{5}({\lambda^{*}})$ &$ \frac{\lambda ^*}{\sqrt{6}}$ & $\lambda ^*$& $0$ & $0$ & $\frac{1}{2} \left(\lambda ^{*^2}-6\right)$ & $\frac{1}{2} \left(\lambda ^{*^2}-6\right)$ &$ \lambda ^{*^2}-2$ & $-\lambda ^* f'\left(\lambda ^*\right) $& Eq.~\eqref{A1},  Eq.~\eqref{A1b}, Eq.~\eqref{A1c}\\\hline
$C_{6}({\lambda^{*}})$ & $\frac{\lambda ^*}{\sqrt{6}}$ & $\lambda ^*$ & $0 $& $-\pi$  & $\frac{1}{2} \left(\lambda ^{*^2}-6\right)$ &$ \frac{1}{2} \left(\lambda ^{*^2}-6\right)$ & $\lambda ^{*^2}-2$ &$ -\lambda ^* f'\left(\lambda ^*\right)$& Eq.~\eqref{A1},  Eq.~\eqref{A1b}, Eq.~\eqref{A1c} \\\hline 
$C_{7}({\lambda^{*}})$ &$ \frac{\lambda ^*}{\sqrt{6}}$ & $\lambda ^*$ &$ 0$ & $\pi$  &$ \frac{1}{2} \left(\lambda ^{*^2}-6\right) $& $\frac{1}{2} \left(\lambda ^{*^2}-6\right)$ & $\lambda ^{*^2}-2$ &$ -\lambda ^* f'\left(\lambda ^*\right)$ & Eq.~\eqref{A1},  Eq.~\eqref{A1b}, Eq.~\eqref{A1c}\\\hline
$C_{8}({\lambda^{*}})$& $-1 $& $\lambda ^* $& $0 $&$ 0$ &$ 4$ &$ 0$ & $\sqrt{6} \lambda ^*+6$ &$ \sqrt{6} f'\left(\lambda ^*\right)$ & Eq.~\eqref{A2}, Eq.~\eqref{A2b}, Eq.~\eqref{A2c} \\\hline
$C_{9}({\lambda^{*}})$ & $-1$ & $\lambda ^*$& $0$ & $-\pi $ & $4$ &$0$ &$ \sqrt{6} \lambda ^*+6$ & $\sqrt{6} f'\left(\lambda ^*\right)$& Eq.~\eqref{A2}, Eq.~\eqref{A2b}, Eq.~\eqref{A2c} \\\hline
$C_{10}({\lambda^{*}}) $& $-1$ &$ \lambda ^*$ &$ 0$ & $\pi$  &$ 4$ & $0$ & $\sqrt{6} \lambda ^*+6 $& $\sqrt{6} f'\left(\lambda ^*\right)$& Eq.~\eqref{A2}, Eq.~\eqref{A2b}, Eq.~\eqref{A2c}  \\\hline
$C_{11}({\lambda^{*}})$ &$ 1$ & $\lambda ^*$ & $0$ & $0$ & $4$ &$ 0$ &$ 6-\sqrt{6} \lambda ^*$ & $-\sqrt{6} f'\left(\lambda ^*\right) $& Eq.~\eqref{A2}, Eq.~\eqref{A2b}, Eq.~\eqref{A2c} \\\hline
$C_{12}({\lambda^{*}})$ & $1$ & $\lambda ^* $& $0$ & $-\pi$  &$4$ & $0 $&$ 6-\sqrt{6} \lambda ^*$ & $-\sqrt{6} f'\left(\lambda ^*\right)$ & Eq.~\eqref{A2}, Eq.~\eqref{A2b}, Eq.~\eqref{A2c} \\\hline
$C_{13}({\lambda^{*}}) $& $1$ & $\lambda ^*$ & $0$ &$ \pi$  &$ 4$ &$ 0 $&$ 6-\sqrt{6} \lambda ^*$ &$ -\sqrt{6} f'\left(\lambda ^*\right)$ & Eq.~\eqref{A2}, Eq.~\eqref{A2b}, Eq.~\eqref{A2c} \\\hline
$C_{14}$ & $x_c$ &$ \lambda_c$ & $1$ & $-\frac{\pi }{2}$ & $0$ & $0$ & $0$ & $0$ & Eq.~\eqref{CASE-A}, Eq.~\eqref{CASE-Ab}, Eq.~\eqref{CASE-Ac} \\\hline
$C_{15}$ & $x_c$ & $\lambda_c$ & $1$ &$ \frac{\pi }{2}$ & $0$ &$ 0$ & $0$ &$ 0$ & Eq.~\eqref{CASE-A}, Eq.~\eqref{CASE-Ab}, Eq.~\eqref{CASE-Ac} \\\hline
$C_{16} $& $0$ & $0$ & $1$ & $-\frac{\pi }{2}$ &$ 0$ & $0$ & $0$ & $0$ & $ e^{H_0\left(t-t_U\right)}, H_0, \phi_0$\\\hline
$C_{17}$& $0$ &$ 0$ & $1 $& $\frac{\pi }{2}$ & $0$ & $0$ &$ 0$ &$ 0$ & $ e^{H_0\left(t-t_U\right)}, H_0, \phi_0$ \\\hline
$C_{18}$ & $-\frac{1}{\sqrt{3}}$ & $-\sqrt{2}$ &$ 1$ & $-\frac{\pi }{2}$ & $0$ &$ 0$& $0$ & $0$ & Eq.~\eqref{A3}, Eq.~\eqref{A3b}, Eq.~\eqref{A3c}\\\hline
$C_{19}$ &$ -\frac{1}{\sqrt{3}} $& $-\sqrt{2}$ &$ 1$ & $\frac{\pi }{2}$ & $0$ & $0$ &$ 0$ & $0$ & Eq.~\eqref{A3}, Eq.~\eqref{A3b}, Eq.~\eqref{A3c}\\\hline
$C_{20}$ & $\frac{1}{\sqrt{3}} $& $\sqrt{2}$ & $1$ & $-\frac{\pi }{2}$ & $0$ & $0$ &$ 0$ &$ 0$ & Eq.~\eqref{A3}, Eq.~\eqref{A3b}, Eq.~\eqref{A3c}\\\hline
$C_{21}$ &$ \frac{1}{\sqrt{3}}$ & $\sqrt{2}$ &$ 1$ & $\frac{\pi }{2}$ & $0$ & $0$ & $0 $& $0$ & Eq.~\eqref{A3}, Eq.~\eqref{A3b}, Eq.~\eqref{A3c}\\\hline
    \end{tabular}}
    \label{tab:V}
\end{table}
In Table~\ref{tab:V}, the equilibrium points of system Eq.~\eqref{(eq:99)} and Eq.~\eqref{(eq:103)} are presented.

These equilibrium points and the stability conditions are summarized as follows.

 $C_{1,2}({\lambda^{*}}):\left(\frac{\lambda ^*}{\sqrt{6}}, \lambda ^*, 0, \mp\cos ^{-1}\left(\mp\frac{2}{\sqrt{\lambda ^{*^4}-12 \lambda ^{*^2}+40}}\right)\right)$ exist for $-\sqrt{6}\leq \lambda^{*}\leq \sqrt{6}$, and are saddles. The scale factor has an asymptotic form Eq.~\eqref{A1}. Since $\frac{{\phi_{ck}}'}{\phi_{ck}}= \frac{1}{2} |\lambda ^{*^2}-6| $, the amplitude of super-horizon Sasaki-Mukhanov variable is exponentially increasing.
Using the procedures of section \ref{Section-6.3.3}, that is, under the transformation, 
\begin{equation}
    \label{transform3}
  \varphi_{ck} =  a^{-1}v_k,
\end{equation} we obtain the equation,  
\begin{equation}
\label{Bessel3}
    \frac{d^2 v_k}{d\eta^2} +    v_k\left(k^2 + \eta ^{-2} \left| 1-\frac{{\lambda ^{*2}}}{2}\right|^{-1}\right)=0 \,\,\text{,}
\end{equation}
with solution,
\begin{equation}
    \label{sol3}
 v_k(\eta)=  C_+ \sqrt{\eta } J_{\nu}(k \eta )+C_- \sqrt{\eta
   } Y_{\nu}(k \eta ) \,\,\text{,}
\end{equation}
where, 
\begin{equation}\label{nu3}
 \nu= \frac{1}{2} \sqrt{1-4 \left| 1-\frac{\lambda ^{*2}}{2}\right|^{-1}} \,\,\text{.}
\end{equation}

 $C_{3,4}({\lambda^{*}}):\left(\frac{\lambda ^*}{\sqrt{6}}, \lambda ^*, 0, \mp\cos ^{-1}\left(\pm \frac{2}{\sqrt{\lambda ^{*^4}-12 \lambda ^{*^2}+40}}\right)\right)$ exist for $-\sqrt{6}\leq \lambda^{*}\leq \sqrt{6}$. They are saddles for $f'(\lambda^*)<0, -\sqrt{2}<\lambda^*<0$, or $f'(\lambda^*)>0, 0<\lambda^*<\sqrt{2}$ or $2 <\lambda^{*^2}<6$ or ${\lambda^{*}} f'({\lambda^{*}})<0$. They are non-hyperbolic otherwise. The scale factor has an asymptotic form Eq.~\eqref{A1}. Since $\frac{{\phi_{ck}}'}{\phi_{ck}}=-\frac{1}{2} |\lambda ^{*^2}-6| $, the amplitude of super-horizon Sasaki-Mukhanov variable is exponentially decreasing. Introducing the transformation Eq.~\eqref{transform3}, we acquire the Bessel equation in Eq.~\eqref{Bessel3} with solution in Eq.~\eqref{sol3} where the parameter $\nu$ is defined by Eq.~\eqref{nu3}, and $C_+$ and $C_-$ are complex constants depending on $k$. 

 $C_{5}({\lambda^{*}}): \left(\frac{\lambda ^*}{\sqrt{6}}, \lambda ^*, 0, 0\right)$, $C_{6}({\lambda^{*}}):\left(\frac{\lambda ^*}{\sqrt{6}}, \lambda ^*, 0,-\pi\right)$ and $C_{7}({\lambda^{*}}):\left(\frac{\lambda ^*}{\sqrt{6}}, \lambda ^*,  0, \pi\right)$, exist for $-\sqrt{6}\leq \lambda^{*}\leq \sqrt{6}$. They are sinks for $f'(\lambda^*)<0, -\sqrt{2}<\lambda^*<0$, or $f'(\lambda^*)>0, 0<\lambda^*<\sqrt{2}$. They are saddles for $2 <\lambda^{*^2}<6$ or ${\lambda^{*}} f'({\lambda^{*}})<0$. They are non-hyperbolic otherwise. The scale factor has an asymptotic form Eq.~\eqref{A1}. Since $\frac{{\phi_{ck}}'}{\phi_{ck}}=0$, the amplitude of super-horizon Sasaki-Mukhanov variable is frozen. Introducing the transformation Eq.~\eqref{transform3}, we acquire the Bessel equation in Eq.~\eqref{Bessel3} with solution in Eq.~\eqref{sol3} where the parameter $\nu$ is defined by Eq.~\eqref{nu3}, and $C_+$ and $C_-$ are complex constants depending on $k$. 

 $C_{8}({\lambda^{*}}):\left(-1, \lambda ^*, 0, 0\right)$, $C_{9}({\lambda^{*}}):\left(-1, \lambda ^*, 0, -\pi\right)$ and $C_{10}({\lambda^{*}}):\left(-1, \lambda ^*, 0, \pi\right)$ are non-hyperbolic with a three-dimensional unstable manifold for  $\lambda^{*}>-\sqrt{6}$ and $f'({\lambda^{*}})>0$. The scale factor has an asymptotic form Eq.~\eqref{A2}. Since $\frac{{\phi_{ck}}'}{\phi_{ck}}=0$, the amplitude of super-horizon comoving curvature perturbation is frozen. 

 Using the procedures of section \ref{Section-6.3.3}, that is, under the transformation given in Eq.~\eqref{transform3}, we obtain the equation,  
\begin{equation}
\label{Bessel3b}
    \frac{d^2 v_k}{d\eta^2} +    v_k\left(k^2 + \frac{1}{2 \eta ^2}\right)=0 \,\,\text{,}
\end{equation}
with solution,
\begin{equation}
    \label{sol3b}
 v_k(\eta)=  C_+ \sqrt{\eta } J_{\frac{i}{2}}(k \eta )+C_- \sqrt{\eta } Y_{\frac{i}{2}}(k \eta) \,\,\text{,}
\end{equation}
and $C_+$ and $C_-$ are complex constants depending on $k$.
  
 $C_{11}({\lambda^{*}}):\left(1, \lambda ^*, 0, 0\right)$, $C_{12}({\lambda^{*}}):\left(1, \lambda ^* ,0, -\pi\right)$ and $C_{13}({\lambda^{*}}):\left(1, \lambda ^*, 0, \pi\right)$ are non-hyperbolic with a three-dimensional unstable manifold for  $\lambda^{*}<\sqrt{6}$ and $f'({\lambda^{*}})<0$.  The scale factor has an asymptotic form Eq.~\eqref{A2}. Since $\frac{{\phi_{ck}}'}{\phi_{ck}}=0$, the amplitude of super-horizon Sasaki-Mukhanov variable is frozen.  Introducing the transformation Eq.~\eqref{transform3}, we acquire the Bessel equation given in Eq.~\eqref{Bessel3b} with solution given in Eq.~\eqref{sol3b}, and $C_+$ and $C_-$ are complex constants depending on $k$.

As we commented,  the invariant set $\bar{Z}=1$ is spanned by a family of heteroclinic cycles with constant $x, \lambda$. They are denoted by       $C_{14}$  and $C_{15}$ in Table~\ref{tab:V}. Due to their physical importance, we have distinguished some special points from these sets of equilibrium points ($C_{16}(\lambda^*)$  to $C_{21}$). The eigenvalues of the linearisation of system Eq.~\eqref{(eq:103)} are  $0 0, 0, 0$ at those equilibrium points. Therefore, they are non-hyperbolic.

$C_{14}({\lambda^{*}}):\left(x_c, \lambda ^*, 1, -\frac{\pi }{2}\right)$, with $-1\leq x_c\leq 1$.   The scale factor has the asymptotic form  Eq.~\eqref{CASE-A}. For  ${x_c}=0$ we have a de Sitter expansion with $a(t)= e^{H_0 \left(t-t_U\right)}$. Since $\frac{{\phi_{ck}}'}{\phi_{ck}}\rightarrow-\infty$, the amplitude of sub-horizon Sasaki-Mukhanov variable quickly decays. 
 Introducing the transformation given in Eq.~\eqref{transform3}, we obtain the Bessel equation, 
\begin{align}\label{Bessel3c}
 \frac{d^2 v_k}{d\eta^2}+    v_k  \left(k^2+\eta ^{-2} |3 x_c^2-1|^{-1}\right)=0,
 \end{align} where $x_c \neq0$, with the solution, 
\begin{align}\label{sol3c} 
  v_k(\eta )= C_+ \sqrt{\eta } J_{\nu}(k \eta )+C_- \sqrt{\eta } Y_{\nu}(k \eta ) \,\,\text{,}
\end{align}
where, 
 \begin{equation}\label{nu3c}
   \nu=\frac{1}{2} \sqrt{1- {4}{|3 x_c^2-1|^{-1}}} \,\,\text{.}
 \end{equation}
 $C_+$ and $C_-$ are complex constants depending on $k$. 

 $C_{15}({\lambda^{*}}):\left(x_c, \lambda ^*, 1, \frac{\pi }{2}\right)$, with $-1\leq x_c\leq 1$. The scale factor has the asymptotic form  Eq.~\eqref{CASE-A}. For  ${x_c}=0$ we have a de Sitter expansion with $a(t)= e^{H_0 \left(t-t_U\right)}$.  Since $\frac{{\phi_{ck}}'}{\phi_{ck}}\rightarrow\infty$, the amplitude of sub-horizon Sasaki-Mukhanov variable quickly diverges.   Introducing the transformation Eq.~\eqref{transform3}, we acquire the Bessel equation given in Eq.~\eqref{Bessel3c} with solution in  Eq.~\eqref{sol3c} where the parameter $\nu$ is defined in Eq.~\eqref{nu3c}. 

 $C_{16}:\left(0, 0, 1, -\frac{\pi }{2}\right)$. The scale factor has the asymptotic form $a(t)= e^{H_0 \left(t-t_U\right)}$, which corresponds to de Sitter expansion. Since $\frac{{\phi_{ck}}'}{\phi_{ck}}\rightarrow-\infty$, the amplitude of sub-horizon Sasaki-Mukhanov variable quickly decays. 

 $C_{17}:\left(0, 0, 1, \frac{\pi }{2}\right)$. The scale factor has the asymptotic form $a(t)= e^{H_0 \left(t-t_U\right)}$, which corresponds to de Sitter expansion. Since $\frac{{\phi_{ck}}'}{\phi_{ck}}\rightarrow\infty$, the amplitude of sub-horizon Sasaki-Mukhanov variable quickly diverges. 

 $C_{18}:\left(-\frac{1}{\sqrt{3}}, -\sqrt{2}, 1, -\frac{\pi }{2}\right)$ with $f(-\sqrt{2})=0$. The scale factor has an asymptotic form  Eq.~\eqref{A3}. Since $\frac{{\phi_{ck}}'}{\phi_{ck}}\rightarrow-\infty$, the amplitude of sub-horizon Sasaki-Mukhanov variable quickly decays. 
Introducing the new variable Eq.~\eqref{2eq:Uggla-special}, equation Eq.~\eqref{eq:Uggla-special} becomes Eq.~\eqref{3eq:Ugglaspecial} with solution Eq.~\eqref{4eq:Uggla-special} where  $C_+$ and $C_-$ are complex constants depending on $k$.  

 $C_{19}:\left( -\frac{1}{\sqrt{3}},-\sqrt{2}, 1, \frac{\pi }{2}\right)$  with $f(-\sqrt{2})=0$. The scale factor has an asymptotic form Eq.~\eqref{A3}. Since $\frac{{\phi_{ck}}'}{\phi_{ck}}\rightarrow\infty$, the amplitude of sub-horizon Sasaki-Mukhanov variable quickly diverges.  
Introducing the new variable Eq.~\eqref{2eq:Uggla-special}, equation Eq.~\eqref{eq:Uggla-special} becomes Eq.~\eqref{3eq:Ugglaspecial} with solution Eq.~\eqref{4eq:Uggla-special} where  $C_+$ and $C_-$ are complex constants depending on $k$.  

 $C_{20}:\left(\frac{1}{\sqrt{3}},\sqrt{2}, 1, -\frac{\pi }{2}\right)$  with $f(\sqrt{2})=0$. The scale factor has an asymptotic form  Eq.~\eqref{A3}. Since $\frac{{\phi_{ck}}'}{\phi_{ck}}\rightarrow-\infty$, the amplitude of sub-horizon Sasaki-Mukhanov variable quickly decays. 
Introducing the new variable Eq.~\eqref{2eq:Uggla-special}, equation Eq.~\eqref{eq:Uggla-special} becomes Eq.~\eqref{3eq:Ugglaspecial} with solution Eq.~\eqref{4eq:Uggla-special} where  $C_+$ and $C_-$ are complex constants depending on $k$.   

 $C_{21}:\left(\frac{1}{\sqrt{3}}, \sqrt{2}, 1, \frac{\pi }{2}\right)$  with $f(\sqrt{2})=0$. The scale factor has an asymptotic form Eq.~\eqref{A3}. Since $\frac{{\phi_{ck}}'}{\phi_{ck}}\rightarrow\infty$, the amplitude of sub-horizon Sasaki-Mukhanov variable quickly diverges. 
Introducing the new variable Eq.~\eqref{2eq:Uggla-special}, equation Eq.~\eqref{eq:Uggla-special} becomes Eq.~\eqref{3eq:Ugglaspecial} with solution Eq.~\eqref{4eq:Uggla-special} where  $C_+$ and $C_-$ are complex constants depending on $k$.

\section{Dynamical system analysis of matter perturbations on top of equilibrium points}
\label{sect:new}
To complement our analysis, we investigate cosmological perturbations with two matter components, e.g., a perfect fluid and a scalar field. A widespread practice in the literature focuses on a particular cosmological epoch in which only one matter component is dominant. In that sense, even though not generic, our subsequent analysis remains relevant when the Universe is scalar-field-dominated, e.g., during the early inflationary epoch or the late-time acceleration. 

In a general non-interacting scenario, which includes dust matter and dynamical dark energy, the scalar perturbations in the Newtonian gauge are determined by the equations \cite{Ma:1995ey}: 
\begin{subequations}
\begin{eqnarray}
&&\dot{\delta}_m+\frac{\theta_m}{a}=0 \,\,\text{,}\;  \label{eq:line1} \\
&&\dot{\delta}_\phi+(1+{{w_\phi})\frac{\theta_\phi}{a}+3H(c_{\mathrm{eff}}^{2}-w_\phi)%
\delta_\phi=0 \,\,\text{,}\;}  \label{eq:line2} \\
&&\dot{\theta}_m+H\theta_m-\frac{k^{2}\Phi }{a}=0 \,\,\text{,}\;  \label{eq:line3} \\
&&\dot{\theta}_\phi+H\theta_\phi-\frac{k^{2}c_{\mathrm{eff}}^{2}\delta_\phi}{(1+%
{{w_\phi})a}}-\frac{k^{2}\Phi }{a}=0 \,\,\text{.}\;  \label{eq:line4}
\end{eqnarray}%
Where $k$ is the wavenumber of Fourier modes, and $\Phi$ is 
the scalar metric perturbation assuming zero anisotropic stress, and the dot means derivative with respect to time. Additionally, $\delta_{i} \equiv \delta \rho_i/\rho_i$, $i\in\{m, \phi\}$ are the densities perturbations and $\theta_i$,  $i\in\{m, \phi\}$ are the velocity perturbations \cite{Ma:1995ey}. Furthermore, $c_{\mathrm{eff}}^{2}$ is the effective sound speed of the dark energy perturbations (the corresponding quantity for matter is zero in the dust case), which determines the amount of
dark-energy clustering. Note that the above equations can be simplified by considering the Poisson equation,
which in sub-horizon scales becomes~\cite{Ma:1995ey}:
\end{subequations}
\begin{equation}
-\frac{k^{2}}{a^{2}}\Phi =\frac{3}{2}H^{2}\left[\Omega_m\delta_{m}+\left(1+3c_{\mathrm{%
eff}}^{2}\right)\Omega_\phi\delta_\phi\right]  \,\,\text{.} \label{eq:poisson}
\end{equation}
In the above equations, we have introduced the density parameters, 
\begin{equation}
\Omega_i=\frac{\rho_i}{3 H^2} \,\,\text{.}
\end{equation}
The amount of DE clustering depends on the magnitude of its effective sound
speed $c_{\mathrm{eff}}^{2}$, and for $c_{\mathrm{eff}}^{2}=0$, the dark energy clusters in
a manner similar to dark matter. However, due to the presence of the dark energy
pressure, one may expect that the amplitude of the DE perturbations is
relatively low with respect to that of dark matter. Notice that in the following we
set $c_{\mathrm{eff}}^{2}=0$.

In the current work, we treat DE as a perfect fluid, which implies that the
effective sound speed coincides with the adiabatic sound speed, 
\begin{equation}
c_{\mathrm{a}}^{2}={{w_\phi}-\frac{a d{w_\phi}/da}{3(1+{w_\phi})} \,\,\text{.}\;}
\label{eq:c_a}
\end{equation} 

Now eliminating $\theta $ from the system of equations Eq.~\eqref{eq:line1}, Eq.~\eqref{eq:line2}, Eq.~\eqref{eq:line3} and Eq.~\eqref{eq:line4}, and using $\frac{d}{dt}=aH \frac{d}{da}$, $\frac{d}{dt}=H\frac{d}{d\ln a},\frac{d}{da}=\frac{1}{a}\frac{d}{d\ln a}$, $\frac{d^{2}}{da^{2}}=\frac{d}{da}\left(\frac{1}{a}\frac{d}{d\ln a}\right)=-\frac{1}{a^{2}}\frac{d}{d\ln a}+\mathrm{\;}\frac{1}{a^{2}}\frac{d^{2}}{d\ln a^{2}}=
\frac{1}{a^{2}}\left(\frac{d^{2}}{d\ln a^{2}}-\frac{d}{d\ln a}\right)$, we obtain after some calculations the following second order
differential equations which describe the evolution of matter and dark energy
perturbations respectively: 
\begin{subequations}
\begin{eqnarray}
&&\delta_m^{\prime \prime }+(A_{m}-1)\delta_m^{\prime }+B_{m}\delta_m=\frac{3%
}{2}(\Omega_m\delta_{\mathrm{m }}+\Omega_\phi\delta_\phi) \,\,\text{,}  \label{perteq1} \\
&& \delta_\phi^{\prime \prime } +(A_\phi-1)\delta_\phi^{\prime}+B_\phi\delta_\phi=%
\frac{3}{2}(1+{{w_\phi} )(\Omega_m\delta _m+\Omega_\phi\delta_\phi)} \,\,\text{,}
\label{perteq2}
\end{eqnarray}
where the coefficients are 
\end{subequations}
\begin{eqnarray}
A_m &=&\frac{3}{2}(1-\Omega_\phi{w_\phi}) \,\,\text{,} \;  
B_m =0 \,\,\text{,} \notag \\
A_\phi &=& -3{w_\phi}-\frac{{{w_\phi^{\prime }}}}{1+{{w_\phi}}}+\frac{3}{2}%
(1-\Omega_\phi{{w_\phi})} \,\,\text{,} \; 
B_\phi = -{{w_\phi^{\prime }}+\frac{{w_\phi^{\prime }}{w_\phi}}{1+{w_\phi}}-%
\frac{1}{2}{w_\phi}(1-3\Omega_\phi{w_\phi})} \,\,\text{,}
\end{eqnarray}
where the prime means derivative with respect to $\ln a$.

\subsection{Application to  $\Lambda$CDM}
Let us first examine the simple $\Lambda$CDM example. In this case, and
assuming dust matter, the background equations are 
\begin{align}
& H^{2}=\frac{1}{3}(\rho_m+\Lambda) \,\,\text{,} \\
& {H}'=-\frac{\rho_m}{2H} \,\,\text{,} \\
& {\rho}_m'+3\rho_m=0 \,\,\text{,} 
\end{align}%
and the equation for the matter perturbations is
\begin{eqnarray}
&&\delta_m^{\prime \prime }+ \left(2-\frac{3}{2} \Omega_m \right)\delta_m^{\prime }=\frac{3%
}{2} \Omega_m\delta_{\mathrm{m }} \,\,\text{,}  \label{Aperteq1} 
\end{eqnarray}
where the prime means derivative with respect to $N=\ln a$. 
Using the above equations, and defining the the ratio $U_m=\frac{\delta_m^{\prime }(N)}{\delta_m}$, we obtain the autonomous system:
\begin{subequations}
\label{LCDM-perts}
\begin{align}
&\Omega_m'= 3 (\Omega_{m}-1)\Omega_{m} \,\,\text{,}\quad
U_m'= \frac{3}{2}(U_{m}+1) \Omega_{m}-U_m (U_m+2) \,\,\text{.} 
\end{align}
\end{subequations}
The system is integrable, depending on $N$ and  $\Omega_{m0}$ and $U_{m0}$ are the values of $\Omega_m$ and $U_m$ today ($N=0, a=1$).

 As we will discuss in more detail later, one can think of $U_{m}$ as the phase of the matter
perturbation. If $U_{m}>0$ during the evolution, it follows by
definition that the perturbations $\delta_{m}$ are growing with time
(since $\delta_{m}>0$ and $\delta'_{m}>0$ or $\delta_{m}<0$ and $%
\delta_{m}'<0$), while if $U_{m}<0$ at that time, the
perturbation is decaying. 
Now, we can analyze the stability of the equilibrium points in the plane $(\Omega_m, U_m)$. In this extended phase space, we can see both the background equations' stability and the perturbations' stability.  

There are identified the equilibrium points of the system Eq.~\eqref{LCDM-perts} are:
\begin{enumerate}
\item $P_1: (\Omega_m, U_m)=(0, -2)$. The phase of the perturbations $U_m$ is negative; therefore, the perturbation $\delta_m$ is decaying with time. The eigenvalues of the linear matrix evaluated at the equilibrium point are $\{-3,2\}$; that is, the equilibrium point is a saddle. That corresponds to the Universe dominated by the cosmological constant, with matter perturbations scaling as $\delta_m\propto e^{-2 N}=a^{-2}$ when the equilibrium point is approached along the stable direction.

\item $P_2: (\Omega_m, U_m)=\left(1, -\frac{3}{2}\right)$. The phase of the perturbations $U_m$ is negative; therefore, the perturbation $\delta_m$ is decaying with time. 
 The eigenvalues of the linear matrix evaluated at the equilibrium point are $\left\{3,\frac{5}{2}\right\}$. That is, the equilibrium point is a source. That corresponds to the matter-dominated Universe, with matter perturbations scaling as $\delta_m\propto e^{-\frac{3}{2} N}=a^{-\frac{3}{2}}$ as $N\rightarrow -\infty$ ($a \rightarrow 0$).  

\item $P_3: (\Omega_m, U_m)=(0, 0)$.  The phase of the perturbations $U_m$  is zero. Therefore, the perturbation $\delta_m$ remains constant. The eigenvalues of the linear matrix evaluated at the equilibrium point are $\{-3,-2\}$. Then, this equilibrium point is a sink. That corresponds to the Universe dominated by the cosmological constant with $\delta_m=\text{const.}$ as $N\rightarrow +\infty$ ($a \rightarrow +\infty$).   It is stable in the extended phase space.

\item $P_4:  (\Omega_m, U_m)=(1, 1)$. The phase of the perturbations $U_m$ is positive; therefore, the perturbation $\delta_m$ is growing with time. The eigenvalues of the linear matrix evaluated at the equilibrium point are $\left\{3,-\frac{5}{2}\right\}$. Then, it is a saddle. That corresponds to a matter-dominated Universe, with matter perturbations scaling as $\delta_m\propto e^{-\frac{5}{2} N}=a^{-\frac{5}{2}}$ when the equilibrium point is approached along the stable direction.

\end{enumerate}

\begin{figure*}[t!]
    \centering
    \begin{subfigure}{(a)}
        \centering
        \includegraphics[scale=0.7]{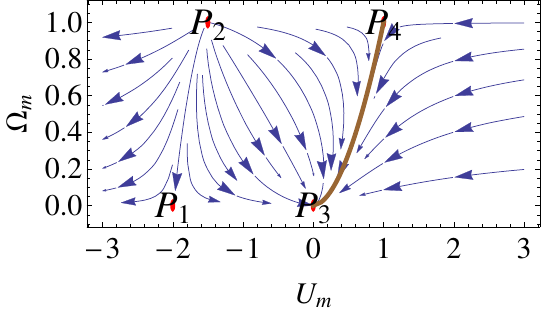}
				           \end{subfigure}%
    ~ 
    \begin{subfigure}{(b)}
        \centering
        \includegraphics[scale=0.7]{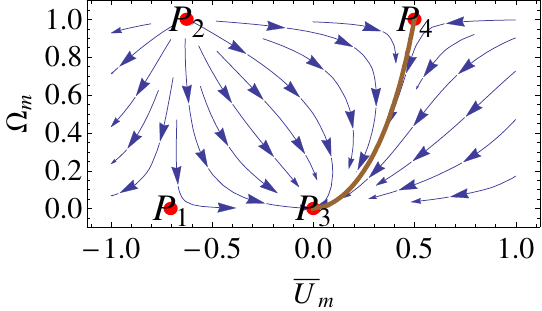}
            \end{subfigure}
    \caption{	\label{fig:LCDM_perts} Phase plane for the system Eq.~\eqref{LCDM-perts} showing both the stability of the background equations as well as the stability of the perturbations for the $\Lambda$CDM model. (a) in the original fractional energy density $\Omega_m$ vs the phase of matter perturbations $U_m$. (b) Using the compact variables $\Omega_m$ vs $\frac{2}{\pi}\arctan(U_m)$.}
\end{figure*}

Fig.~\ref{fig:LCDM_perts} presents a flow for the system Eq.~\eqref{LCDM-perts} showing both the stability of the background equations and the stability of the perturbations for the $\Lambda$CDM model. (a) in the original fractional energy density $\Omega_m$ vs the phase of matter perturbations $U_m$. (b) Using the compact variables $\left(\Omega_m, \frac{2}{\pi}\arctan(U_m)\right)$ 
we see that there are no equilibrium points at infinity.

This numerical elaboration suggests that the unstable manifold of $P_4$ connects $P_4$ and $P_3$, which is stable. To find an estimate of this solution, we proceed as follows. We propose the polytropic law $U_m=\Omega_m^\Gamma$.  Using the equations in Eq.~\eqref{LCDM-perts} we obtain the equation,
\begin{equation}
3 \Gamma (\Omega_m-1) \Omega_m^\Gamma+\left(\Omega_m^\Gamma+2\right) \Omega_m^\Gamma-\frac{3}{2}\Omega_m \left(\Omega_m^\Gamma+1\right)=0 \,\,\text{.}
\end{equation}
It is required that the solution passes through $P_4$, therefore, expanding the above equation in Taylor series around $\Omega_m=1$ we obtain the approximation,
\begin{equation}
-\left(\frac{11 \Gamma}{2}-3\right) (1-\Omega_m)+O\left((1-\Omega_m)^2\right)=0 \,\,\text{.}
\end{equation}
This implies $\Gamma=\frac{6}{11}$. That is, $U_m\simeq \Omega_m^{\frac{6}{11}}$ for the matter dominated Universe. 
This line is represented in the Fig.~\ref{fig:LCDM_perts} by a thick (brown) line and incidentally coincides with the stable manifold of the matter-dominated solution $P_4$.

\begin{figure*}[t!]
    \centering
		\includegraphics[width=1.00\textwidth]{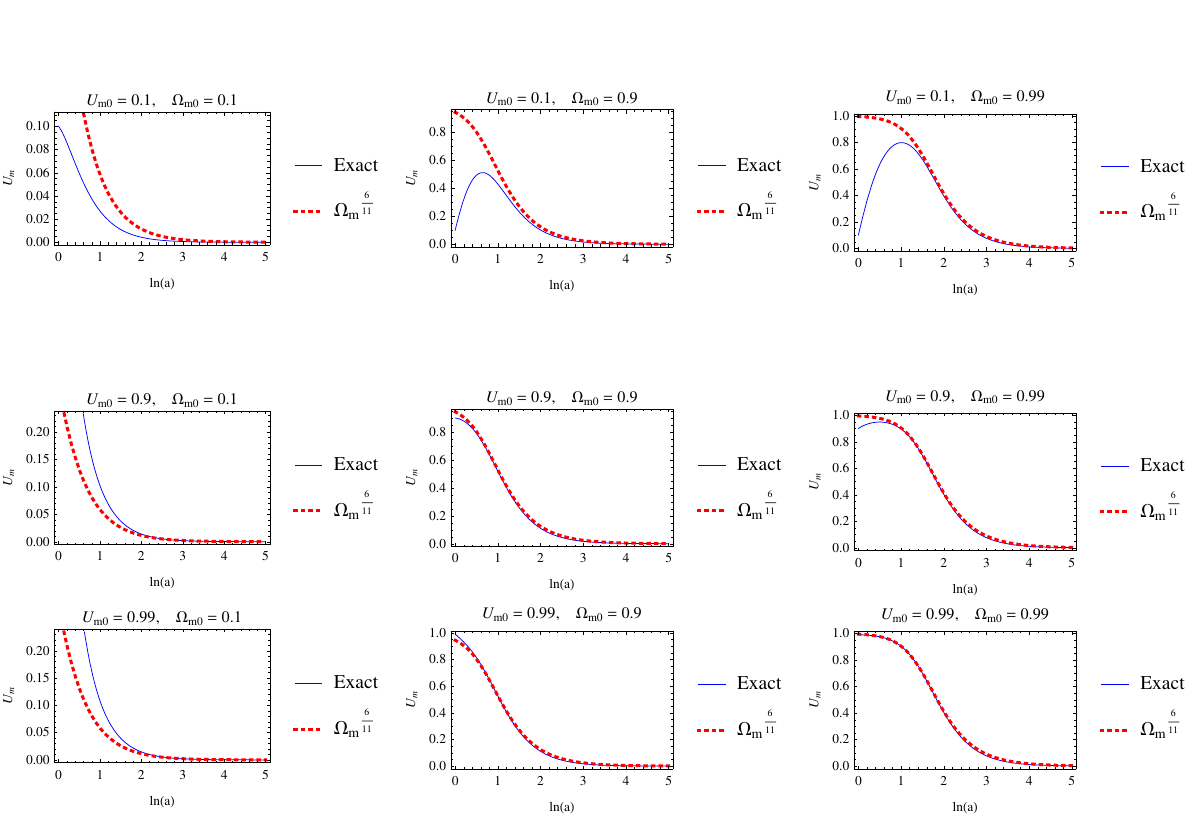}
	\caption{	\label{fig:Exact-vs-App} $U_m$ vs $\ln a$. The plot shows the exact solution  (solid blue line) and the approximated solution $U_m=\Omega_m^{\frac{6}{11}}$ (the red dotted line) for different values of the initial conditions $U_{m0}, \Omega_{m0}$. The closer the initial conditions to the equilibrium point $P_4$, the more accurate the approximation will be.}
\end{figure*}

In the Fig.~\ref{fig:Exact-vs-App} $U_m$ vs $\ln a$ are presented the exact solution (solid blue line) and the approximated solution $U_m=\Omega_m^{\frac{6}{11}}$ (the red dotted line) for different values of the initial conditions $U_{m0}, \Omega_{m0}$. The closer the initial conditions to the equilibrium point $P_4$, the more accurate the approximation will be.

\subsection{Application to Quintessence}

As a second example, let us examine the simple quintessence example. In this case, and
assuming dust matter, the background equations are,

\begin{align}
&H^{2} =\frac{1}{3}(\rho_m+\rho_\phi) \,\,\text{,} \\
&  \dot{H}=-\frac{1}{2} (\rho_m+\rho_\phi+p_\phi) \,\,\text{,} \\
&\dot{\rho}_m+3H\rho_m =0 \,\,\text{,}  \\
&\dot{\rho}_\phi+3H(1+w_\phi)\rho_\phi =0 \,\,\text{,} 
\end{align}
where 
\begin{eqnarray}
\rho_\phi=\frac{\dot{\phi}^2}{2}+V(\phi) \,\,\text{,}  \quad 
p_\phi=\frac{\dot{\phi}^2}{2}-V(\phi) \,\,\text{,}  \quad 
w_\phi=\frac{p_\phi}{\rho_\phi} \,\,\text{,} 
\end{eqnarray}
while the perturbation equations are (\ref{perteq1}),(\ref{perteq2}). Let us
take for simplicity the usual case, 
\begin{eqnarray}
V(\phi)=V_0 e^{-\lambda \phi} \,\,\text{.} 
\end{eqnarray}

We introduce the following auxiliary variables $(x,y)$ given by Eq.~\eqref{(9)} \cite{COPE}
such that, 
\begin{equation}
\Omega_\phi= x^2+y^2, \quad \Omega_m=1-\Omega_\phi, \quad w_\phi=\frac{x^2-y^2}{%
x^2+y^2} \,\,\text{.} 
\end{equation}
As before, the background equations are recast as  Eq.~\eqref{(18)}-\eqref{(19)}.

The equations of matter perturbations become, 
\begin{subequations}
\begin{align}
\delta_{m}^{\prime \prime }&= \frac{1}{2} \delta_{m}^{\prime }\left(1-3
x^2+3 y^2\right)+\frac{3}{2} \delta_\phi \left(x^2+y^2\right)-\frac{3}{2}
\delta_{m} \left(x^2+y^2-1\right) \,\,\text{,}   \label{eqn.27} \\
\delta_\phi^{\prime \prime }&=\frac{1}{2} \delta_\phi^{\prime }\left(x^2
\left(\frac{24}{x^2+y^2}+3\right)+\frac{2 \sqrt{6} \lambda y^2}{x}-3
y^2-19\right)  \notag \\
& +\frac{1}{2} \delta_\phi \left(\frac{2 \left(7-6 x^2\right) x^2}{x^2+y^2}%
+15 x^2+\frac{y^2 \left(2 \sqrt{6} \lambda -3 x\right)}{x}-13\right)  \nonumber \\
& + 3 \delta_{m} x^2 \left(\frac{1}{x^2+y^2}-1\right) \,\,\text{.}   \label{eqn.28}
\end{align}
\end{subequations}
The stability of the equilibrium point in the plane $(x,y)$ was examined in 
\cite{COPE}.

\begin{enumerate}
\item At the equilibrium point $A: (x,y)=(0,0)$, the equation for the matter
perturbation is written as, 
\begin{equation}  \label{perts1}
{\delta_m}^{\prime }+3 {\delta_m}=2 {\delta_m}^{\prime \prime } \,\,\text{,} 
\end{equation}
where prime means derivative with respect to  $N=\ln a$. \newline
Hence, the perturbations evolve as, 
\begin{equation}  \label{solperts1}
\delta_{m}(N )= \frac{1}{5} {\delta_m}_{0} e^{-N } \left(2 e^{5 N
/2}+3\right)+\frac{2}{5} {\delta_m^{\prime }}_{0} e^{-N } \left(e^{5 N
/2}-1\right) \,\,\text{,} 
\end{equation}
for the initial conditions ${\delta_m}_{0}={\delta_m}|_{N=0}, {%
\delta_m^{\prime }}_{0}={\delta_m}^{\prime }|_{N=0}$.

For $x^2+y^2=0$, we have no scalar field such that the perturbations $%
\delta_\phi $ are not defined.

Defining the ratio $U_m=\frac{\delta_m^{\prime }(N)}{\delta_m}$%
, we have the equation, 
\begin{equation}
U_m^{\prime }=\frac{1}{2} \left(-2 U_m^2+U%
_m+3\right) \,\,\text{.} 
\end{equation}

The equilibrium points of the above system are, 

\begin{enumerate}
\item $A_1: U_m=-1$ and 
\item $A_2: U_m=\frac{3}{2}$. 
 \end{enumerate}

Since, 
\begin{equation}
\lambda(U_m):= \frac{d U_m^{\prime }}{d U_m}=%
\frac{1}{2} (1-4 U_m) \,\,\text{,}  \quad \lambda(-1)=\frac{5}{2} \,\,\text{,} \quad
\lambda(3/2)=-\frac{5}{2} \,\,\text{.} 
\end{equation}
That is, $U_m=-1$ is unstable, such that the decaying mode $%
\delta_m\propto e^{-N}$ is important as $N\rightarrow -\infty$.
Besides $U_m=\frac{3}{2}$ is stable, such that the growing mode $%
\delta_m\propto e^{\frac{3}{2}N}$ is the dominant at late times, that is,
as $N\rightarrow +\infty$. Using this qualitative analysis, we can
anticipate the result that is expected from Eq.~\eqref{solperts1} that matters
perturbations decays as $\delta_m\propto e^{-N}$ in the past, but they
grows as $\delta_m\propto e^{\frac{3}{2}N}$ latter on the evolution,
without having solved the original equation Eq.~\eqref{perts1}. The aim of 
this section is to formulate the standard quintessence model as an extended
dynamical system (in scale-invariant variables), with the lower dimension as
possible, that comprised both the background equations as well as the equations of the perturbations. 
Inspired by the analysis
done in Chapter 14 of Wainwright \& Ellis book \cite{Ellis} we extent 
this analysis for scalar field cosmologies could also be extended to modified gravity theories. 
The method consists in defining
$U_m=\frac{\delta_m^{\prime }(N)}{\delta_m}, \quad U_d=%
\frac{\delta_\phi^{\prime }(N)}{\delta_\phi}$ 
(and some other variables with physical interpretation).
In the standard method to write second-order equations like Eq.~\eqref{eqn.27} or Eq.~\eqref{eqn.28} is to define $(X_{(n)},Y_{(n)})=\left(%
\delta_{(n)}, \delta^{\prime }_{(n)}\right)$ as variables. Hence, we should
regard this $U_{(n)}$ as $\tan\theta_{(n)}$ where this $%
\theta_{(n)}$ is the usual polar angle in the $\left(\delta_{(n)},
\delta^{\prime }_{(n)}\right)$-plane with $0\leq \theta_{(n)}<2\pi$.
Therefore, one can think of $U_{(n)}$ as the phase of the
perturbation. If $U_{(n)}>0$ during the evolution, it follows by
definition that the perturbations $\delta_{(n)}$ are growing with time
(since $\delta_{(n)}>0$ and $\delta'_{(n)}>0$ or $\delta_{(n)}<0$ and $%
\delta_{(n)}'<0$), while if $U_{(n)}<0$ at that time, the
perturbation is decaying. If an orbit of the flow is asymptotic to an
equilibrium point, then the perturbation approaches a stationary state,
decaying to zero if $U_{(n)}<0$ or growing if $U_{(n)}>0$. If the orbits are asymptotic to a periodic solution, then the perturbation
propagates as sound waves.

\item At the equilibrium points $B, C: (x,y)=(\pm 1, 0)$,  the evolution of perturbations
is: 
\begin{align}  \label{perts2}
{\delta_m}^{\prime \prime }=\frac{3 {\delta_\phi}}{2}-{\delta_m}^{\prime },
\quad {\delta_\phi}^{\prime \prime }=4 {\delta_\phi}^{\prime }+2 {\delta_\phi} \,\,\text{.} 
\end{align}
Hence, the perturbations evolve as,
\begin{subequations}
\begin{align}
& \delta_m=\frac{1}{8} {\delta_\phi}_{0} e^{\left(2-\sqrt{6}\right) N }
\left(e^{\left(\sqrt{6}-3\right) N } \left(24 e^{N }+\left(9 \sqrt{6}%
-22\right) e^{\left(3+\sqrt{6}\right) N }+20\right)-9 \sqrt{6}-22\right) 
\notag \\
& +\frac{1}{8} {\delta_\phi^{\prime }}_{0} e^{\left(2-\sqrt{6}\right) N }
\left(e^{\left(\sqrt{6}-3\right) N } \left(-6 e^{N }+\left(5-2 \sqrt{6}%
\right) e^{\left(3+\sqrt{6}\right) N }-4\right)+2 \sqrt{6}+5\right)\nonumber \\
&  +{%
\delta_m}_{0}  +{\delta_m^{\prime }}_{0} (\sinh (N )-\cosh (N )+1) \,\,\text{,}  \\
&\delta_\phi= \frac{{\delta_\phi^{\prime }}_{0} e^{2 N } \sinh \left(\sqrt{6}
N \right)}{\sqrt{6}}-\frac{1}{3} {\delta_\phi}_{0}e^{2 N } \left(\sqrt{6}
\sinh \left(\sqrt{6} N \right)-3 \cosh \left(\sqrt{6} N \right)\right) \,\,\text{,} 
\end{align}
\end{subequations}
for the initial conditions ${\delta_m}_{0}={\delta_m}|_{N=0}, {\delta_\phi}%
_{0}={\delta_\phi}|_{N=0}, {\delta_m^{\prime }}_{0}={\delta_m}^{\prime
}|_{N=0} \,\,\text{,} {\delta_\phi^{\prime }}_{0}={\delta_\phi}^{\prime }|_{N=0}$ \,\,\text{.}

With the above perturbation equations Eq.~\eqref{perts2}, we construct a system
of differential equations for the quantities,
\begin{equation}
\quad V_m=\frac{%
\delta_m^{\prime }(N)}{\delta_\phi}, \quad U_d=\frac{%
\delta_\phi^{\prime }(N)}{\delta_\phi} \,\,\text{,} 
\end{equation}
as given by,
\begin{subequations}
\begin{align}
& V_m^{\prime }=\frac{3}{2}-(U_d+1) V_m\,\,\text{,}  \\
& U_d^{\prime }= 2-(U_d-4) U_d \,\,\text{.} 
\end{align}
\end{subequations}

We now try to research its stability and integrability in the reduced phase plane  $(V_m, U_d)$. 
We obtain the equilibrium points: 
\begin{enumerate}
\item $B_1: (V_m,U_d)=\left(\frac{1}{2} \left(3+\sqrt{6}\right),2-\sqrt{6}\right)$ 
The eigenvalues are $\left\{2 \sqrt{6},\sqrt{6}-3\right\}$. Therefore, the equilibrium point is a saddle (unstable). 
The phase of the matter perturbation at the equilibrium point is  $U_m^*=U_d^*=2-\sqrt{6}$, which are both negative; therefore, the perturbations $\delta_m$ and $\delta_\phi$ decay with time. 
\item $B_2: (V_m,U_d)=\left(\frac{1}{2} \left(3-\sqrt{6}\right),2+\sqrt{6}\right)$. 
The eigenvalues are $\left\{-3-\sqrt{6},-2 \sqrt{6}\right\}$. Therefore, the equilibrium point is stable. The phase of the matter perturbation at the equilibrium point is  $U_m^*=U_d^*=2+\sqrt{6}$, which are both positive; therefore, the perturbations $\delta_m$ and $\delta_\phi$ are growing with time. 
\end{enumerate}

Introducing the compact variables,
\begin{equation}
\bar{V}_m=\frac{2}{\pi}\arctan\left(V_m
\right)
\,\,\text{,} \quad \bar{U}_d=\frac{2}{\pi}\arctan\left(U_d\right) \,\,\text{.} 
\end{equation}
We obtain the dynamical system,
\begin{subequations}
\label{SCF_perts_3_4}
\begin{align}
&\bar{V}_m'=\frac{\sin (\pi  \bar{V}_m) \left(3 \cot \left(\frac{\pi  \bar{V}_m}{2}\right)-2 \left(\tan \left(\frac{\pi  \bar{U}_d}{2}\right)+1\right)\right)}{2 \pi } \,\,\text{,} \\
&\bar{U}_d'=\frac{4 \sin (\pi  \bar{U}_d)+3 \cos (\pi  \bar{U}_d)+1}{\pi} \,\,\text{.} 
\end{align}
\end{subequations}

\begin{figure}[]
\centering
        \includegraphics[scale=0.9]{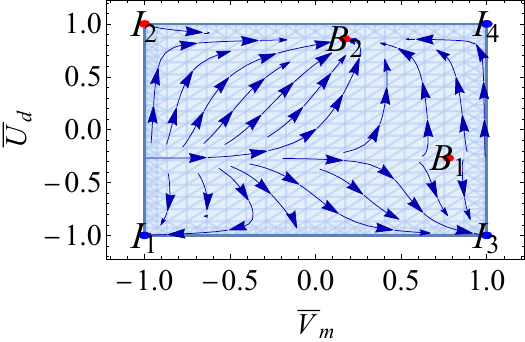}
      
    \caption{	\label{fig:SCF_perts_3_4} Phase plane  for the system Eq.~\eqref{SCF_perts_3_4} showing both the stability of the background equations as well as the stability of the perturbations for the Kinetic dominated points for the quintessence model using the compact variables $\bar{V}_m=\frac{2}{\pi}\arctan\left(V_m
\right)$ vs. $\bar{U}_d=\frac{2}{\pi}\arctan\left(U_d\right)$.}
\end{figure}

In Fig.~\ref{fig:SCF_perts_3_4}, it is presented a flow for the system Eq.~\eqref{SCF_perts_3_4} showing both the stability of the background equations as well as the stability of the perturbations for the  Kinetic dominated points $(x,y)=(\pm 1, 0)$ for the quintessence model using the compact variables $\bar{V}_m=\frac{2}{\pi}\arctan\left(V_m
\right)$ vs. $\bar{U}_d=\frac{2}{\pi}\arctan\left(U_d\right)$. There are some configurations $I_{1,2}:(\bar{V}_m, \bar{U}_d)=(-1, \pm 1)$ and $I_{3,4}:(\bar{V}_m, \bar{U}_d)=(1, \pm 1)$ which are not equilibrium points due to $\bar{U}_d'=-\frac{2}{\pi}\neq 0$ at the fixed points but they are essential for the dynamics at infinity. The figure suggests that the attractor at the infinity region are $I_1$ and $I_3$, and the other points at infinity are saddle points. 
Indeed, if we choose $\bar{V}_m=\pm 1$, then $\bar{U}_d'= -\frac{2}{\pi} \implies \bar{U}_d=-\frac{2}{\pi} N+c$, but by definition $-1< \bar{U}< 1$, therefore any solution starting on the line $\bar{V}_m=\pm 1$ hits the boundary $\bar{U}_d=+1$ in a finite time-lapse to the past and the boundary $\bar{U}_d=-1$ in a finite time-lapse to the future.

\item For the equilibrium points $D: (x,y)=\left(\frac{\lambda }{\sqrt{6}}, \sqrt{1-%
\frac{\lambda^2}{6}}\right)$, we have 
\begin{small}
\begin{align}  \label{perts3}
\delta_{m}^{\prime \prime }= \frac{1}{2} \left(3 {\delta_\phi}-\left(\lambda
^2-4\right) {\delta_m}^{\prime }\right) \,\,\text{,} \quad {\delta_\phi}^{\prime \prime
}= \frac{1}{2} \left(3 \lambda ^2-10\right) {\delta_\phi}^{\prime }-\frac{1}{6}
\left(\lambda ^4-10 \lambda ^2+12\right) {\delta_\phi} \,\,\text{.}
\end{align}
\begin{align}
&\delta_m={\delta_m}_{0}+\frac{2 {\delta_m^{\prime }}_{0} \left(1-e^{-\frac{1%
}{2} (\lambda^2 -4) N }\right)}{\lambda^2 -4}  \notag \\
& +9{\delta_\phi}_{0} \Big(\frac{e^{N_{-}} (k (\lambda^2 ((401-43 \lambda^2 )
\lambda^2 -1348)+1644)+(\lambda^2 (17 \lambda^2 -113)+198) (\lambda^2 (19 \lambda^2
-100)+204))}{2 ((\lambda^2 -10) \lambda^2 +12) (\lambda^2 (7 \lambda^2 -55)+96)
(\lambda^2 (19 \lambda^2 -100)+204)}  \notag \\
& +\frac{e^{N_{+}} (k (\lambda^2 (\lambda^2 (43 \lambda^2
-401)+1348)-1644)+(\lambda^2 (17 \lambda^2 -113)+198) (\lambda^2 (19 \lambda^2
-100)+204))}{2 ((\lambda^2 -10) \lambda^2 +12) (\lambda^2 (7 \lambda^2 -55)+96)
(\lambda^2 (19 \lambda^2 -100)+204)}  \notag \\
&+\frac{2 (2 \lambda^2 -7) e^{-\frac{1}{2} (\lambda^2 -4) N }}{(\lambda^2 -4)
(\lambda^2 (7 \lambda^2 -55)+96)}+\frac{10-3 \lambda^2 }{(\lambda^2 -4) ((\lambda^2
-10) \lambda^2 +12)}\Big)  \notag \\
& +\frac{9{\delta_\phi^{\prime }}_{0}}{(\lambda^2 -4) ((\lambda^2 -10) \lambda^2 +12)}
\Big(2+ \frac{e^{-\frac{\lambda^2 N }{2}} }{(\lambda^2 (7 \lambda^2 -55)+96)
(\lambda^2 (19 \lambda^2 -100)+204)} \times  \notag \\
& \Big\{-(\lambda^2 -4) (k ((103-16 \lambda^2 ) \lambda^2 -186)+3 (2 \lambda^2 -7)
(\lambda^2 (19 \lambda^2 -100)+204)) e^{\frac{\lambda^2 N }{2}+N_{-}}  \notag
\\
& -(\lambda^2 -4) (k (\lambda^2 (16 \lambda^2 -103)+186)+3 (2 \lambda^2 -7) (\lambda^2
(19 \lambda^2 -100)+204)) e^{\frac{\lambda^2 N }{2}+N_{+}}  \notag \\
& -2 ((\lambda^2 -10) \lambda^2 +12) (\lambda^2 (19 \lambda^2 -100)+204) e^{2 N }%
\Big\}\Big) \,\,\text{,} \\
&\delta_\phi=\frac{{\delta_\phi}_{0}\left(k (3 \lambda^2 -10)
\left(e^{N_{+}}-e^{N_{-}}\right)+(\lambda^2 (19 \lambda^2 -100)+204)
\left(e^{N_+}+e^{N_{-}}\right)\right)}{2 (\lambda^2 (19 \lambda^2
-100)+204)}  \nonumber \\
& -\frac{2 {\delta_\phi^{\prime }}_{0} \left(e^{N_{+}} -e^{N _{-}}\right)%
}{\sqrt{\frac{1}{3} \lambda^2 (19 \lambda^2 -100)+68}} \,\,\text{,}
\end{align}
\end{small}
for the initial conditions ${\delta_m}_{0}={\delta_m}|_{N=0}, {\delta_\phi}%
_{0}={\delta_\phi}|_{N=0}, {\delta_m^{\prime }}_{0}={\delta_m}^{\prime
}|_{N=0}, {\delta_\phi^{\prime }}_{0}={\delta_\phi}^{\prime }|_{N=0}$, where 
$N_+=-\frac{1}{12} \left(-9 \lambda^2 +30+k\right) N$, $N_-=-\frac{1}{%
12} \left(-9 \lambda^2 +30-k\right) N$, \newline $k=\sqrt{57 \lambda^4-300 \lambda^2
+612}$.

With the above perturbation equations given in Eq.~\eqref{perts3}, we construct a system
of differential equations for the quantities 
\begin{equation}
V_m=\frac{%
\delta_m^{\prime }(N)}{\delta_\phi} \,\,\text{,} \quad U_d=\frac{%
\delta_\phi^{\prime }(N)}{\delta_\phi} \,\,\text{,}
\end{equation}
as given by 
\begin{subequations}
\begin{align}
&Q^{\prime }=-Q (U_d-U_m) \,\,\text{,} \\
&U_m^{\prime }= \frac{1}{2} \left(\frac{3}{Q}-U_m
\left(\lambda ^2+2 U_m-4\right)\right) \,\,\text{,} \\
&U_d^{\prime }= -\frac{\lambda ^4}{6}+\frac{5 \lambda ^2}{3}+\frac{%
3 \lambda ^2 U_d}{2}-U_d(U_d+5)-2 \,\,\text{.}
\end{align}
\end{subequations}
For further simplification, we define $%
V_m={U_m}{Q}$, so that we acquire the reduced
dynamical system 
\begin{subequations}
\label{systemC}
\begin{align}
& V_m^{\prime }=\frac{1}{2} \left(3-V_{m} \left(\lambda
^2+2 U_d-4\right)\right) \,\,\text{,}\\
&U_d^{\prime }= -\frac{\lambda ^4}{6}+\frac{5 \lambda ^2}{3}+\frac{%
3 \lambda ^2 U_d}{2}-U_d(U_d+5)-2 \,\,\text{.}
\end{align}
\end{subequations}

For arbitrary $\lambda$, we obtain the equilibrium points: 
\begin{enumerate}
\item $D_1: (V_m,U_d)=\left(-\frac{18}{-15 \lambda ^2+\sqrt{57 \lambda ^4-300 \lambda ^2+612}+54}, \frac{1}{12} \left(9 \lambda ^2-\sqrt{57 \lambda ^4-300 \lambda ^2+612}-30\right)\right)$. For $-\sqrt{\frac{1}{14} \left(55+\sqrt{337}\right)}<\lambda <\sqrt{\frac{1}{14} \left(55+\sqrt{337}\right)}\approx 2.28907$, both phases of the perturbations are negative, implying that the perturbations $\delta_m$ and $\delta_\phi$ are decaying with time.  The eigenvalues  of the linear matrix are \newline $\left\{\frac{\sqrt{19 \lambda ^4-100 \lambda ^2+204}}{2 \sqrt{3}},\frac{1}{12} \left(-15 \lambda ^2+\sqrt{3} \sqrt{19 \lambda ^4-100 \lambda ^2+204}+54\right)\right\}$. For the above interval, the point is a source. For $\lambda ^2 >{\frac{1}{14} \left(55+\sqrt{337}\right)}$  it is a saddle. 

\item $D_2: (V_m,U_d)=\left(\frac{18}{15 \lambda ^2+\sqrt{57 \lambda ^4-300 \lambda ^2+612}-54}, \frac{1}{12} \left(9 \lambda ^2+\sqrt{57 \lambda ^4-300 \lambda ^2+612}-30\right)\right)$.  For  $-\sqrt{5-\sqrt{13}}<\lambda <\sqrt{5-\sqrt{13}}\approx 1.18087$, both phases of the perturbations are negative, implying that the perturbations $\delta_m$ and $\delta_\phi$ are decaying with time. The eigenvalues  of the linear matrix are \newline
$\left\{-\frac{\sqrt{19 \lambda ^4-100 \lambda ^2+204}}{2 \sqrt{3}},\frac{1}{12} \left(-15 \lambda ^2-\sqrt{3} \sqrt{19 \lambda ^4-100 \lambda ^2+204}+54\right)\right\}$. 

When $-\sqrt{\frac{55}{14}-\frac{\sqrt{337}}{14}}<\lambda <\sqrt{\frac{55}{14}-\frac{\sqrt{337}}{14}}$ the point is a saddle. It is stable (sink) when $\lambda <-\sqrt{\frac{55}{14}-\frac{\sqrt{337}}{14}}\lor \lambda >\sqrt{\frac{55}{14}-\frac{\sqrt{337}}{14}}$. In this region, both phases of the perturbations are positive, implying that the perturbations $\delta_m$ and $\delta_\phi$ are growing with time. Therefore, one of the scalar field-dominated solutions can have stable phases of the perturbations, but the perturbations $\delta_m$ and $\delta_\phi$ themselves are, in this case growing with time.  
\end{enumerate}

For $\lambda=0$, the points above correspond to the de Sitter point (exact $\Lambda$CDM model) that has $(x,y)=(0,1)$.
That is,  the de Sitter solution is represented in the extended phase space $(V_m, U_d)$ by two equilibrium points:
\begin{enumerate}
\item $D_1: (V_m,U_d)=\left( -\frac{18}{54+6 \sqrt{17}},  \frac{1}{12} \left(-30-6 \sqrt{17}\right)\right)$.  The eigenvalues of the linear matrix are $\left\{\frac{1}{2} \left(9+\sqrt{17}\right),\sqrt{17}\right\}$. Hence, the point is unstable (source). 
\item $D_2: (V_m,U_d)=\left(\frac{18}{6 \sqrt{17}-54}, \frac{1}{12} \left(6\sqrt{17}-30\right)\right)$. The eigenvalues of the linear matrix are  $\left\{-\sqrt{17},\frac{1}{2} \left(9-\sqrt{17}\right)\right\}$ and the point is a saddle point. 
\end{enumerate}
For these equilibrium points, the  phases of the perturbations $U_m$ and $U_d$ 
are negative therefore, the perturbations $\delta_m$ and $\delta_\phi$ are decaying with time. That is expected for the exact $\Lambda$CDM model with $\delta_m\rightarrow 0, \delta_\phi=0$. 
Although in the phase plane, $(x,y)$, the de Sitter solution is stable, when the directions along the ``phase of the perturbations'' are considered in the dynamics, the de Sitter solution (and its two representations) becomes unstable. The other becomes saddle in the phase plane  $(V_m, U_d)$. Summarizing, on top of the fixed point $D$, the phases of the perturbations are not stable, although the perturbations $\delta_m$ and $\delta_\phi$ are decaying with time.  

Introducing the compact variables
\begin{equation}
\quad \bar{V}_m=\frac{2}{\pi}\arctan\left(V_m
\right)
,\quad \bar{U}_d=\frac{2}{\pi}\arctan\left(U_d\right),
\end{equation}
we obtain the dynamical system
\begin{subequations}
\label{SCF_perts}
\begin{align}
&\bar{V}_m'=-\frac{\cos ^2\left(\frac{\pi  \bar{V}_m}{2}\right) \left(\tan \left(\frac{\pi  \bar{V}_m}{2}\right) \left(2 \tan \left(\frac{\pi  \bar{U}_d}{2}\right)+\lambda^2-4\right)-3\right)}{\pi } \,\,\text{,}\\
&\bar{U}_d'=-\frac{\left(30-9 \lambda ^2\right) \sin \left(\pi  \bar{U}_d\right)+\left(\lambda ^4-10 \lambda ^2+6\right) \cos \left(\pi  \bar{U}_d\right)+\lambda
   ^4-10 \lambda ^2+18}{6 \pi } \,\,\text{.}
\end{align}
\end{subequations}

Fig.~\ref{fig:SFC}: it is presented a flow  for the system Eq.~\eqref{SCF_perts} showing  both the stability of the background equations as well as the stability of the perturbations for the scalar field dominated point  $(x,y)=\left(\frac{\lambda }{\sqrt{6}}, \sqrt{1-%
\frac{\lambda^2}{6}}\right)$ for the quintessence model with exponential potential with $\lambda=0, 1, 2, 3$ using the compact variables $\bar{V}_m=\frac{2}{\pi}\arctan\left(V_m
\right)$ vs. $\bar{U}_d=\frac{2}{\pi}\arctan\left(U_d\right)$. There are some configurations $I_{1,2}:(\bar{V}_m, \bar{U}_d)=(-1, \pm 1)$ and $I_{3,4}:(\bar{V}_m, \bar{U}_d)=(1, \pm 1)$ which are not equilibrium points due to $\bar{U}_d'=-\frac{2}{\pi}\neq 0$ at the fixed points but they are important for the dynamics at infinity. The figure suggests that the attractor at the infinity region are $I_1$ and $I_3$, and the other points at infinity are saddle points. 
Indeed, if we choose $\bar{V}_m=\pm 1$, then $\bar{U}_d'= -\frac{2}{\pi} \implies \bar{U}_d=-\frac{2}{\pi} N+c$, but by definition $-1< \bar{U}< 1$, therefore any solution starting on the line $\bar{V}_m=\pm 1$ hits the boundary $\bar{U}_d=+1$ in a finite time-lapse to the past and the boundary $\bar{U}_d=-1$ in a finite time-lapse to the future.  

\begin{figure*}
	\centering
		\includegraphics[scale=0.65]{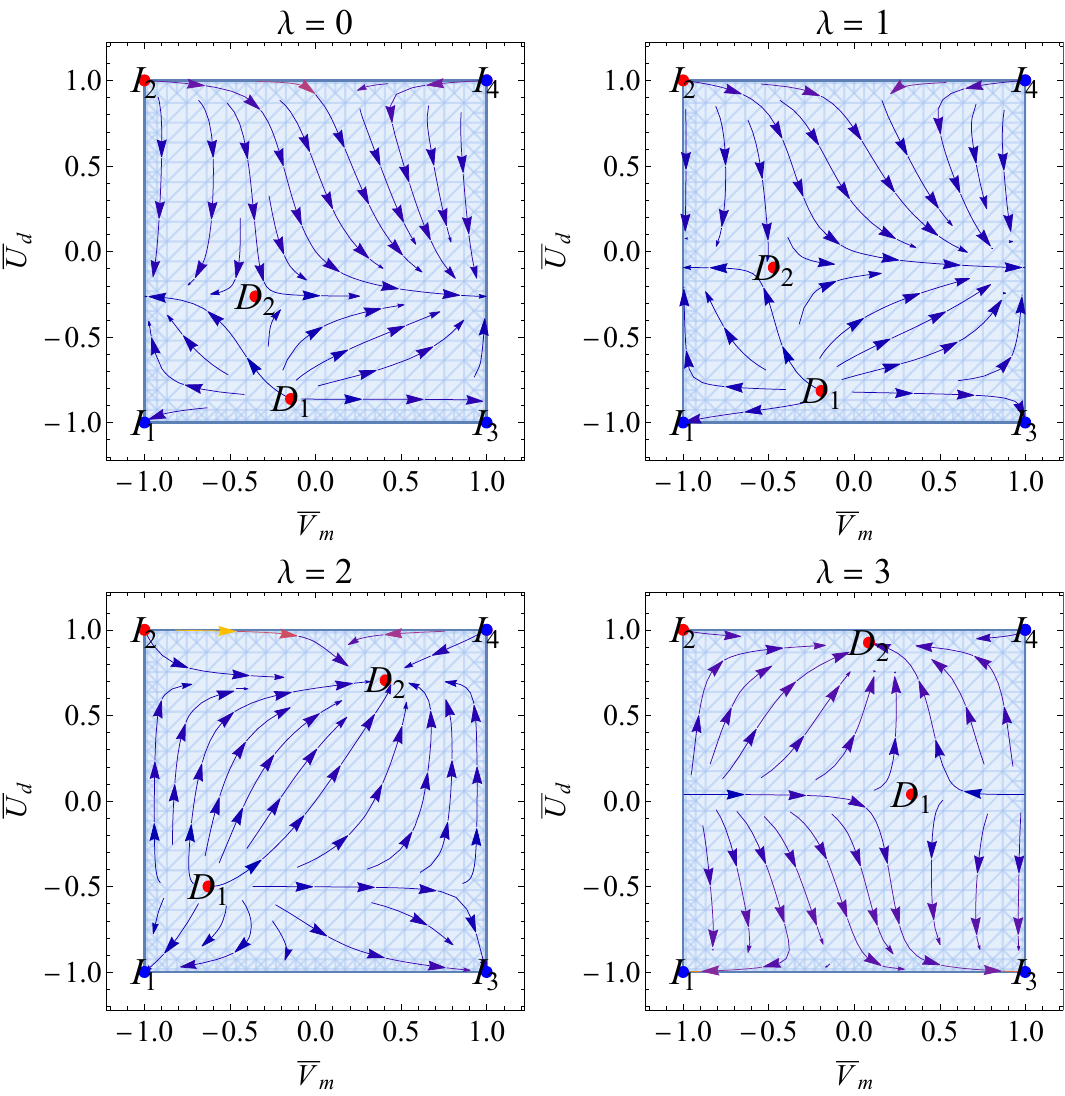}
	\caption{\label{fig:SFC} Phase plane  for the system Eq.~\eqref{SCF_perts} showing  both the stability of the background equations as well as the stability of the perturbations for the scalar field dominated point  $(x,y)=\left(\frac{\lambda }{\sqrt{6}}, \sqrt{1-%
\frac{\lambda^2}{6}}\right)$ for the quintessence model with exponential potential with $\lambda=0, 1, 2, 3$ using the compact variables $\bar{V}_m=\frac{2}{\pi}\arctan\left(V_m
\right)$ vs. $\bar{U}_d=\frac{2}{\pi}\arctan\left(U_d\right)$.}
\end{figure*}

\item For the equilibrium points $E:= (x,y)=\left(\frac{\sqrt{\frac{3}{2}}}{\lambda }, 
\sqrt{\frac{3}{2 \lambda^2}}\right)$, we have 
\begin{align}  \label{perts4}
{\delta_m}^{\prime \prime }= \frac{9 {\delta_\phi}+\lambda ^2 {\delta_m}%
^{\prime }+3 \left(\lambda ^2-3\right) {\delta_m}}{2 \lambda ^2} \,\,\text{,}\quad {%
\delta_\phi}^{\prime \prime }= \frac{-\lambda ^2 {\delta_\phi}^{\prime }+9 {%
\delta_\phi}+3 \left(\lambda ^2-3\right) {\delta_m}}{2 \lambda ^2} \,\,\text{.}
\end{align}

With the above perturbation equations are given by Eq.~\eqref{perts4}, we construct a system
of differential equations for the quantities 
\begin{equation}
Q=\frac{\delta_m}{\delta_\phi} \,\,\text{,} \quad V_m=\frac{%
\delta_m^{\prime }(N)}{\delta_\phi} \,\,\text{,} \quad U_d=\frac{%
\delta_\phi^{\prime }(N)}{\delta_\phi} \,\,\text{,}
\end{equation}
as given by 
\begin{align}
& V_m'=\frac{\lambda ^2 (3 Q-2 U_\phi V_{m}+ V_{m})-9 Q+9}{2 \lambda ^2} \,\,\text{,} \\
& U_d'= \frac{3 \left(\lambda ^2-3\right) Q-\lambda ^2 U_\phi (2 U_\phi+1)+9}{2 \lambda ^2} \,\,\text{,}\\
& Q'=V_{m}-Q  U_\phi \,\,\text{.}
\end{align}

In this case, the real-valued equilibrium points are 
\begin{enumerate}
\item $E_1: (V_m, U_d, Q)= \left( 0,  0,  -\frac{3}{\lambda ^2-3}\right)$
and 
\item $E_2: (V_m, U_d, Q)=$ \newline
$\Bigg( \frac{4 \Delta ^{2/3} \left(8 \lambda ^2-27\right) \lambda ^2+\sqrt[3]{3} \Delta 
   \left(7 \lambda ^2-18\right)+2 \sqrt[6]{3} \sqrt[3]{\Delta } \left(29 \sqrt{3} \lambda ^2+\sqrt{-25 \lambda ^4-729 \lambda ^2+2187}-90 \sqrt{3}\right) \lambda ^4+49 \sqrt[3]{3} \lambda
   ^8}{36 \Delta ^{2/3} \lambda ^2 \left(\lambda ^2-3\right)}$,\\
   $ \frac{\Delta ^{2/3}+7 \sqrt[3]{3} \lambda ^4}{2\ 3^{2/3} \sqrt[3]{\Delta } \lambda ^2},$\\
   $\frac{3 \Delta ^{2/3}
   \left(14 \lambda ^2+\sqrt[3]{81 \lambda ^6+6 \left(\sqrt{-75 \lambda ^4-2187 \lambda ^2+6561}-81\right) \lambda ^4}-54\right) \lambda ^2+3^{2/3} \Delta ^{4/3}+21\ 3^{2/3} \sqrt[3]{\Delta
   } \lambda ^6+147 \sqrt[3]{3} \lambda ^8}{54 \Delta ^{2/3} \lambda ^2 \left(\lambda ^2-3\right)}\Bigg)$, \\
where $\Delta =27 \lambda ^6+2 \left(\sqrt{-75 \lambda ^4-2187 \lambda ^2+6561}-81\right) \lambda ^4$.
\end{enumerate}
The first point $E_1$ has constant densities perturbations with $\delta_m=-\frac{3}{\lambda ^2-3}\delta_\phi$. They remain constant with evolution.
The eigenvalues of $E_1$ are $\left\{\frac{ k_{1}}{2 \lambda ^2 \left(\lambda ^2-3\right)},\frac{k_{2}}{2 \lambda ^2 \left(\lambda ^2-3\right)},\frac{ k_{3}}{2 \lambda ^2
   \left(\lambda ^2-3\right)}\right\}$, where $k_{1,2,3}$ are the three roots of the polynomial $P(k)=-6 \lambda ^{12}+90 \lambda ^{10}-486 \lambda ^8+1134 \lambda ^6-972 \lambda ^4+k^3+k \left(-7 \lambda ^8+42 \lambda ^6-63 \lambda ^4\right)$. 
   
   \begin{figure}[]
\centering
        \includegraphics[scale=0.8]{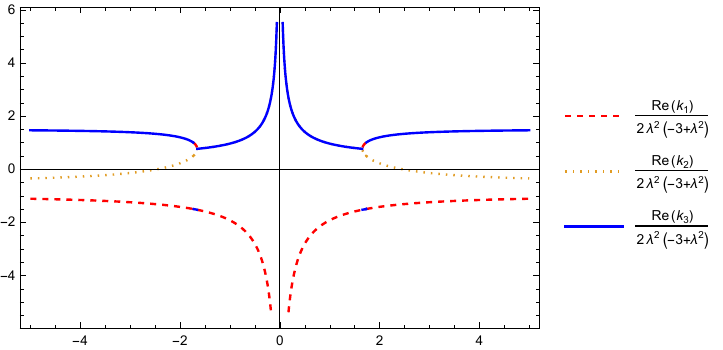}
      
    \caption{	\label{fig:EigenE1} Real parts of the eigenvalues of  $E_1$. }
\end{figure}
   Fig.~\eqref{fig:EigenE1} presents the real parts of the eigenvalues of $E_1$. At least two eigenvalues have different signs for all the values of $\lambda$. Therefore, $E_1$ is a saddle. 

For $E_2$, we  have \newline $\delta_m=\frac{3 \Delta ^{2/3}
   \left(14 \lambda ^2+\sqrt[3]{81 \lambda ^6+6 \left(\sqrt{-75 \lambda ^4-2187 \lambda ^2+6561}-81\right) \lambda ^4}-54\right) \lambda ^2+3^{2/3} \Delta ^{4/3}+21\ 3^{2/3} \sqrt[3]{\Delta
   } \lambda ^6+147 \sqrt[3]{3} \lambda ^8}{54 \Delta ^{2/3} \lambda ^2 \left(\lambda ^2-3\right)} \delta_\phi$. Moreover,  $U_d>0$  implies that the perturbation  $\delta_\phi$ (therefore, $\delta_m$) is growing with time. 
   
      \begin{figure}[]
\centering
        \includegraphics[scale=0.8]{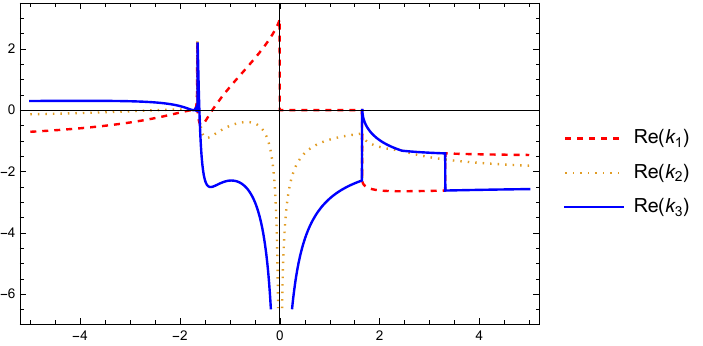}
          \caption{\label{fig:EigenE2} Real parts of the eigenvalues of  $E_2$. }
\end{figure}

Fig.~\ref{fig:EigenE2} presents the real parts of the eigenvalues of $E_2$. For $\lambda \gtrsim  1.65014$ or $- 1.65014 \lesssim \lambda \lesssim -1.35169$, $E_2$ is a sink; otherwise, it is a saddle. 

\begin{figure}[]
	\centering
		\includegraphics[scale=0.65]{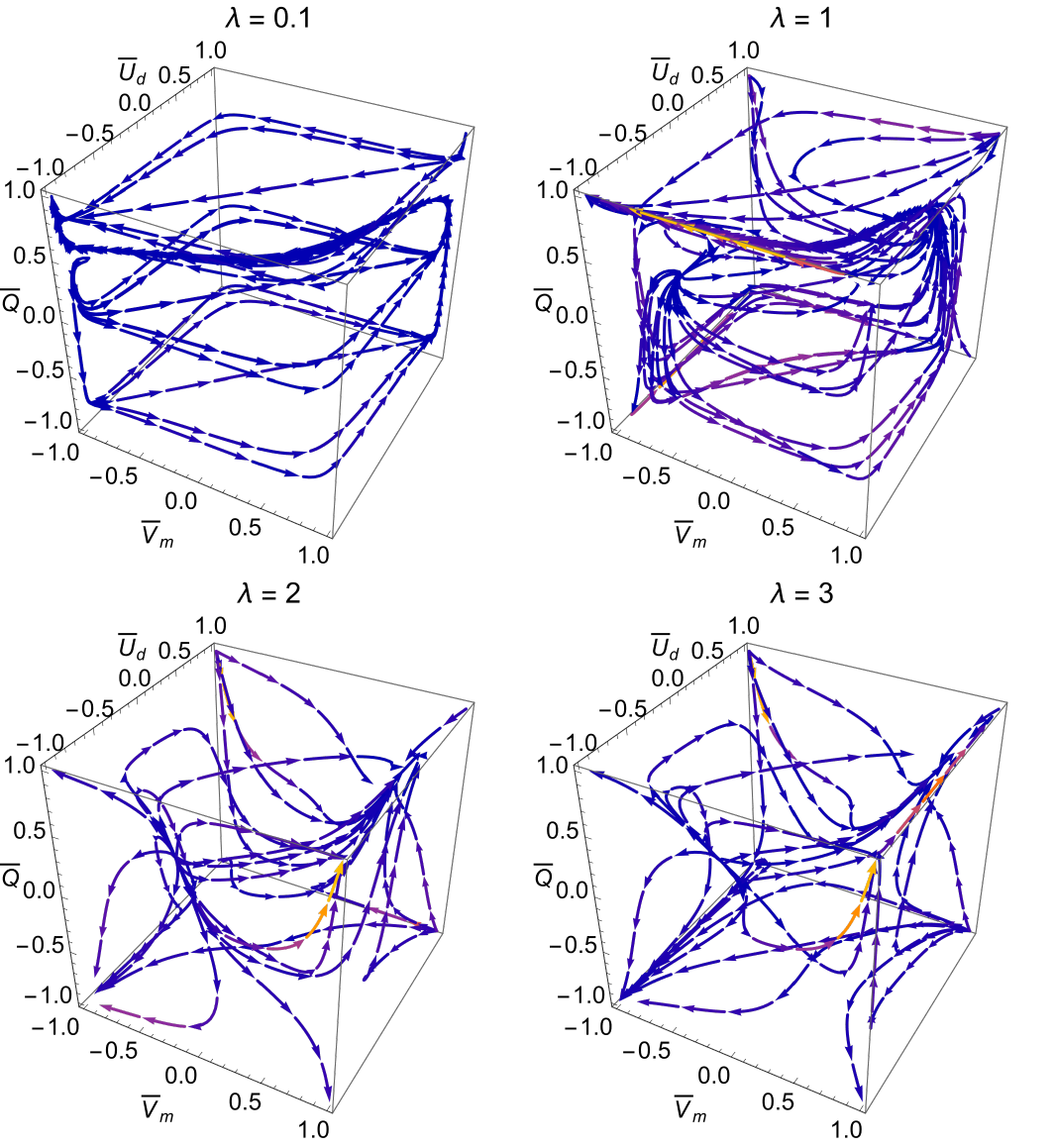}
	\caption{\label{fig:SFC3D} Phase space for the system Eq.~\eqref{SCF_perts3D} showing  both the stability of the background equations as well as the stability of the perturbations for the scalar field dominated point  $(x,y)=\left(\frac{\lambda }{\sqrt{6}}, \sqrt{1-%
\frac{\lambda^2}{6}}\right)$ for the quintessence model with exponential potential with $\lambda=0.1, 1, 2, 3$ using the compact variables $\bar{V}_m=\frac{2}{\pi}\arctan\left(V_m\right),\quad \bar{U}_d=\frac{2}{\pi}\arctan\left(U_d\right), \quad \bar{Q}=\frac{2}{\pi}\arctan\left(\frac{ \delta_m%
}{\delta_\phi}\right)$.}
\end{figure}

Introducing the compact variables
\begin{equation}
\bar{V}_m=\frac{2}{\pi}\arctan\left(V_m
\right)
\,\,\text{,}\quad \bar{U}_d=\frac{2}{\pi}\arctan\left(U_d\right) \,\,\text{,} \quad \bar{Q}=\frac{2}{\pi}\arctan\left(\frac{ \delta_m%
}{\delta_\phi}\right) \,\,\text{,}
\end{equation}
we obtain the dynamical system
\begin{subequations}
\label{SCF_perts3D} 
\begin{align}
&\bar{V}_m'=\frac{\cos ^2\left(\frac{\pi  \bar{V}_m}{2}\right) \left(\lambda ^2 \left(1-2 \tan \left(\frac{\pi  \bar{U}_d}{2}\right)\right) \tan \left(\frac{\pi  \bar{V}_m}{2}\right)+3
   \left(\lambda ^2-3\right) \tan \left(\frac{\pi  \bar{Q}}{2}\right)+9\right)}{\pi  \lambda ^2} \,\,\text{,}\\
&\bar{U}_d'=\frac{\cos ^2\left(\frac{\pi  \bar{U}_d}{2}\right) \left(\lambda ^2 \left(-\tan
   \left(\frac{\pi  \bar{U}_d}{2}\right)\right) \left(2 \tan \left(\frac{\pi  \bar{U}_d}{2}\right)+1\right)+3 \left(\lambda ^2-3\right) \tan \left(\frac{\pi 
   \bar{Q}}{2}\right)+9\right)}{\pi  \lambda ^2} \,\,\text{,}\\
&\bar{Q}'=\frac{2 \cos ^2\left(\frac{\pi  \bar{Q}}{2}\right) \left(\tan \left(\frac{\pi  \bar{V}_m}{2}\right)-\tan \left(\frac{\pi 
   \bar{Q}}{2}\right) \tan \left(\frac{\pi  \bar{U}_d}{2}\right)\right)}{\pi } \,\,\text{.}
\end{align}
\end{subequations}

\end{enumerate}

Fig.~\ref{fig:SFC3D} shows a phase space for the system Eq.~\eqref{SCF_perts3D} showing  both the stability of the background equations and the stability of the perturbations for the scalar field dominated point  $(x,y)=\left(\frac{\lambda }{\sqrt{6}}, \sqrt{1-%
\frac{\lambda^2}{6}}\right)$ for the quintessence model with exponential potential with $\lambda=0.1, 1, 2, 3$ using the compact variables $\bar{V}_m=\frac{2}{\pi}\arctan\left(V_m\right),\quad \bar{U}_d=\frac{2}{\pi}\arctan\left(U_d\right), \quad \bar{Q}=\frac{2}{\pi}\arctan\left(\frac{ \delta_m}{\delta_\phi}\right)$.
Both plots show that there are late-time attractors at $(\bar{V}_m, \bar{U}_d, \bar{Q})= (-1, -1, + 1)$ (top panels) and at $(\bar{V}_m, \bar{U}_d, \bar{Q})= (-1, -1, + 1)$, or at $(\bar{V}_m, \bar{U}_d, \bar{Q})= (-1, -1, - 1)$, or both (bottom panels). Some orbits are the bottom panels approaches $(\bar{V}_m, \bar{U}_d, \bar{Q})= (+1, -1, -1)$. 

\section{Concluding Remarks}
\label{sect:6}
This chapter extensively analyzed dynamical systems in cosmological models involving a scalar field with an arbitrary potential. Our research covers a broad class of potentials and examines the system's behavior at the background and perturbation levels. We utilized the $f$-deviser method to study a non-interacting scalar field cosmology with arbitrary potential. This approach facilitates a comprehensive analysis of the system's dynamics, extending the findings of the existing literature. We analysed the background quantities using Hubble-normalized variables. We provide two examples: the monomial potential and the double exponential potential, which includes hyperbolic cosine, exponential potential, and cosmological constant. These two classes of potentials, monomial and double exponential, comprise the asymptotic behaviour of several classes of scalar field potentials. Therefore, they provide the skeleton for the typical behaviour of arbitrary potentials.

As mentioned earlier, technical papers such as \cite{Alho:2020cdg} have derived a new regular dynamical system on a three-dimensional compact state space. This system describes linear scalar perturbations of spatially flat RW geometries for relativistic models with a minimally coupled scalar field with exponential potential. This allows them to create a global solution space where known solutions reside on specific invariant sets. Their dynamical systems approach has obtained new findings about comoving and uniform density curvature perturbations. Additionally, they have extended this approach to more general scalar field potentials, resulting in state spaces where the models' exponential potential state space appears as invariant boundary sets, demonstrating their significance as building blocks in a hierarchy of increasingly complex cosmological models. 

Our chapter also employs a similar approach as in \cite{Alho:2020cdg} by assuming the matterless scenario for simplicity. Using dynamical systems methods, we analysed the dynamics of linear scalar cosmological perturbations for a generic scalar field model. We focused on three scalar perturbations: the evolution of the Bardeen potentials, the comoving curvature perturbation, and the Sasaki-Mukhanov variable (the scalar field perturbation in uniform curvature gauge). To achieve this, we created three autonomous nonlinear first-order ordinary differential equations with a product structure for the state space $S=B\times P$, a product space of the background state space $B$ that describes the dynamics of a FLRW background, and $P$ which contains Fourier decomposed gauge invariant variables that describe linear cosmological perturbations. Our investigation employed methodologies to explore scalar field theories at the background level for exact spacetimes. Specifically, an exhaustive dynamical system analysis for each scalar perturbation was presented, and we have integrated the different subsystems numerically. It has also been shown that one can generalize this gauge invariant argument \cite{Mishra:2025} to modified gravity as well. These are powerful tools for investigating homogeneous scalar field cosmologies with arbitrary potential.

To finish this section, we compare the procedures presented in \cite{Alho:2020cdg} for the generic potential with the method of $f$-devisers used here. 
Following the notations of \cite{Alho:2020cdg}, we are defining two quantities for an arbitrary potential as follows, 
\begin{equation}
    \lambda_\phi = - \frac{\sqrt{6}}{6} \frac{d \ln V(\phi)}{d \phi } \,\,\text{,} \quad \Upsilon_\phi= \frac{V''(\phi)}{6 V(\phi)} \,\,\text{.}
\end{equation}
Compared with Eq.~\eqref{sdef} and Eq.~\eqref{fdef}, we have 
\begin{equation}
  \lambda= \sqrt{6}  \lambda_\phi \,\,\text{,} \quad f = 6(\Upsilon_\phi-\lambda_\phi^2) \,\,\text{.}  \label{(eq190)}
\end{equation}
Therefore, the conditions upon the scalars   $\lambda_\phi, \Upsilon_\phi$ to satisfy 
\begin{equation}
    \lim_{\phi \rightarrow \pm \infty} \lambda_\phi(\phi)= \lambda_{\pm} \,\,\text{,} \quad  \lim_{\phi \rightarrow \pm \infty}  \Upsilon_\phi (\phi)= \lambda_{\pm}^2 \,\,\text{,}
\end{equation}
are translated in our scenario to  
\begin{equation}
    \lim_{\phi \rightarrow \pm \infty} \lambda(\phi)= \sqrt{6} \lambda_{\pm} \,\,\text{,} \quad    \lim_{\phi \rightarrow \pm \infty} f (\phi)= 0 \,\,\text{.}
\end{equation}
Both cases reduce to a Potential that is asymptotically an exponential one, such as $\phi \rightarrow \pm \infty$. These asymptotically exponential potentials can be studied using the formalism of \cite{Alho:2020cdg} or the procedures previously introduced in \cite{Foster:1998sk}. Following the nomenclature and formalism
introduced in \cite{Foster:1998sk}, let $V:\mathbb{R}\rightarrow \mathbb{R}$ be a $C^2$ non-negative
function. Let there exist some $\phi_0>0$ for which $V(\phi)> 0$
 for all $\phi>\phi_0$ and some number $N$ such that the function
$W_V:[\phi_0,\infty)\rightarrow \mathbb{R},  W_V(\phi)=\frac{V^{\prime}(\phi)}{ V(\phi)} - N $
 satisfies $\lim_{\phi\rightarrow\infty}W_V(\phi)=0$.
Then we say that $V$ is Well Behaved at Infinity (WBI) of
exponential order $N$.

Assume that there are $\phi_{0}> 0$, and a coordinate transformation $\varphi=h(\phi)$,
with inverse $h^{(-1)}(\varphi)$, which maps the interval $[\phi_{0},\infty)$ onto
$(0, \delta]$, where $\delta=h(\phi_{0})$, satisfying $\lim
_{\phi\rightarrow+\infty}h(\phi)=0$, and has the following additional properties:
\begin{enumerate}
\item $h$ is $C^{k+1}$ and strictly decreasing,
\item
\begin{equation}
\overline{h^{\prime}}(\varphi)=\left\{
\begin{array}
[c]{cc}%
h^{\prime}(h^{(-1)}(\varphi))\,\,\text{,} & \varphi>0 \,\,\text{,}\\
\lim_{\varphi\rightarrow\infty} h^{\prime}(\varphi) \,\,\text{,} & \varphi=0 \,\,\text{,}
\end{array}
\right.  \label{eq23}%
\end{equation}
is $C^{k}$ on the closed interval $[0, \delta]$ and
\item $\frac{d \overline{h^{\prime}}}{d \varphi}(0)$ and the higher derivatives
$\frac{d^{m}\overline{h^{\prime}}}{d \varphi^{m}}(0)$ satisfy
\begin{equation}
\frac{d \overline{h^{\prime}}}{d \varphi}(0)=\frac{d^{m}\overline{h^{\prime}}}{d \varphi^{m}%
}(0)=0 \,\,\text{.}
\end{equation}
\end{enumerate}

\begin{table}
\begin{center}
\bigskip
\caption{ Simple examples of WBI behavior at large $\phi$. $n$ and $\lambda$ are arbitrary constants.}
\begin{tabular}{|l|l|l|l|l|}
\hline
 $V(\phi)$&$W_V(\phi)$ & $\varphi=h(\phi)$&$\overline{W_V}(\varphi)$ &$\overline{h^{\prime}}(\varphi)$ \\ \hline
$\left|\frac{\lambda}{n}\right|\phi^{n}$&$n\phi^{-1}$
 &$\phi^{-\frac{1}{2}}$
 &$n \varphi^2$&
$-\frac{1}{2}\varphi^3$\\[3pt]
$e^{\lambda\phi}$&0&$\phi^{-1}$&0&$-\varphi^2$\\[3pt]
$2e^{\lambda\sqrt{\phi}}$ &$\lambda\phi^{-\frac{1}{2}}$
&$\phi^{-\frac{1}{4}}$ &$\lambda \varphi^2$&$-\frac{1}{4}\varphi^5$\\[3pt]
$\left(A+(\phi-B)^2\right)e^{-\mu\phi}$ &
$\frac{2\left(\phi-B\right)}{A+(B-\phi)^2}$&$\phi^{-\frac{1}{2}}$
 &$-\frac{2\varphi^2\left(B\varphi^2-1\right)}{A \varphi^4+\left(B\varphi^2-1\right)^2}$&
$-\frac{1}{2}\varphi^3$\\[3pt]
$\left(1-e^{-\lambda^2\phi}\right)^2$ & $ -\frac{2 \lambda
^2}{1-e^{\lambda ^2 \phi }}$
&$\phi^{-1}$& $-\frac{2 \lambda ^2}{1-e^{\frac{\lambda ^2}{\varphi }}}$&$-\varphi^2$\\[3pt]
 $\ln{\phi}$&$(\phi\ln\phi)^{-1}$&$(\ln\phi)^{-1}$&$\varphi e^{-\frac{1}{\varphi}}$&$-\varphi e^{-\frac{2}{\varphi}}$\\[3pt]
$\phi^2\ln{\phi}$&$2\phi^{-1} +(\phi\ln\phi)^{-1}$&
$(\ln\phi)^{-1}$&$(2+\varphi)e^{-\frac{1}{\varphi}}$&$-\varphi e^{-\frac{2}{\varphi}}$\\[3pt]
$\begin{cases}
   V_0(\phi^4+M^4)   &  \textrm{if} \quad \phi<0 \\
   \frac{V_0 M^8}{\phi^4+M^4}  & \textrm{if} \quad \phi \geq 0
\end{cases}$ & $ -\frac{4|\phi|^3}{(\phi^4+M^4)}$ & $|\phi|^{-1}$ & $  -\frac{4|\varphi|}{(1+M^4\varphi^4)}  $ & $- \text{sgn}(\phi) \varphi^2 $\\[3pt]
\hline
\end{tabular}
\label{WBItransf}
\end{center}
\end{table}
Table \ref{WBItransf} displays simple examples of WBI behaviour at large $\phi.$
We see that, as $|\phi|\rightarrow \infty$, a potential function $V(\phi)$ satisfying Eq.~\eqref{(eq190)} is a function Well-Behaved at Infinity (WBI)  of exponential order $N_\pm= -\sqrt{6} \lambda_{\pm}$. 
Hence, we have to transform the system Eq.~\eqref{(18)}, Eq.~\eqref{(19)} and Eq.~\eqref{(20)} to a system well suited for the analysis at infinity  \cite{Foster:1998sk,Leon:2008de,Fadragas:2014mra}.  Then, using the procedures of \cite{Foster:1998sk,Leon:2008de,Fadragas:2014mra}
for $\phi>0$, taking the scalar field transformation  $\varphi= h(\phi)$ (satisfying conditions 1, 2, and 3 before) and replacing 
\begin{equation}
    \lambda \mapsto -\left(\overline{W_V} +N\right) \,\,\text{,} \label{lambda_varphi}
\end{equation} we obtain
\begin{align}
\frac{d x}{{d \bar{N}} }& =- (1-x^2)\left(1-\bar{Z}\right) \left[3 x + \sqrt{\frac{3}{2}} \left(\overline{W_V} +N\right)\right] \,\,\text{,} \label{(57)}\\
\frac{d\varphi}{{d \bar{N}} }& = \sqrt{6} x \overline{h^{\prime}}(\varphi) \left(1-\bar{Z}\right) \,\,\text{,} \label{(59)}\\
 \frac{d\bar{Z}}{d \bar{N}} & = 2\left(3x^2 - 1\right)\bar{Z}\left(1-\bar{Z}\right)^2 \,\,\text{.}
\end{align}
With the restriction $x^2+y^2=1$ and considering one of the following perturbation equations for $\theta$. 
From Eq.~\eqref{(eq:99b)}, 
\begin{align}
&\frac{d\theta}{d\bar{N}} = - \Bigg[\sin^2\theta + \left(7-3x^{2}- \sqrt{6}\left(\overline{W_V}(\varphi) +N\right)\left(\frac{1-x^2}{x}\right)\right) \sin\theta\cos\theta
    \nonumber\\
   & \qquad\qquad +\left(6\left(1-x^2\right)-\sqrt{\frac{3}{2}}\left(\overline{W_V} +N\right)\left(\frac{1-x^2}{x}\right)\right)\cos^2\theta\Bigg]\left(1-\bar{Z}\right)  - \bar{Z}\cos^2\theta \,\,\text{.}
\end{align}

From Eq.~\eqref{eq100}, the comoving curvature perturbation evolves as  
\begin{align}
& \frac{d\theta}{d\bar{N}} = - \left[\sin^2\theta - \sqrt{6}\left(\overline{W_V} +N\right)\left(\frac{1-x^2}{x}\right)\sin\theta\cos\theta\right]\left(1-\bar{Z}\right) - \bar{Z}\cos^2 \theta \,\,\text{.}
\end{align}
From Eq.~\eqref{(eq:103)}, by using the transformation Eq.~\eqref{lambda_varphi}, and 
\begin{equation}
f(\lambda) \mapsto \bar{f}(\varphi)=\left\{\begin{array}{cc}
f\left(-\left(\overline{W_V}(\varphi) +N\right)\right) \,\,\text{,} & \varphi>0 \,\,\text{,}\\
0 \,\,\text{,} & \varphi=0 \,\,\text{,}\end{array}\right. 
\end{equation}  and we obtain the following. 
\begin{align}
& \frac{d\theta}{d\bar{N}} =- \Bigg[ \sin^2\theta + 3\left(1-x^2\right)\sin\theta\cos\theta  \nonumber \\
& + 18\left(1-x^2\right) \left(  \frac{\bar{f}(\varphi)}{6}+\left(x+\frac{\left(\overline{W_V}(\varphi) +N\right)}{\sqrt{6}}\right)^2 \right)\cos^2\theta\Bigg]\left(1-\bar{Z}\right)   -\bar{Z} \cos^2\theta \,\,\text{,}  
\end{align}
defined in the phase-space $B\times P$, modulo $n\pi, n\in\mathbb{Z}$, where  the background space is 
\begin{equation}
B =  \left\{ (x, \varphi, \bar{Z})\in [-1,1] \times [0, h(\phi_{0})] \times [0,1]\right\}, 
\end{equation}
and the perturbation space is 
\begin{equation}
  P =  \left\{ \theta\in  [-\pi, \pi]\right\} \,\,\text{.}
\end{equation}
The three possible dynamical systems presented here have the same asymptotic behavior as those investigated in Sect. \ref{sect:4}. To investigate the potential $V_* \left(e ^ {\kappa \beta (1 - \tanh{(\phi/\beta)})}-1\right)$, $\beta>0$ of \cite{Alho:2020cdg}, we relax the condition $\frac{d \overline{h^{\prime}}}{d \varphi}(0)=0$, and define $h(\phi)=1-\tanh \left(\frac{\phi }{\beta }\right)$  to obtain   $\overline{W_V}(\varphi)=\frac{e^{\beta  \kappa  \varphi} (\beta  \kappa  (\varphi-2) \varphi+2)-2}{\beta  \left(e^{\beta  \kappa  \varphi}-1\right)}$, $\overline{h^{\prime}}(\varphi)=\frac{(\varphi-2) \varphi}{\beta }$ and $N=-2/\beta$ as $\phi\rightarrow \infty$. 

Finally, we have investigated cosmological perturbations in the presence of two matter components, e.g. a perfect fluid and a scalar field with exponential potential. As a drawback of this approach, we must concentrate on a particular cosmological epoch when only one matter component is dominant. In that sense, even though not generic, our subsequent analysis is still relevant when the Universe is a scalar field-dominated, e.g., during the early inflationary epoch or the late-time acceleration.  

Our future aim is to evaluate the viability of cosmological models by utilizing observational data from various sources such as Supernovae Ia, Cosmic Chronometers, baryon acoustic oscillations, and cosmic microwave background. We will comprehensively analyze beyond the traditional linear stability approach, exploring various gravitational and cosmological models. This analysis will include studying multiple-scale, slow-fast dynamics, averaging theory, and non-smooth dynamical systems.



\chapter{Concluding Remarks and Future Perspectives} 

\label{Chapter7} 

\lhead{Chapter 7. \emph{Concluding Remarks and Future Perspectives}} 

 \clearpage

\epigraph{``We have to remember that what we observe is not nature herself, but nature exposed to our method of questioning.''}{--- Werner Heisenberg, \textit{Physics and Philosophy} (1958), Ch.~3}

This thesis has been devoted to a systematic investigation of the nature of dark energy, dark matter, and cosmic acceleration within a unified, multi-layered framework. By combining modified gravity, scalar field dynamics, non-canonical field theories, and model-independent data reconstruction techniques, the overarching objective has been to explore whether the observed late-time acceleration of the Universe, along with the dark sector phenomenology, can be consistently explained beyond the standard $\Lambda$CDM paradigm while remaining compatible with current observational constraints and fundamental theoretical principles.

At a foundational level, the thesis demonstrates that modified gravity theories based on non-metricity, particularly within the $f(Q)$ framework, provide a robust geometric arena in which dark energy and dark matter phenomena can be modeled in a unified manner. By exploiting the coincident gauge and the simplicity of the non-metric scalar $Q=6H^2$ in a flat FLRW background, the theory offers a transparent extension of GR while preserving its well-tested limits. Throughout the thesis, consistency with Einstein gravity has been maintained by recovering the STEGR limit $f(Q) = -Q$, ensuring that deviations arise only where they are observationally motivated.

A central result of this work is the demonstration that Chaplygin gas-type equations of state emerge naturally across multiple physical realizations, ranging from scalar and DBI-type field constructions to BEC dark matter and effective fluid descriptions within modified gravity. This universality strongly supports the Chaplygin gas as a powerful phenomenological bridge between dark matter-dominated and dark energy-dominated epochs. Within the interacting dark sector scenario developed in this thesis, the parameter $m$ plays a decisive role: values of $m \approx -5$ correspond to dark energy domination driven by modified gravity effects, while $m \approx 2$ corresponds to BEC-like dark matter behavior. The existence of these two observationally viable branches highlights a nontrivial degeneracy in late-time cosmology, one that cannot be resolved without precise data analysis and null diagnostics.

Through a comprehensive observational study employing Hubble data, Pantheon+SH0ES supernovae, and MCMC techniques, it has been shown that both branches are phenomenologically consistent with $\Lambda$CDM at the background level. However, model selection criteria such as $\Delta$AIC and $\Delta$BIC consistently favor the BEC-dominated interacting scenario, indicating that dark matter-to-dark energy conversion plays a statistically significant role. The analysis of statefinder and Om diagnostics further confirms that the models pass through a transient phantom regime before asymptotically approaching a de Sitter phase, in perfect agreement with late-time cosmic acceleration.

A second major pillar of this thesis is the model-independent reconstruction of dark energy dynamics using Gaussian processes. By reconstructing $H(z)$, the equation of state $\omega(z)$, the density parameter $\Omega(z)$, and the DBI scalar field potential directly from observational datasets (including 32 Cosmic Chronometers, 26 BAO points, and DESI data), this work avoids the theoretical biases inherent in rigidly parametric cosmological models. The reconstructed equation of state $\omega_\phi \approx -0.95$ provides a strong phenomenological justification for DBI-type scalar fields, which naturally approach $\omega \to -1$ in the slow-roll regime. The subsequent MCMC fitting of reconstructed potentials reveals that a wide class of scalar field models---including exponential, power-law, free-field (quadratic), and Higgs-like potentials---remain compatible with observations, provided a nonzero vacuum energy term is present. This recurrent appearance of a term $\mathcal{V}_0$ reinforces the conclusion that vacuum energy, whether fundamental or emergent, remains an indispensable ingredient of late-time cosmology.

Importantly, the GP-based reconstruction also opens a highly relevant pathway toward testing quantum gravity consistency conditions, such as the Swampland conjectures proposed within string theory. Because the GP reconstruction is strictly non-parametric, it provides a purely data-driven arena in which theoretical bounds on scalar potentials (such as the de Sitter and refined de Sitter conjectures) can be empirically examined without imposing specific ultraviolet assumptions. This makes the approach particularly valuable in light of the ongoing tensions between observable de Sitter-like cosmology and string-theoretic expectations.

The dynamical systems analyses presented in the later chapters further strengthen the theoretical foundations of the thesis. By studying scalar field cosmologies in $f(Q)$ gravity using phase-space techniques, it has been shown that viable cosmic histories---including stiff-fluid, matter-dominated, and de Sitter phases arise naturally for well-motivated choices of the gravitational function and scalar potential. Exponential potentials consistently yield more robust and realistic evolutionary sequences than power-law potentials, although both classes can successfully reproduce late-time acceleration. The inclusion of DBI scalar fields within this framework demonstrates that non-canonical kinetic terms do not obstruct cosmic viability; rather, they enrich the structure of fixed points and transition dynamics.

Extending beyond background evolution, the thesis has also addressed linear scalar cosmological perturbations using a fully dynamical systems approach. By constructing a product state space that unifies background and perturbation variables, the analysis provides a global picture of the evolution of gauge-invariant quantities such as the Bardeen potentials, comoving curvature perturbation, and the Sasaki-Mukhanov variable. The use of the $f$-deviser method enables a systematic treatment of arbitrary scalar field potentials, showing that asymptotically exponential potentials act as universal attractors in the space of models. This result establishes exponential-like behavior as a structural backbone of scalar field cosmology rather than a merely phenomenological choice.

\section*{Chapter-by-Chapter Summary}

To provide a granular view of the theoretical and observational milestones achieved in this work, the core findings are summarized chapter by chapter as follows:

\textbf{Chapter~\ref{Chapter1}: Introduction} \\
This opening chapter established the theoretical and observational foundations of the thesis. We reviewed the remarkable successes of the standard $\Lambda$CDM model alongside its persistent theoretical challenges, including the cosmological constant problem, the unknown fundamental nature of dark matter, and emerging observational crises like the $H_0$ tension. This chapter motivated the exploration of alternative cosmological frameworks, specifically introducing symmetric teleparallel $f(Q)$ gravity, non-canonical scalar fields, and the mathematical machinery of dynamical systems and Gaussian processes utilized in subsequent chapters.

\textbf{Chapter~\ref{Chapter2}: Observational Constraints on Dissipative Chaplygin Gas Cosmology in Coincident $f(Q)$ Gravity} \\
In this chapter, we investigate a unified dark sector framework by confronting a dissipative Chaplygin gas model with recent observational data. By incorporating an interaction term between dark matter and dark energy, we performed robust MCMC analyses using CC and Pantheon+SH0ES datasets. We demonstrated that the model exhibits two distinct viable branches: a modified gravity-driven dark energy phase ($m \approx -5$) and a BEC-like dark matter phase ($m \approx 2$). Crucially, statistical information criteria ($\Delta$AIC and $\Delta$BIC) favored the interacting BEC scenario, which successfully navigates late-time cosmic acceleration while passing rigorous statefinder and Om diagnostics.

\textbf{Chapter~\ref{Chapter3}: Reconstruction of Dark Energy Using Hubble and DESI Data for Dirac-Born-Infeld Scalar Field via Gaussian Process} \\
Transitioning to a purely data-driven methodology, this chapter utilized Gaussian Processes to reconstruct the cosmological expansion history without relying on rigid parametric assumptions. By extracting $H(z)$, $\omega(z)$, and the DBI scalar field potential directly from 32 Cosmic Chronometers, 26 BAO points, and DESI data, we found compelling phenomenological support for slow-rolling DBI fields ($\omega_\phi \approx -0.95$). Subsequent analysis of various physically motivated potentials confirmed that a nonzero vacuum energy ($\mathcal{V}_0$) is strictly required to match observations, simultaneously providing a non-parametric testing ground for quantum gravity Swampland conjectures.

\textbf{Chapter~\ref{Chapter4}: Dynamical System Analysis of Scalar Field Cosmology in Coincident $f(Q)$ Gravity} \\
Here, we shifted our focus to the rigorous mathematical framework of autonomous dynamical systems. By casting the cosmological equations of scalar field $f(Q)$ gravity into a phase-space representation, we mapped the evolutionary trajectories of the Universe. The analysis proved that observationally viable cosmic histories -- smoothly transitioning from a stiff-fluid epoch to a matter-dominated phase, and finally arriving at a stable de Sitter attractor --emerge naturally. We established that exponential potentials provide significantly more robust and stable evolutionary sequences toward late-time acceleration compared to power-law alternatives.

\textbf{Chapter~\ref{Chapter5}: Dynamical System Analysis of Dirac-Born-Infeld Scalar Field Cosmology in Coincident $f(Q)$ Gravity} \\
Building upon the previous chapter, we extended the phase-space methodology to encompass the non-canonical DBI scalar field. We demonstrated that the introduction of non-canonical kinetic terms, governed by the DBI Lorentz-like factor, dynamically enriches the phase space without destroying the viability of the cosmic evolution. The interplay between the non-metricity of spacetime and the relativistic limits of the scalar field was shown to generate novel stable attractors, further cementing the DBI field as a highly versatile and theoretically sound dark energy candidate.

\textbf{Chapter~\ref{Chapter6}: Scalar Field Evolution at Background and Perturbation Levels For a Broad Class of Potentials} \\
Moving beyond homogeneous background dynamics, this chapter provided a comprehensive analysis of dynamical systems incorporating linear cosmological perturbations. By constructing a unified product state space, we systematically tracked the evolution of crucial gauge-invariant quantities such as the Bardeen potentials and the Sasaki-Mukhanov variable. Using the $f$-deviser method, we analyzed a broad spectrum of arbitrary scalar field potentials and mathematically proved that asymptotically exponential potentials function as universal attractors. This key result confirmed that exponential-like late-time behavior is a fundamental structural backbone of scalar field cosmology.

\textbf{Chapter~\ref{Chapter7}: Conclusion} \\
This final chapter (the present one) synthesizes the overarching narrative of the thesis, tying together the phenomenological successes and theoretical insights derived from modified gravity, scalar field dynamics, observational reconstructions, and dynamical phase-space analyses.

Along with these seven chapters, there are four additional appendices that give all the theoretical details regarding $f(Q)$ gravity, which we could not fit in the main chapter of the thesis.

\textbf{Appendix~\ref{AppendixA}: Field Equations for $f(Q)$ Gravity} \\
This appendix outlines the variational derivation of the gravitational field equations within the coincident $f(Q)$ framework. We expand the non-metricity scalar and define the relevant deformation tensor and superpotential in the gauge where the affine connection vanishes globally. By varying the action with respect to the metric, we recover the general covariant field equations that provide the mathematical basis for the derivations used throughout this thesis.

\textbf{Appendix~\ref{AppendixB}: Cosmology in $f(Q)$ gravity} \\
Here, we apply the general $f(Q)$ field equations to a spatially flat FLRW background. After calculating the non-vanishing components of the non-metricity tensor and the superpotential, we show that the scalar reduces to $Q=6H^2$. These results are then used to formulate the modified Friedmann equations, which govern the cosmological dynamics and observational constraints analyzed in the main text.

\textbf{Appendix~\ref{AppendixC}: Degrees of Freedom} \\
This appendix addresses the counting of physical degrees of freedom in the STEGR. Using Hamiltonian constraint analysis via the ADM formalism and gauge-invariant linear perturbations, we demonstrate that the requirement of a flat, torsion-free connection does not introduce extra dynamical modes. The analysis confirms that STEGR propagates only two degrees of freedom, consistent with the massless spin-2 graviton of standard General Relativity.

\textbf{Appendix~\ref{AppendixD}: Linear Approximation} \\
We derive the STEGR action by enforcing consistency with the weak-field limit of General Relativity. Starting from a general quadratic Lagrangian built from the five independent non-metricity contractions, we linearize the theory around a Minkowski background in the coincident gauge. By matching the result to the Fierz-Pauli action for a free massless spin-2 field, we fix the coupling coefficients and establish STEGR as a ghost-free teleparallel equivalent of linearized gravity.

\begin{figure}[H]
{\includegraphics[width=8cm,height=5cm]{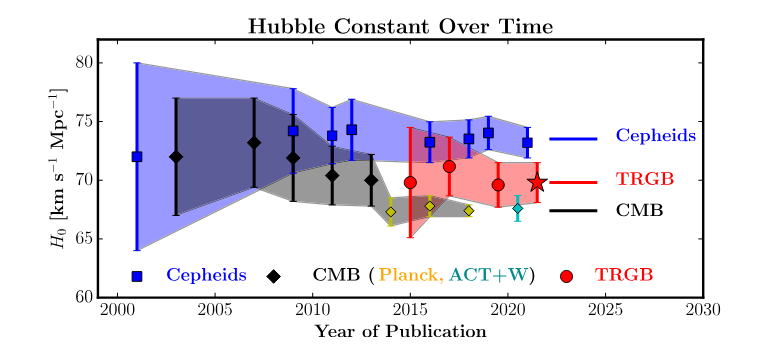}}
{\includegraphics[width=7.5cm,height=7cm]{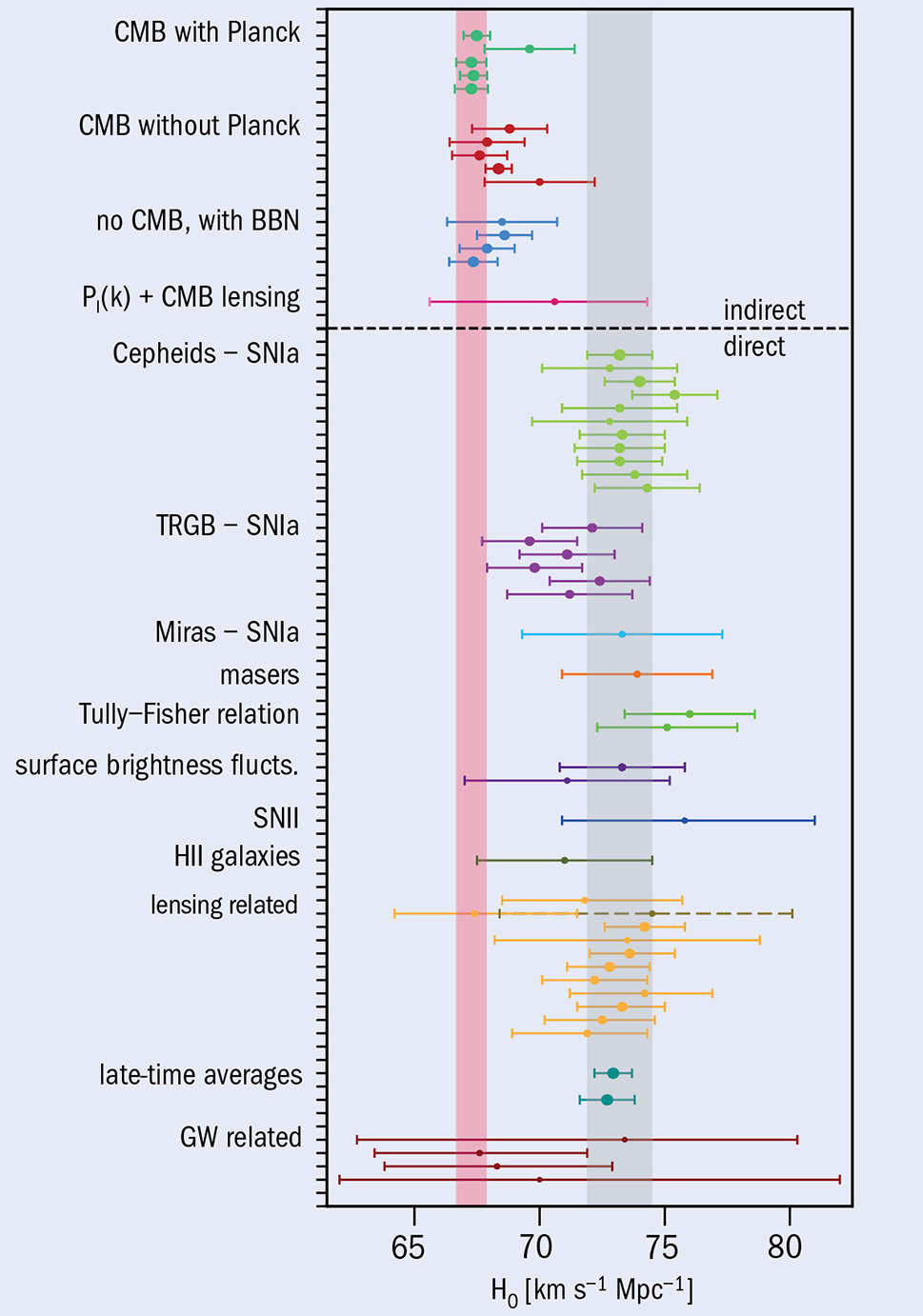}}
\caption{On the left side, we show roughly year-by-year progression of the tension via three independent methods of measuring $H_0$ \cite{Freedman:22}, and on the right, we show the rough status of tension with respect to various observations \cite{DiValentino:2021izs}. }\label{fig_7.1} 
\end{figure}

\section*{Overarching Synthesis and Future Outlook}

Taken together, the results of this thesis strongly suggest that a unified description of the dark sector is not only possible but theoretically well-motivated when modified gravity, scalar field dynamics, and interacting fluids are treated within a coherent mathematical framework. The convergence of independent approaches, observational reconstruction, phase-space analysis, null diagnostics, and model selection lends exceptional robustness to the conclusions drawn here. As illustrated in Fig.~ \ref{fig_7.1}, persistent discrepancies in the Hubble constant ($H_0$) and the structure growth parameter ($\sigma_8$) underscore the need for such extensions to the standard cosmological model \cite{Freedman:22, DiValentino:2021izs}.

The current era of precision cosmology finds itself in a state that Charles Dickens in the ``Tale of two cities" might have recognized: ``It was the best of times, it was the worst of times, it was the age of wisdom, it was the age of foolishness, it was the epoch of belief, it was the epoch of incredulity, it was the season of light, it was the season of darkness, it was the spring of hope, it was the winter of despair''. Record-breaking observational capabilities now coexist with persistent theoretical tensions—the $H_0$ and $\sigma_8$ discrepancies exemplify a field where the standard model is simultaneously exquisitely validated and fundamentally challenged. Yet this duality is not a paralysis but a catalyst. As Ya.~B.~Zel'dovich observed, ``the problem of cosmological analysis is to derive the observed present-day situation and structure of the Universe from certain plausible assumptions about its early behavior \cite[p.~115]{Zee1982}. This principle anchors the central thesis of this work: late-time dark sector phenomenology cannot be disentangled from primordial dynamics. By unifying modified gravity, non-canonical scalar fields, and interacting fluids within a coherent framework, we have shown that resolving contemporary anomalies requires tracing their imprints to the early Universe. The convergence of next-generation surveys, high-precision CMB constraints, and the emerging cosmological collider program will ultimately test whether these early-universe assumptions can naturally evolve into the late-time cosmos we observe today.

Looking ahead, several natural and exciting extensions arise from this work. First, the formalisms developed here can be seamlessly extended to more general modified gravity frameworks, such as $f(Q,T)$, $f(Q,B)$, or $f(Q,T,L_m)$, to investigate whether the phenomenological successes of the $f(Q)$ dark sector hold universally when matter-geometry couplings or boundary terms are introduced. Second, the incorporation of radiation and baryonic matter at the perturbative level will be essential for testing these models against the highly precise Cosmic Microwave Background (CMB) power spectra. Third, the utilization of forthcoming, unprecedentedly high-precision datasets from next-generation observatories like DESI, Euclid, and the Vera C. Rubin Observatory (LSST) will be crucial. These datasets will not only tighten constraints on GP-reconstructed potentials but could conclusively resolve existing anomalies such as the Hubble tension \cite{Freedman:22} and the $\sigma_8$ tension observed in weak lensing surveys \cite{planck, Heymans:2021}.

Finally, a highly promising frontier lies in linking these non-canonical scalar field and modified gravity frameworks to the rapidly growing field of cosmological collider physics. Although this thesis has focused on late-time dynamics, the scalar degrees of freedom responsible for dark energy may also leave imprints during inflation. By analyzing the primordial non-Gaussianities and oscillatory signatures left by massive particles, future work can probe the ``cosmological collider'' regime. Building on the foundational formalism of Maldacena regarding primordial fluctuations \cite{Maldacena:2003}, and the specific cosmological collider framework established by Arkani-Hamed and Maldacena \cite{ArkaniHamed:2015}, recent advances by Ghoshal et al.\ \cite{Ghoshal:2020} have shown how to identify specific particle signatures in the squeezed limit. Future research could extract these new physics signatures, thereby providing a direct observational bridge between the late-time scalar dynamics explored in this thesis and the ultra-high-energy physics of the early Universe. Such a connection would elevate the dark sector from a phenomenological fix to a fundamental component of high-energy particle physics.

In conclusion, this thesis provides a comprehensive and internally consistent exploration of late-time cosmology beyond $\Lambda$CDM, demonstrating that modified gravity combined with advanced scalar field dynamics and data-driven reconstruction offers a flexible, robust, and highly promising framework for understanding the dark Universe.

\appendix

\chapter{Field equation for $f(Q)$ gravity} 

\label{AppendixA} 

\lhead{Appendix A. \emph{Field Equation}} 

\section{Geometric preliminaries}
We work on a manifold endowed with a metric \(g_{\mu\nu}\) and a general affine connection \(\Gamma^\alpha{}_{\mu\nu}\).
The non-metricity tensor and its traces are defined by,
\begin{align}
Q_{\alpha\mu\nu} &\equiv \nabla_{\alpha} g_{\mu\nu}\,\, , \\
Q_\alpha &\equiv Q_{\alpha}{}^{\mu}{}_{\mu}\,\, , \qquad
\widetilde Q_\alpha \equiv Q^{\mu}{}_{\alpha\mu}\,\, .
\end{align}
The affine connection can be decomposed as,
\begin{equation}
\Gamma^\alpha{}_{\mu\nu} = \{^{\;\alpha}{}_{\mu\nu}\} + K^\alpha{}_{\mu\nu} + L^\alpha{}_{\mu\nu}\,\, ,
\end{equation}
where \(\{^{\;\alpha}{}_{\mu\nu}\}\) is the Levi--Civita connection, \(K^\alpha{}_{\mu\nu}\) the contortion (torsion part) and \(L^\alpha{}_{\mu\nu}\) the dis formation (non-metricity part). In the symmetric teleparallel setting, we have the following,
\begin{equation}
R^\lambda{}_{\beta\mu\nu}(\Gamma)=0,\qquad T^\alpha{}_{\mu\nu}=0 \,\, ,
\end{equation}
so \(K^\alpha{}_{\mu\nu}=0\) and the disformation tensor is expressed in terms of the non-metricity,
\begin{equation}\label{eq:disformation}
L^\alpha{}_{\mu\nu}=\frac{1}{2} g^{\alpha\rho}\big(-Q_{\mu\rho\nu}-Q_{\nu\rho\mu}+Q_{\rho\mu\nu}\big) \,\, .
\end{equation}
From the definition of $Q_{\alpha\mu\nu}$ we can form two independent traces are defined as,
\begin{align}
Q_{\alpha} &= {Q}_{\alpha}\,^{\mu}\,_{\mu} = g^{\mu\nu}Q_{\alpha\mu\nu}\,\, , \\
\tilde{Q}_{\alpha} &= Q^{\mu}\,_{\alpha \mu} = g^{\mu\nu}Q_{\mu\alpha\nu}\,\, ,
\end{align}
The super potential $P$ is defined as,
\begin{equation}
    P^{\alpha}\,_{\mu\nu}=\frac{1}{2}\frac{\partial Q}{\partial Q_{\alpha}\,^{\mu\nu}}=-\frac{1}{4}Q^{\alpha}\,_{\mu\nu}+\frac{1}{2}Q_{(\mu}\,^{\alpha}\,_{\nu)}+\frac{1}{4}g_{\mu\nu}Q^{\alpha}-\frac{1}{4}\left(g_{\mu\nu}\tilde{Q}^{\alpha}+\delta^{\alpha}\,_{(\mu}Q_{\nu)}\right) \,\, .
\end{equation}
Or more explicitly as follows,
\begin{equation}
\boxed{P^{\alpha}_{\phantom{\alpha}\mu\nu} = \frac{1}{4}\left( -Q^{\alpha}_{\phantom{\alpha}\mu\nu} + 2Q_{(\mu\phantom{\alpha}\nu)}^{\phantom{(\mu}\alpha} + Q^{\alpha}g_{\mu\nu} - \tilde{Q}^{\alpha}g_{\mu\nu} - \delta^{\alpha}_{(\mu}Q_{\nu)} \right)}
\end{equation}
From this we can get,
\begin{equation}
P^{\alpha\gamma\delta}=g^{\mu\gamma}g^{\nu\delta}P^{\alpha}\,_{\mu\nu}=\frac{1}{4}\left(-Q^{\alpha\gamma\delta}+2Q^{(\gamma|\alpha|\delta)}+Q^{\alpha}g^{\gamma\delta}-\tilde{Q}^{\alpha}g^{\gamma\delta}-\frac{1}{2}(g^{\alpha\gamma}Q^{\delta}+g^{\alpha\delta}Q^{\gamma})\right) \,\, .
\end{equation}
The non-metricity scalar $Q$ is expressed via the superpotential as,
\begin{equation}
\boxed{Q = -Q_{\alpha\mu\nu}P^{\alpha\mu\nu}=-Q^{\alpha\mu\nu}P_{\alpha\mu\nu}}
\end{equation}
Explicitly expanded, the scalar is as follows,
\begin{align}
    Q&=-Q_{\alpha\beta\gamma}g^{\beta\mu}g^{\gamma\nu}\frac{1}{4}\left( -Q^{\alpha}_{\phantom{\alpha}\mu\nu} + 2Q_{(\mu\phantom{\alpha}\nu)}^{\phantom{(\mu}\alpha} + Q^{\alpha}g_{\mu\nu} - \tilde{Q}^{\alpha}g_{\mu\nu} - \delta^{\alpha}_{(\mu}Q_{\nu)} \right)\\
    &=\frac{1}{4}[Q_{\alpha\mu\nu}Q^{\alpha\mu\nu}-Q_{\alpha\mu\nu}Q^{\mu\alpha\nu}-Q_{\alpha\mu\nu}Q^{\nu\alpha\mu}-Q_{\alpha\mu\nu}Q^\alpha g^{\mu\nu}+Q_{\alpha\mu\nu}\tilde{Q}^{\alpha}g^{\mu\nu}+Q_{\alpha\mu\nu}\delta^{\alpha(\mu}Q^{\nu)}] \nonumber \\
    &=\frac{1}{4}Q_{\alpha\mu\nu}Q^{\alpha\mu\nu}-\frac{1}{2}Q_{\alpha\mu\nu}Q^{\mu\alpha\nu}-\frac{1}{4}Q_{\alpha}Q^{\alpha}+\frac{1}{4}Q_{\alpha}\tilde{Q}^{\alpha}+\frac{1}{8}[\tilde{Q}_{\nu}Q^{\nu}+\tilde{Q}_{\mu}Q^{\mu}] \nonumber \\
    &= \frac{1}{4}Q_{\alpha\mu\nu}Q^{\alpha\mu\nu}-\frac{1}{2}Q_{\alpha\mu\nu}Q^{\mu\alpha\nu}-\frac{1}{4}Q_{\alpha}Q^{\alpha}+\frac{1}{2}Q_{\alpha}\tilde{Q}^{\alpha} \nonumber \,\, .
\end{align}
So we get,
\begin{equation}
\boxed{Q = \frac{1}{4}Q_{\alpha\mu\nu}Q^{\alpha\mu\nu} - \frac{1}{2}Q_{\alpha\mu\nu}Q^{\mu\alpha\nu} - \frac{1}{4}Q_{\alpha}Q^{\alpha} + \frac{1}{2}Q_{\alpha}\tilde{Q}^{\alpha}} 
\end{equation}
\section{Variation of each term}
\subsection{Variation of $Q_{\alpha\mu\nu}$}
\begin{equation}
    \delta Q_{\alpha\mu\nu}=\delta \partial_{\alpha}g_{\mu\nu}=\partial_{\alpha}\delta g_{\mu\nu} \,\, .
\end{equation}
\subsection{Variation of $Q^{\alpha\mu\nu}$}
\begin{align*}
\delta Q^{\alpha\mu\nu}
&= \delta\!\left(g^{\alpha\beta} g^{\mu\gamma} g^{\nu\delta} Q_{\beta\gamma\delta}\right) \\[6pt]
&= g^{\alpha\beta} g^{\mu\gamma} g^{\nu\delta} \, \delta Q_{\beta\gamma\delta}
   + Q_{\beta\gamma\delta} \, \delta\!\left(g^{\alpha\beta} g^{\mu\gamma} g^{\nu\delta}\right) \\[6pt]
&= g^{\alpha\beta} g^{\mu\gamma} g^{\nu\delta} \, \partial_\beta \delta g_{\gamma\delta}
   + Q_{\beta}\,^{\mu\nu} \, \delta g^{\alpha\beta}
   + Q^{\alpha}\,_{\beta}\,^{\nu} \, \delta g^{\mu\beta}
   + Q^{\alpha\mu}\,_{\beta} \, \delta g^{\nu\beta} \\[6pt]
&= g^{\alpha\beta} g^{\mu\gamma} g^{\nu\delta} \, \partial_\beta \delta g_{\gamma\delta}
   - \left(
       g^{\gamma\alpha} Q^{\delta\mu\nu}
       + g^{\gamma\mu} Q^{\alpha\delta\nu}
       + g^{\gamma\nu} Q^{\alpha\mu\delta}
     \right) \delta g_{\gamma\delta} \,\, .
\end{align*}
Where we have used the fact,
\begin{align*}
    g^{\alpha\gamma} g_{\gamma\beta} &=\delta^\alpha_\beta\\
    \delta g^{\alpha\gamma} \cdot g_{\gamma\beta} + g^{\alpha\gamma} \cdot \delta g_{\gamma\beta} &= 0\\
    \delta g^{\alpha\gamma} \delta_\gamma^\delta + g^{\alpha\gamma} g^{\beta\delta} \delta g_{\gamma\beta} &= 0\\
    \delta g^{\alpha\delta} &= -g^{\alpha\gamma} g^{\beta\delta} \delta g_{\gamma\beta} \,\, .
\end{align*}
\subsection{Variation of $Q_{\alpha}$}
\begin{align*}
\delta Q_{\alpha}
&= \delta(g^{\mu\nu} Q_{\alpha\mu\nu}) \\[6pt]
&= g^{\mu\nu} \delta Q_{\alpha\mu\nu} + Q_{\alpha\mu\nu} \delta g^{\mu\nu} \\[6pt]
&= g^{\mu\nu} \partial_\alpha \delta g_{\mu\nu} - Q_{\alpha\mu\nu} (g^{\mu\rho} g^{\nu\sigma} \delta g_{\rho\sigma}) \\[6pt]
&= g^{\mu\nu} \partial_\alpha \delta g_{\mu\nu} - Q_{\alpha}{}^{\rho\sigma} \delta g_{\rho\sigma}\\
&=g^{\gamma\delta}\partial_{\alpha}\delta g_{\gamma\delta}-g^{\mu\gamma}g^{\nu\delta}Q_{\alpha\mu\nu}\delta g_{\gamma\delta}\,\, .
\end{align*}
\subsection{Variation of $Q^{\alpha}$}
\begin{align*}
    \delta Q^{\alpha}= \delta (g^{\mu\nu}Q^{\alpha}\,_{\mu\nu})&=\delta(g^{\mu\nu}g^{\alpha\lambda}Q_{\lambda\mu\nu})\\
    &=-g^{\mu\lambda}g^{\nu\delta}Q^{\alpha}\,_{\mu\nu}\delta g_{\lambda\sigma}+g^{\mu\nu}g^{\alpha\lambda}\partial_{\lambda}\delta g_{\mu\nu}+g^{\mu\nu}Q_{\lambda\mu\nu}\delta g^{\alpha\lambda}\\
    &= -g^{\mu\lambda}g^{\nu\delta}Q^{\alpha}\,_{\mu\nu}\delta g_{\lambda\sigma}-g^{\mu\nu}Q_{\lambda\mu\nu}g^{\alpha\beta}g^{\lambda\sigma}\delta g_{\beta\sigma}+g^{\mu\nu}Q_{\lambda\mu\nu}\delta g^{\alpha\lambda}\\
    &= -g^{\mu\lambda}g^{\nu\delta}Q^{\alpha}\,_{\mu\nu}\delta g_{\lambda\sigma}-Q_{\lambda}g^{\alpha\mu}g^{\lambda\nu}\delta g_{\mu\nu}+g^{\mu\nu}g^{\alpha\lambda}\partial_{\lambda}\delta g_{\mu\nu}\\
    &= g^{\gamma\delta}g^{\alpha\lambda}\partial_{\lambda}\delta g_{\gamma\delta}-Q^{\alpha\gamma\delta}\delta g_{\gamma\delta}-Q_{\lambda}g^{\alpha\gamma}g^{\gamma\delta}\delta g_{\gamma\delta} \,\, .
\end{align*}
\subsection{Variation of $\tilde{Q}^{\alpha}$}
\begin{align*}
    \delta \tilde{Q}^{\alpha}= \delta(g^{\mu\nu}Q_{\mu\nu}\,^{\alpha})=\delta(g^{\mu\nu}g^{\alpha\lambda}Q_{\mu\nu\alpha})&= g^{\alpha\lambda}Q_{\mu\nu\lambda}\delta g^{\mu\nu}+g^{\mu\nu}Q_{\mu\nu\lambda}\delta g^{\alpha\lambda}+g^{\mu\nu}g^{\alpha\lambda}\partial_{\mu}\delta g_{\nu\lambda}\\
    &= -g^{\gamma\mu}g^{\delta\nu}Q_{\mu\nu}\,^{\alpha}\delta g_{\gamma\delta} -\tilde{Q}_{\lambda}g^{\gamma\alpha}g^{\delta\lambda}\delta g^{\gamma\delta}+g^{\mu\nu}g^{\alpha\lambda}\partial_{\mu}\delta g_{\nu\lambda}\\
    &=-Q^{\gamma\delta\alpha}\delta g_{\gamma\delta}-\tilde{Q}^{\delta}g^{\gamma\alpha}\delta g_{\gamma\delta}+g^{\mu\gamma}g^{\alpha\delta}\partial_{\mu}\delta g_{o\gamma\delta} \,\, .
\end{align*}
\subsection{Variation of $\tilde{Q}_{\alpha}$}
\begin{align*}
    \delta \tilde{Q}_{\alpha}=\delta(g^{\mu\nu}Q_{\mu\nu\alpha})&=\delta g^{\mu\nu}Q_{\mu\nu\alpha}+g^{\mu\nu}\partial_{\mu}\delta g_{\nu\alpha}\\
    &=-g^{\gamma\mu}g^{\delta\nu}Q_{\mu\nu\alpha}\delta g_{\gamma\delta}+g^{\mu\nu}\partial_{\mu} \delta g_{\nu\alpha}\\
    &=-Q^{\gamma\delta}\,_{\alpha}\delta g_{\gamma\delta}+g^{\mu\gamma}\partial_{\mu}\delta g_{\gamma\alpha} \,\, .
\end{align*}
\subsection{Miscellaneous identities that would be used}
In order to derive the full field equation from the variational principle, here are some of the identities we have used in various steps,
\begin{equation*}
    g^{\alpha\gamma}Q^{\delta}\partial_{\alpha}\delta g_{\gamma\delta} = \frac{1}{2} (g^{\alpha\gamma}Q^{\delta}\partial_{\alpha}\delta g_{\gamma\delta}+g^{\alpha\gamma}Q^{\delta}\partial_{\alpha}\delta g_{\delta\gamma})
    = \frac{1}{2}(g^{\alpha\gamma}Q^{\delta}+g^{\alpha\delta}Q^{\gamma})\partial_{\alpha}\delta g_{\gamma\delta} \,\, .
\end{equation*}
\begin{equation*}
    -g_{\beta\alpha}\delta^{\beta}\,_{(\mu}Q_{\nu)}=\frac{1}{2}g_{\mu\alpha}Q_{\nu}-\frac{1}{2}g_{\alpha\nu}Q_{\mu} \,\, .
\end{equation*}
\begin{equation*}
    g^{\mu\gamma}g^{\nu\delta}\delta^{\alpha}\,_{(\mu}Q_{\nu)}= \frac{1}{2}(g^{\alpha\gamma}g^{\nu\delta}Q_{\nu}+g^{\mu\gamma}g^{\alpha\delta}Q_{\mu})
    = \frac{1}{2}(g^{\alpha\gamma}Q^{\delta}+g^{\alpha\delta}Q^{\gamma}) \,\, .
\end{equation*}
Also, we have used these facts,
\begin{align*}
    P^{\alpha}\,_{\lambda\sigma}g_{\gamma\mu}g_{\delta\nu}(-g^{\sigma\delta}Q_{\alpha}\,^{\lambda\sigma}-g^{\lambda\gamma}Q_{\alpha}\,^{\sigma\delta})&=-P^{\alpha}\,_{\lambda\sigma}g_{\gamma\mu}g_{\delta\nu} g^{\sigma\delta}Q_{\alpha}\,^{\lambda\gamma}-P^{\alpha}\,_{\lambda\sigma}g_{\gamma\nu}g_{\delta\mu}g^{\lambda\gamma}Q_{\alpha}\,^{\sigma\delta}\\
    &= -2P^{\alpha}\,_{\lambda\sigma}g_{\gamma\nu}g_{\delta\mu}g^{\sigma\delta}Q_{\alpha}\,^{\lambda\gamma} \nonumber \,\, ,
    \end{align*}
and,
\begin{equation*}
        \partial_{\alpha}g^{\lambda\gamma}g^{\sigma\delta}+g^{\lambda\gamma}\partial_{\alpha}g^{\sigma\delta}=-g^{\sigma\delta}Q_{\alpha}\,^{\lambda\gamma}-g^{\lambda\gamma}Q_{\alpha}\,^{\sigma\delta} \,\, ,
\end{equation*}
and,
\begin{align*}
P^{\alpha}\,_{\lambda\sigma}g_{\gamma\nu}g_{\delta\mu}g^{\lambda\nu}Q_{\alpha}\,^{\sigma\delta}&=P^{\alpha}\,_{\lambda\sigma}g_{\delta\mu}g_{\gamma\nu}g^{\lambda\nu}Q_{\alpha}\,^{\sigma\delta}\\
    &=P^{\alpha}\,_{\sigma\lambda}g_{\gamma\mu}g_{\delta\nu}g^{\lambda\delta}Q_{\alpha}\,^{\sigma\nu}  \nonumber \\
    &=P^{\alpha}\,_{\lambda\sigma}g_{\gamma\mu}g_{\delta\nu}g^{\sigma\delta}Q_{\alpha}\,^{\lambda\gamma} \,\, , \nonumber
\end{align*}
\section{Variation of $Q$}
We note that,
\begin{equation}
    Q=-Q_{\alpha\mu\nu}P^{\alpha\mu\nu}=-Q^{\alpha\mu\nu}P_{\alpha\mu\nu} \,\, .
\end{equation}

So, the variation of $Q$ is given by,
\begin{align*}
    \delta Q&=-\delta Q^{\alpha\mu\nu}P_{\alpha\mu\nu}-Q^{\alpha\mu\nu}\delta P_{\alpha\mu\nu}\\
            &= -\delta Q^{\alpha\mu\nu}P_{\alpha\mu\nu}- P^{\alpha\mu\nu}\delta Q_{\alpha\mu\nu}\\
            &= -P_{\alpha\mu\nu}(g^{\alpha\beta}g^{\mu\gamma}g^{\nu\delta}\partial_{\beta}\delta g_{\gamma\delta}-(g^{\gamma\alpha}Q^{\delta \mu\nu}+g^{\gamma\mu}Q^{\alpha \delta\nu}+g^{\gamma\nu}Q^{\alpha \mu\delta})\delta g_{\gamma\delta})-P^{\alpha\mu\nu}\partial_{\alpha}\delta g_{\mu\nu}\\
            &= -P^{\alpha\mu\nu}\partial_{\alpha}\delta g_{\gamma\delta}+(P^\gamma\,_{\mu\nu} Q^{\delta \mu\nu}+P_\alpha\,^\gamma\,_{\nu}Q^{\alpha \delta\nu}+P_{\alpha\mu}\,^{\gamma}Q^{\alpha \mu\delta})\delta g_{\gamma\delta}-P^{\alpha\gamma\delta}\partial_{\alpha}\delta g_{\gamma\delta}\\
            &=- 2P^{\alpha\mu\nu}\partial_{\alpha}\delta g_{\gamma\delta}+(P^{\gamma}\,_{\mu\nu}Q^{\delta\mu\nu}+2P_{\mu\nu}\,^{\gamma}Q^{\mu\nu\delta}) \,\, .
\end{align*}
Where in second line we have used this useful identity given as follows,
\begin{alignat*}{2}
Q^{\alpha\mu\nu} \delta P_{\alpha\mu\nu} &= Q^{\alpha\mu\nu} \frac{1}{4} \delta \left[ -Q_{\alpha\mu\nu} + 2 Q_{(\mu|\alpha|\nu)} + Q_\alpha g_{\mu\nu} - \tilde{Q}_\alpha g_{\mu\nu} - \frac{1}{2} g_{\mu\alpha} Q_\nu - \frac{1}{2} g_{\alpha\nu} Q_\mu \right] \nonumber \\
&= Q^{\alpha\mu\nu} \left[ -\partial_\alpha \delta g_{\mu\nu} + \partial_\mu \delta g_{\alpha\nu} + \partial_\nu \delta g_{\alpha\mu} + \delta Q_\alpha g_{\mu\nu} + Q_\alpha \delta g_{\mu\nu} - \delta \tilde{Q}_\alpha g_{\mu\nu} - \tilde{Q}_\alpha \delta g_{\mu\nu} \right. \nonumber \\
&\quad \left. - \frac{1}{2} Q_{\nu}\delta g_{\mu\alpha} - \frac{1}{2} g_{\mu\alpha} \delta Q_\nu - \frac{1}{2}  Q_\mu \delta g_{\alpha\nu} - \frac{1}{2} g_{\alpha\nu} \delta Q_\mu \right] \nonumber \\
&= - \left[ Q^{\alpha\gamma\delta} \partial_\alpha \delta g_{\gamma\delta} - Q^{\gamma\alpha\delta} \partial_\alpha \delta g_{\gamma\delta} - Q^{\gamma\delta\alpha} \partial_\alpha \delta g_{\gamma\delta} - Q^\alpha g^{\gamma\delta} \partial_\alpha \delta g_{\gamma\delta} + Q^\alpha g^{\mu\gamma} g^{\nu\delta}Q_{\alpha\mu\nu}\delta g_{\gamma\delta}  \right. \nonumber \\
&\quad \left. -Q^{\alpha\gamma\delta}Q_{\alpha}\delta g_{\mu\nu} + Q^\alpha \left( -Q^{\gamma\delta}_{\alpha}\delta g_{\gamma\delta}+g^{\mu\gamma}\partial_{\mu}\delta g_{\gamma\alpha}\right) + Q^{\alpha\mu\nu} \tilde{Q}_\alpha \delta g_{\mu\nu} + \frac{1}{2} Q^{\alpha\mu\nu} Q_{\nu} \delta g_{\mu\alpha} \right. 
\nonumber \\
&\quad \left.  + \frac{1}{2} \tilde{Q}^{\alpha} \left( g^{\gamma\delta} \partial_{\alpha}\delta g_{\gamma\delta} - Q_{\alpha}^{\gamma\delta} \delta g_{\gamma\delta} \right)+\frac{1}{2}Q^{\gamma\mu\delta}Q_{\mu}\delta g_{\gamma\delta}+\frac{1}{2} \tilde{Q}^{\alpha}(g^{\gamma\delta}\partial_{\alpha}\delta g_{\gamma\delta} -Q_{\alpha}^{\gamma\delta}\delta g_{\gamma\delta}) \right] \nonumber \\
&= - \left[ Q^{\alpha\gamma\delta} \partial_\alpha \delta g_{\gamma\delta} - Q^{\gamma\alpha\delta} \partial_\alpha \delta g_{\gamma\delta} - Q^{\delta\alpha\gamma} \partial_\alpha \delta g_{\gamma\delta} - Q^\alpha g^{\gamma\delta} \partial_\alpha \delta g_{\gamma\delta} + Q^{\delta} g^{\alpha\gamma} \partial_\alpha \delta g_{\gamma\delta}  \right. 
\nonumber \\
&\quad \left.  + \frac{1}{2}\tilde{Q}^{\alpha}g^{\gamma\delta}\partial_{\alpha}\delta g_{\gamma\delta}  +\frac{1}{2}\tilde{Q}^{\alpha}g^{\gamma\delta}\partial_{\alpha}\delta g_{\gamma\delta} -Q^{\alpha}Q^{\gamma\delta}_{\alpha}\delta g_{\gamma\delta} + \tilde{Q}_{\alpha}Q^{\alpha\gamma\delta}\delta g_{\gamma\delta} +\frac{1}{2} Q^{\delta\gamma\alpha} Q_{\alpha} \delta g_{\gamma\delta} \right. 
\nonumber \\
&\quad \left. - \frac{1}{2} \tilde{Q}_{\alpha} Q^{\alpha\gamma\delta} \delta g_{\gamma\delta} + \frac{1}{2} Q^{\gamma\alpha\delta} Q_{\alpha} \delta g_{\gamma\delta} - \frac{1}{2} \tilde{Q}_{\alpha} Q^{\alpha\gamma\delta} \delta g_{\gamma\delta}  \right] \nonumber \\
&= -\frac{1}{4} \left[ Q^{\alpha\gamma\delta}- 2Q^{(\gamma|\alpha|\delta)} - g^{\gamma\delta} Q^{\alpha} + g^{\gamma\delta} Q^{\alpha} + g^{\alpha\gamma} Q^{\delta}+\frac{1}{2}(g^{\alpha\gamma}Q^{\delta}+g^{\alpha\delta}Q^{\gamma}) \right] \partial_\alpha \delta g_{\gamma\delta}
\nonumber \\
&\quad -\frac{1}{4}[-\frac{1}{2}Q_{\alpha}Q^{\gamma\delta\alpha}+\frac{1}{2}\tilde{Q}_{\alpha}Q^{\alpha\gamma\delta}+\frac{1}{2}Q^{\delta\gamma\alpha}Q_{\alpha}-\frac{1}{2}\tilde{Q}_{\alpha}Q^{\alpha\gamma\delta}]\delta g_{\gamma\delta} \nonumber \\
&= -\frac{1}{4} \left[ Q^{\alpha\gamma\delta}- 2Q^{(\gamma|\alpha|\delta)} - g^{\gamma\delta} Q^{\alpha} + g^{\gamma\delta} Q^{\alpha} + g^{\alpha\gamma} Q^{\delta}+\frac{1}{2}(g^{\alpha\gamma}Q^{\delta}+g^{\alpha\delta}Q^{\gamma}) \right] \partial_\alpha \delta g_{\gamma\delta}-\frac{1}{4}[0]\delta g_{\gamma\delta} \nonumber \\
&= P^{\alpha\tau\sigma} \partial_\alpha \delta g_{\tau\sigma} = P^{\alpha\mu\nu} \delta Q_{\alpha\mu\nu} \\
&\Rightarrow \boxed{ Q^{\alpha\mu\nu} \delta P_{\alpha\mu\nu} = P^{\alpha\mu\nu} \delta Q_{\alpha\mu\nu} }
\end{alignat*}
\section{Full variation to get the field equation}
Now we do the variation of the full action as follows, 
 \begin{align*}
 & \delta\int\left(\frac{1}{2}f(Q)+\mathcal{L}_m\right)\sqrt{-g}
= 0 \\
\implies& \int\frac{1}{2}f(Q)\delta\sqrt{-g}
+ \frac{1}{2}\sqrt{-g}f_Q\delta Q
+ \delta(\sqrt{-g}\mathcal{L}_m)
= 0 \\
\implies&\frac{1}{2}\int \left(\frac{1}{2}f\sqrt{-g}g^{\gamma\delta}\right)\delta g_{\gamma\delta}+\frac{1}{2}\int \sqrt{-g} f_{Q}(P^{\gamma}\,_{\mu\nu}Q^{\delta\mu\nu}+2P_{\mu\nu}\,^{\gamma}Q^{\mu\nu\delta})\delta g_{\gamma\delta} \\ &\hspace{6cm} -\frac{1}{2}\int 2\sqrt{-g}f_{Q}P^{\alpha\gamma\delta}\partial_{\alpha}\delta g_{\gamma\delta} +\int \delta(\sqrt{-g}L_m)=0\\
\implies&\frac{1}{2}\int \left(\frac{1}{2}f\sqrt{-g}g^{\gamma\delta}\right)\delta g_{\gamma\delta}+\frac{1}{2}\int \sqrt{-g} f_{Q}(P^{\gamma}\,_{\mu\nu}Q^{\delta\mu\nu}+2P_{\mu\nu}\,^{\gamma}Q^{\mu\nu\delta})\delta g_{\gamma\delta} \\ &\hspace{6cm} +\frac{1}{2}2\int \partial_{\alpha}(\sqrt{-g}f_Q P^{\alpha\gamma\delta})\delta g_{\gamma\delta} +\int \delta(\sqrt{-g}L_m)=0\\
\implies& \left(\frac{1}{2}f\sqrt{-g}g^{\gamma\delta}\right)+\sqrt{-g} f_{Q}(P^{\gamma}\,_{\mu\nu}Q^{\delta\mu\nu}+2P_{\mu\nu}\,^{\gamma}Q^{\mu\nu\delta})+2 \partial_{\alpha}(\sqrt{-g}f_Q P^{\alpha\gamma\delta})=-2\frac{\partial(\sqrt{-g}L_m)}{\partial g_{\gamma\delta}}\\
\implies& \left(\frac{1}{2}f\sqrt{-g}g^{\gamma\delta}\right)+\sqrt{-g} f_{Q}(P^{\gamma}\,_{\mu\nu}Q^{\delta\mu\nu}+2P_{\mu\nu}\,^{\gamma}Q^{\mu\nu\delta})+2 \partial_{\alpha}(\sqrt{-g}f_Q P^{\alpha}\,_{\lambda\sigma}g^{\lambda\gamma}g^{\sigma\delta})=-\sqrt{-g}T^{\gamma\delta}\\
\implies&  g_{\mu\gamma}g_{\nu\delta}\left(\frac{1}{2}f\sqrt{-g}g^{\gamma\delta}\right)+\sqrt{-g}f_{Q}(P_{\mu\alpha\beta}Q_{\nu}\,^{\alpha\beta}+2P_{\alpha\beta\mu}Q^{\alpha\beta}\,_{\nu})+2g_{\gamma\mu}g_{\delta\nu}g^{\lambda\gamma}g^{\sigma\delta}\partial_{\alpha}(\sqrt{-g}f_{Q}P^{\alpha}\,_{\lambda\sigma}) \\ &\hspace{6cm} +2\sqrt{-g}f_{Q}P^{\alpha}\,_{\lambda\sigma}g_{\gamma\mu}g_{\delta\nu}\partial_{\alpha}(g^{\lambda\gamma} g^{\sigma\delta})=-g_{\mu\gamma}g_{\nu\delta}\sqrt{-g}T^{\gamma\delta}\\
\implies& g_{\mu\gamma}g_{\nu\delta}\left(\frac{1}{2}f\sqrt{-g}g^{\gamma\delta}\right) + \sqrt{-g} f_{Q}(P_{\mu\alpha\beta}Q_{\nu}\,^{\alpha\beta}+2P_{\alpha\beta\mu}Q^{\alpha\beta}\,_{\nu})+2 \partial_{\alpha}(\sqrt{-g}f_{Q} P^{\alpha}\,_{\mu\nu})=-\sqrt{-g}T_{\mu\nu}\\
\implies& \frac{1}{2}f\sqrt{-g}g_{\mu\nu}+\sqrt{-g}f_{Q}(P_{\mu\alpha\beta}Q_{\nu}\,^{\alpha\beta}+2P_{\alpha\beta\mu}Q^{\alpha\beta}\,_{\nu})+2 \partial_{\alpha}(\sqrt{-g}f_{Q} P^{\alpha}\,_{\mu\nu})=-\sqrt{-g}T_{\mu\nu}\\
\implies& \frac{f}{2}g_{\mu\nu}+f_{Q}(P_{\mu\alpha\beta}Q_{\nu}\,^{\alpha\beta}+2P_{\alpha\beta\mu}Q^{\alpha\beta}\,_{\nu})+\frac{2}{\sqrt{-g}}\partial_{\alpha}(\sqrt{-g}f_{Q} P^{\alpha}\,_{\mu\nu})=-T_{\mu\nu}\\
\implies&  \frac{f}{2}g_{\mu\nu}+f_{Q}(P_{\mu\alpha\beta}Q_{\nu}\,^{\alpha\beta}+2P_{\alpha\beta\mu}Q^{\alpha\beta}\,_{\nu})+2P^{\alpha}\,_{\mu\nu}f_{QQ}\partial_{\alpha}Q+\frac{2}{\sqrt{-g}}f_{Q}\partial_{\alpha}(\sqrt{-g}P^{\alpha}\,_{\mu\nu})=-T_{\mu\nu} \,\, .
\end{align*}
Using the standard definition $T_{\mu\nu}\equiv-\frac{2}{\sqrt{-g}}\frac{\delta(\sqrt{-g}\mathcal{L}_m)}{\delta g^{\mu\nu}}$ and the relation $\delta g^{\alpha\beta}=-g^{\alpha\mu}g^{\beta\nu}\delta g_{\mu\nu}$, the variation with respect to the covariant metric yields
\begin{equation}
    \frac{\delta(\sqrt{-g}\mathcal{L}_m)}{\delta g_{\mu\nu}} = -g^{\mu\alpha}g^{\nu\beta}\frac{\delta(\sqrt{-g}\mathcal{L}_m)}{\delta g^{\alpha\beta}} = -g^{\mu\alpha}g^{\nu\beta}\left(-\frac{1}{2}\sqrt{-g}\,T_{\alpha\beta}\right) = \boxed{\frac{1}{2}\sqrt{-g}\,T^{\mu\nu}}\,
\end{equation}
which explicitly produces the \textit{positive} contravariant energy-momentum tensor.

Where we first note that,
\begin{equation*}
P^{\lambda\beta}{}_{\nu} = -P^{\beta\lambda}{}_{\nu}\,\, , \qquad Q_{\lambda\beta\mu} = Q_{\beta\lambda\mu} \,\, .
\end{equation*}
Due to the antisymmetric property of the super potential.\\
From this we note that,
\begin{align*}
-2Q_{\lambda\beta\mu}P^{\lambda\beta}{}_{\nu}
&= -2Q_{\beta\lambda\mu}(-P^{\beta\lambda}{}_{\nu}) \nonumber\\
&= +2Q_{\beta\lambda\mu}P^{\beta\lambda}{}_{\nu} \nonumber\\
&= +2Q_{\alpha\beta\mu}P^{\alpha\beta}{}_{\nu} \nonumber\\
&= +2P_{\alpha\beta\mu}Q^{\alpha\beta}{}_{\nu} \,\, .
\end{align*}
So, we identify,
\begin{equation}
\therefore\quad P_{\mu\lambda\beta}Q_{\nu}{}^{\lambda\beta} - 2Q_{\lambda\beta\mu}P^{\lambda\beta}{}_{\nu}
= P_{\mu\alpha\beta}Q_{\nu}{}^{\alpha\beta} + 2P_{\alpha\beta\mu}Q^{\alpha\beta}{}_{\nu} \,\, .
\end{equation}
Also, using the coincident gauge identity $\nabla_\alpha = \partial_\alpha$ and $\frac{1}{\sqrt{-g}}\partial_\lambda(\sqrt{-g}V^\lambda) = \nabla_\lambda V^\lambda$:
\begin{align}
\frac{2}{\sqrt{-g}}\nabla_\lambda(\sqrt{-g}f_Q P^\lambda{}_{\mu\nu})
&= \frac{2}{\sqrt{-g}}\partial_\lambda(\sqrt{-g}f_Q P^\lambda{}_{\mu\nu}) \nonumber\\
&= \frac{2f_Q}{\sqrt{-g}}\partial_\lambda(\sqrt{-g}P^\lambda{}_{\mu\nu}) + 2f_{QQ}(\partial_\lambda Q)P^\lambda{}_{\mu\nu} \nonumber\\
&= \frac{2}{\sqrt{-g}}f_Q\partial_\alpha(\sqrt{-g}P^\alpha{}_{\mu\nu}) + 2P^\alpha{}_{\mu\nu}f_{QQ}\partial_\alpha Q \,\, .
\end{align}
So, using that, we get the full field equation given as,
\begin{equation}
\boxed{\frac{2}{\sqrt{-g}}\nabla_{\lambda}(\sqrt{-g}f_{Q}P^{\lambda}_{\phantom{\lambda}\mu\nu}) + \frac{1}{2}g_{\mu\nu}f + f_{Q}\left(P_{\mu\lambda\beta}Q_{\nu}^{\phantom{\nu}\lambda\beta} - 2Q_{\lambda\beta\mu}P^{\lambda\beta}_{\phantom{\lambda\beta}\nu}\right) = -T_{\mu\nu}}
\end{equation}
Which exactly matches with the form given in \cite{Harko/2018} and \cite{Lazkoz/2019}.



\subsection{Generalized Bianchi identity and curvature decomposition}
Here, we give an alternative derivation based on \cite{Capozziello/2020}. Here, we do not consider the coincident gauge, so some formulas and sign conventions are different as we have not chosen a gauge, but the overall conclusion would be exactly the same.\\
Start from the second Bianchi identity for a general connection,
\begin{equation}\label{eq:second-bianchi}
\nabla_\lambda R^\beta{}_{\mu\nu\alpha}(\Gamma)
+ \nabla_\nu R^\beta{}_{\lambda\mu\alpha}(\Gamma)
+ \nabla_\mu R^\beta{}_{\nu\lambda\alpha}(\Gamma)
=0 \,\, .
\end{equation}
Decompose the total Riemann tensor of \(\Gamma\) into the Levi--Civita Riemann plus curvature-like terms constructed from the disformation tensor,
\begin{equation}\label{eq:R-decomp}
R^\beta{}_{\mu\nu\alpha}(\Gamma)
= R^{\circ\,\beta}{}_{\mu\nu\alpha}
+ \mathcal{R}^\beta{}_{\mu\nu\alpha}(L) \,\,  ,
\end{equation}
where (schematically) we have,
\begin{equation}
\mathcal{R}^\beta{}_{\mu\nu\alpha}(L)
= \nabla^\circ_\mu L^\beta{}_{\nu\alpha} - \nabla^\circ_\nu L^\beta{}_{\mu\alpha}
+ L^\sigma{}_{\mu\alpha} L^\beta{}_{\nu\sigma} - L^\sigma{}_{\nu\alpha} L^\beta{}_{\mu\sigma} \,\, .
\end{equation}
Because in STG / STEGR the total curvature vanishes, \(R(\Gamma)=0\), we have the follwing,
\begin{equation}\label{eq:riemann-cancel}
R^{\circ\,\beta}{}_{\mu\nu\alpha} = -\,\mathcal{R}^\beta{}_{\mu\nu\alpha}(L) \,\,  .
\end{equation}
Substituting \eqref{eq:R-decomp} into \eqref{eq:second-bianchi} and using \(\nabla\to\nabla^\circ\) in the coincident gauge (or equivalently expressing covariant derivatives with \(\nabla^\circ\) plus \(L\)-terms) yields an identity of the form,
\begin{equation}\label{eq:generalized-bianchi}
\nabla^\circ_\lambda R^{\circ\,\beta}{}_{\mu\nu\alpha}
+ \nabla^\circ_\nu R^{\circ\,\beta}{}_{\lambda\mu\alpha}
+ \nabla^\circ_\mu R^{\circ\,\beta}{}_{\nu\lambda\alpha}
+ \big(\nabla^\circ_\lambda\mathcal{R}^\beta{}_{\mu\nu\alpha} + \text{cyclic}\big)=0 \,\,  ,
\end{equation}
which is the generalized Bianchi identity used in the following.

\subsection{Contraction and split field equations}
We contract \eqref{eq:generalized-bianchi} in the standard way (first set \(\lambda=\alpha\) and sum, then further contractions with the metric). The Levi--Civita pieces produce the usual contracted Bianchi identity structure, while the \(L\)-dependent pieces collect into contractions of \(\mathcal{R}(L)\). After straightforward index manipulations, one finds a split of the form,
\begin{equation}\label{eq:split-eq}
G^{\circ}{}_{\mu\nu} \;=\; -\big( \mathcal{L}_{\mu\nu} - \tfrac{1}{2} g_{\mu\nu}\,\mathcal{L}\big) \,\, ,
\end{equation}
where \(G^{\circ}_{\mu\nu}=R^{\circ}{}_{\mu\nu}-\tfrac12 g_{\mu\nu}R^{\circ}\) is the Einstein tensor built with the Levi--Civita connection, and we have defined the disformation-contracted quantities,
\begin{equation}
\mathcal{L}_{\mu\nu} \equiv \mathcal{R}^\alpha{}_{\mu\alpha\nu}(L), \qquad
\mathcal{L} \equiv g^{\mu\nu}\mathcal{L}_{\mu\nu} \,\, . 
\end{equation}
Equation \eqref{eq:split-eq} shows that the Levi--Civita Einstein tensor is balanced by purely non-metricity (disformation) contributions. In STEGR, these RHS terms can be interpreted as an effective geometric energy-momentum of the non-metricity sector; in the vacuum, the RHS vanishes when the disformation satisfies \(\mathcal{L}_{\mu\nu}-\tfrac12 g_{\mu\nu}\mathcal{L}=0\).

\subsection{Expression in terms of non-metricity and the superpotential}
Using \eqref{eq:disformation} one can expand \(\mathcal{L}_{\mu\nu}\) and \(\mathcal{L}\) explicitly in terms of \(Q_{\alpha\mu\nu}\) and \(P^{\alpha}{}_{\mu\nu}\). After algebraic rearrangement 
one obtains identities of the schematic form,
\begin{align}
\mathcal{L}_{\mu\nu}
&= \nabla^\circ_\alpha L^\alpha{}_{\mu\nu} + \frac{1}{2}\nabla^\circ_\nu Q_\mu
- \frac{1}{2}Q_\alpha L^\alpha{}_{\mu\nu} + \mathcal{Q}_{\mu\nu} \,\, , \label{eq:L-munu-expanded}\\
\mathcal{L}
&= \nabla^\circ_\alpha\big(Q^\alpha-\widetilde Q^\alpha\big) - Q \,\, , \label{eq:L-expanded}
\end{align}
where \(Q\equiv Q_{\alpha\mu\nu}P^{\alpha\mu\nu}\) is the non-metricity scalar and \(\mathcal{Q}_{\mu\nu}\) denotes quadratic combinations of \(Q_{\alpha\mu\nu}\).
Using these explicit forms and grouping total divergences into the superpotential \(P^{\alpha}{}_{\mu\nu}\) yields the field equation in the form customary for STEGR and its generalization \(f(Q)\).

\subsection{Final form: \texorpdfstring{$f(Q)$}{f(Q)} field equations}
If the gravitational action is,
\begin{equation}
S=\frac{1}{2}\int\!d^4x\,\sqrt{-g}\,f(Q) + S_m \,\,
\end{equation}
variation with respect to the metric gives the standard \(f(Q)\) field equations, 
\begin{equation}\label{eq:fQ-eqs}
\frac{2}{\sqrt{-g}}\nabla^\circ_{\alpha}\!\big(\sqrt{-g}\,f_Q\,P^{\alpha}{}_{\mu\nu}\big)
+\tfrac12 g_{\mu\nu} f
+ f_Q\!\left(P_{\mu\alpha\beta}Q_{\nu}{}^{\alpha\beta}-2Q_{\alpha\beta\mu}P^{\alpha\beta}{}_{\nu}\right)
= - T_{\mu\nu} \,\, ,
\end{equation}
where \(f_Q\equiv df/dQ\) and \(T_{\mu\nu}\) is the matter energy–momentum tensor. Equation \eqref{eq:fQ-eqs} reduces to the STEGR (equivalent to GR) case for \(f(Q)=Q\) (up to the boundary term relating \(Q\) to the Ricci scalar).
\section{Derivation of $f(Q)$ action based Schrödinger's motivation}
The foundation of Symmetric Teleparallel Gravity rests on the observation that the Einstein-Hilbert Lagrangian differs from a first-derivative bulk term only by a total divergence. Following Schrödinger's seminal splitting procedure \cite{Schrodinger1950}, the second-derivative content of the Ricci scalar is isolated into a boundary term, leaving a purely quadratic connection piece as the dynamical core. When the connection is constrained to be flat and torsion-free, this quadratic piece becomes the non-metricity scalar $Q$. This exact equivalence not only reproduces General Relativity but also provides the rigorous justification for $f(Q)$ gravity: replacing $Q$ with an arbitrary function $f(Q)$ yields a well-posed, second-order theory. Unlike $f(R)$ gravity, the field equations remain second-order in the metric because the connection is auxiliary and can be eliminated globally via the coincident gauge $\Gamma^\lambda_{\mu\nu}=0$, thereby avoiding Ostrogradsky instabilities while enabling rich cosmological phenomenology.

We introduce the Levi-Civita connection $\mathring{\Gamma}^\lambda_{\mu\nu}$ and a symmetric teleparallel connection $\Gamma^\lambda_{\mu\nu}$ ($R^\lambda{}_{\rho\mu\nu}=0,\ T^\lambda_{\mu\nu}=0$). Their difference defines the disformation tensor $L^\lambda_{\mu\nu} \equiv \Gamma^\lambda_{\mu\nu} - \mathring{\Gamma}^\lambda_{\mu\nu}$, which relates to the non-metricity tensor $Q_{\alpha\mu\nu} \equiv \nabla_\alpha g_{\mu\nu}$ via $Q_{\alpha\mu\nu} = -L_{\alpha\mu\nu} - L_{\alpha\nu\mu}$. The independent traces are $Q_\alpha \equiv Q_{\alpha\mu}{}^\mu$ and $\tilde{Q}_\alpha \equiv Q^\mu{}_{\alpha\mu}$.

Schrödinger's standard splitting of the Levi-Civita Ricci scalar reads,
\begin{equation}
\sqrt{-g}\,\mathring{R} = \partial_\mu\!\left[\sqrt{-g}\left(g^{\alpha\beta}\mathring{\Gamma}^\mu_{\alpha\beta} - g^{\mu\alpha}\mathring{\Gamma}^\beta_{\alpha\beta}\right)\right] + \sqrt{-g}\,g^{\mu\nu}\left(\mathring{\Gamma}^\lambda_{\lambda\rho}\mathring{\Gamma}^\rho_{\mu\nu} - \mathring{\Gamma}^\lambda_{\nu\rho}\mathring{\Gamma}^\rho_{\mu\lambda}\right) \,\,.
\label{eq:schrodinger_split}
\end{equation}
Substituting $\mathring{\Gamma}^\lambda_{\mu\nu} = \Gamma^\lambda_{\mu\nu} - L^\lambda_{\mu\nu}$ and expanding, the pure-$\Gamma$ terms reconstruct $\sqrt{-g}\,R(\Gamma)$, which vanishes identically by flatness. The remaining cross and quadratic contributions reorganize as,
\begin{equation}
\sqrt{-g}\,\mathring{R} = \sqrt{-g}\,g^{\mu\nu}\left(L^\lambda_{\lambda\rho}L^\rho_{\mu\nu} - L^\lambda_{\nu\rho}L^\rho_{\mu\lambda}\right) - \partial_\mu\!\left[\sqrt{-g}\left(g^{\alpha\beta}L^\mu_{\alpha\beta} - g^{\mu\alpha}L^\beta_{\alpha\beta}\right)\right] \,\, .
\end{equation}
Expressing $L$ in terms of $Q_{\alpha\mu\nu}$ and contracting indices, the quadratic bulk piece defines the non-metricity scalar,
\begin{equation}
Q \equiv -\frac{1}{4}Q_{\alpha\mu\nu}Q^{\alpha\mu\nu} + \frac{1}{2}Q_{\alpha\mu\nu}Q^{\mu\alpha\nu} + \frac{1}{4}Q_\alpha Q^\alpha - \frac{1}{2}Q_\alpha \tilde{Q}^\alpha \,\, ,
\end{equation}
while the divergence term simplifies to $\partial_\mu[\sqrt{-g}(Q^\mu - \tilde{Q}^\mu)]$. This yields the fundamental decomposition,
\begin{equation}
\sqrt{-g}\,\mathring{R} = \sqrt{-g}\,Q + \partial_\mu\!\left[\sqrt{-g}\,(Q^\mu - \tilde{Q}^\mu)\right] \,\, . \label{eq:STEGR_identity}
\end{equation}
Integrating over spacetime, the boundary term vanishes under standard variational conditions, establishing the action equivalence $S_{\text{EH}} = S_Q$. Consequently, the $f(Q)$ action $S_{f(Q)} = \frac{1}{2}\int d^4x \sqrt{-g} f(Q)$ inherits this structure: it reduces to GR at linear order ($f(Q)\to Q$) and defines a consistent, second-order modification where the connection remains non-dynamical and fully gauge-fixable \cite{NEST,Jimenez/JCAP2018}.


\section{Concluding remarks}
This appendix provides a complete, self-contained derivation of the $f(Q)$ gravitational field equations from first principles. Starting from the geometric foundations of symmetric teleparallel gravity, we systematically computed the metric variation of the non-metricity tensor, its independent traces, and the superpotential, leading to the explicit variation of the non-metricity scalar $Q$. By applying these results to the $f(Q)$ action, we derived the full second-order field equations and cross-verified their structure through the generalized Bianchi identity and the curvature decomposition induced by the disformation tensor. Additionally, we presented an alternative, historically motivated derivation based on Schrödinger’s splitting procedure, which rigorously establishes the equivalence between the Einstein-Hilbert Lagrangian and the non-metricity scalar up to a boundary term. Together, these steps confirm the mathematical consistency, geometric transparency, and auxiliary nature of the connection in $f(Q)$ gravity, providing a robust foundation for its further application in cosmological and astrophysical contexts.


\chapter{Cosmology in $f(Q)$ gravity} 

\label{AppendixB} 

\lhead{Appendix B. \emph{Cosmology in $f(Q)$ gravity}} 

Here, we provide a detailed description of the Friedmann equation on $f(Q)$ gravity with the FLRW metric.

\section{FLRW METRIC}
 Here we take the FLRW metric in the  Cartesian coordinates, 
$ds^2= -dt^2+a(t)^2(dx^2+dy^2+dz^2)$, \\
$g_{\mu\nu}= diag(-1,a^2,a^2,a^2)$ and $g^{\mu\nu}= diag \left(-1,\frac{1}{a^2},\frac{1}{a^2},\frac{1}{a^2}\right)$\\
We have taken the co-incident gauge choice, so, \\
\begin{equation}
    Q_{\lambda\mu\nu}=\frac{\partial g_{\mu\nu}}{\partial x^{\lambda}} \,\, , 
\end{equation}
so 
\begin{equation}
    Q_{011}=\frac{\partial a^2}{\partial t} = 2a\dot{a} \,\, .
\end{equation}
So, by symmetry, we conclude that,
\begin{equation}
    \boxed{Q_{011}=Q_{022}=Q_{033}=2a\dot{a}\,\, .}
\end{equation}
Now, the deformation tensor becomes,
\begin{equation}
    L^{\lambda}\,_{\mu\nu}=\frac{1}{2}g^{\lambda\sigma}(-Q_{\mu\sigma\nu}-Q_{\nu\sigma\mu}+Q_{\sigma\mu\nu})\,\, .
    \end{equation}
By simplifying, we obtain,
\begin{align}
    L^{\lambda}\,_{\mu\nu}&=\frac{1}{2}g^{\lambda\sigma}(-Q_{\mu\sigma\nu}-Q_{\nu\sigma\mu}+Q_{\sigma\mu\nu})\\
    &=\frac{1}{2}(-Q_{\mu}\,^{\lambda}\,_{\nu} -Q_{\nu}\,^{\lambda}\,_{\mu}+Q^{\lambda}\,_{\mu\nu})\nonumber\\
    &=\frac{1}{2}Q^{\lambda}\,_{\mu\nu}-Q_{(\nu}\,^{\lambda}\,_{\mu)}\nonumber \,\, .
\end{align}
Now we can see that the non-zero components of $L$ would come from,
 \begin{align}
     L^{0}\,_{11}&=\frac{1}{2}g^{0\sigma}(-Q_{1\sigma1}-Q_{1\sigma1}+Q_{\sigma11})\\
     &= \frac{1}{2}g^{00}(Q_{011}) \nonumber\\
     &=\frac{1}{2}(-1)(2a\dot{a})\nonumber\\
     &=-a\dot{a} \,\, .
 \end{align}
By symmetry, we can argue, 
\begin{equation}
  \boxed{L^{0}\,_{11}=L^{0}\,_{22}=L^{0}\,_{33}=-a\dot{a}}
\end{equation}
Similarly,
\begin{align}
     L^{1}\,_{01}&=\frac{1}{2}g^{1\sigma}(-Q_{0\sigma1}-Q_{1\sigma0}+Q_{\sigma01})\\
     &= \frac{1}{2}g^{11}(-Q_{011}-Q_{110}+Q_{101})\nonumber\\
     &=\frac{1}{2}\frac{1}{a^2}(-2a\dot{a})\nonumber\\
     &=-\frac{\dot{a}}{a} \,\, .
 \end{align}
 Again, by symmetry, we can get, 
 \begin{equation}
     \boxed{L^{1}\,_{01}= L^{2}\,_{02}= L^{3}\,_{03}=-\frac{\dot{a}}{a}= L^{1}\,_{10}= L^{2}\,_{20}= L^{3}\,_{30}}
 \end{equation}
 Now we can take two different traces: one is $Q_\alpha=Q_{\alpha}\,^{\nu}\,_{\nu}=g^{\nu\mu}Q_{\alpha\mu\nu}$ and the other is $\tilde{Q}_\alpha=Q^{\nu}\,_{\alpha\nu}=g^{\nu\mu}Q_{\mu\alpha\nu}$.\\
 So we can get $Q^{\lambda}=g^{\lambda\alpha}g^{\beta\nu}Q_{\alpha\beta\nu}$\\
 and $\tilde{Q}^{\lambda}=g^{\lambda\beta}g^{\alpha\nu}Q_{\alpha\beta\nu}$ \,\, .\\
 Now, the definition of the superpotential is given by,
 \begin{align}
     P^{\lambda}\,_{\mu\nu}&=\frac{1}{4}\left(-2\left(\frac{1}{2}Q^{\lambda}\,_{\mu\nu}-Q_{(\mu}\,^{\lambda}\,_{\nu)}\right)+Q^{\lambda}g_{\mu\nu}-\tilde{Q}^{\lambda}g_{\mu\nu}-\delta^{\lambda}_{(\mu}Q_{\nu)}\right)\\
     &= -\frac{1}{2}L^{\lambda}\,_{\mu\nu}+\frac{1}{4}(Q^{\lambda}g_{\mu\nu}-\tilde{Q}^{\lambda}g_{\mu\nu})-\frac{1}{8}(\delta^{\lambda}_{\mu}Q_{\nu}+\delta^{\lambda}_{\nu}Q_{\mu}) \,\, .
 \end{align}
 So we get,
 \begin{align}
     Q^{\lambda}&=g^{\lambda\alpha}g^{\beta\nu}Q_{\alpha\beta\nu}\\
     Q^0&=g^{00}g^{\beta\nu}Q_{0\beta\nu} \nonumber\\
     &= (-1)3\left(\frac{1}{a^2}\right)(2a\dot{a})\nonumber\\
     &=-6H \,\, .
 \end{align}
 So we get, 
    \begin{equation}
        \boxed{Q^0=-6H}
    \end{equation}
One can show similarly,
\begin{equation}
    \boxed{Q^1=Q^2=Q^3=0}
\end{equation}

Similarly, we get,
\begin{align}
    \tilde{Q}^{\lambda}&=g^{\lambda\beta}g^{\alpha\nu}Q_{\alpha\beta\nu}\\
    \tilde{Q}^{0}&=g^{00}g^{\alpha\nu}Q_{\alpha0\nu}\nonumber\\
    &=0 \,\, .
\end{align}
One can show similarly that,
\begin{equation}
    \boxed{\tilde{Q}^1=\tilde{Q}^2=\tilde{Q}^3=0}
\end{equation}
So, the first non-zero combination of the superpotential is given by,
\begin{align}
    P^0\,_{11}&=-\frac{1}{2}L^0\,_{11}+\frac{1}{4}(Q^0-\tilde{Q}^0)g_{11}-\frac{1}{8}(\delta^{\lambda}_{\mu}Q_{\nu}+\delta^{\lambda}_{\nu}Q_{\mu})\\
    &= -\frac{1}{2}(-a\dot{a})+\frac{1}{4}(-6H)a^2-0 \nonumber\\
    &=\frac{a\dot{a}}{2}-\frac{3}{2}(a\dot{a}) \nonumber\\
    &=-a\dot{a} \nonumber \,\, .
\end{align}
So, using the symmetry, we can show,
\begin{equation}
    \boxed{P^0\,_{11}=P^0\,_{22}=P^0\,_{33}=-a\dot{a}}
\end{equation}

Now, for the second set, we get to see that from the following, 
\begin{align}
     P^1\,_{01}&=-\frac{1}{2}L^1\,_{01}+\frac{1}{4}(-6H-0)g_{01}-\frac{1}{8}(\delta^{1}_{0}Q_{1}+\delta^{1}_{1}Q_{0})\\
     &=-\frac{1}{2}\left(-\frac{\dot{a}}{a}\right)+0-\frac{1}{8}g_{00}Q^0  \nonumber\\
     &=\frac{1}{2}\left(\frac{\dot{a}}{a}\right)+\frac{1}{8}(-6H) \nonumber\\
     &=-\frac{1}{4}H \,\, . \nonumber
\end{align}
 Again, using symmetry, we can show that,
\begin{equation}
    \boxed{P^1\,_{01}=P^2\,_{02}=P^3\,_{03}=-\frac{1}{4}H=P^1\,_{10}=P^2\,_{20}=P^3\,_{30}}
\end{equation}

Now we use  the fact that $P^{\lambda\mu\nu}=P^{\lambda}\,_{\alpha\beta}g^{\alpha\mu}g^{\beta\nu}$.\\
Using the above, we get,
\begin{align}
    P^{011}&=P^0\,_{11}g^{11}g^{11}\\
    &=(-a\dot{a})\left(\frac{1}{a^2}\right)\left(\frac{1}{a^2}\right) \nonumber\\
    &=-\frac{\dot{a}}{a^3} \,\, . \nonumber
\end{align}
So, by symmetry, we get,
\begin{equation}
    \boxed{P^{011}=P^{022}=P^{033}=-\frac{\dot{a}}{a^3}}
\end{equation}
We recall that $Q_{011}=Q_{022}=Q_{033}=2a\dot{a}$, so the non-metricity tensor becomes,
\begin{align}
    Q&=-Q_{\lambda\mu\nu}P^{\lambda\mu\nu}\\
    &=-(Q_{011}P^{011}+Q_{022}P^{022}+Q_{033}P^{033}) \nonumber\\
    &=-3(-\frac{\dot{a}}{a^3})(2a\dot{a}) \nonumber \\
    &=6H^2 \,\, . \nonumber
\end{align}
So we finally get,
\begin{equation}
    \boxed{Q=6H^2}
\end{equation}
\subsection{Modified Friedmann equations for $f(Q)$ gravity}
In this subsection, we will completely derive the modified Friedmann equation by putting the (flat) FLRW metric in the original field equation.
\section{Field equation}
We recall the field equation is given by,
\begin{equation}
\frac{2}{\sqrt{-g}}\nabla_{\lambda}(\sqrt{-g}f_{Q}P^{\lambda}_{\phantom{\lambda}\mu\nu}) + \frac{1}{2}g_{\mu\nu}f + f_{Q}\left(P_{\mu\lambda\beta}Q_{\nu}^{\phantom{\nu}\lambda\beta} - 2Q_{\lambda\beta\mu}P^{\lambda\beta}_{\phantom{\lambda\beta}\nu}\right) = -T_{\mu\nu} \,\, ,
\end{equation}
Here, because of the metric's symmetry, we split the calculation into two parts. 
\subsection{$00$ component}
Here we take $\mu=\nu=0$, and noting that $T_{00}=\rho$ we get,
\begin{align}
    \frac{2}{a^3}\nabla_\lambda(a^3f_QP^\lambda_{00})+\frac{1}{2}(-1)f+f_Q(P_{0\lambda\beta}Q_0\,^{\lambda\beta}-2Q_{\lambda\beta0}P^{\lambda\beta}\,_0)&=-T_{00}\\
    -\frac{f}{2}+f_Q(g_{00}P^0\,_{\lambda\beta} \,g^{\lambda\alpha}g^{\beta\delta}Q_{0\alpha\delta})&=-T_{00} \nonumber \\
     -\frac{f}{2}+3f_Q(g_{00}P^0\,_{11}g^{11}g^{11}Q_{011})&=-T_{00} \nonumber\\
     -\frac{f}{2}+3f_Q((-1)\left(-a\dot{a}\frac{1}{a^4}\right)(2a\dot{a}))&=-T_{00} \nonumber\\
     3H^2&=\frac{1}{2f_Q}\left(-\rho+\frac{f}{2}\right) 
\end{align}

\subsection{$11, 22, 33$ components}
As we can see in the Cartesian coordinate, the FLRW metric takes a rather simple form, so we can just take ``$11$" component and by symmetry of the metric we can be sure that the ``$22$" and ``$33$" components would be exactly the same,
\begin{equation}
\frac{2}{\sqrt{-g}}\nabla_{\lambda}(\sqrt{-g}f_{Q}P^{\lambda}_{\phantom{\lambda}\mu\nu}) + \frac{1}{2}g_{\mu\nu}f + f_{Q}(P_{\mu\lambda\beta}Q_{\nu}^{\phantom{\nu}\lambda\beta} - 2Q_{\lambda\beta\mu}P^{\lambda\beta}_{\phantom{\lambda\beta}\nu}) = -T_{\mu\nu} \,\,. 
\end{equation}
Here, if we put $\mu=\nu=1$ and $T_{11}=pa^2$ we get, 

\begin{align}
    \frac{2}{a^3}\nabla_0(a^3f_QP^0\,_{11})+\frac{1}{2}g_{11}f+f_Q(P_{1\lambda\beta}Q_1\,^{\lambda\beta}-2Q_{\lambda\beta 1}P^{\lambda\beta}\,_1)&=-T_{11} \\
    \frac{2}{a^3}\frac{\partial}{\partial t}(a^3f_Q(-a\dot{a}))+\frac{1}{2}(a^2)f+f_Q(P_{110}Q_1\,^{10}+P_{101}Q_1\,^{01}-2Q_{011}P^{01}\,_1)&=-T_{11}\nonumber \\
    \frac{2}{a^3}\frac{\partial}{\partial t}(a^3f_Q(-a\dot{a}))+\frac{1}{2}(a^2)f+f_Q(-2Q_{011}P^{01}\,_1)&=-T_{11}\nonumber \\
    -\frac{2}{a^3}\frac{\partial}{\partial t}(a^4\dot{a}f_Q)+\frac{1}{2}(a^2)f+f_Q(-2(2a\dot{a})g^{11}P^0\,_{11})&=-T_{11}\nonumber\\
    -\frac{2}{a^3}(4a^3\dot{a}^2f_Q+a^4 \Ddot{a}f_Q +a^4\dot{a}f_{QQ}\dot{Q})+\frac{1}{2}a^2f +f_Q(4a\dot{a})\left(\frac{1}{a^2}\right)(a\dot{a})&=-T_{11} \nonumber\\
   -8\dot{a}^2f_Q-2a\Ddot{a}f_Q-2a\dot{a}f_{QQ}\dot{Q}+\frac{1}{2}a^2f+4f_Q\dot{a}^2 &=-pa^2 \nonumber\\
   4\dot{a}^2 f_{Q}+2a\Ddot{a}f_Q+2a\dot{a}\dot{f_Q}-\frac{1}{2}a^2f &=pa^2 \nonumber\\
   2\left(\frac{\dot{a}}{a}\right)\dot{f}_Q+2f_Q\left(\frac{\Ddot{a}}{a}+\frac{\dot{a}}{a}^2\right)+4f_QH^2&=p+\frac{f}{2}\nonumber\\
   2\dot{f}_QH+2f_Q(\dot{H}+3H^2)&=p+\frac{f}{2} \nonumber\\
   \dot{H}+3H^2+\frac{\dot{f}_Q}{f_Q}H=\frac{1}{2f_Q}\left(p+\frac{f}{2}\right) 
\end{align}

The calculations for the other two special components can be done exactly the same way.\\
So, we get the two Friedmann equations (also colloquially known as ``the field equation for cosmology in $f(Q)$ gravity") are given as follows, 
\begin{equation}\label{f1}
    \boxed{3H^2 =\frac{1}{2f_Q}\left(-\rho+\frac{f}{2}\right) }
\end{equation}
and,
\begin{equation}\label{f2}
    \boxed{\dot{H}+3H^2+\frac{\dot{f}_Q}{f_Q}H=\frac{1}{2f_Q}\left(p+\frac{f}{2}\right)\, .}
\end{equation}
\section{Concluding remarks}
In this appendix, we have provided a concise, step-by-step derivation of the modified Friedmann equations in $f(Q)$ gravity using the flat FLRW metric in the coincident gauge. By explicitly computing the non-metricity tensor, deformation tensor, and superpotential components, we showed that the non-metricity scalar reduces to $Q=6H^2$, and substituting into the general field equations yields the cosmological equations \eqref{f1}--\eqref{f2}, consistent with \cite{Harko/2018, Lazkoz/2019}. This compact derivation serves as a self-contained reference for the background dynamics in $f(Q)$ cosmology.

\chapter{Degrees of freedom} 

\label{AppendixC} 

\lhead{Appendix C. \emph{Degrees of freedom}} 

\section{Motivation}
In this appendix, we provide two complementary derivations that the Symmetric Teleparallel Equivalent of General Relativity (STEGR) propagates exactly two physical degrees of freedom per spacetime point, corresponding to the massless spin-2 graviton polarizations. The first method employs Hamiltonian (ADM) constraint analysis, while the second uses linear perturbation theory with gauge-invariant variables. Both approaches confirm that despite the reformulation in terms of non-metricity, STEGR remains dynamically equivalent to General Relativity.

\subsection{Hamiltonian (ADM) constraint analysis}

\subsubsection{Motivation and setup}
The STEGR action is constructed from the non-metricity scalar $\mathbb{Q}$ rather than the Ricci scalar $R$ of General Relativity. Despite this geometric reformulation, STEGR must propagate the same physical content as GR to remain viable. The key question is whether the independent affine connection introduces additional dynamical degrees of freedom or remains purely gauge. In the coincident gauge where the connection vanishes ($\Gamma^\alpha_{\mu\nu} = 0$), the STEGR action reduces to the Einstein-Hilbert action up to a boundary term \cite{Jimenez/JCAP2018}. However, to rigorously count degrees of freedom, we must work with the full theory before gauge fixing and analyze the constraint structure using the Dirac-Bergmann algorithm.

\subsubsection{Configuration variables}
We begin by counting the fundamental fields in STEGR:
\begin{itemize}
    \item Metric tensor $g_{\mu\nu}$: 10 independent components (symmetric $4\times4$ tensor)
    \item Connection via Stueckelberg fields $\xi^\alpha$: 4 components (parameterizing the flat, torsion-free connection)
    \item Total configuration variables: $10 + 4 = 14$
\end{itemize}
The connection in STEGR is constrained to be flat and torsion-free, which allows it to be written as, 
\begin{equation}
\Gamma^\alpha_{\mu\nu} = \frac{\partial x^\alpha}{\partial \xi^\lambda}\partial_\mu\partial_\nu\xi^\lambda \,\, .
\end{equation}
In the coincident gauge $\xi^\alpha = x^\alpha$, we recover $\Gamma^\alpha_{\mu\nu} = 0$ globally.

\subsubsection{ADM decomposition}
We perform a $3+1$ decomposition of spacetime with metric,
\begin{equation}
ds^2 = -N^2 dt^2 + \gamma_{ij}(dx^i + N^i dt)(dx^j + N^j dt) \,\, , 
\end{equation}
where $N$ is the lapse function, $N^i$ is the shift vector, and $\gamma_{ij}$ is the spatial metric. The ADM variables contribute:
\begin{itemize}
    \item Lapse $N$: 1 component (non-dynamical, acts as Lagrange multiplier)
    \item Shift $N^i$: 3 components (non-dynamical, act as Lagrange multipliers)
    \item Spatial metric $\gamma_{ij}$: 6 components (symmetric $3\times3$ tensor, potentially dynamical)
    \item Stueckelberg fields $\xi^\alpha$: 4 components
\end{itemize}

\subsubsection{Primary constraints}
The momenta conjugate to certain variables vanish identically, yielding primary constraints,
\begin{equation}
\pi_N \approx 0 \,\, ,  \qquad \pi_i \approx 0 \,\, , \qquad p_\alpha \approx 0 \,\, ,
\end{equation}
where $\pi_N$ is the momentum conjugate to the lapse, $\pi_i$ are momenta conjugate to the shift, and $p_\alpha$ are momenta conjugate to the Stueckelberg fields. This gives 8 primary constraints total ($1 + 3 + 4 = 8$). The vanishing of $p_\alpha$ reflects that the connection has no kinetic term---no $\dot{\Gamma}$ appears in the STEGR action. The connection is algebraic rather than dynamical \cite{NEST}.

\subsubsection{Secondary constraints}
Preservation of the primary constraints under time evolution generates secondary constraints. The Poisson bracket of the primary constraints with the Hamiltonian yields,
\begin{equation}
\mathcal{H}_\perp \approx 0 \quad \text{(Hamiltonian constraint)}, \qquad 
\mathcal{H}_i \approx 0 \quad \text{(Momentum constraints)}.
\end{equation}
This gives 4 secondary constraints ($1 + 3 = 4$). The Hamiltonian constraint generates time reparametrizations while the momentum constraints generate spatial diffeomorphisms. The consistency condition for $p_\alpha \approx 0$ is automatically satisfied because the Stueckelberg fields enter only through gauge-invariant connection combinations; thus, no additional secondary constraints arise.

\subsubsection{Constraint classification}
A careful count reveals that while 12 constraint equations are formally written (8 primary + 4 secondary), only \textbf{8 independent first-class constraints} exist. The 4 primary constraints $p_\alpha \approx 0$ are absorbed into the diffeomorphism generators via the Stueckelberg reparameterization symmetry and do not produce independent secondary constraints. The remaining 8 constraints ($\pi_N, \pi_i, \mathcal{H}_\perp, \mathcal{H}_i$) weakly commute with each other on the constraint surface. Their algebra closes and is isomorphic to the hypersurface deformation algebra of General Relativity. This confirms they are all first-class, generating spacetime diffeomorphisms and connection gauge transformations, and that STEGR possesses the exact same gauge symmetry structure as GR despite the different geometric formulation.

\subsubsection{Motivation for the constraint hierarchy}
The appearance of primary and secondary constraints is a direct consequence of the singular nature of the STEGR Lagrangian. Following Dirac's systematic treatment of constrained systems \cite{Dirac1964,Henneaux:1992ig}, we distinguish their physical origins:
\begin{itemize}
    \item \textbf{Primary constraints} arise immediately upon Legendre transformation when the Hessian matrix $\partial^2 \mathcal{L}/\partial \dot{q}^A \partial \dot{q}^B$ is singular. In STEGR, $N$, $N^i$, and $\xi^\alpha$ enter without time derivatives, so their conjugate momenta vanish identically. These are not dynamical equations but rather relations defining the physical submanifold of phase space, signaling gauge redundancies rather than independent degrees of freedom.
    \item \textbf{Secondary constraints} emerge from consistency requirements: primary constraints must remain valid under time evolution, $\dot{\phi} \approx \{\phi, H_T\} \approx 0$. In generally covariant theories, this procedure generates the Hamiltonian and momentum constraints, which encode invariance under spacetime diffeomorphisms. The hierarchy ensures that all unphysical directions are systematically identified and removed before counting propagating modes \cite{Teitelboim1973,Golovnev:2018ocj}.
\end{itemize}

\subsubsection{Detailed degree of freedom counting}
With the constraint structure established, we rigorously count the physical degrees of freedom using Dirac's recipe for singular Hamiltonian systems \cite{Dirac1964,Henneaux:1992ig}:
\begin{enumerate}
    \item \textbf{Configuration space dimension:} Excluding the Lagrange multipliers $N$ and $N^i$, the dynamical configuration variables are the spatial metric $\gamma_{ij}$ (6 independent components) and the Stueckelberg fields $\xi^\alpha$ (4 components). Thus, $N_{\text{conf}} = 10$.
    
    \item \textbf{Phase space dimension:} Each configuration variable has a conjugate momentum, giving a naive phase space of dimension $2N_{\text{conf}} = 20$.
    
    \item \textbf{Independent first-class constraints:} The system possesses 8 independent first-class constraints. While the formal counting yields 12 constraint equations (8 primary + 4 secondary), the 4 primary constraints $p_\alpha \approx 0$ are not independent generators; they combine with the diffeomorphism constraints to form a closed algebra of exactly 8 independent first-class generators.
    
    \item \textbf{Phase space reduction:} Each independent first-class constraint removes two dimensions from phase space: one from the constraint equation itself, and one from the associated gauge-fixing condition required to select a representative from each gauge orbit. Therefore, the physical phase space dimension is given by,
    \begin{equation}
    \dim(\Gamma_{\text{phys}}) = 2N_{\text{conf}} - 2N_{\text{1st}} = 20 - 2\times 8 = 4 \,\, .
    \end{equation}
    
    \item \textbf{Physical configuration degrees of freedom:} Since each physical degree of freedom corresponds to a conjugate pair in phase space, the number of propagating modes are given by, 
    \begin{equation}
    N_{\text{DoF}} = \frac{1}{2}\dim(\Gamma_{\text{phys}}) = \frac{1}{2}\left(2N_{\text{conf}} - 2N_{\text{1st}}\right) = 10 - 8 = 2 \,\, .
    \end{equation}
\end{enumerate}
This counting matches the configuration-space formula $N_{\text{DoF}} = N_{\text{conf}} - N_{\text{1st}}$ exactly. The two remaining physical degrees of freedom correspond to the transverse-traceless tensor polarizations $h_+$ and $h_\times$ of the massless spin-2 graviton. The non-metricity formulation of STEGR, despite its distinct geometric variables and constraint hierarchy, reproduces the identical physical content as curvature-based General Relativity \cite{Golovnev:2018ocj,Jimenez/JCAP2018}.

\subsubsection{Connection to Dirac's formalism and BRST quantization}
The constraint analysis follows Dirac's classification, where first-class constraints generate gauge transformations rather than restricting dynamics. In the path integral formulation, this redundancy requires gauge fixing. BRST quantization \cite{Becchi:1974xu,Tyutin:1975qk,Henneaux:1992ig} promotes each first-class constraint to a cohomological operator by pairing it with Faddeev-Popov ghosts, ensuring unitarity and gauge independence. For example, in $U(1)$ gauge theory, the first-class constraints $\pi^0 \approx 0$ and $\partial_i \pi^i \approx 0$ generate electromagnetic gauge transformations. BRST introduces ghost $c$, antighost $\bar{c}$, and auxiliary field $B$, constructing a nilpotent charge $Q_{\text{BRST}}$ such that physical states satisfy $Q_{\text{BRST}}|\text{phys}\rangle = 0$, projecting out unphysical modes and leaving exactly two transverse polarizations. STEGR follows identical logic: its 8 first-class constraints generate diffeomorphism and connection gauge symmetries, and BRST quantization would introduce 8 ghost-antighost pairs to consistently isolate the two physical graviton degrees of freedom in the quantum theory.

\subsection{Linear perturbation theory and gauge-invariant counting}

\subsubsection{Motivation}
While Hamiltonian analysis establishes the DoF count non-perturbatively, linear perturbation theory provides complementary insight into which modes propagate and which are constrained or pure gauge. This is particularly important for understanding why the non-metricity tensor does not introduce additional scalar or vector modes beyond the standard graviton polarizations.

\subsubsection{SVT decomposition and gauge rules}
We consider linear perturbations around the Minkowski background: $g_{\mu\nu} = \eta_{\mu\nu} + h_{\mu\nu}$. Adopting the active diffeomorphism convention $\delta h_{\mu\nu} = \partial_\mu \xi_\nu + \partial_\nu \xi_\mu$ with generator $\xi^\mu = (\xi_0, \partial_i \xi + \xi_i^\perp)$ ($\partial^i \xi_i^\perp = 0$), the scalar-vector-tensor (SVT) decomposition reads,
\begin{align}
h_{00} &= -2\phi \,\, , \\
h_{0i} &= \partial_i B + B_i\,\, , \qquad (\partial^i B_i = 0) \,\, , \\
h_{ij} &= -2\psi\delta_{ij} + 2\left(\partial_i\partial_j - \frac{1}{3}\delta_{ij}\nabla^2\right)E + 2\partial_{(i} E_{j)} + h_{ij}^{\text{TT}} \,\, ,
\end{align}
where $\partial^i E_i = 0$ and $h_{ij}^{\text{TT}}$ is transverse-traceless. Substituting the gauge transformation and matching components yields the consistent rules,
\begin{align}
\delta\phi &= -\dot{\xi}_0 \,\, , \qquad &\delta\psi &= -\frac{1}{3}\nabla^2\xi \,\, , \\
\delta B &= \xi_0 + \dot{\xi} \,\, , \qquad &\delta E &= \xi \,\, , \\
\delta B_i &= \dot{\xi}_i^\perp \,\, , \qquad &\delta E_i &= \xi_i^\perp \,\, , \\
\delta h_{ij}^{\text{TT}} &= 0 \,\, .
\end{align}
The gauge freedom consists of 4 functions ($\xi_0$, $\xi$, $\xi_i^\perp$), which can be used to eliminate 4 components of the perturbation.

\subsubsection{Gauge-invariant variables}
Following standard constructions \cite{Bardeen:1980kt,Mukhanov}, we define combinations invariant under the above rules,
\begin{equation}
\Phi \equiv \phi + \dot{B} - \ddot{E} \,\, , \qquad \Psi \equiv \psi + \frac{1}{3}\nabla^2 E \,\, , \qquad \mathcal{V}_i \equiv B_i - \dot{E}_i \,\, .
\end{equation}
Direct substitution verifies $\delta\Phi = \delta\Psi = \delta\mathcal{V}_i = 0$. The tensor sector $h_{ij}^{\text{TT}}$ is already gauge-invariant.

\subsubsection{Field equations and constraint analysis}
The linearized STEGR field equations are equivalent to the linearized Einstein equations $G_{\mu\nu}^{(1)} = 0$. Here, we analyze each sector:
\textbf{Vector Sector:} The vector modes $B_i$ and $E_i$ do not contain independent kinetic terms; their time derivatives enter only through total derivatives or couple to constraints. The momentum constraint $\mathcal{H}_i = 0$ yields,
\begin{equation}
\nabla^2 \mathcal{V}_i = 0 \quad \Rightarrow \quad \mathcal{V}_i = 0 \quad \text{(in vacuum with appropriate boundary conditions)}.
\end{equation}
Thus vector modes do not propagate---they are either pure gauge or constrained to vanish.
\textbf{Scalar Sector:} The Hamiltonian constraint $\mathcal{H}_\perp = 0$ and momentum constraints relate to scalar potentials. In vacuum STEGR,
\begin{equation}
\Phi = \Psi = 0 \,\, ,
\end{equation}
i.e., no scalar gravitational waves propagate. The scalars are determined instantaneously by matter sources (when present) rather than evolving dynamically.
\textbf{Tensor Sector:} The transverse-traceless modes satisfy the wave equation,
\begin{equation}
\Box h_{ij}^{\text{TT}} = 0 \,\, ,
\end{equation}
where $\Box = -\partial_t^2 + \nabla^2$. These are the only propagating degrees of freedom.

\subsubsection{Final counting}
We can now count the physical modes systematically:
\begin{itemize}
    \item Initial components: 10
    \item Gauge freedom removes: 4 (via choice of $\xi^\mu$)
    \item Constraint equations remove: 4 (Hamiltonian + momentum constraints)
    \item Remaining physical modes: $10 - 4 - 4 = 2$
\end{itemize}
Only the tensor polarizations $h_{ij}^{\text{TT}}$ remain, corresponding to the $+$ and $\times$ graviton modes.

\subsection{There are no extra modes in non-metricity}
A critical question is why the non-metricity tensor $Q_{\alpha\mu\nu}$ does not introduce additional degrees of freedom beyond standard GR. The answer lies in three key properties of STEGR:
\begin{enumerate}
    \item \textbf{Connection is Pure Gauge:} The flat, torsion-free connection in STEGR is parameterized entirely by the Stueckelberg fields $\xi^\alpha$, which represent a coordinate choice rather than independent dynamics \cite{Jimenez/JCAP2018}.
    \item \textbf{No Kinetic Term:} The STEGR action contains no terms with time derivatives of the connection ($\dot{\Gamma}$). The connection appears algebraically and can be eliminated by its own field equation, yielding the coincident gauge $\Gamma^\alpha_{\mu\nu} = 0$.
    \item \textbf{Equivalent to GR:} The STEGR Lagrangian differs from the Einstein-Hilbert Lagrangian only by a boundary term \cite{Jimenez/JCAP2018},
    \begin{equation}
    \mathcal{L}_{\text{STEGR}} = \mathcal{L}_{\text{EH}} + \nabla_\mu\left(Q^\mu - \tilde{Q}^\mu\right) \,\, ,
    \end{equation}
    where $Q^\mu$ and $\tilde{Q}^\mu$ are specific non-metricity vectors. Boundary terms do not affect the equations of motion or the degree of freedom count.
\end{enumerate}
This contrasts sharply with general metric-affine gravity theories, where the connection carries independent kinetic terms and typically propagates additional modes (often ghost-like) \cite{Hehl1995}. STEGR avoids this pathology by constraining the connection to be flat and torsion-free from the outset.

\subsection{Concluding remarks}
Both the Hamiltonian constraint analysis and the linear perturbation theory rigorously confirm that STEGR propagates exactly two physical degrees of freedom, corresponding to the massless spin-2 graviton polarizations. The non-metricity formulation provides a geometrically distinct but dynamically equivalent description of gravity. By recasting curvature into a boundary term and shifting the bulk dynamics to the non-metricity scalar, STEGR demonstrates that the physical content of General Relativity is robust against changes in the underlying geometric variables, provided curvature and torsion are strictly constrained. This equivalence not only validates STEGR as a viable classical theory but also establishes a clean, second-order foundation for exploring modified gravity extensions such as $f(Q)$ theories.


\chapter{Linear Approximation} 

\label{AppendixD} 

\lhead{Appendix D. \emph{Linear Approximation}} 

\section{Introduction}
In this appendix, we determine the specific coefficients of the most general quadratic non-metricity Lagrangian by requiring consistency with General Relativity in the weak-field limit. Although the geometric identity $R =- Q + \text{(boundary)}$ uniquely fixes the theory, it is instructive to derive the coefficients $(c_1,c_2,c_3,c_4,c_5)$ from first principles using perturbation theory. This approach demonstrates that STEGR is the unique theory of a massless spin-2 field constructed from non-metricity.

\subsection{The general quadratic Lagrangian}
Instead of deriving the STEGR action from the curvature identity relating the Ricci scalar $R$ and the non-metricity scalar $Q$, we construct the theory by starting with the most general quadratic Lagrangian in non-metricity and demanding consistency with General Relativity in the weak-field limit.

The non-metricity tensor is defined as the covariant derivative of the metric,
\begin{equation}
Q_{\alpha\mu\nu}\equiv\nabla_\alpha g_{\mu\nu} \,\, , 
\end{equation}
and it satisfies $Q_{\alpha\mu\nu}=Q_{\alpha\nu\mu}$ because $g_{\mu\nu}=g_{\nu\mu}$. The independent quadratic scalars that one can form from two copies of $Q$ by contracting the indices with the inverse metric are
\begin{enumerate}
    \item $Q_{\alpha\mu\nu}Q^{\alpha\mu\nu}$ \,\, ,
    \item $Q_{\alpha\mu\nu}Q^{\mu\alpha\nu}$\,\, ,
    \item $Q_\alpha Q^\alpha$ where $Q_\alpha\equiv Q_{\alpha\ \beta}^{\ \beta}$ \,\, ,
    \item $\bar Q_\alpha\bar Q^\alpha$ where $\bar Q_\alpha\equiv Q^{\beta}_{\ \beta\alpha}$ \,\,,
    \item $Q_\alpha\bar Q^\alpha$ \,\, .
\end{enumerate}
There are no other independent quadratic contractions, given the index symmetries.

Therefore, the most general quadratic non-metricity Lagrangian is given by, 
\begin{equation}\label{eq:general_lagrangian}
\mathcal{L} \;=\; c_1\,Q_{\alpha\mu\nu}Q^{\alpha\mu\nu}
+ c_2\,Q_{\mu\alpha\nu}Q^{\alpha\mu\nu}
+ c_3\,Q_\mu Q^\mu
+ c_4\,\bar Q_\mu\bar Q^\mu
+ c_5\,Q_\mu\bar Q^\mu \,\, ,
\end{equation}
with dimensionless coefficients $c_i$ to be determined by physical consistency.

\subsection{Linearization in the coincident gauge}
To study the weak-field limit, we work in the \emph{coincident gauge} where the (flat, torsion-free) affine connection is set to zero locally,
\[
\Gamma^\alpha{}_{\mu\nu}=0 \,\, .
\]
In this gauge, the covariant derivative reduces to the partial derivative and given by,
\begin{equation}
Q_{\alpha\mu\nu}=\partial_\alpha g_{\mu\nu}.
\end{equation}
Expand the metric about Minkowski spacetime,
\[
g_{\mu\nu}=\eta_{\mu\nu}+h_{\mu\nu},\qquad |h_{\mu\nu}|\ll 1 \,\, , 
\]
with $\eta_{\mu\nu}=\mathrm{diag}(-1,+1,+1,+1)$ and $h:=\eta^{\mu\nu}h_{\mu\nu}$. To first order in $h$,
\begin{equation}\label{eq:Q_linear}
Q_{\alpha\mu\nu}=\partial_\alpha h_{\mu\nu}+ \mathcal{O}(h^2) \,\, ,
\qquad
Q^{\alpha\mu\nu}=\partial^\alpha h^{\mu\nu}+\mathcal{O}(h^2) \,\, ,
\end{equation}
where the indices are raised and lowered with $\eta^{\mu\nu}$ in linear order.

The two traces reduce to the following,
\begin{align}
Q_\alpha &= Q_{\alpha\ \beta}^{\ \beta} = \partial_\alpha h \,\, , \label{eq:trace1}\\
\bar Q_\alpha &= Q^\beta_{\ \beta\alpha} = \partial^\beta h_{\beta\alpha}\,\, . \label{eq:trace2}
\end{align}

\subsection{Quadratic building blocks}
Introduce the following quadratic derivative building blocks of $h_{\mu\nu}$:
\begin{align}
A &:= \partial_\alpha h_{\mu\nu}\partial^\alpha h^{\mu\nu}\,\, , \label{eq:blockA}\\
B &:= \partial_\alpha h_{\mu\nu}\partial^\mu h^{\alpha\nu} \,\, , \label{eq:blockB}\\
C &:= \partial_\alpha h\,\partial^\alpha h \,\, , \label{eq:blockC}\\
D &:= \partial_\lambda h^\lambda{}_{\mu}\,\partial_\rho h^{\rho\mu} = (\partial\!\cdot\! h)_\mu(\partial\!\cdot\! h)^\mu \,\, , \label{eq:blockD}\\
E &:= \partial_\mu h\,\partial_\lambda h^{\lambda\mu} = (\partial h)\cdot(\partial\!\cdot\! h) \,\, . \label{eq:blockE}
\end{align}
Using \eqref{eq:Q_linear}--\eqref{eq:trace2}, the five quadratic invariants map to these blocks at quadratic order:
\begin{align}
Q_{\alpha\mu\nu}Q^{\alpha\mu\nu} &= A \,\, , \\
Q_{\mu\alpha\nu}Q^{\alpha\mu\nu} &= B  \,\, , \\
Q_\mu Q^\mu &= C\,\, , \\
\bar Q_\mu\bar Q^\mu &= D \,\, , \\
Q_\mu\bar Q^\mu &= E\,\,.
\end{align}
Thus, the quadratic action \eqref{eq:general_lagrangian} becomes (up to total derivatives),
\begin{equation}\label{eq:lagrangian_blocks}
\mathcal{L}^{(2)} = c_1 A + c_2 B + c_3 C + c_4 D + c_5 E + \text{(total derivatives)}\,\, .
\end{equation}

\subsection{The Fierz--Pauli action}
The unique gauge-invariant kinetic Lagrangian for a massless spin-2 field in flat spacetime (Fierz--Pauli), which reproduces the quadratic expansion of the Einstein--Hilbert action, is the following,
\begin{equation}\label{eq:FP}
\mathcal{L}_{\mathrm{FP}}
=\tfrac{1}{4}\partial_\lambda h_{\mu\nu}\partial^\lambda h^{\mu\nu}
-\tfrac{1}{2}\partial_\mu h^{\mu\nu}\partial^\lambda h_{\lambda\nu}
+\tfrac{1}{2}\partial_\mu h^{\mu\nu}\partial_\nu h
-\tfrac{1}{4}\partial_\lambda h\,\partial^\lambda h \,\, .
\end{equation}
In terms of the blocks \eqref{eq:blockA}--\eqref{eq:blockE}, this reads,
\begin{equation}\label{eq:FP_blocks}
\mathcal{L}_{\mathrm{FP}}=\tfrac{1}{4}A - \tfrac{1}{2}D + \tfrac{1}{2}E - \tfrac{1}{4}C \,\,.
\end{equation}
Matching to Fierz--Pauli ensures that the theory propagates exactly the two graviton polarizations and is invariant under linearized diffeomorphisms,
\[
\delta h_{\mu\nu}=\partial_\mu\xi_\nu+\partial_\nu\xi_\mu \,\, .
\]

\subsection{Matching conditions from gauge invariance}
At the linear level, blocks $B$ and $D$ are equivalent up to integration by parts. Indeed, integrating by parts twice gives,
\begin{align}
B &= \partial_\alpha h_{\mu\nu}\partial^\mu h^{\alpha\nu} \nonumber \\
  &= -\,h_{\mu\nu}\,\partial_\alpha\partial^\mu h^{\alpha\nu} + \text{(boundary)} \nonumber \\
  &= +\,(\partial^\mu h_{\mu\nu})(\partial_\alpha h^{\alpha\nu}) + \text{(boundary)} \nonumber \\
  &= D + \text{(boundary)} \,\, .
\end{align}
Hence in the action, we may replace $B\to D$ (up to total derivatives), and \eqref{eq:lagrangian_blocks} becomes,
\begin{equation}\label{eq:reduced_lagrangian}
\mathcal{L}^{(2)} = c_1 A + (c_2+c_4)D + c_3 C + c_5 E \,\, .
\end{equation}
Requiring $\mathcal{L}^{(2)}$ to reproduce the Fierz--Pauli Lagrangian \eqref{eq:FP_blocks}, we identify the coefficients term-by-term:
\begin{align}
c_1 &= \tfrac{1}{4}\,\,, \label{eq:match1}\\
c_2+c_4 &= -\tfrac{1}{2}\,\, , \label{eq:match2}\\
c_3 &= -\tfrac{1}{4}\,\, , \label{eq:match3}\\
c_5 &= \tfrac{1}{2}\,\, . \label{eq:match4}
\end{align}
Linear analysis leaves the combination $c_2+c_4$ fixed, but does not determine $c_2$ and $c_4$ separately.

\subsection{Nonlinear consistency and the Bianchi identity}
To fix the remaining freedom one must require nonlinear consistency: the metric field equations obtained from the full action must be divergence-free (so that $\nabla_\mu M^{\mu\nu}=0$ and matter energy--momentum is conserved). Varying the full action built from \eqref{eq:general_lagrangian} and taking the covariant divergence produces terms that place additional constraints on the coefficients so the Bianchi identity holds off-shell. A detailed nonlinear analysis \cite{Jimenez/2018, Heisenberg/2024} leads to,
\begin{equation}\label{eq:bianchi_constraints}
c_2 = -2c_1\,\, ,\qquad c_4 = 0\,\, .
\end{equation}
Using $c_1=\tfrac{1}{4}$ from \eqref{eq:match1} gives,
\begin{equation}
c_2 = -\tfrac{1}{2}\,\, ,\qquad c_4=0\,\, .
\end{equation}
Note that this satisfies the linear matching condition \eqref{eq:match2}: $c_2+c_4 = -\tfrac{1}{2}+0 = -\tfrac{1}{2}$.

\subsection{Determined coefficients}
Collecting all results, the coefficients are uniquely fixed as:
\[
c_1=\tfrac{1}{4}\,\, ,\quad c_2=-\tfrac{1}{2}\,\,,\quad c_3=-\tfrac{1}{4}\,\,,\quad c_4=0,\quad c_5=\tfrac{1}{2} \,\, .
\]

\begin{table}[h]
\centering
\begin{tabular}{c c l}
\toprule
\textbf{Coefficient} & \textbf{Value} & \textbf{Physical meaning} \\
\midrule
$c_1$ & $\tfrac{1}{4}$ & Normalizes the main kinetic term $Q_{\alpha\mu\nu}Q^{\alpha\mu\nu}$ \\
$c_2$ & $-\tfrac{1}{2}$ & Ensures gauge invariance and correct tensor structure \\
$c_3$ & $-\tfrac{1}{4}$ & Cancels the trace-mode kinetic term (no scalar graviton) \\
$c_4$ & $0$ & Required by the Bianchi identity (no $\bar Q_\mu\bar Q^\mu$ term) \\
$c_5$ & $\tfrac{1}{2}$ & Ensures the correct divergence structure \\
\bottomrule
\end{tabular}
\caption{STEGR coefficients determined from perturbative analysis and the Bianchi identity.}
\label{tab:coefficients}
\end{table}

Thus, the STEGR Lagrangian reads,
\begin{equation}\label{eq:STEGR_final}
\boxed{\;
\mathcal{L}_{\mathrm{STEGR}}
=\tfrac{1}{4}Q_{\alpha\mu\nu}Q^{\alpha\mu\nu}
-\tfrac{1}{2}Q_{\mu\alpha\nu}Q^{\alpha\mu\nu}
-\tfrac{1}{4}Q_\mu Q^\mu
+\tfrac{1}{2}Q_\mu\bar Q^\mu
\;}
\end{equation}
or, equivalently, defining the non-metricity scalar $Q$ (with the sign convention used in the STEGR literature),
\begin{equation}\label{eq:Q_scalar}
Q = -\tfrac{1}{4}Q_{\alpha\mu\nu}Q^{\alpha\mu\nu}
+\tfrac{1}{2}Q_{\alpha\mu\nu}Q^{\mu\alpha\nu}
+\tfrac{1}{4}Q_\alpha Q^\alpha
-\tfrac{1}{2}Q_\alpha\bar Q^\alpha \,\, ,
\end{equation}
so that with the action normalization $S=-\frac{1}{2\kappa}\int d^4x\sqrt{-g}\,Q$ one reproduces the Einstein--Hilbert action up to a boundary term.

\subsection{Equivalence to general relativity}
The fundamental identity relating the Ricci scalar computed from the Levi-Civita connection and the non-metricity scalar is given by,
\begin{equation}\label{eq:R_Q_identity}
R = Q + \nabla_\mu\bigl(Q^\mu-\bar Q^\mu\bigr) \,\, ,
\end{equation}
so that the difference between the Einstein--Hilbert Lagrangian and the STEGR Lagrangian is a total divergence. Hence,
\[
S_{\mathrm{STEGR}} = -\frac{1}{2}\int d^4x\sqrt{-g}\,Q
= S_{\mathrm{EH}} + \text{(boundary term)} \,\, .
\]
The field equations are therefore identical to Einstein's equations. The degree-of-freedom count is exactly two (a massless spin-2 graviton).

\section*{Concluding remarks}
By matching the linearized quadratic non-metricity action to the standard Fierz--Pauli structure and imposing nonlinear consistency (Bianchi identity/divergence-free field equations), we uniquely fix the coefficients in the general quadratic non-metricity Lagrangian and recover the STEGR action, which is dynamically equivalent to General Relativity.

\addtocontents{toc}{\vspace{2em}} 

\backmatter


\label{References}
\lhead{\emph{References}}


\begin{thebibliography}{100}




\bibitem{harari:2014} 
Y. N. Harari, \textit{Sapiens: A Brief History of Humankind}, Harvill Secker, London (2014).

\bibitem{marchant:2020} 
J. Marchant, \textit{The Human Cosmos: Civilization and the Stars}, Little, Brown Spark, New York (2020).

\bibitem{rappengluck:2008} 
M. Rappenglück, \textit{Insights into the Rock Art of Europe}, vol. 1, p. 112 (2008).

\bibitem{piccardi:2007} 
L. Piccardi and W. B. Masse, \textit{Myth and Geology}, Geological Society of London, London (2007).

\bibitem{usscher:1654} 
J. Ussher, \textit{The Annals of the World}, vol. 1, p. 1 (1654).

\bibitem{dalley:1998}
S. Dalley (Trans.), \textit{Myths from Mesopotamia: Creation, the Flood, Gilgamesh, and Others}, Oxford University Press, Oxford (1998).

\bibitem{yang:2005}
L. Yang and D. An, \textit{Handbook of Chinese Mythology}, ABC-CLIO, Santa Barbara (2005).

\bibitem{griaule:1986}
M. Griaule and G. Dieterlen, \textit{The Pale Fox}, translated by S. C. Infantino, Continuum Foundation, Arizona (1986). 

\bibitem{idowu:1962}
E. B. Idowu, \textit{Olodumare: God in Yoruba Belief}, Longmans, London (1962).

\bibitem{lewis-williams:2002}
J. D. Lewis-Williams, \textit{The Mind in the Cave: Consciousness and the Origins of Art}, Thames and Hudson, London (2002).

\bibitem{rigveda:ancient} 
\textit{Rig Veda}, (Trans. R. T. H. Griffith), Motilal Banarsidass, Delhi (1973).

\bibitem{aryabhata:499} 
Aryabhata, \textit{Aryabhatiya}, (Ed. \& Trans. K. S. Shukla), Indian National Science Academy, New Delhi (1976).

\bibitem{sarma:1997} 
K. V. Sarma, \textit{A History of the Kerala School of Hindu Astronomy}, Vishveshvaranand Institute, Hoshiarpur (1997).

\bibitem{bag:1979} 
A. K. Bag, \textit{Aryabhata's Contributions to Mathematics and Astronomy}, Aryabhata Publications, Calcutta (1979).

\bibitem{brahmagupta:628} 
Brahmagupta, \textit{Brahmasphu\d{t}asiddhanta}, (Ed. S. Dwivedi), Vishveshvaranand Vedic Research Institute, Hoshiarpur (1996).

\bibitem{albiruni:1000} 
Al-Biruni, \textit{Kitab al-Jamahir}, (Trans. C. E. Sachau), Kegan Paul, Trench, Trübner \& Co., London (1910).

\bibitem{bhaskara:1150} 
Bhaskara II, \textit{Siddhanta Siroma\d{n}i} (Goladhyaya), (Trans. H. T. Colebrooke), John Murray, London (1817).

\bibitem{ibnsina:1020} 
Ibn Sina, \textit{Kitab al-Shifa}, (Ed. A. M. Goichon), Desclée de Brouwer, Paris (1951).

\bibitem{copernicus:1543} 
N. Copernicus, \textit{De revolutionibus orbium coelestium}, Johannes Petreius, Nuremberg (1543).

\bibitem{galileo:1610} 
G. Galilei, \textit{Sidereus Nuncius}, Thomas Baglioni, Venice (1610).

\bibitem{kepler:1609} 
J. Kepler, \textit{Astronomia nova}, Ex officina typographica, Prague (1609).

\bibitem{singh:2005} 
S. Singh, \textit{Big Bang: The Most Important Scientific Discovery of All}, HarperCollins, London (2005).

\bibitem{newton:1687} 
I. Newton, \textit{Philosophi{\ae} Naturalis Principia Mathematica}, Jussu Societatis Regi{\ae}, London (1687).

\bibitem{hutton:1788} 
J. Hutton, \textit{Trans. Roy. Soc. Edinburgh} \textbf{1}, 209 (1788).

\bibitem{lyell:1830} 
C. Lyell, \textit{Principles of Geology}, vol. 1, John Murray, London (1830).

\bibitem{kelvin:1862} 
W. Thomson, \textit{Trans. Roy. Soc. Edinburgh} \textbf{23}, 157 (1862).

\bibitem{kelvin:1897} 
W. Thomson, \textit{Popular Lectures and Addresses} \textbf{1}, 350 (1897).

\bibitem{rutherford:1905} 
E. Rutherford, \textit{Phil. Mag.} \textbf{10}, 202 (1905).

\bibitem{eddingtion:1920} 
A. S. Eddington, \textit{The Internal Constitution of the Stars}, Cambridge University Press, Cambridge (1920).

\bibitem{leavitt:1912} 
H. S. Leavitt, \textit{Ann. Harvard Coll. Obs.}, \textbf{60}, 87 (1912).

\bibitem{shapley:1918} 
H. Shapley, \textit{Astrophys. J.} \textbf{47}, 1 (1918).

\bibitem{curtis:1920} 
H. D. Curtis, \textit{Pub. Allegheny Obs.} \textbf{11}, 1 (1920).

\bibitem{slipher:1917} 
V. M. Slipher, \textit{Proc. Amer. Phil. Soc.} \textbf{56}, 403 (1917).

\bibitem{lemaitre:1927} 
G. Lemaître, \textit{Ann. Soc. Sci. Brux.} \textbf{47}, 49 (1927).

\bibitem{hubble:1925} 
E. Hubble, \textit{Astrophys. J.} \textbf{62}, 409 (1925).

\bibitem{Hubble:1929} 
E. Hubble, \textit{Proc. Nat. Acad. Sci.} \textbf{15}, 168 (1929).













\bibitem{kiessling/2012} M. K.-H. Kiessling, \textit{Studies in History and Philosophy of Modern Physics} \textbf{43}, 28 (2012).
\bibitem{Scali2025} F. Scali, \textit{Philos. Trans. R. Soc. A} \textbf{383}, 20230292 (2025).
\bibitem{kant:1755} I. Kant \textit{Allgemeine Naturgeschichte und Theorie des Himmels}, Johann Friedrich Petersen, 1 (1755).
\bibitem{weyl:1918} H. Weyl, \textit{Raum, Zeit, Materie}, Julius Springer, 1 (1918).
\bibitem{herschel:1785} W. Herschel, \textit{Philosophical Transactions of the Royal Society} \textbf{75}, 213 (1785).
\bibitem{einstein:1905} A. Einstein, \textit{Annalen der Physik} \textbf{322}, 891 (1905).
\bibitem{einstein:1915} A. Einstein, \textit{Sitzungsberichte der Koeniglich Preussischen Akademie der Wissenschaften}, \textbf{1915}, 844 (1915).
\bibitem{Einstein:1917} A. Einstein \textit{Sitzungsberichte der Koeniglich Preussischen Akademie der Wissenschaften} \textbf{1917}, 142 (1917).
\bibitem{gamow:1970} G. Gamow, \textit{My World Line: An Informal Autobiography}, Viking Press (1970).
\bibitem{eddington:1930} A. S. Eddington, \textit{Mon. Not. R. Astron. Soc.} \textbf{90}, 668 (1930).
\bibitem{desitter:1917} W. de Sitter, \textit{Mon. Not. R. Astron. Soc.} \textbf{78}, 3 (1917).
\bibitem{friedmann:1922} A. Friedmann, \textit{Zeitschrift für Physik} \textbf{10}, 377 (1922).
\bibitem{robertson:1935} H. P. Robertson \textit{Astrophysical Journal} \textbf{82}, 284 (1935).
\bibitem{walker:1937} A. G. Walker \textit{Proceedings of the London Mathematical Society} \textbf{42}, 90 (1937).
\bibitem{hawking:1973} S. W. Hawking and G. F. R. Ellis \textit{The Large Scale Structure of Space-Time}, Cambridge University Press (1973).
\bibitem{islam:2001} J. N. Islam \textit{An Introduction to Mathematical Cosmology} Cambridge University Press (2001).
\bibitem{planck} Planck Collaboration, \textit{A}\&\textit{A} 
\textbf{641 A(6)}, 1-67 (2020). 
\bibitem{sandage1970} A. R. Sandage, \textit{Physics Today} \textbf{23}, 34--41 (1970).
\bibitem{ZEL} Y. B. Zeldovich et al., \textit{Sov. phys. Usp.} \textbf{11}, 209-230 (1968).
\bibitem{Efstathiou:1988} G. Efstathiou, R. S. Ellis and B. A. Peterson, \textit{Mon. Not. R. Astron. Soc.} \textbf{232}, 431-444 (1988).
\bibitem{Efstathiou:1990} G. Efstathiou, W. J. Sutherland and S. J. Maddox, \textit{Nature} \textbf{348}, 705-707 (1990).

\bibitem{Loh:1986} E. D. Loh and E. J. Spillar, \textit{Astrophys. J.} \textbf{303}, L15 (1986).
\bibitem{late1} A.G. Riess et al., \textit{ Astron. J.} \textbf{116}, 1009 (1998).
\bibitem{late2} S. Perlmutter et al., \textit{Astrophys. J.} \textbf{517}, 565 (1999).
\bibitem{Eisenstein:2005}
Eisenstein, D. J., et al. (2005). Detection of the Baryon Acoustic Peak in the Large-Scale Correlation Function of SDSS Luminous Red Galaxies. \textit{The Astrophysical Journal} 633, 560.
\bibitem{Crittenden:1996}
R. G. Crittenden and N. Turok, \textit{Phys. Rev. Lett.} \textbf{76}, 575 (1996)
\bibitem{WENB} S. Weinberg, \textit{Rev. Mod. Phys.} \textbf{61}, 1 (1989).
\bibitem{CANT} E. N. Saridakis et al., \textit{Modified Gravity and Cosmology: An Update by the CANTATA Network}, Springer (2021).
\bibitem{COSI} E. Abdalla et al., \textit{JHEAp} \textbf{34}, 49-211 (2022).
\bibitem{NEST} J. M. Nester and H. J. Yo, \textit{Chin. J. Phys.} \textbf{37}, 113 (1999).
\bibitem{Jimenez/2018} J. B. Jimenez, L. Heisenberg and T. Koivisto, \textit{Phys. Rev. D} \textbf{98}, 044048 (2018).
\bibitem{Harko/2018} T. Harko, T. S. Koivisto, F. S. N. Lobo, G. J. Olmo and D. Rubiera-Garcia, \textit{Phys. Rev. D} \textbf{98}, 084043 (2018).
\bibitem{Lazkoz/2019} R. Lazkoz, F. S. N. Lobo, M. Ortiz-Baños and V. Salzano, \textit{Phys. Rev. D} \textbf{100}, 104027 (2019).
\bibitem{Capozziello/2020} W. G. Boskoff and S. Capozziello, \textit{A Mathematical Journey to Relativity: Deriving Special and General Relativity with Basic Mathematics}, Springer (2020).
\bibitem{BARR} B. J. Barros, T. Barreiro, T. Koivisto, and N. J. Nunes, \textit{Phys. Dark Univ.} \textbf{30}, 100616 (2020).

\bibitem{ANAG} F. K. Anagnostopoulos, S. Basilakos, E. N. Saridakis, \textit{Phys. Lett. B} \textbf{822}, 136634 (2021).

\bibitem{ARORA}  O. Sokoliuk, S. Arora, S. Praharaj, A. Baransky, and P. K. Sahoo, \textit{Mon. Not. R. Astron. Soc.} \textbf{522}, 252-267 (2023).

\bibitem{NUNES} J. Ferreira, T. Barreiro, J. Mimoso, and N. J. Nunes, \textit{Phys. Rev. D} \textbf{105}, 123531 (2022).

\bibitem{RODR}  J. T. S. S. Junior and M. E. Rodrigues, \textit{Eur. Phys. J. C} \textbf{83}, 475 (2023).

\bibitem{LAVI-1} F. D'Ambrosio, S. D. B. Fell, L. Heisenberg and S. Kuhn, \textit{Phys. Rev. D} \textbf{105}, 024042 (2022).

\bibitem{NEOM} L. Atayde and N. Frusciante, \textit{arXiv}:2306.03015 (2023).

\bibitem{CAPE-1} F. Bajardi and S. Capozziello, \textit{Eur. Phys. J. C} \textbf{83}, 531 (2023).

\bibitem{PALIA}  N. Dimakis, A. Paliathanasis and T. Christodoulakis, \textit{Class. Quantum Grav.} \textbf{38}, 225003 (2021).

\bibitem{GADB} G. N. Gadbail, A. Kolhatkar, S. Mandal and P. K. Sahoo, \textit{Eur. Phys. J. C} \textbf{83}, 595 (2023)

\bibitem{CAPE-2} S. Capozziello and M. Shokri, \textit{Phys. Dark Univ.} \textbf{37}, 101113 (2022).

\bibitem{ANDER} A. Lymperis, \textit{JCAP} \textbf{11}, 018 (2022).

\bibitem{SNEHA} S. Pradhan, S. Mandal and P. K. Sahoo, \textit{Chin. Phys. C} \textbf{47}, 055103 (2023).

\bibitem{Hassan/2022} Z. Hassan, S. Ghosh, P. K. Sahoo and K. Bamba, \textit{Eur. Phys. J. C} \textbf{82}, 1116 (2022)

\bibitem{Hassan/2023} Z. Hassan, S. Ghosh, P. K. Sahoo and V. S. H. Rao, \textit{Gen. Rel. Grav.} \textbf{55}, 90 (2023)

\bibitem{ET} S. Sahlu and E. Tsegaye, \textit{arXiv}:2206.02517 (2022).

\bibitem{ANAG-2}  F. K. Anagnostopoulos, V. Gakis, E. N. Saridakis and S. Basilakos, \textit{Eur. Phys. J. C} \textbf{83}, 58 (2023).

\bibitem{DE-1} G. Subramaniam, A. De, T. H. Loo and Y. K. Goh, \textit{Fortschr. Phys.} \textbf{2023}, 2300038 (2023).

\bibitem{DE-2} G. Subramaniam, A. De, T. H. Loo and Y. K. Goh, \textit{Phys. Dark Univ.} \textbf{41}, 101243 (2023).

\bibitem{PALIA-2} N. Dimakis, M. Roumeliotis, A. Paliathanasis, P. S. Apostolopoulos and T. Christodoulakis,  \textit{Phys. Rev. D} \textbf{106}, 123516 (2022).

\bibitem{HOH} M. Hohmann, \textit{Phys. Rev. D} \textbf{104}, 124077 (2021).

\bibitem{WOM-2} W. Khyllep, A. Paliathanasis and J. Dutta, \textit{Phys. Rev. D} \textbf{103}, 103521 (2021).

\bibitem{Buchdahl} H. A. Buchdahl, \textit{Mon.\ Not.\ Roy.\ Astron.\ Soc.} \textbf{150}, 1 (1970).
\bibitem{star}  A. A. Starobinsky , \textit{Phys. Lett. B} \textbf{91}, 99-102 (1980).
\bibitem{Cartan}
E. Cartan, \textit{Ann.\ Ec.\ Norm.\ Sup.} \textbf{40}, 325 (1923).

\bibitem{Sciama}
D. W. Sciama, \textit{Rev.\ Mod.\ Phys.} \textbf{36}, 463 (1964).

\bibitem{Einstein-Tele}
A. Einstein,
\textit{Sitzungsber.\ Preuss.\ Akad.\ Wiss.} 543 (1930).
\bibitem{Bengochea/2009}
G.~R.~Bengochea and R.~Ferraro,
\textit{Phys.\ Rev.\ D} \textbf{79}, 124019 (2009).

\bibitem{Baojiu/2011}
B.~Li, T.~P.~Sotiriou, and J.~D.~Barrow,
\textit{Phys.\ Rev.\ D} \textbf{83}, 064035 (2011).

\bibitem{Bohmer/2011}
C.~G.~Böhmer, A.~Mussa, and N.~Tamanini,
\textit{Class.\ Quant.\ Grav.} \textbf{28}, 245020 (2011).

\bibitem{Jarv2020}L.~Jarv, M.~Hohmann, and M.~Saal, \textit{Eesti Vabariigi preemiad}, 57 (2020).
\bibitem{Jimenez/2019} J. B. Jimenez, L. Heisenberg and T. S. Koivisto, \textit{Universe} \textbf{5}, 173 (2019).
\bibitem{Heisenberg/2024} L. Heisenberg, \textit{Phys. Rep.} \textbf{1066}, 1 (2024).
\bibitem{KUHN} F. D'Ambrosio, L. Heisenberg and S. Kuhn, \textit{Class. Quantum Grav.} \textbf{39}, 025013 (2021).
\bibitem{BAHA} S. Bahamonde, C. G. Bohmer, S. Carloni, E. J. Copeland, W. Fang and N. Tamanini, \textit{Phys. Rept.} \textbf{775}, 1-122 (2018).
\bibitem{Pati:2023} L.~Pati, D.~Blixt and M.-J.~Guzm\'{a}n, \textit{Phys.\ Rev.\ D} \textbf{107}, 044071 (2023).
\bibitem{Guzman:2024} M.-J.~Guzm\'{a}n, L.~J\"{a}rv and L.~Pati,
\textit{Phys.\ Rev.\ D} \textbf{110}, 124013 (2024).
\bibitem{Hu:2022} K.~Hu, T.~Katsuragawa and T.~Qiu, \textit{Phys.\ Rev.\ D} \textbf{106}, 044025 (2022).
\bibitem{Bardeen:1980kt} J. M. Bardeen, \textit{Phys. Rev. D} \textbf{22}, 1882--1905 (1980).
\bibitem{Mukhanov} V. F. Mukhanov, H. A. Feldman and R. H. Brandenberger, \textit{Phys. Rept.} \textbf{215}, 203--333 (1992).
\bibitem{Hinterbichler:2011tt} K.~Hinterbichler, \textit{Rev.\ Mod.\ Phys.} \textbf{84}, 671 (2012).
\bibitem{Fierz:1939ix} M.~Fierz and W.~Pauli, \textit{Proc.\ Roy.\ Soc.\ Lond.\ A} \textbf{173}, 211 (1939).
\bibitem{2df} J. Einasto et. al., \textit{A}\&\textit{A} \textbf{462}, 397 - 410 (2006).
\bibitem{Hehl1995} F.~W.~Hehl, J.~D.~McCrea, E.~W.~Mielke, and Y.~Ne'eman,
\textit{Phys.\ Rept.} \textbf{258}, 1 (1995).
\bibitem{JIM-2} J. B. Jimenez, L. Heisenberg, T. S. Koivisto, and S. Pekar, \textit{Phys. Rev. D} \textbf{101}, 103507 (2020).
\bibitem{ALAN} A. H. Guth, \textit{Phys. Rev. D} \textbf{23}, 347-356 (1981).
\bibitem{Linde} A. D. Linde, \textit{Phys. Lett. B} \textbf{129}, 177-181 (1983).
\bibitem{Ratra:1988} B. Ratra and P.J.E. Peebles, \textit{Phys. Rev. D} \textbf{37}, 3406 (1988).
\bibitem{quint2} C. Rubano and P. Scudellaro, \textit{Gen. Rel. Grav} \textbf{34}, 307-328 (2002).
\bibitem{quintom1} Y. F. Cai, E. N. Saridakis, M. R. Setare and J. Q. Xia, \textit{Phys. Rept.} \textbf{493}, 1-60 (2010).
\bibitem{Guo/2005} Z. K. Guo, Y. S. Piao, X. M. Zhang and Y. Z. Zhang \textit{Phys. Lett. B} \textbf{608}, 177 (2005).
\bibitem{mult} E. Elizalde, S. Nojiri and S. D. Odintsov, \textit{Phys. Rev. D} \textbf{70}, 043539 (2004).
\bibitem{Riess} A. G. Riess, S. Casertano, W. Yuan, L. M. Macri and D. Scolnic, \textit{Astrophys. J.} \textbf{876}, 85 (2019).
\bibitem{fr} A. De Felice and S. Tsujikawa, \textit{Living Rev. Relativ.} \textbf{13}, 1-161 (2010).
\bibitem{ft} M. Gonzalez-Espinoza and G. Otalora, \textit{Eur. Phys.J.C} \textbf{81}, 480 (2021).
\bibitem{fg} A.D. Millano, G.Leon and A. Paliathanasis, \textit{Phys. Rev. D} \textbf{108}, 023519 (2023).
\bibitem{ratrareview} P. J. E. Peebles and B. Ratra, \textit{Rev. Mod. Phys.} \textbf{75}, 559 (2003).
\bibitem{rajame} G.Leon et. al. \textit{Fortschr. Phys.}, \textbf{71}, 2300006 (2023).

\bibitem{tension} S. Mandal, S.S Mishra and P.K Sahoo, \textit{Nucl. Phys. B} \textbf{993},  116285 (2023).
\bibitem{green} Green, Schwarz and Witten, \textit{Superstring Theory}, (First edition) Cambridge University Press (1988).
\bibitem{polchinski} J. Polchinski, \textit{String Theory}, (First edition) Cambridge University Press (2005).
\bibitem{mazumdar} A. Mazumdar, S. Panda and A. Perez-Lorenzana, \textit{Nucl. Phys. B} \textbf{614}, 101-116 (2001).
\bibitem{sen1} A. Sen, \textit{Modern Physics Letters A} \textbf{17}, 1797-1804 (2002).
\bibitem{sen2} A. Sen, \textit{JHEP} \textbf{04}, 048 (2002).
\bibitem{sen3} A. Sen, \textit{JHEP} \textbf{07}, 065 (2002).
\bibitem{paddy} T. Padmanabhan et al., \textit{Phys.Rev. D} \textbf{66}, 021301 (2002).
\bibitem{gibbons1} G. W. Gibbons, \textit {Phys. Lett. B} \textbf{537}, 1-4 (2002).
\bibitem{gibbons2} G. W. Gibbons, \textit{Class.Quant.Grav.} \textbf{20}, S321-S346 (2003).
\bibitem{bhagla} J. Bhagla et al., \textit{Phys.Rev. D} \textbf{67}, 063504 (2003).
\bibitem{gorini} V. Gorini et al., \textit{Phys.Rev. D} \textbf{69}, 123512 (2004).
\bibitem{paddy2} T. Padmanabhan et al., \textit{Phys.Rev. D} \textbf{66}, 081301 (2002).

\bibitem{copeland1} E. J. Copeland et al., \textit{Phys.Rev. D} \textbf{71}, 043003 (2005).

\bibitem{aguirregabiria} Aguirregabiria et al.,\textit{Phys.Rev. D} \textbf{69}, 123502 (2004).
\bibitem{fang2010a} W. Fang and H. Q. Lu, \textit{Eur. Phys. J. C} \textbf{68}, 567-572 (2010).
\bibitem{Quiros} Quiros et al., \textit{Class. Quant. Grav.} \textbf{27}, 215021 (2010).
\bibitem{guoexp} Z. K. Guo et al., \textit{Phys. Rev. D} \textbf{71}, 023501 (2005).

\bibitem{tong} E. Silverstein and D. Tong, \textit{Phys.Rev. D} \textbf{70}, 103505 (2004).


\bibitem{R15} Timothy Clifton et al., \textit{Physics Reports}, \textbf{513}: 1-189 (2012).
\bibitem{R18} M. Hohmann et al, \textit{Phys. Rev. D}, \textbf{99}: 024009 (2019).






\bibitem{R24} S. Capozziello, V. De Falco and C. Ferrara, \textit{Eur. Phys. J. C.}, \textbf{82}: 865 (2022).

\bibitem{R25} D. Zhao, \textit{Eur. Phys. J. C}, \textbf{82}: 303 (2022).

\bibitem{R26} A. De and L.T. How, \textit{Phys. Rev. D}, \textbf{106}: 048501 (2022).

\bibitem{R27} N. Frusciante, \textit{Phys. Rev. D}, \textbf{103}: 0444021 (2021).
\bibitem{R29} M. Calza and L. Sebastiani, \textit{Eur. Phys. J. C.}, \textbf{83}: 247 (2023).
\bibitem{Zwicky1933}
F.~Zwicky, \textit{Helvetica Physica Acta} \textbf{6}, 110 (1933).

\bibitem{Rubin1980}
V.~C.~Rubin, W.~K.~Ford~Jr. and N.~Thonnard, 
\textit{Astrophysical Journal} \textbf{238}, 471 (1980).

\bibitem{Yeager2022}
A.~J.~Yeager, \textit{Bright Galaxies, Dark Matter and Beyond}, 
MIT Press (2022).
\bibitem{Boehmer/2007} C. G. Boehmer and T. Harko \textit{JCAP} \textbf{06}, 025 (2007).
\bibitem{Sahni/2008} V. Sahni, A. Shafieloo and A. A. Starobinsky, 
\textit{Phys. Rev. D} \textbf{78}, 103502 (2008).
\bibitem{Sahni/2003} V. Sahni, T. D. Saini, A. A. Starobinsky and U. Alam, 
\textit{JETP Lett.} \textbf{77}, 201 (2003) .
\bibitem{Rasmussen2005} C. Rasmussen and C. Williams, Gaussian Processes for Machine Learning, The MIT Press, (2006).
\bibitem{Seikel2012} M. Seikel, C. Clarkson and  M. Smith \textit{JCAP} \textbf{06}, 036 (2012).
\bibitem{Seikel2013} M. Seikel and C. Clarkson, \textit{arXiv}: 1311.6678 (2013).
\bibitem{Mehrabi2021} A. Mehrabi and M. Rezaei \textit{Astrophys. J.} \textbf{923}, 274 (2021).
\bibitem{Busti:2014} V. C. Busti, C. Clarkson and M. Seikel, \textit{Mon. Not. R. Astron. Soc. Lett.} \textbf{441}, L11 (2014).
\bibitem{Sun:2021} W. Sun, K. Jiao and T.-J. Zhang, \textit{Astrophys. J.} \textbf{915}, 123 (2021).
\bibitem{vafa} C. Vafa, \textit{arXiv}: 0509212 (2005).
\bibitem{elizaldeswampland1} E. Elizalde and M. Khurshudyan, \textit{Phys. Rev. D} \textbf{99}, 103533  (2019).
\bibitem{elizaldeswampland2} E. Elizalde and M. Khurshudyan, \textit{Eur. Phys. J. C } \textbf{82}, 811 (2022).
































\bibitem{Simpson/2025} G. Simpson, K. Bolejko and S. Walters, \textit{Class. Quantum Grav.} \textbf{42}, 145001 (2025).
\bibitem{Kazantzidis/2018} L. Kazantzidis and L. Perivolaropoulos, \textit{Phys. Rev. D} \textbf{97}, 103503 (2018).
\bibitem{Das/2013} U. Das and B. Mukhopadhyay, \textit{Phys. Rev. Lett.} \textbf{110}, 071102 (2013).

\bibitem{Pun/2008} C. S. J. Pun, Z. Kovács and T. Harko, \textit{Phys. Rev. D} \textbf{78}, 024043 (2008).



\bibitem{Jimenez/JCAP2018} J. B. Jimenez, L. Heisenberg and T. S. Koivisto, \textit{JCAP} \textbf{08}, 039 (2018).



\bibitem{Mohanty/2024} D. Mohanty, S. Ghosh and P. K. Sahoo, \textit{Ann. Phys.} \textbf{463}, 169636 (2024).

\bibitem{Mohanty/2025} D. Mohanty, S. Ghosh and P. K. Sahoo, \textit{Int. J. Mod. Phys. D} \textbf{34}, 2550041 (2025).

\bibitem{Amendola/2000} L. Amendola, \textit{Phys. Rev. D} \textbf{62}, 043511 (2000).
\bibitem{FarrarPeebles/2004} G. R. Farrar and P. J. E. Peebles, \textit{Astrophys. J.} \textbf{604}, 1 (2004).
\bibitem{ZimdahlPavon/2007} W. Zimdahl and D. Pavon, \textit{Class. Quantum Grav.} \textbf{24}, 5461 (2007).
\bibitem{Bolotin/2015} Yu. L. Bolotin, A. Kostenko, O. A. Lemets and D. A. Yerokhin, \textit{Int. J. Mod. Phys. D} \textbf{24}, 1530007 (2015).
\bibitem{Wang/2016} B. Wang, E. Abdalla, F. Atrio-Barandela and D. Pav\'on, \textit{Rep. Prog. Phys.} \textbf{79}, 096901 (2016).
\bibitem{Bhat/2024} A. Bhat, R. Solanki and P. K. Sahoo, \textit{Gen. Relativ. Gravit.} \textbf{56}, 63 (2024).

\bibitem{Kamenshchik/2001} A. Kamenshchik, U. Moschella and V. Pasquier, \textit{Phys. Lett. B} \textbf{511}, 265 (2001).
\bibitem{Bento/2002} M. C. Bento, O. Bertolami and A. Sen, \textit{Phys. Rev. D} \textbf{66}, 043507 (2002).

\bibitem{Kamenshchik/2000} A. Kamenshchik, U. Moschella and V. Pasquier, \textit{Phys. Lett. B} \textbf{487}, 7 (2000).
\bibitem{Kama/1998} S. K. Kama, \textit{Phys. Lett. B} \textbf{424}, 39 (1998).

\bibitem{Novello/2005} M. Novello, M. Makler, L. S. Werneck and C. A. Romero, \textit{Phys. Rev. D} \textbf{71}, 043515 (2005).
\bibitem{Benaoum/2022} H. Benaoum, \textit{Universe} \textbf{8}, 340 (2022).
\bibitem{Carturan/2003} D. Carturan and F. Finelli, \textit{Phys. Rev. D} \textbf{68}, 103501 (2003).

\bibitem{Bertolami/2004} O. Bertolami, A. A. Sen, S. Sen and P. T. Silva, \textit{Mon. Not. R. Astron. Soc.} \textbf{353}, 329 (2004).

\bibitem{Makler/2003} M. Makler, S. Quinet de Oliveira and I. Waga, \textit{Phys. Lett. B} \textbf{555}, 1 (2003).

\bibitem{Bento/2003} M. C. Bento, O. Bertolami and A. A. Sen, \textit{Phys. Lett. B} \textbf{575}, 172 (2003).

\bibitem{Dev/2003} A. Dev, D. Jain and J. S. Alcaniz, \textit{Phys. Rev. D} \textbf{67}, 023515 (2003).

\bibitem{Alcaniz/2003} J. S. Alcaniz, D. Jain and A. Dev, \textit{Phys. Rev. D} \textbf{67}, 043514 (2003).

\bibitem{Cunha/2004} J. V. Cunha, J. A. S. Lima and J. S. Alcaniz, \textit{Phys. Rev. D} \textbf{69}, 083501 (2004).


\bibitem{Randall/1999} L. Randall and R. Sundrum, \textit{Phys. Rev. Lett.} \textbf{83}, 3370 (1999).

\bibitem{Israel:1966} W. Israel, \textit{Nuovo Cimento B} \textbf{44}, 1 (1966).

\bibitem{Shiromizu:2000} T. Shiromizu, K. Maeda and M. Sasaki, \textit{Phys. Rev. D} \textbf{62}, 044021 (2000).

\bibitem{Bradley/1995} C. C. Bradley, C. A. Sackett, J. J. Tollett and R. G. Hulet, \textit{Phys. Rev. Lett.} \textbf{75}, 1687 (1995).

\bibitem{Harko/2011} T. Harko, \textit{Phys. Rev. D} \textbf{83}, 123515 (2011).
\bibitem{Harko/2015} T. Harko and F. S. N. Lobo, \textit{Phys. Rev. D} \textbf{92}, 043011 (2015).
\bibitem{Mambrini/2016} Y. Mambrini, S. Profumo and F. S. Queiroz, \textit{Phys. Lett. B} \textbf{760}, 807 (2016).
\bibitem{Hooper/2007} D. Hooper and S. Profumo, \textit{Phys. Rep.} \textbf{453}, 29 (2007).
\bibitem{Mishra/2020} S. S. Mishra and V. Sahni, \textit{JCAP} \textbf{04}, 007 (2020).
\bibitem{Odintsov/2019} S. D. Odintsov and V. K. Oikonomou, \textit{Phys. Rev. D} \textbf{99}, 104070 (2019).
\bibitem{Das/2018} S. Das and R. K. Bhaduri, \textit{arXiv}:1812.07647 [gr-qc] (2018).
\bibitem{Das/2023} S. Das and S. Sur, \textit{Phys. Dark Univ.} \textbf{42}, 101331 (2023).

\bibitem{Mahichi/2021} E. Mahichi, A. Amani and M. A. Ramzanpour, \textit{Can. J. Phys.} \textbf{99}, 11 (2021).
\bibitem{Mahichi/2022} E. Mahichi, A. Amani and M. A. Ramzanpour, \textit{Mod. Phys. Lett. A} \textbf{37}, 2250228 (2022).
\bibitem{Mahichi/2023} E. Mahichi and A. Amani, \textit{Phys. Dark Univ.} \textbf{39}, 101167 (2023).
\bibitem{Setare/2009} M. R. Setare, J. Sadeghi and A. R. Amani, \textit{Phys. Lett. B} \textbf{673}, 241 (2009).
\bibitem{Chimento/2003} L. P. Chimento, A. S. Jakubi, D. Pavon and W. Zimdahl, \textit{Phys. Rev. D} \textbf{67}, 083513 (2003).


\bibitem{Moresco/2016} M. Moresco, \textit{Mon. Not. Roy. Astron. Soc.} \textbf{463}, L6 (2016).
\bibitem{Solanki2021} R. Solanki, S. K. J. Pacif, A. Parida and P.K. Sahoo, \textit{Phys. Dark Univ.} \textbf{32}, 100820 (2021).
\bibitem{Solanki2022} R. Solanki, A. De, S. Mandal and P. K. Sahoo, \textit{Phys. Dark Universe} \textbf{36}, 101053 (2022).
\bibitem{Kowalski/2008} M. Kowalski et al., \textit{Astrophys. J.} \textbf{686}, 749 (2008).

\bibitem{Amanullah/2010} R. Amanullah et al., \textit{Astrophys. J.} \textbf{716}, 712 (2010).

\bibitem{Suzuki/2012} N. Suzuki et al., \textit{Astrophys. J.} \textbf{746}, 85 (2012).

\bibitem{Betoule/2014} M. Betoule et al., \textit{Astron. Astrophys.} \textbf{568}, A22 (2014).

\bibitem{Scolnic/2018} D. M. Scolnic et al., \textit{Astrophys. J.} \textbf{859}, 101 (2018).

\bibitem{Scolnic/2022} D. M. Scolnic et al., \textit{Astrophys. J.} \textbf{938}, 113 (2022).
\bibitem{Akaike/1974} H. Akaike, \textit{IEEE Trans. Autom. Control} \textbf{19}, 716 (1974)

\bibitem{Schwarz/1978} G. Schwarz, \textit{Ann. Statist.} \textbf{6}, 461 (1978).

\bibitem{Liddle/2007} A. R. Liddle, \textit{Mon. Not. Roy. Astron. Soc.} \textbf{377}, L74 (2007).






\bibitem{cmb} A. A. Penzias and R. W. Wilson \textit{Astrophys. J.} \textbf{142}, 419-421 (1965).
\bibitem{gamow} R. A. Alpher, H. Bethe and G. Gamow, \textit{Phys. Rev.} \textbf{73}, 803 (1948).





\bibitem{mincouple} A. Mohammadi and F. Kheirandish, \textit{Phys.Dark Univ.} \textbf{42}, 101362 (2023).
\bibitem{nonmincouple} A. I. Keskin, \textit{Mod. Phys. Lett. A} \textbf{33}, 1850199 (2018).
\bibitem{medbi} S. Ghosh, R. Solanki and P. K. Sahoo, \textit{Chin. Phys. C } \textbf{48}, 095102 (2024).
\bibitem{Jimenez2002} R. Jimenez and A. Loeb \textit{Astrophys. J.} \textbf{573}, 37-42 (2002).
\bibitem{Moresco2016} M. Moresco et. al. \textit{JCAP} \textbf{05}, 14 (2016).
\bibitem{Gaztanaga2009} A. G. Sanchez et. al. \textit{Mon. Not. Roy. Astron. Soc.} \textbf{400}, 1643 (2009).
\bibitem{Alam2017} S. Alam et. al. \textit{Mon. Not. Roy. Astron. Soc.} \textbf{470}, 2617--2652 (2017).
\bibitem{gaurav1} G. Gadbail, S. Mandal, P. K. Sahoo \textit{Astrophys. J.} \textbf{972}, 172 (2024).
\bibitem{desi} A.G. Adame et. al. \textit{JCAP} \textbf{2015} \textbf{(01)}, 124 (2025).
\bibitem{desi1} A.G. Adame et. al. \textit{JCAP} \textbf{2015} \textbf{(04)}, 012 (2025).

\bibitem{sahnireconstruction} V. Sahni and A. Starobinsky, \textit{Int. J. Mod. Phys. D} \textbf{15}, 2105-2132 (2006).
\bibitem{lindergaussian} A. Shafieloo, A. G. Kim and E. V. Linder, \textit{Phys. Rev. D} \textbf{85}, 123530  (2012).

\bibitem{jesusgaussian} J. F. Jesus, R. Valentim, A. A. Escobal, S. H. Pereira and D. Benndorf, \textit{JCAP} \textbf{11}, 037 (2022).

\bibitem{niugaussian} J. Niu, K. Jiao, P. He and T. J. Zhang, \textit{Astrophys. J.} \textbf{972}, 1-14 (2024).



\bibitem{gaurav2} G. Gadbail, S. Mandal, P. K. Sahoo and K. Bamba \textit{Phys. Lett. B} \textbf{860}, 139232 (2025).
\bibitem{genetic} R. E. Ouardi, A. Bouali, I. E. Bojaddaini, A. Errahmani and T. Ouali, \textit{Chin. Phys. C} (2025).
\bibitem{gaussianom} P. Mukherjee and A. A. Sen, \textit{Phys. Rev. D} \textbf{110}, 123502 (2024).
\bibitem{hao} J. Hao and X. Li, \textit{Phys. Rev. D} \textbf{68}, 083514 (2003).
\bibitem{chingangbam} P. Chingangbam and T. Qureshi, \textit{Int. J. Mod. Phys. A} \textbf{20}, 6083 (2005).
\bibitem{expnentiallavinia} L. Heisenberg, M. Bartelmann, R. Brandenberger and A. Refregier, \textit{Phys. Rev. D} \textbf{98}, 123502 (2018).
\bibitem{Peebles:1988} P.J.E. Peebles and B. Ratra, \textit{Astrophys. J. Lett.} \textbf{325}, L17 (1988).




\bibitem{freefieldlopez} L.A. Urena-Lopez and M.J. Reyes-Ibarra, \textit{Int. J. Mod. Phys. D} \textbf{18}, 621 (2009).
\bibitem{higgsamin} M. A. Amin, J. Fan, K. D. Lozanov and Matthew Reece, \textit{Phys. Rev. D} \textbf{99}, 035008 (2019). 
\bibitem{higgs} P. W. Higgs, \textit{Phys. Rev. Lett.} \textbf{13}, 508 (1964).



















\bibitem{phant} L. A. Urena-Lopez, \textit{JCAP} \textbf{09}, 013 (2005).












\bibitem{COPE} E. J. Copeland, A. R. Liddle and D. Wands, \textit{Phys. Rev. D} \textbf{57}, 4686-4690 (1998).

\bibitem{DE-3} H. Shabani, Avik De and T. H. Loo, \textit{Eur. Phys. J. C} \textbf{88}, 535 (2023).

\bibitem{WOM} W. Khyllep, J. Dutta, E. N. Saridakis, and K. Yesmakhanova, \textit{Phys. Rev. D} \textbf{107}, 044022 (2023).

\bibitem{Mishra-1} L. K. Duchaniya, S. V. Lohakare, B. Mishra and S. K. Tripathy, \textit{Eur. Phys. J. C} \textbf{82}, 448 (2022).

\bibitem{Mishra-2} S. A. Kadam, B. Mishra and J. L. Said, \textit{Eur. Phys. J. C} \textbf{82}, 680 (2022).

\bibitem{HAMID} H. Shabani and M. Farhoudi, \textit{Phys. Rev. D} \textbf{88}, 044048 (2013).

\bibitem{APLL} A. Paliathanasis, \textit{Phys. Dark Univ.} \textbf{41}, 101255 (2023). 







\bibitem{LAS} L. A. Salo and J. Haro, \textit{Eur. Phys. J. C } \textbf{81}, 105 (2021).

\bibitem{TBR} T. Barreiro, E. J. Copeland and N. J. Nunes, \textit{Phys. Rev. D} \textbf{61}, 127301 (2000).

\bibitem{DBL} D. Baumann and  L. McAllister, \textit{Inflation and String Theory} (Cambridge Monographs on Mathematical Physics),  Cambridge University Press (2015).

\bibitem{JYJ} J. Yearsley and J. D Barrow, \textit{Class. Quantum Grav.} \textbf{13}, 2693 (1996).

\bibitem{Perko:2001} L. Perko, Differential Equations and Dynamical Systems, \textit{Springer}, New York (2001).

\bibitem{Wiggins:2003} S. Wiggins, Introduction to Applied Nonlinear Dynamical Systems and Chaos, \textit{Springer}, New York (2003).

\bibitem{Carr:2012} J. Carr, Applications of Center Manifold Theory, \textit{Springer}, New York (2012).






\bibitem{born} W. Fang, H. Q. Lu and Z. G. Huang, \textit{Int J Mod Phys A.} \textbf{22}, 2173-2195 (2007).

\bibitem{LRA} L. R. Abramo and F. Finelli, \textit{Phys. Lett. B} \textbf{575}, 165-171 (2003). 



\bibitem{Carroll:2000} S. M. Carroll, \textit{Living Rev. Rel.} \textbf{4}, 1 (2001). 













\bibitem{OO1} S. Ghosh, R. Solanki and P. K. Sahoo, \textit{Phys. Scr.}, \textbf{99}: 055021 (2024).




\bibitem{effective} A. Baldazzi, O. Melichev and R. Percacci, \textit{Ann. Phys.}, \textbf{438}: 168757 (2022).









\bibitem{Parsons:1995kt} P. Parsons and J. D. Barrow, \textit{Class. Quant. Grav.} \textbf{12}, 1715--1721 (1995).

\bibitem{Barrow:2016qkh} J. D. Barrow and A. Paliathanasis, \textit{Phys. Rev. D} \textbf{94}, 083518 (2016).
\bibitem{Feng:2004ff} B. Feng, M. Li, Y.-S. Piao and X. Zhang, \textit{Phys. Lett. B} \textbf{634}, 101--105 (2006).
\bibitem{Zhang:2005eg} X.-F. Zhang, H. Li, Y.-S. Piao and X.-M. Zhang, \textit{Mod. Phys. Lett. A} \textbf{21}, 231--242 (2006).
\bibitem{Zhang:2005kj} X. Zhang, \textit{Commun. Theor. Phys.} \textbf{44}, 762--768 (2005).
\bibitem{Lazkoz:2006pa} R. Lazkoz and G. Leon, \textit{Phys. Lett. B} \textbf{638}, 303--309 (2006).
\bibitem{Lazkoz:2007mx} R. Lazkoz, G. Leon and I. Quiros, \textit{Phys. Lett. B} \textbf{649}, 103--110 (2007).
\bibitem{Setare:2008pc} M. R. Setare and E. N. Saridakis, \textit{Phys. Lett. B} \textbf{671}, 331--338 (2009).

\bibitem{Setare:2008pz} M. R. Setare and E. N. Saridakis, \textit{Phys. Lett. B} \textbf{668}, 177--181 (2008).
\bibitem{Leon:2012vt} G. Leon, Y. Leyva and J. Socorro, \textit{Phys. Lett. B} \textbf{732}, 285--297 (2014).
\bibitem{Leon:2018lnd} G. Leon, A. Paliathanasis and J. L. Morales-Mart\'\i{}nez, \textit{Eur. Phys. J. C} \textbf{78}, 753 (2018).
\bibitem{Mishra:2018dzq} S. Mishra and S. Chakraborty, \textit{Eur. Phys. J. C} \textbf{78}, 917 (2018).
\bibitem{Marciu:2020vve} M. Marciu, \textit{Eur. Phys. J. C} \textbf{80}, 894 (2020).
\bibitem{Dimakis:2020tzc} N. Dimakis and A. Paliathanasis, \textit{Class. Quant. Grav.} \textbf{38}, 075016 (2021).
\bibitem{Chervon:2013btx} S. V. Chervon, \textit{Quant. Matt.} \textbf{2}, 71--82 (2013).
\bibitem{Paliathanasis:2020wjl} A. Paliathanasis, \textit{Class. Quant. Grav.} \textbf{37}, 195014 (2020).
\bibitem{Elizalde:2008yf} E. Elizalde, S. Nojiri, S. D. Odintsov, D. Saez-Gomez and V. Faraoni, \textit{Phys. Rev. D} \textbf{77}, 106005 (2008).
\bibitem{Paliathanasis:2018vru} A. Paliathanasis, G. Leon and S. Pan, \textit{Gen. Rel. Grav.} \textbf{51}, 106 (2019).
\bibitem{DiValentino:2021izs} E. Di Valentino \textit{et al.}, \textit{Class. Quant. Grav.} \textbf{38}, 153001 (2021). 
\bibitem{Efstathiou:2021ocp} G. Efstathiou, \textit{Mon. Not. Roy. Astron. Soc.} \textbf{505}, 3866--3872 (2021). 
\bibitem{Banerjee:2020xcn} A. Banerjee, H. Cai, L. Heisenberg, E. Ó. Colgáin, M. M. Sheikh-Jabbari and T. Yang, \textit{Phys. Rev. D} \textbf{103}, L081305 (2021).
\bibitem{Lee:2022cyh} B.-H. Lee, W. Lee, E. Ó. Colgáin, M. M. Sheikh-Jabbari and S. Thakur, \textit{JCAP} \textbf{04}, 004 (2022). 
\bibitem{Capozziello:2005tf} S. Capozziello, S. Nojiri and S. D. Odintsov, \textit{Phys. Lett. B} \textbf{632}, 597--604 (2006).
\bibitem{Nojiri:2005pu} S. Nojiri and S. D. Odintsov, \textit{Gen. Rel. Grav.} \textbf{38}, 1285--1304 (2006).
\bibitem{Briscese:2006xu} F. Briscese, E. Elizalde, S. Nojiri and S. D. Odintsov, \textit{Phys. Lett. B} \textbf{646}, 105--111 (2007). 
\bibitem{Nojiri:2006ww} S. Nojiri and S. D. Odintsov, \textit{Phys. Lett. B} \textbf{637}, 139--148 (2006). 
\bibitem{Capozziello:2005mj} S. Capozziello, S. Nojiri and S. D. Odintsov, \textit{Phys. Lett. B} \textbf{634}, 93--100 (2006). 
\bibitem{Astashenok:2012kb} A. V. Astashenok, S. Nojiri, S. D. Odintsov and R. J. Scherrer, \textit{Phys. Lett. B} \textbf{713}, 145--153 (2012). 
\bibitem{Astashenok:2012tv} A. V. Astashenok, S. Nojiri, S. D. Odintsov and A. V. Yurov, \textit{Phys. Lett. B} \textbf{709}, 396--403 (2012). 
\bibitem{Ito:2011ae} Y. Ito, S. Nojiri and S. D. Odintsov, \textit{Entropy} \textbf{14}, 1578--1605 (2012). 
\bibitem{Frampton:2011rh} P. H. Frampton, K. J. Ludwick, S. Nojiri, S. D. Odintsov and R. J. Scherrer, \textit{Phys. Lett. B} \textbf{708}, 204--211 (2012). 
\bibitem{Bamba:2014daa} K. Bamba, S. Nojiri and S. D. Odintsov, \textit{Phys. Lett. B} \textbf{737}, 374--378 (2014). 
\bibitem{Odintsov:2018zai} S. D. Odintsov and V. K. Oikonomou, \textit{EPL} \textbf{126}, 20002 (2019). 
\bibitem{Odintsov:2018uaw} S. D. Odintsov and V. K. Oikonomou, \textit{Phys. Rev. D} \textbf{98}, 024013 (2018). 
\bibitem{verhulst2006method} F. Verhulst, \textit{Methods and Applications of Singular Perturbations}, Springer (2006).
\bibitem{fenichel} N. Fenichel, \textit{J. Differ. Equ.} \textbf{31}, 53--98 (1979).
\bibitem{Fusco1989SlowmotionMD} G. Fusco and J. K. Hale, \textit{J. Dyn. Diff. Eq.} \textbf{1}, 75--94 (1989).
\bibitem{Berglund} N. Berglund and B. Gentz, \textit{Noise-Induced Phenomena in Slow-Fast Dynamical Systems: A Sample-Paths Approach}, Springer-Verlag (2006).
\bibitem{holmes2012introduction} M. H. Holmes, \textit{Introduction to Perturbation Methods}, Springer (2012).
\bibitem{kevorkian2013perturbation} J. Kevorkian and J. D. Cole, \textit{Perturbation Methods in Applied Mathematics}, Springer (2013).

\bibitem{Rendall:2006cq} A. D. Rendall, \textit{Class. Quant. Grav.} \textbf{24}, 667--678 (2007).
\bibitem{Alho:2015cza} A. Alho, J. Hell and C. Uggla, \textit{Class. Quant. Grav.} \textbf{32}, 145005 (2015).
\bibitem{Alho:2019pku} A. Alho, V. Bessa and F. C. Mena, \textit{J. Math. Phys.} \textbf{61}, 032502 (2020).
\bibitem{Fajman:2020yjb} D. Fajman, G. Hei\ss{}el and M. Maliborski, \textit{Class. Quant. Grav.} \textbf{37}, 135009 (2020).
\bibitem{Fajman:2021cli} D. Fajman, G. Hei\ss{}el and J. W. Jang, \textit{Class. Quant. Grav.} \textbf{38}, 085005 (2021).

\bibitem{Leon:2021lct} G. Leon \textit{et al.}, \textit{Eur. Phys. J. C} \textbf{81}, 414 (2021). [Erratum: \textbf{81}, 1097 (2021)].

\bibitem{Leon:2021rcx} G. Leon \textit{et al.}, \textit{Eur. Phys. J. C} \textbf{81}, 489 (2021). [Erratum: \textbf{81}, 1100 (2021)].
\bibitem{Chakraborty:2021vcr} S. Chakraborty, E. González, G. Leon and B. Wang, \textit{Eur. Phys. J. C} \textbf{81}, 1039 (2021). 
\bibitem{Leon:2020pvt} G. Leon, E. González, A. D. Millano and F. O. F. Silva, \textit{Class. Quant. Grav.} \textbf{39}, 115003 (2022).
\bibitem{Leon:2019iwj} G. Leon and F. O. F. Silva, \textit{arXiv}:1912.09856 (2019).
\bibitem{Leon:2020pfy} G. Leon and F. O. F. Silva, \textit{Class. Quant. Grav.} \textbf{37}, 245005 (2020).
\bibitem{Leon:2020ovw} G. Leon and F. O. F. Silva, \textit{Class. Quant. Grav.} \textbf{38}, 015004 (2021).
\bibitem{Leon:2021hxc} G. Leon \textit{et al.}, \textit{Eur. Phys. J. C} \textbf{81}, 867 (2021). [Erratum: \textbf{81}, 1096 (2021)].
\bibitem{Llibre:2012zz} J. Llibre and C. Vidal, \textit{J. Math. Phys.} \textbf{53}, 012702 (2012).
\bibitem{Leon:2008de} G. Leon, \textit{Class. Quant. Grav.} \textbf{26}, 035008 (2009).
\bibitem{Giambo:2009byn} R. Giambo and J. Miritzis, \textit{Class. Quant. Grav.} \textbf{27}, 095003 (2010).
\bibitem{Tzanni:2014eja} K. Tzanni and J. Miritzis, \textit{Phys. Rev. D} \textbf{89}, 103540 (2014). [Addendum: \textit{Phys. Rev. D} \textbf{89}, 129902 (2014)]
\bibitem{Paliathanasis:2015cza} A. Paliathanasis, S. Pan, and S. Pramanik, \textit{Class. Quant. Grav.} \textbf{32}, 245006 (2015).
\bibitem{Paliathanasis:2021egx} A. Paliathanasis, G. Leon, W. Khyllep, J. Dutta and S. Pan, \textit{Eur. Phys. J. C} \textbf{81}, 607 (2021).
\bibitem{Escobar:2013js} D. Escobar, C. R. Fadragas, G. Leon and Y. Leyva, \textit{Astrophys. Space Sci.} \textbf{349}, 575--602 (2014).
\bibitem{delCampo:2013vka} S. del Campo et.al. \textit{Phys. Rev. D} \textbf{88}, 023532 (2013).
\bibitem{Wetterich:1987fm} C. Wetterich, \textit{Nucl. Phys. B} \textbf{302}, 668--696 (1988).

\bibitem{Sahni:1999qe} V. Sahni and L. M. Wang, \textit{Phys. Rev. D} \textbf{62}, 103517 (2000).
\bibitem{Sahni:1999gb} V. Sahni and A. A. Starobinsky, \textit{Int. J. Mod. Phys. D} \textbf{9}, 373--444 (2000).
\bibitem{Urena-Lopez:2000ewq} L. A. Ureña-López and T. Matos, \textit{Phys. Rev. D} \textbf{62}, 081302 (2000).
\bibitem{Matos:2000ng} T. Matos and L. A. Ureña-López, \textit{Class. Quant. Grav.} \textbf{17}, L75--L81 (2000).
\bibitem{Cardenas:2002np} R. Cardenas, T. Gonzalez, Y. Leiva, O. Martin and I. Quiros, \textit{Phys. Rev. D} \textbf{67}, 083501 (2003).
\bibitem{Matos:2009hf} T. Matos, J. R. Luevano, I. Quiros, L. A. Ureña-López and J. A. Vázquez, \textit{Phys. Rev. D} \textbf{80}, 123521 (2009).
\bibitem{Copeland:2009be} E. J. Copeland, S. Mizuno and M. Shaeri, \textit{Phys. Rev. D} \textbf{79}, 103515 (2009).
\bibitem{Leyva:2009zz} Y. Leyva, D. Gonzalez, T. Gonzalez, T. Matos and I. Quiros, \textit{Phys. Rev. D} \textbf{80}, 044026 (2009).
\bibitem{Lidsey:2001nj} J. E. Lidsey, T. Matos and L. A. Ureña-López, \textit{Phys. Rev. D} \textbf{66}, 023514 (2002).
\bibitem{Pavluchenko:2003ge} S. A. Pavluchenko, \textit{Phys. Rev. D} \textbf{67}, 103518 (2003).

\bibitem{Gonzalez:2007hw} T. Gonzalez, R. Cardenas, I. Quiros and Y. Leyva, \textit{Astrophys. Space Sci.} \textbf{310}, 13--18 (2007).
\bibitem{Gonzalez:2006cj} T. Gonzalez, G. Leon and I. Quiros, \textit{Class. Quant. Grav.} \textbf{23}, 3165--3179 (2006).
\bibitem{Saridakis:2009pj} E. N. Saridakis, \textit{Nucl. Phys. B} \textbf{819}, 116--126 (2009). 
\bibitem{Saridakis:2009ej} E. N. Saridakis, \textit{Nucl. Phys. B} \textbf{830}, 374--389 (2010). 
\bibitem{Leon:2009dt} G. Leon and E. N. Saridakis, \textit{Phys. Lett. B} \textbf{693}, 1--10 (2010).
\bibitem{Chang:2013cba} H. Y. Chang and R. J. Scherrer, \textit{Phys. Rev. D} \textbf{88}, 083003 (2013). 
\bibitem{Skugoreva:2013ooa} M. A. Skugoreva, S. V. Sushkov and A. V. Toporensky, \textit{Phys. Rev. D} \textbf{88}, 083539 (2013). 
\bibitem{Pavlov:2013nra} A. Pavlov, S. Westmoreland, K. Saaidi and B. Ratra, \textit{Phys. Rev. D} \textbf{88}, 123513 (2013).

\bibitem{Vernov:2019ubo} S. Yu. Vernov, V. R. Ivanov and E. O. Pozdeeva, \textit{Phys. Part. Nucl.} \textbf{51}, 744--749 (2020). 


\bibitem{Brandenberger:1992dw} R. H. Brandenberger, H. Feldman, V. F. Mukhanov and T. Prokopec, in \textit{The Origin of Structure in the Universe} (1992).

\bibitem{Brandenberger:1993zc} R. H. Brandenberger, H. Feldman and V. F. Mukhanov, \textit{37th Yamada Conference: Evolution of the Universe and its Observational Quest}, 19--30 (1993).

\bibitem{Brandenberger:1992qj} R. H. Brandenberger, H. Feldman and V. F. Mukhanov, in \textit{International Conference on Gravitation and Cosmology} (1992).
\bibitem{dunsby:1997} P. K. S. Dunsby, in \textit{Dynamical Systems in Cosmology}, eds. J. Wainwright and G. F. R. Ellis (Cambridge Univ. Press, 1997), pp. 287--304.
\bibitem{amendola_tsujikawa_2010} L. Amendola and S. Tsujikawa, \textit{Dark Energy: Theory and Observations}, Cambridge University Press (2010).
\bibitem{Amendola:1999dr} L. Amendola, \textit{Mon. Not. Roy. Astron. Soc.} \textbf{312}, 521 (2000).
\bibitem{Basilakos:2019dof} S. Basilakos, G. Leon, G. Papagiannopoulos and E. N. Saridakis, \textit{Phys. Rev. D} \textbf{100}, 043524 (2019).
\bibitem{Alho:2019jho} A. Alho, C. Uggla and J. Wainwright, \textit{JCAP} \textbf{09}, 045 (2019).
\bibitem{Alho:2020cdg} A. Alho, C. Uggla and J. Wainwright, \textit{Class. Quant. Grav.} \textbf{37}, 225011 (2020).

\bibitem{Uggla:2011jn} C. Uggla and J. Wainwright, \textit{Class. Quant. Grav.} \textbf{28}, 175017 (2011).
\bibitem{Uggla:2011hs} C. Uggla and J. Wainwright, \textit{Class. Quant. Grav.} \textbf{29}, 105002 (2012).
\bibitem{Uggla:2012gg} C. Uggla and J. Wainwright, \textit{Gen. Rel. Grav.} \textbf{45}, 643--674 (2013).
\bibitem{Uggla:2013paa} C. Uggla and J. Wainwright, \textit{Gen. Rel. Grav.} \textbf{45}, 1467--1492 (2013).
\bibitem{Uggla:2013kya} C. Uggla and J. Wainwright, \textit{Class. Quant. Grav.} \textbf{31}, 105008 (2014).
\bibitem{Uggla:2014hva} C. Uggla and J. Wainwright, \textit{Phys. Rev. D} \textbf{90}, 043511 (2014).
\bibitem{Uggla:2018fiy} C. Uggla and J. Wainwright, \textit{Class. Quant. Grav.} \textbf{36}, 035004 (2019).
\bibitem{Uggla:2018cct} C. Uggla and J. Wainwright, \textit{Phys. Rev. D} \textbf{98}, 103534 (2018).
\bibitem{Uggla:2019rho} C. Uggla and J. Wainwright, \textit{Phys. Rev. D} \textbf{100}, 023544 (2019).
\bibitem{Landim:2019lvl} R. G. Landim, \textit{Eur. Phys. J. C} \textbf{79}, 889 (2019).
\bibitem{Uggla:2019zdm} C. Uggla and J. Wainwright, \textit{JCAP} \textbf{06}, 021 (2019).
\bibitem{Khyllep:2021wjd} W. Khyllep, J. Dutta, S. Basilakos and E. N. Saridakis, \textit{Phys. Rev. D} \textbf{105}, 043511 (2022).
\bibitem{Khyllep:2022spx} W. Khyllep, J. Dutta, E. N. Saridakis and K. Yesmakhanova, \textit{Phys. Rev. D} \textbf{107}, 044022  (2023)
\bibitem{Sharma:2021ivo} M. K. Sharma and S. Sur, \textit{Int. J. Mod. Phys. D} \textbf{31}, 2250017 (2022).

\bibitem{Sharma:2021ayk} M. K. Sharma and S. Sur, \textit{arXiv}:2112.14017 (2021).

\bibitem{Tot:2022dpr} J. Tot, B. Yildirim, A. Coley and G. Leon, \textit{Phys. Dark Univ.} \textbf{39}, 101155 (2023).
\bibitem{Kodama:1984ziu} H. Kodama and M. Sasaki, \textit{Prog. Theor. Phys. Suppl.} \textbf{78}, 1--166 (1984).
\bibitem{Mukhanov:1988jd} V. F. Mukhanov, \textit{Sov. Phys. JETP} \textbf{67}, 1297--1302 (1988).
\bibitem{Arefeva:2009tkq} I. Ya. Aref’eva, N. V. Bulatov and S. Yu. Vernov, \textit{Theor. Math. Phys.} \textbf{163}, 788--803 (2010).
\bibitem{Arefeva:2004odl} I. Ya. Aref'eva, A. S. Koshelev and S. Yu. Vernov, \textit{Theor. Math. Phys.} \textbf{148}, 895--909 (2006). 
\bibitem{Barrow:1995xb} J. D. Barrow and P. Parsons, \textit{Phys. Rev. D} \textbf{52}, 5576--5587 (1995).
\bibitem{Burd:1988ss} A. B. Burd and J. D. Barrow, \textit{Nucl. Phys. B} \textbf{308}, 929--945 (1988).
\bibitem{tavakol_1997} R. Tavakol, \textit{Dynamical Systems in Cosmology}, eds. J. Wainwright and G. F. R. Ellis (Cambridge Univ. Press, 1997), pp. 84--104.
\bibitem{Parsons:1995ew} P. Parsons and J. D. Barrow, \textit{Phys. Rev. D} \textbf{51}, 6757--6763 (1995).
\bibitem{Coley:2003mj} A. A. Coley, \textit{Dynamical Systems and Cosmology} (Kluwer, Dordrecht, 2003).
\bibitem{Leon2012CosmologicalDS} G. Leon and C. R. Fadragas, \textit{Cosmological Dynamical Systems} (LAP LAMBERT Acad. Publ., Saarbrücken, 2012).
\bibitem{Gong:2006sp} Y. Gong, A. Wang and Y. Z. Zhang, \textit{Phys. Lett. B} \textbf{636}, 286--292 (2006).
\bibitem{Setare:2008sf} M. R. Setare and E. N. Saridakis, \textit{Phys. Rev. D} \textbf{79}, 043005 (2009)
\bibitem{Chen:2008ft} X. M. Chen, Y. G. Gong and E. N. Saridakis, \textit{JCAP} \textbf{04}, 001 (2009).
\bibitem{Gupta:2009kk} G. Gupta, E. N. Saridakis and A. A. Sen, \textit{Phys. Rev. D} \textbf{79}, 123013 (2009).
\bibitem{Farajollahi:2011ym} H. Farajollahi, A. Salehi, F. Tayebi and A. Ravanpak, \textit{JCAP} \textbf{05}, 017 (2011).
\bibitem{Urena-Lopez:2011gxx} L. A. Ureña-López, \textit{JCAP} \textbf{03}, 035 (2012).
\bibitem{Escobar:2011cz} D. Escobar, C. R. Fadragas, G. Leon and Y. Leyva, \textit{Class. Quant. Grav.} \textbf{29}, 175005 (2012).
\bibitem{Escobar:2012cq} D. Escobar, C. R. Fadragas, G. Leon and Y. Leyva, \textit{Class. Quant. Grav.} \textbf{29}, 175006 (2012).
 \bibitem{Xu:2012jf} C. Xu, E. N. Saridakis and G. Leon, \textit{JCAP} \textbf{07}, 005 (2012).
\bibitem{Leon:2013qh} G. Leon, J. Saavedra and E. N. Saridakis, \textit{Class. Quant. Grav.} \textbf{30}, 135001 (2013).
\bibitem{Lucchin:1984yf} F. Lucchin and S. Matarrese, \textit{Phys. Rev. D} \textbf{32}, 1316 (1985).
\bibitem{Kitada:1992uh} Y. Kitada and K. Maeda, \textit{Class. Quant. Grav.} \textbf{10}, 703--734 (1993).
\bibitem{Halliwell:1986ja} J. J. Halliwell, \textit{Phys. Lett. B} \textbf{185}, 341 (1987).
\bibitem{Coley:1997nk} A. A. Coley, J. Ibanez and R. J. van den Hoogen, \textit{J. Math. Phys.} \textbf{38}, 5256--5271 (1997).
\bibitem{Wands:1993zm} D. Wands, E. J. Copeland and A. R. Liddle, in \textit{16th Texas Symposium on Relativistic Astrophysics} (1993), pp. 647--652.
\bibitem{Ma:1995ey} C. P. Ma and E. Bertschinger,  \textit{Astrophys. J.} \textbf{455}, 7--25 (1995).
\bibitem{Ellis} J. Wainwright and G. F. R. Ellis,  \textit{Dynamical Systems in Cosmology} (Cambridge University Press, 1997).
\bibitem{Mishra:2025} S. K. Mishra, J. L. Said and B. Mishra, \textit{Phys. Rev. D} \textbf{112}, 064019 (2025).
\bibitem{Foster:1998sk} S. Foster, \textit{Class. Quant. Grav.} \textbf{15}, 3485--3504 (1998).
\bibitem{Fadragas:2014mra} C. R. Fadragas and G. Leon, \textit{Class. Quant. Grav.} \textbf{31}, 195011 (2014).
















































\bibitem{Zee1982} A. Zee, \textit{Unity of Forces in the Universe}, World Scientific (1982).
\bibitem{Freedman:22} W. L. Freedman and B. F. Madore, \textit{Proc. Int. Astron. Union} \textbf{18}, 1 (2022).
\bibitem{Heymans:2021} L. Heymans et al., \textit{Astronomy \& Astrophysics} \textbf{646}, A140 (2021).
\bibitem{Maldacena:2003} J. M. Maldacena, \textit{Journal of High Energy Physics} \textbf{0305}, 013 (2003).
\bibitem{ArkaniHamed:2015} N. Arkani-Hamed and J. Maldacena, \textit{arXiv}:1503.08043 (2015).
\bibitem{Ghoshal:2020} A. Ghoshal et al., \textit{JCAP} \textbf{07}, 051 (2020).







\bibitem{Schrodinger1950} E. Schr\"odinger \textit{Space-Time Structure} (Cambridge University Press, 1950).


\bibitem{Dirac1964} P. A. M. Dirac \textit{Lectures on Quantum Mechanics} (Yeshiva University, 1964).
\bibitem{Henneaux:1992ig} M. Henneaux and C. Teitelboim \textit{Quantization of Gauge Systems} (Princeton University Press, 1992).
\bibitem{Teitelboim1973} C. Teitelboim \textit{Annals Phys.} \textbf{79}, 542--557 (1973).
\bibitem{Golovnev:2018ocj} A. Golovnev and T. Koivisto \textit{Phys. Rev. D} \textbf{97}, 104050 (2018).
\bibitem{Becchi:1974xu} C. Becchi, A. Rouet and R. Stora \textit{Commun. Math. Phys.} \textbf{42}, 127--132 (1975).
\bibitem{Tyutin:1975qk} I. V. Tyutin \textit{Lebedev Preprint FIAN} \textbf{39} (1975).
\end{thebibliography}
\cleardoublepage
\pagestyle{fancy}

\label{Publications}
\lhead{\emph{List of Publications}}

\chapter{List of Publications}
\section*{Thesis Publications}
\begin{enumerate}
 \item \textbf{Sayantan Ghosh}, Raja Solanki, Pradyumn Kumar Sahoo, \textit{Observational Constraints on Dissipative Chaplygin Gas Cosmology in the Framework of Coincident $f(Q)$ Gravity}, \textcolor{blue}{Chinese Physics C}, \textbf{50}, 065110 (2026). DOI: 10.1088/1674-1137/ae50e5.
\item \textbf{Sayantan Ghosh}, Gaurav N. Gadbail, P.K. Sahoo, Kazuharu Bamba \textit{Reconstruction of a dark energy model for the Dirac-Born-Infeld scalar field with the Hubble and DESI dataset via Gaussian process.}, (manuscript submitted),	\textcolor{blue}{arXiv:2607.09731}.

\item \textbf{Sayantan Ghosh}, Raja Solanki, P.K. Sahoo, \textit{Dynamical system analysis of scalar field cosmology in coincident $f(Q)$ gravity}, \textcolor{blue}{Physica Scripta}, \textbf{99} (5), 055021 (2024). 

\item \textbf{Sayantan Ghosh}, Raja Solanki, P.K. Sahoo, \textit{Dynamical system analysis of Dirac–Born–Infeld scalar field cosmology in coincident $f(Q)$ gravity}, \textcolor{blue}{Chinese Physics C}, \textbf{48} (9), 095102 (2024). 

\item Genly Leon, Saikat Chakraborty, \textbf{Sayantan Ghosh}, Raja Solanki, P.K. Sahoo, Esteban González, \textit{Scalar Field Evolution at Background and Perturbation Levels for a Broad Class of Potentials}, \textcolor{blue}{Fortschritte der Physik}, \textbf{71} (10–11), 2300006 (2023). 

\end{enumerate}
\section*{Other Publications}
\begin{enumerate}
\item Zinnat Hassan, \textbf{Sayantan Ghosh}, P.K. Sahoo, Kazuharu Bamba, \textit{Casimir wormholes in modified symmetric teleparallel gravity}, \textcolor{blue}{European Physical Journal C}, \textbf{82} (12), 1116 (2022). DOI: 10.1140/epjc/s10052-022-11107-0.
\item Zinnat Hassan, \textbf{Sayantan Ghosh}, P.K. Sahoo, V. Sree Hari Rao, \textit{GUP corrected Casimir wormholes in $f(Q)$ gravity}, \textcolor{blue}{General Relativity and Gravitation}, \textbf{55} (8), 90 (2023). DOI: 10.1007/s10714-023-03139-y.
\item Moreshwar Tayde, \textbf{Sayantan Ghosh}, P.K. Sahoo, \textit{Non-exotic static spherically symmetric thin-shell wormhole solution in $f(Q,T)$ gravity}, \textcolor{blue}{Chinese Physics C}, \textbf{47} (7), 075102 (2023). DOI: 10.1088/1674-1137/acd2b7.
\item Debasmita Mohanty, \textbf{Sayantan Ghosh}, P.K. Sahoo, \textit{Study of charged gravastar model in $f(Q)$ gravity}, \textcolor{blue}{Annals of Physics}, \textbf{463}, 169636 (2024). DOI: 10.1016/j.aop.2024.169636.

\item Debasmita Mohanty, \textbf{Sayantan Ghosh}, P.K. Sahoo, \textit{Charged gravastar model in noncommutative geometry under $f(T)$ gravity}, \textcolor{blue}{Physics of the Dark Universe}, \textbf{46}, 101692 (2024). DOI: 10.1016/j.dark.2024.101692.
\item Debasmita Mohanty, \textbf{Sayantan Ghosh}, P.K. Sahoo, \textit{Impact of Chaplygin gas model on the characteristics of gravastar in $f(Q,T)$ gravity}, \textcolor{blue}{International Journal of Modern Physics D}, \textbf{34} (11), 2550041 (2025). DOI: 10.1142/S0218271825500415.
\item \textbf{Sayantan Ghosh}, Sneha Pradhan, Pradyumn Kumar Sahoo, \textit{Maximum Mass Limit of Generalized MIT Compact Star: A Theoretical Study of Thin Shell Dynamics and Gravitational Lensing}, \textcolor{blue}{Fortschritte der Physik}, \textbf{73} (8), e70020 (2025). DOI: 10.1002/prop.70020.
\item \textbf{Sayantan Ghosh}, Debasmita Mohanty, P.K. Sahoo, \textit{Scalar perturbations of gravastar model in the presence of global monopole}, \textcolor{blue}{The European Physical Journal C}, \textbf{86} (6), 666 (2026). DOI: 10.1140/epjc/s10052-026-15965-w.

\end{enumerate}
\cleardoublepage
\pagestyle{fancy}
\lhead{\emph{Biography}}

\chapter{Biography}

\section*{Brief Biography of the Supervisor:}
\noindent\textbf{Prof. Pradyumn Kumar Sahoo} is a Professor in the Department of Mathematics at BITS Pilani, Hyderabad Campus, India. He has more than two decades of research experience in Applied Mathematics, Theoretical Cosmology, General Relativity, Modified Theories of Gravity, and Astrophysical Objects. He obtained his Ph.D. in Mathematics from Sambalpur University, Odisha, India, in January 2004.

In 2009, he joined BITS Pilani, Hyderabad Campus, as an Assistant Professor and is currently serving as a Professor. He also served as the Head of the Department of Mathematics from 2020 to 2024. His research focuses mainly on theoretical cosmology, modified gravity theories, and compact astrophysical objects.

Prof. Sahoo has published more than 270 research articles in reputed national and international journals. According to a survey conducted by researchers from Stanford University, he has been ranked among the top 2\% of scientists worldwide in the field of Nuclear and Particle Physics. He is also the recipient of the \textit{Prof. S. Venkateswaran Faculty Excellence Award} (2022) from BITS Pilani.

He has been awarded a Visiting Professor Fellowship at Transilvania University of Brașov, Romania, and is also an Associate Member of the Inter-University Centre for Astronomy and Astrophysics (IUCAA), Pune. As a visiting scientist, he has also visited the European Organization for Nuclear Research (CERN), Geneva.

Prof. Sahoo has actively participated in numerous national and international conferences and has delivered several invited talks. He is an active member of COST Action CA21136, which focuses on addressing observational tensions in cosmology.

He has supervised or co-supervised 21 Ph.D. students (12 awarded and 9 ongoing) and guided several M.Sc. dissertations. He has also contributed to several sponsored research projects funded by UGC, CSIR, NBHM, ANRF–DST, and DAAD (RISE) Worldwide. In addition, he serves as an expert reviewer for research projects of ANRF-DST, and UGC, and is an editorial board member of several reputed international journals.

\section*{Brief Biography of the Candidate:}
\textbf{Mr. Sayantan Ghosh} obtained his Bachelor’s degree in Mathematics from the Chennai Mathematical Institute (CMI), followed by a Master’s degree in Mathematics from the Indian Institute of Science (IISc), Bengaluru. He pursued doctoral research for one year at the Raman Research Institute (RRI), Bengaluru, and subsequently joined BITS Pilani, Hyderabad Campus, to continue his Ph.D. studies in Theoretical Physics.\\
He has qualified national-level examinations such as CSIR-NET and JEST. In recognition of his research contributions, he received the Young Relativist Award from the Tensor Society in 2023. His research has been disseminated through presentations at several prestigious international conferences, including those held in Japan and Thailand.

\end{document}